\documentclass[iop,revtex4]{emulateapj}

 \bibpunct{(}{)}{;}{a}{,}{,}
 \usepackage{lscape}
  \usepackage{color}
\usepackage{mathrsfs}
\usepackage[caption=false]{subfig}

\newcommand{\um}{$\mu$m}

\newcommand{\kms}{km s$^{-1}$}
\newcommand{\msun}{\mbox{M$_\odot$}}

\newcommand{\td}{\mbox{$T_d$}}
\newcommand{\jyb}{\mbox{Jy beam$^{-1}$}}
\newcommand{\mask}{\mbox{\textit{labelmask}}}
\newcommand{\hii}{\mbox{\ion{H}{2}}}
\newcommand{\bolocat}{\textit{Bolocat}}
\newcommand{\ammonia}{\mbox{{\rm NH}$_3$}}

\shorttitle{The Bolocam Galactic Plane Survey. XI.}
\shortauthors{Merello et al.}

\begin{document}

\title{The Bolocam Galactic Plane Survey. XI. \\Temperatures and Substructure of Galactic Clumps Based on 350 \um\ Observations}

\author{Manuel Merello$^{1,2}$, Neal J. Evans II$^1$, Yancy L. Shirley$^3$, Erik
Rosolowsky$^4$, Adam Ginsburg$^5$, John Bally$^6$, Cara Battersby$^7$, Michael
M. Dunham$^7$}

\affil{$^1$The University of Texas at Austin, Department of Astronomy, 2515
Speedway, Stop C1400, Austin, Texas 78712-1205}
\affil{$^2$Istituto di Astrofisica e Planetologia Spaziali-INAF, Via Fosso del Cavaliere 100, I-00133 Roma, Italy}
\affil{$^3$Steward Observatory, University of Arizona, 933 North Cherry Avenue,
Tucson, AZ 85721, USA}
\affil{$^4$Department of Physics, 4-181 CCIS, University of Alberta, Edmonton,
AB T6G 2E1, Canada}
\affil{$^5$European Southern Observatory, ESO Headquarters,
Karl-Schwarzschild-Strasse 2, D-95748 Garching bei MŸnchen, Germany}
\affil{$^6$CASA, University of Colorado, 389-UCB, Boulder, CO 80309, USA}
\affil{$^7$Harvard-Smithsonian Center for Astrophysics, 60 Garden Street, MS
78, Cambridge, MA 02138, USA }

\email{ \href{mailto:manuel.merello@iaps.inaf.it}{manuel.merello@iaps.inaf.it}}

\begin{abstract}
We present 107 maps of continuum emission at 350 \um\ from Galactic molecular 
clumps. Observed sources were mainly selected from the Bolocam Galactic Plane Survey (BGPS) catalog, with 3 additional maps covering star forming regions in the outer Galaxy. The higher resolution of the
SHARC-II images (8.5\arcsec\ beam) compared with the 1.1 mm images from BGPS
(33\arcsec\ beam) allowed us to identify a large population of smaller
substructures within the clumps. A catalog is presented for the 1386
sources extracted from the 350 \um\ maps.  The color
temperature distribution of clumps based on the two wavelengths has a median of
13.3 K and mean of $16.3 \pm 0.4$ K, assuming an opacity law index
of 1.7. For the structures with good determination of color temperatures,
the mean ratio of gas temperature, determined from \ammonia\ observations,
to dust color temperature is $0.88$ and the median ratio is 0.76. 
About half the clumps have more than two substructures and 22 clumps 
have more than 10. The fraction of the mass in dense substructures seen at 350 \um\  compared to the mass of their parental clump is $\sim$0.19, and the surface densities
of these substructures are, on average, 2.2 times
those seen in the clumps identified at 1.1 mm. For a well-characterized
sample, 88 structures (31\%) exceed a surface density of 0.2 g cm$^{-2}$,
and 18 (6\%) exceed 1.0 g cm$^{-2}$, thresholds for massive star formation
suggested by theorists.

\end{abstract}



\section{Introduction}

Although high-mass stars play a key role in the structure and evolution of our
Galaxy, there are still many questions that need to be answered before we have a
paradigm for their formation process~\citep{sta00,bal05,zinn07,tan14}. Adopting
the nomenclature of
\citet{williams2000}, star clusters form in dense
clumps ($M\sim50-500$ \msun, sizes $\sim0.3-3$ pc) of giant molecular clouds
while individual stars form in cores ($M\sim0.5-5$ \msun, sizes $\sim0.03-0.2$
pc). An important
step in the observational study of massive star formation is the identification
and characterization of clumps and their resolution into cores.
Studies of compact structures in our Galaxy
will give fundamental ``ground truth'' for the analysis
of nearby galaxies where individual clumps and
cores cannot be resolved, and provide source information for observations of
the Milky Way with a new generation of instruments, such as the Atacama Large
Millimeter/Submillimeter Array.

There are complementary approaches to the study of massive star forming
regions. 
We may consider obtaining molecular line observations,
which are a good diagnostic of line-of-sight motions in a cloud, but
variations in
tracer abundances due to depletion and complex chemical processing, excitation
conditions and the impacts of radiation fields and shocks
add uncertainties~\citep{eva99}. We may
also proceed by mapping the continuum emission toward these sources. Newly formed
stars heat the dust
of their parental molecular cores and this dust emits in the far infrared and
submillimeter wavelengths~\citep{gar99}. Continuum emission observations at
(sub) millimeter wavelengths offer a reliable tracer of the column density
and clump masses due to the low optical depth of the
dust~\citep[e.g.][]{joh06}, and
they have made an increasingly significant contribution to the understanding of
star formation. 
Reddening and extinction of starlight provide also a measure of column densities~\citep{boh78},
and mid-IR large scale surveys have identified dark patches obscuring the bright-background, which are associated with 
dense molecular gas~\citep[e.g.,][]{ega98,sim06}. Studies of the physical properties of these regions, called Infrared Dark Clouds, suggest they are the birthplaces of high-mass stars and stellar clusters~\citep[e.g.,][]{rat06,rat10,bat10}. In addition, near infrared extended emission has also been considered a tracer of the column density of dense dark clouds, interpreted as scattered ambient starlight observed from the outer parts of dense cores~\citep[``cloudshine", e.g.][]{fos06,pad06}.

Several studies have made progress toward
the characterization of massive star forming regions, but they are usually biased because target selection required
existing signposts of newly form high-mass stars, including maser emission, far infrared
emission, or the presence of \hii\ regions \citep{chu90,ces91,plu92}.
Therefore, the identification of
clumps with a less biased survey will improve our understanding of the early
stages of massive stars. Examples include large scale surveys of
particular regions, such as
the study of the W3 giant molecular cloud with Submillimetre Common-User
Bolometer Array (SCUBA) by~\cite{moo07} or the Cygnus-X molecular cloud complex
region with MAMBO receiver by \cite{mot07}.
Large-scale blind surveys of the Galactic Plane at far-IR, submm and
mm wavelengths have recently been completed. The ATLASGAL survey mapped the Southern
part of the plane at 870 \micron\ using the 12 m APEX
telescope~\citep{sch09,con13}.
Space borne missions, such as {\textit{WISE}}~\citep{wri10} and
{\textit{Spitzer}}/GLIMPSE~\citep{ben03} and MIPSGAL~\citep{car09}, have mapped
most of the Galactic plane at a resolution of a few to 10\arcsec, at
wavelengths between 3.6 and 24 \um. The {\textit{Herschel}}/Hi-GAL 
survey~\citep{mol10} covered the entire Galactic plane
with resolution between 5\arcsec\ and 36\arcsec, in five bands at 70, 160, 250,
350 and 500 \um. In addition, the 15-m James Clerk Maxwell Telescope is using
SCUBA-2 850 \um\ and 450 \um\ emission to map the Galactic plane and several
nearby Gould Belt clouds. These surveys have provided a robust tracer of dust
properties (column density and mean grain temperature), for different phases of
the interstellar medium.

The Bolocam Galactic Plane Survey~\citep[BGPS;][]{agu11} is one of the first
ground-based surveys to map the continuum emission at
1.1 mm over a large region of the northern Galactic plane. The BGPS coverage
region extended from
$l=-10\arcdeg$ to $+86.5\arcdeg$ for $b=\pm 0.5\arcdeg$, at a resolution of
$\sim$30\arcsec, and it identified 8454 compact sources throughout the
Galaxy~\citep{ros09}.
Depending on their distance, millimeter features may be cores, clumps, or clouds~\citep{mck07}, although BGPS sources are mostly clumps~\citep{dun11}.
An improved reduction and
a new version (V2.0) of the BGPS maps and catalog was released
\citep{gin13}, including additional $\sim$20 square degree coverage area in the
third and fourth quadrant, and $\sim$2 square degree in the 1st quadrant. This
new reduction shows an improved flux calibration with respect to the former
version (V1.0), in agreement with other data surveys on the Galaxy. The
characterization of the angular transfer function of the Bolocam pipeline shows
that the flux recovery is above 90\% for scales between 33\arcsec\ and
80\arcsec, with a steep drop for scales above $\sim$100\arcsec. While version
2.0 of the catalog contains 8559 sources, 35 sources were unintentionally
excluded, and they were added in a new version (V2.1), for a
total of 8594 sources. 
Figure~\ref{fig:bgps_sharc_distr} shows the distribution of
40\arcsec\ aperture flux density of V2.1 sources.
Although the basic statistical properties of the catalogs remains unaltered, there are some effects on the source extraction when considering the new map versions with improved quality.  Comparing the source extraction in the V1.0 and V2.1 overlap region, V2.1 recovers fewer compact sources (8004), with only $\sim70\%$ of those having a clear V1.0 match, and these are considered more reliable and stable identified sources~\citep[see Fig. 10 in][]{gin13}. Fig. 11 and 12 in~\cite{gin13} show visual examples of the ÒlostÓ sources, which are generally small sources on the shoulders of bright ones that have now been included in the bright source in the catalog.

We selected a sample of sources from the BGPS catalog, initially from V1.0 and
later from V2.1, towards the most
crowded and brightest portions of the Galactic Plane including the Galactic
Center, the Molecular Ring near $l = 30$\arcdeg\ and Cygnus-X, and
used the Submillimeter High Angular Resolution Camera II~\citep[SHARC-II;][]{dow03} to
obtain continuum maps at 350 $\mu$m of them. The high resolution of the
SHARC-II
observations (FWHM beam size of 8.5\arcsec) is in particular very useful when
we observe small structures at
great distances in the Galaxy, and the comparison of flux densities of sources
at 350 \um\  and 1.1 mm will give us constraints on their dust temperatures,
hence improving the determination of masses and column densities of the
molecular
clumps.

SHARC-II has been used before for studying low-mass dense cores from the
\textit{Spitzer} Legacy Program ``From Molecular Cores to Planet-Formation
Disks" \citep[][A. Suresh et al. in preparation]{wu07}, concluding that observations
with this instrument are better to distinguish between starless and
protostellar cores than observations at longer wavelengths. 
While \citet{mueller02} studied dense clumps selected for having
massive star formation using SHARC at 350 \um, there has not been a
comprehensive study of BGPS-selected sources with SHARC-II.
A recent study by \cite{rag13} shows how ground-based
observations at 350 \um\ continuum emission of regions of high-mass and cluster
formation can give information on substructures within them. For a set of 11
nearby Infrared-dark clouds (IRDCs), they obtained data on 350 \um\ continuum
emission with the SABOCA instrument at the APEX telescope, 
finding a large population of small core-like
structures. Higher resolution observations at submm wavelengths 
allow the study of the complex process of massive star formation from 
fragmentation of molecular clumps to individual
substructures, such as cold and hot cores.

We present a set of 107 continuum emission maps at 350 \um\ obtained with
SHARC-II. These include 104 maps of clump-like BGPS V2.1 sources, and 3 maps 
of outer Galaxy sources not included in the catalog. This paper is
organized as follows. Section 2 describes the selection of targets from the
BGPS catalog and the SHARC-II instrument. Section 3 describes the data
reduction and calibration methods for our set of submm continuum maps. Section
4 presents the results of the extraction of sources in the images, including
source recovery tests, a catalog of these sources, the procedure for
association between sources in the 350 \um\ maps and the 1.1 mm sources,
and the comparison between SHARC-II and {\textit{Herschel}} at 350 \um.
Section 5 describes the determination of color temperatures for BGPS sources found in our set of maps,comparing them with temperature determination from spectral energy distribution (SED) fitting from Hi-GAL and other continuum surveys. Correlations between our dust temperature estimations and high-density molecular tracers are also presented. In Section 6, we explore consequences from our results for mass and surface density estimations of millimeter sources. In Section 7, we present conclusions from our analysis..

\section{Observations}

Observations were taken in several runs between June 2006 and September 2012 at
the Caltech Submillimeter Observatory (CSO\footnote{The Caltech Submillimeter Observatory
was operated by the California Institute of Technology, until 2013
March 31 under cooperative agreement with the National Science Foundation
(AST-0838261)}).
The observations required
excellent weather conditions, with a measured optical depth at 225 GHz ranging
between 0.030 and 0.094 (measured at zenith position), with an average value of $\tau_{225\ GHz}=0.057$,
corresponding to an optical depth at 350 \um\ (857 GHz) of $\tau_{350\ \mu
m}=1.41$. In general, we observed our targets when they were culminating, and
we avoided observing sources below an elevation of 30\arcdeg\ or above
80\arcdeg.

\subsection {Target selection}\label{sec:target_selection}

The targets were initially selected from sources in the BGPS V1.0 catalog. We
chose sources with strong emission at 1.1 mm (typically with peak emission
above 1 \jyb) toward densely populated regions, many of them known as active
star forming regions, such as the Galactic Center, $l=30\arcdeg$, Cygnus-X
region, W3 and W5 molecular clouds, GemOB1 region, and others. Several of those
targets were extended and they were likely to have unresolved substructure at a
resolution of 30\arcsec.
Most of the time the observation fields were centered directly on a V1.0
source, but in some cases we tried to cover several sources in the same field.
Starting in December 2009, we added to our list of targets bright sources
selected from new Bolocam maps that were later incorporated in BGPS V2.0.
Considering that almost all targets from V1.0 catalog are incorporated in V2.1
\citep{gin13}, we consider the latest version of the catalog for our analysis.
In just one of our maps, L136.52+1.24, there are three sources marginally
detected in V1.0 but not included in subsequent V2.0 or V2.1
catalog versions.
We also obtained 350 \um\ maps of the outer Galaxy star forming regions SH2-209, SH2-307 and
IRAS 03233+5833~\citep[e.g.,][]{kle05,ric12}, which were not covered by the BGPS survey. The maps of these regions are 
L151.61$-$0.24, L234.57+0.82, and L142.01+1.77, respectively.
These observations are included for completeness, but they are not included in analysis that requires BGPS data.

\begin{figure}[h] 
   \centering
      \includegraphics[width=0.6\textwidth]{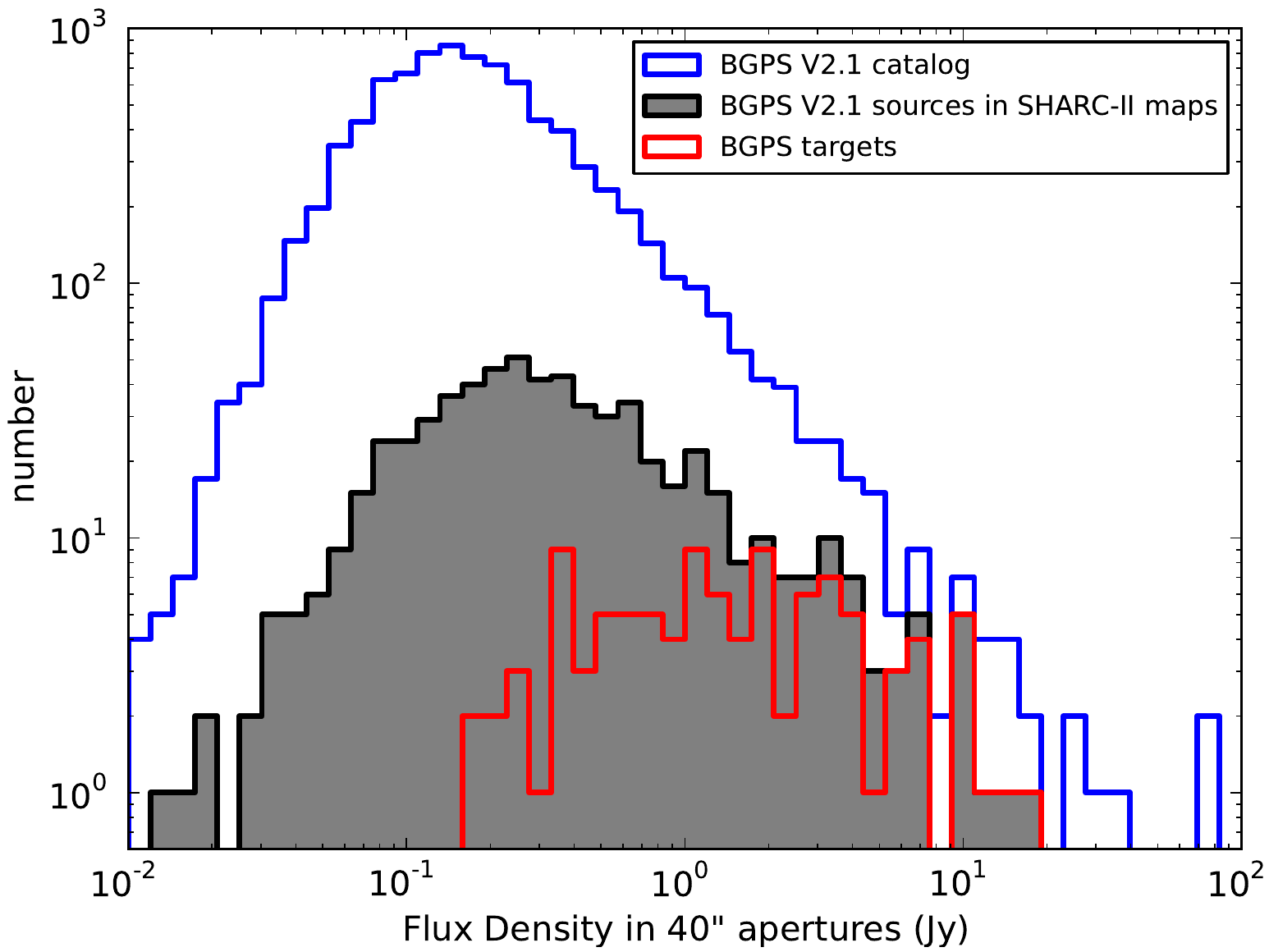}

   \caption{Distribution of 40\arcsec\ aperture flux density of the complete sample of 8594 sources from the BGPS V2.1 catalog (blue line), along with those 619 sources contained in the SHARC-II maps (shaded black line). The red line shows the distribution of the 104 sources considered as representative targets (see Section~\ref{sec:target_selection}).}
  \label{fig:bgps_sharc_distr}
\end{figure}

For a better characterization of our targets with respect to the complete
catalog, we consider the brightest V2.1 source in each map as the
representative target for that map. In some cases, where adjacent maps share
the brightest source, we consider the second brightest source as a
representative target. From the 104 maps at 350 \um\ with BGPS sources,
we have then a sub-sample of V2.1 sources considered as representative
targets. Figure~\ref{fig:bgps_sharc_distr} shows in red the distribution of
flux density in 40\arcsec\ apertures for our sample of 104 representative targets.
The distribution has an average of 2.63 Jy and a median of 1.37 Jy,
much higher than the values for the overall BGPS catalog.
However, weaker sources were covered in the same fields, so the 
final flux density distribution of BGPS sources with 350 \um\ data is closer
to that of the full catalog (see black line in 
Figure~\ref{fig:bgps_sharc_distr}).
We found a total of 619 V2.1 sources in our sample of SHARC-II maps, with an average and median flux density of 
0.79 Jy and 0.29 Jy, respectively, 
about twice the values for
the complete V2.1 catalog (average of 0.35 Jy and median of 0.16
Jy for 40\arcsec\ apertures). 

\subsection{Instrument and Observational Method}

SHARC-II is a background-limited facility camera operating at 350 \um\ or 450 \um, located at the CSO on Mauna Kea,
Hawaii. It consists of a 32$\times$12 array of doped silicon
`pop-up' bolometers and it has a mean beam size of $8.5\arcsec$ at 350 \um. The
array has a full size of $2.59\arcmin\times0.97\arcmin$, which makes the
instrument start to lose sensitivity to emission on scales larger than
$\sim$1\arcmin.
We used the box scan mode of SHARC-II without chopping\footnote{See
http://www.submm.caltech.edu/$\sim$sharc/} for mapping regions with
sizes significantly larger than the size of the array. In the box scan mode,
the scan pattern moves across the rectangular field of view, starting in the
center of the array and going straight until the edge of the bounding box,
where it changes trajectory emulating the bounce of a ball on a billiard table.
The advantage of this scan mode is that it provides better isolation against
1/\textit{f} noise, with more uniformly distributed noise across the field of
view in comparison with the alternative sweep mode with Lissajous scan pattern.
Different box sizes were used in our data acquisition, between
2.5\arcmin$\times$2.5\arcmin\ and 18\arcmin$\times$18\arcmin, although the bulk of
our observations were obtained with 10\arcmin$\times$10\arcmin\ boxes. Column 3
in Table \ref{tbl:observation_information} shows the size for each map.

For all our observations we used the Dish Surface Optimization System
(DSOS)\footnote{See http://www.cso.caltech.edu/dsos/DSOS\_MLeong.html} to
correct the dish-surface figure for imperfections and gravitational
deformations as the dish moves in elevation during observations.

\section{Data reduction and calibration}

\subsection{Data reduction}\label{sec:reduction}

The 350 $\mu$m raw data were reduced using the Comprehensive Reduction Utility
for SHARC-II~\citep[CRUSH;][]{kov06}. CRUSH is a Java-based software which solves
a series of iterative models, attempting to reproduce atmospheric and
instrumental effects on the observations, removing the sky emission common to
all bolometers at first order~\citep[e.g., ][]{bee06}. We used version 2.10
of CRUSH, publicly accessible from the CSO webpage.

\begin{figure}[h] 
   \centering
	\includegraphics[width=0.6\textwidth]{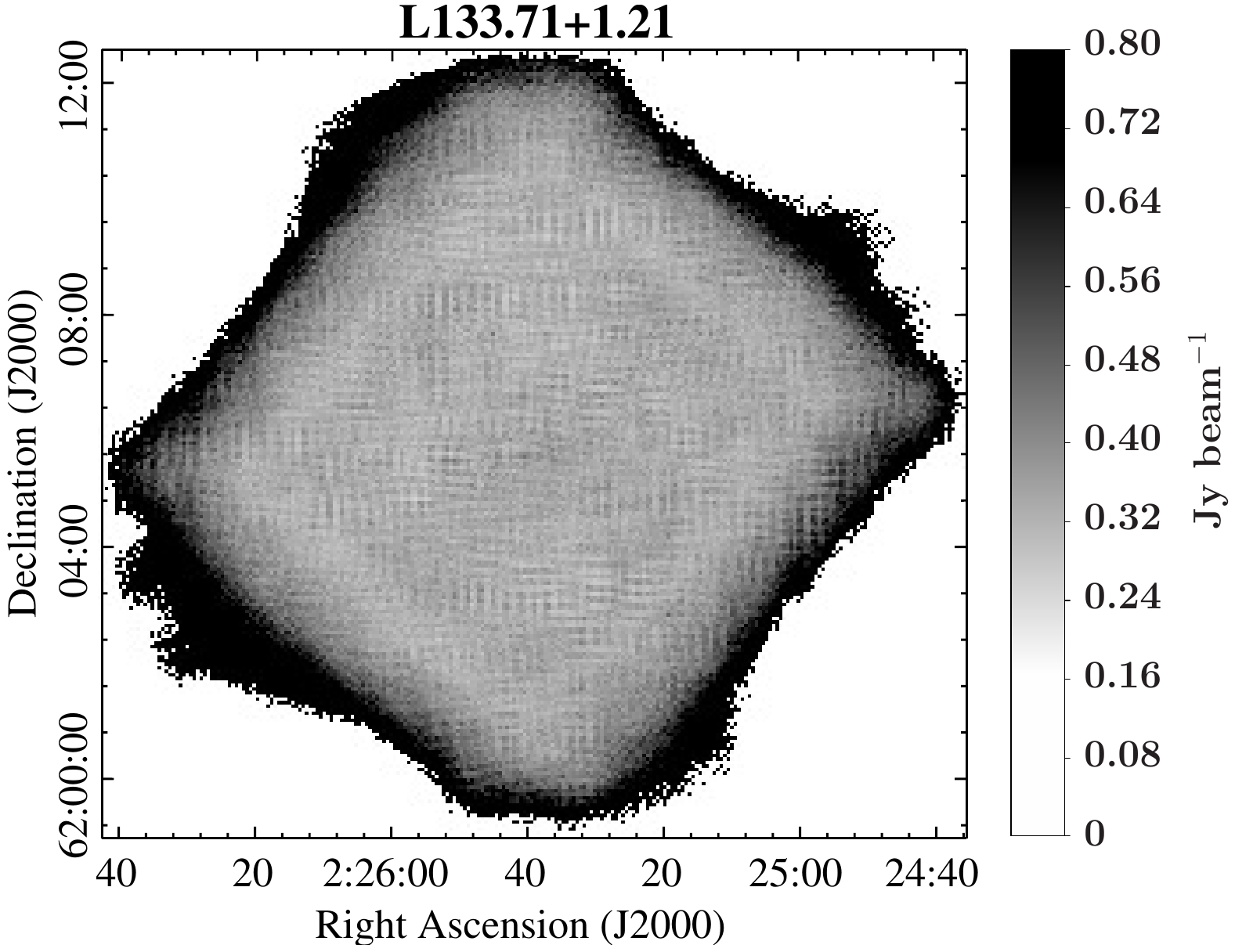}
   \caption{Error map obtained from CRUSH reduction for L133.71+1.21. The image unit is Jy beam$^{-1}$ (beam size of 8.5\arcsec). The measured noise for the L133.71+1.21 map is 0.338 Jy beam$^{-1}$. The average
representative noise for the complete sample of maps is 478 mJy beam$^{-1}$}
  \label{fig:example_errormap}
\end{figure}

Figure~\ref{fig:example_errormap} shows an example of the error or noise maps
that reduction with CRUSH produces for each observed region. In general, all
SHARC-II maps are much
noisier toward the edges by a factor of $\sim$4 with 
respect to the measured average
noise of the image. At a distance of $\sim$20\arcsec\ (twice the beam size)
from the edge, the noise is only increased a factor of $\sim$2. Also, when
taking multiple scans for a single map at different runs, the orientation of
the field of view rotates, and the regions that do not overlap in the
integration of the source show a significant increase in the noise. The rest
of the map shows some residuals of the pattern used in the box scan mode, but
in general the inner parts of these maps are well represented by uniform noise.

\subsection{Calibration}

The maps were reduced in instrumental units of $\mu$V, and we estimated the
flux
conversion factors $C_\theta$ in order to measure the flux densities of the
sources in units of Jy. On our analysis, we followed the same method used on
SHARC-II maps by~\cite{wu07}, and on SCUBA maps by~\cite{shi00}, based on
aperture calibrations over a given angular size $\theta$. Flux estimation
obtained from aperture corrected photometry are 
less sensitive to uncertainties due to effects of sidelobes.
The observed flux density for
an aperture of diameter $\theta$ is $S_\nu(\lambda , \ \theta)\ =\ C_\theta \
V(\lambda, \ \theta)$, with $V(\lambda, \ \theta)$ the voltage measured at
wavelength $\lambda$ in an aperture of diameter $\theta$.
We used the fluxes from Mars, Uranus and Neptune to obtain the flux conversion factors
for two different apertures, 20$\arcsec$ and 40$\arcsec$. We restricted our flux
estimation to 40$\arcsec$ considering the limited sensitivity of the
instrument to large scale emission. The bulk of the compact structures
identified on the maps have sizes below this limit.

To compare the 350 \um\ data with the 1.1 mm images from BGPS survey, 
we also convolved the  350 \um\ to match the
$33\arcsec$ FWHM beam from Bolocam. Therefore, we also estimated the flux
conversion factor C$_{40}^{\ 33\arcsec}$ for an aperture of 40\arcsec\ using
the images of the calibration sources convolved to a 33\arcsec\ beam. The flux
densities of
sources and calibrators in instrumental units were
measured using Starlink's $aperadd$ package.

In addition, we estimated the flux conversion factor for one beam,
$C_{beam}$, which we used to express the peak flux of each
source and the 1 $\sigma$ rms of the maps in units Jy beam$^{-1}$.
$C_{beam}$ is obtained from the integrated flux density of the
calibrators in one beam and the peak pixel of the map in instrument units.
We have then an average value for the conversion $\mu$V to Jy pixel$^{-1}$ =
32.23 Jy beam$^{-1}$, for the beam size and pixel scale of the instrument.
Therefore, the value of $C_{\theta}$ will depend of the number of pixels
considered. From the maps, the average value of $C_{beam}=8.27$ Jy beam$^{-1}$
$\mu$V$^{-1}$, and then
\begin{eqnarray}
\nonumber
1\ \mathrm{Jy\ beam}^{-1}	 &=& 503.8\ \mathrm{MJy\ sr}^{-1}\nonumber \cr
					 &=& 0.031\ \mathrm{Jy\ pixel}^{-1} \cr
					 &=& 0.121\ \mu \mathrm{V}^{-1}\ .
\label{eq:conversion}
\end{eqnarray}

For the calibration of the images, only the observations of planets were
considered. While the secondary calibrators mentioned in 
section~\ref{sec:map_description} are useful to
check the pointing when planets are not available, they are not as bright and
their flux densities are not as well known as planets. Therefore, we avoided
using them to minimize uncertainties in the estimation of
calibration factors.

Table~\ref{tbl:calibrators} in the Appendix section gives the observation dates
(column 1), the planet
observed (column 2), the observed opacity at 225 GHz at that date (column 3),
the computed values of $C_{beam}$ (column 4) and $C_\theta$ at different
apertures (columns 5 and 6). Column 7 gives the calibration factor for an
aperture of 40$\arcsec$ obtained for the planet images convolved to a beam of
33$\arcsec$. Table~\ref{tbl:calibrators_average} shows the average values of
$C_\theta$ for the different observation epochs. According to these results,
the uncertainties in the calibration of the maps are between 15-20$\%$,
consistent with previous observations and calibration estimations for the
SHARC-II instrument \citep{wu07}.

While some sources may have significant contributions to the broad-band
flux density from line emission, it is usually a small fraction of the 
dust continuum emission (e.g., \citealt{groesbeck95}).
The continuum flux measured with SHARC-II
(780 - 910 GHz for the 350 \um\ filter) could in principle be 
contaminated by the CO$(7-6)$ line.
\cite{hatchell09} estimated a contribution less than 100 mJy beam$^{-1}$
in the SHARC-II 350 \um\ band for a CO$(7-6)$ integrated intensity of 100 K
\kms, which is smaller that the typical sensitivity found in our maps, and they considered that
the CO line may contribute $\sim$20\% of the continuum emission in the 350 \um\ band. We adopted that limit in this work as a conservative value for possible line contamination.

\section{Results}

\subsection{Map description}\label{sec:map_description}

\begin{figure}[h] 
   \centering
  \includegraphics[width=0.7\textwidth]{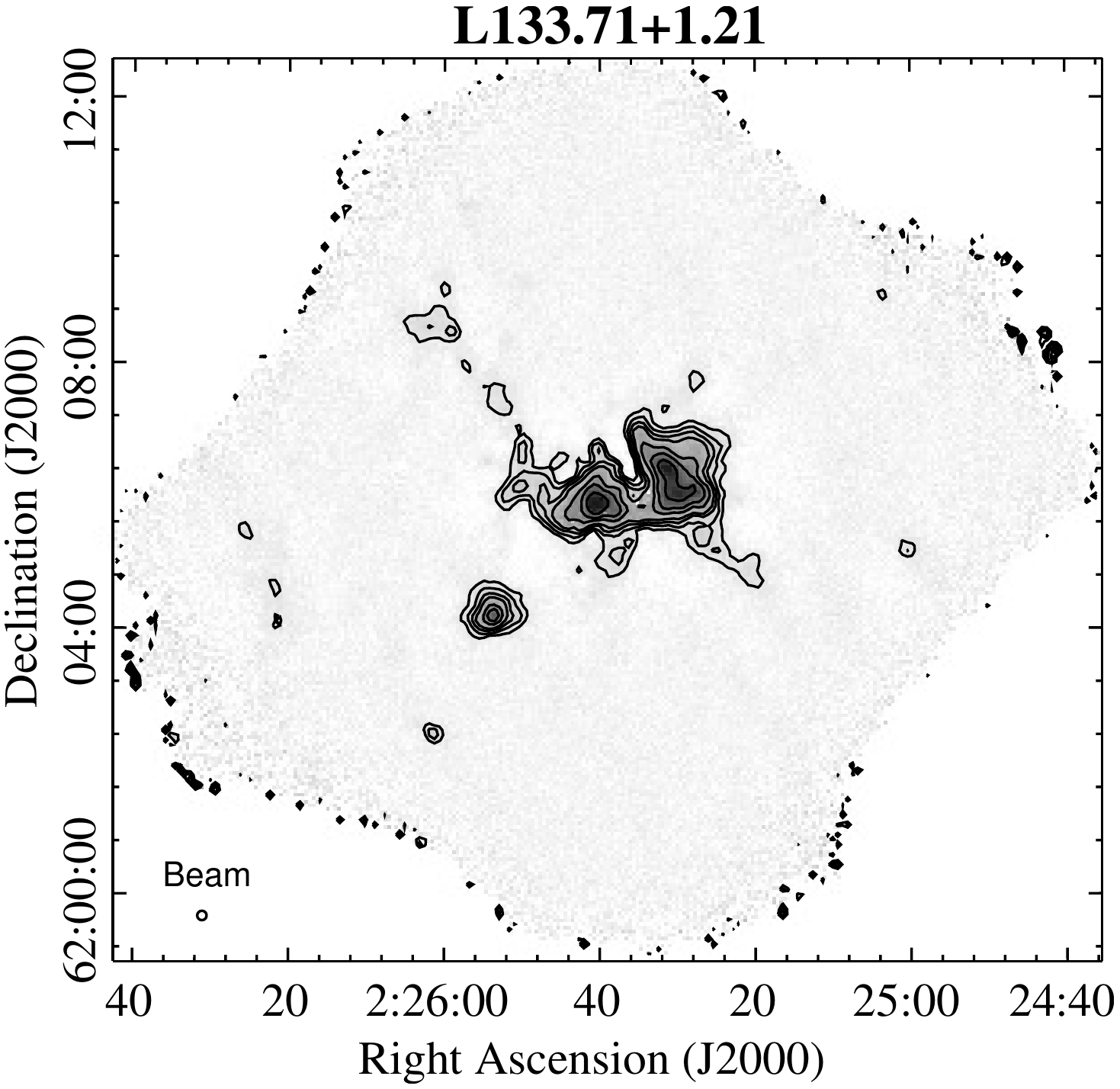}
   \caption{Map at 350 \um\ obtained with SHARC-II toward L133.71+1.21, corresponding to the W3 Main star forming region. Beam size of image is shown in the bottom left corner. Contour levels represent 3$\sigma$, 6$\sigma$, 10$\sigma$ ,15$\sigma$, 30$\sigma$, 50$\sigma$ and 100$\sigma$, with a rms noise $\sigma$ =  338 mJy beam$^{-1}$.}
  \label{fig:example_map}
\end{figure}

Figure~\ref{fig:example_map} shows as an example the 350 \um\ continuum map of
L133.71+1.21, corresponding to the
W3 Main region \cite[see][and references therein]{riv13,meg08}.
Besides the two
bright infrared sources IRS5 and IRS4~\citep{wyn72}, the small beam size of the
350 \um\ maps allows us to identify several sources near them, and some
elongated, filamentary-like structures and scattered faint sources. These types
of features are found recurrently in the 350 \um\ maps and we characterize
their correspondence with 1.1 mm sources later in section
\ref{sec:correlation}.

We obtained a total sample of 107 maps, with an average
noise of 744 mJy beam$^{-1}$, and a standard deviation of 443 mJy beam$^{-1}$. These maps are presented in Figure~\ref{fig:sharcmaps} in the Appendix.
Because the noise increases at the edges of the images
(see Section~\ref{sec:reduction}), we estimated a more representative value of the mean rms of
each map on a region toward the central area. In the
10\arcmin$\times$10\arcmin\ maps, the considered area has a diameter of
7\arcmin. Column 7 in Table \ref{tbl:observation_information} shows the
representative rms noise of each map. The average 1$\sigma$ noise value
estimated this way is 478 mJy beam$^{-1}$, with a median of 375 mJy beam$^{-1}$
and a standard deviation of 298 mJy beam$^{-1}$. Each map has an angular scale
of 1.618 arc-seconds per pixel.
Table~\ref{tbl:observation_information} lists, for each map observed, the
configuration used, the size and center of each map, the date on which the map was
taken, and the 1 $\sigma$ noise of the map in units of mJy beam$^{-1}$.
Integration time was obtained in blocks of $\sim$ 14 minutes. The
pointing was checked on planets such as Mars, Uranus and Neptune, and with
secondary objects when the planets were not available, such as IRAS 16293-2422,
K350, G34.3, W75N and CRL 618. The blind pointing uncertainty varied between
1.8$\arcsec$ and 2.1$\arcsec$ for azimuth and between 0.8$\arcsec$ and
1.6$\arcsec$ for zenith angle.

\subsection{Source Extraction}\label{sec:extraction}

To facilitate comparison with the BGPS sources, we used the same algorithm for
source extraction as was used for the BGPS
catalogs, \bolocat~\citep{ros09}. \bolocat\ identifies sources based on their
significance with respect to the local estimate of the noise in a map,
subdividing regions with high significance into individual sources based on
local maxima inside that region. Each pixel is assigned to an individual source
using a seeded watershed, in a similar way to other source extraction
algorithms (\textit{Clumpfind}, \citealt{wil94}; \textit{SExtractor}, \citealt{ber96}). The
source identification process is determined by three parameters: $P_{amp}$, the
amplitude of the signal compared to a local estimation of the noise
$\sigma$ (noise estimation as $\alpha,\delta$); $P_{base}$, the base level of emission at which the
identified region is expanded; and the deblending parameter $P_{deb}$, used in
decomposing regions of emission with multiple local maxima. First, \bolocat\ masks all data above $P_{amp}$ of a particular image, and extends a region to
include all connected regions of emission above
$P_{base}$=1$\sigma(\alpha,\delta)$ since areas with marginal significance
adjacent to regions of emission are probably real. Second, each identified
region is examined and subdivided according to the level of contrast between
local maxima. For each pair of local maxima in a region, the amplitudes of
emission, $I_1$ and $I_2$, are compared with the highest contour of emission
containing the local maxima pair, $I_{crit}$, and if any $I_1$ or $I_2$ is less
than $P_{deb}$ above $I_{crit}$, that local maximum is discarded as a
subregion. \bolocat\ also avoids recovering sources (and sub-sources from the
deblending process) with sizes less than the beam FWHM. A detailed description
of the algorithm is presented in \citet{ros09}. BGPS catalogs V1.0 and V2.1
used as parameters for the extraction of sources $P_{amp}$=2$\sigma(l,b)$,
$P_{base}$=1$\sigma(l,b)$ and $P_{deb}$=0.5$\sigma(l,b)$, with $\sigma(l,b)$ the
local noise estimation, as a function of Galactic coordinates, on the BGPS
maps.

We tested \bolocat\ in the SHARC-II maps varying the $P_{amp}$, $P_{base}$ and
$P_{deb}$ parameters, checking which of them reproduce a ``by eye" extraction
of sources in those maps. Using the same parameters as BGPS, spurious low
brightness sources were recovered across the maps, with several of them toward
the noisy edges of the maps, and bright extended sources are sub-divided until
the routine reaches small areas, comparable to the beam size of the maps. 
Better results were
obtained considering $P_{amp}$=3$\sigma(\alpha,\delta)$,
$P_{base}$=1$\sigma(\alpha,\delta)$ and $P_{deb}$=1$\sigma(\alpha,\delta)$, and
these values were used for all maps (inclusive those outside the BGPS coverage) and the following catalog of BPGS substructures at 350 \um.

\begin{figure}[h] 
   \centering
  \includegraphics[angle=0,width=1\textwidth]{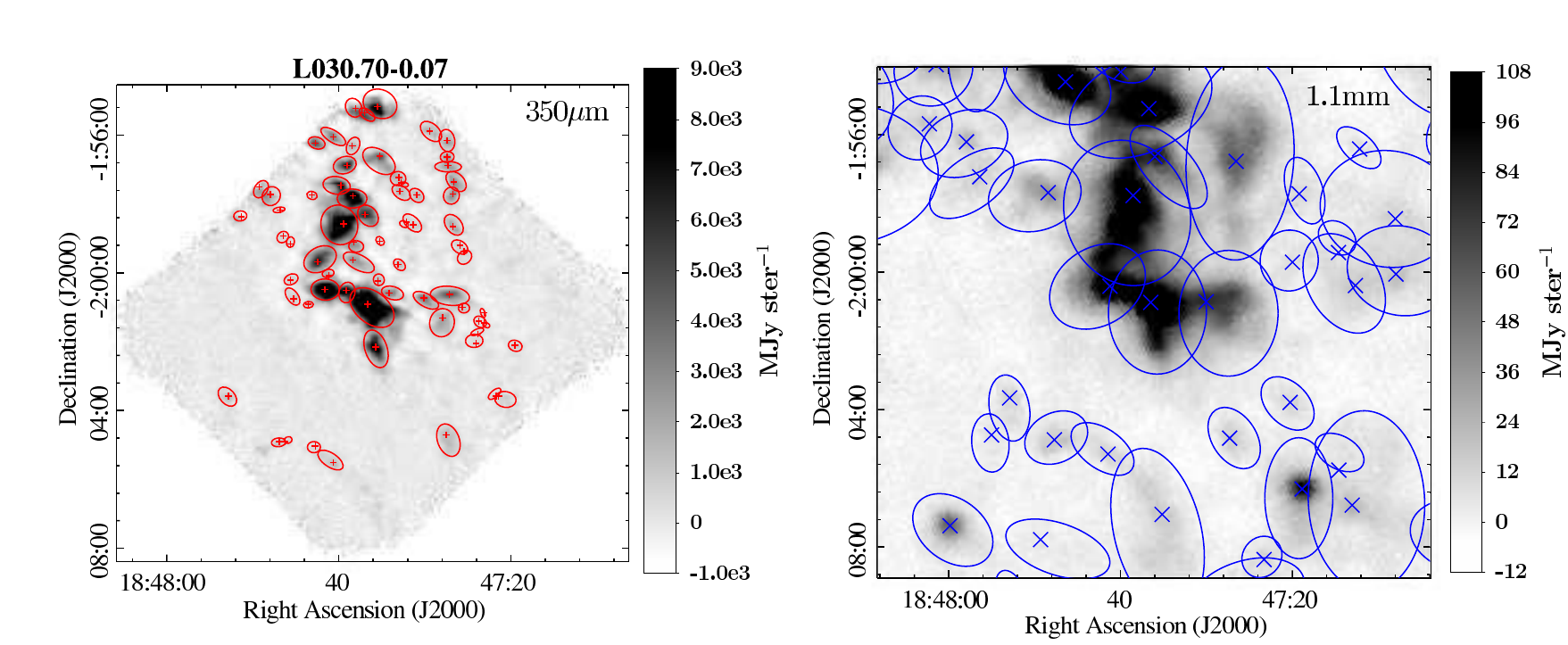}
   \caption{Example of the extraction of sources with \bolocat. Left: Map of L030.70-0.07 with SHARC-II at 350 \um\ (FWHM beam of 8.5\arcsec). Right: The same region mapped with Bolocam at 1.1 mm (FWHM beam of 33\arcsec). Some structures of Bolocam have several substructures mapped at the better resolution of SHARC-II. Also, the figure shows BGPS sources with no counterpart in the 350 \um\ map (see Section~\ref{sec:extraction} for a description of \bolocat\ parameters used on the extraction of sources). }
  \label{fig:sharccat_bolocat}
\end{figure}

Figure~\ref{fig:sharccat_bolocat} shows an example of one SHARC-II map
and the corresponding image at 1.1 mm obtained from BGPS. The angular scales of
SHARC-II and BGPS images are 1.6 arc-seconds per pixel and and 7.2 arc-seconds
per pixel, respectively. The better angular resolution of SHARC-II reveals
substructures within the clumps identified by the BGPS catalogue. Crowded
fields and extended clumps identified at 1.1 mm show small components and
filaments when they are observed at 8.5$\arcsec$ resolution.

\subsection{Source recovery experiments}\label{sec:tests}

We performed similar tests to the ones done for BGPS to assess catalog
properties compared with the real distribution of emission on the sky. For
these tests, we inserted sources in selected maps that do not have detected
sources, or that have just a couple of compact identified sources.
The selected maps are those of L030.15+0.00 (2 identified sources),
L078.92$-$0.19 (0), L079.62+0.49 (0), L079.11$-$0.35 (1), L080.86+0.38 (2),
L111.26$-$0.77 (2), L111.79+0.71 (0), L136.52+1.24 (2), and L137.69+1.46 (0). The
median rms noise of these maps is 667 mJy beam$^{-1}$, larger than 83$\%$ of
the estimated noise in the whole set of maps.
This set of maps will be considered as a conservative
representation of an emission free SHARC-II map.
For those maps that already have identified sources, we flagged
the detection of those sources and any input sources near them, and therefore
the recovered properties of input sources will not be affected by the emission
prior to the test.
After the artificial Gaussian objects were added, the maps were reduced in the
same standard way as the rest of the SHARC-II maps.
Although Gaussian brightness profiles do not quite represent all structures
observed in the 350 \um\ maps, they are still good models of compact sources.
Thus, the bulk of sources will be well represented in these tests.

\begin{figure}[h] 
   \centering
  \includegraphics[width=0.5\textwidth]{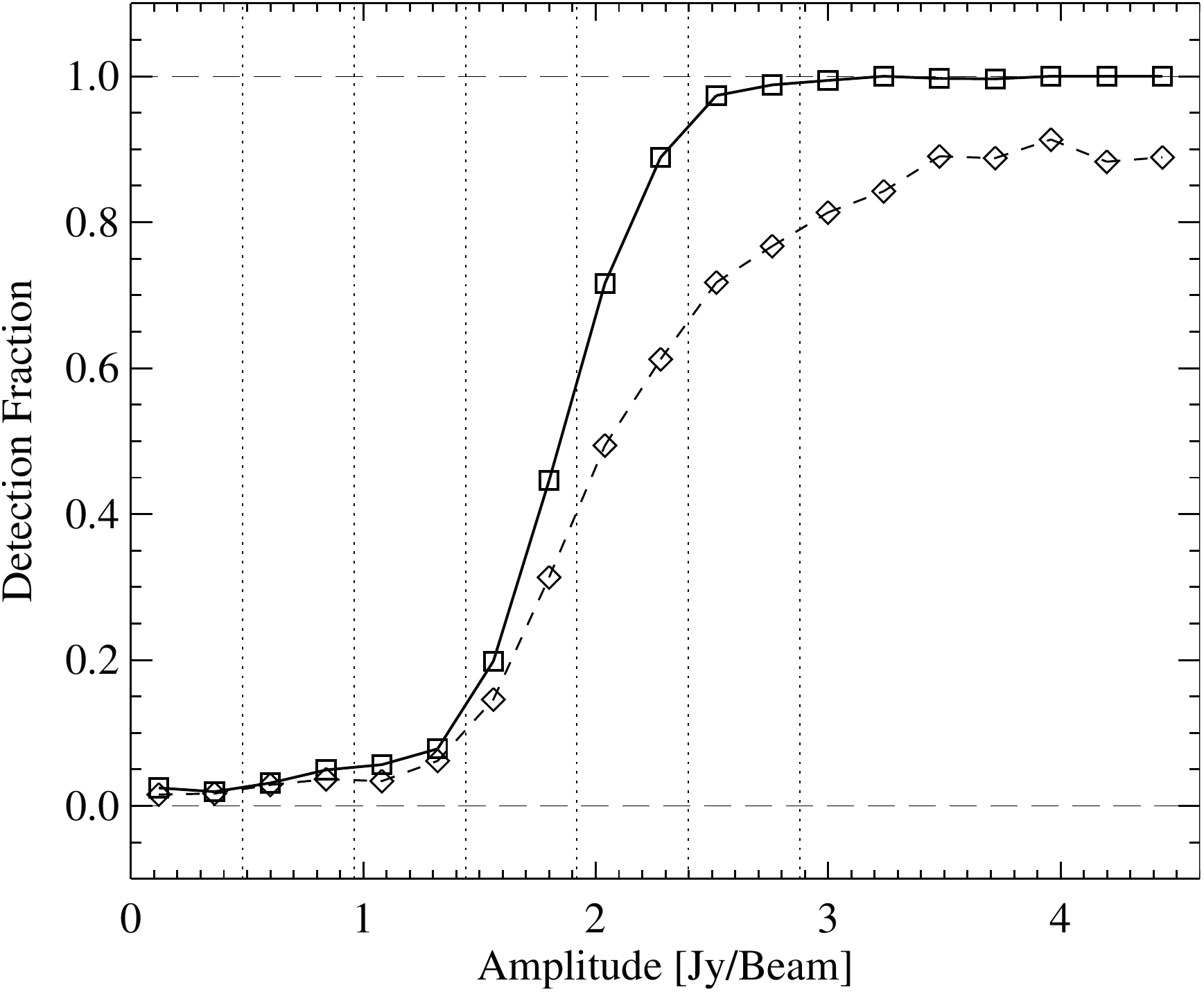}
   \caption{Completeness fraction derived from fake source tests. The tests were performed in nine SHARC-II maps with different noise levels. The plot shows the results of those detections scaled to the mean noise of the catalog maps (0.48 Jy beam$^{-1})$. The fraction of sources recovered is plotted as a function of input source intensity. The vertical dotted lines indicate \{1, 2, 3, 4, 5, 6\}$\sigma$. The dashed line with diamonds corresponds to fake sources across the whole map, while the continuous line with squares  corresponds to sources at a distance $<$3.5\arcmin\ from the center of the map. The results indicate that the catalog is complete at the $>$99\% limit for sources with flux densities $>6\sigma$.}
  \label{fig:test_completeness}
\end{figure}

We examined the degree of completeness of the catalog of substructures at different
flux density limits. For the nine test maps, we input sources with
FWHM of 9\arcsec\ and amplitudes uniformly ranging between 0.1$\sigma_{rms}$
and 15$\sigma_{rms}$, with $\sigma_{rms}$ the noise of each map. For $\sigma
_{rms}=667$ mJy beam$^{-1}$, this range correspond to sources with flux
densities between 0.19 Jy to 28.5 Jy. Taking into account that the edges of the
maps have a larger noise than the central area of the map with uniform noise,
we consider the detection fraction of sources across the whole map, and sources
inside a central area of 7\arcmin\ in diameter.
Figure~\ref{fig:test_completeness} shows the results of the test of
completeness. The vertical lines represent 1 to 6 times the average
$\sigma_{rms}$ value of the complete set of SHARC-II maps. For sources inside
the central area of the map, the detection is $>$99\% complete at the
6$\sigma_{rms}$ level. For sources with
amplitude less than 3$\sigma_{rms}$, the detection fraction is less than
$15\%$. For the rest of our test, we only considered sources inside a central
area of radius 3.5\arcmin\ to avoid noisy edge effects.

\begin{figure}[h] 
   \centering

 \includegraphics[width=0.5\textwidth]{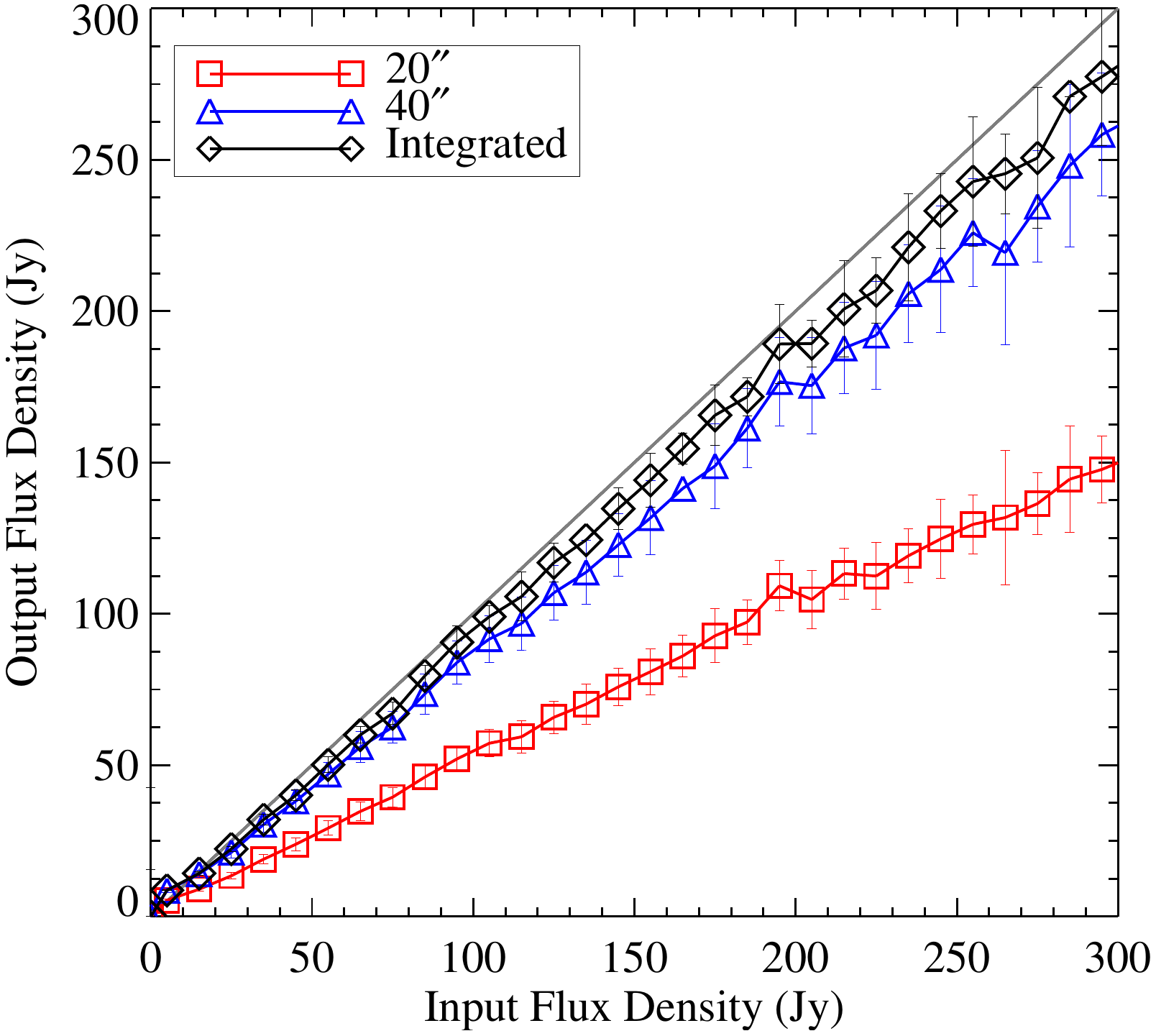}
       
   \caption{Recovery of source flux density for simulated resolved sources (FHWM $=2.7\theta_{beam}$ $=$ 23\arcsec) in SHARC-II maps. Results of the calibrated flux recovery are shown in units of Jy. Small differences are found between the 40\arcsec\ aperture and the integrated flux, with an under estimation of the integrated flux of $\sim10\%$. }

  \label{fig:test_flux}
\end{figure}

The next test we performed was examining the properties recovered by the
extracting algorithm for a distribution of input sources.
We compared input and extracted
flux densities for objects with FWHM equal to 23\arcsec\ (2.7 times the beam size of the
350 \um\ maps). Initially, we tested the flux recovery of artificial sources on apertures of 20\arcsec\ and
40\arcsec, and the total integrated flux, in units of $\mu$V. 
The integrated flux is estimated from the area of significant emission defined by the watershed extraction.
At an aperture smaller than the source size, we only recovered part
of the total intensity as expected, and the flux recovery at a larger aperture
as 40\arcsec\ is almost equivalent to the integrated flux. All the sources
extracted in our catalog have a recovered flux density less than 3000 $\mu$V ($\sim300$ Jy). Figure~\ref{fig:test_flux} shows results of
this flux recovery test. For the integrated emission, calibration 
was done using the C$_{beam}$ conversion factor, and flux recovery at
20\arcsec\ and 40\arcsec\ was calibrated using C$_{20}$ and C$_{40}$,
respectively. The image indicates that the 40\arcsec\ aperture flux density
could be underestimating the amount of flux recovered for input sources, with
a difference with respect to the curve of the integrated flux of $\sim10$\%. 
That
difference between integrated and aperture flux density recovery gets bigger
for input sources with larger size. For example, for input sources with FWHM
equal to 4 times the beam size ($\sim$30\arcsec), the difference between a 
40\arcsec\ aperture and the integrated flux is around 35\%.

\begin{figure}[h] 
   \centering
  \includegraphics[width=0.5\textwidth]{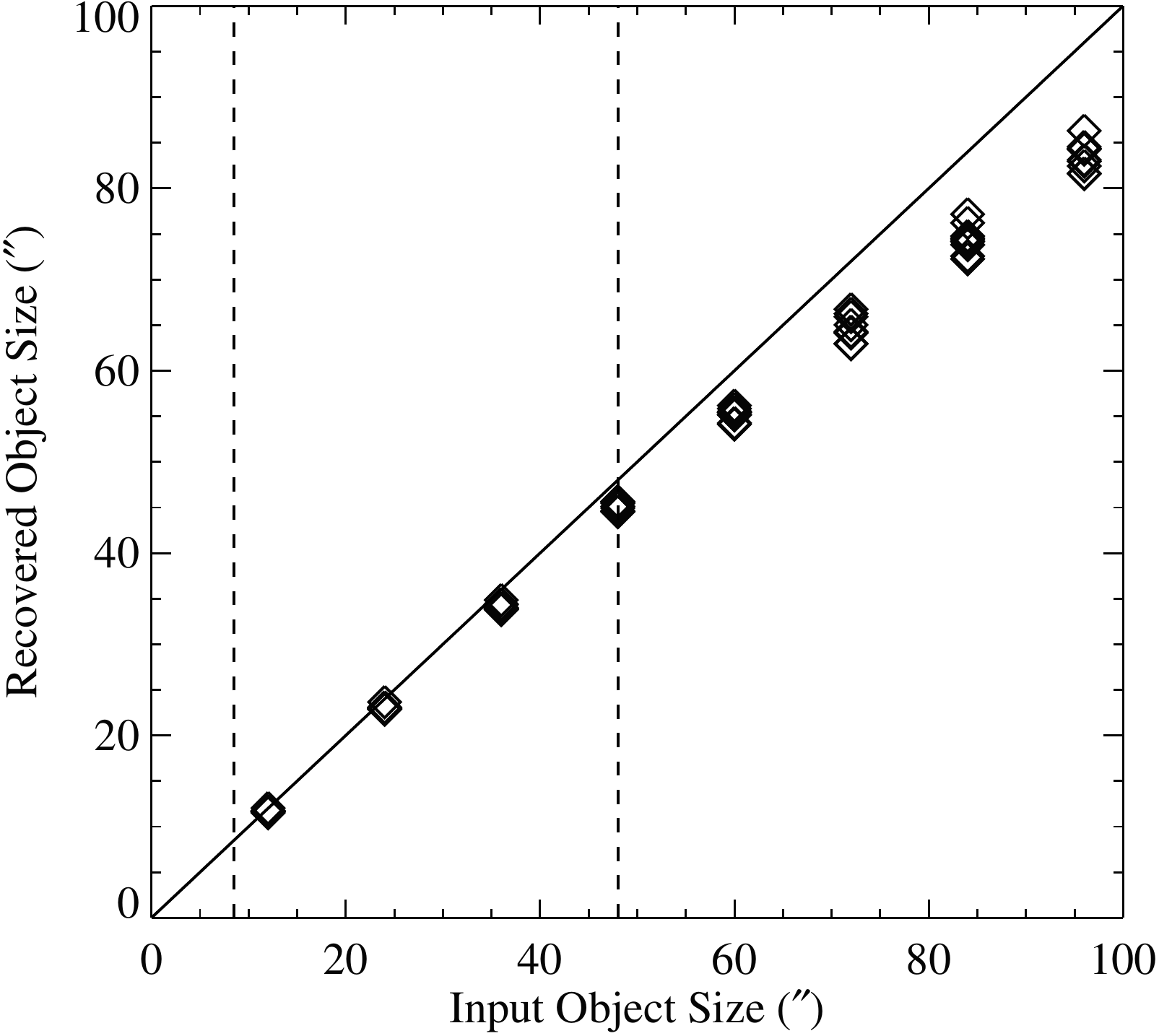}
   \caption{Recovery of source size for simulated observations in SHARC-II maps. The vertical lines at 8.5\arcsec\ and 48\arcsec\ indicate, respectively, the beam size of the images, and the largest size recovered by \bolocat\ for identified objects in our sample of maps. The largest major-axis recovered for identified sources is 65\arcsec.}
  \label{fig:test_size}
\end{figure}

The size recovered with \bolocat\ for input sources with a peak signal of 50$\sigma_{rms}$ is shown in Figure~\ref{fig:test_size}. The sizes of small
sources are well recovered by the algorithm, but they become underestimated
for sources
larger than $\sim60$\arcsec. These results are similar to those found in the
BGPS maps, where the radii of recovered sources become underestimated for
radii $\ge 200$\arcsec.  
The largest source that we found in the 350 \um\ maps has a size of $\sim$48\arcsec, with a major-axis of 65\arcsec.

\begin{figure}[h] 
   \centering
  \includegraphics[width=0.5\textwidth]{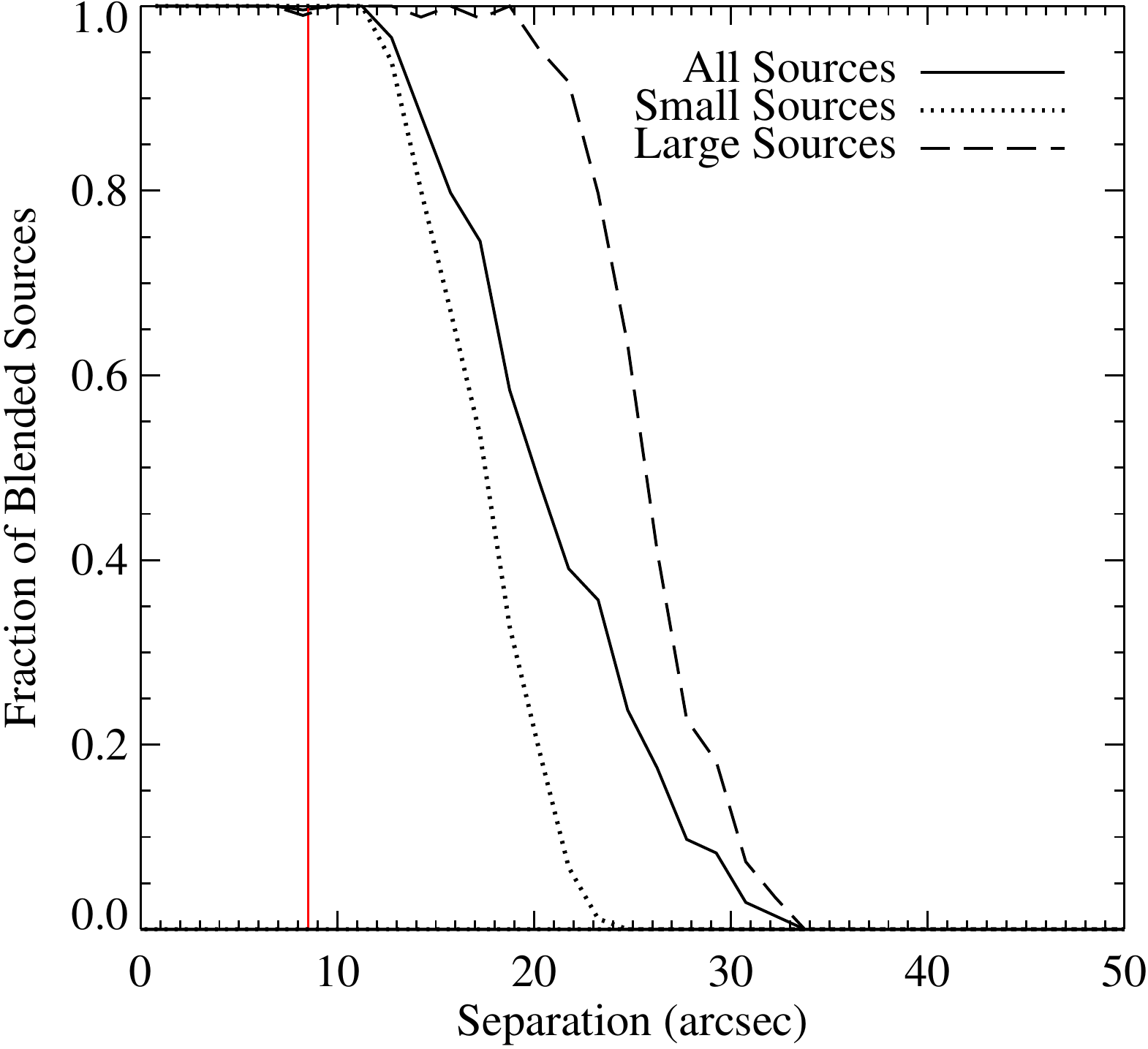}
   \caption{Fraction of blended fake objects as a function of source pair separation (solid line). The vertical red line shows the FWHM beam size of the 350 \um\ maps. Smaller input sources (15\arcsec$-$24\arcsec\ in size) are represented by the dotted line, and large input sources (24\arcsec$-$34\arcsec\ in size) are represented by segmented line.  }
  \label{fig:test_deblending}
\end{figure}

Finally, we inspected how the algorithm decomposes sources into individual
substructures. We tested how blended sources are identified as a single or
as a couple of individual objects, considering pairs of input objects in the
set of test maps. Each source of the pair of input fake sources have sizes
between 14\arcsec\ and 34\arcsec, and peak flux density of 50$\sigma_{rms}$.
Figure~\ref{fig:test_deblending} shows the fraction of blended sources as a
function of the separation between pairs of input sources. In general, sources
at a distance less than the beam FWHM cannot be recovered individually and they
are assigned to the same source in the recovered catalog, and more than 50\% of
the pairs of input sources are recovered individually at distances larger than
20\arcsec\ (2.4 times the beam size). Figure~\ref{fig:test_deblending} also
shows that pairs of input sources with sizes less than 24\arcsec\ are better resolved
at shorter distance than pairs with larger sizes than that. Pairs of sources
are resolved more than 50\% at separations of 16\arcsec\ for small input 
sources, and at 25\arcsec\ for larger sources.

\subsection{Catalog of sources in the 350 \um\ maps}

We have recovered from the \bolocat\ extraction 1386 sources in the 350 \um\
maps. We name our sources in a similar way as described in the BGPS catalog,
using the peak position of the source in galactic coordinates:
SHARC\_G$lll.llll\pm bb.bbbb$. An additional digit with respect to the BGPS
catalog was required to account for the better resolution and smaller pixel
size in the SHARC-II maps. Table~\ref{tbl:sources_detection} presents the
properties recovered for the extracted sources. The positions of 
the recovered sources in the SHARC-II maps are shown in 
Figure~\ref{fig:sharcmaps} in the Appendix.

Figure~\ref{fig:hist_sharc_flux} shows the distribution of recovered flux for
these sources using three methods from our catalog: photometry in 20\arcsec\
and 40\arcsec\ apertures, and the integrated flux. 
For the 20\arcsec\ aperture, the
average and median value of the flux density are 23.15$\pm$1.59 Jy and 8.95 Jy,
respectively. For the 40\arcsec\ aperture, the average and median values are
45.11$\pm$2.82 Jy and 16.63 Jy, resp. For the integrated flux density, 
the average and median values are 59.59$\pm$5.19 Jy and 12.06 Jy, resp.

Figure~\ref{fig:hist_sharc_radius} shows the deconvolved radii and the aspect
ratio distribution.
The deconvolved radius distribution
extends to 46\arcsec, and the catalog has 257 sources ($\sim19$\% of the
total) with radii not resolved. These small sources are usually faint, with an
average value of their peak signal-to-noise ratio of 4.4. The average and
median of the radius distribution are 15.0\arcsec\ and 14.0\arcsec,
respectively. The aspect ratio is defined as $\sigma_{maj}/\sigma_{min}$, and
its average and median values are 1.53 and 1.45, respectively.
A similar median aspect ratio is found in the BGPS V2.1 catalog, and this could
be a common feature of millimeter and submillimeter sources, or an 
artifact of the extraction algorithm.

\subsection{Correlation with BGPS sources}\label{sec:correlation}

We performed a spatial position matching analysis to get a correlation between
sources found in the 350 \um\ maps and the BGPS V2.1 sources contained in them. Therefore, we excluded from this analysis the 3 maps of outer Galaxy regions outside the coverage area of BGPS.
A simple criterion to match sources from catalogs at different resolutions is
to take the position of maximum intensity for sources of one catalog, and find
which sources in the second catalog are 1 beam distant from that position. For
example, \cite{con13} performed this kind of matching between sources of the
ATLASGAL survey (870 \um, 19.2\arcsec\ FWHM beam size) and BGPS V1.0 sources,
considering that sources between these catalogs are associated if their
peak positions are less than 40\arcsec\ in angular distance, finding around
$\sim$3000 likely matches. One of the issues with this approach is that it
does not yield a one-to-one association.
This has special importance in catalogs based on sources extracted from maps at different resolutions.
Even more, \cite{gin13} found that for the same
1.1 mm data at 33\arcsec\ FWHM beam resolution, re-processed V2.1 maps with
better spatial filtering recover in some cases more than the one source that
previous V1.0 maps found, and in others V2.1 maps recover a single structure where V1.0 recovered many. Then, sources required more than simple matching at
beam distance to compare different versions of the catalog.

Considering the different beam size of the 350 \um\ maps (8.5\arcsec) and 1.1
mm maps (33\arcsec), we expect to resolve extended millimeter sources in some
cases into multiple smaller components, and we put the constraint that compact
sources identified in the 350 \um\ maps must be associated with a single 1.1 mm
source. For the following analysis, we will consider as a ``parental clump" a
1.1 mm source from the BGPS catalog harboring one or more associated matched
structures on the 350 \um\ maps, which we will refer to as ``high-resolution
sources" or just simply as ``substructures".

We made use of one of the sub-products of the \bolocat\ algorithm, the mask
produced for each map in the identification of individual significant emission.
These \mask\ maps give information on which particular position in a map
contains emission, and associates that position with a
single source from the catalog produced for that map. Our approach was taking
the position of maximum intensity from the 350 \um\ sources, and if that peak is within the \mask\ of a BGPS V2.1 source, it is considered associated.

For the sample of 104 SHARC-II maps containing BGPS sources, there are 1374 high-resolution sources.
Only 24 (2\%)  of those sources have their peak position not associated to a 
BGPS V2.1 source from the \mask\ maps. The association between 350
\um\ high resolution sources and 1.1 mm sources suffers from intrinsic, noise driven uncertainty in the peak flux density position, and the algorithm
uncertainties in the \mask\ area assigned to a specific source. We improved the matching of
sources considering also the \mask\ maps of the 350 \um\ emission, overlapping
both masks and estimating how much of the substructure mask area is associated
with the parental source in the BGPS \mask\ area. Two of the 24 sources
have most of
their mask area associated with a parental source and therefore are re-considered
as matched sources. For the sample 350 \um\ sources with a match, there are 17
sources with their mask area shared between two parent sources, and for them we
made a visual inspection to see the most likely correct parental association.
Figure~\ref{fig:sharc_bolo_regions} shows a couple of examples for association
between 350 \um\ and 1.1 mm \mask\ regions.

We obtained a total of 1352 high-resolution sources associated with 349 different
parental sources.
The results of the previously described spatial matching of 350 \um\ SHARC-II
and 1.1 mm BGPS sources are shown in Figure~\ref{fig:hist_match}. Half of the
parental clumps have only one or two associated high-resolution sources, but
toward several of the parental clumps we found a large multiplicity of sources,
indicating possible fragmentation. The clumps with the largest multiplicity are
BGPSv2\_G213.705-12.603 (34 substructures), BGPSv2\_G034.256+00.154 (27),
BGPSv2\_G133.716+01.220 (25), BGPSv2\_G000.014-00.017 (23),
BGPSv2\_G029.916-00.045 (22), BGPSv2\_G081.477+00.020 (20), and
BGPSv2\_G029.958-00.017 (19). There are 22 parental millimeter sources with
more than 10 substructures, containing a total of 350 high-resolution
sources (26\% of the total).

These parental sources show some clear strong compact sources in them, but also
some elongated, filamentary structures are present.
We explored whether the large number of substructures in parental sources is
due to spurious low brightness sources, and/or algorithm fragmentation of large
sources. We consider the peak signal-to-noise ratio
identified for each source by \bolocat, and compare the number of identified sources under different ranges of this ratio.
Table~\ref{tbl:large_frag} presents the number of substructures with amplitude $P_{amp}$ (described in section~\ref{sec:extraction}) above
6$\sigma_{rms}$, 10$\sigma_{rms}$ and 20$\sigma_{rms}$ for the 22 BGPS sources
with large multiplicity. For these, 66\% of 350 \um\ sources have $P_{amp}$ of
6$\sigma_{rms}$, 42\% of 10$\sigma_{rms}$, and 19\% of 20$\sigma_{rms}$.
Figure~\ref{fig:sharc_bolo_regions} shows two examples of the distribution at
different peak signal-to-noise levels for two of these parental sources. While
compact, strong sources have $P_{amp}$ above 10$\sigma_{rms}$, detections
below this limit recover not only isolated low emission objects, but also filament-like
features in chains of sources, and weak detections produced in the
decomposition process of the algorithm. We conclude then that sources with
$P_{amp}$ above 10$\sigma_{rms}$ (``compact substructure") are more related
with possible dense core-like structures in the interior of millimeter clumps, and
sources below this limit (``faint substructure") trace a more diffuse medium in
addition to weak sources. For the total number of high-resolution sources
recovered from the SHARC-II maps, only 437 ($\sim$32\%) are considered compact
substructures. Column 14 in Table~\ref{tbl:sources_detection} indicates if a
source recovered in the 350 \um\ maps is ``compact" or ``faint".

\subsection{Flux densities of BGPS sources at 350 \um}\label{sec:flux_bgps_350}

We determined the flux densities at 350 \um\ of BGPS V2.1 sources by
measuring the flux density in an aperture of 40\arcsec\ on our sample of SHARC-II maps. To match the 33\arcsec\ FWHM effective BGPS beam at 1.1 mm, the 350 \um\ images, (1.618\arcsec\ per pixel) were convolved with gaussian
kernels (using IRAF task GAUSS) having $\sigma$ = 13.5\arcsec\ (8.33 pixels), i.e., convolving with a gaussian FWHM$=\sigma\sqrt{8\,\mathrm{ln}\,2}=31.8$\arcsec.

From the total of 619 BGPS V2.1 sources in the 350 \um\ maps, there are 82 sources
that appear in more than one map; we preferred the flux determination from the
maps where those sources suffer fewer sampling artifacts, such as high noise due
to map edge proximity, or negative bowls around the source. In case there
are not evident problems in the source flux estimation, we just considered the
averaged result of the flux density. In addition, there are 45 BGPS sources with negative integrated fluxes on the 350 \um\ maps due to proximity to noisy edges or negative bowls surrounding areas of strong emission, and therefore these sources are not considered as reliable measurements.

Results of the estimated values of 350 \um\ continuum emission from BGPS sources are presented in Table~\ref{tbl:bgps_sources_temperatures}. Column 1 gives the name of the source in the V2.1 catalog. Column 2 gives
the flux density integrated in an aperture of 40$\arcsec$ centered on the peak
position of the 1.1 mm source. The values of the flux density are corrected by
the factor 1.46 suggested by~\cite{agu11} for the 40\arcsec\ aperture flux
obtained in BGPS catalog. Column 3 gives the integrated flux density for
aperture photometry in the same previous position, but this time in the 350
\um\ convolved maps.

\subsection{Comparison between SHARC-II and {\textit{Herschel}} images}\label{sec:comparison_sharc_herschel_images}

From~\cite{mol10}, we used the Hi-GAL $2\arcdeg\times2\arcdeg$ images at 350
\um\ obtained during science demonstration phase centered toward $l=30$\arcdeg,
$b=0$\arcdeg\ and compared the results of flux recovery between the
\textit{Herschel}/SPIRE image and the 10 SHARC-II maps contained in that area.
The \textit{Herschel} image has a FWHM beam resolution of 24.9\arcsec, and
details in the reduction process are found in detail in~\cite{tra11}. 
Figure~\ref{fig:bgps_herschel1} in
the Appendix shows 1.1 mm BGPS images, and their corresponding Hi-GAL
images at 350 $\mu$m toward $l=30$\arcdeg, $b=0$\arcdeg, for regions mapped with SHARC-II.
While the 1.1 mm images in this figure show the position of the sources
from the BGPS V2.1 catalog, the 350 \um\ images show in blue ($< 10 \sigma$) 
or red ($>10 \sigma$) the
objects identified in our high-resolution source catalog.  There
are 213 of these substructures found, with only one left out of the following analysis
due to noisy edge effects.

First, we compared the flux obtained on the sources recovered directly by
\bolocat\ in this set of SHARC-II images, with emission measured in the \textit{Herschel} map.
Upper panel of Figure~\ref{fig:fluxrec_herschel_sharc} shows the ratio between the density
flux at 40\arcsec\ aperture $F_{SHARC}$ and the recovered flux at the same
aperture, centered on the peak position of the high-resolution sources,
obtained from the \textit{Herschel} image, $F_{Herschel}$. For faint substructures, the
flux ratio is in general below 0.3, with a cut around $F_{Herschel}\sim100$ Jy,
which suggests that the \textit{Herschel} image is recovering additional flux
from diffuse, large scale emission that SHARC-II maps do not recover.

Uniform background emission ranging between 2700 - 3300 MJy sr$^{-1}$ in
the {\textit{Herschel}} image would give a flux density into a 40\arcsec\ 
aperture of 80 - 100 Jy. 
Figure~\ref{fig:images_sharc_herschel} shows one of the SHARC-II maps toward $l=30$\arcdeg, L029.95-0.05, and the {\textit{Herschel}} emission in the same region. This figure also presents the SHARC-II map convolved to a resolution 
of 24.9\arcsec, to match the beam size of Hi-GAL data at the same wavelength.
The emission level at
17$\sigma$ in the {\textit{Herschel}} data corresponds to $\sim$2790 MJy sr$^{-1}$, 
and the morphology of emission recovered above that contour level in the {\textit{Herschel}} map is similar to the structures detected above 3$\sigma$ level in the SHARC-II images.
A ``background emission" level  for the {\textit{Herschel}} maps would
lie at about 2460 MJy sr$^{-1}$, corresponding to the 15$\sigma$ level 
(white contour in Fig.~\ref{fig:images_sharc_herschel}), and a 40\arcsec\ aperture integrated flux 
density of $\sim$73 Jy. The
beam-matched SHARC-II image shows that the 1$\sigma$ emission level has a good
resemblance with that background emission level.

Inspecting Figure~\ref{fig:bgps_herschel1}, most of the faint substructures
do not look like well-defined entities at a 10$\sigma$ contour, but instead
they seem to be immersed in background extended emission. For the compact
substructures recovered in the SHARC-II maps, the average and
median values for the flux ratio are 0.56$\pm$0.03 and 0.53, respectively. For
those high-emission compact sources with peak signal-to-noise above 50,
represented by green points in Figure~\ref{fig:fluxrec_herschel_sharc}, flux
recovery in the SHARC-II maps and the \textit{Herschel} map are nearly the same
(average flux ratio of 1.00$\pm$0.11).

We performed a similar analysis for the 102 BGPS sources found in this region.
We first convolved the 350 \um\ \textit{Herschel} image to match the 33\arcsec\
resolution of Bolocam, and compare later the 40\arcsec\ aperture flux emission
obtained toward the peak position of BGPS sources. Results are shown in
bottom panel of Figure~\ref{fig:fluxrec_herschel_sharc}. Errors in the flux ratio consider
conservative uncertainties of 20\% in the flux from the {\textit{Herschel}}
image, and 30\% in fluxes from SHARC-II data. Sources without any
substructure associated have low mm emission and flux ratio
$F_{SHARC}/F_{Herschel}$ on average lower than 0.1. For those 1.1 mm
parental clumps associated with compact substructures (shown in red in the
figure), the average and median values of the flux ratio are 0.76$\pm$0.08 and
0.69, respectively.

We emphasize that the differences arise from the better angular resolution of SHARC-II compared to {\textit{Herschel}}/SPIRE data, and not properly to the background subtraction; SHARC-II images are able to better disentangle the very dense and inner parts of clumps from the surrounding  diffuse emission.


\section{Analysis}

\subsection{Temperature determination}\label{sec:temperature_determination}

The data at 350 $\mu$m and 1.1 mm can be used to define a color temperature.
We will refer to this quantity through this analysis as the ``dust temperature", although  
variations in temperature across and along the line of sight
make this color temperature only a rough guide to the actual dust temperatures. 

The observed intensity of continuum emission is given
by $S_\nu \ = \ \Omega_{beam}[1-e^{-\tau_{\nu d}}]B_{\nu}(\td)$,  where
$\Omega_{beam}$ is the beam solid angle (33\arcsec\ beam) for both the 1.1 mm data and the convolved,
beam matched, 350 $\mu$m observations,
$\tau_{\nu d}$ is the optical depth of the emitting dust at each frequency and
$B_\nu (\td)$ is the Planck function at the dust temperature \td.
The optical depth in the
submm to mm regime is considered proportional to $\nu^\beta$. 
In the optically thin limit, the temperature
can be estimated
according to:

\begin{equation}
R = \frac{S_{350 \mu m }}{S_{1.1mm}} = \frac{\nu^{3+\beta}_{350 \mu m}
[\mathrm{exp}( h\nu_{1.1mm}/ k T_d)-1] }{\nu^{3+\beta}_{1.1mm} [\mathrm{exp}(
h\nu_{350 \mu m}/ k T_d)-1] }\ .
\label{eq:ratio}
\end{equation}

$S_{350 \mu m}$ and $S_{1.1mm}$ are the integrated flux densities obtained
in 40\arcsec\ apertures in the convolved SHARC-II maps and in the BGPS maps,
respectively.
The distribution of
the ratio $S_{350 \mu m}/S_{1.1\ mm}$ for the 574 V2.1 sources with reliable
flux values is
shown in Figure~\ref{fig:hist_flux_ratio_low}. The ratio distribution has a
maximum of 181.0, with a median value of 19.7.

Equation~\ref{eq:ratio} cannot be solved analytically for \td.
The adjustment of the pair of parameters \td\ and $\beta$ has been a recurrent
issue in the study of the interstellar medium and star formation.
Frequently, a value of $\beta$$\sim$2 in the millimeter to submillimeter regime
is assumed, referring to studies of dust grains composed by graphite and
silicate~\citep{dra84}, but models with variations of ice mantles due to
coagulation give lower values~\citep[e.g.,][estimated $\beta = 1.8$ for their
OH5 model]{oss94}. Values near $\beta\sim$1 may be appropriate for
circumstellar disks~\citep[e.g.,][]{bec91}.
\cite{shi05} summarize different opacity models in the submillimeter-millimeter
regime for low-mass pre-protostellar cores, with values of $\beta$ 
between 1.3 and 2.3. 
For the present analysis, we obtained dust temperatures considering three
different values of $\beta$: 1.0, 1.7, and 2.0. Nevertheless, considering that
most BGPS sources properties are related with dust clumps of dense material,
$\beta=1.7$ should be the closest model of the true nature of these structures.
Figure~\ref{fig:temp_plot} shows the fitted values for temperatures as a
function of the ratio $S_{350 \mu m}$ / $S_{1.1 mm}$ for the three models
previously described. For $\beta=1.7$, uncertainties in the determination of
temperatures are shown in the figure, considering an error of the flux ratio of
10\%.

The large uncertainty for high temperatures is unavoidable with these
data.  
At high temperatures ($>50$ K, see below), both 1.1 mm and 350 \um\ flux densities
approach the Rayleigh-Jeans limit, and therefore the ratio 
becomes nearly constant, increasing significantly the uncertainty
in the determination of \td.

The computed temperatures using three different values of $\beta$ are presented in Table~\ref{tbl:bgps_sources_temperatures}.
Columns 4 and 5 gives the source color temperature determined
from equation \ref{eq:ratio}, using a emissivity index $\beta=1.0$, and the
upper and lower limit for that temperature. Values of temperatures and 
upper and lower limits for a spectral index of $\beta=1.7$ and $\beta=2.0$ are
given in Columns 6-7 and 8-9, respectively.

From equation~\ref{eq:ratio}, sources with flux ratio $S_{350 \mu m}$ /
$S_{1.1\ mm}$ below 70.2 have temperatures lower than 1000 K for $\beta=1.7$.
Only 32 V2.1 sources ($\sim6\%$ of sample) have estimated temperatures
above 1000 K, most of them (24) found toward the Galactic Center region. In
only one source, the large flux ratio is the consequence of low emission at 1.1
mm and comparable noise level, and almost half of the rest have $S_{1.1\
mm}>1$ Jy, which included well-known regions such as G034.256+00.154,
G029.958-00.017, G030.702-00.067, and the ``Brick" IRDC~\citep{lon12}. 
These very high temperatures reflect flux density ratios near those
expected in the Rayleigh-Jeans limit, and the temperatures are not
constrained by our data. A centrally heated region will produce strong
350 \um\ emission; using a ratio with the 1.1 mm emission, which traces
a larger region of cold dust, can produce an artificially high color
temperature.

At a limiting temperature of 50 K, the corresponding flux ratio is 52.7 for
$\beta
= 1.7$, and a variation of 10\% of this ratio will give an increase of 40\% in
the estimated temperature. Therefore, to avoid large uncertainties we consider
as a ``good fit" a determined
temperature less or equal than 50 K, and a similar limiting temperature value
is found for $\beta=1.0$ and $\beta=2.0$. There are 512 sources in our sample
below this limit for $\beta=1.7$, and 30 sources with temperatures 
between 50 and 290 K with quite large uncertainties.

Figure~\ref{fig:hist_bgps_temp_limit} shows the temperature distribution
for the 512 sources, assuming $\beta=1.7$, with good fits for
temperature. The median value is 13.3 K, with an average of 16.3$\pm$0.4 K. The
difference between the median and the average is explained by the positive
skewness of the distribution. The figure also shows the distribution of sources
with different models for the spectral index. Considering a value of
$\beta=1.0$, the sample of V2.1 sources tends to have larger values of
temperature, and then the number of these sources with $T_d\le50$ K is 322, with
a median value of 17.3 K and an average of 19.6$\pm$0.5 K. For $\beta=2.0$, the
distribution of fitted temperatures has lower values, with 543 sources below 50
K, and with median and average temperatures of 11.2 K and 13.3$\pm$0.3 K,
respectively.

Figure~\ref{fig:fluxbolo_temp} plots temperature versus flux density
at 1.1 mm for a fixed value $\beta=1.7$. For faint mm sources,
uncertainties in the flux dominate the uncertainties in the temperature.
In contrast, temperatures above 30-40 K are uncertain because of the
weak constraints as the Rayleigh-Jeans limit is approached.
The figure also shows the presence of 8 sources with high flux density ($>4$ Jy at 1 mm) and low estimated temperatures ($<20$ K). These are 
BGPSv2\_G359.946-00.045 on the Galactic center region, 
BGPSv2\_G075.834+00.400 and 
BGPSv2\_G076.384-00.622 on Cygnus-X, 
BGPSv2\_G133.716+01.220 on W3 main, 
BGPSv2\_G031.411+00.307, and the sources
BGPSv2\_G206.557-16.361 and 
BGPSv2\_G206.534-16.356 on NGC 2024, toward the Orion B South molecular cloud. In the case of BGPSv2\_G133.716+01.220, corresponding to W3 East,~\cite{riv13} recovered a higher temperature ($\sim30$ K for $\beta=2.0$) for this source using {\textit{Herschel}} data from the HOBYS program~\citep{mot10}, difference that is explained in part for the underestimation of the flux at 350 \um\ due to a shift on the peak of the BGPS source (and therefore the integration aperture), and the peak of the source on the SHARC-II map. In fact, there is a better agreement on the sources W3 West (BGPSv2\_G133.698+01.216; $T_d=28$ K, $\beta=2.0$) and W3 SE (BGPSv2\_G133.747+01.198; $T_d=25$ K, $\beta=2.0$), and the values estimated by~\citeauthor{riv13} for these clumps ($\sim$25 K and $\sim$21 K, respectively). 

In the case of the clump associated to the G31.41+0.31 hot core~\citep{ces94}, an isothermal temperature of $T=29$ K was estimated for this source by~\cite{mueller02} from continuum data modeling, which could indicate that the low temperature is the result of unreliability on the measurement of derived fluxes.

For the two clumps on the Orion B cloud,~\cite{joh06b} considered for both sources a temperature of 50 K, though they were unable to obtain color temperatures for them from SCUBA 450 \um\ and 850 \um\ continuum data due to high flux ratios, inconsistent with modified blackbody emission.

There are other 22 sources with $S_{1.1mm}>1$ Jy and $T_d<15$
K. These sources could be dense, prestellar clumps, but further observations
and analysis, particularly in molecular line observations of high-density
tracers, are necessary to test their nature.

\subsection{Comparison of temperatures with other surveys}\label{sec:comp_surveys}

We found in general a good agreement between our determination of temperatures and  estimations of temperature from other surveys of high-mass star forming regions. In this section, we describe details of these comparisons.

\cite{fau04} performed an analysis
of physical properties of 146 continuum emission structures detected at 1.2 mm
with SIMBA toward IRAS sources associated with CS(2$-$1) detections. The
temperature and spectral index of these sources were obtained from SED fitting
using additional IRAS four band fluxes, obtaining values for $\beta$ ranging
between 1.5 and 2.5, with average temperature for their sample of $\sim$32 K.
Temperatures found for their millimeter structures are larger than
ours because their sample was biased toward IRAS sources that are intrinsically warm. Their sample has strong emission at 1.2 mm,
with all their sources (except one) above 1 Jy and with an average flux 
density of 16.3$\pm$3.2 Jy.
Considering the emission at 1.1 mm from our sources that are, within
uncertainties and corrections, comparable with their sample, the average
temperature for the 124 V2.1 sources with 40\arcsec\ aperture flux at 1.1 mm
above 1 Jy is $\sim$38 K for $\beta=1.7$. 
Similarly, \citet{mueller02}
modeled SEDs of a group of 51 massive star forming clumps, originally selected
by having water masers, and found that a
temperature of $29\pm9$ K best characterized the sources.
Cross-matching their sample with BGPS sources inside SHARC-II maps,
considering one beam (33\arcsec) in distance associations between samples,
we found 13 BGPS clumps with good determination of dust temperature associated with
the sources in \citet{mueller02}. These sources have average and median values for their 40\arcsec\ aperture integrated flux of 8.5 Jy and 7 Jy, respectively, with only one source with integrated flux below 1 Jy. One of these sources corresponds to BGPSv2\_G031.411+00.307, for which its fitted dust temperature is considered underestimated (see Section~\ref{sec:temperature_determination}). Excluding this source, the fitted dust temperatures of the rest of the clumps have an average of 27 K, and a median of 29 K.
Thus, the parameters that we found in
sources with strong ($>$1 Jy) integrated emission at 1.1 mm are similar to those found with SED fitting,
giving some confidence to our values based on only two wavelengths.
In addition, BGPS clumps associated with massive star forming regions generally have
strong emission at mm and submm wavelengths, with warm temperatures $\sim$30 K,
lying above 89\% of the distribution of fitted temperatures of BGPS sources with T$\le$50 K.

\cite{bat11} used Hi-GAL 70 to 500 \um\ data to identify significant dust
continuum emission in two 2\arcdeg$\times$2\arcdeg\ regions centered at
$l=30$\arcdeg\ and $l=59$\arcdeg, and obtained temperature and column density
maps for them by fitting spectral energy distributions for each pixel. While
the temperature is a free parameter for their fitting model, the spectral index
is fixed to 1.75 following~\cite{oss94}. From our sample, we have 91 BGPS V2.1
sources with determined temperatures in the $l=30$\arcdeg\ map, and none in the
$l=59$\arcdeg\ map. Considering $T_{SHARC}$ the fitted temperature obtained
from equation~\ref{eq:ratio}, the median and average temperature for those
sources from our analysis is 18 K and 34$\pm$4 K for $\beta=1.7$. We estimated
a value of the temperature of these sources from the maps presented
by~\citeauthor{bat11}, $T_{Herschel}$, averaging the temperature values of each
pixel on an aperture of 40\arcsec\ centered on the peak position of each
source. Figure~\ref{fig:herschel_temp} shows the ratio $T_{SHARC}$ /
$T_{Herschel}$ as a function of the flux density at 1.1 mm for the sample of
BGPS sources. The average ratio from this figure, without considering those
sources with $T_{SHARC}>50$ K for which the uncertainties are much larger, is
0.83$\pm$0.05, while the weighted mean is 0.50$\pm$0.01.

To identify dense material associated with the formation of massive stars and clusters, \cite{bat11} performed iterative removal of cirrus cloud emission on Hi-GAL images. The spectral emission distribution fitted for each point across the map results in a median value for temperature of  T$=$23 K for a spectral index of $\beta=$1.7 on $l=30$\arcdeg. The region considered as source emission is roughly the contour at $6.5\sigma$ ($\sim$1070 MJy sr$^{-1}$) in the 350 \um\ SPIRE band, and hence SHARC-II maps still miss extended emission that is considered to be source emission by~\citeauthor{bat11}. From their temperature map, we estimated a temperature for this extended background emission, masking the region above 15$\sigma$ from the 350 \um\ SPIRE map, and also masking circular areas of 80\arcsec\ diameter centered on each BGPS V2.1 source on the map. We found a temperature of ~25 K, for a fixed $\beta=$1.75 for the emission considered for SED fitting by~\citeauthor{bat11} but filtered out in SHARC-II maps.

We tested then the estimation of color temperature on compact sources, if additional background emission were considered. A simple model of a source emitting uniformly as a modified blackbody, characterized by a temperature $T_{source}$ and a spectral index $\beta_{source}$, combined with background emission, also modeled as a modified blackbody, with T$_{back}$ and $\beta_{back}$. The measured  fluxes of the composite emission, $S_{350 \mu m}$ and $S_{1.1mm}$, correspond  to the sum of individual contributions at those wavelengths, and following equation~\ref{eq:ratio}, differences between $T_{source}$ and $T_{back}$, and $\beta_{source}$ and $\beta_{back}$, will recover a new set of measured values $T_{M}$ and $\beta_{M}$. For a field near $l=30\arcdeg$, where the very dense structures are likely spread over diffuse emission with $T_{back}=25$ K and $\beta_{back}=1.7$, we simulated the recovery of properties for input sources with a range of parameters characteristic of clump-like features, $5<T_{source}<50$ K, and $1.5<\beta_{source}<2.5$. The ratio $T_{source}/T_{source\,+\,bg}$ of the input to the determined temperatures, as a function of the measured 1.1mm flux, are shown in the lower panel of Fig~\ref{fig:herschel_temp}. The results are qualitatively similar to the comparison between $T_{SHARC}$ and $T_{Herschel}$, with low-flux, cold sources having low ratios of temperatures; warm, strong mm emission clumps have ratios of temperatures slightly above 1.0; and with sources with temperatures of 10-20 K (the bulk of BGPS clumps in $l=30$\arcdeg) with temperature ratios in the range 0.4-1.0. 

A more direct comparison can be performed with compact sources extracted on {\textit{Herschel}} images using the CUTEX method, designed for detection of sources on intense and highly variable fore/background~\citep{mol11}. Using a similar method as presented by~\cite{eli13}, a catalog of extracted compact sources and their associated physical parameters, using the five bands of Hi-GAL, have been obtained for the inner Galaxy ($-70\arcdeg<l<67\arcdeg$) (D. Elia et al., in preparation). These sources are extracted from background emission and therefore considered to represent the very dense clump-core features across molecular clouds. From a simple spatial match performed to the 102 BGPS sources toward $l=30$\arcdeg\ that have estimated color-temperatures, 76 of these clumps have a counterpart Hi-GAL compact source, and preliminary results produce much better agreement with the temperatures derived in this paper from SHARC and BGPS
data; the ratio between these temperatures have a mean of 1.05 and a median of 0.97. In both cases, a fixed value of the spectral index $\beta=2.0$ was considered. Several considerations need to be assessed, such as different beam resolutions between observation bands, and a more complete analysis between the presented sample of SHARC-II maps and properties of extracted sources with Hi-GAL, including substructures and association with emission at 70 \um, will be the subject of a forthcoming work.

\subsection{Comparison between gas temperature from NH$_3$ and color temperature}

Figure~\ref{fig:nh3_dust} 
shows the ratio of gas to dust temperatures for a sample of BGPS sources.
The gas temperatures were determined from
NH$_3$ observations~\citep[][and BGPS team in prep.]{dun10,wie12}. For those observations not centered on peak positions of BGPS sources, we associated the NH$_3$ detections to their closest BGPS source within one beam distance.
Only the 49 clumps with  dust temperatures below 50 K and gas temperatures
below 30 K are plotted. The ratio $T_{gas}/T_{dust}$ has a weighted 
average of 0.88 and a median of 0.76. There are few sources with equal
temperatures, but instead a group with lower $T_{gas}$ and a group
with higher $T_{gas}$.

\subsection{Comparison with tracers of dense gas}

\cite{sch11} presented spectroscopic observations of HCO$^+$ and N$_2$H$^+$
(3-2) for 1882
sources from BGPS. The selection of sources was made between Galactic
longitudes $10\arcdeg\le l\le90\arcdeg$, considering all sources with
integrated flux in a 40\arcsec\ aperture greater than
S$_{1.1mm}\sim$0.4 Jy, and from bins of logarithmically spaced flux of 0.1 for
sources with S$_{1.1mm}=0.1-0.4$ Jy. A new, complete version of this catalog
by~\cite{shi13} includes all 6194 sources in the BGPS
V1.0 catalog between $7.5\arcdeg\le l \le194\arcdeg$ of this pair of
molecular transitions. 
After considering a one-to-one match between sources
from V1.0 and V2.1 catalogs, with a spatial matching of one beam of distance
(33\arcsec), we checked which of those sources are present in our sample
of SHARC-II maps. There are 300 V2.1 sources with spectroscopy data, 250
of them with HCO$^+$ detections and estimated dust temperatures from our
analysis, and 199 with N$_2$H$^+$ detections with dust temperatures.
Figure~\ref{fig:spec_cat_bgps} shows the peak main beam temperature versus our fitted temperature for that group of sources. The points scatter widely about the plot. Considering only the sources with dust temperatures less than 50 K (169 for N$_2$H$^+$ and 223 for HCO$^+$), the Spearman's rank correlation coefficient of the distributions of log($T^{pk}_{mb}$) and log($T_{dust}$) are 0.51 for sources with N$_2$H$^+$ detections, and 0.33 for sources with HCO$^+$ detections.
We tested the null hypothesis of no significant correlation between log($T^{pk}_{mb}$) and log($T_{dust}$) for these distributions. For a sample size $n$, the Student's $t$ value for the Pearson correlation coefficient $r(n)$ is estimated from $t=r/s_r$, with $s_r=\sqrt{(1-r^2)/(n-2)}$. For a two-tailed probability value $p=0.05$, our sample requires $t=2$, corresponding to $r(169)=0.151$ and $r(223)=0.131$. The estimated values for the correlation coefficients of our sample are $r=0.508$ and $r=0.332$ for N$_2$H$^+$ and HCO$^+$, respectively, rejecting the null hypothesis at a 95\% confidence. Therefore, although the correlation is weak, in both cases it is statistically significant for the detected peak main beam temperature and the fitted dust temperature. In the case of N$_2$H$^+$, the correlation can be fitted as log($T^{pk}_{mb}$)\ =\ B$\times$log($T_d$)\ +\ A, with fitted values A=$-$1.55 and B=1.04$\pm$0.13.
For HCO$^+$, the fitted parameters are A=$-$0.59 and B=0.52$\pm$0.08.

\cite{shi13}\ suggested  that BGPS sources with a higher N$_2$H$^+/$HCO$^+$ ratio could be  clumps with  dense core-like structures in them, as a consequence of cold, CO-depleted gas within those cores. Nevertheless, we did not find a clear correlation between the ratio of high density tracers N$_2$H$^+/$HCO$^+$, and number of substructures or estimated dust temperature in our source sample. Parental clumps with large multiplicity of substructures and high N$_2$H$^+/$HCO$^+$ ratios represent  good candidates for future interferometric observations to study core properties in regions of clustered massive-star formation. 

\section{Discussion}

\subsection{Mass and surface density of substructures}\label{sec:mass_surface_substructures}

We use  the following equation to calculate the mass:

\begin{equation}
M_d=\frac{S_{\nu}D^2}{\kappa_{\nu}B_{\nu}(T_d)}\ ,
\end{equation}

\noindent where $S_\nu$ is the flux density, $D$ is the distance to the source, and $\kappa_{\nu}$ is the dust mass opacity coefficient.

We use a dust opacity of 1.14 cm$^2$g$^{-1}$ of dust at 1.1 mm
\citep{oss94} for a model with thin ice mantles (OH5). Assuming a
dust-to-gas mass ratio of $R=M_d/M_g=0.01$, the mass of gas and dust
will be given by:

\begin{eqnarray}\nonumber
M_{1.1 mm}\ &=&\ 14.26\left(\frac{S_{\nu}}{Jy}\right)\left(\frac{\mathrm{1.14\: cm^2\:g^{-1}}}{\kappa_{1.1 mm}}\right)\left(\frac{D}{\mathrm{kpc}}\right)^2\\ &
&\times\left\{\mathrm{exp}\left(\frac{13.01}{T_d}\right)-1\right\}\ M_\odot\ .
\label{eq:mass_bgps}
\end{eqnarray}

From our sample of 512 BGPS sources with temperatures $\le50$ K, we can give
some estimates of how the values of masses will change when a fixed
temperature is used. The average ratio between the clump mass $M(T_{d})$, with
$T_d$ the fitted dust temperature from continuum observations between 350 \um\ and
1.1 mm, and the mass estimated at a fixed temperature, $M(T_{fix})$, is
obtained from:

\begin{equation}
\left<\ \frac{ M(T_{d})}{M(T_{fix})}\right> \ = \ \left<
\frac{\mathrm{exp}(13.01/
T_{d})-1}{\mathrm{exp}(13.01/ T_{fix})-1}\right>
\end{equation}

Figure~\ref{fig:model_ratio_mass} shows the results of the above equation for
different values of $T_{fix}$. It is common to use for simplicity a fixed
value of 20 K in determination of masses from millimeter 
surveys of Galactic star forming regions~\cite[e.g.][]{mot07,sch11,urq13}.
Using $T_{fix}=20$ K, the average value of the ratio is 2.43$\pm$0.11, with a
median value of 1.83. Therefore, assuming a fiducial value for the temperature
of 20 K for our sample of clumps, the
estimation of masses would be underestimated on average, with respect to the
mass obtained from our analysis of color temperatures.

From \cite{ell13,ell15}, kinematic distances were determined for a large number of BGPS sources. Only a fraction of our sample of BGPS clumps have estimated distances, so we cannot calculate masses for the entire group. However, we can estimate  the contribution of masses from substructures detected at 350 um with respect to the parental clump, because
both substructures and clump are located at the same distance. 
We use the flux density at 350 \um\ to compute masses of the substructures.
Similar to equation~\ref{eq:mass_bgps}, but with appropriate change in
the exponential and with a dust opacity of 11 cm$^2$g$^{-1}$ (OH5), 
the mass measured at 350 \um\ is as follows:

 \begin{eqnarray}\nonumber
M_{350 \mu m}\ &=&\ 0.053\left(\frac{S_\nu}{Jy}\right)\left(\frac{\mathrm{11\: cm^2\:
g^{-1}}}{\kappa_{350 \mu m}}\right)\left(\frac{D}{\mathrm{kpc}}\right)^2\\ &
&\times\left\{\mathrm{exp}\left(\frac{41.14}{T_d}\right)-1\right\}\ M_\odot
\label{eq:mass_sharc}
\end{eqnarray}

From our sample
of 349 BGPS parental clumps associated with 1352 high-resolution sources, the
fraction of the total mass estimated for structures at 350 \um\ with respect to the
mass of the parental 1.1 mm clump is determined by:

 \begin{eqnarray}\nonumber
\frac{\sum \left(M_{substructures}\right)}{M_{clump}}\ &=\
&3.75\times10^{-3}\left(\frac{S^{*}_{350\mu m}}{S^{*}_{1.1mm}}\right)\\& &\
\times\left(\frac{\mathrm{exp}(41.14/
T_{d})-1}{\mathrm{exp}(13.01/ T_{d})-1}\right)\ ,
\end{eqnarray}

\noindent where $S^{*}_{1.1mm}$ is
the emission of the parental BGPS clump and $S^{*}_{350\mu m}$ represent the total
emission of substructures inside that clump. We consider the same dust temperature $T_d$, fitted from equation~\ref{eq:ratio}, for substructures and their parental clump. Here the integrated emission
recovered by \bolocat\ for each clump is taken into account, which estimates the total flux for the
area delimited by the \mask\ region (see Section~\ref{sec:correlation}). In
that way, the contribution of each substructure will be associated with a single
parental clump, with both considered at the same dust temperature.

Figure~\ref{fig:ratio_mass_sharc_bgps} shows the ratio between masses obtained
for high-resolution sources at 350 \um\ and their parental BGPS clump, as a
function of the dust temperature and as a function of the integrated flux at
1.1 mm. Black points represent the mass ratio considering the flux contribution
of all high-resolution sources, and red points only consider emission from
strong, compact sources. The average and median values for the mass ratio
$\sum (M_{substructures})/M_{clump}$ when all substructures are considered, are
0.48 and 0.40, respectively, with a mean weighted by errors of 0.22$\pm$0.01. 
Parental clumps that are more massive and cold have
smaller uncertainties, so the mean mass ratio weighted inversely by uncertainty is smaller.
When only
compact substructures (see Section~\ref{sec:correlation})
are considered, the average and median of the
distribution are 0.39 and 0.37, respectively, with a mean weighted by errors of
0.19$\pm$0.01. 
These results for the mass fraction are larger than the results
from \cite{rag13}, who fit SEDs and obtained a value of $\sim14$\% for the mass
contribution of core-like features observed at a resolution of 7.8\arcsec\
compared to the total mass of their harboring IRDCs. Possible explanations of
this difference are our larger sample of sources (\citeauthor{rag13} consider 11
IRDCs, harboring a total of 83 substructures), and also that, contrary to BGPS sources, IRDCs are not limited to compact structures. Besides, our sample of
clumps has a broader range of evolutionary stages, 
not necessarily limited to the coldest, initial stages of 
massive star formation, represented by the IRDC phase.

Assuming a uniform temperature will also introduce errors in
the mass determination. The parental molecular clump will be affected by both
embedded stars and the external radiation field, and it is
likely that substructures will have a different temperature from their
surrounding medium. In addition, SHARC-II and Bolocam are sensitive to
different spatial scale emission (see Section~\ref{sec:tests}), and therefore the recovered flux at 1.1 mm
could include faint, extended background emission not
detected in 350 \um\ continuum emission.  A better analysis of the mass ratio
between parental clumps and their internal structures (compact, core-type 
sources, and faint, filamentary emission) will require a comparison between SHARC-II and {\textit{Herschel}} maps at 350 \um, or modeling these sources
using radiative transfer codes.

Figure~\ref{fig:surface_density} shows the surface density $\Sigma_{350 \mu m}$
estimated for the substructures found in the 350 \um\ map, as a function of
the surface density of their parental clump, $\Sigma_{1.1mm}$. We considered
the deconvolved radius of resolved substructures at 8.5\arcsec and resolved
clumps at 33\arcsec. Identified substructures have in general larger surface
densities than their parental BGPS source. For the sample of spatially
resolved sources, there are 418 compact substructures, 286 of which have
estimated temperatures $T_d\le50$ K. These are represented by red points in the
figure. A linear fit of the compact sources indicate that they have
$\Sigma_{350 \mu m}$ $\sim$ 2.2$\times\Sigma_{1.1mm}$. 
Eighteen of them (6\%) have  $\Sigma_{350 \mu m} \ge 1.0$ g cm$^{-2}$, a
theoretically predicted threshold for the formation
of massive stars~\citep{kru08}. 
More recently a lower value of 0.2 g cm$^{-2}$ was proposed by~\cite{but12} 
as the initial condition of massive star formation in the Galaxy;
88 (31\%) of our sources meet that criterion.

Thus, for our sample of high-resolution structures observed at 350 \um, 
there are probably many dense cores, whose physical parameters 
suggest on-going or imminent formation of massive stars.

\subsection{Uncertainties on determined physical parameters of clumps}

Following equation~\ref{eq:ratio}, the presence of additional integrated flux densities observed at 350 \um\ and 1.1 mm not related to the dust emission of a given clump-like source, such as emission from the surrounding environment of the compact source, or contamination from molecular lines on those bands, can introduce dissimilarities between the recovered and real flux ratio and therefore in subsequent derived physical parameters. As seen in Section~\ref{sec:comp_surveys}, the temperature estimated for a source embedded in a background with different temperature (or spectral index) may differ from the actual source temperature.
While the data-reduction pipeline of the Hi-GAL survey is coded to remove low frequency noise, which imply that the 350 \um\ Herschel images recover faint diffuse emission~\citep{tra11}, SHARC-II maps filter out most of that emission and recover the very dense, compact structures, and BGPS maps are not sensitive to extended emission at scales larger than 2\arcmin-3\arcmin~\citep{gin13}. The precise amount of additional flux per band is determined by the temperature and spectral index of the background extended emission (see Fig.~\ref{fig:herschel_temp}).

In the case that observations of a particular clump recover some background emission in the 1.1 mm BGPS but not in the 350 \um\ SHARC-II maps, their overall effect would consist of lower measured color temperature, and higher mass estimation. Contamination of CO$(7-6)$ lines by the  350 \um\ emission is not expected for the bulk of clumps in this work. The $J=7$ level of CO is 155 K above ground, and even though this line has been seen in sources with kinetic gas temperature of $\sim$25 K~\citep{kru89}, emission is expected only in the warmer and denser clumps. No line contamination is possible at 1.1 mm since the passband of Bolocam excludes the strong CO$(2-1)$ line.

To quantify these uncertainties, we consider two hypothetical cases of clumps with equal spectral index $\beta=1.7$ and temperatures $T_{source}$ of 20 K and 40 K, respectively. These sources have an associated value $R_{source}$ of the ratio of integrated flux densities observed at 350 \um\ and 1.1 mm, and additional emission in these bands will produce variation of the measured value of the flux ratio, $R_{M}$. For the case $T_{source}=20$ K, differences of less than 20\% between $R_{M}$ and the intrinsic flux ratio of the source $R_{source}$, will imply a measured temperature $T_{M}$ also 20\% different from the source temperature $T_{source}$. From equation~\ref{eq:mass_bgps}, the uncertainty in the determination of the clump mass and surface density will be~$\sim$30\%.

For warmer sources, even small changes to the flux ratio will result on large uncertainties in the measured $T_{M}$. For $T_{source}=40$ K, differences of 10\% between $R_{M}$ and $R_{source}$ will result in differences of 30\% of $T_{M}$ with respect to $T_{source}$, and clump mass and surface density will also have uncertainties~$\sim$30\%. In the case of differences of 20\% between $R_{M}$ and $R_{source}$, measured temperatures could be 80\% different from the real $T_{source}$, and clump mass and surface density will have additional uncertainties of $\sim$50\%.

\section{Summary and Conclusions}

We presented a set of 107 SHARC-II 350 \um\ maps toward star forming clumps. Most of maps are observed toward sources from the BGPS V2.1 catalog, with additional maps covering three outer Galaxy star forming regions. The maps have a resolution of 8.5\arcsec, improving 
upon the resolution of the BGPS maps at 1.1 mm (33\arcsec) and revealing a 
population of substructures.

We used \bolocat\ to identify significant emission in the SHARC-II
maps, and we obtained a catalog of 1386 high-resolution sources, with average
fluxes of 23.15$\pm$1.59 Jy and average radius of 15\arcsec. We found that
$\sim$32\% of these features have peak emission above 10$\sigma_{rms}$, and we
consider these ``compact substructures" as core-like sources immersed in 
parental clumps. Below
this limit, recovered sources are called ``faint substructures", and they are related
with fragmentation of filamentary features, isolated low intensity sources,
and residuals from the source extraction algorithm.

We found 619 BGPS V2.1 sources in our set of 350 \um\ maps, only 56\%
of which are associated with 1352 high-resolution substructures. While most of
the parental clumps have only one or two associated substructures, we found
significant multiplicity in some of them, with 22 clumps having more than 10
substructures in them.

We compared the 350 \um\ continuum emission from SHARC-II maps with
\textit{Herschel} images from the Hi-GAL survey toward the 
$l=30\arcdeg$ region at
same wavelengths. Strong emission detected in SHARC-II maps is in general in
good agreement with the \textit{Hershel} fluxes, but faint sources from the
SHARC-II maps are immersed in background emission.

We estimated temperatures for different models of the
spectral index. For the 512 sources with ``good fits" in the determination of temperatures
($T_d\le50$ K), the median and average values of their temperatures are 13.3 K
and 16.3$\pm$0.4 K, respectively, in reasonable agreement
with other temperature determination methods based on SED analysis.

Comparison to gas temperatures derived from \ammonia\ observations 
reveals sources with ratios of $T_{gas}/T_{dust}$ both larger and smaller
than unity, but with mean value of 0.88.

Sources with N$_2$H$^+$ and HCO$^+$ detections present weak correlations between their observed peak main beam temperature and their fitted dust temperature. There is no clear correlation between the ratio T$^{pk}_{mb}$(N$_2$H$^+$)$/$T$^{pk}_{mb}$(HCO$^+$) and the number of substructures found on parental clumps.

The fraction of the mass contained in substructures compared
with the total parental clump mass has an average of 0.48, but this value
decreases to 0.22 when weighted by uncertainties. Considering only the compact
structures, likely to be core sources, the mass fraction between substructures
and parental clumps is 0.19, higher than other studies of high angular
resolution at 350 \um.

Small, but significant fractions of substructures satisfy theoretical 
thresholds for massive star formation of 0.2 (31\%) or 1.0 (6\%) g
cm$^{-2}$.
\acknowledgments
We thank the anonymous referee for helpful comments and suggestions that improved the clarity of this article. We would like to thank all the people involved on obtaining our observations, including D. Dowell, students and researchers from the University of Texas, and staff from the CSO.
The BGPS project was supported in part by the National Science Foundation through NSF grant AST-0708403. The first observing runs for BGPS were supported by travel funds provided by NRAO. Support for the development of Bolocam was provided by NSF grants AST-9980846 and AST-0206158. NJE and MM were supported by NSF grant AST-1109116. MM was also supported by a Fulbright Fellowship. 
This material is based upon work at the Caltech Submillimeter Observatory, which was operated by the California Institute of Technology under cooperative agreement with the National Science Foundation (AST-0838261).
Herschel Hi-GAL data processing, maps production and source catalogue generation from the Hi-GAL Team have been possible thanks to Contracts I/038/080/0 and I/029/12/0 from ASI, Agenzia Spaziale Italiana.
Herschel is an ESA space observatory with science instruments provided by European-led Principal Investigator consortia and with important participation from NASA. 
SPIRE has been developed by a consortium of institutes led by Cardiff Univ. (UK) and including Univ. Lethbridge (Canada); NAOC (China); CEA, LAM (France); IAPS, Univ. Padua (Italy); IAC (Spain); Stockholm Observatory (Sweden); Imperial College London, RAL, UCL-MSSL, UKATC, Univ. Sussex (UK); Caltech, JPL, NHSC, Univ. Colorado (USA). This development has been supported by national funding agencies: CSA (Canada); NAOC (China); CEA, CNES, CNRS (France); ASI (Italy); MCINN (Spain); Stockholm Observatory (Sweden); STFC (UK); and NASA (USA).




\clearpage

\clearpage

\begin{deluxetable}{lcccc}
\tabletypesize{\scriptsize}
\tablecaption{Average Calibration Factors}
\tablewidth{0pt}
\tablehead{
\colhead{Date} & \colhead{C$_{beam}$} & \colhead{C$_{20}$} &
\colhead{C$_{40}$}  & \colhead{C$_{40}^{\ 33\arcsec}$}	\cr
		&\colhead{(Jy beam$^{-1}\ \mu V^{-1}$)}&\colhead{(Jy$ \ \mu V^{-1}$)}&\colhead{(Jy$ \ \mu V^{-1}$)} & \colhead{(Jy$ \ \mu V^{-1}$)} 
}
\startdata

2006 Jun	 &	9.84$\pm$0.68	& 0.35$\pm$0.03	& 0.28$\pm$0.02	& 0.42$\pm$0.03 \cr
2007 Jul	 &	7.01$\pm$0.21		& 0.23$\pm$0.01	& 0.20$\pm$0.01	& 0.29$\pm$0.01 \cr
2007 Oct	 &	8.08$\pm$0.32		& 0.27$\pm$0.01	& 0.23$\pm$0.01	& 0.34$\pm$0.01 \cr
2009 Sep	 &	6.05$\pm$0.25		& 0.21$\pm$0.02	& 0.17$\pm$0.01	& 0.26$\pm$0.02 \cr
2009 Dec	 &	5.89$\pm$0.11		&0.22$\pm$0.01	& 0.17$\pm$0.01  & 0.25$\pm$0.01 \cr
2010 Jul\tablenotemark{a}	 &	7.31$\pm$0.25		& 0.24$\pm$0.01	&0.20$\pm$0.01  & 0.30$\pm$0.01 \cr

2010 Jul 24	&	17.83$\pm$2.50		&	0.67$\pm$0.06	& 0.51$\pm$0.06  & 0.77$\pm$0.09 \cr

2010 Dec	 &	6.69$\pm$0.16		& 0.23$\pm$0.01	& 0.19$\pm$0.01  & 0.28$\pm$0.01 \cr
2011 Dec  &	10.82$\pm$0.39		& 0.38$\pm$0.01	& 0.31$\pm$0.01  & 0.46$\pm$0.02 \cr
2012 Sep	 &	15.59$\pm$0.85		& 0.52$\pm$0.03	& 0.43$\pm$0.02  & 0.65$\pm$0.04

\enddata
\label{tbl:calibrators_average}
\tablenotetext{a}{Does not include calibration from July 24, 2010.}
\end{deluxetable}

\clearpage
{\LongTables
\begin{deluxetable}{lcccccc}
\tabletypesize{\scriptsize}
\tablecaption{Observing Information}
\tablewidth{0pt}
\tablehead{
\colhead{Map} 	&\colhead{Config.\tablenotemark{a}}&\colhead{Size}	&	\multicolumn{2}{c}{Map Center}			&	\colhead{Obs. Date} &\colhead{1 $\sigma$ Noise\tablenotemark{b}}\cr
			&					&				& \colhead{R.A. (J2000)} & \colhead{Dec. (J2000)}  &                                            & \colhead{(mJy beam$^{-1}$)} 
}
\startdata
L359.85+0.00\tablenotemark{c}	&	A	&	$10\arcmin	\times	10\arcmin$	&	17:45:15.356	&	-29:03:48.64	&	2006Jun	&	994	\cr	
L000.00+0.00\tablenotemark{c}	&	A	&	$10\arcmin	\times	10\arcmin$	&	17:45:38.222	&	-28:56:15.83	&	2006Jun	&	1453	\cr	
L000.15+0.00\tablenotemark{c}	&	A	&	$10\arcmin	\times	10\arcmin$	&	17:45:59.055	&	-28:48:38.10	&	2006Jun	&	1438	\cr	
L000.30+0.00\tablenotemark{c}	&	A	&	$10\arcmin	\times	10\arcmin$	&	17:46:20.056	&	-28:40:51.99	&	2006Jun	&	1086	\cr	
L029.95-0.05	&	A	&	$10\arcmin	\times	10\arcmin$	&	18:46:10.078	&	-02:40:45.19	&	2006Jun	&	514	\cr	
L030.00+0.00	&	A	&	$10\arcmin	\times	10\arcmin$	&	18:46:05.764	&	-02:36:40.25	&	2006Jun	&	462	\cr	
L030.15+0.00	&	A	&	$10\arcmin	\times	10\arcmin$	&	18:46:21.926	&	-02:28:41.39	&	2006Jun	&	426	\cr	
L030.30+0.00	&	A	&	$10\arcmin	\times	10\arcmin$	&	18:46:38.373	&	-02:20:39.29	&	2006Jun	&	496	\cr	
L030.45+0.00	&	A	&	$10\arcmin	\times	10\arcmin$	&	18:46:54.711	&	-02:12:38.81	&	2006Jun	&	603	\cr	
L030.60+0.00	&	A	&	$10\arcmin	\times	10\arcmin$	&	18:47:11.158	&	-02:04:35.09	&	2006Jun	&	888	\cr	
L030.70-0.07	&	A	&	$10\arcmin	\times	10\arcmin$	&	18:47:36.576	&	-02:01:32.42	&	2006Jun	&	510	\cr	
L030.80-0.05	&	A	&	$10\arcmin	\times	10\arcmin$	&	18:47:43.784	&	-01:55:09.17	&	2006Jun	&	529	\cr	
L030.88+0.13	&	A	&	$10\arcmin	\times	10\arcmin$	&	18:47:13.872	&	-01:45:57.32	&	2006Jun	&	367	\cr	
L034.26+0.15	&	A	&	$10\arcmin	\times	10\arcmin$	&	18:53:19.759	&	+01:14:36.57	&	2006Jun	&	395	\cr	
L076.16+0.09	&	A	&	$10\arcmin	\times	10\arcmin$	&	20:23:52.185	&	+37:36:41.20	&	2006Jun	&	249	\cr	
L077.93+0.02	&	A	&	$10\arcmin	\times	10\arcmin$	&	20:29:23.675	&	+39:00:58.88	&	2006Jun	&	389	\cr	
L078.14-0.32	&	A	&	$10\arcmin	\times	10\arcmin$	&	20:31:26.817	&	+38:58:53.21	&	2006Jun	&	408	\cr	
L078.96+0.37	&	A	&	$10\arcmin	\times	10\arcmin$	&	20:31:03.924	&	+40:02:54.24	&	2006Jun	&	445	\cr	
L079.28+0.30	&	A	&	$10\arcmin	\times	10\arcmin$	&	20:32:20.042	&	+40:15:45.21	&	2006Jun	&	558	\cr	
L080.92-0.11	&	A	&	$10\arcmin	\times	10\arcmin$	&	20:39:16.364	&	+41:19:41.83	&	2006Jun	&	724	\cr	
L081.45+0.04	&	A	&	$10\arcmin	\times	10\arcmin$	&	20:40:23.200	&	+41:50:33.25	&	2006Jun	&	274	\cr	
L081.68+0.54	&	A	&	$10\arcmin	\times	10\arcmin$	&	20:39:01.659	&	+42:19:37.11	&	2006Jun	&	802	\cr	
L082.55+0.14	&	A	&	$10\arcmin	\times	10\arcmin$	&	20:43:35.922	&	+42:46:07.20	&	2006Jun	&	255	\cr	
L075.76+0.41	&	A	&	$10\arcmin	\times	10\arcmin$	&	20:21:24.076	&	+37:27:55.95	&	2007Jul	&	652	\cr	
L076.12-0.24	&	B	&	$6\arcmin	\times	6\arcmin$	&	20:25:07.547	&	+37:23:14.34	&	2007Jul	&	388	\cr	
L076.35-0.58	&	B	&	$6\arcmin	\times	6\arcmin$	&	20:27:10.982	&	+37:22:29.34	&	2007Jul	&	430	\cr	
L078.92-0.19	&	A	&	$10\arcmin	\times	10\arcmin$	&	20:33:18.191	&	+39:41:24.34	&	2007Jul	&	689	\cr	
L078.17-0.31	&	A	&	$10\arcmin	\times	10\arcmin$	&	20:31:29.785	&	+39:00:20.75	&	2007Jul	&	800	\cr	
L079.62+0.49	&	A	&	$10\arcmin	\times	10\arcmin$	&	20:32:36.698	&	+40:39:16.34	&	2007Jul	&	1489	\cr	
L079.11-0.35	&	A	&	$10\arcmin	\times	10\arcmin$	&	20:34:33.491	&	+39:44:46.95	&	2007Jul	&	1226	\cr	
L080.86+0.38	&	A	&	$10\arcmin	\times	10\arcmin$	&	20:37:00.649	&	+41:34:35.34	&	2007Jul	&	1874	\cr	
L110.11+0.05	&	C	&	$4\arcmin	\times	4\arcmin$	&	23:05:11.371	&	+60:14:41.67	&	2007Oct	&	288	\cr	
L111.62+0.38	&	C	&	$4\arcmin	\times	4\arcmin$	&	23:15:32.196	&	+61:07:30.77	&	2007Oct	&	342	\cr	
L134.28+0.86	&	C	&	$4\arcmin	\times	4\arcmin$	&	02:29:02.834	&	+61:33:28.83	&	2007Oct	&	275	\cr	
L134.83+1.31	&	C	&	$4\arcmin	\times	4\arcmin$	&	02:34:45.364	&	+61:46:15.77	&	2007Oct	&	296	\cr	
L136.38+2.27	&	C	&	$4\arcmin	\times	4\arcmin$	&	02:50:08.515	&	+61:59:54.27	&	2007Oct	&	257	\cr	
L136.83+1.07	&	C	&	$4\arcmin	\times	4\arcmin$	&	02:49:04.352	&	+60:43:23.95	&	2007Oct	&	250	\cr	
L111.28-0.66	&	A	&	$10\arcmin	\times	10\arcmin$	&	23:16:04.681	&	+60:02:06.00	&	2009Sep-2009Dec	&	286	\cr	
L111.54+0.78	&	A	&	$10\arcmin	\times	10\arcmin$	&	23:13:44.302	&	+61:28:10.18	&	2009Sep	&	496	\cr	
L133.71+1.21	&	A	&	$10\arcmin	\times	10\arcmin$	&	02:25:41.066	&	+62:05:42.68	&	2009Sep-2009Dec	&	338	\cr	
L133.95+1.06	&	A	&	$10\arcmin	\times	10\arcmin$	&	02:27:03.912	&	+61:52:14.05	&	2009Sep-2009Dec	&	338	\cr	
L111.26-0.77	&	A	&	$10\arcmin	\times	10\arcmin$	&	23:16:11.144	&	+59:55:27.74	&	2009Dec	&	538	\cr	
L111.78+0.59	&	A	&	$10\arcmin	\times	10\arcmin$	&	23:16:13.500	&	+61:22:51.01	&	2009Dec	&	412	\cr	
L111.79+0.71	&	D,A	&	$10\arcmin	\times	10\arcmin$	&	23:15:52.220	&	+61:30:02.26	&	2009Dec-2012Sep	&	307	\cr	
L111.88+0.82	&	D,A	&	$10\arcmin	\times	10\arcmin$	&	23:16:15.419	&	+61:37:42.69	&	2009Dec	&	473	\cr	
L136.52+1.24	&	D	&	$10\arcmin	\times	10\arcmin$	&	02:47:25.531	&	+61:00:34.20	&	2009Dec	&	260	\cr	
L136.85+1.14	&	E	&	$3\arcmin	\times	3\arcmin$	&	02:49:28.155	&	+60:47:02.96	&	2009Dec	&	337	\cr	
L136.95+1.09	&	D	&	$10\arcmin	\times	10\arcmin$	&	02:50:02.727	&	+60:41:52.32	&	2009Dec	&	475	\cr	
L137.69+1.46	&	H	&	$11\arcmin	\times	11\arcmin$	&	02:56:47.527	&	+60:41:21.90	&	2009Dec	&	667	\cr	
L138.30+1.56	&	D,A	&	$10\arcmin	\times	10\arcmin$	&	03:01:34.013	&	+60:29:10.72	&	2009Dec-2011Dec-2012Sep	&	205	\cr	
L138.48+1.63	&	D,A	&	$10\arcmin	\times	10\arcmin$	&	03:03:09.052	&	+60:27:39.61	&	2009Dec-2011Dec-2012Sep	&	266	\cr	
L173.14+2.36	&	A	&	$10\arcmin	\times	10\arcmin$	&	05:37:57.556	&	+36:00:18.63	&	2009Dec	&	777	\cr	
L173.17+2.35	&	G	&	$4\arcmin	\times	4\arcmin$	&	05:37:59.420	&	+35:58:27.54	&	2009Dec	&	320	\cr	
L173.47+2.43	&	F	&	$2.5\arcmin	\times	2.5\arcmin$	&	05:39:07.589	&	+35:46:02.82	&	2009Dec	&	487	\cr	
L173.57+2.44	&	F	&	$2.5\arcmin	\times	2.5\arcmin$	&	05:39:24.823	&	+35:40:55.71	&	2009Dec	&	421	\cr	
L173.62+2.81	&	D	&	$10\arcmin	\times	10\arcmin$	&	05:41:07.430	&	+35:50:21.17	&	2009Dec	&	359	\cr	
L173.72+2.70	&	G	&	$4\arcmin	\times	4\arcmin$	&	05:40:52.684	&	+35:41:45.31	&	2009Dec	&	387	\cr	
L173.76+2.67	&	G	&	$4\arcmin	\times	4\arcmin$	&	05:40:51.995	&	+35:38:53.79	&	2009Dec	&	301	\cr	
L188.79+1.03	&	A	&	$10\arcmin	\times	10\arcmin$	&	06:09:06.309	&	+21:50:45.79	&	2009Dec	&	259	\cr	
L188.95+0.88	&	D,A	&	$10\arcmin	\times	10\arcmin$	&	06:08:52.987	&	+21:38:19.76	&	2009Dec	&	268	\cr	
L189.03+0.78	&	D,A	&	$10\arcmin	\times	10\arcmin$	&	06:08:39.856	&	+21:31:11.38	&	2009Dec	&	290	\cr	
L189.12+0.64	&	A	&	$10\arcmin	\times	10\arcmin$	&	06:08:20.114	&	+21:22:04.54	&	2009Dec	&	403	\cr	
L189.68+0.19	&	D,A	&	$10\arcmin	\times	10\arcmin$	&	06:07:47.653	&	+20:39:28.06	&	2009Dec	&	155	\cr	
L189.85+0.39	&	I	&	$18\arcmin	\times	18\arcmin$	&	06:08:53.475	&	+20:36:24.00	&	2009Dec	&	505	\cr	
L190.17+0.74	&	A	&	$10\arcmin	\times	10\arcmin$	&	06:10:51.620	&	+20:29:49.63	&	2009Dec	&	184	\cr	
L192.60-0.16	&	D	&	$10\arcmin	\times	10\arcmin$	&	06:12:28.348	&	+17:56:17.12	&	2009Dec	&	340	\cr	
L192.60-0.05	&	D,A	&	$10\arcmin	\times	10\arcmin$	&	06:12:53.279	&	+17:59:28.55	&	2009Dec-2010Dec	&	271	\cr	
L192.72+0.04	&	A	&	$10\arcmin	\times	10\arcmin$	&	06:13:28.147	&	+17:55:38.60	&	2009Dec	&	335	\cr	
L192.81+0.11	&	D	&	$10\arcmin	\times	10\arcmin$	&	06:13:55.135	&	+17:53:09.02	&	2009Dec	&	328	\cr	
L192.98+0.14	&	D	&	$10\arcmin	\times	10\arcmin$	&	06:14:21.502	&	+17:45:02.52	&	2009Dec	&	383	\cr	
L196.42-1.66	&	A	&	$10\arcmin	\times	10\arcmin$	&	06:14:36.606	&	+13:52:04.40	&	2009Dec	&	470	\cr	
L203.23+2.06	&	A	&	$10\arcmin	\times	10\arcmin$	&	06:41:00.779	&	+09:33:56.58	&	2009Dec	&	269	\cr	
L203.35+2.03	&	A	&	$10\arcmin	\times	10\arcmin$	&	06:41:07.741	&	+09:26:52.53	&	2009Dec	&	314	\cr	
L213.71-12.62	&	D	&	$10\arcmin	\times	10\arcmin$	&	06:07:42.663	&	-06:23:27.15	&	2009Dec	&	421	\cr	
L217.37-0.07	&	A	&	$10\arcmin	\times	10\arcmin$	&	06:59:17.374	&	-03:59:14.04	&	2009Dec	&	704	\cr	
L234.57+0.82	&	D	&	$10\arcmin	\times	10\arcmin$	&	07:35:28.482	&	-18:45:34.36	&	2009Dec	&	1051	\cr	
L001.10-0.07	&	A	&	$10\arcmin	\times	10\arcmin$	&	17:48:28.655	&	-28:01:44.77	&	2010Jul	&	849	\cr	
L023.31-0.26	&	A	&	$10\arcmin	\times	10\arcmin$	&	18:34:39.947	&	-08:40:36.21	&	2010Jul	&	379	\cr	
L023.43-0.22	&	A	&	$10\arcmin	\times	10\arcmin$	&	18:34:45.046	&	-08:32:55.27	&	2010Jul	&	375	\cr	
L024.50-0.08	&	A	&	$10\arcmin	\times	10\arcmin$	&	18:36:14.747	&	-07:31:49.77	&	2010Jul	&	306	\cr	
L024.65-0.13	&	A	&	$10\arcmin	\times	10\arcmin$	&	18:36:44.447	&	-07:25:23.71	&	2010Jul	&	357	\cr	
L024.78+0.12	&	A	&	$10\arcmin	\times	10\arcmin$	&	18:36:04.446	&	-07:11:28.75	&	2010Jul	&	363	\cr	
L025.40-0.18	&	A	&	$10\arcmin	\times	10\arcmin$	&	18:38:16.246	&	-06:47:09.20	&	2010Jul	&	482	\cr	
L030.61+0.16	&	A	&	$10\arcmin	\times	10\arcmin$	&	18:46:39.147	&	-01:59:36.24	&	2010Jul	&	349	\cr	
L031.28+0.05	&	A	&	$10\arcmin	\times	10\arcmin$	&	18:48:14.651	&	-01:26:51.80	&	2010Jul	&	303	\cr	
L081.11-0.16	&	A	&	$10\arcmin	\times	10\arcmin$	&	20:40:05.498	&	+41:26:56.24	&	2010Jul	&	341	\cr	
L081.28+1.01	&	A	&	$10\arcmin	\times	10\arcmin$	&	20:35:39.600	&	+42:17:38.79	&	2010Jul	&	250	\cr	
L081.39+0.73	&	A	&	$10\arcmin	\times	10\arcmin$	&	20:37:12.963	&	+42:12:38.77	&	2010Jul	&	341	\cr	
L081.48+0.00	&	A	&	$10\arcmin	\times	10\arcmin$	&	20:40:37.896	&	+41:50:26.77	&	2010Jul	&	353	\cr	
L081.76+0.60	&	A	&	$10\arcmin	\times	10\arcmin$	&	20:39:01.567	&	+42:25:36.73	&	2010Jul	&	329	\cr	
L031.41+0.31	&	A	&	$10\arcmin	\times	10\arcmin$	&	18:47:34.152	&	-01:12:46.30	&	2010Jul24	&	791	\cr	
L081.88+0.77	&	A	&	$10\arcmin	\times	10\arcmin$	&	20:38:39.559	&	+42:37:37.68	&	2010Jul24	&	717	\cr	
L183.40-0.58	&	A	&	$10\arcmin	\times	10\arcmin$	&	05:51:16.720	&	+25:43:32.50	&	2010Dec	&	255	\cr	
L189.78+0.33	&	A	&	$10\arcmin	\times	10\arcmin$	&	06:08:32.990	&	+20:38:33.34	&	2010Dec	&	294	\cr	
L202.58+2.42	&	A	&	$10\arcmin	\times	10\arcmin$	&	06:41:06.051	&	+10:18:59.80	&	2010Dec	&	370	\cr	
L142.01+1.77	&	A	&	$10\arcmin	\times	10\arcmin$	&	03:27:29.910	&	+58:43:55.31	&	2011Dec	&	305	\cr	
L151.61-0.24	&	A	&	$10\arcmin	\times	10\arcmin$	&	04:11:07.205	&	+51:09:21.54	&	2011Dec	&	266	\cr	
L154.37+2.58	&	A	&	$10\arcmin	\times	10\arcmin$	&	04:36:18.153	&	+51:11:02.11	&	2011Dec	&	266	\cr	
L169.18-0.89	&	A	&	$10\arcmin	\times	10\arcmin$	&	05:13:26.754	&	+37:27:38.75	&	2011Dec	&	232	\cr	
L172.88+2.27	&	A	&	$10\arcmin	\times	10\arcmin$	&	05:36:53.085	&	+36:10:29.56	&	2011Dec	&	315	\cr	
L211.53-19.27	&	A	&	$10\arcmin	\times	10\arcmin$	&	05:39:57.873	&	-07:27:48.36	&	2011Dec	&	254	\cr	
L111.42+0.76	&	A	&	$10\arcmin	\times	10\arcmin$	&	23:12:52.873	&	+61:24:31.27	&	2012Sep	&	375	\cr	
L111.88+0.99	&	A	&	$10\arcmin	\times	10\arcmin$	&	23:15:45.489	&	+61:47:37.40	&	2012Sep	&	226	\cr	
L134.20+0.75	&	A	&	$10\arcmin	\times	10\arcmin$	&	02:28:05.671	&	+61:29:25.54	&	2012Sep	&	525	\cr	
L189.85+0.50	&	A	&	$10\arcmin	\times	10\arcmin$	&	06:09:19.407	&	+20:39:39.18	&	2012Sep	&	506	\cr	
L189.94+0.34	&	A	&	$10\arcmin	\times	10\arcmin$	&	06:08:53.761	&	+20:30:07.72	&	2012Sep	&	590	\cr	
L206.60-16.37	&	A	&	$10\arcmin	\times	10\arcmin$	&	05:41:47.488	&	-01:57:59.97	&	2012Sep	&	563	

\enddata
\tablenotetext{a}{A: BOX SCAN  571.429 600 40.0 45. B: BOX SCAN  345.600 360 30.0 45. C: BOX SCAN  228.571 240 20.0 45. D: BOX SCAN 606.1 636.4 60.0 45. E: BOX SCAN 169.7 167.5 60.0 45. F: BOX SCAN 144.6 148.0 60.0 45. G: BOX SCAN 144.6 148.0 60.0 45. H: BOX SCAN 707.1 669.9 60.0 45. I: BOX SCAN 1060.7 1157.1 60.0 45. The configuration BOX SCAN X Y R A indicates that the map has a size of X(arcsec)$\times$Y(arcsec), a scan rate R (arcsec/sec), and a scanning angle A (deg).}
\tablenotetext{b}{Representative noise is estimated on the central area of each map, avoiding edges (see Section~\ref{sec:map_description}).}
\tablenotetext{c}{Additional SHARC-II 350 \um\ continuum data on the Galactic center were presented by~\cite{bal10}. }
\label{tbl:observation_information}
\end{deluxetable}
}

\clearpage
\begin{landscape}
\clearpage
\begin{deluxetable}{llcccccccccccc}
\tabletypesize{\scriptsize}
\tablecaption{Properties of sources recovered in the 350 $\mu$m maps}
\tablewidth{0pt}
\tablehead{
\colhead{No.}&\colhead{Name}&\colhead{$l_{max}$}&\colhead{$b_{max}$}&\colhead{$l$}&\colhead{$b$}&\colhead{$\sigma_{maj}$}&\colhead{$\sigma_{min}$}&\colhead{P.A.}&\colhead{$\Theta_R$}&\colhead{$S_{20}$}&\colhead{$S_{40}$}&\colhead{$S$}&\colhead{Type}\cr
	&	&\colhead{(\arcdeg)}&\colhead{(\arcdeg)}&\colhead{(\arcdeg)}&\colhead{(\arcdeg)}&\colhead{(\arcsec)}&\colhead{(\arcsec)}&\colhead{(\arcdeg)}&\colhead{(\arcsec)}&\colhead{(Jy)}&\colhead{(Jy)}&\colhead{(Jy)}&\cr
\colhead{(1)}&\colhead{(2)}&\colhead{(3)}&\colhead{(4)}&\colhead{(5)}&\colhead{(6)}&\colhead{(7)}&\colhead{(8)}&\colhead{(9)}&\colhead{(10)}&\colhead{(11)}&\colhead{(12)}&\colhead{(13)}&\colhead{(14)}
 }
\startdata
1&SHARC\_G000.0002-00.0200&0.0002&-0.0200&359.9995&-0.0175&8&6&36&15&42.91$\pm$4.51&118.56$\pm$11.11&100.41$\pm$6.94&F \cr
2&SHARC\_G000.0006-00.0246&0.0006&-0.0246&359.9994&-0.0253&7&6&64&13&55.84$\pm$5.86&122.38$\pm$11.47&96.41$\pm$6.66 &C \cr
3&SHARC\_G000.0008-00.0304&0.0008&-0.0304&0.0002&-0.0305&5&4&102&7&26.85$\pm$2.82&55.23$\pm$5.18&26.29$\pm$1.82 &F \cr
4&SHARC\_G000.0013-00.0784&0.0013&-0.0784&0.0008&-0.0780&5&4&142&7&13.62$\pm$1.43&23.03$\pm$2.17&12.71$\pm$0.89 &F \cr
5&SHARC\_G000.0065-00.0557&0.0065&-0.0557&0.0058&-0.0575&13&8&153&22&49.25$\pm$5.17&104.71$\pm$9.81&146.57$\pm$10.13 &C \cr
6&SHARC\_G000.0067-00.0146&0.0067&-0.0146&0.0052&-0.0139&6&5&104&9&30.57$\pm$3.21&89.52$\pm$8.39&37.29$\pm$2.58 &F \cr
7&SHARC\_G000.0075-00.0199&0.0075&-0.0199&0.0058&-0.0208&11&6&67&17&68.11$\pm$7.15&162.68$\pm$15.24&163.46$\pm$11.30 &C \cr
8&SHARC\_G000.0089+00.0034&0.0089&0.0034&0.0092&0.0035&5&3&17&\nodata&7.53$\pm$0.80&13.34$\pm$1.26&5.00$\pm$0.35 &F \cr
9&SHARC\_G000.0120-00.0201&0.0120&-0.0201&0.0129&-0.0182&12&8&47&21&80.66$\pm$8.47&188.93$\pm$17.70&237.90$\pm$16.44 &C \cr
10&SHARC\_G000.0121-00.0513&0.0121&-0.0513&0.0113&-0.0511&9&7&34&17&60.03$\pm$6.30&121.66$\pm$11.40&122.43$\pm$8.46 &C \cr
11&SHARC\_G000.0130+00.0064&0.0130&0.0064&0.0127&0.0065&9&4&164&10&15.16$\pm$1.60&34.25$\pm$3.21&21.75$\pm$1.51 &F \cr
12&SHARC\_G000.0135-00.0217&0.0135&-0.0217&0.0156&-0.0233&7&5&165&10&78.04$\pm$8.19&161.42$\pm$15.12&102.54$\pm$7.09 &C \cr
13&SHARC\_G000.0142+00.0364&0.0142&0.0364&0.0148&0.0359&10&8&111&19&26.86$\pm$2.82&48.54$\pm$4.55&55.72$\pm$3.86 &F \cr
14&SHARC\_G000.0145+00.0127&0.0145&0.0127&0.0150&0.0125&3&3&169&\nodata&5.00$\pm$0.54&7.71$\pm$0.74&3.05$\pm$0.22 &F \cr
15&SHARC\_G000.0188+00.0040&0.0188&0.0040&0.0186&0.0072&7&5&20&11&26.13$\pm$2.75&59.59$\pm$5.58&39.78$\pm$2.75 &F \cr
16&SHARC\_G000.0195-00.0067&0.0195&-0.0067&0.0184&-0.0077&12&6&62&19&32.01$\pm$3.36&56.13$\pm$5.26&63.15$\pm$4.37 &F \cr
17&SHARC\_G000.0196-00.0501&0.0196&-0.0501&0.0199&-0.0511&12&10&30&24&73.00$\pm$7.66&143.62$\pm$13.46&208.13$\pm$14.38 &C \cr
18&SHARC\_G000.0197+00.0017&0.0197&0.0017&0.0197&0.0016&6&5&165&10&27.63$\pm$2.90&58.82$\pm$5.51&34.29$\pm$2.37 &F\cr
19&SHARC\_G000.0238+00.0359&0.0238&0.0359&0.0237&0.0357&11&6&78&16&15.41$\pm$1.62&33.58$\pm$3.15&33.12$\pm$2.30 &F\cr
20&SHARC\_G000.0247+00.0032&0.0247&0.0032&0.0252&0.0024&7&5&166&11&19.90$\pm$2.09&39.84$\pm$3.74&26.18$\pm$1.81 &F\cr
21&SHARC\_G000.0279-00.0533&0.0279&-0.0533&0.0279&-0.0538&7&6&14&13&29.31$\pm$3.08&79.58$\pm$7.46&55.47$\pm$3.84 &F\cr
22&SHARC\_G000.0281-00.0576&0.0281&-0.0576&0.0278&-0.0587&8&4&87&10&23.22$\pm$2.44&55.01$\pm$5.16&32.71$\pm$2.26 &F\cr
23&SHARC\_G000.0308+00.0211&0.0308&0.0211&0.0303&0.0218&11&7&18&18&35.87$\pm$3.77&65.86$\pm$6.17&71.75$\pm$4.96 &F\cr
24&SHARC\_G000.0315-00.0512&0.0315&-0.0512&0.0304&-0.0507&5&4&35&7&26.45$\pm$2.78&65.31$\pm$6.12&26.94$\pm$1.86 &F\cr
25&SHARC\_G000.0331+00.0054&0.0331&0.0054&0.0337&0.0061&8&4&90&8&15.37$\pm$1.62&15.11$\pm$1.43&16.67$\pm$1.16 &F

\enddata
\tablecomments{(1) Running source number. (2) Name derived from Galactic coordinates of the maximum intensity in the object. (3), (4) Galactic coordinates of maximum
intensity in the catalog object. (5), (6) Galactic coordinates of emission centroid. (7)$-$(9) Major and minor axis 1/e widths and position angle of source. (10) Deconvolved angular size of source. (11), (12) Flux densities derived for 20 and 40 apertures. (13) Integrated flux density in the object. (14) Type of substructure: ``C$=$"compact or ``F"$=$faint, as defined in section~\ref{sec:correlation}.\\
(This table is available in its entirety in a machine-readable form in the online journal. A portion is shown here for guidance regarding its form and content.)}
\label{tbl:sources_detection}
\end{deluxetable}
\clearpage
\end{landscape}

\clearpage

\begin{deluxetable}{lcccc}
\tabletypesize{\scriptsize}
\tablecaption{Sources from BGPS catalog with large number of high-resolution associated substructures.}
\tablewidth{0pt}
\tablehead{
\colhead{BGPS V2.1 name}&\multicolumn{4}{c}{Peak emission of 350 \um\ substructures}\cr
\cline{2-5}
&\colhead{Total\tablenotemark{a}}&\colhead{$>6\sigma_{rms}$}&\colhead{$>10\sigma_{rms}$}&\colhead{$>20\sigma_{rms}$}
}
\startdata

BGPSv2\_G213.705-12.603 &	34 &	24 &	15 &	12	\cr
BGPSv2\_G034.256+00.154 &	27 &	18 &	15 &	8	\cr
BGPSv2\_G133.716+01.220 &	25 &	15 &	9 &	4	\cr
BGPSv2\_G000.014-00.017 &	23 &	9 &	5 &	1	\cr
BGPSv2\_G029.916-00.045 &	22 &	18 &	8 &	3	\cr
BGPSv2\_G029.958-00.017 &	19 &	13 &	6 &	4	\cr
BGPSv2\_G359.867-00.083 &	16 &	12 &	8 &	4	\cr
BGPSv2\_G081.477+00.020 &	16 &	11 &	6 &	0	\cr
BGPSv2\_G359.982-00.069 &	14 &	12 &	7 &	1	\cr
BGPSv2\_G359.946-00.045 &	14 &	5 &	1 &	0	\cr
BGPSv2\_G030.751-00.051 &	14 &	8 &	7 &	5	\cr
BGPSv2\_G081.753+00.593 &	13 &	9 &	7 &	5	\cr
BGPSv2\_G076.359-00.600 &	13 &	11 &	7 &	1	\cr
BGPSv2\_G030.786-00.025 &	13 &	12 &	11 &	3	\cr
BGPSv2\_G203.223+02.076 &	12 &	8 &	1 &	0	\cr
BGPSv2\_G192.598-00.049 &	12 &	7 &	5 &	1	\cr
BGPSv2\_G203.320+02.058 &	11 &	7 &	6 &	4	\cr
BGPSv2\_G023.437-00.183 &	11 &	5 &	2 &	1	\cr
BGPSv2\_G023.273-00.211 &	11 &	7 &	1 &	0	\cr
BGPSv2\_G081.721+00.573 &	10 &	9 &	8 &	7	\cr
BGPSv2\_G030.690-00.043 &	10 &	4 &	3 &	0	\cr
BGPSv2\_G024.493-00.039 &	10 &	7 &	6 &	3	

\enddata

\tablenotetext{a}{Total number of associated substructures (compact and faint) found on SHARC-II maps per each BGPS parental clump. See Section~\ref{sec:correlation} for details.}

\label{tbl:large_frag}
\end{deluxetable}

{\tabletypesize{\normalsize}
\renewcommand{\arraystretch}{1.6}
\begin{deluxetable}{lcc|rl|rl|rl}
\tabletypesize{\normalsize}
\tablecaption{BGPS V2.1 sources fluxes and fitted dust temperature.}
\tablewidth{0pt}
\tablehead{
\colhead{BGPS source}&\colhead{1.1 mm}&\colhead{Convolved 350 $\mu$m\tablenotemark{a}}&\multicolumn{6}{c}{Temperature}\cr
	&\colhead{Flux 40$\arcsec$}&\colhead{Flux 40$\arcsec$}&\multicolumn{2}{c}{$\beta=1.0$}&\multicolumn{2}{c}{$\beta=1.7$}&\multicolumn{2}{c}{$\beta=2.0$}\cr
	&\colhead{(Jy)}&\colhead{(Jy)}&\multicolumn{2}{c}{(K)}&\multicolumn{2}{c}{(K)}&\multicolumn{2}{c}{(K)}\	
}
\startdata
BGPSv2\_G024.743+00.179	&	0.51$\pm$0.11	&	17.82$\pm$0.64	&	$>$1000.0	&	$^{>1000.0}_{103.1}$	&	23.5	&	$^{31.1}_{18.3}$	&	16.7	&	$^{19.9}_{14.1}$	\cr
BGPSv2\_G024.745+00.161	&	0.60$\pm$0.12	&	18.42$\pm$0.66	&	351.7	&	$^{>1000.0}_{58.2}$	&	20.2	&	$^{24.8}_{16.6}$	&	15.1	&	$^{17.3}_{13.1}$	\cr
BGPSv2\_G024.757+00.091	&	1.36$\pm$0.12	&	61.75$\pm$2.19	&	$>$1000.0	&	$^{>1000.0}_{>1000.0}$	&	34.9	&	$^{42.9}_{29.2}$	&	21.2	&	$^{23.6}_{19.1}$	\cr
BGPSv2\_G024.759+00.065	&	0.43$\pm$0.09	&	8.47$\pm$0.38	&	33.3	&	$^{52.7}_{23.4}$	&	14.2	&	$^{16.2}_{12.3}$	&	11.6	&	$^{12.9}_{10.4}$	\cr
BGPSv2\_G024.760+00.163	&	0.34$\pm$0.13	&	4.80$\pm$0.18	&	20.8	&	$^{31.9}_{14.2}$	&	11.7	&	$^{14.0}_{9.5}$	&	10.0	&	$^{11.5}_{8.4}$\cr
BGPSv2\_G024.773+00.125	&	0.22$\pm$0.09	&	1.67$\pm$0.07	&	12.9	&	$^{16.2}_{9.9}$	&	8.9	&	$^{10.3}_{7.5}$	&	8.0	&	$^{8.9}_{6.8}$	\cr
BGPSv2\_G024.791+00.083	&	10.07$\pm$0.64	&	556.65$\pm$19.73	&	$>$1000.0	&	$^{>1000.0}_{>1000.0}$	&	58.6	&	$^{78.9}_{46.3}$	&	27.1	&	$^{30.2}_{24.4}$	\cr
BGPSv2\_G024.795+00.101	&	3.14$\pm$0.22	&	142.57$\pm$4.87	&	$>$1000.0	&	$^{>1000.0}_{>1000.0}$	&	34.8	&	$^{41.0}_{30.1}$	&	21.2	&	$^{23.0}_{19.5}$	\cr
BGPSv2\_G024.807+00.039	&	0.32$\pm$0.09	&	5.84$\pm$0.23	&	29.1	&	$^{49.6}_{19.4}$	&	13.5	&	$^{16.0}_{11.3}$	&	11.2	&	$^{12.7}_{9.7}$	\cr
BGPSv2\_G024.815+00.189	&	0.10$\pm$0.09	&	-0.75$\pm$0.05	&	\nodata	&	$^{--}_{--}$	&	\nodata	&	$^{--}_{--}$	&	\nodata	&	$^{--}_{--}$	\cr
BGPSv2\_G024.817+00.129	&	0.48$\pm$0.09	&	6.69$\pm$0.27	&	20.7	&	$^{25.5}_{17.0}$	&	11.6	&	$^{12.8}_{10.5}$	&	9.9	&	$^{10.7}_{9.1}$	\cr
BGPSv2\_G024.824+00.181	&	0.17$\pm$0.09	&	1.89$\pm$0.07	&	16.5	&	$^{25.0}_{10.7}$	&	10.3	&	$^{12.7}_{7.9}$	&	9.0	&	$^{10.7}_{7.2}$\cr
BGPSv2\_G024.850+00.085	&	0.90$\pm$0.11	&	36.60$\pm$1.32	&	$>$1000.0	&	$^{>1000.0}_{>1000.0}$	&	28.9	&	$^{35.9}_{23.7}$	&	19.0	&	$^{21.5}_{16.8}$	\cr
BGPSv2\_G024.863+00.145	&	0.16$\pm$0.09	&	1.08$\pm$0.04	&	12.2	&	$^{16.4}_{8.4}$	&	8.6	&	$^{10.3}_{6.7}$	&	7.7	&	$^{9.0}_{6.1}$	\cr
BGPSv2\_G025.329-00.196	&	0.56$\pm$0.12	&	3.95$\pm$0.16	&	12.4	&	$^{13.9}_{10.9}$	&	8.7	&	$^{9.4}_{8.0}$	&	7.8	&	$^{8.3}_{7.2}$	\cr
BGPSv2\_G025.339-00.170	&	0.19$\pm$0.12	&	0.76$\pm$0.04	&	9.5	&	$^{11.7}_{7.0}$	&	7.3	&	$^{8.4}_{5.8}$	&	6.6	&	$^{7.5}_{5.4}$	\cr
BGPSv2\_G025.355-00.156	&	0.19$\pm$0.13	&	1.49$\pm$0.08	&	13.2	&	$^{19.5}_{8.0}$	&	9.1	&	$^{11.3}_{6.4}$	&	8.1	&	$^{9.7}_{5.9}$	\cr
BGPSv2\_G025.355-00.190	&	1.46$\pm$0.15	&	40.94$\pm$1.53	&	116.1	&	$^{738.0}_{61.5}$	&	18.6	&	$^{20.6}_{16.8}$	&	14.2	&	$^{15.3}_{13.2}$	\cr
BGPSv2\_G025.382-00.182	&	3.23$\pm$0.24	&	115.44$\pm$4.01	&	$>$1000.0	&	$^{>1000.0}_{>1000.0}$	&	24.0	&	$^{26.7}_{21.8}$	&	17.0	&	$^{18.1}_{15.9}$	\cr
BGPSv2\_G025.399-00.140	&	4.56$\pm$0.31	&	152.14$\pm$5.34	&	$>$1000.0	&	$^{>1000.0}_{405.7}$	&	22.1	&	$^{24.1}_{20.3}$	&	16.1	&	$^{17.0}_{15.1}$	\cr
BGPSv2\_G025.405-00.256	&	0.31$\pm$0.15	&	3.14$\pm$0.11	&	15.8	&	$^{23.1}_{10.6}$	&	10.1	&	$^{12.3}_{7.9}$	&	8.8	&	$^{10.4}_{7.1}$	\cr
BGPSv2\_G025.413-00.176	&	0.65$\pm$0.12	&	17.80$\pm$0.64	&	98.2	&	$^{>1000.0}_{43.3}$	&	18.2	&	$^{21.6}_{15.4}$	&	14.0	&	$^{15.8}_{12.4}$	\cr
BGPSv2\_G025.455-00.210	&	1.64$\pm$0.17	&	75.41$\pm$2.61	&	$>$1000.0	&	$^{>1000.0}_{>1000.0}$	&	35.9	&	$^{46.0}_{29.1}$	&	21.5	&	$^{24.4}_{19.1}$	\cr
BGPSv2\_G025.467-00.126	&	0.12$\pm$0.12	&	0.33$\pm$0.06	&	8.4	&	$^{11.2}_{--}$	&	6.7	&	$^{8.2}_{--}$	&	6.1	&	$^{7.4}_{--}$	\cr
BGPSv2\_G025.477-00.136	&	0.10$\pm$0.12	&	1.73$\pm$0.06	&	27.0	&	$^{>1000.0}_{--}$	&	13.1	&	$^{25.9}_{--}$	&	10.9	&	$^{17.8}_{--}$

\enddata
\tablenotetext{a}{Fluxes were obtained from SHARC-II maps convolved to match a 33\arcsec\ beam size (see Section~\ref{sec:flux_bgps_350}).}
\tablecomments{(This table is available in its entirety in a machine-readable form in the online journal. A portion is shown here for guidance regarding its form and content.)}
\label{tbl:bgps_sources_temperatures}
\end{deluxetable}
}

\clearpage

\newpage

\newpage
\newpage

\newpage

\newpage

\begin{figure}[h] 
   \centering
    \includegraphics[width=0.6\textwidth]{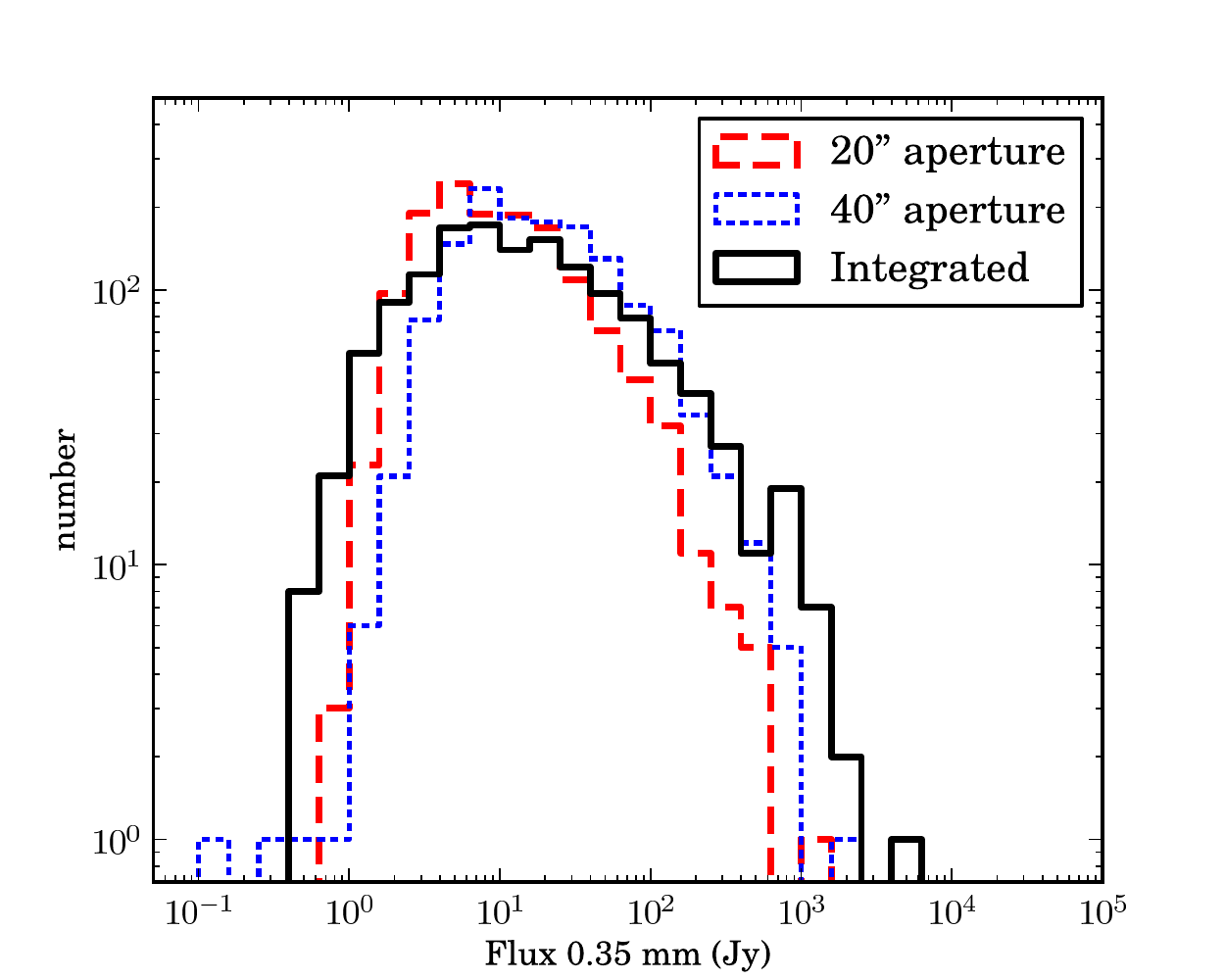}
   \caption{Flux distribution for objects identified in the 107 SHARC-II maps. Three methods of flux recovery are represented in the image: aperture photometry at 20\arcsec (red long dash) and 40\arcsec (blue short dash) in diameter toward the peak, and integrated emission of the source (solid black line). }
  \label{fig:hist_sharc_flux}
\end{figure}

\begin{figure}[h] 
   \centering
  \includegraphics[width=0.46\textwidth]{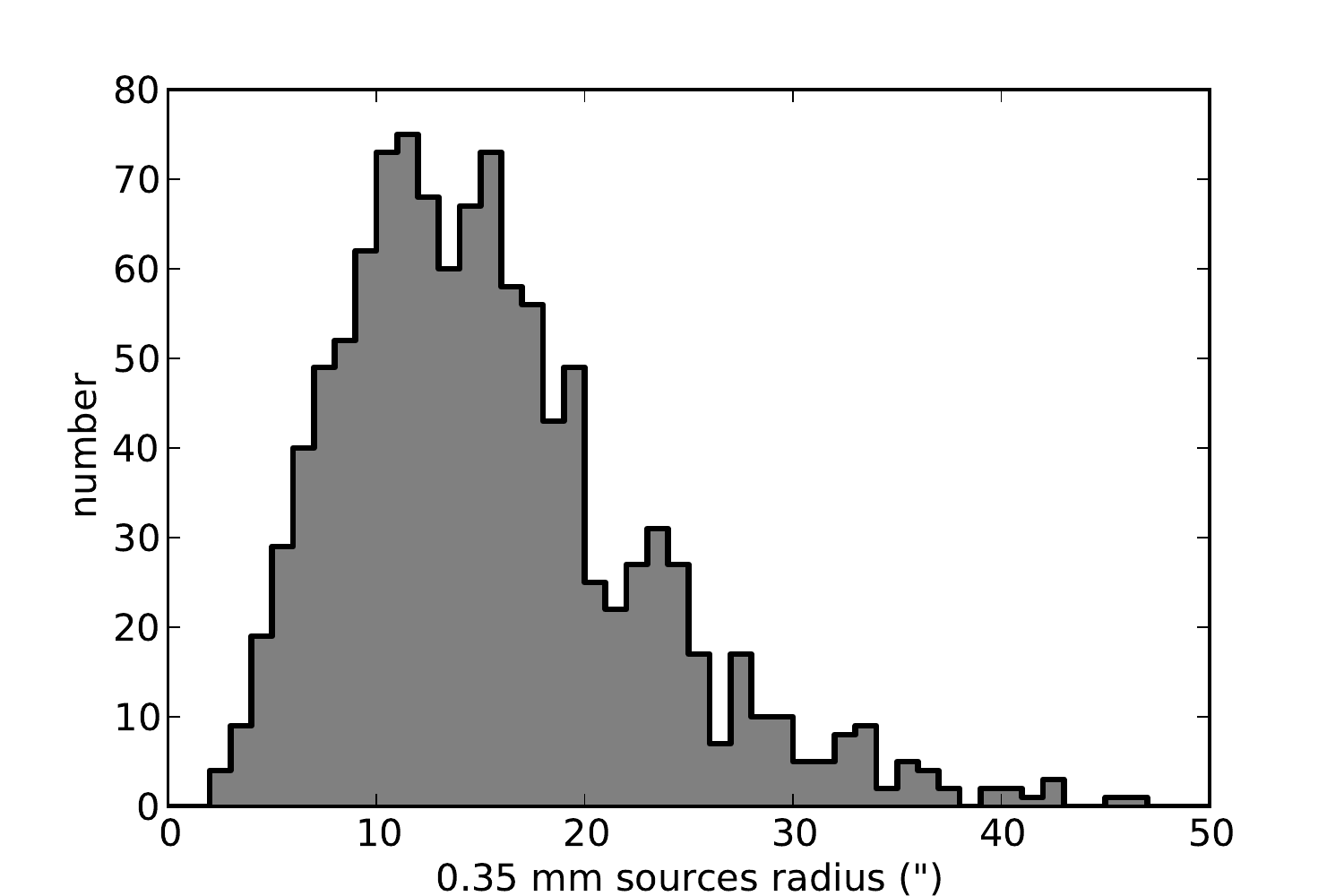}
  \includegraphics[width=0.46\textwidth]{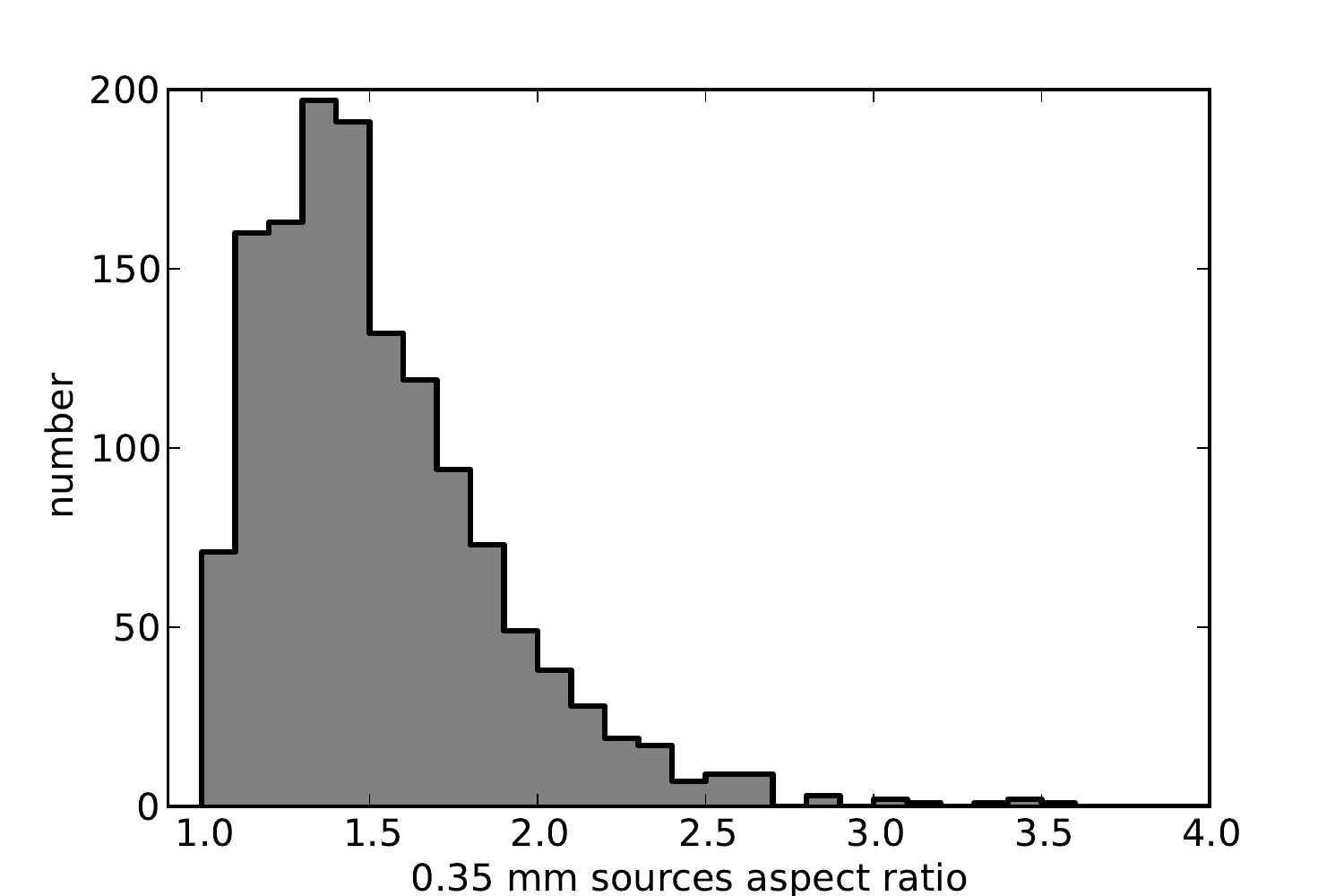}
   \caption{Distribution of the deconvolved radii and aspect ratio of sources in the SHARC-II maps. The average and median values of the radius distribution are 15\arcsec\ and 14\arcsec, respectively. The average and median values of the aspect ratios are respectively 1.53 and 1.45.}
  \label{fig:hist_sharc_radius}
\end{figure}
\newpage

\begin{figure}[h] 
   \centering
  \includegraphics[width=1\textwidth]{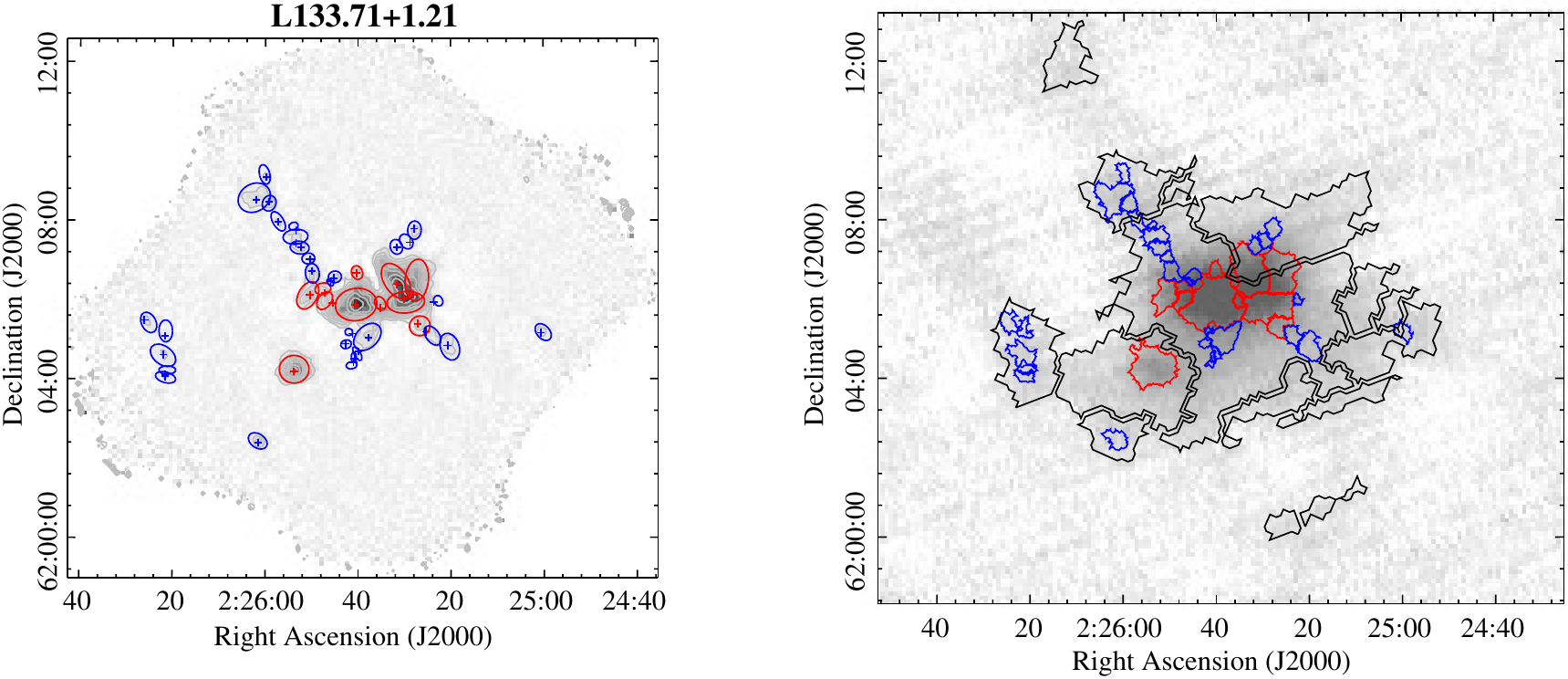}\\
\vspace{4 mm}  
  \includegraphics[width=1\textwidth]{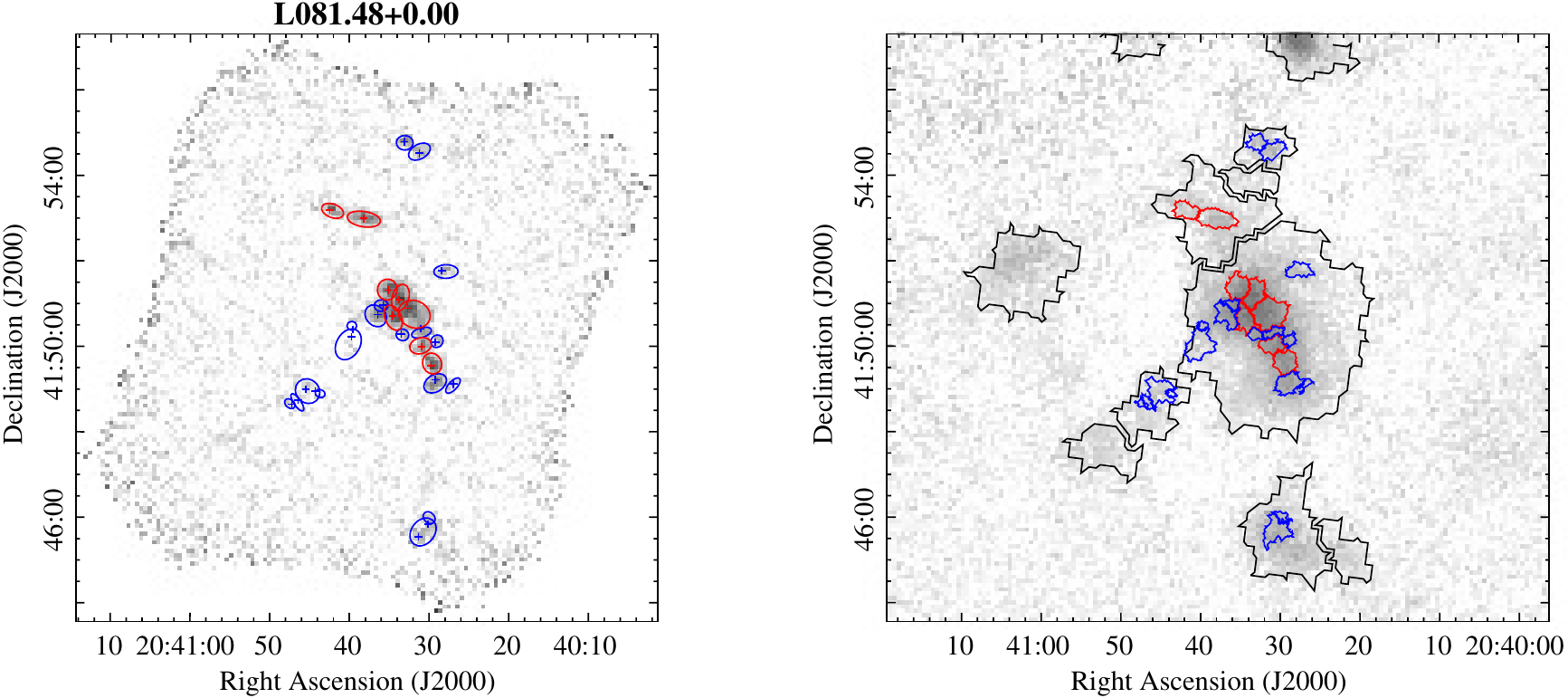}
   \caption{Upper left: Results of Bolocat on the 350 \um\ map of L133.71+1.21, corresponding to the W3 Main region. The different colors indicate the peak signal-to-noise of each source (above $10\sigma$ in red, below that limit in blue). Sources in red are associated with compact and dense core-type structures, the blue sources are related to low emission, field material or possible filamentary structure. Upper right: Grey color scale of the 1.1 mm toward the same region. Black regions show the result of \bolocat\ on the 1.1 mm map, and blue and red regions the results of \bolocat\ on the 350 \um\ map. The central V2.1 source mask shows large number of substructures (both strong and faint). Lower panels are similar to the upper panels for L081.$48+0.00$, in the Cygnus X region. The comparison of source masks reveal a large population of BGPS V2.1 sources located inside the 350 \um\ maps, but not associated with any source at high-resolution. Details on the correlation between 350 \um\ sources and their parental BGPS clump are described in Section~\ref{sec:correlation}.}
  \label{fig:sharc_bolo_regions}
\end{figure}
\newpage

\begin{figure}[h] 
   \centering
  \includegraphics[width=0.47\textwidth]{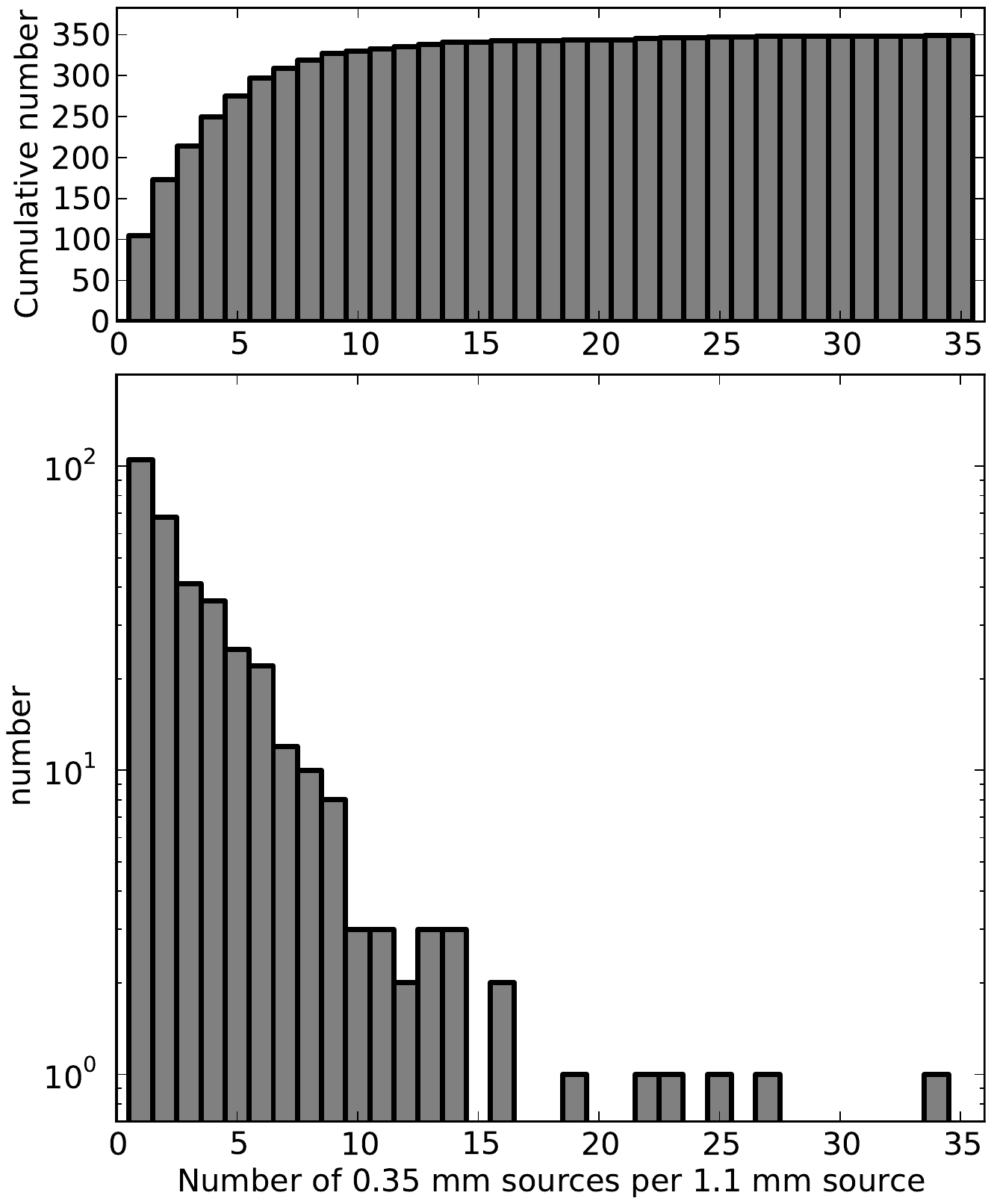}
   \caption{Distribution of 350 \um\ sources per 1.1 mm source from BGPS. The top panel shows the cumulative number. Half of the 1.1 mm parental clumps are associated with one-or-two substructures identified at high-resolution.}
  \label{fig:hist_match}
\end{figure}
\newpage

\begin{figure}[h] 
   \centering
  \includegraphics[width=0.5\textwidth]{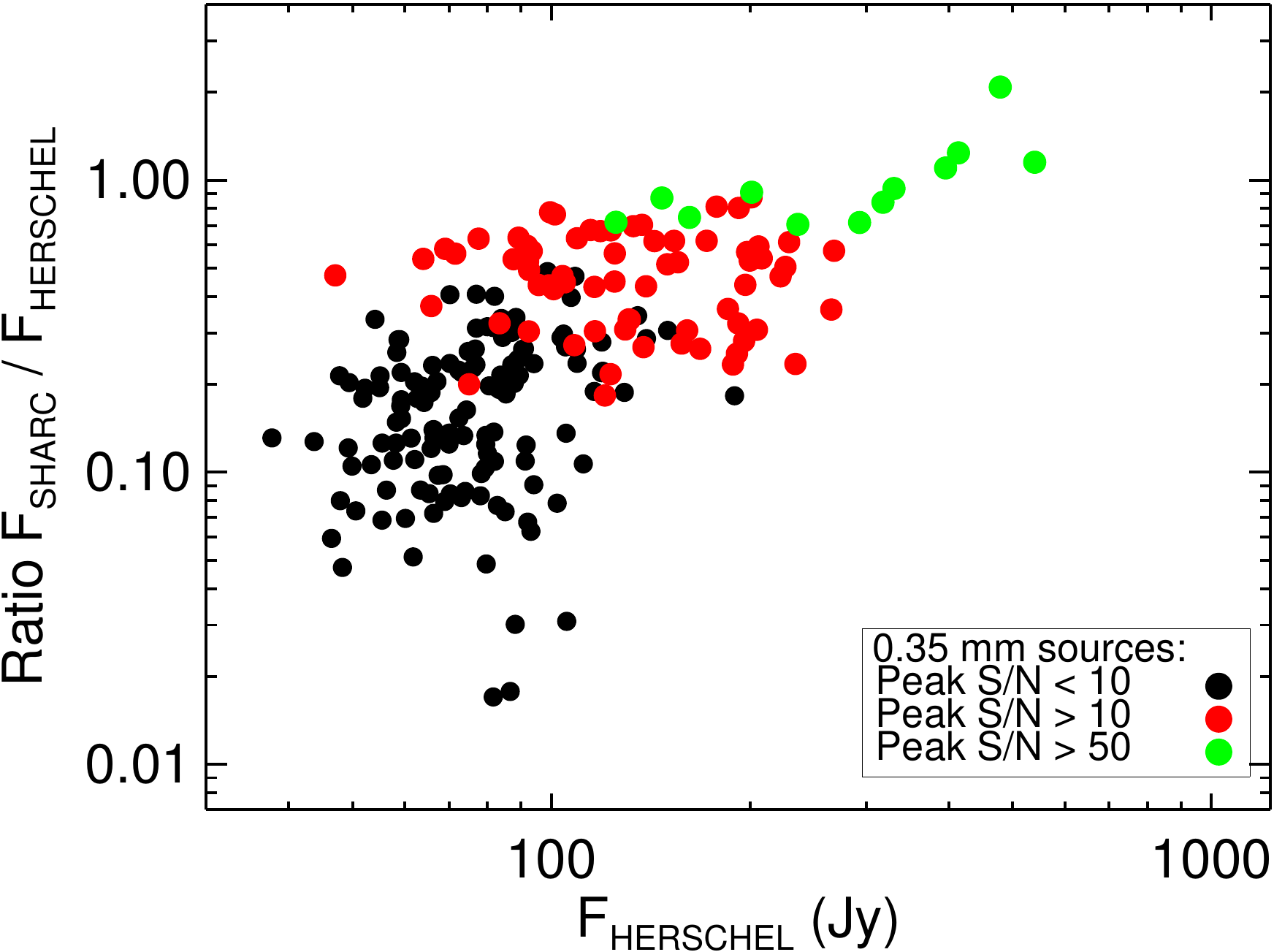}\\
  \vspace{5mm}
  
    \includegraphics[width=0.5\textwidth]{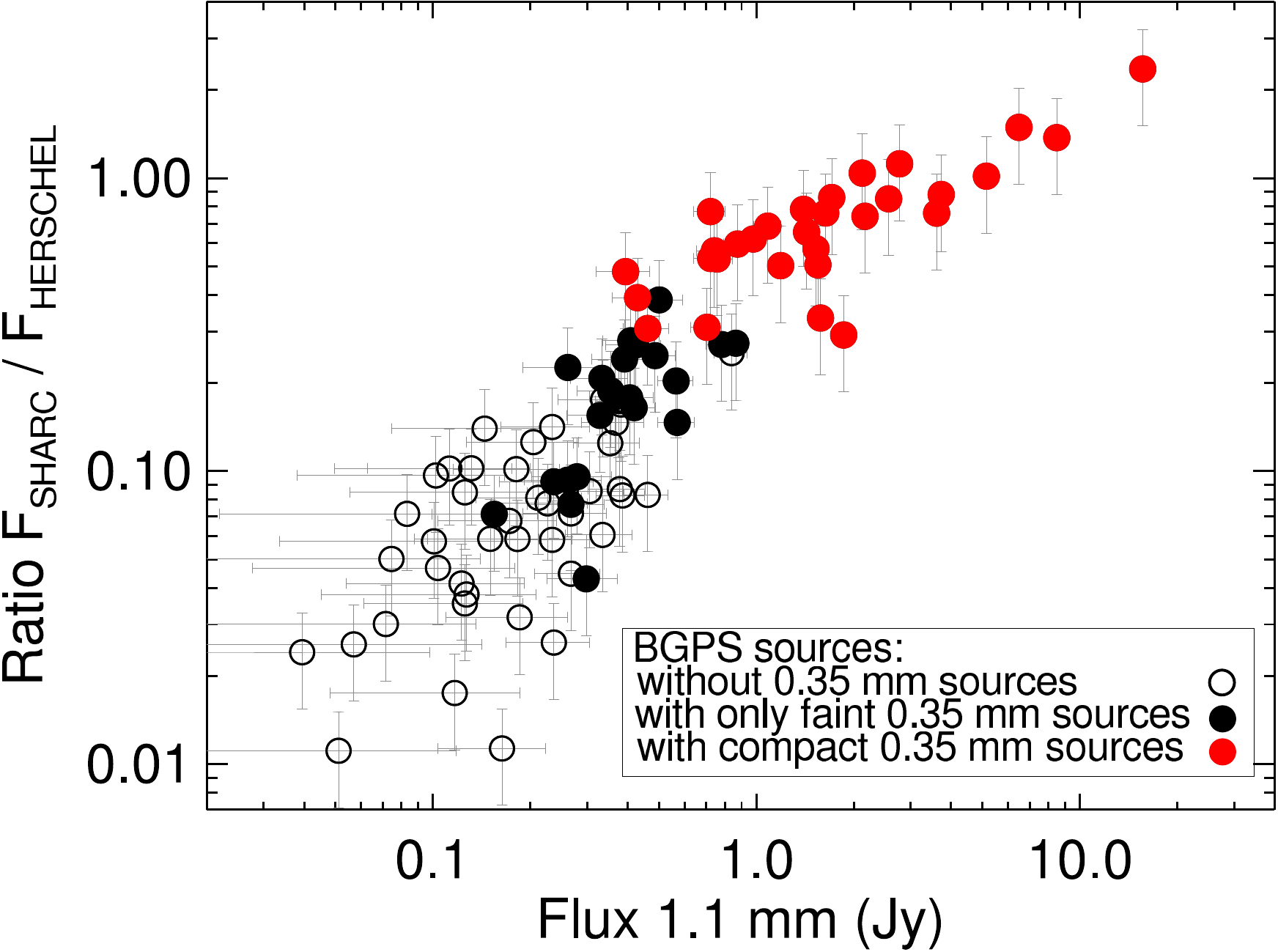}
   \caption{Upper: Comparison between integrated flux in 40\arcsec\ apertures obtained at 350 \um\ on SHARC-II and \textit{Herschel} images toward $l=30\arcdeg$. The points represent the 213 high-resolution sources recovered by \bolocat\ in SHARC-II maps, and the colors show different limits of peak signal-to-noise on those sources. Bottom: Comparison between integrated flux in 40\arcsec\ apertures for 350 \um\ images from SHARC-II and  \textit{Herschel} image on a sample of 102 BGPS sources located toward $l=30\arcdeg$. Both SHARC-II and \textit{Herschel} were convolved to match the resolution of the Bolocam maps (33\arcsec). Unfilled circles represent BGPS sources with no associated source detected at 350 \um, and filled circles show the 1.1 mm parental clumps with only faint objects (black) and with compact substructures (red). The error bars in the flux ratio considered 20\% uncertainties in $F_{Herschel}$, and 30\% in $F_{SHARC}$.}
     \label{fig:fluxrec_herschel_sharc}
\end{figure}
\newpage

\begin{figure}[h] 
   \centering
  \includegraphics[width=1\textwidth]{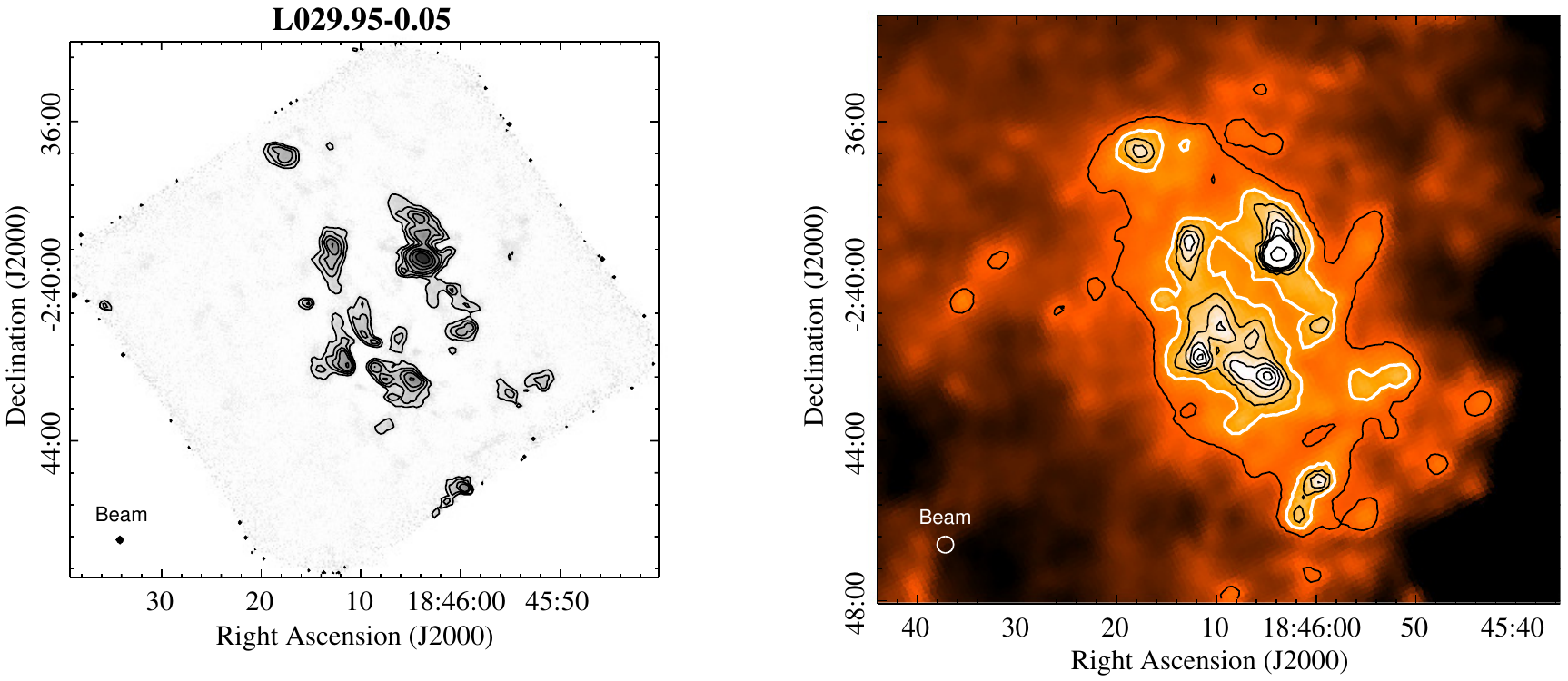}
    \includegraphics[width=0.42\textwidth]{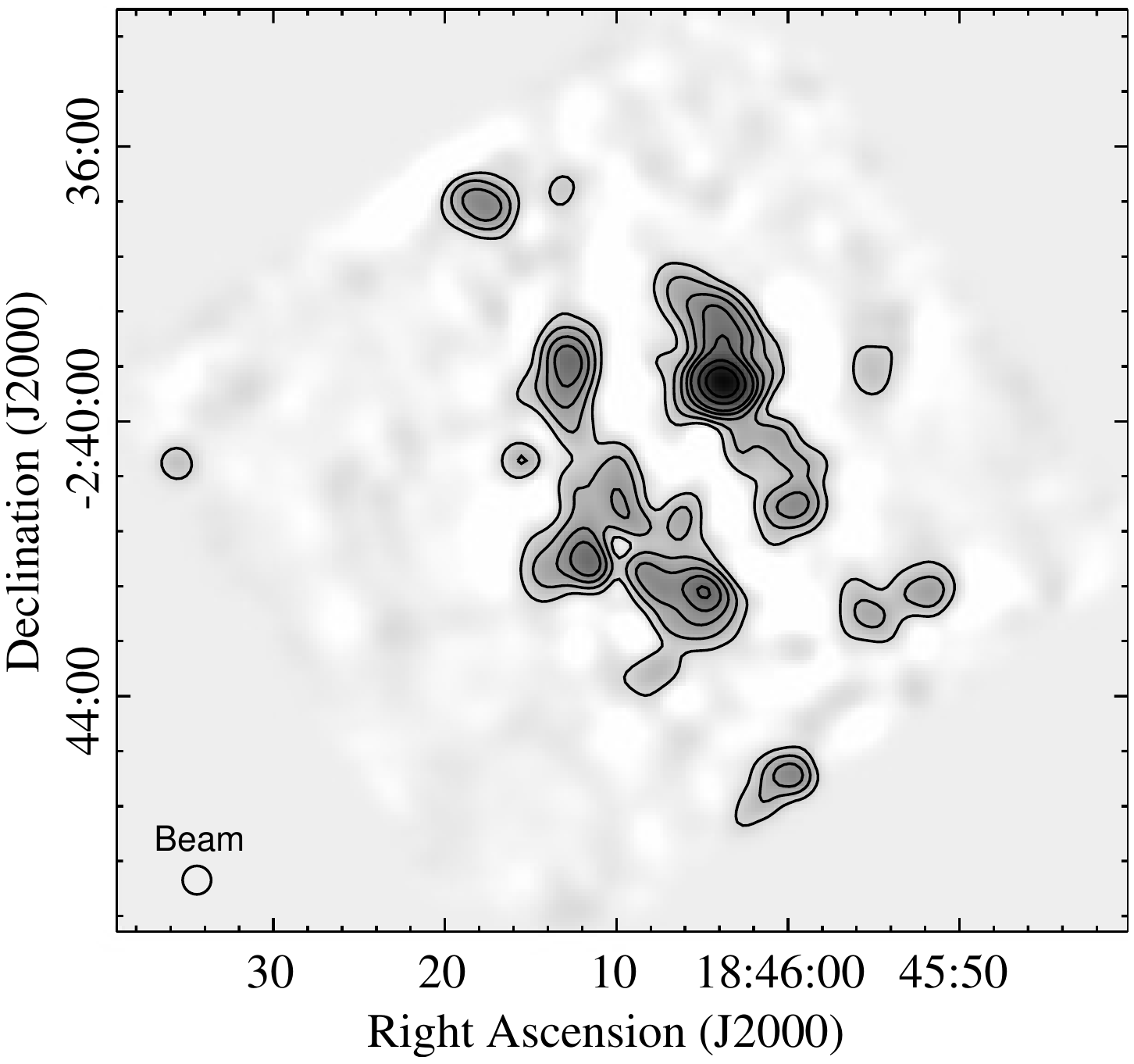}
   \caption{Example of the 350 \um\ continuum emission from SHARC-II (upper left) and {\textit{Herschel}}/SPIRE (upper right) toward the region $l=28.95$\arcdeg, $b=-0.05$\arcdeg. Beam sizes  are shown in the bottom left corner of each map. The contours of the SHARC-II image correspond to 3$\sigma$, 6$\sigma$, 10$\sigma$, 15$\sigma$, 30$\sigma$, 50$\sigma$ and 100$\sigma$, with an rms noise $\sigma=514$ mJy beam$^{-1}$. Contours of the {\textit{Herschel}} image start from 10$\sigma$, with increasing steps of 5$\sigma$ ($\sigma=164$ MJy sr$^{-1}$). The smaller beam size of SHARC-II images allow us to recover structures not identified by {\textit{Herschel}} with lower resolution. 
Bottom: SHARC-II image convolved to match the 24.9\arcsec\ FWHM beam size of {\textit{Herschel}} image.
Contours represent 1, 3, 6, 10, 15, 30 and 50 times the rms noise of the image. The 1$\sigma$ contour level resembles the emission observed at the 15$\sigma$ white contour level of the {\textit{Herschel}} image. The 15$\sigma$ level in the \textit{Herschel} map sets then a limit of the diffuse emission filtered out in the SHARC-II maps (see Section~\ref{sec:comparison_sharc_herschel_images}).}
     \label{fig:images_sharc_herschel}
\end{figure}
\newpage

\begin{figure}[h] 
   \centering
  \includegraphics[width=0.5\textwidth]{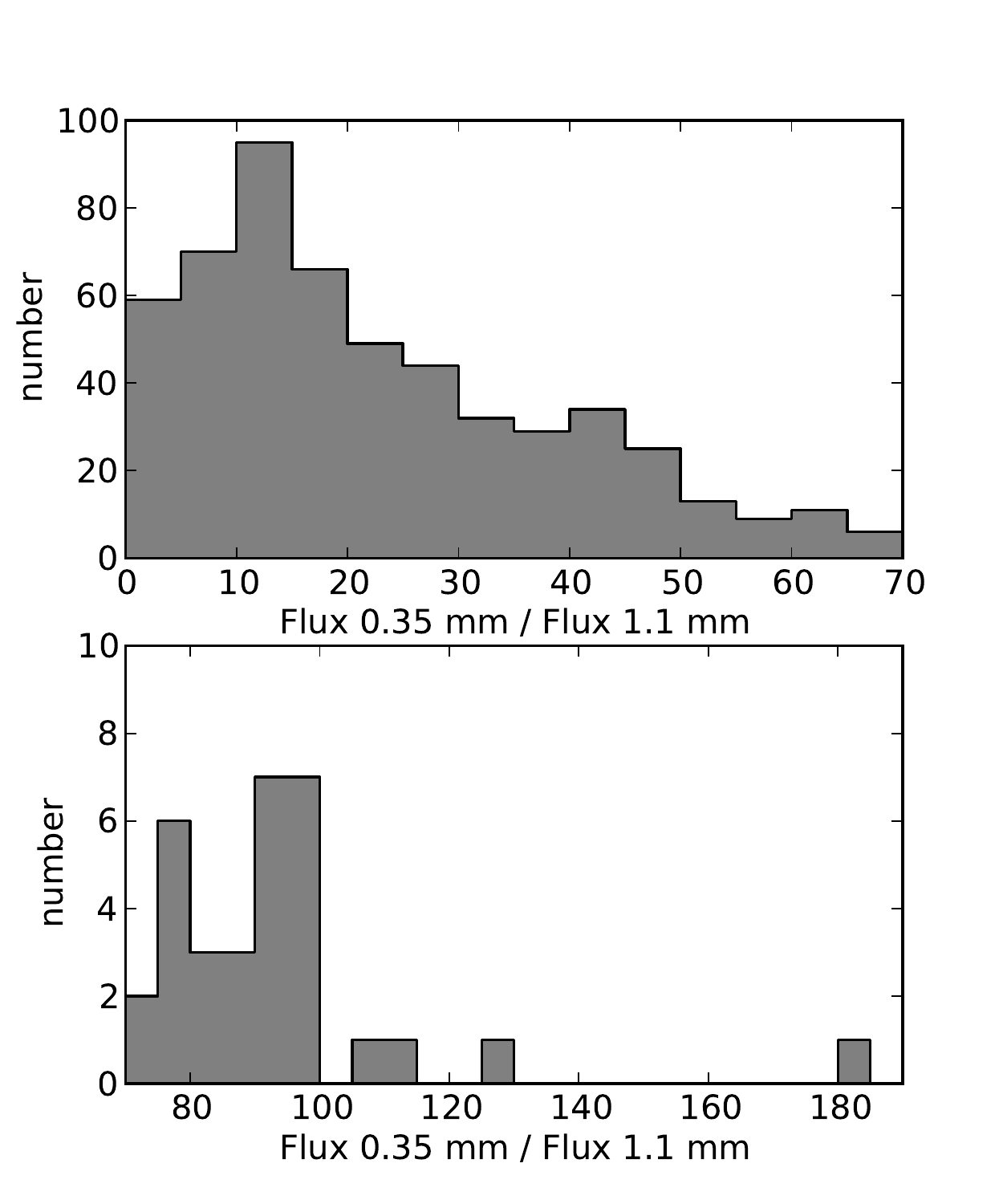}
   \caption{Distribution of the flux ratio between the emission at 0.35 mm and 1.1 mm for our sample of 574 BGPS V2.1 sources with reliable flux values. Nearly 94\% of the sources have a flux ratio lower than 70.2 }
  \label{fig:hist_flux_ratio_low}
\end{figure}

\begin{figure}[h] 
   \centering
  \includegraphics[width=0.45\textwidth]{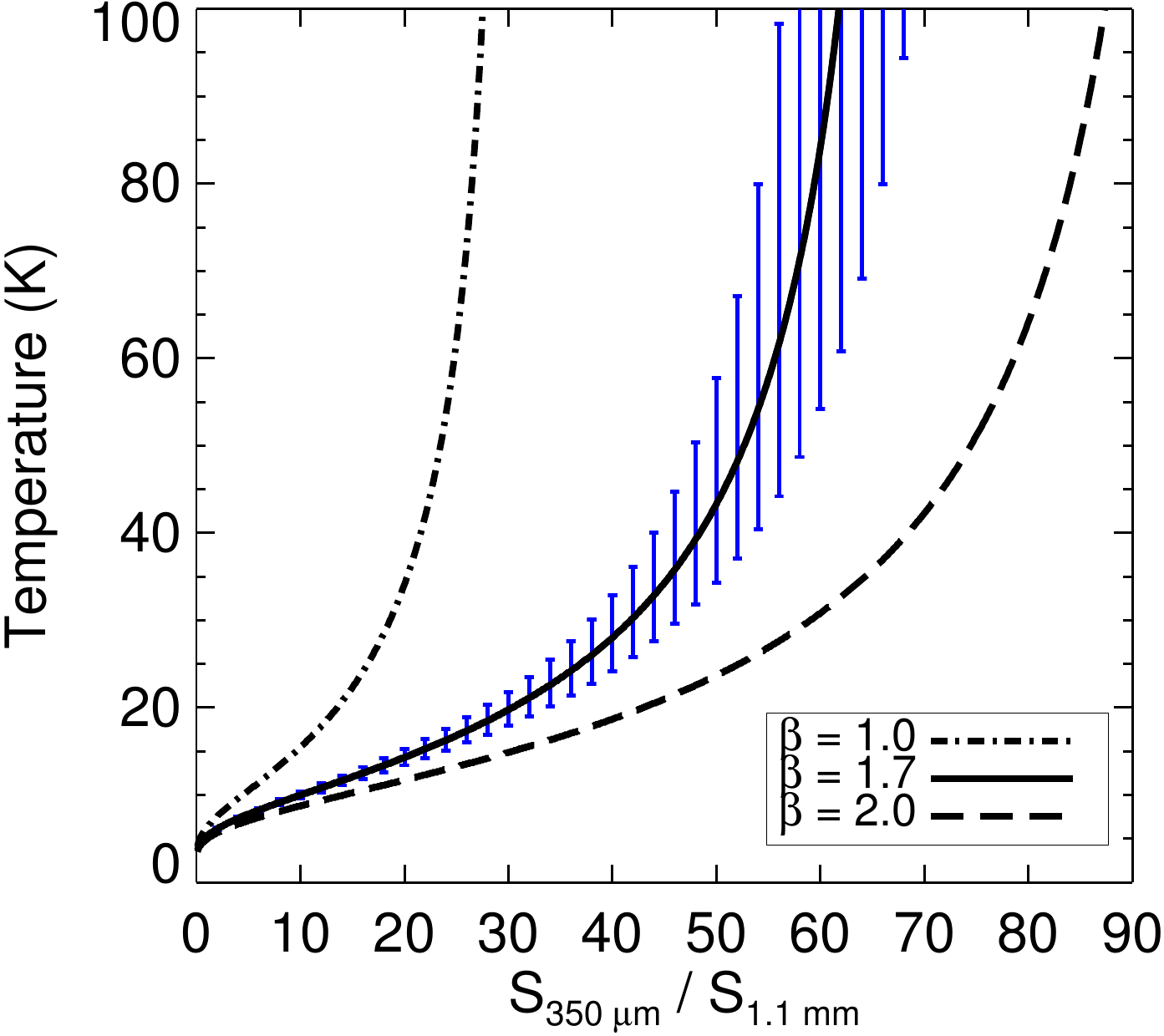}
   \caption{Models of dust temperature as a function of the flux ratio between continuum emission at 350 $\mu$m and 1.1 mm, in the optically thin limit. The three models correspond to different opacity laws, $\beta=1.0$, $\beta=1.7$ and $\beta=2.0$. The error bars for the $\beta=1.7$ model show the expected uncertainty in temperature determination when the flux ratio has an error of 10\%.}
  \label{fig:temp_plot}
\end{figure}
\newpage

\begin{figure}[h] 
   \centering
  \includegraphics[width=0.6\textwidth]{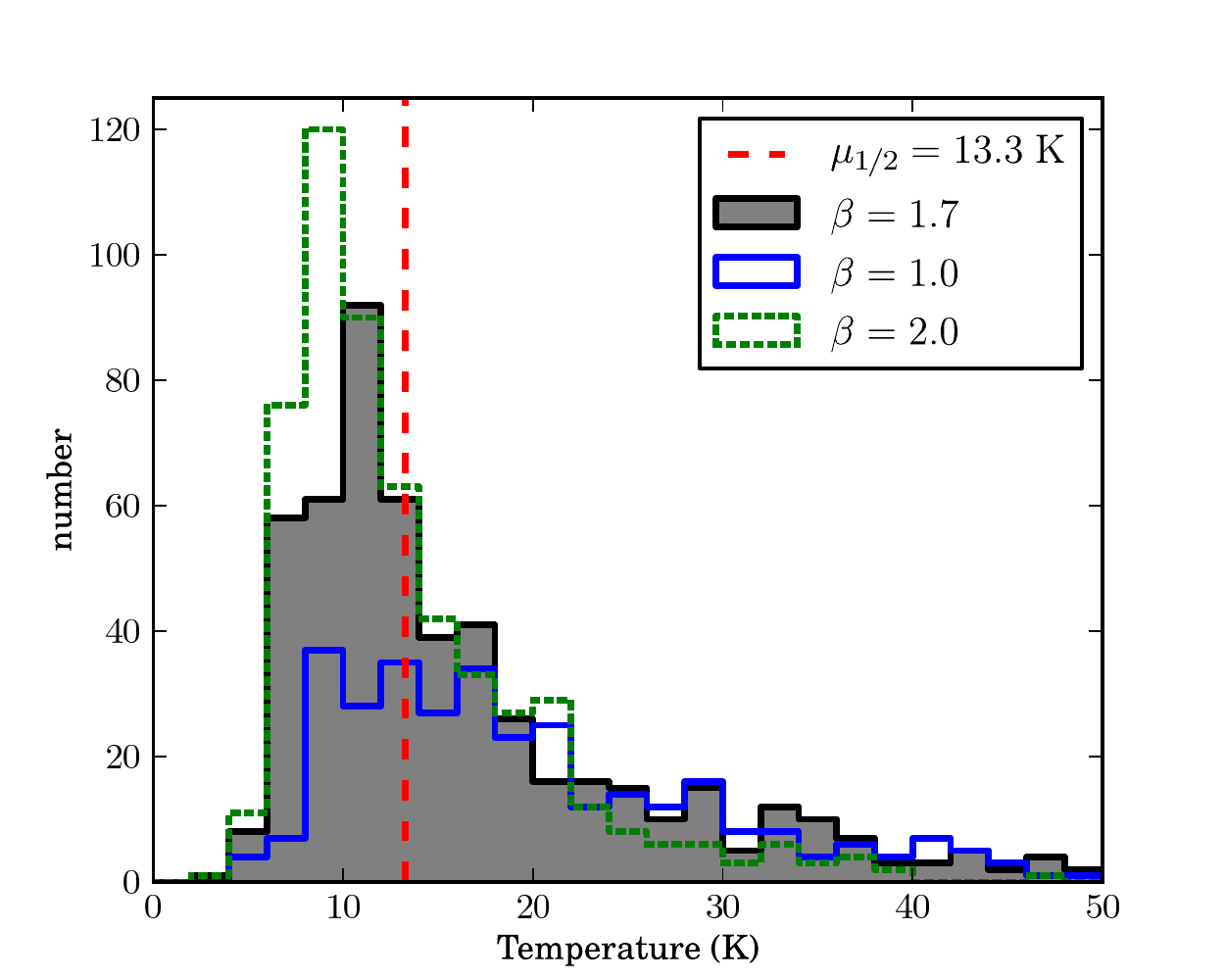}
  
   \caption{Distribution of temperatures of the BGPS sources with good fit in the determination of temperature, using the three different opacity models: spectral index $\beta$=1.0 (blue line), $\beta$=2.0 (green dashed line), and $\beta$=1.7 (shaded black line). The vertical red dashed line at T=13.3 K, represents the median value for the distribution with $\beta$=1.7. }
  \label{fig:hist_bgps_temp_limit}
\end{figure}

\begin{figure}[h] 
   \centering
  \includegraphics[width=0.6\textwidth]{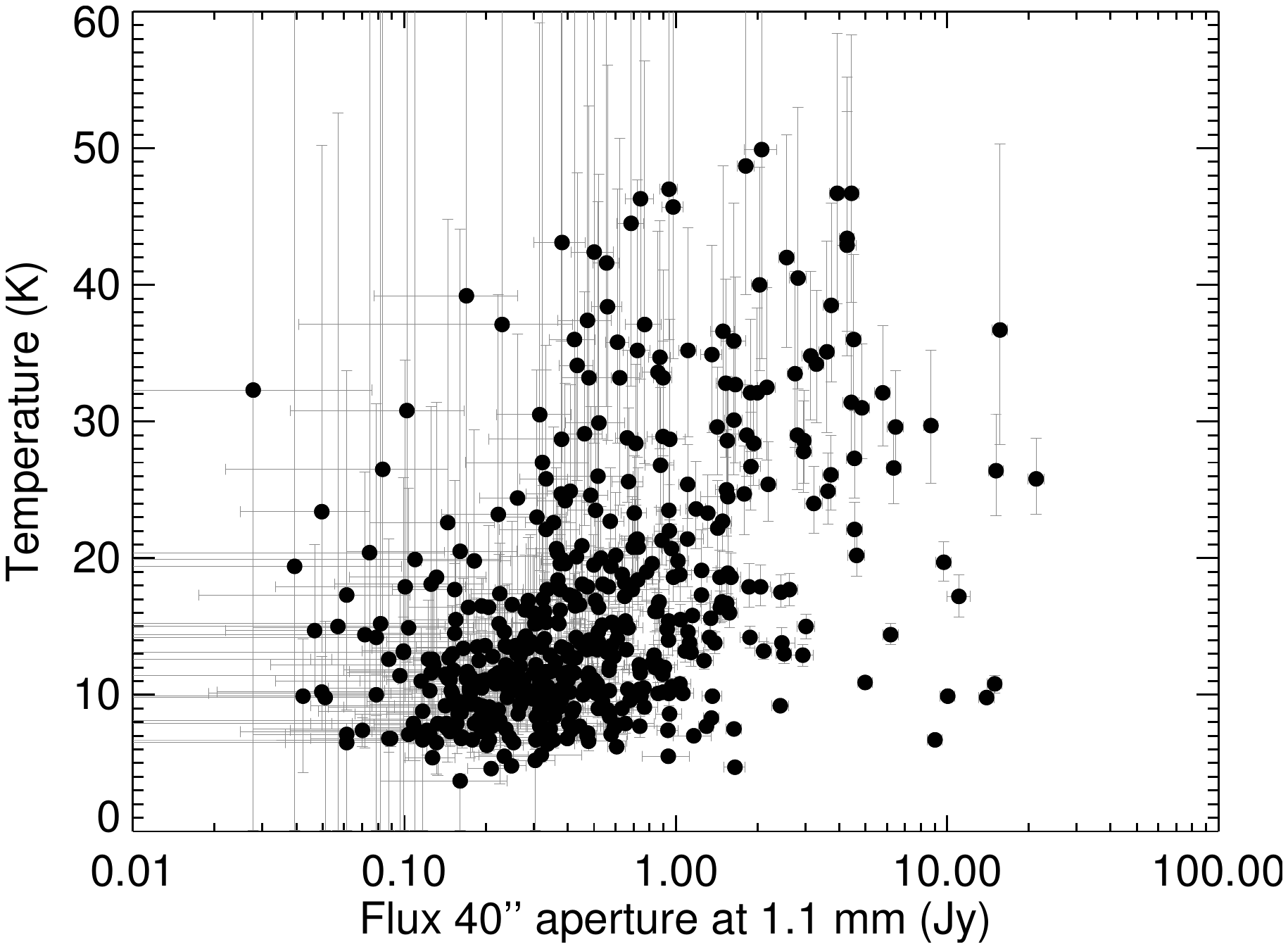}
   \caption{Temperature determined for the BGPS sample of sources as a function of their 40\arcsec\ aperture flux, for a spectral index $\beta=1.7$. Only sources with good fits in the determination of temperature (T$\le50$ K) are shown in this figure. There are few BGPS sources with high flux (F$_{1.1 mm}>4$ Jy) and low temperature (T$<20$ K); a description of them is presented in Section~\ref{sec:temperature_determination}.}
     \label{fig:fluxbolo_temp}
\end{figure}
\newpage

\begin{figure}[h] 
   \centering
  \includegraphics[width=0.47\textwidth]{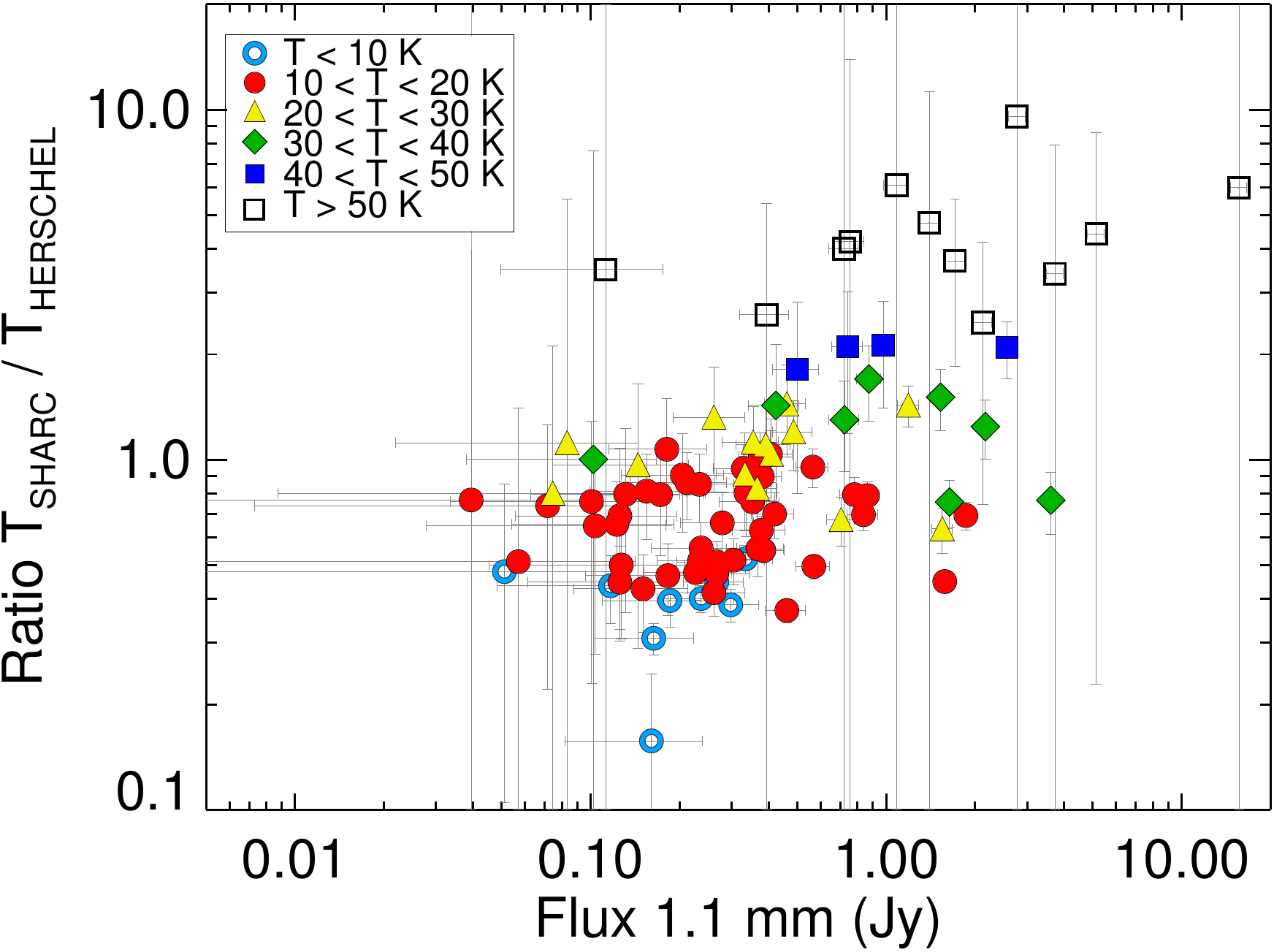}\\
  \includegraphics[width=0.47\textwidth]{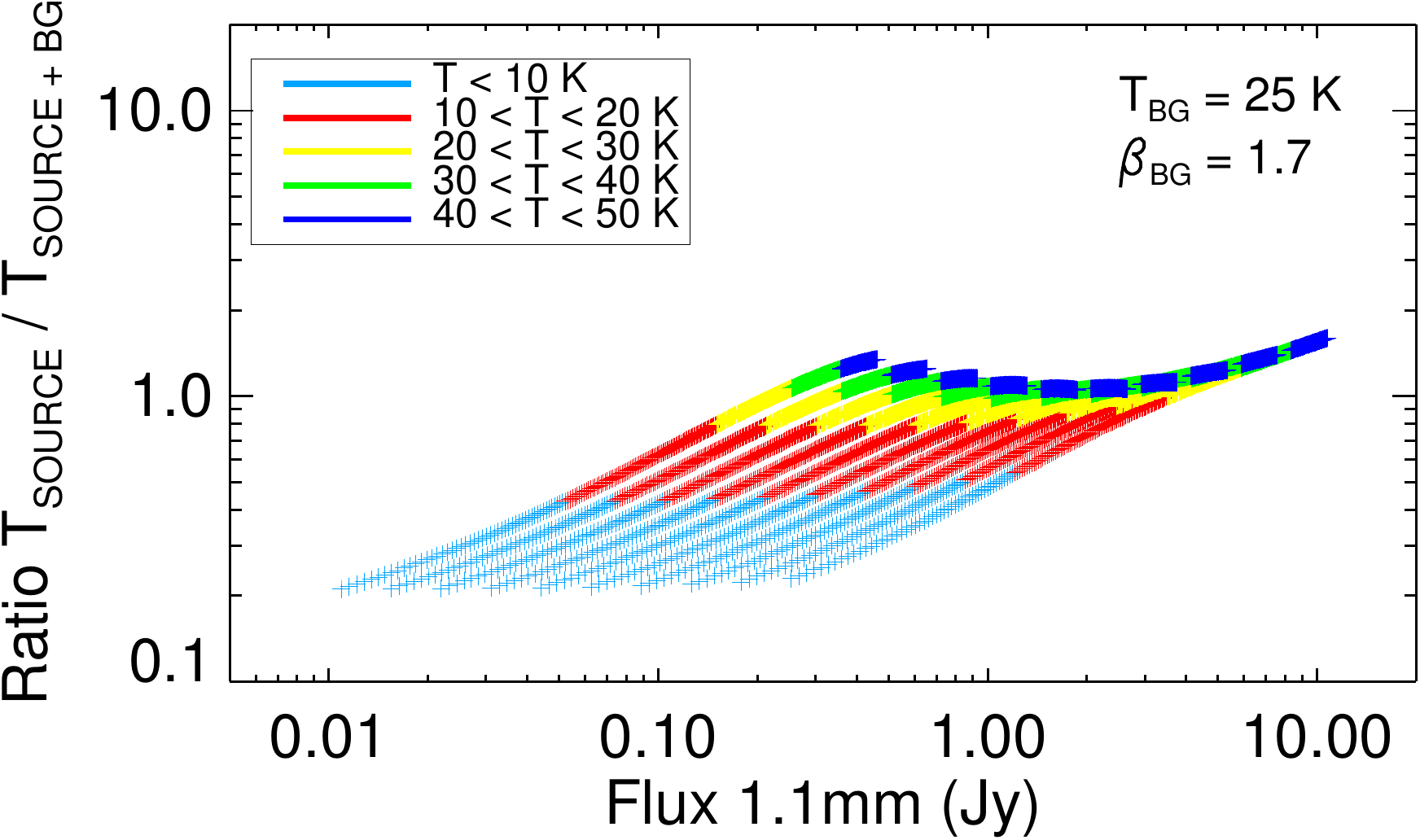}\\

   \caption{Upper: Comparison between estimated temperatures with 350 \um\ and 1.1 mm continuum maps, $T_{SHARC}$, and temperatures obtained from SED fitting by~\cite{bat11}, $T_{Herschel}$, as a function of 40\arcsec\ aperture flux density at 1.1 mm for a sample of 91 BGPS sources toward $l=30$\arcdeg. The colors shown in the legend represent different ranges for $T_{SHARC}$. The mean value of the ratio, weighted by errors, is 0.50$\pm$0.01 for sources with $T_{SHARC}\le50$ K. Bottom: Simulated recovery of sources immersed in a uniform background emitting as a modified blackbody with temperature $T_{bg}$=25 K, and spectral index $\beta=1.7$. The input sources have temperatures and spectral index in the ranges $5<T_{source}<50$ K, and $1.5<\beta_{source}<2.5$, which are characteristic of clump-like features. The temperature ratio $T_{source}/T_{source\,+\,bg}$ of simulated sources immersed in background reproduce qualitatively the observed ratio $T_{SHARC}/T_{Herschel}$ of the upper panel.}
     \label{fig:herschel_temp}
\end{figure}

\begin{figure}[h] 
   \centering
\includegraphics[width=0.46\textwidth]{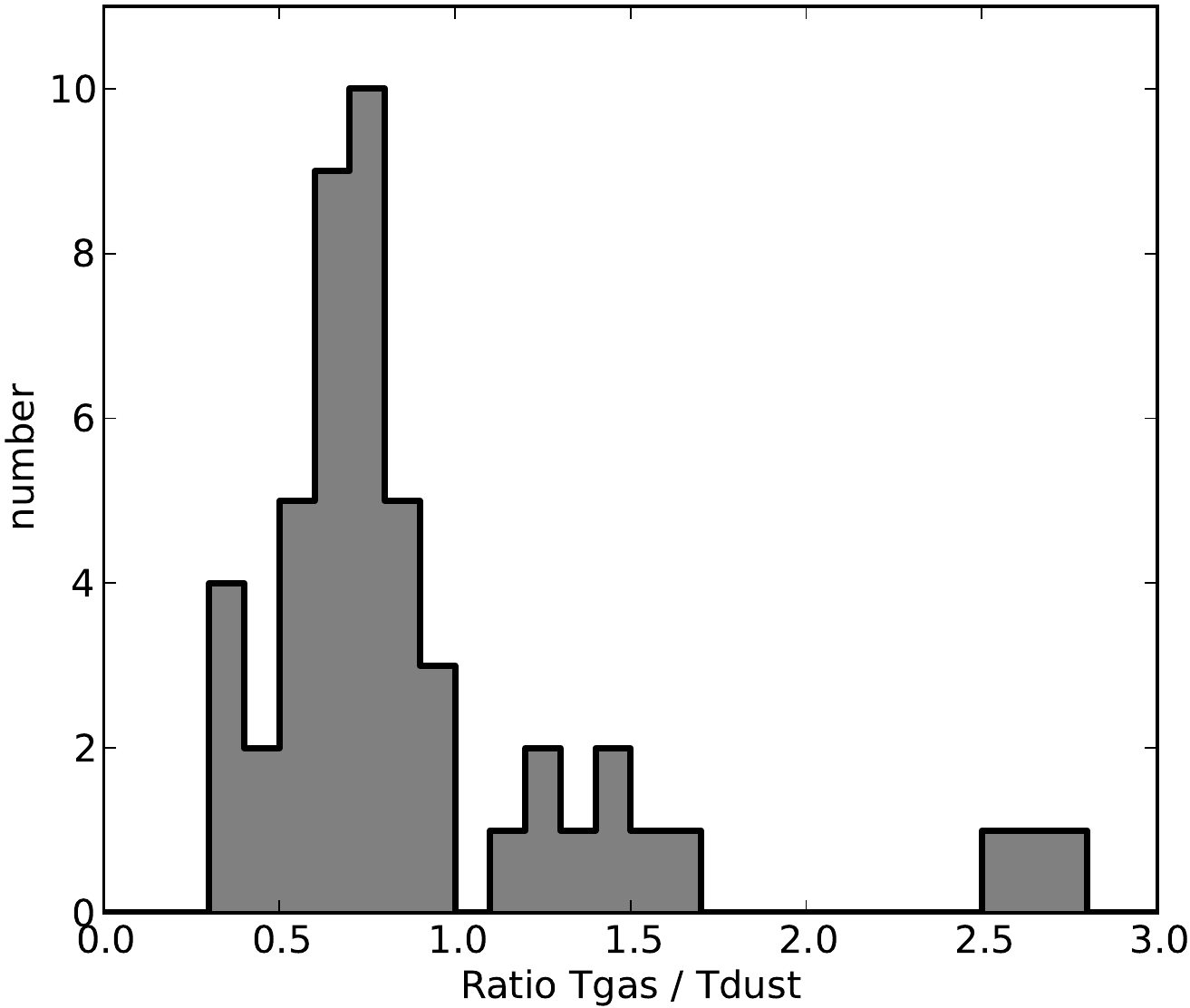}
   \caption{Distribution of the ratio of gas temperature determined from NH$_3$ observations $T_{gas}$, and fitted dust temperatures $T_{dust}$, for the sample of 49 BGPS sources with $T_{gas}<30$ K and $T_{dust}<50$ K, ranges in which both values are relatively well determined. The average and median values of this distribution are 0.88 and 0.76, respectively.}
     \label{fig:nh3_dust}
\end{figure}
\newpage

\begin{figure}[h] 
   \centering
  \includegraphics[width=0.50\textwidth]{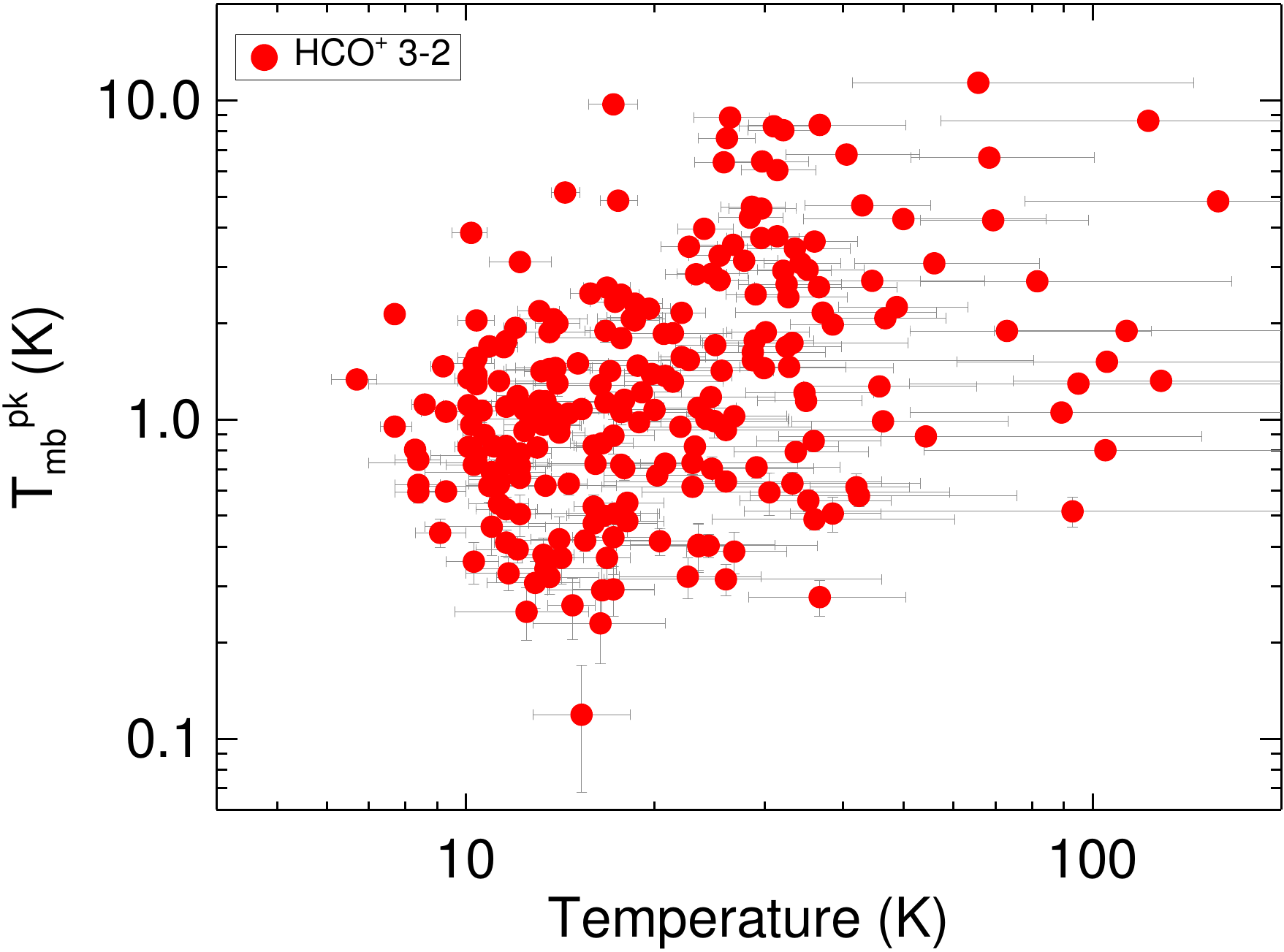}
  \includegraphics[width=0.50\textwidth]{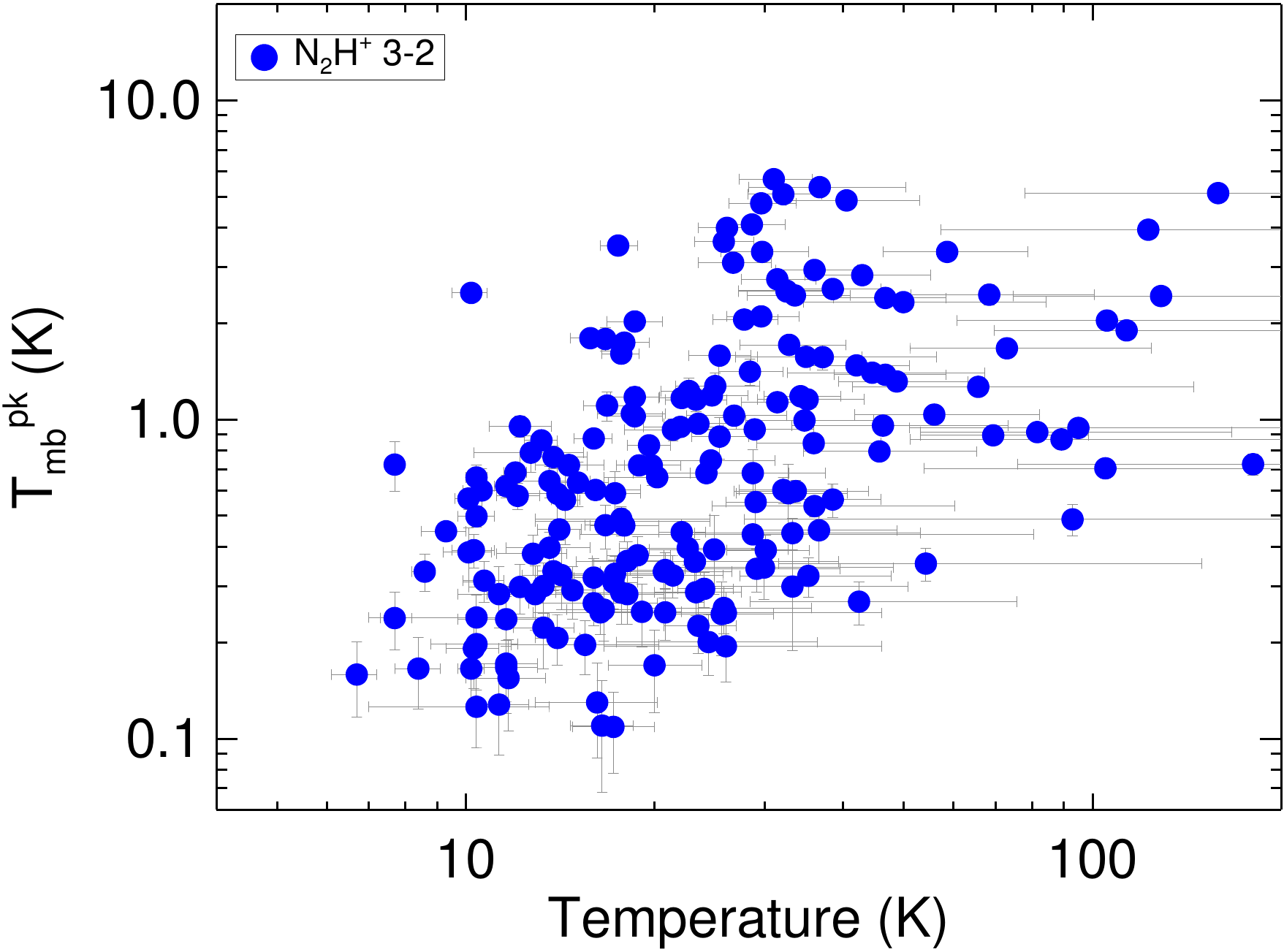}
   \caption{Peak main-beam temperature, as a function of dust temperature, for the sample of BGPS sources from the spectroscopy catalog by~\cite{shi13} that are contained in the SHARC-II maps. The number of sources with HCO$^+$ emission is 250, and the number of sources with N$_2$H$^+$ emission is 199.}
     \label{fig:spec_cat_bgps}
\end{figure}

\begin{figure}[h] 
   \centering
  \includegraphics[width=0.48\textwidth]{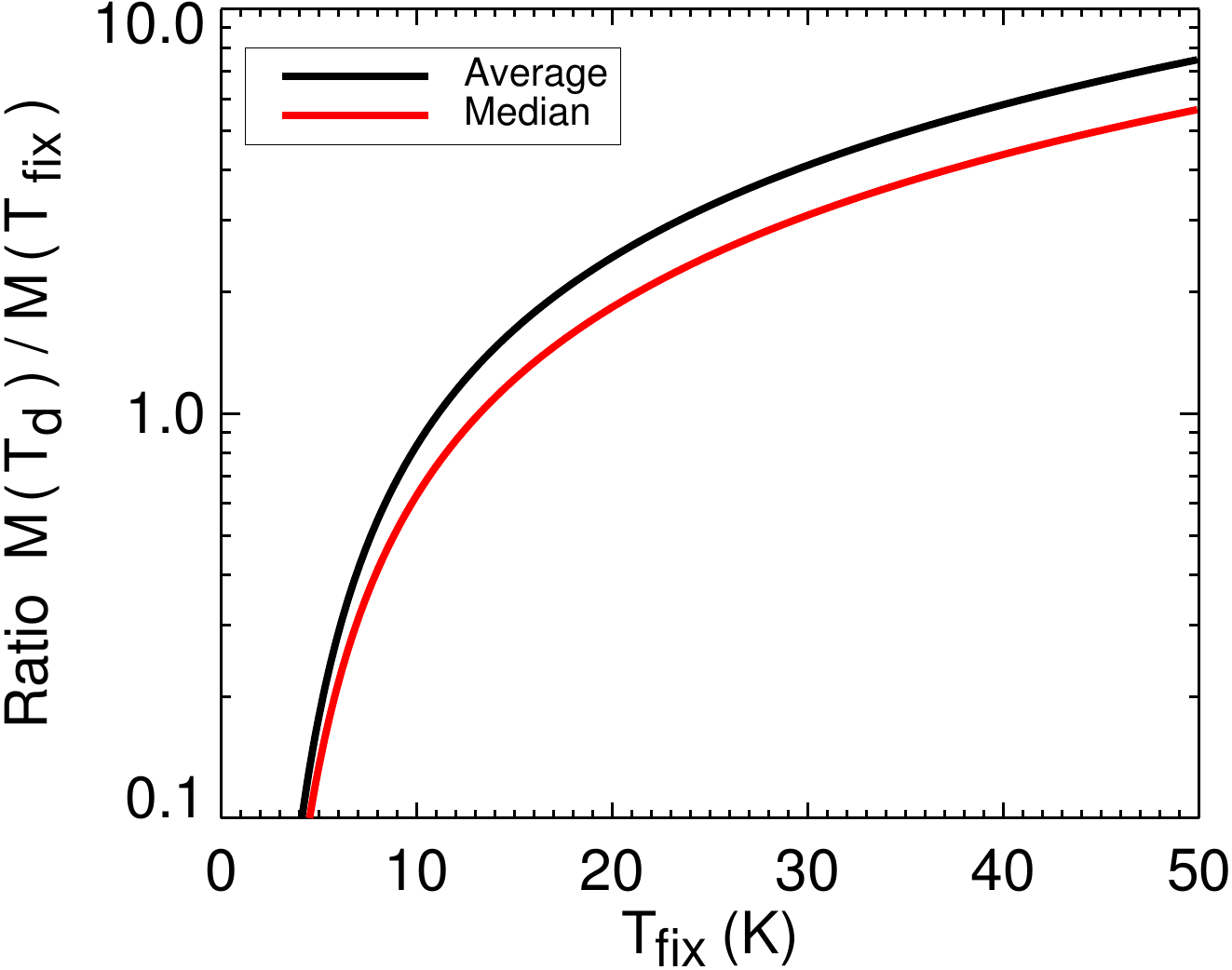}
   \caption{Ratio between the mass estimated using color temperatures determined from 350 \um\ and 1.1 mm images ($T_{clump}$), and the mass estimated using a single fiducial temperature ($T_{fix}$). For the sample of 514 V2.1 BGPS sources with good fits in the determination of temperature, using $T_{fix}=20$ K, the average value of the source masses will be underestimated by a factor of $\sim$2.4.}
     \label{fig:model_ratio_mass}
\end{figure}
\newpage

\begin{figure}[h] 
   \centering
  \includegraphics[width=0.46\textwidth]{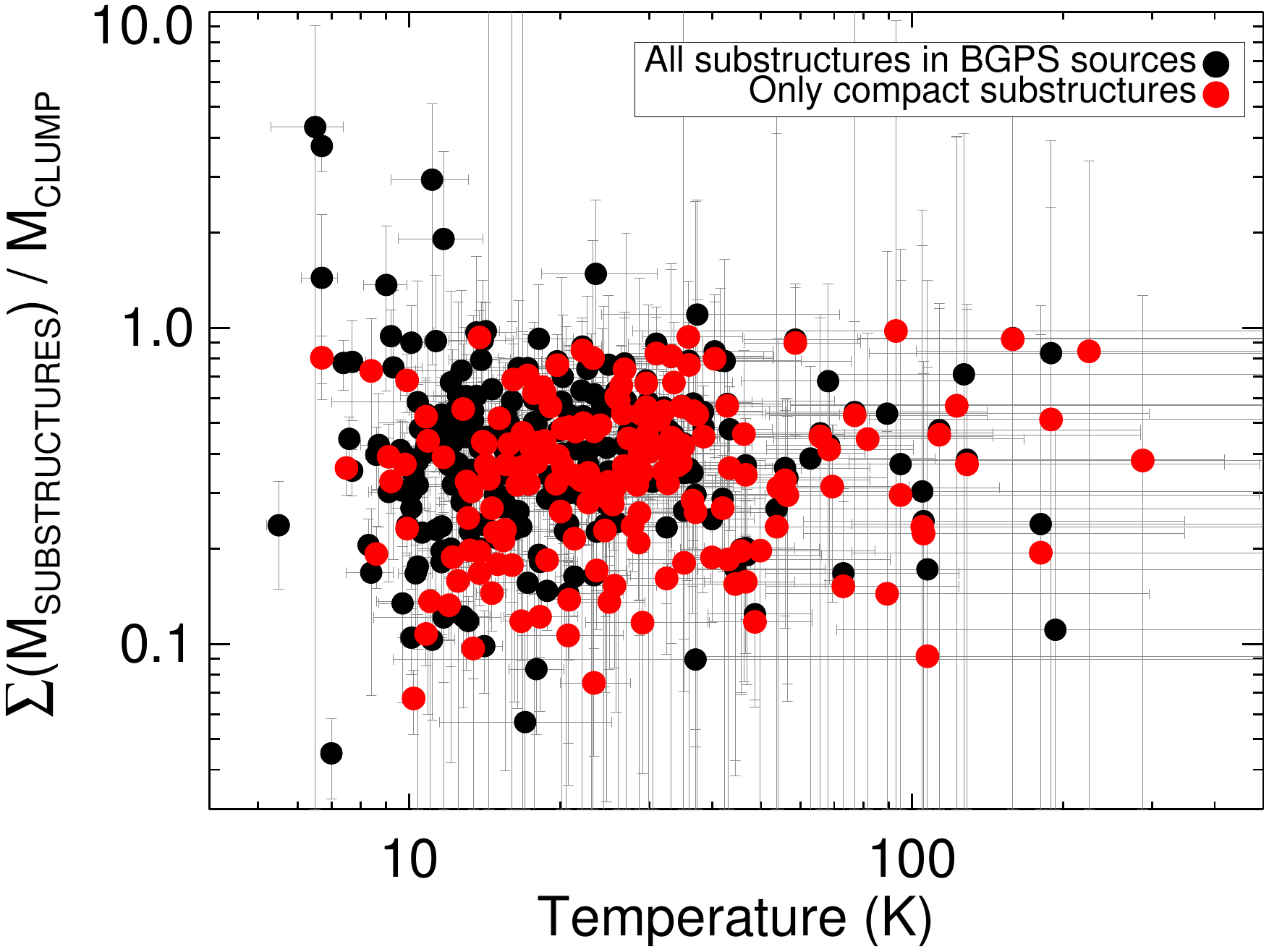} \hspace{4mm}
    \includegraphics[width=0.46\textwidth]{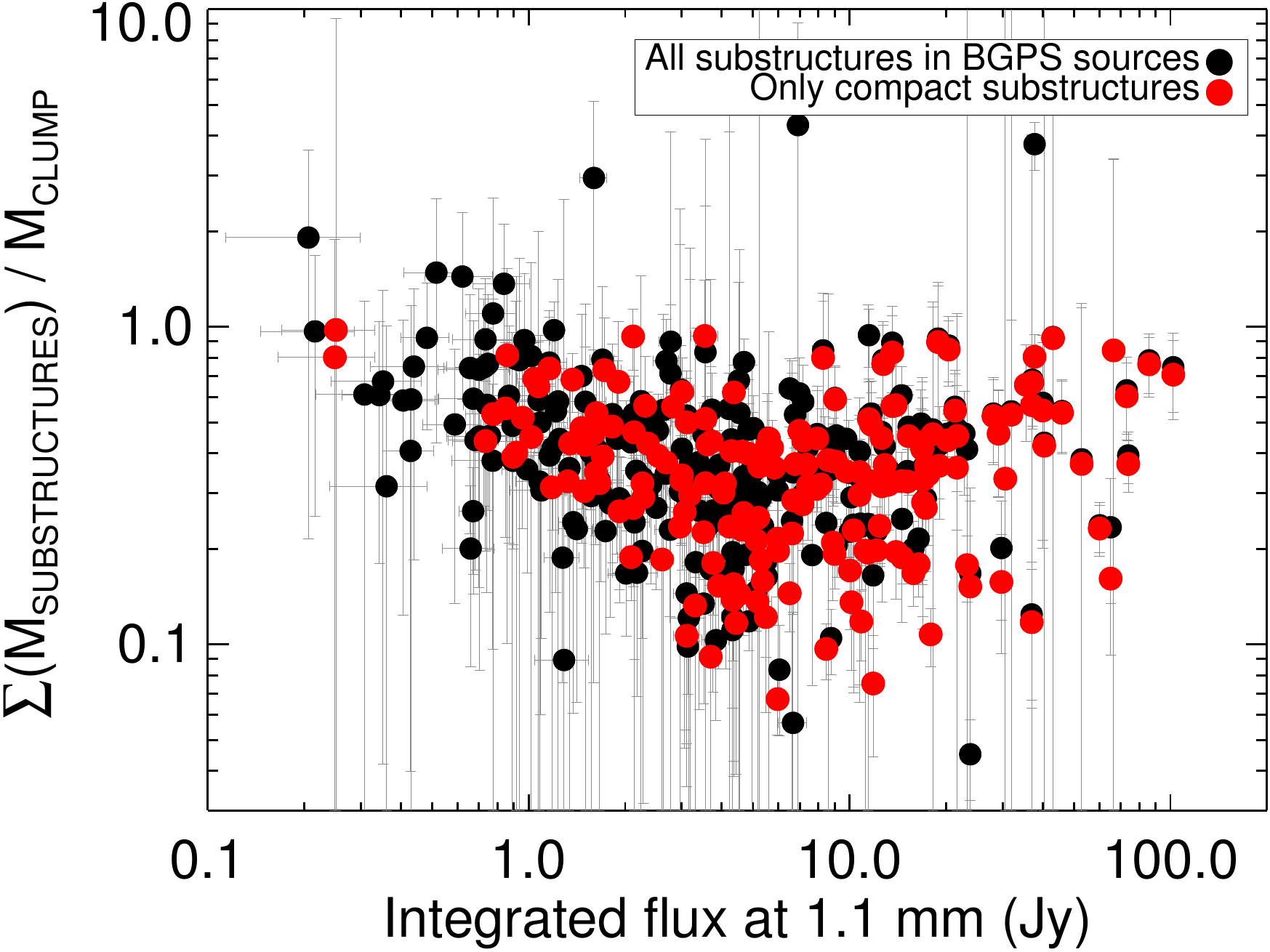}
   \caption{Comparison between the total estimated mass from 350 \um\ substructures in each clump, and the mass estimated at 1.1 mm for that parental BGPS clump, as a function of dust temperature (left) and the integrated flux of the BGPS sources (right). Black points consider for M$_{substructures}$ the flux contribution from all high-resolution sources in the parental clumps, and have a weighted averaged mass fraction $\sum$(M$_{substructures}$)/M$_{clump}$ of 0.22$\pm$0.01. Red points only consider the flux contribution from strong, compact substructures, and the averaged mass fraction, weighted by errors, of this distribution is 0.19$\pm$0.01.}
     \label{fig:ratio_mass_sharc_bgps}
\end{figure}

\begin{figure}[h] 
   \centering

  \includegraphics[width=0.5\textwidth]{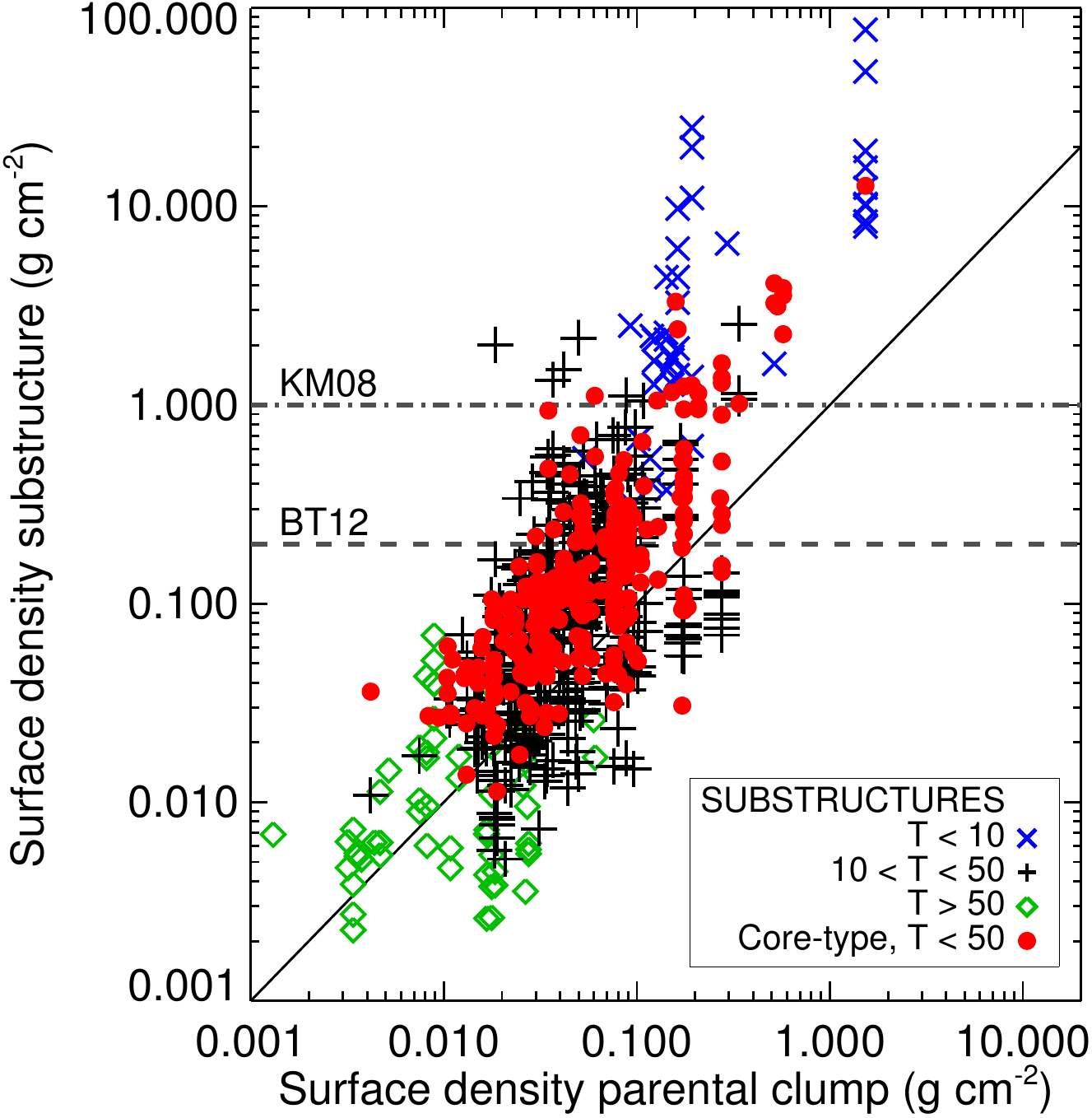}
   \caption{Estimation of surface density between substructures and parental clumps. Solid line represent the locus of equality between the structures. Black crosses show the bulk of the sources, with temperatures between 10$-$50 K. Blue marks show sources with low temperature (T$<10$ K), likely to overestimating the surface density.  Grey marks show sources with high temperature (T$>50$ K). Red points show only those compact, core-type substructures found in the parental clumps, with detections above 10$\sigma$. Most of these compact sources have higher surface 2-3 times larger than their parental clump. Several of these compact sources have surface densities above theoretical thresholds of 0.2 g cm$^{-2}$~\citep{but12} or 1.0 g cm$^{-2}$~\citep{kru08} for massive star formation (see Section~\ref{sec:mass_surface_substructures}).}
     \label{fig:surface_density}
\end{figure}
\newpage

\appendix

\section{Calibrators}

Table~\ref{tbl:calibrators} presents in detail the flux calibration and the observed values of $\tau$ at 225 GHz for each observation run. Flux conversion factors for one beam ($C_{beam}$), 20\arcsec\ and 40\arcsec\ aperture conversion factors on SHARC-II maps ($C_{20}$ and $C_{40}$), and 40\arcsec\ aperture conversion factors for SHARC-II maps convolved to a resolution of 33\arcsec\ are shown in the table.
{\LongTables
\begin{deluxetable}{llccccc}
\tabletypesize{\scriptsize}
\tablecaption{Calibrators}
\tablewidth{0pt}
\tablehead{
\colhead{Date} & \colhead{Calibrator}  & \colhead{$\tau_{225\ GHz}$} & \colhead{C$_{beam}$} & \colhead{C$_{20}$} &
\colhead{C$_{40}$}  & \colhead{C$_{40}^{\ 33\arcsec}$} \cr
 		&	&	&\colhead{(Jy beam$^{-1}\ \mu V^{-1}$)}&\colhead{(Jy$ \ \mu V^{-1}$)}&\colhead{(Jy$ \ \mu V^{-1}$)} & \colhead{(Jy$ \ \mu V^{-1}$)} 
}
\startdata
 
2006 Jun 11	&Mars	 & 0.071	& 10.86	&0.39	&0.31  &0.46   \cr 
			&Mars	 & 0.071	& 10.15	&0.42	&0.30  &0.46   \cr 
			&Mars	 & 0.071	& 11.63	&0.40	&0.33  &0.49   \cr 
2006 Jun 23	&Uranus	 & 0.061	& 8.55	&0.28	&0.24  &0.36   \cr 
			&Uranus	 & 0.062	& 8.02	&0.28	&0.23  &0.34   \cr 
2007 Jul 9 	&Mars	 & 0.068	& 7.56 	&0.27	&0.22  &0.33  \cr 
			&Uranus	 & 0.058	& 6.55	&0.22	&0.18  &0.27 \cr 
			&Uranus	 & 0.068	& 7.57	&0.25	&0.21  &0.31 \cr 
			&Neptune	 & 0.058	& 6.76	&0.22	&0.19  &0.28 \cr 
			&Neptune	 & 0.064	& 6.42	&0.21	&0.18  &0.27 \cr 
			&Neptune	 & 0.060	& 7.18	&0.23	&0.20  &0.30 \cr 
2007 Oct 24	&Uranus	 & 0.053	& 7.80	&0.26	&0.22  & 0.32 \cr 
			&Uranus	 & 0.053	& 8.02	&0.27	&0.23  & 0.34 \cr 
			&Neptune & 0.055	& 9.29	&0.31	&0.26  & 0.39 \cr 
			&Neptune	 & 0.055	& 9.20	&0.30	&0.26  & 0.38 \cr 
2007 Oct 27	&Uranus	 & 0.052	& 7.18	&0.23	&0.20  & 0.30 \cr 
			&Uranus	 & 0.047	& 7.80	&0.25	&0.22  & 0.33 \cr 
			&Uranus	 & 0.047	& 7.29	&0.24	&0.20  & 0.30 \cr 
2009 Sep 10	&Uranus	 & 0.055	& 5.80	&0.19	&0.16  & 0.24 \cr 
			&Mars	 & 0.062	& 6.31	&0.23	&0.18  & 0.27 \cr 
2009 Dec 13	&Uranus	 & 0.050	& 6.34	&0.22	&0.18  & 0.26 \cr 
			&Uranus    & 0.050   & 6.13        &0.23        &0.18  & 0.26 \cr 
			&Mars        & 0.065   & 5.73	&0.20        &0.17  & 0.24 \cr 
			&Mars	 & 0.065   & 5.67	&0.20	&0.16  & 0.24 \cr 
2009 Dec 14	&Uranus    & 0.068   & 6.01       &0.22        &0.17  & 0.25 \cr 
			&Mars	 & 0.058	& 5.28	&0.22	&0.16  & 0.24 \cr 
2009 Dec 29	&Uranus	 & 0.060	& 5.84	&0.20	&0.16  & 0.24 \cr 
			&Uranus	 & 0.052	& 5.85	&0.19	&0.16  & 0.24 \cr 
			&Mars	 & 0.062	& 5.72	&0.22	&0.17  & 0.26 \cr 
			&Mars	 & 0.062	& 5.78	&0.22	&0.17  & 0.26 \cr 
			&Mars	 & 0.056	& 5.80	&0.22	&0.17  & 0.26 \cr 
2009 Dec 30	&Uranus	 & 0.050	& 7.06	&0.24	&0.19  & 0.28 \cr 
			&Uranus	 & 0.051	& 7.08	&0.24	&0.19  & 0.28 \cr 
			&Mars	 & 0.047	& 5.70	&0.22	&0.17  & 0.25 \cr 
			&Mars	 & 0.052	& 5.68	&0.22	&0.17  & 0.25 \cr 
			&Mars	 & 0.040	& 5.66	&0.22	&0.17  & 0.25 \cr 
			&Mars	 & 0.040	& 5.81	&0.24	&0.18  & 0.26 \cr 
2009 Dec 31	&Uranus	 & 0.051	& 6.43	&0.22	&0.18  & 0.27 \cr 
			&Uranus	 & 0.049	& 6.32	&0.22	&0.18  & 0.26 \cr 
			&Uranus	 & 0.050	& 6.26	&0.22	&0.18  & 0.27 \cr 
			&Mars	 & 0.048	& 4.87	&0.19	&0.15  & 0.22 \cr 
			&Mars	 & 0.048	& 4.79	&0.20	&0.15  & 0.22 \cr 
2010 Jan 01	&Mars	 & 0.045	& 5.69	&0.22	&0.17  & 0.25 \cr 
2010 Jul 24	&Neptune	 & 0.063	& 19.30	&0.61	&0.54  & 0.80 \cr 
			&Uranus	 & 0.070	& 26.18	&0.87	&0.72  & 1.07 \cr 
			&Uranus	 & 0.075	& 18.00	&0.76	&0.52  & 0.81 \cr 
			&Uranus	 & 0.056	& 11.68	&0.58	&0.36  & 0.55 \cr 
			&Uranus	 & 0.065	& 13.98	&0.55	&0.39  & 0.61 \cr 
2010 Jul 25	&Neptune	 & 0.077	& 9.63	&0.32	&0.27  & 0.41 \cr 
			&Uranus	 & 0.042	& 6.09	&0.20	&0.17  & 0.25 \cr 
			&Uranus	 & 0.039	& 5.79	&0.19	&0.16  & 0.24 \cr 
			&Uranus	 & 0.046	& 7.12	&0.25	&0.20  & 0.30 \cr 
2010 Jul 28	&Neptune	 & 0.056	& 6.76	&0.21	&0.19  & 0.28 \cr 
			&Uranus	 & 0.056	& 6.57	&0.25	&0.18  & 0.27 \cr 
			&Uranus	 & 0.046	& 6.37	&0.20	&0.18  & 0.26 \cr 
2010 Jul 31	&Neptune	 & 0.050	& 8.58	&0.28	&0.23  & 0.35 \cr 
			&Neptune	 & 0.050	& 7.71	&0.26	&0.21  & 0.32 \cr 
			&Neptune	 & 0.050	& 7.82	&0.27	&0.22  & 0.33 \cr 
			&Neptune	 & 0.050	& 8.22	&0.27	&0.23  & 0.34 \cr 
			&Uranus	 & 0.058	& 6.67	&0.23	&0.19  & 0.28 \cr 
2010 Aug 01	&Neptune	 & 0.036	& 7.42	&0.24	&0.20  & 0.30 \cr 
			&Neptune	 & 0.060	& 7.67	&0.25	&0.21  & 0.31 \cr 
			&Neptune	 & 0.050	& 7.46	&0.25	&0.20  & 0.30 \cr 
			&Uranus	 & 0.053	& 7.06	&0.24	&0.19  & 0.29 \cr 
2010 Dec 05	&Uranus	 & 0.060	& 5.75	&0.20	&0.17  & 0.24 \cr 
			&Uranus	 & 0.030	& 5.96	&0.21	&0.17  & 0.25 \cr 
			&Uranus	 & 0.030	& 6.66	&0.23	&0.19  & 0.27 \cr 
			&Uranus	 & 0.040	& 6.38	&0.21	&0.18  & 0.26 \cr 
2010 Dec 06	&Uranus	 & 0.030	& 7.07	&0.25	&0.19  & 0.29 \cr 
			&Uranus	 & 0.050	& 7.07 	&0.26	&0.20  & 0.30 \cr 
			&Uranus	 & 0.038	& 7.26	&0.24	&0.19  & 0.30 \cr 
			&Uranus	 & 0.038	& 7.04	&0.25	&0.19  & 0.29 \cr 
			&Uranus	 & 0.068	& 6.78	&0.26	&0.19  & 0.29 \cr 
			&Uranus	 & 0.049	& 6.93	&0.23	&0.19  & 0.28 \cr 
2011 Dec 23	&Uranus	 & 0.036	& 11.35	&0.38	&0.32  & 0.49 \cr 
			&Uranus	 & 0.071	& 10.06	&0.37	&0.29  & 0.43 \cr 
			&Uranus	 & 0.070	& 11.06	&0.38	&0.31  & 0.45 \cr 
2012 Sep 14	&Uranus	 & 0.050	& 12.73	&0.42	&0.35  & 0.52 \cr 
			&Uranus	 & 0.057    & 12.88	&0.44	&0.36  & 0.53 \cr 
			&Uranus	 & 0.057	& 12.60	&0.42	&0.35  & 0.52 \cr 
			&Uranus	 & 0.062	& 17.32	&0.57	&0.47  & 0.70 \cr 
			&Uranus	 & 0.065	& 17.37	&0.59	&0.48  & 0.72 \cr 
2012 Sep 21	&Uranus	 & 0.070	& 17.47	&0.56	&0.49  & 0.73 \cr 
			&Uranus	 & 0.094	& 17.79	&0.57	&0.50  & 0.74 \cr 
			&Uranus	 & 0.094	& 16.58	&0.55	&0.47  & 0.70 \cr 

\enddata

\label{tbl:calibrators}
\end{deluxetable}
}

\clearpage
\section{SHARC-II maps}

We present in Figure~\ref{fig:sharcmaps} our complete sample of 107 regions mapped with SHARC-II. For each of them, the image shows grey scale maps with emission contours at 3$\sigma$, 6$\sigma$, 10$\sigma$, 15$\sigma$, 30$\sigma$, 50$\sigma$, and 100$\sigma$, with $\sigma$ the rms noise. The value of $\sigma$ in units of mJy beam$^{-1}$ is stated below each map. The image also shows the high-resolution sources extracted with \bolocat. Red contours represent sources with signal-to-noise values above 10 (compact sources), and blue contours sources below this limit (faint sources).

\renewcommand{\thesubfigure}{\thefigure.\arabic{subfigure}}
\begin{figure} 
\center
\subfloat[L359.85+0.00 map, $\sigma_{rms}=994$ mJy beam$^{-1}$.]{%
\includegraphics[scale=0.43]{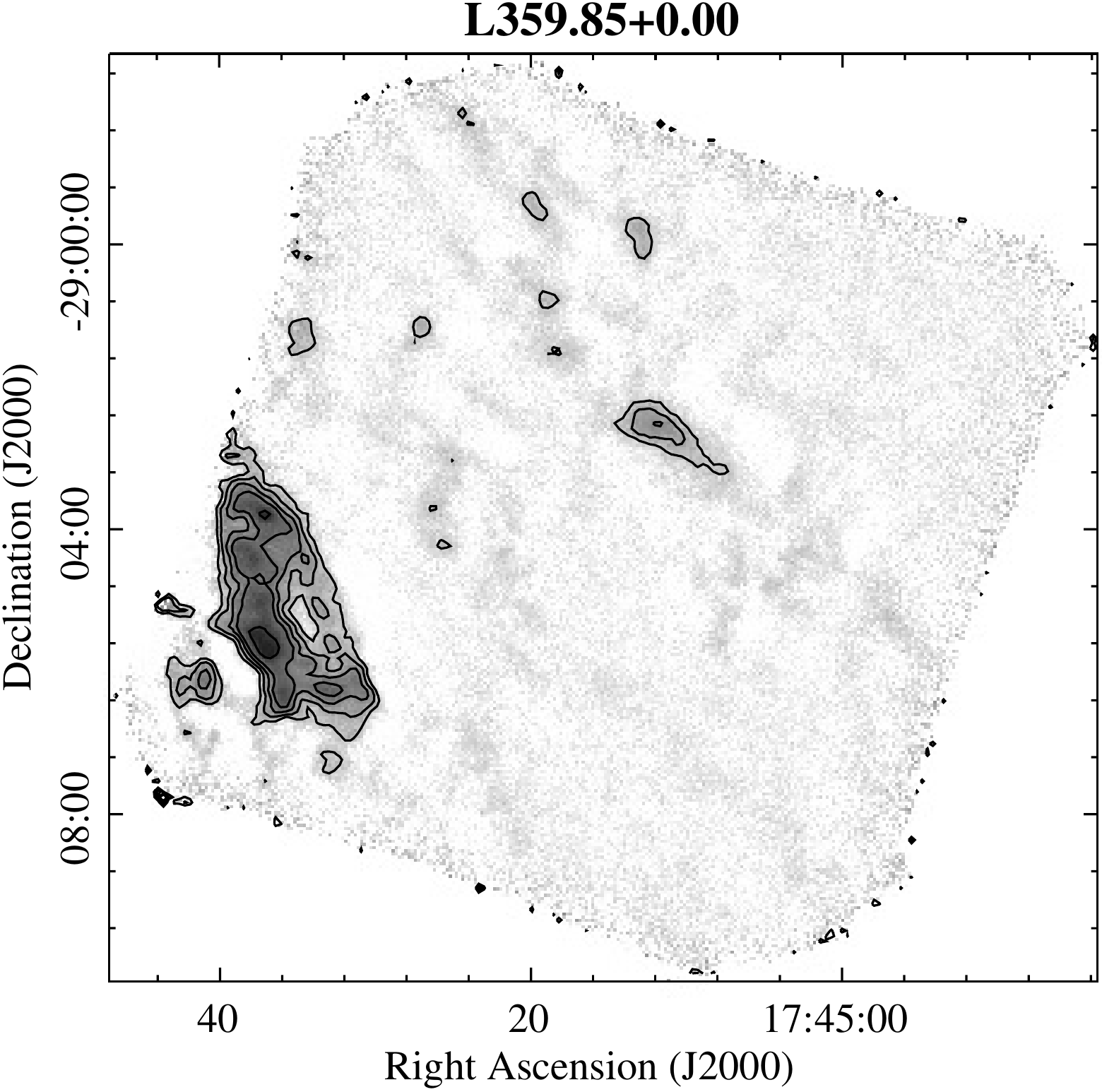}
\includegraphics[scale=0.43]{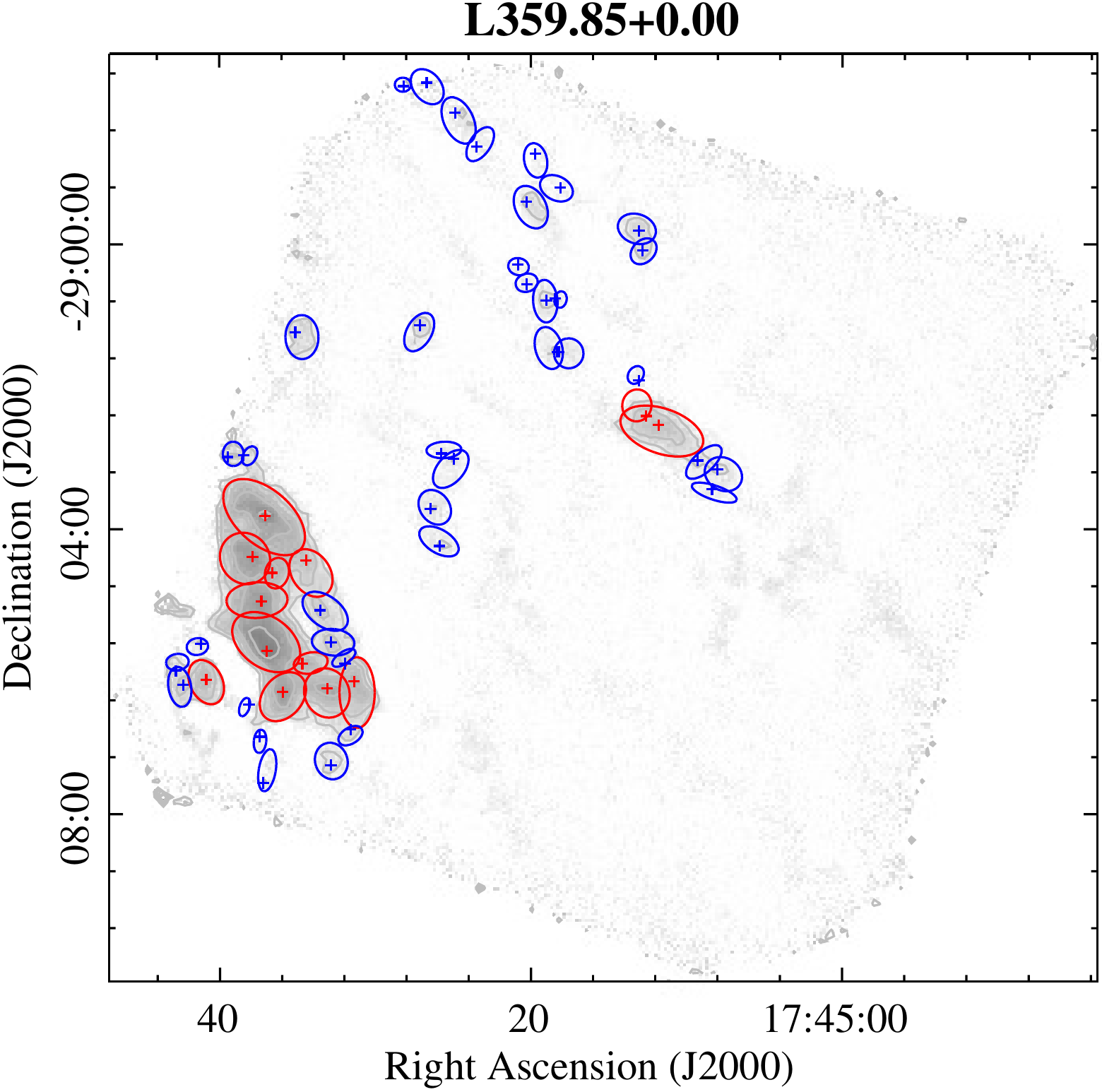}
}\\
\subfloat[L000.00+0.00 map, $\sigma_{rms}=1453$ mJy beam$^{-1}$.]{%
\includegraphics[scale=0.43]{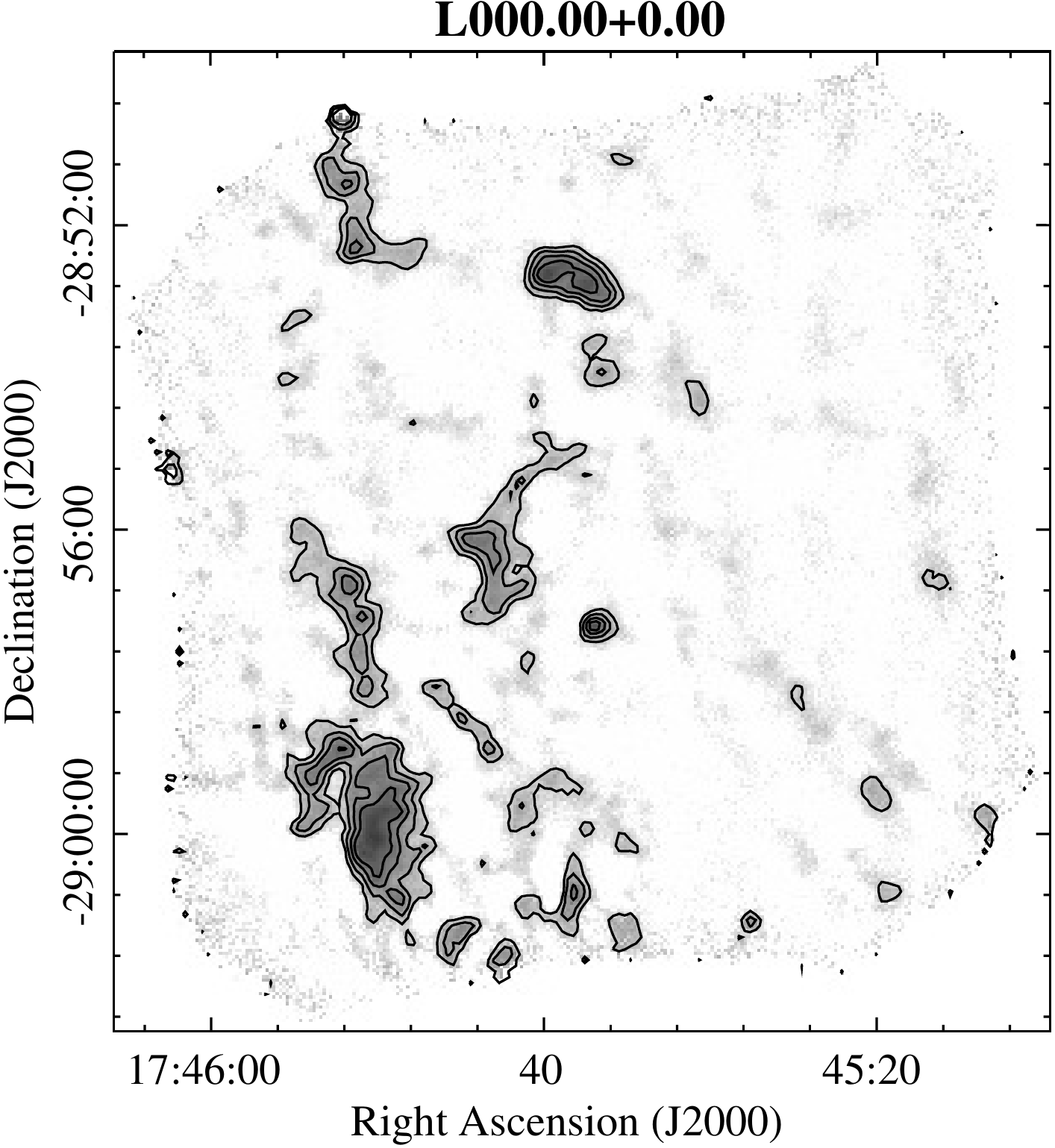}
\includegraphics[scale=0.43]{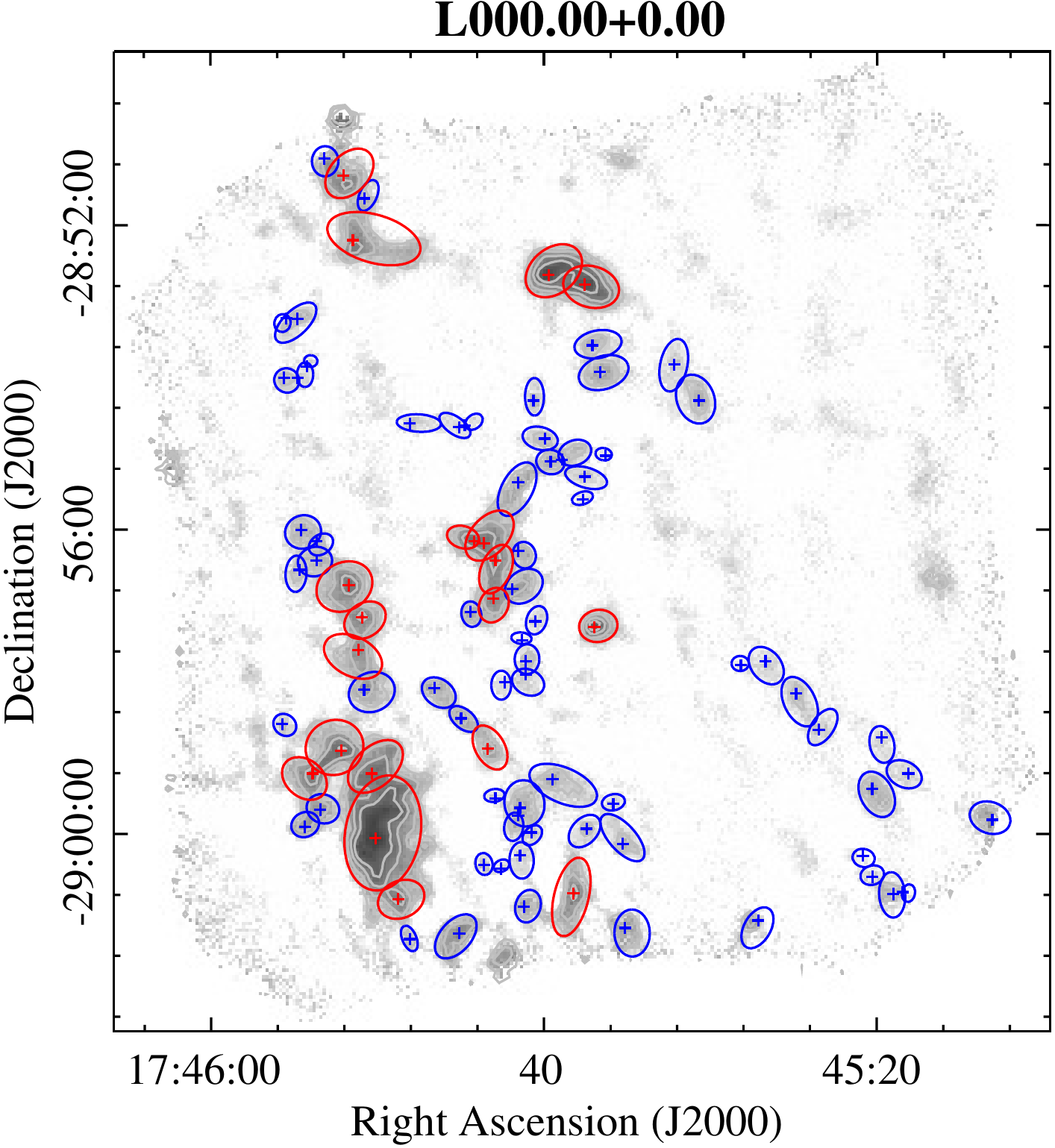}
}\\
\subfloat[L000.15+0.00 map, $\sigma_{rms}=1438$ mJy beam$^{-1}$.]{%
\includegraphics[scale=0.43]{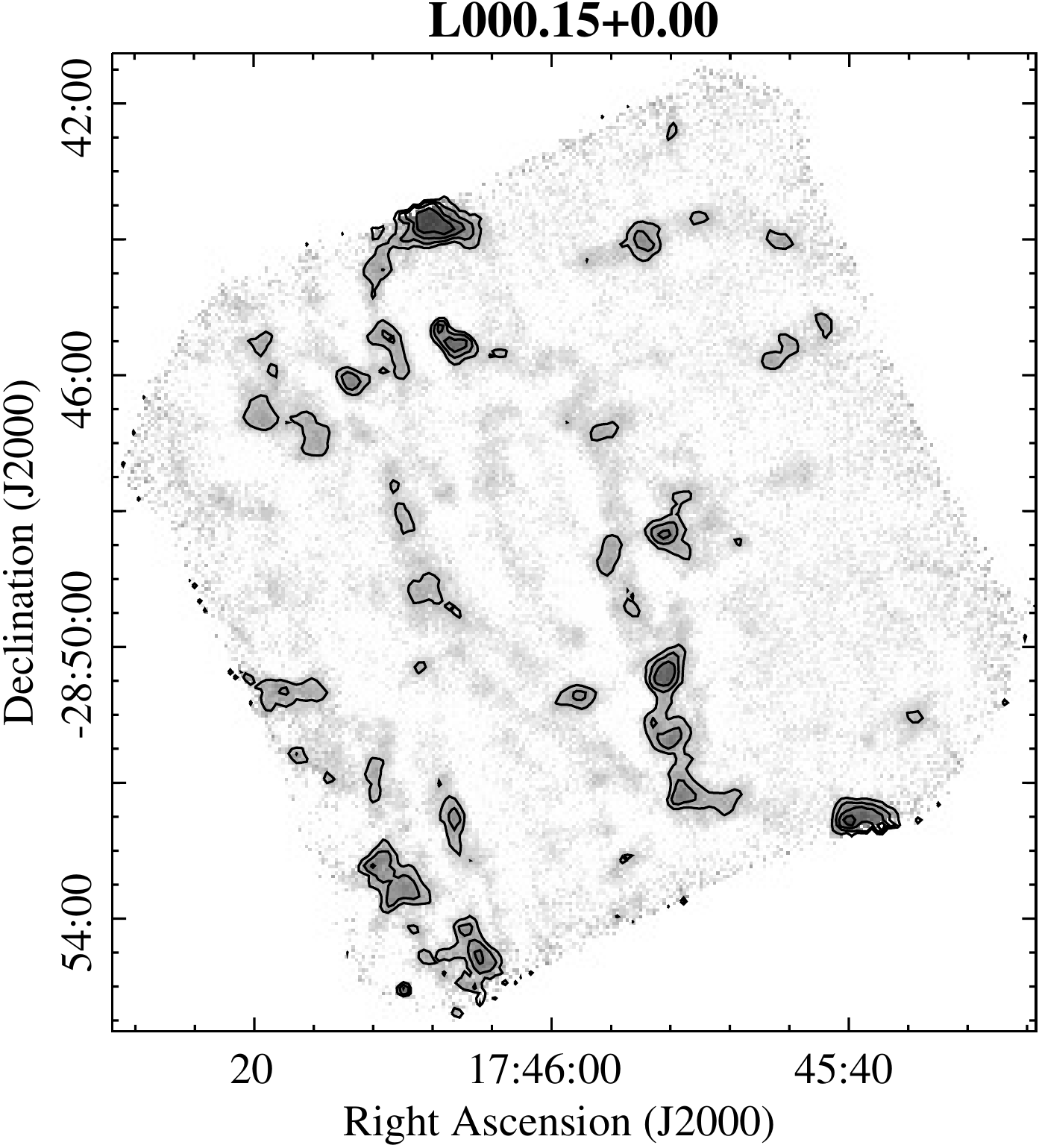}
\includegraphics[scale=0.43]{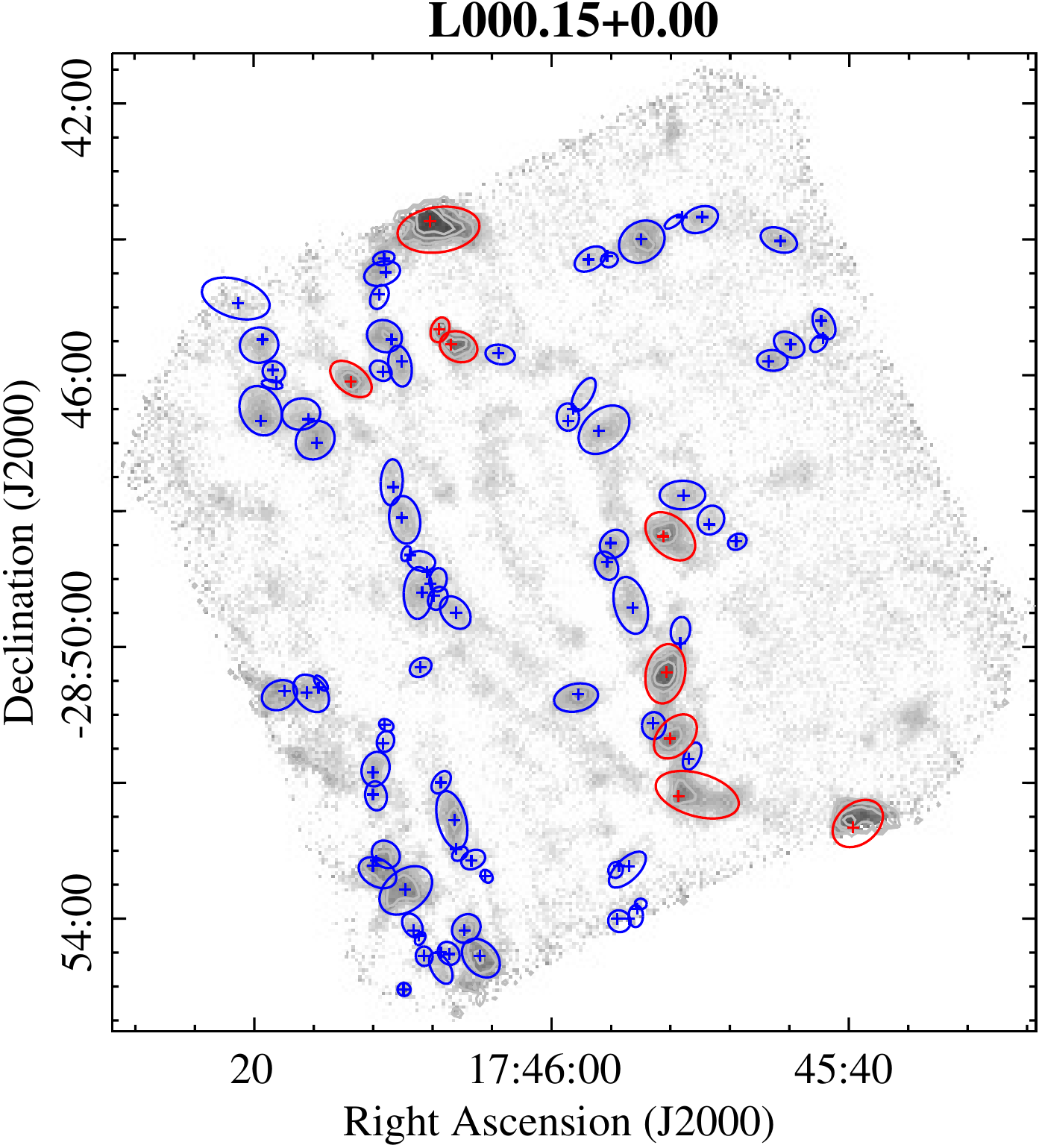}
}\\
	\caption{Continuum maps at 350 $\mu$m of the 107 regions included in our catalog. The beam size of the maps is 8.5\arcsec. Left: Contours are drawn at 3$\sigma$, 6$\sigma$, 10$\sigma$, 15$\sigma$, 30$\sigma$, 50$\sigma$, and 100$\sigma$. Noise level $\sigma_{rms}$ is indicated in the figure for each map. Right: Recovery of substructures with \bolocat\ algorithm. Red and blue regions represent the position of dense, compact cores-type sources and faint substructures, respectively.}

\label{fig:sharcmaps}
\end{figure}

\clearpage
\begin{figure}\ContinuedFloat 
\center
\subfloat[L000.30+0.00 map, $\sigma_{rms}=1086$ mJy beam$^{-1}$.]{
\includegraphics[scale=0.43]{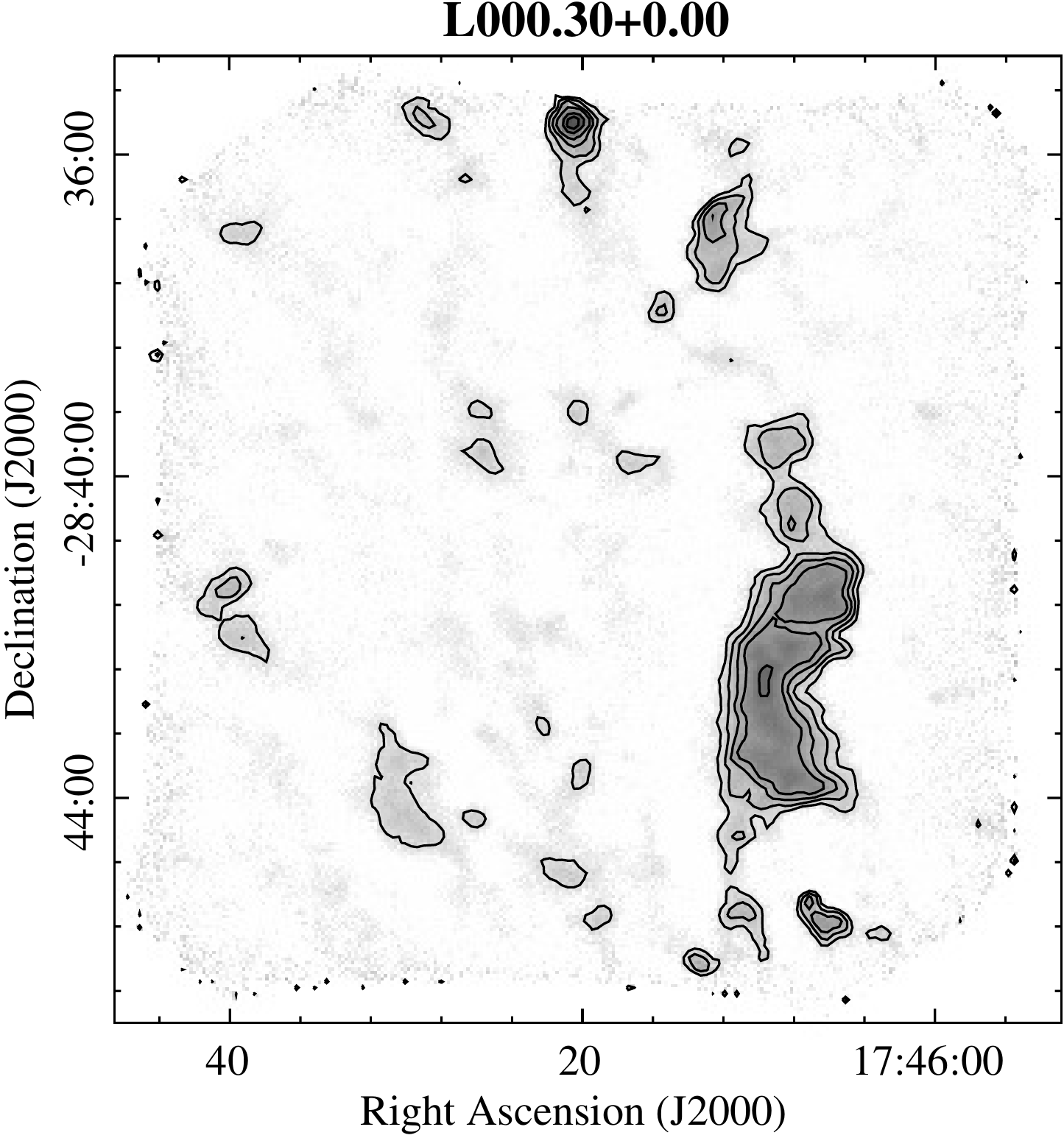}
\includegraphics[scale=0.43]{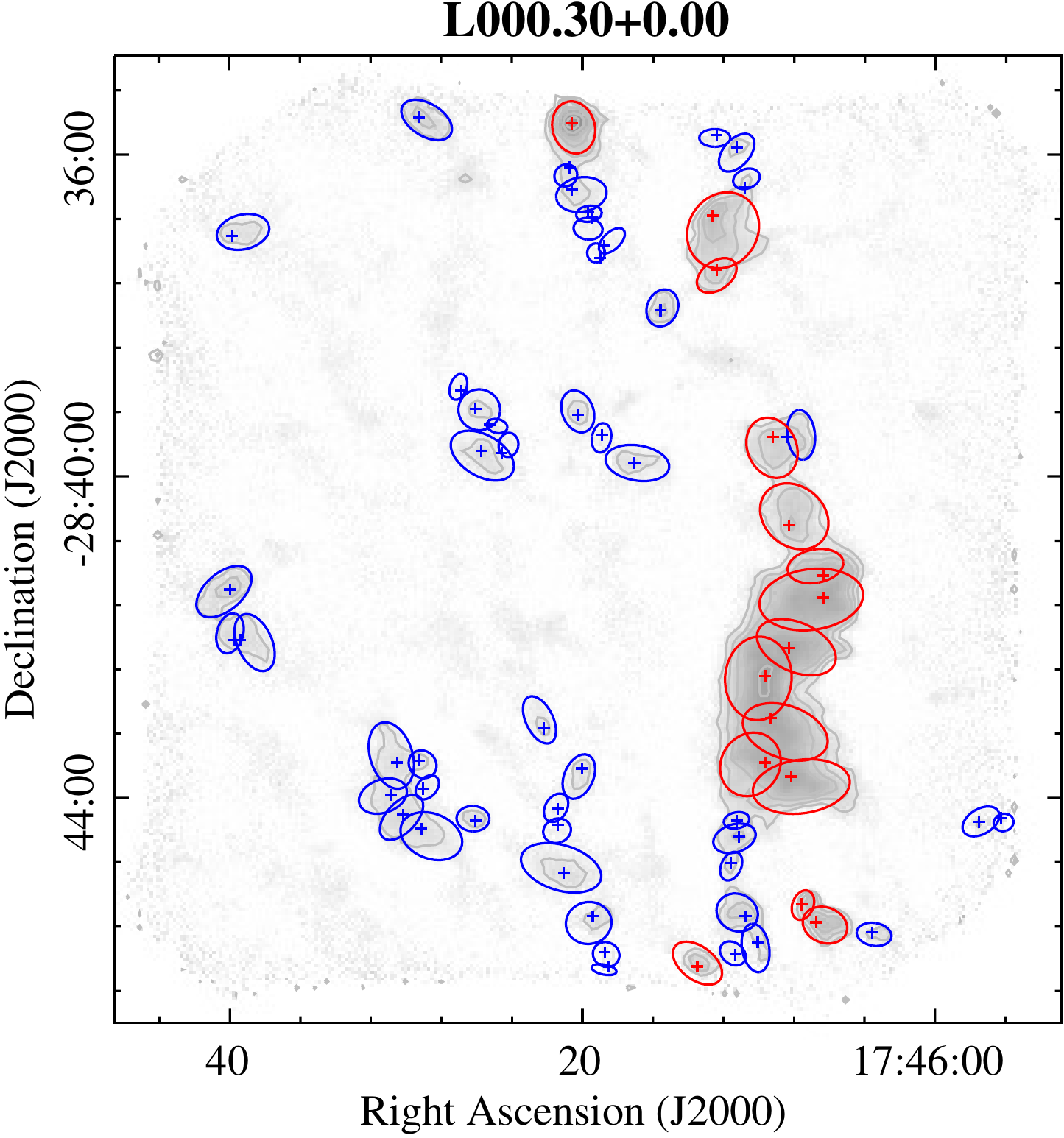}
}\\
\subfloat[L029.95-0.05 map, $\sigma_{rms}=514$ mJy beam$^{-1}$.]{
\includegraphics[scale=0.43]{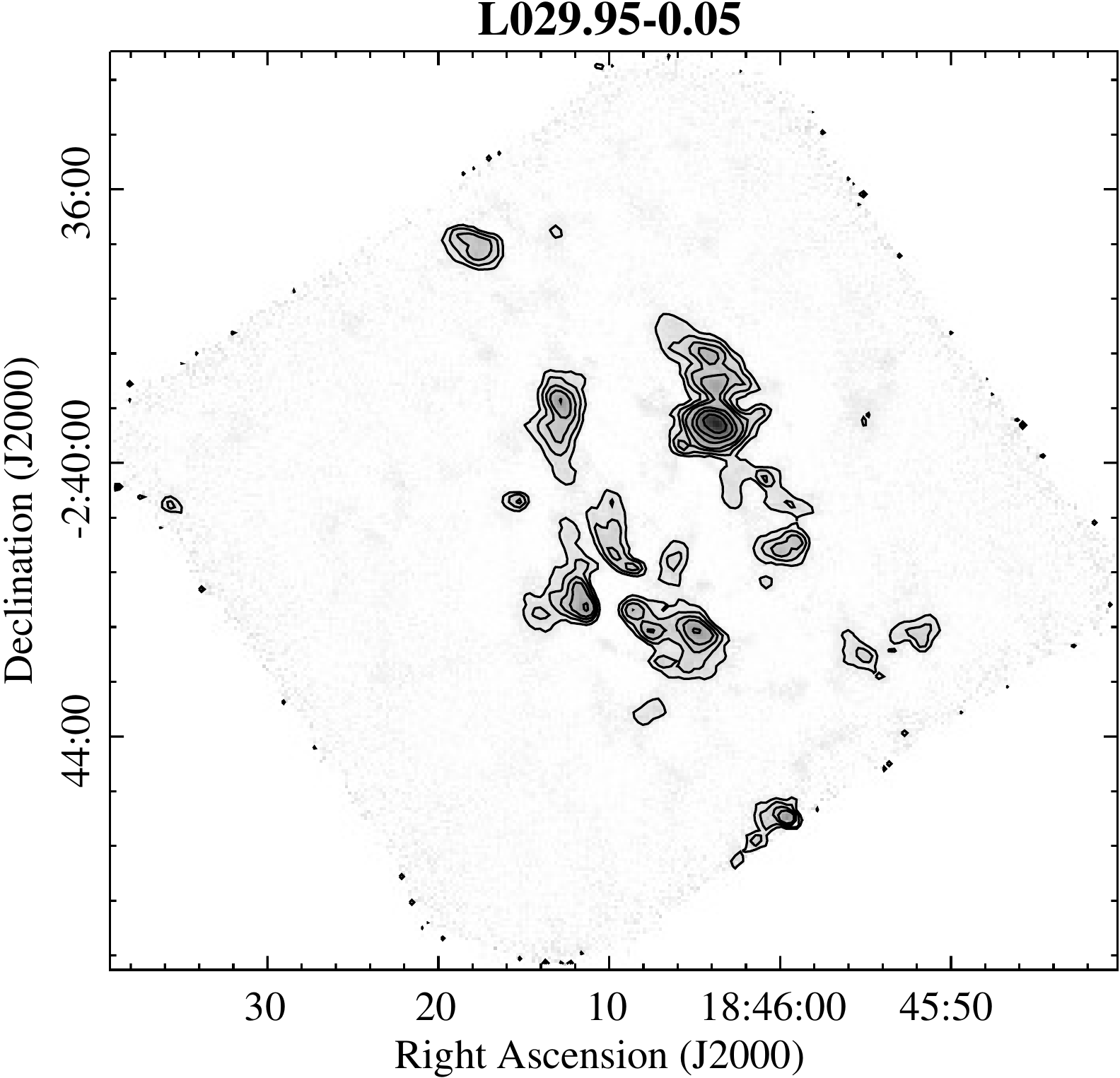}
\includegraphics[scale=0.43]{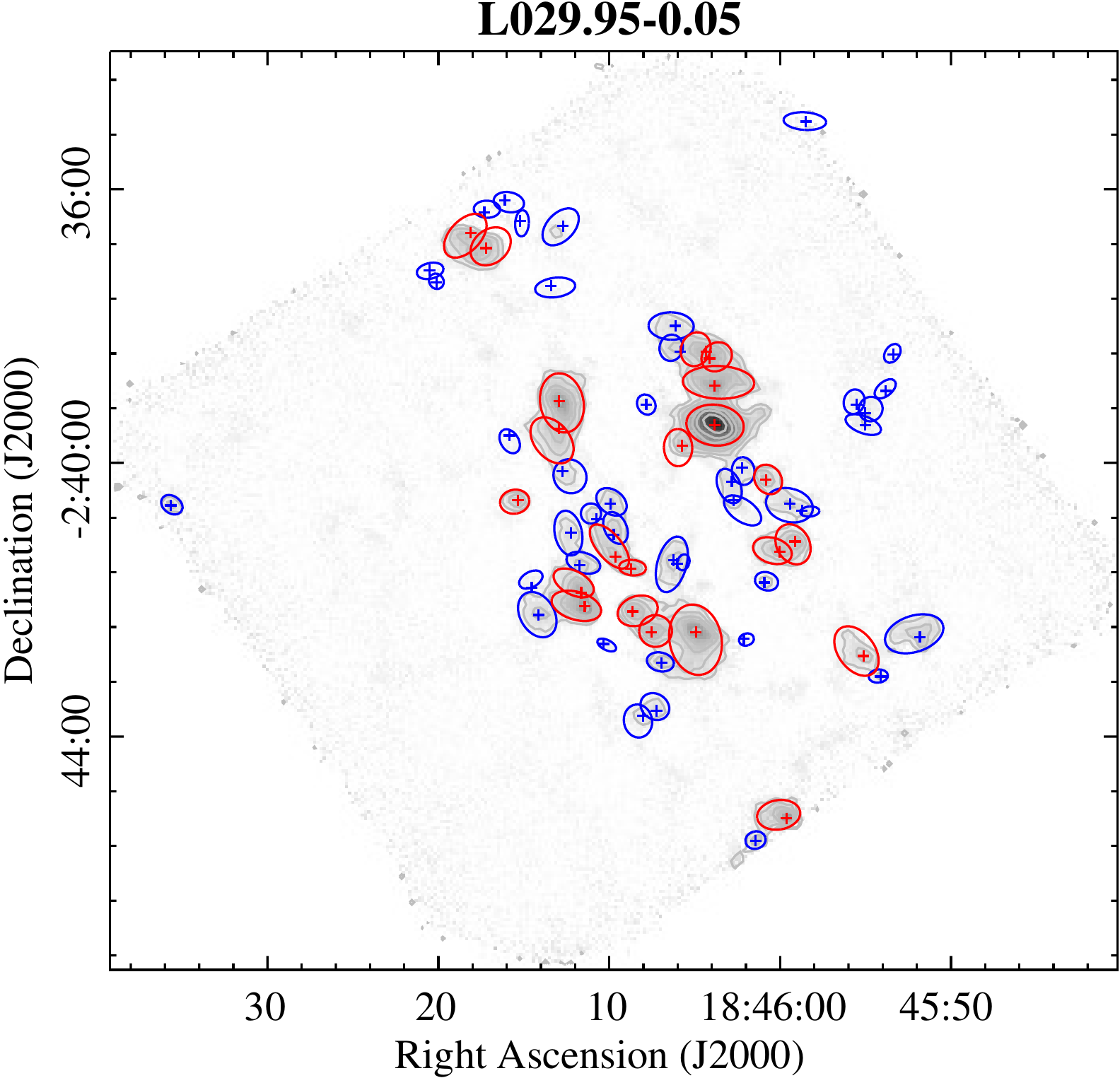}
}\\
\subfloat[L030.00+0.00 map, $\sigma_{rms}=462$ mJy beam$^{-1}$.]{
\includegraphics[scale=0.43]{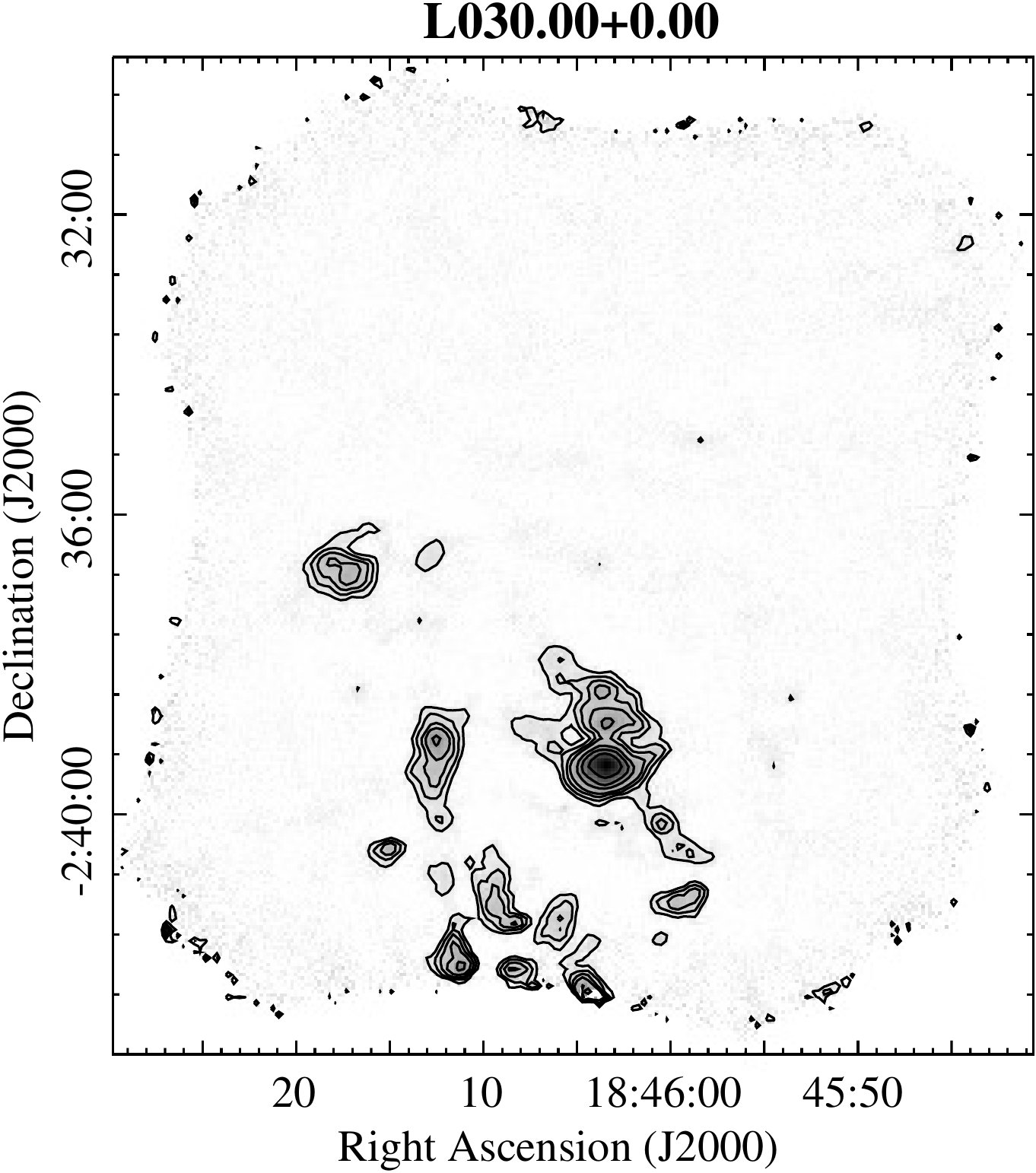}
\includegraphics[scale=0.43]{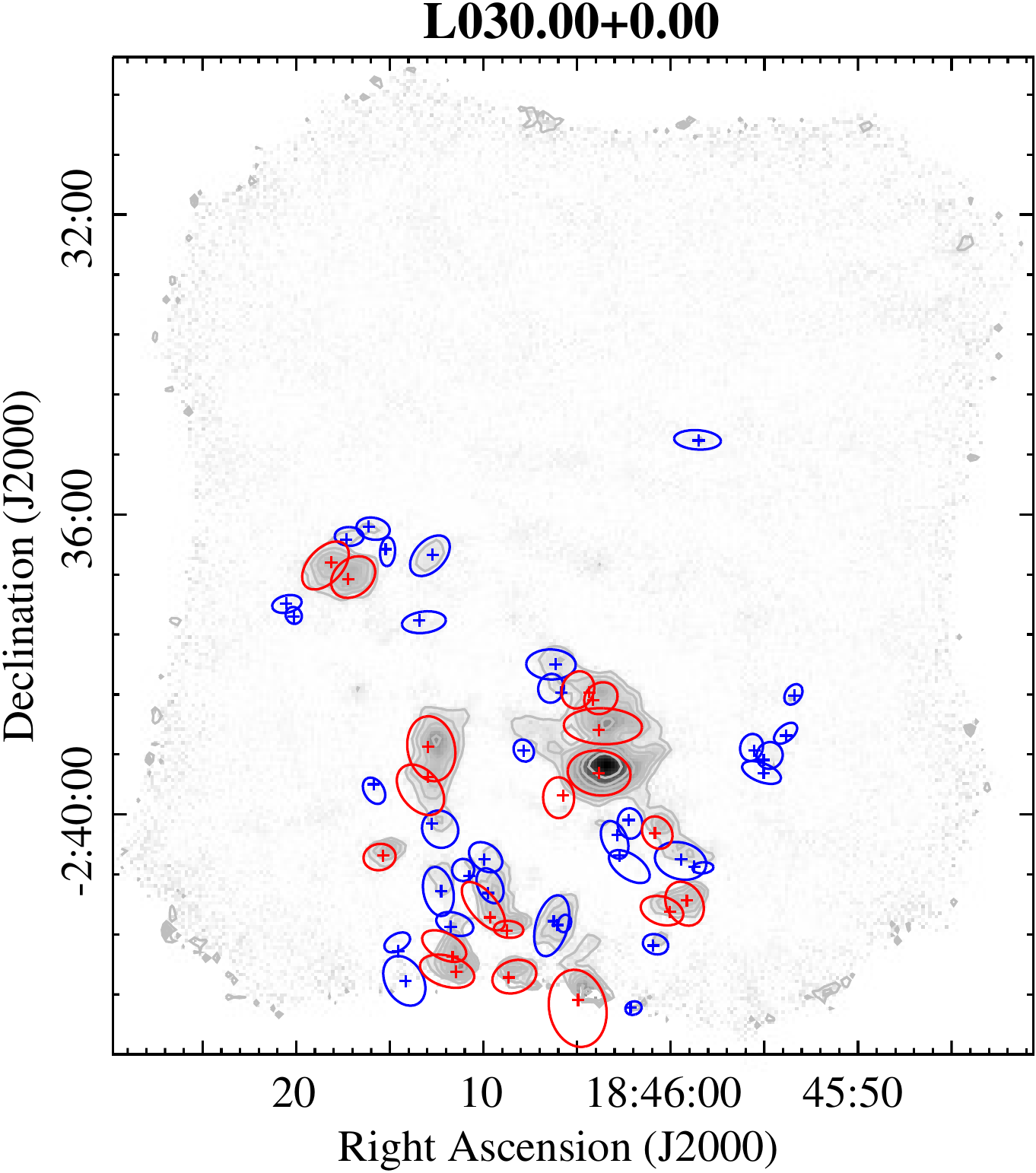}
}\\
\caption{Continuation}
\end{figure}

\clearpage
\begin{figure}\ContinuedFloat 
\center
\subfloat[L030.15+0.00 map, $\sigma_{rms}=426$ mJy beam$^{-1}$.]{
\includegraphics[scale=0.43]{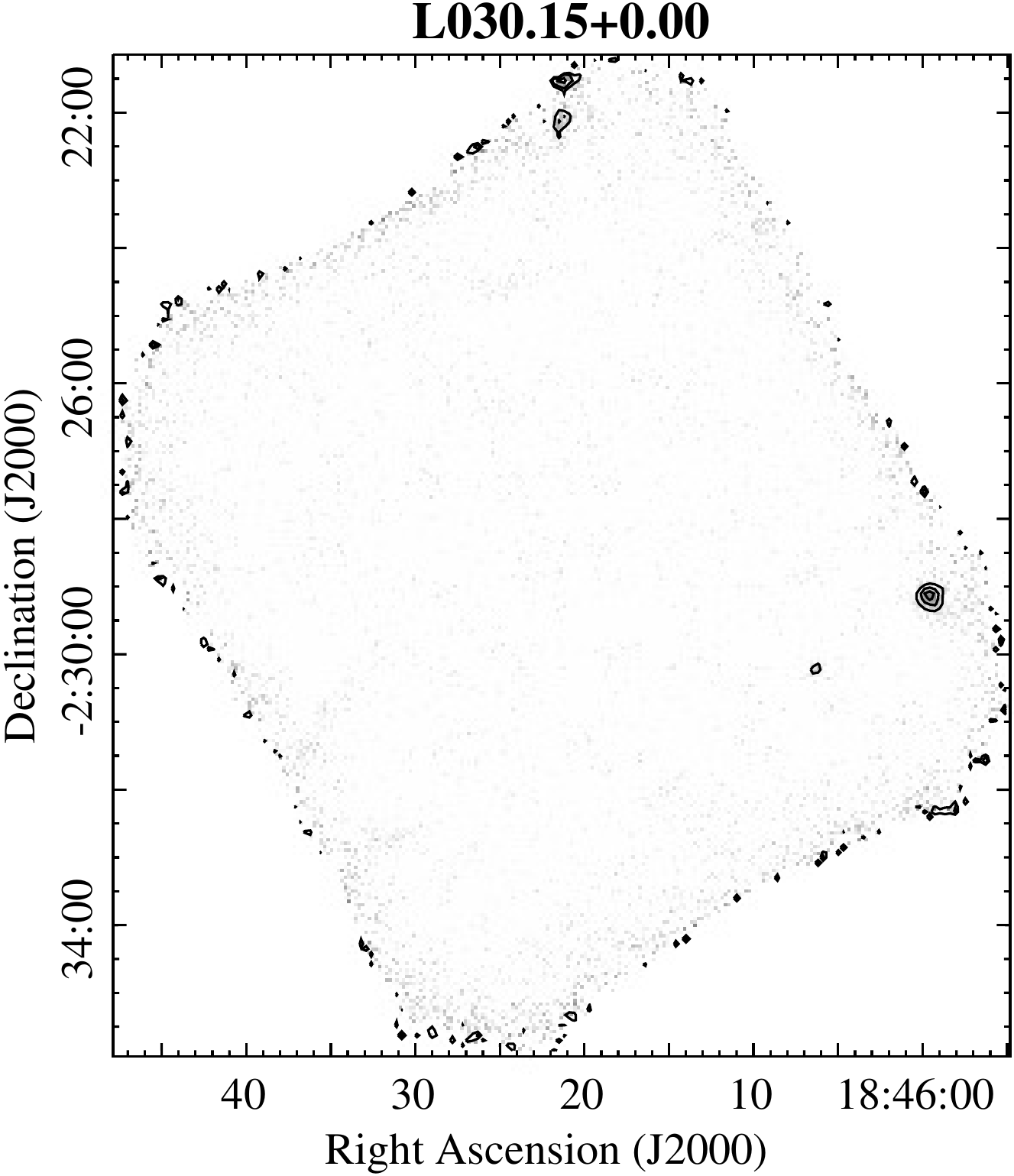}
\includegraphics[scale=0.43]{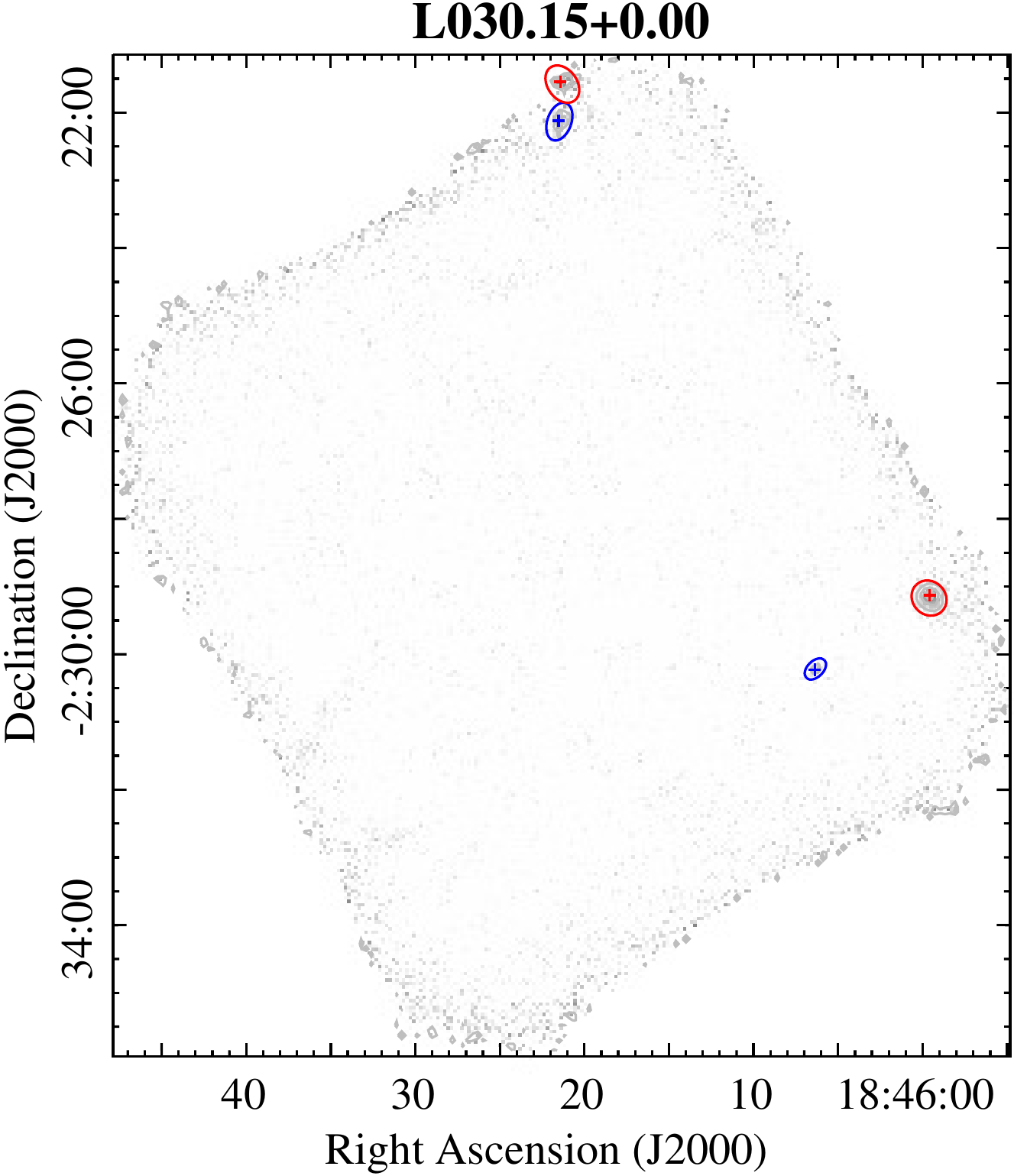}
}\\
\subfloat[L030.30+0.00 map, $\sigma_{rms}=496$ mJy beam$^{-1}$.]{
\includegraphics[scale=0.43]{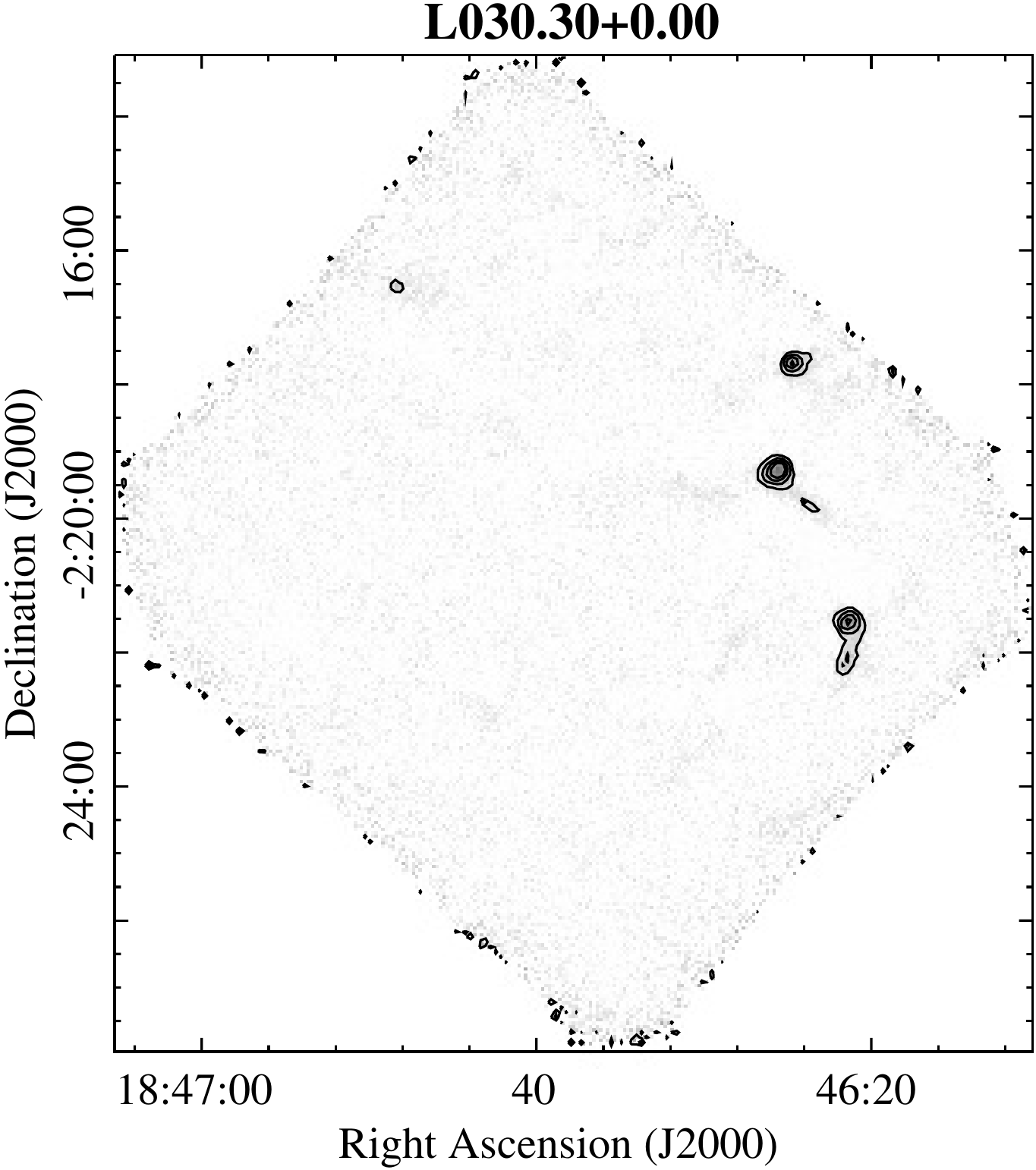}
\includegraphics[scale=0.43]{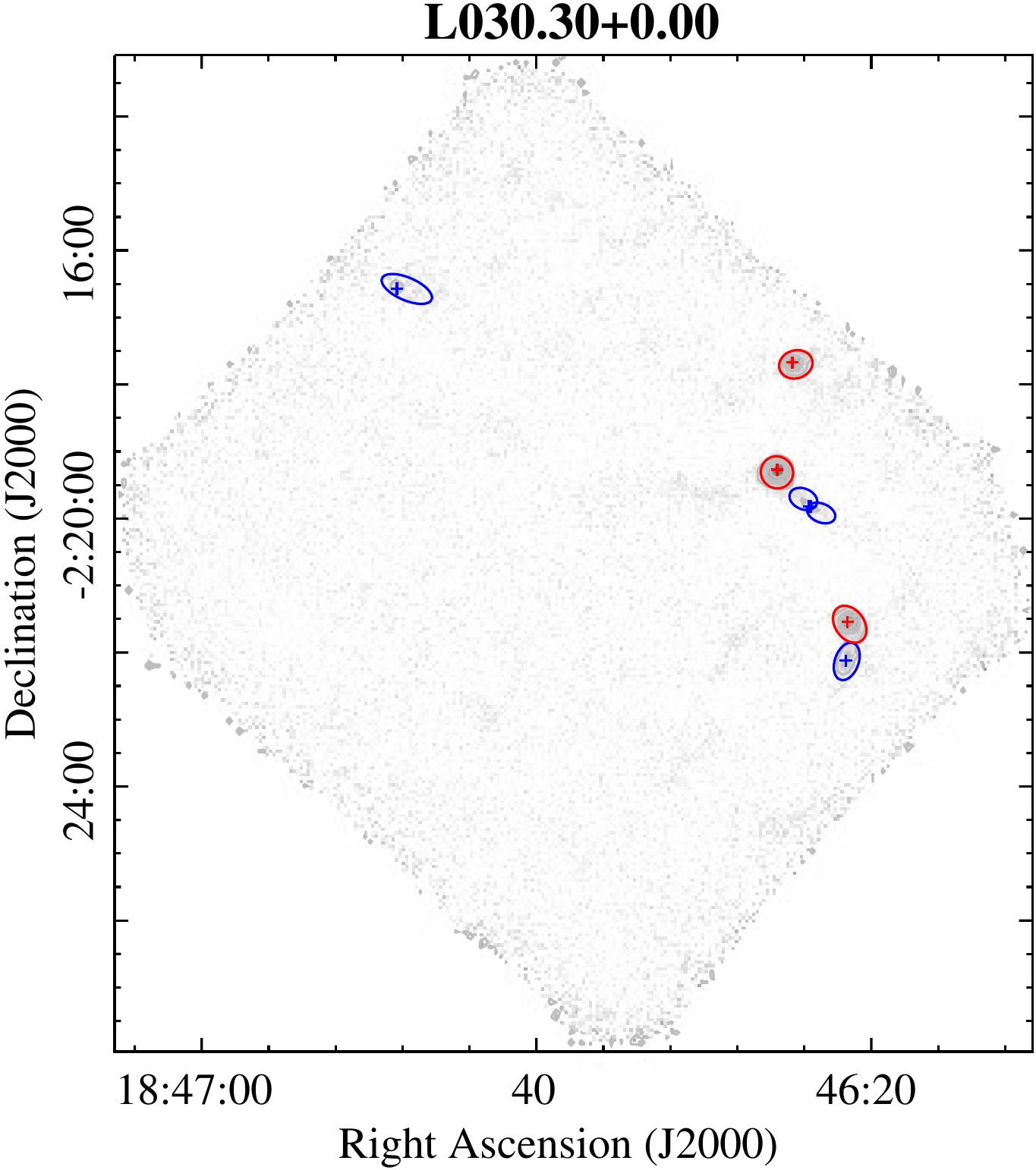}
}\\
\subfloat[L030.45+0.00 map, $\sigma_{rms}=603$ mJy beam$^{-1}$.]{
\includegraphics[scale=0.43]{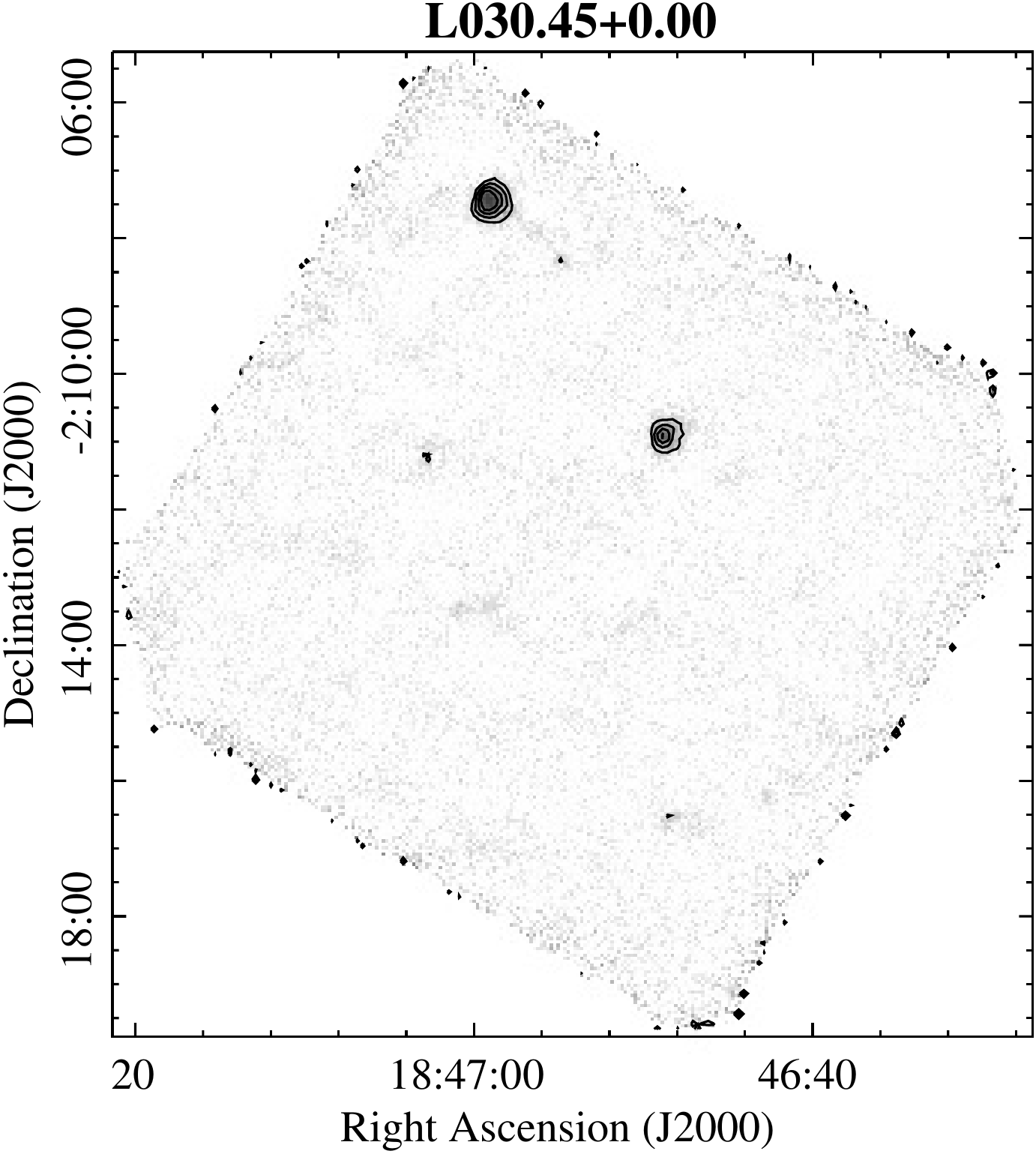}
\includegraphics[scale=0.43]{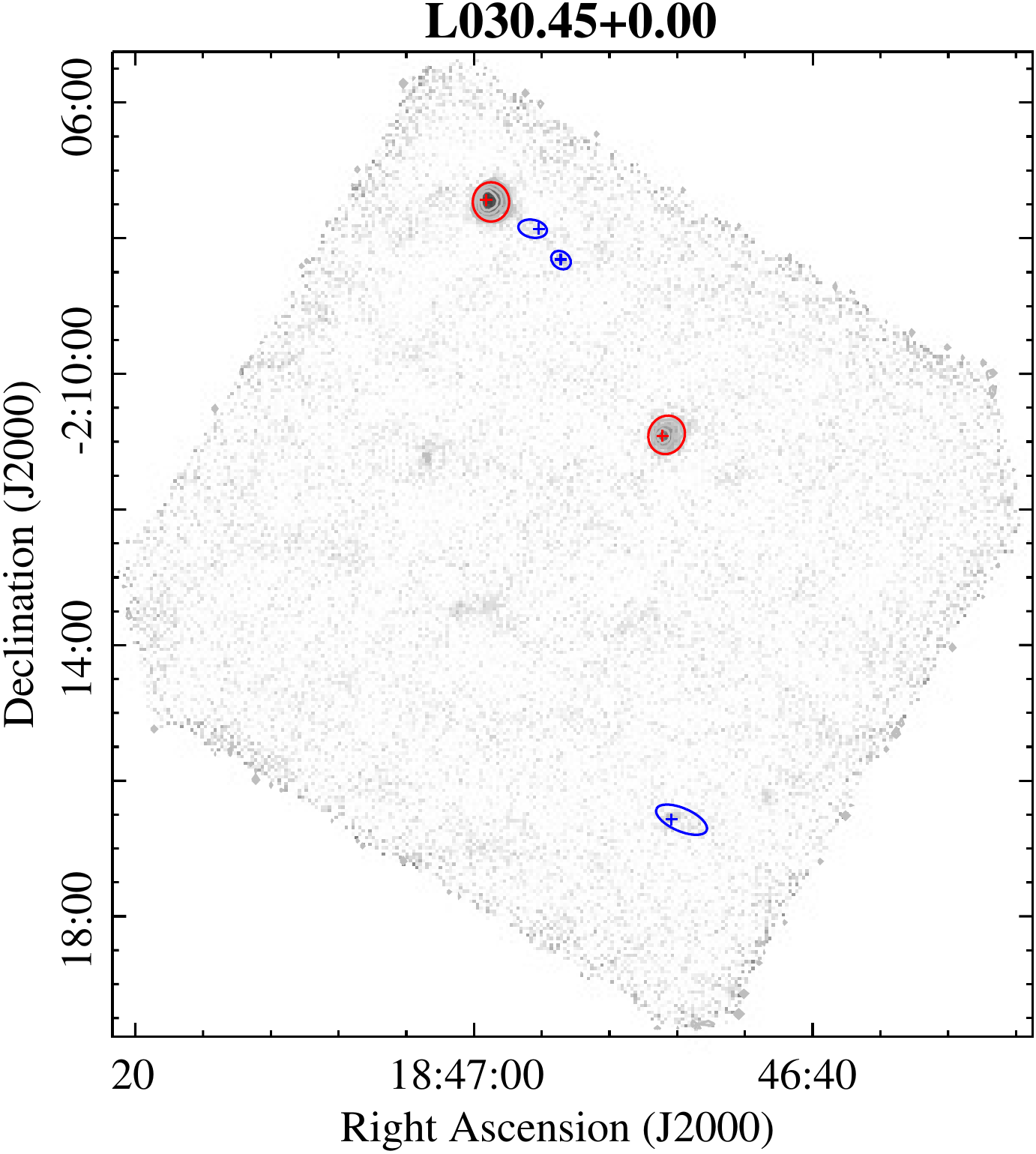}
}\\
\caption{Continuation}
\end{figure}

\clearpage
\begin{figure}\ContinuedFloat 
\center
\subfloat[L030.60+0.00 map, $\sigma_{rms}=888$ mJy beam$^{-1}$.]{
\includegraphics[scale=0.43]{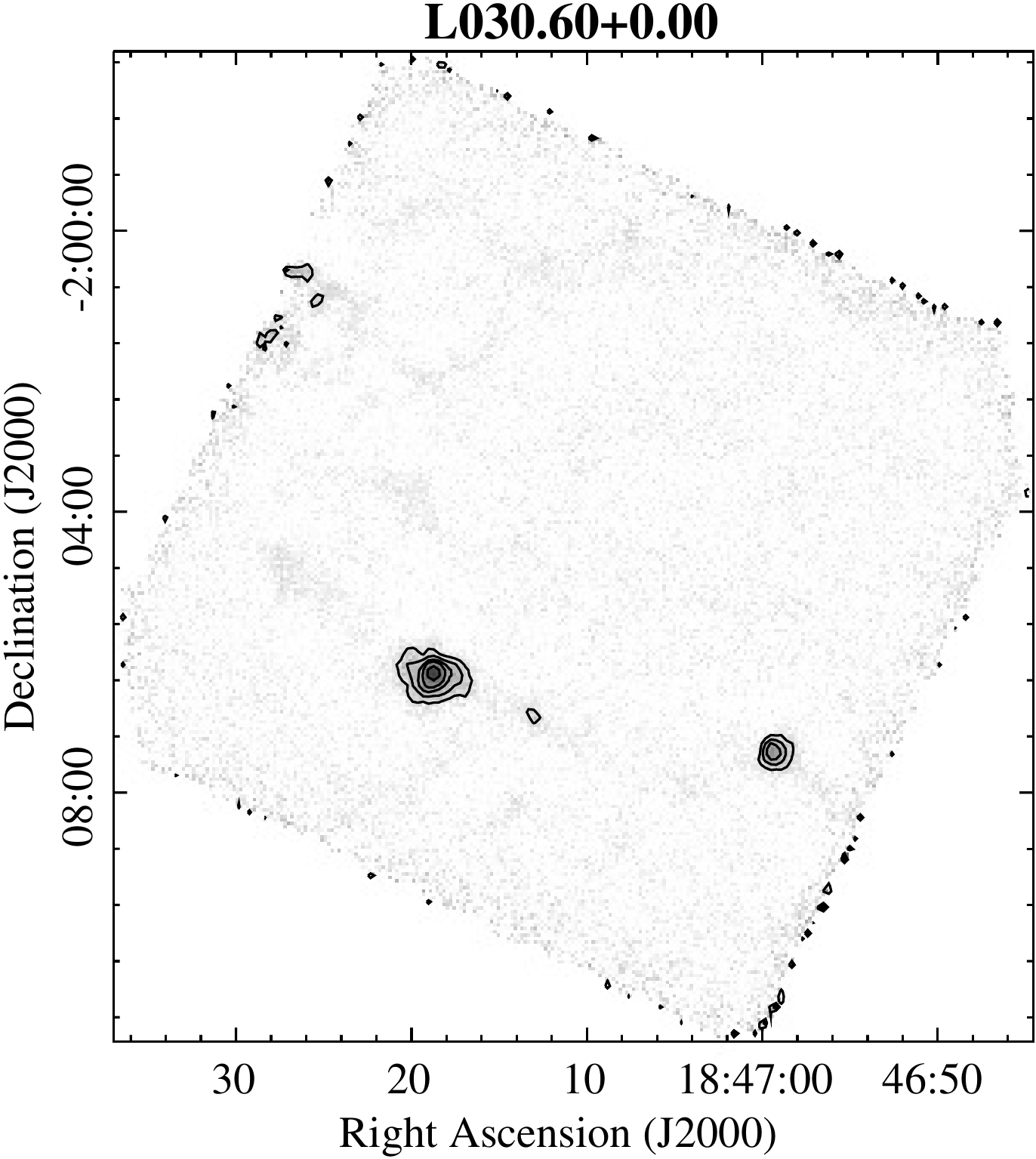}
\includegraphics[scale=0.43]{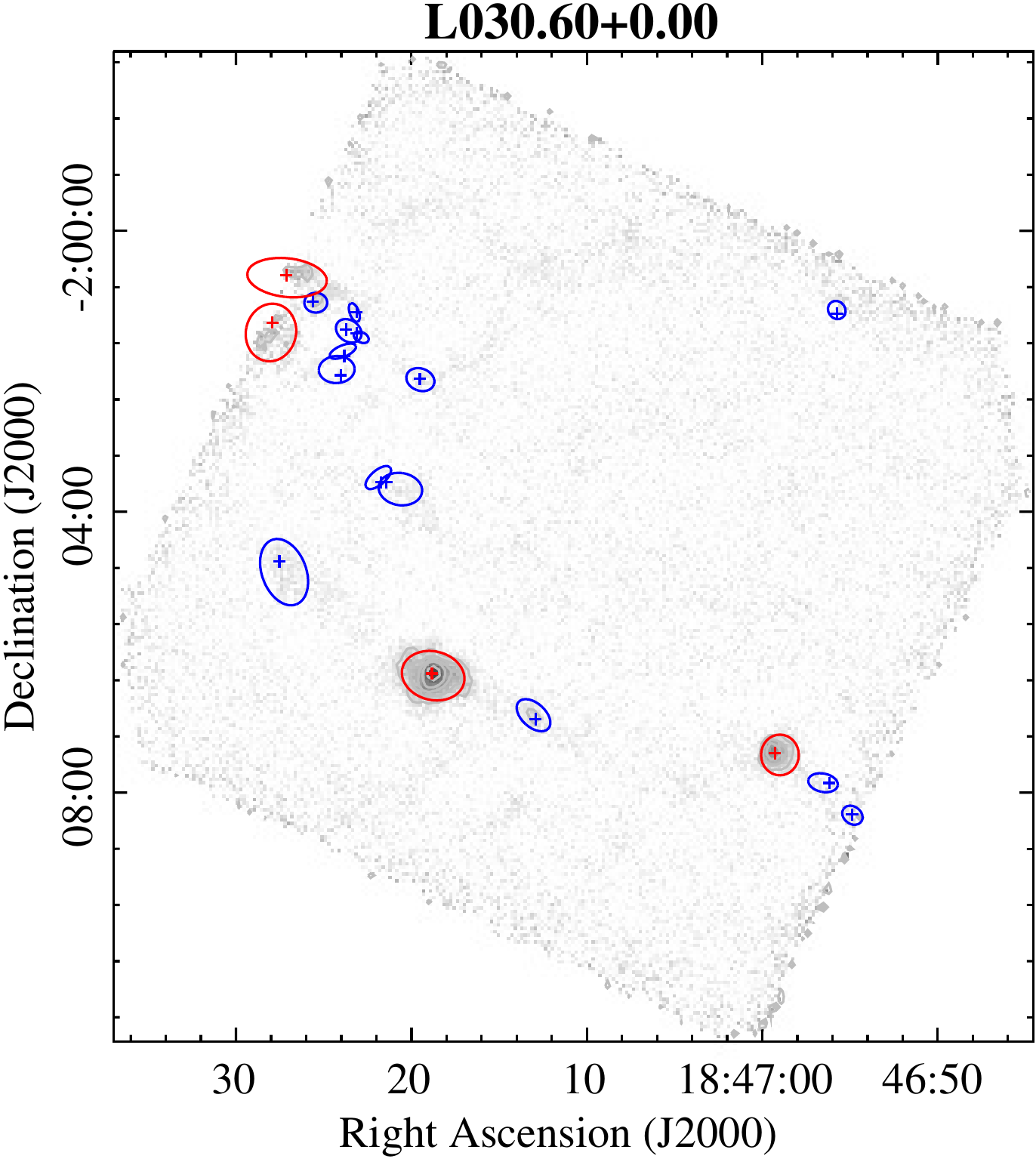}
}\\
\subfloat[L030.70-0.07 map, $\sigma_{rms}=510$ mJy beam$^{-1}$.]{
\includegraphics[scale=0.43]{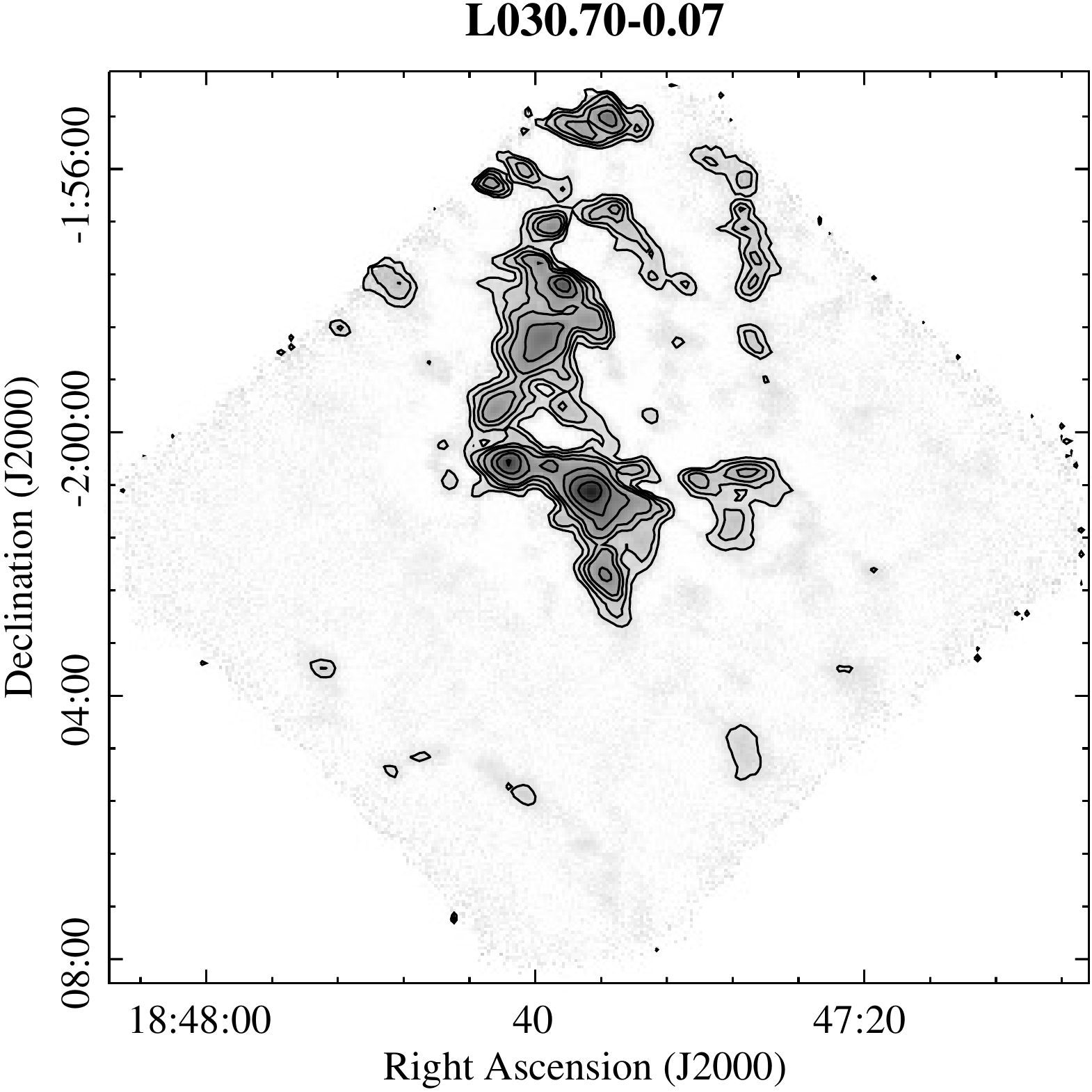}
\includegraphics[scale=0.43]{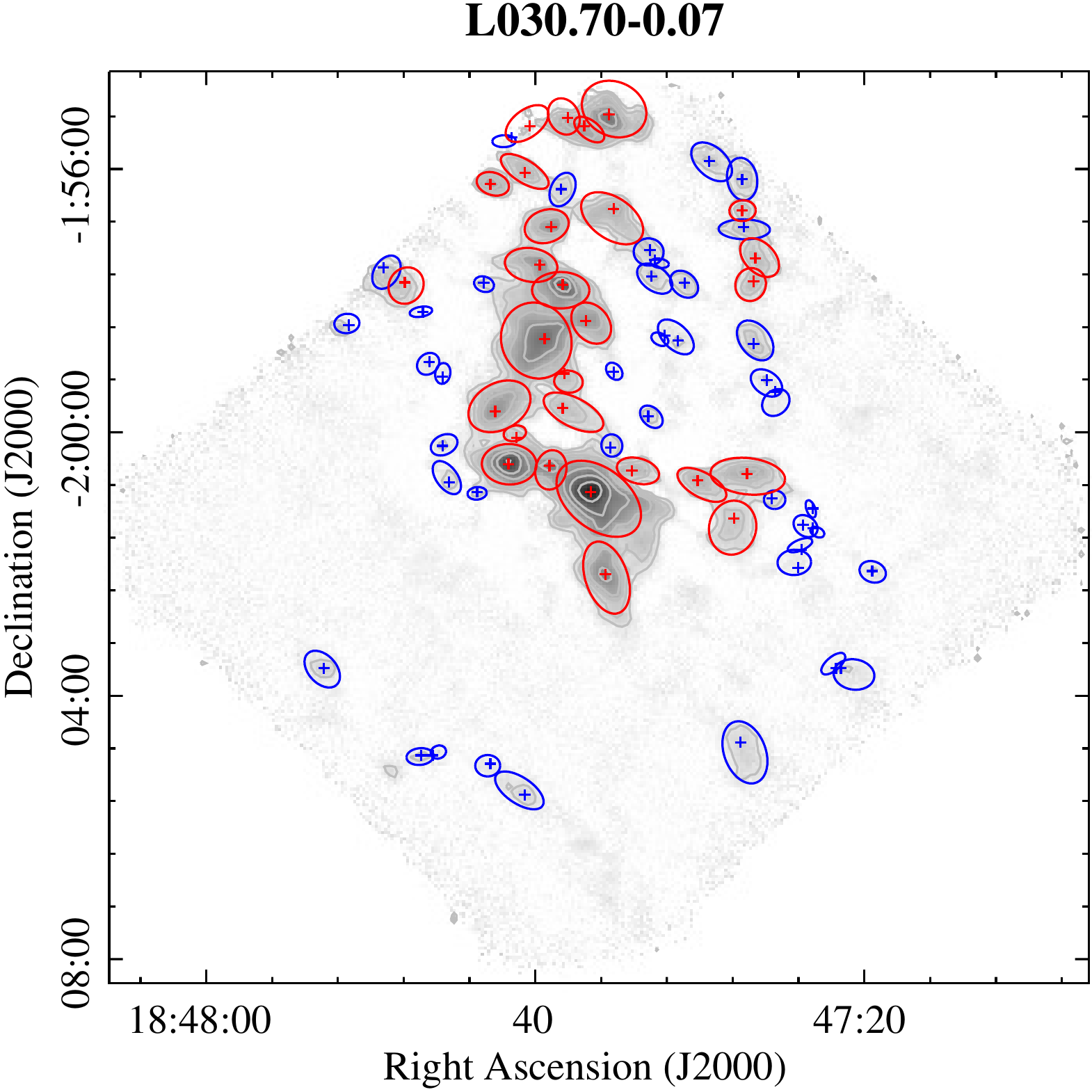}
}\\
\subfloat[L030.80-0.05 map, $\sigma_{rms}=529$ mJy beam$^{-1}$.]{
\includegraphics[scale=0.43]{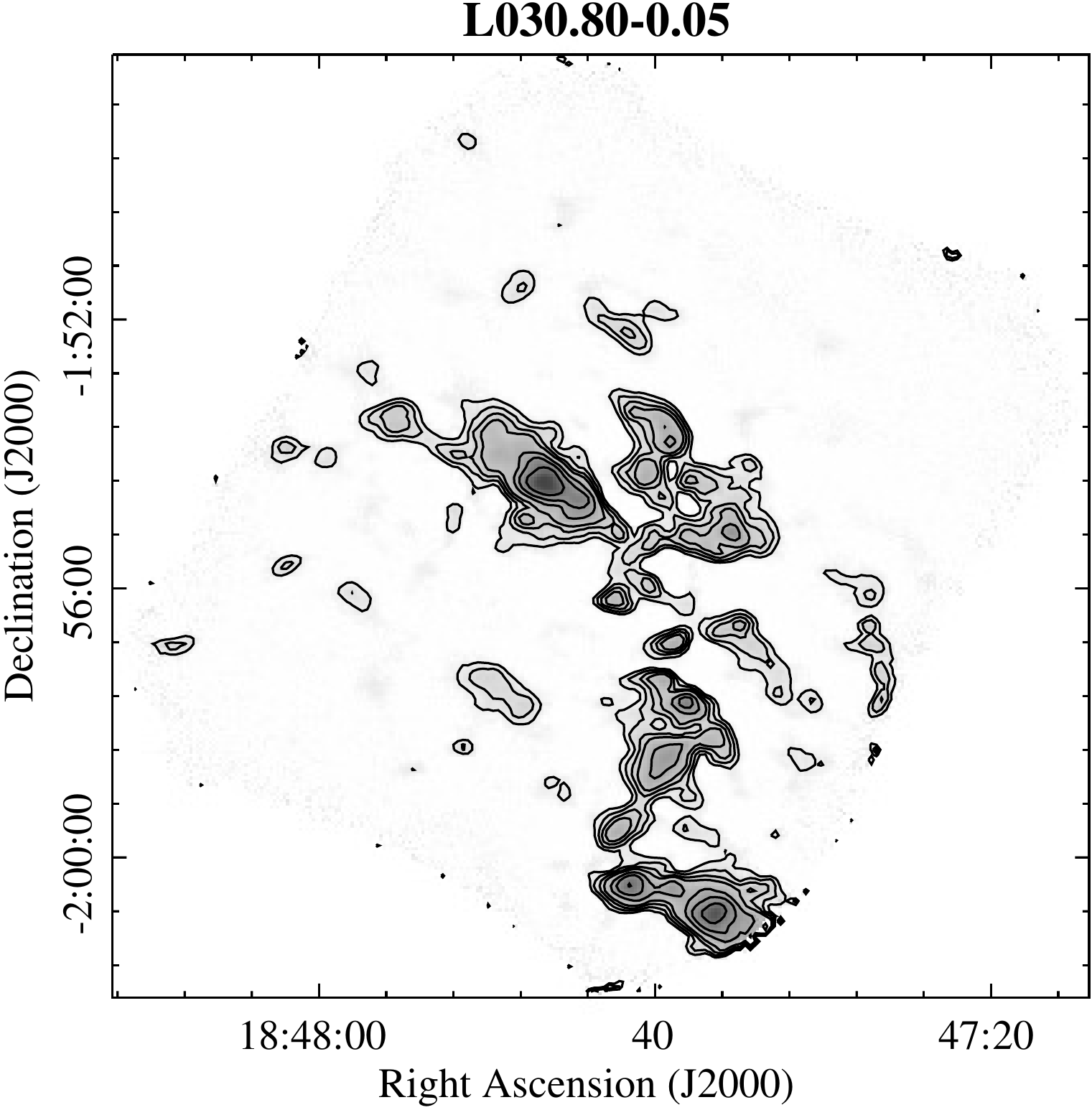}
\includegraphics[scale=0.43]{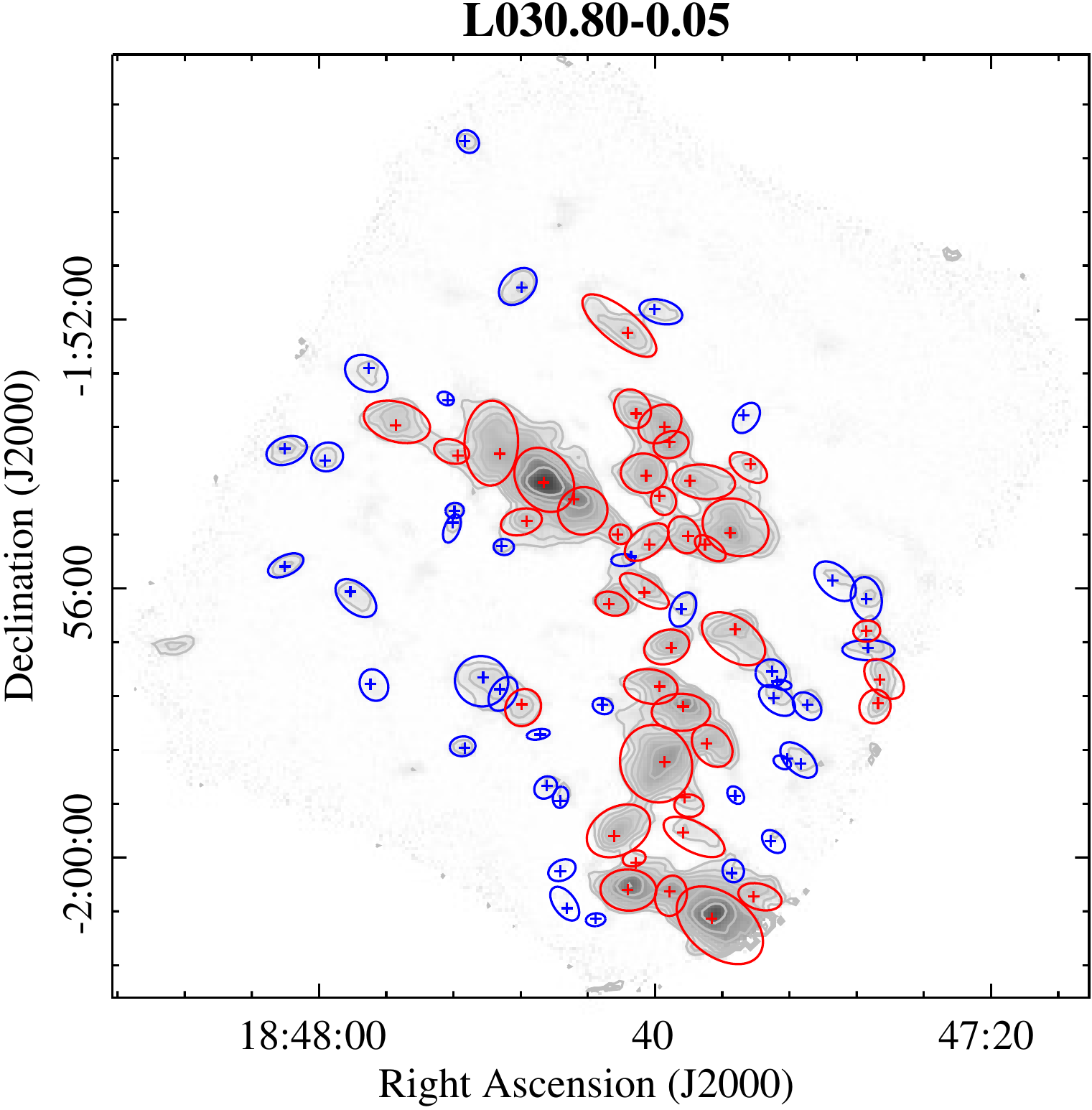}
}\\
\caption{Continuation}
\end{figure}

\clearpage
\begin{figure}\ContinuedFloat 
\center
\subfloat[L030.88+0.13 map, $\sigma_{rms}=367$ mJy beam$^{-1}$.]{
\includegraphics[scale=0.43]{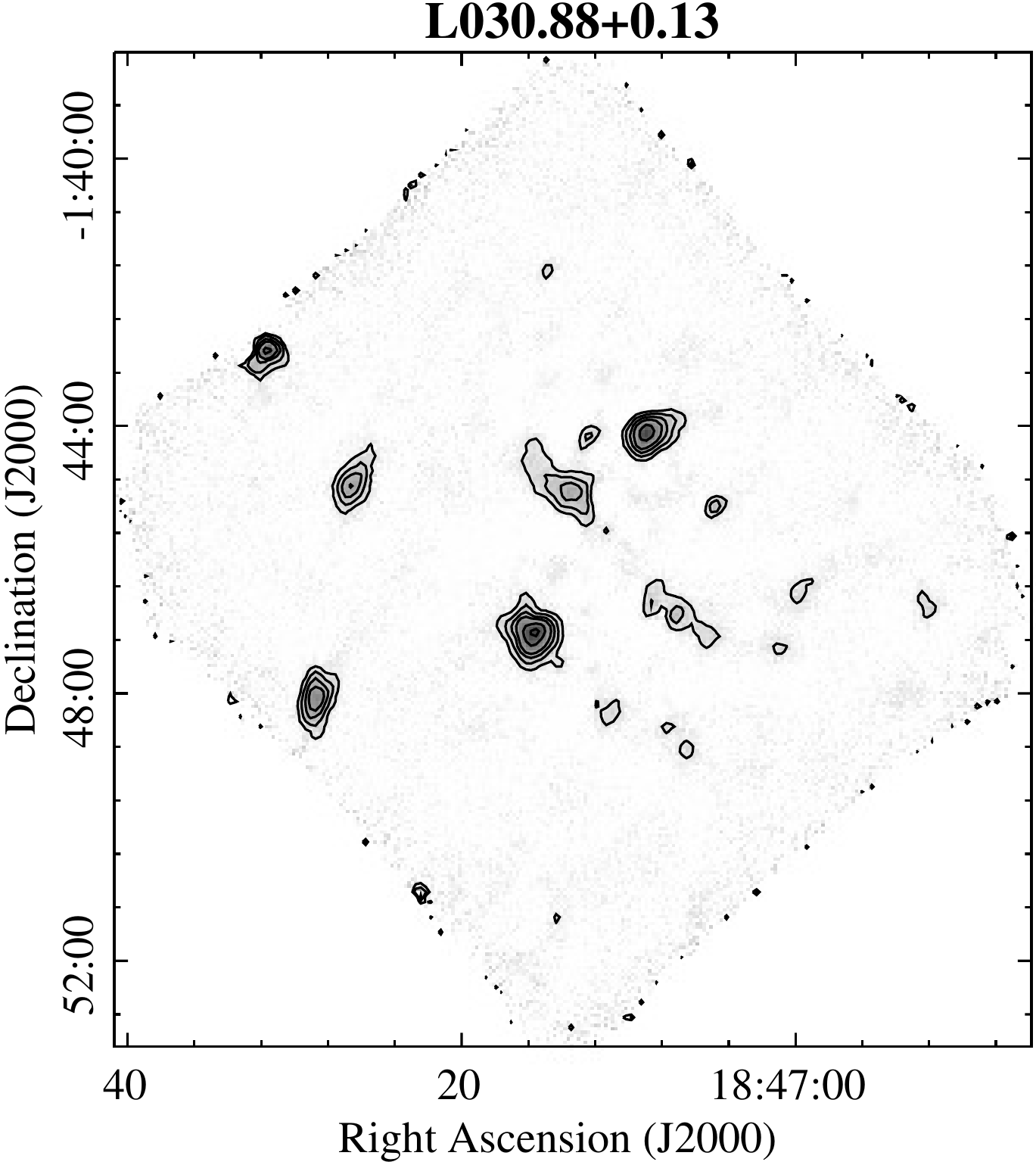}
\includegraphics[scale=0.43]{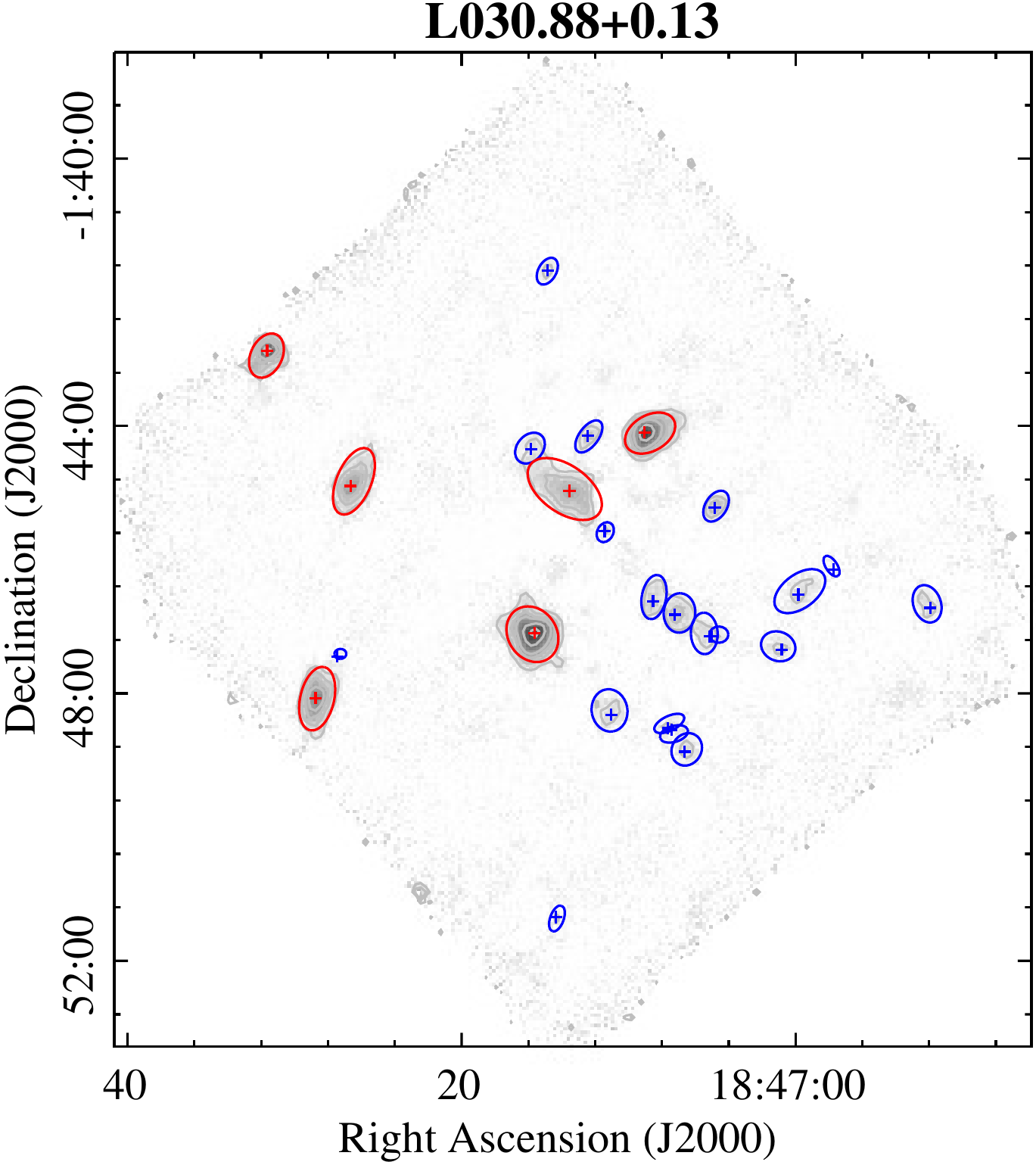}
}\\
\subfloat[L034.26+0.15 map, $\sigma_{rms}=395$ mJy beam$^{-1}$. Additional contour is drawn at 200$\sigma$ and 400$\sigma$.]{
\includegraphics[scale=0.43]{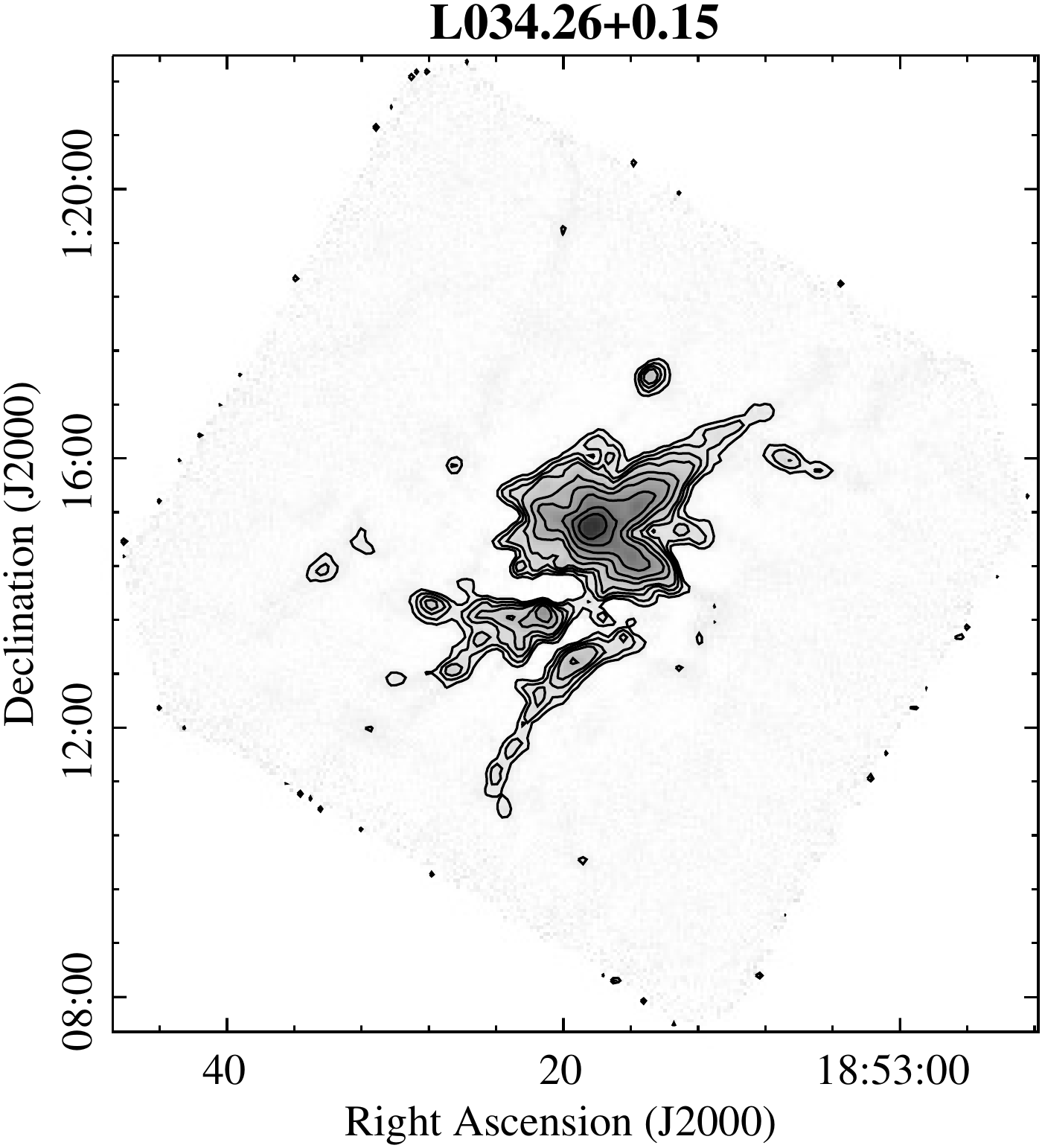}
\includegraphics[scale=0.43]{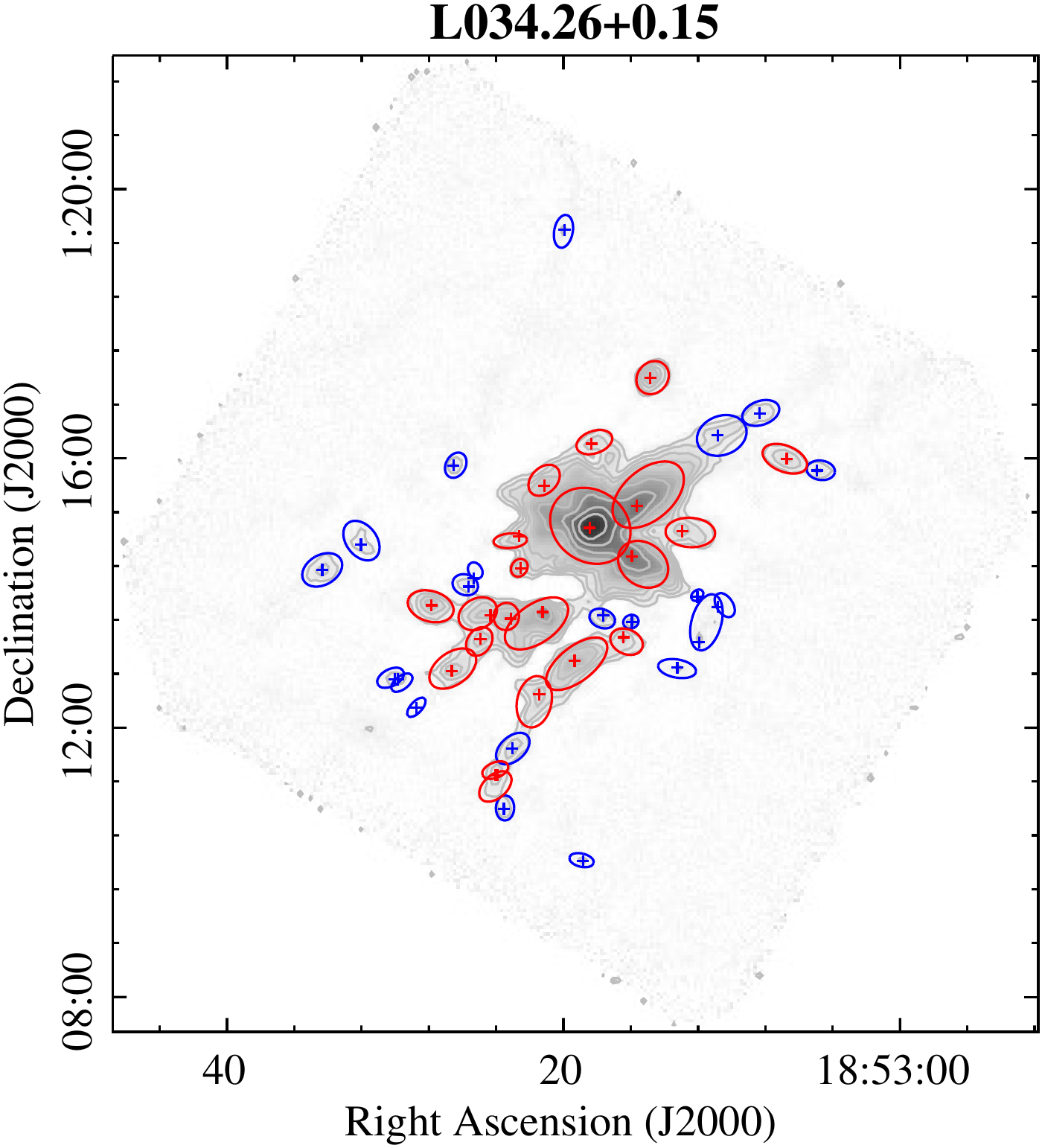}
}\\
\subfloat[L076.16+0.09 map, $\sigma_{rms}=249$ mJy beam$^{-1}$.]{
\includegraphics[scale=0.43]{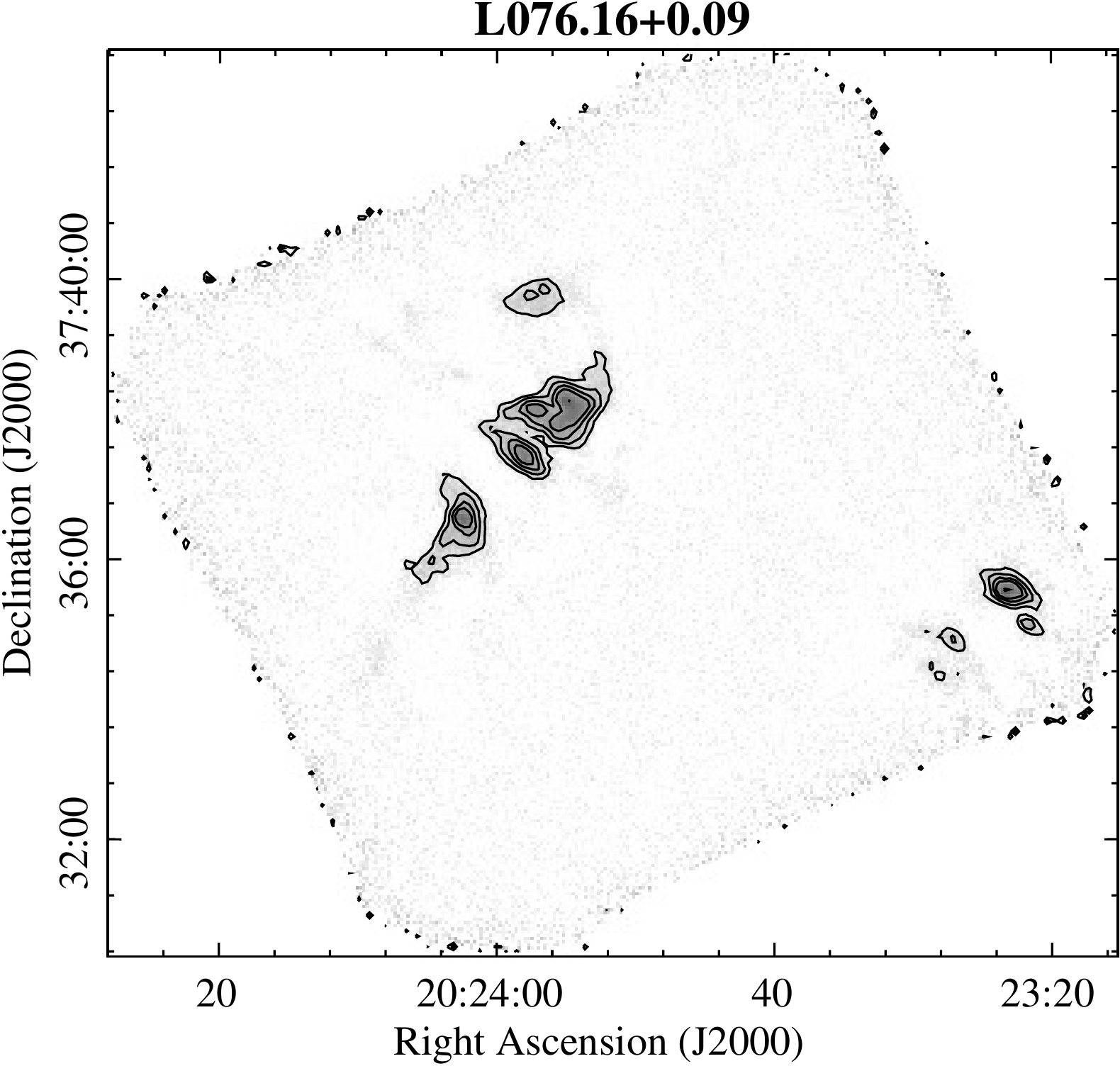}
\includegraphics[scale=0.43]{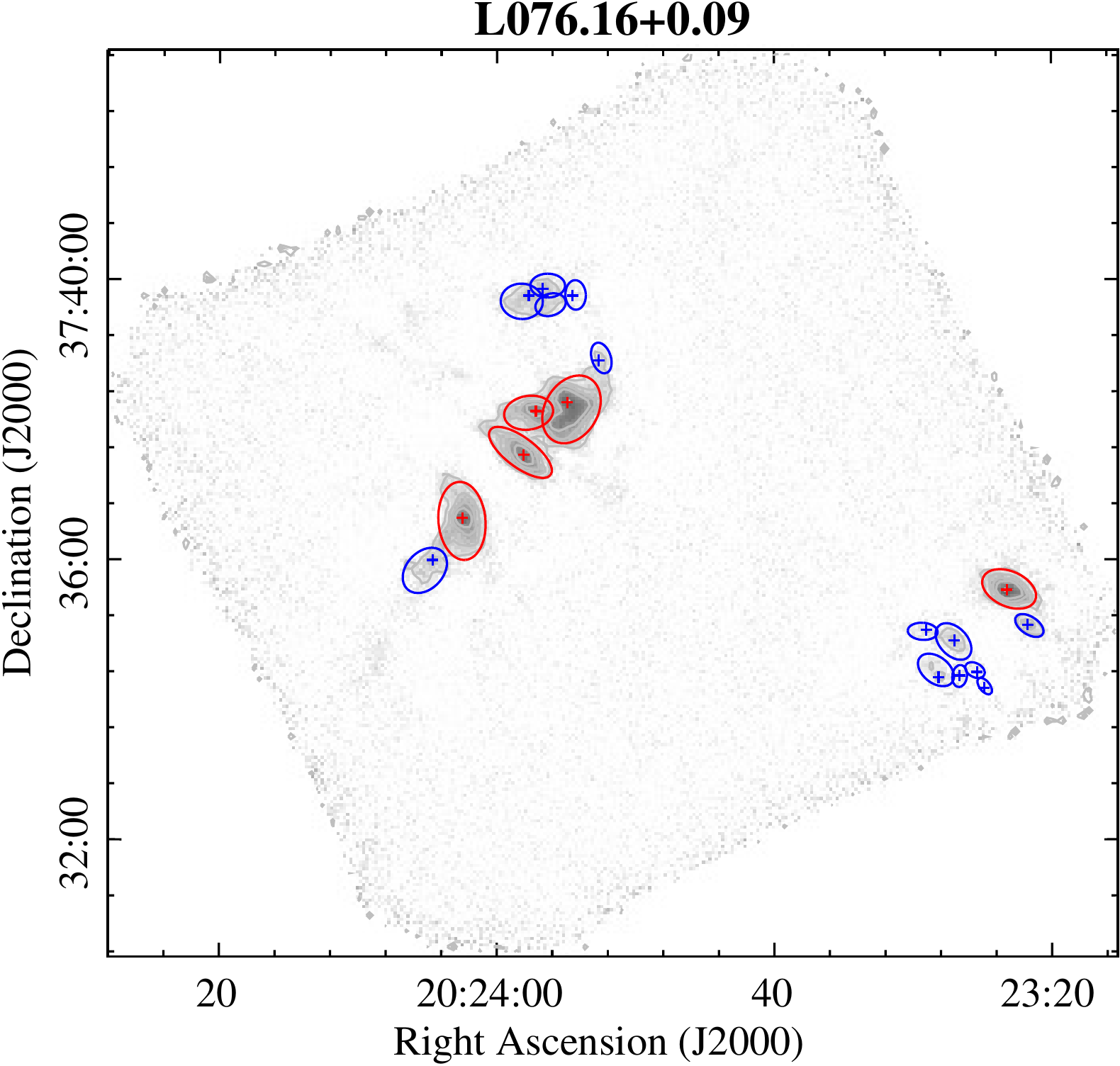}
}\\
\caption{Continuation}
\end{figure}

\clearpage
\begin{figure}\ContinuedFloat 
\center
\subfloat[L077.93+0.02 map, $\sigma_{rms}=389$ mJy beam$^{-1}$.]{
\includegraphics[scale=0.43]{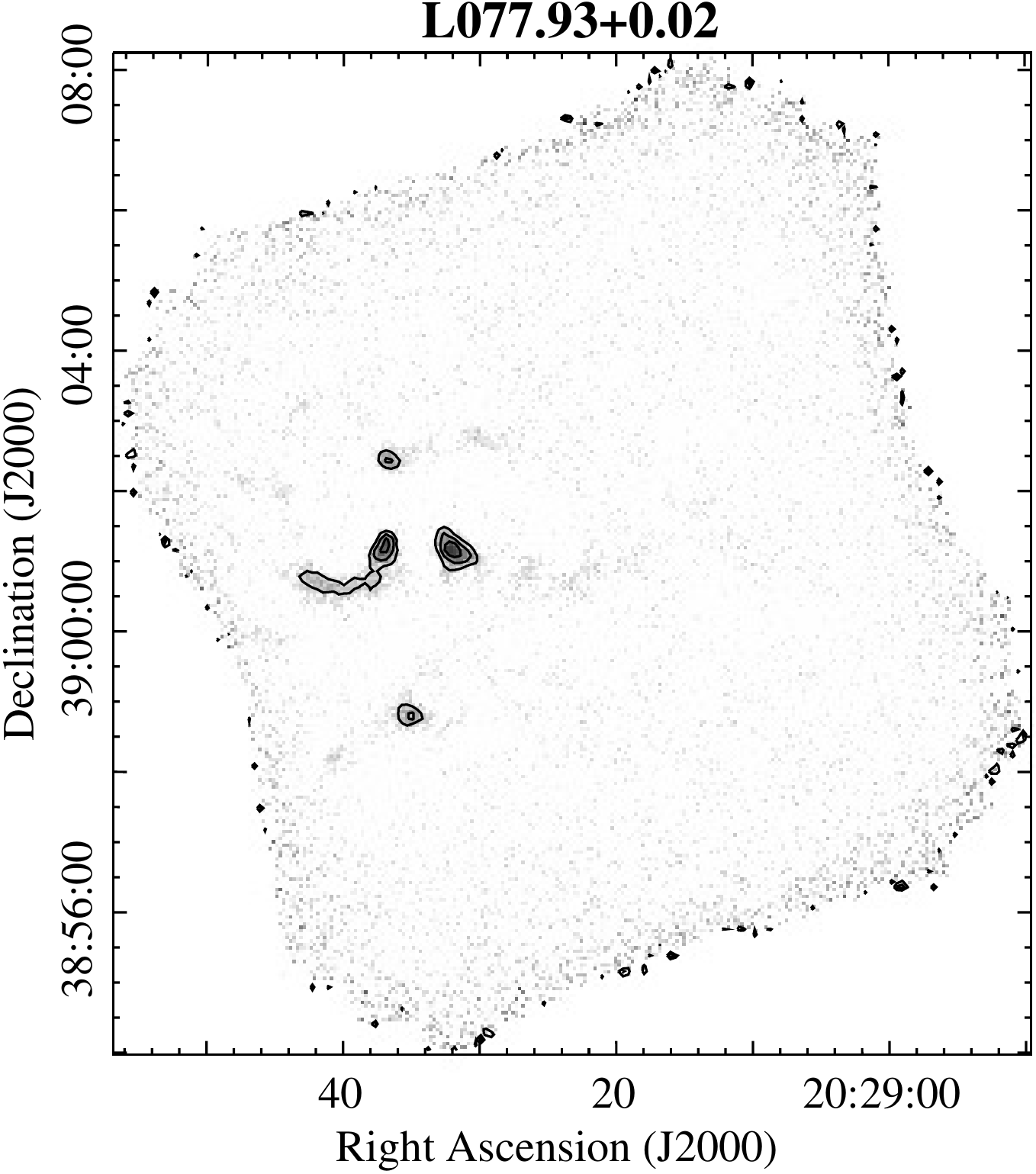}
\includegraphics[scale=0.43]{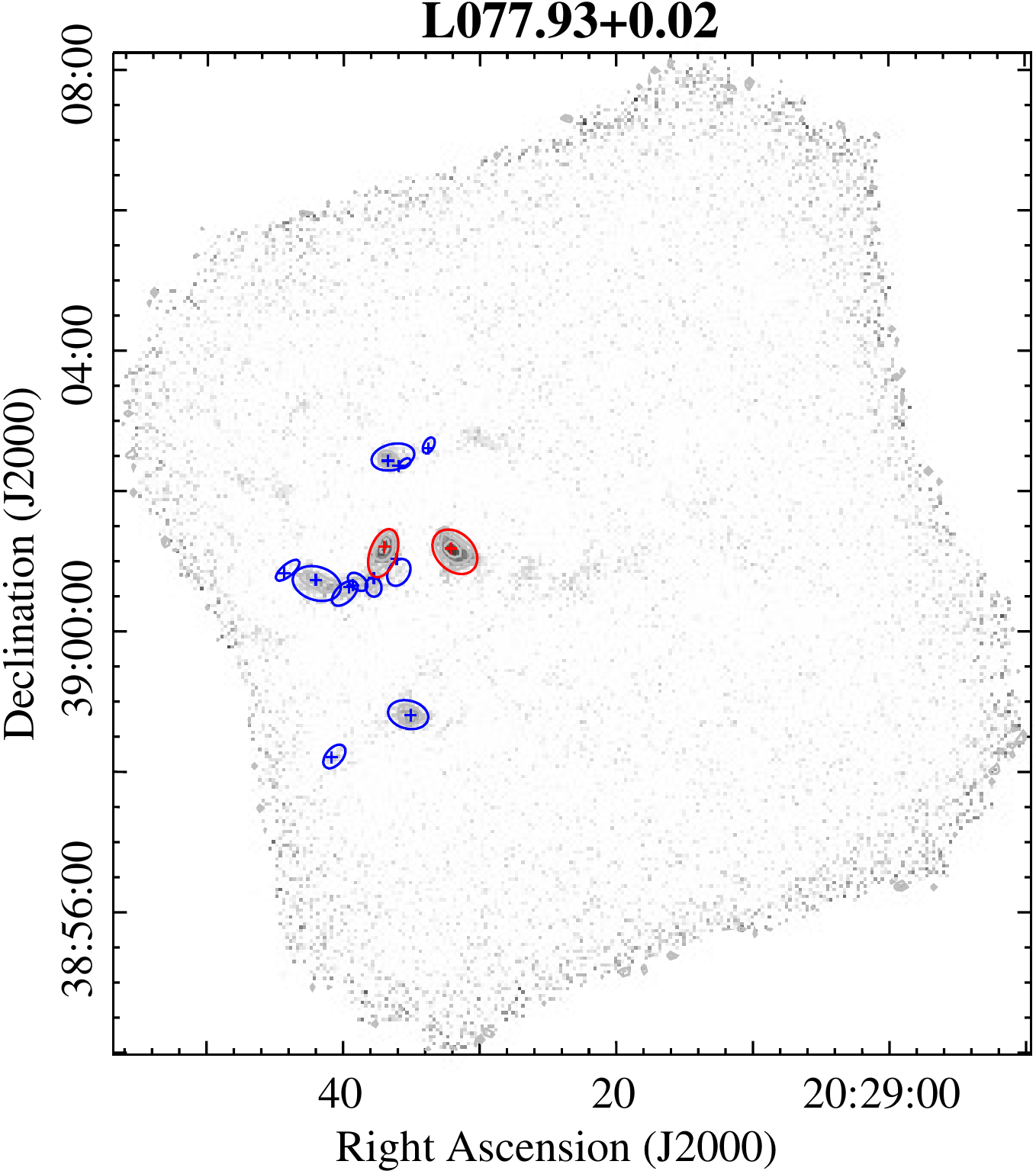}
}\\
\subfloat[L078.14-0.32 map, $\sigma_{rms}=408$ mJy beam$^{-1}$.]{
\includegraphics[scale=0.43]{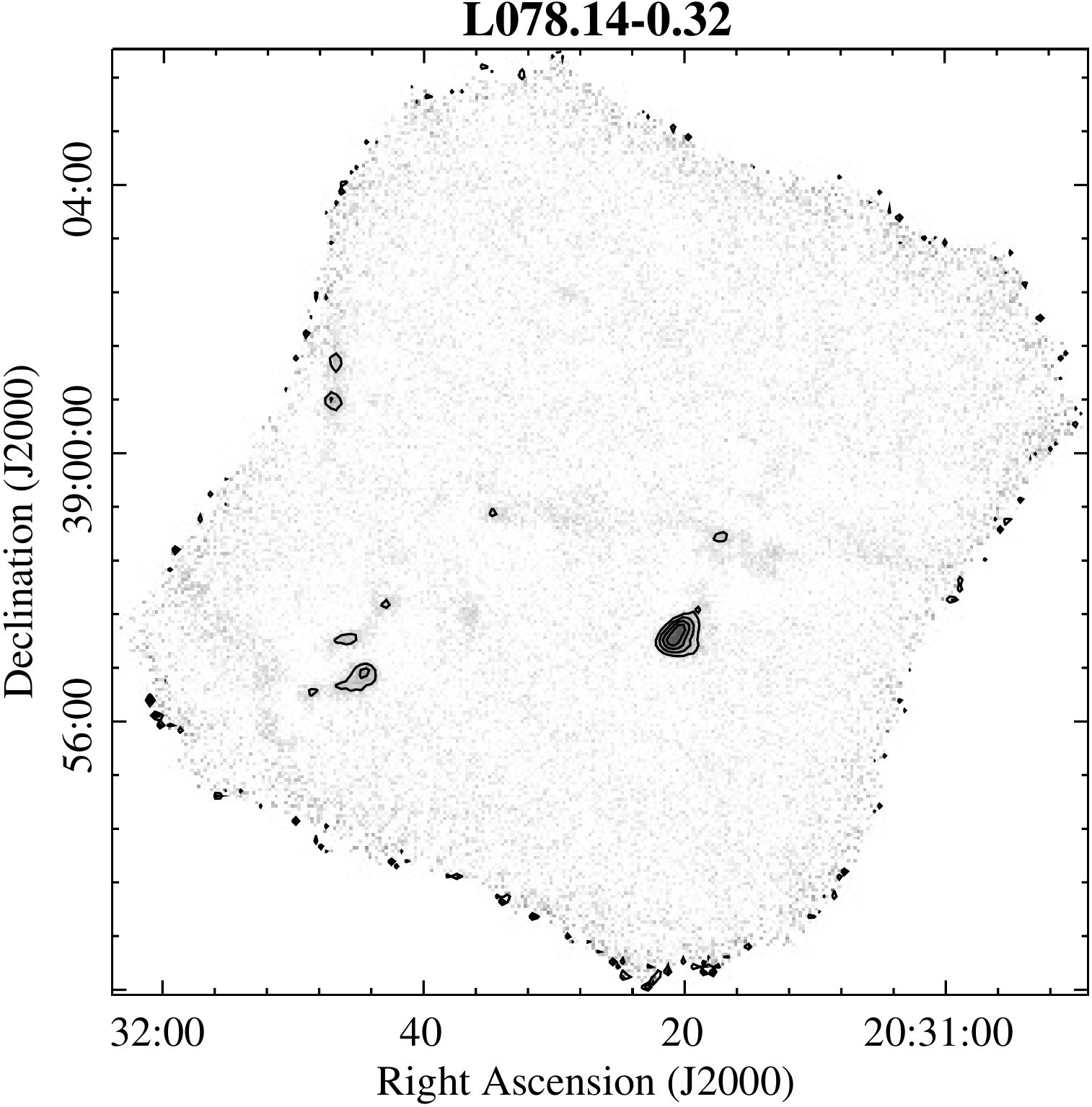}
\includegraphics[scale=0.43]{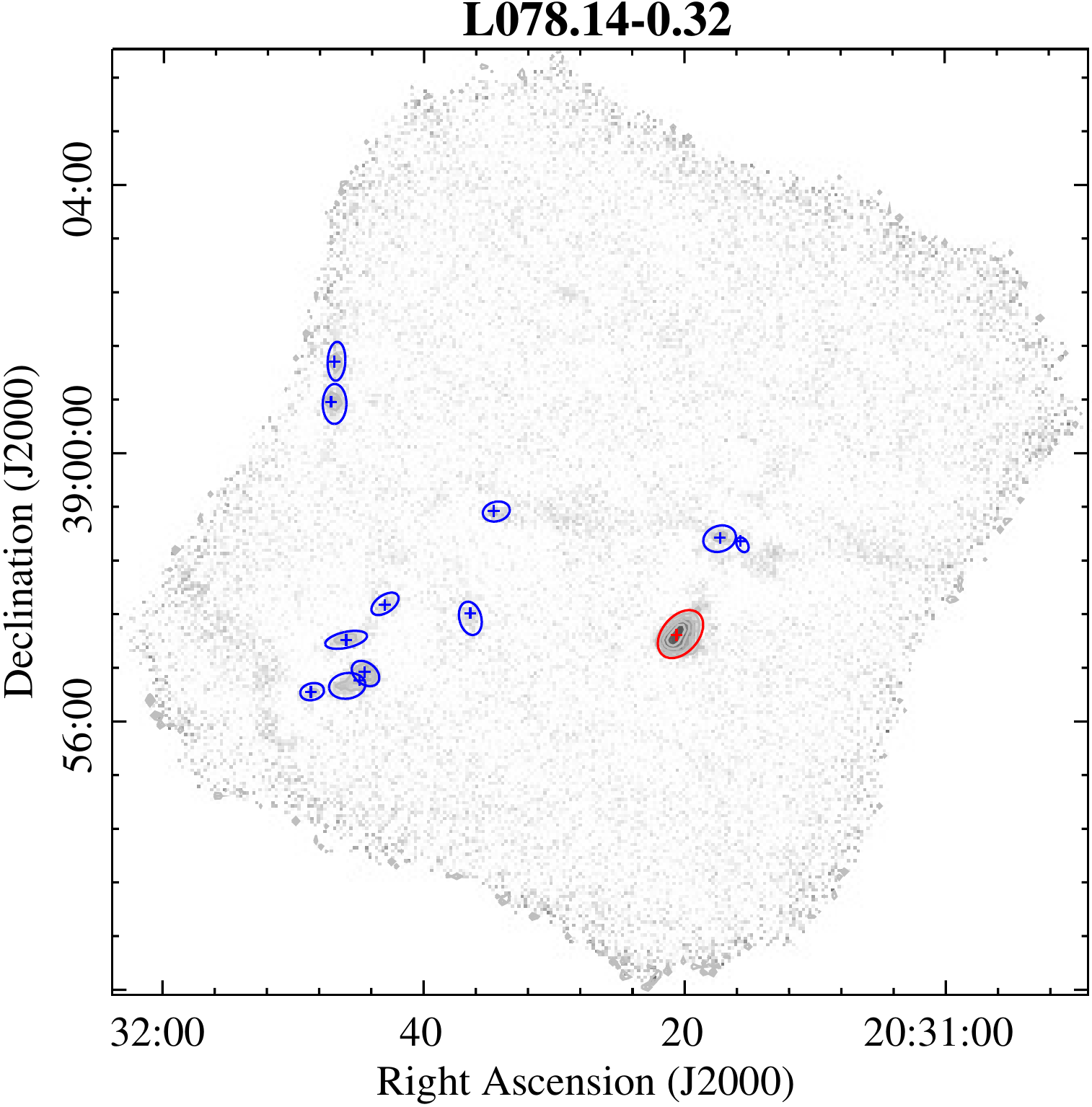}
}\\
\subfloat[L078.96+0.37 map, $\sigma_{rms}=445$ mJy beam$^{-1}$.]{
\includegraphics[scale=0.43]{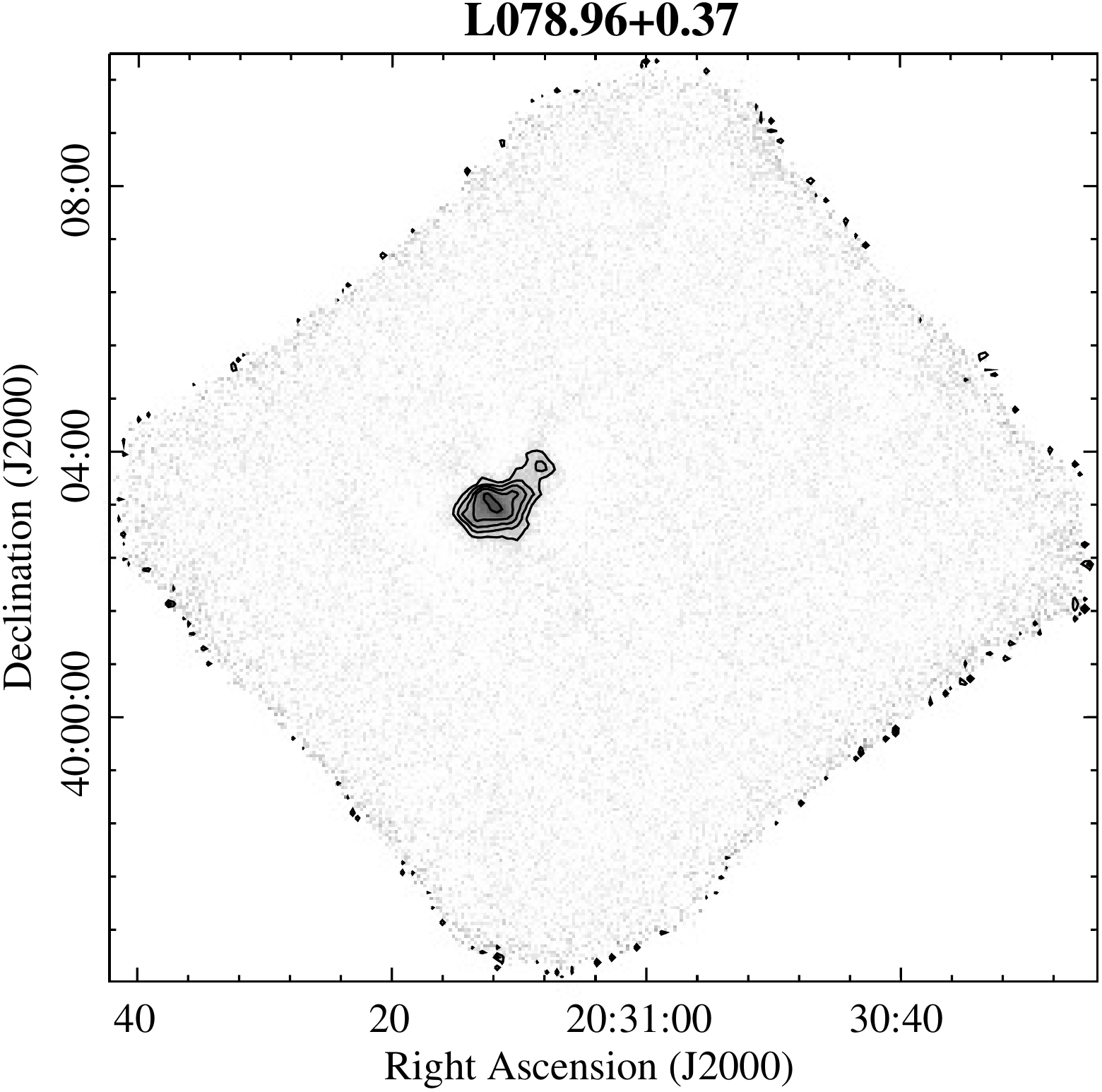}
\includegraphics[scale=0.43]{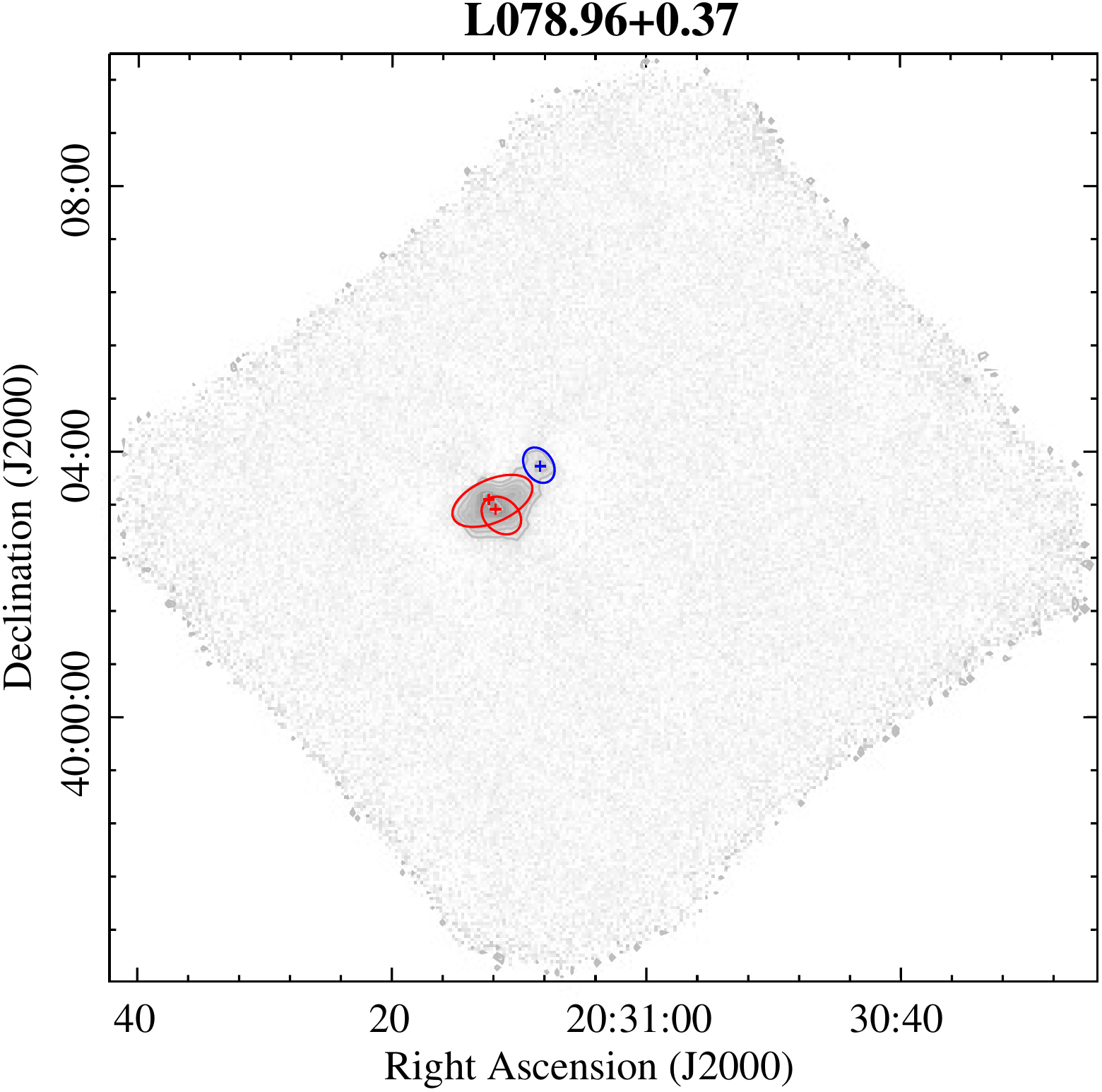}
}\\
\caption{Continuation}
\end{figure}

\clearpage
\begin{figure}\ContinuedFloat 
\center
\subfloat[L079.28+0.30 map, $\sigma_{rms}=558$ mJy beam$^{-1}$.]{
\includegraphics[scale=0.43]{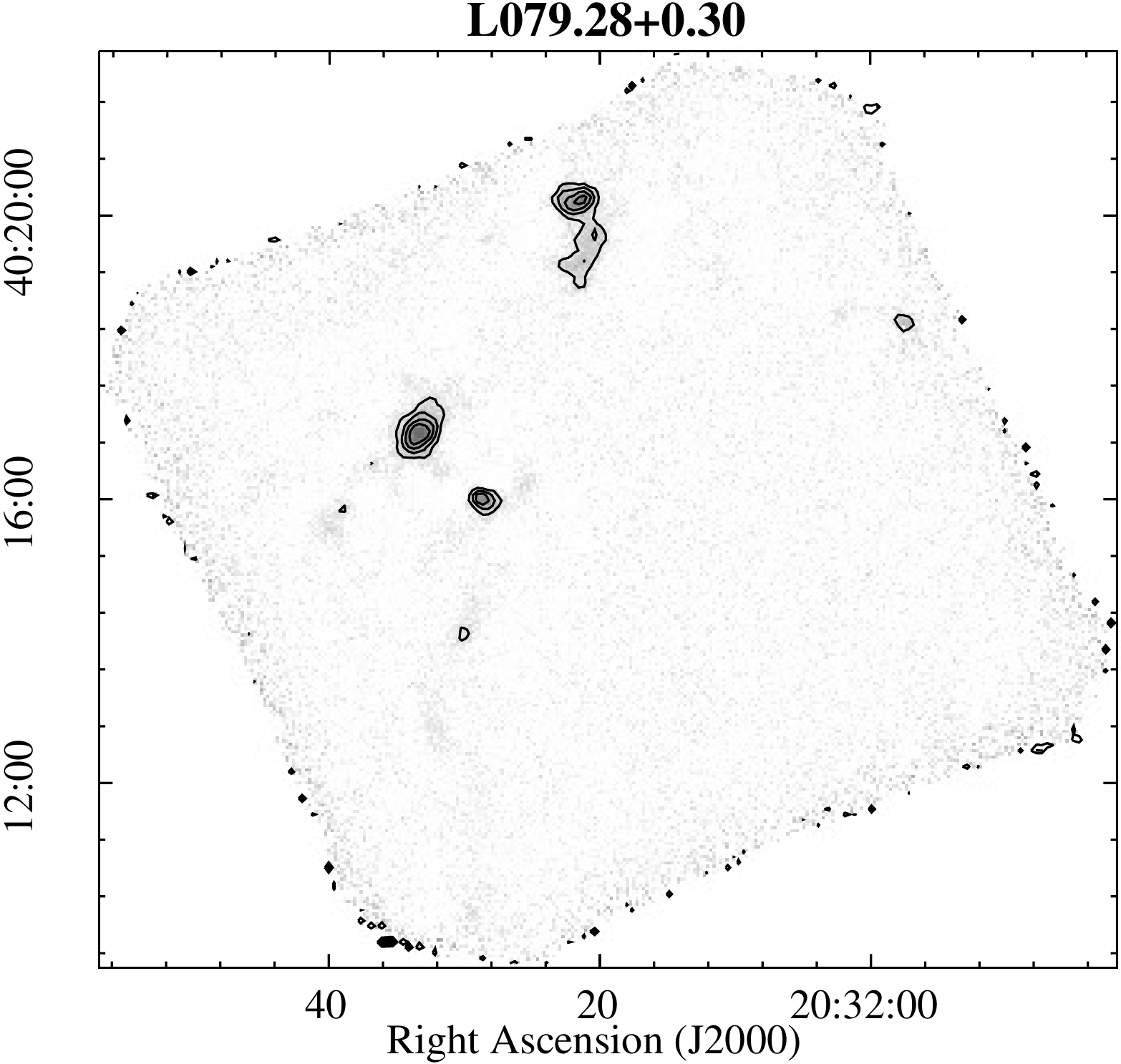}
\includegraphics[scale=0.43]{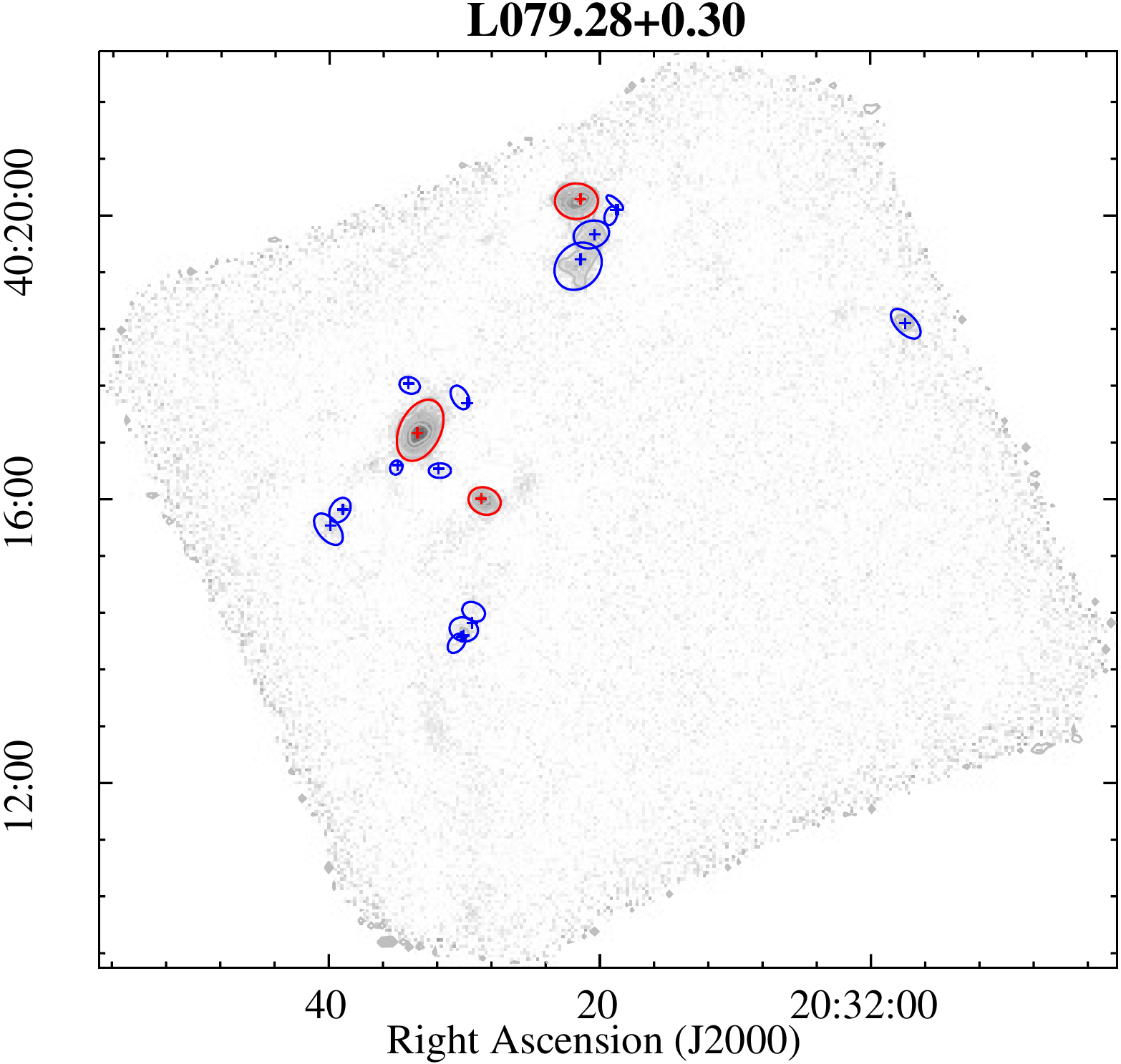}
}\\
\subfloat[L080.92-0.11 map, $\sigma_{rms}=724$ mJy beam$^{-1}$.]{
\includegraphics[scale=0.43]{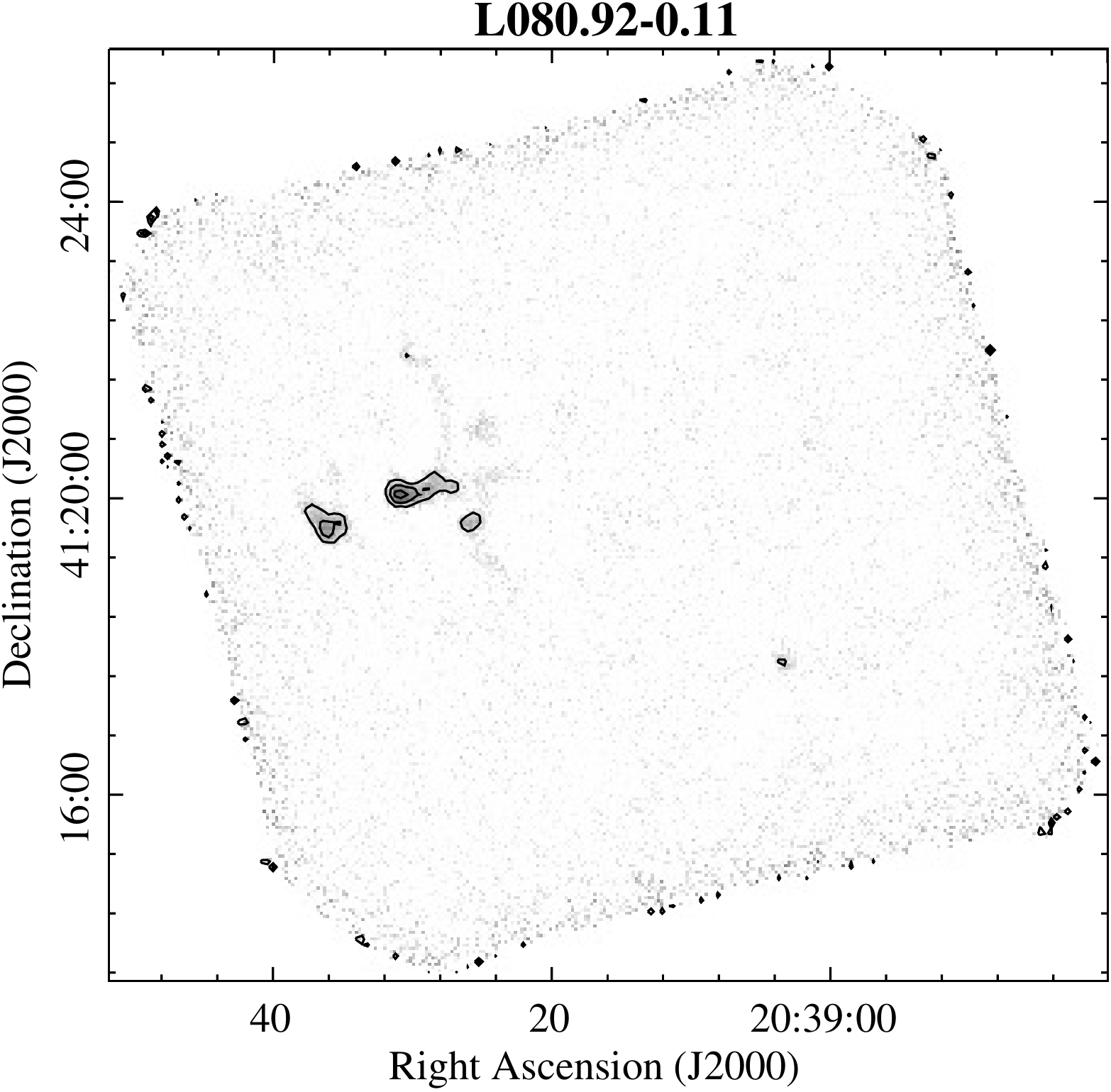}
\includegraphics[scale=0.43]{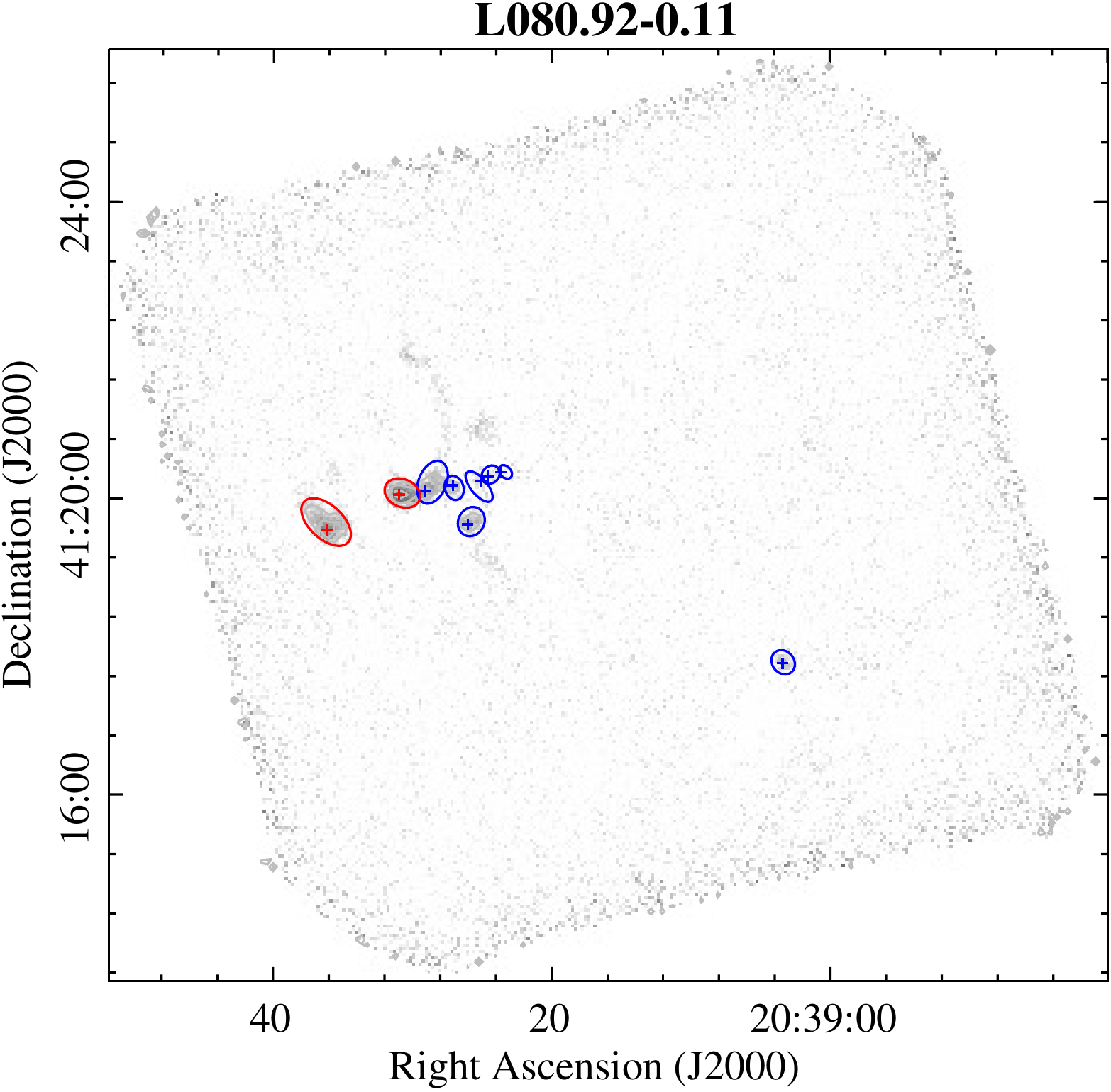}
}\\
\subfloat[L081.45+0.04 map, $\sigma_{rms}=274$ mJy beam$^{-1}$.]{
\includegraphics[scale=0.43]{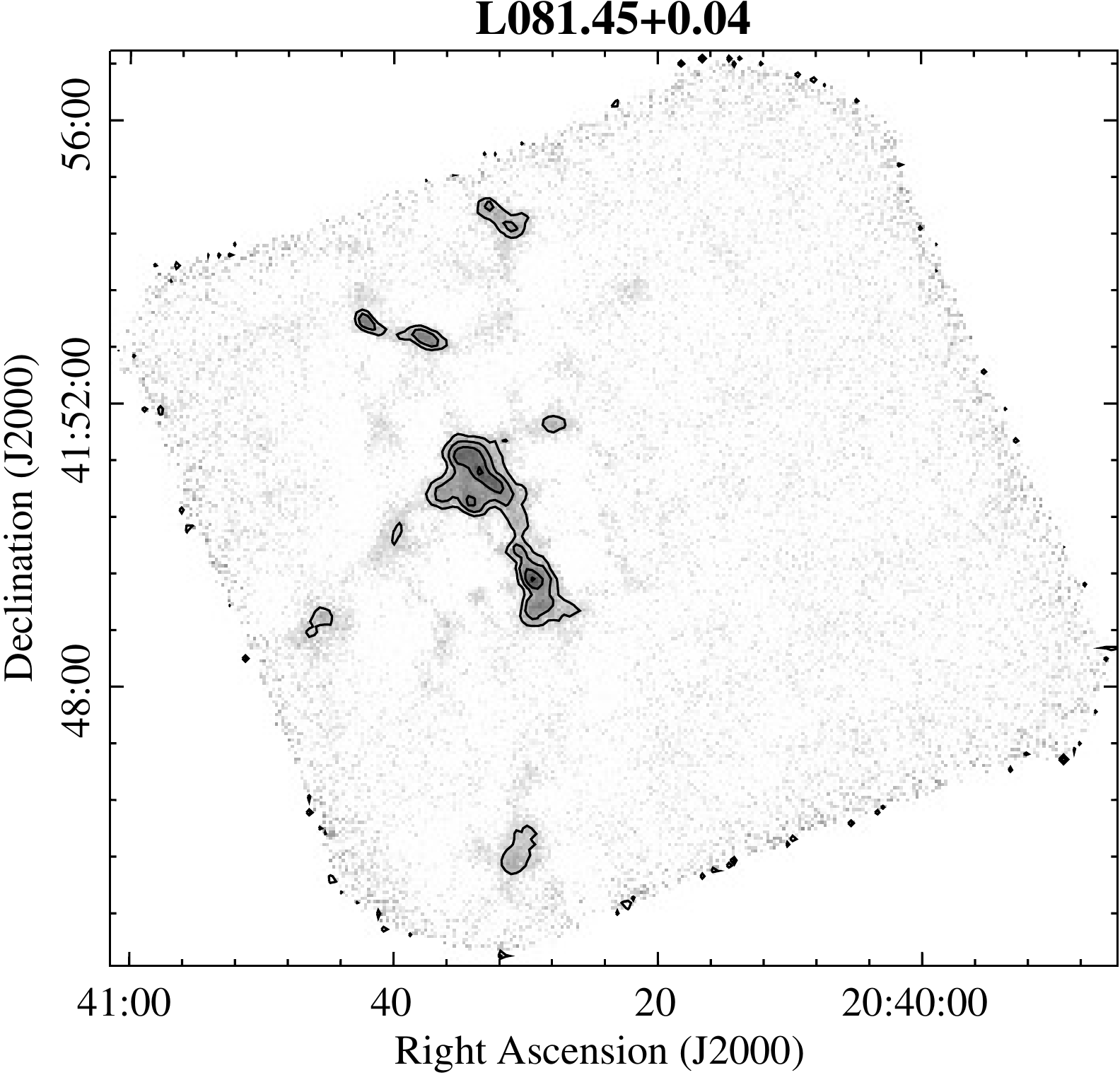}
\includegraphics[scale=0.43]{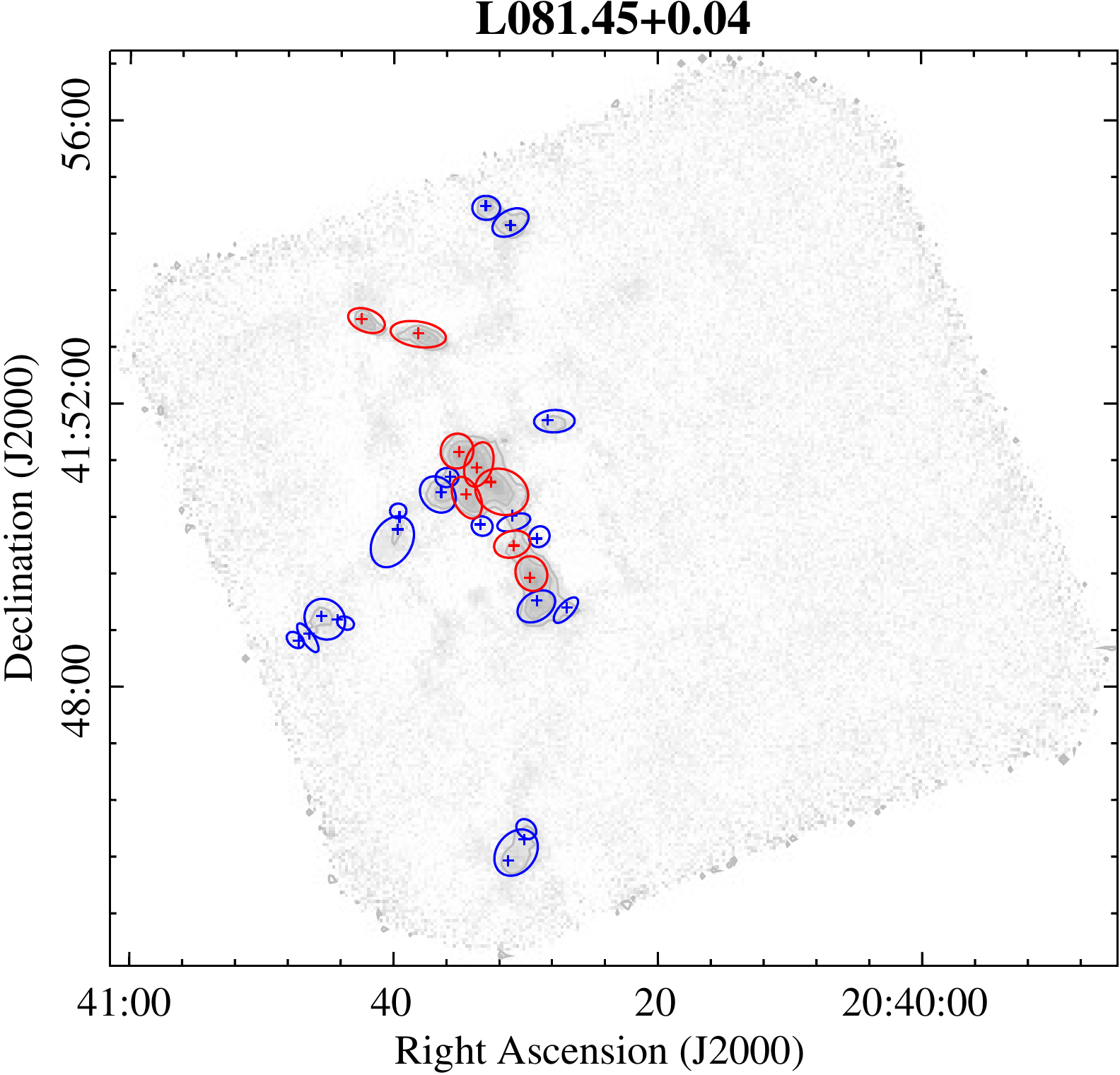}
}\\
\caption{Continuation}
\end{figure}

\clearpage
\begin{figure}\ContinuedFloat 
\center
\subfloat[L081.68+0.54 map, $\sigma_{rms}=802$ mJy beam$^{-1}$.]{
\includegraphics[scale=0.43]{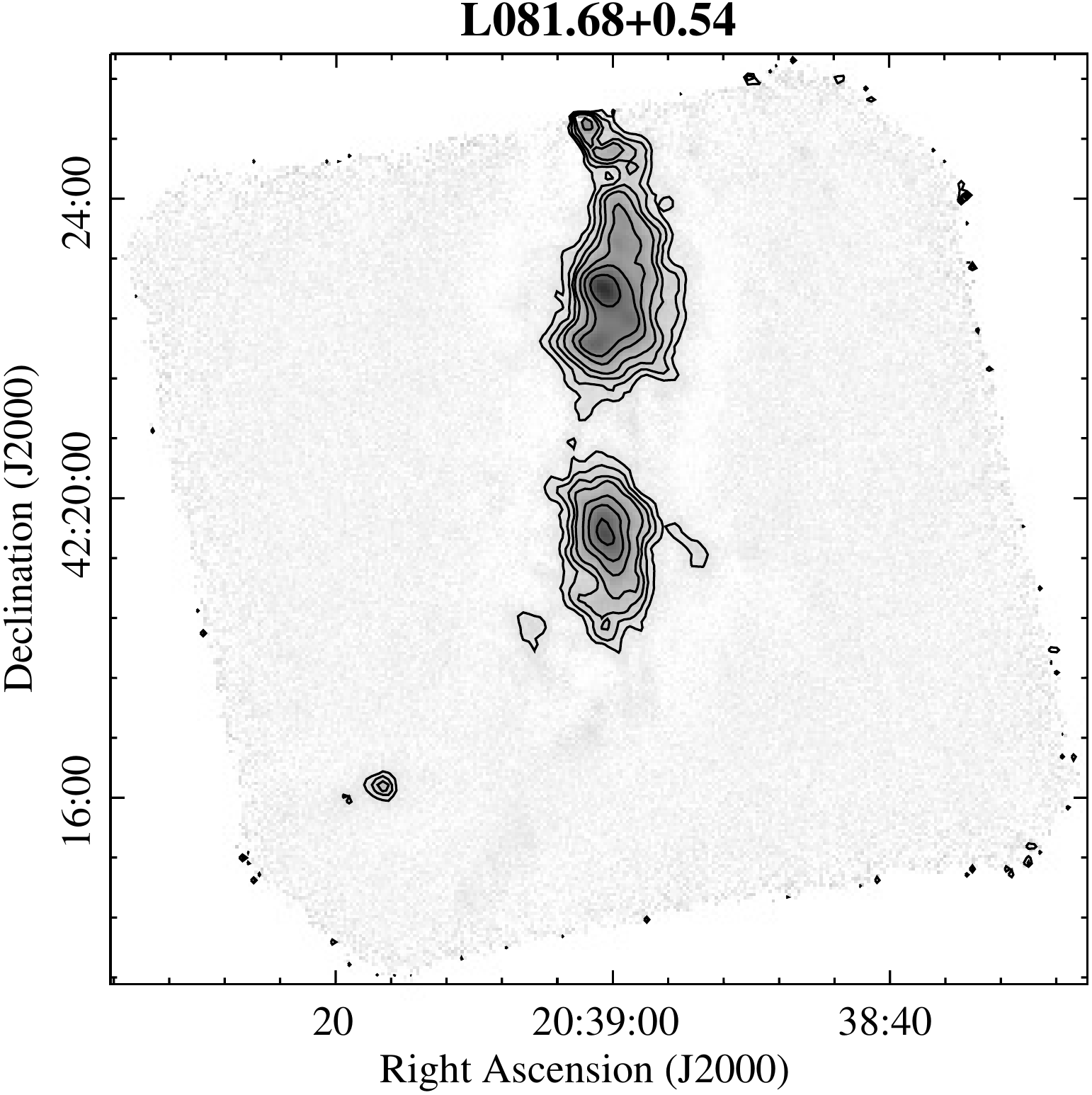}
\includegraphics[scale=0.43]{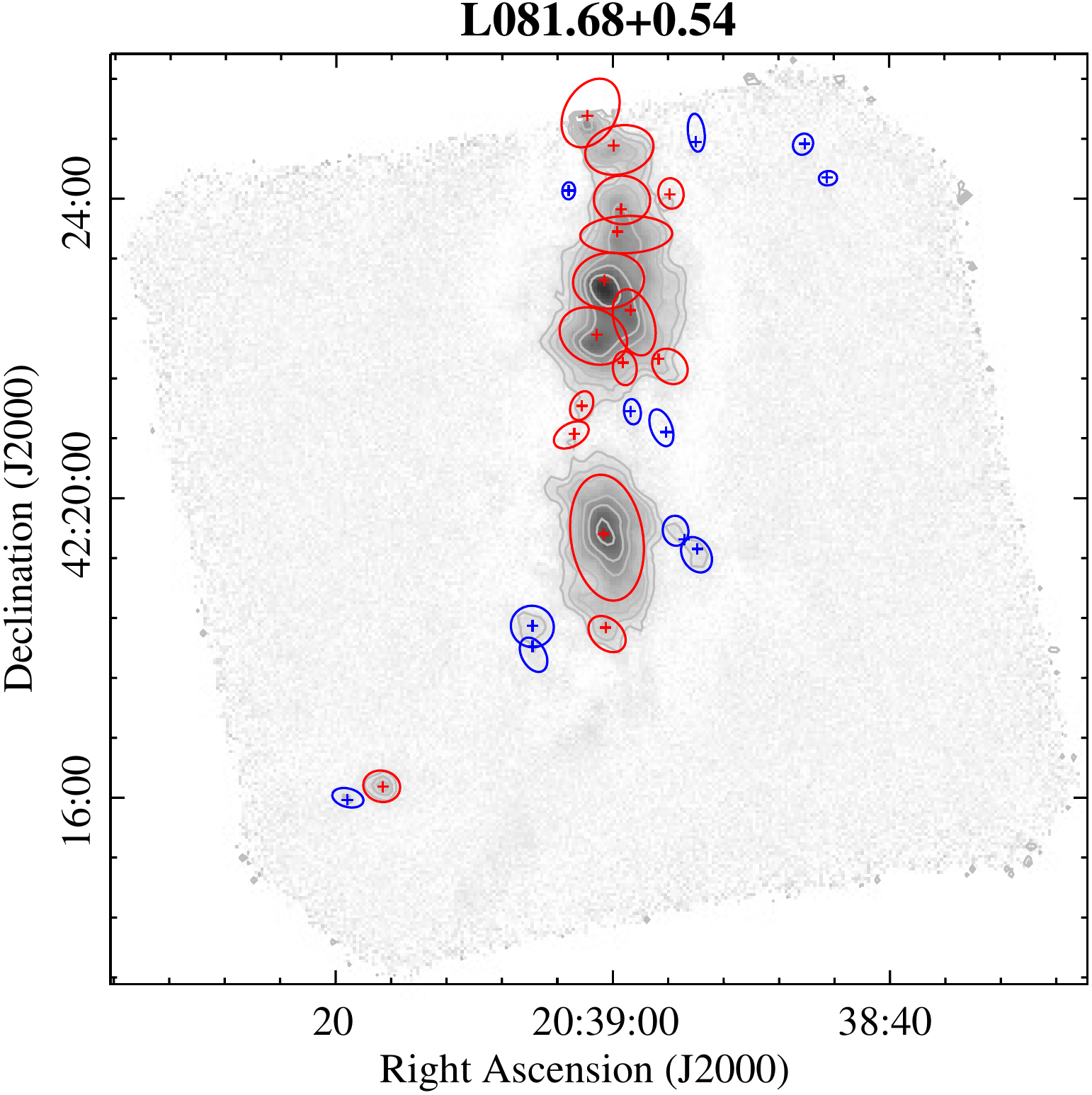}
}\\
\subfloat[L082.55+0.14 map, $\sigma_{rms}=255$ mJy beam$^{-1}$.]{
\includegraphics[scale=0.43]{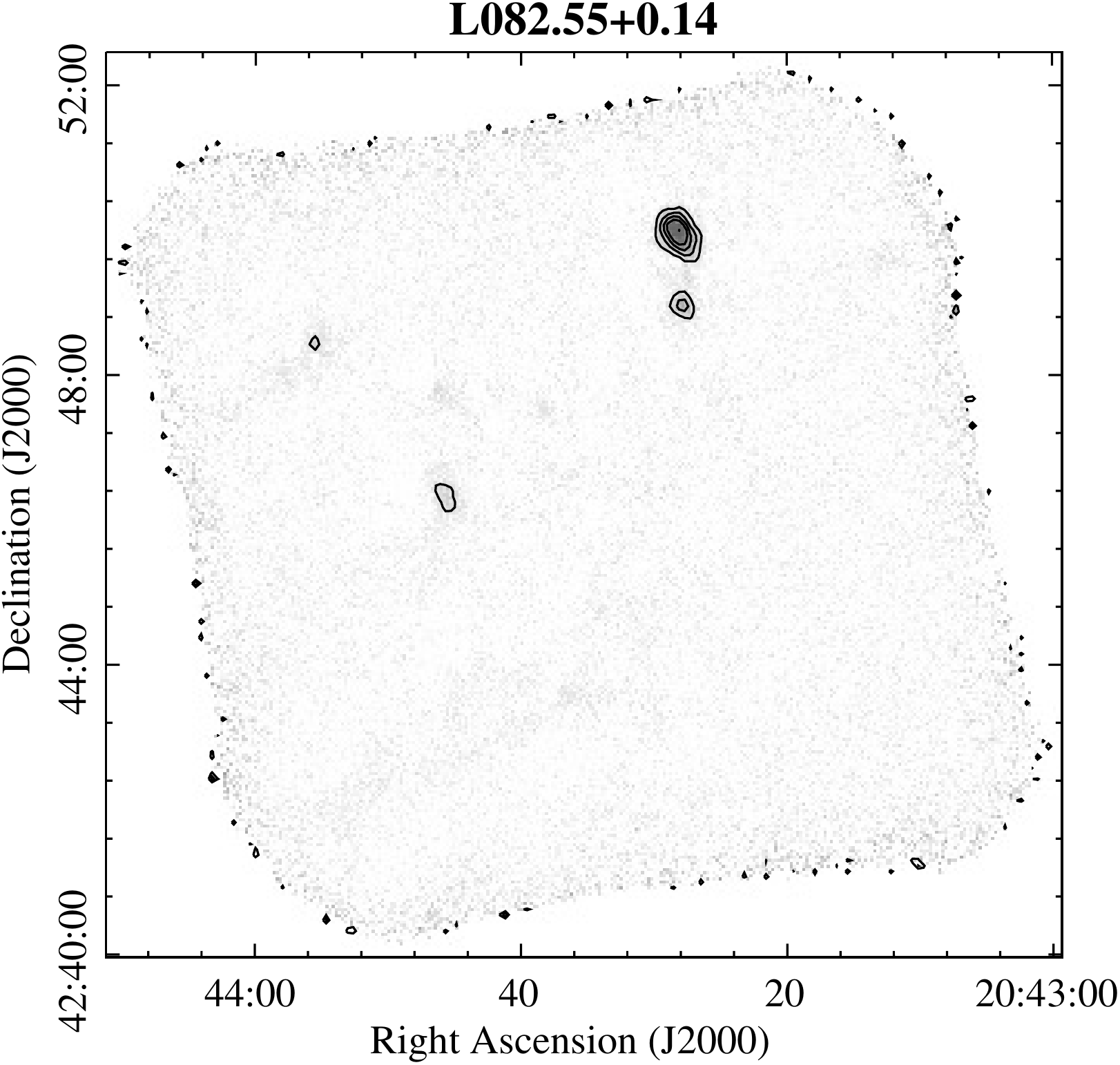}
\includegraphics[scale=0.43]{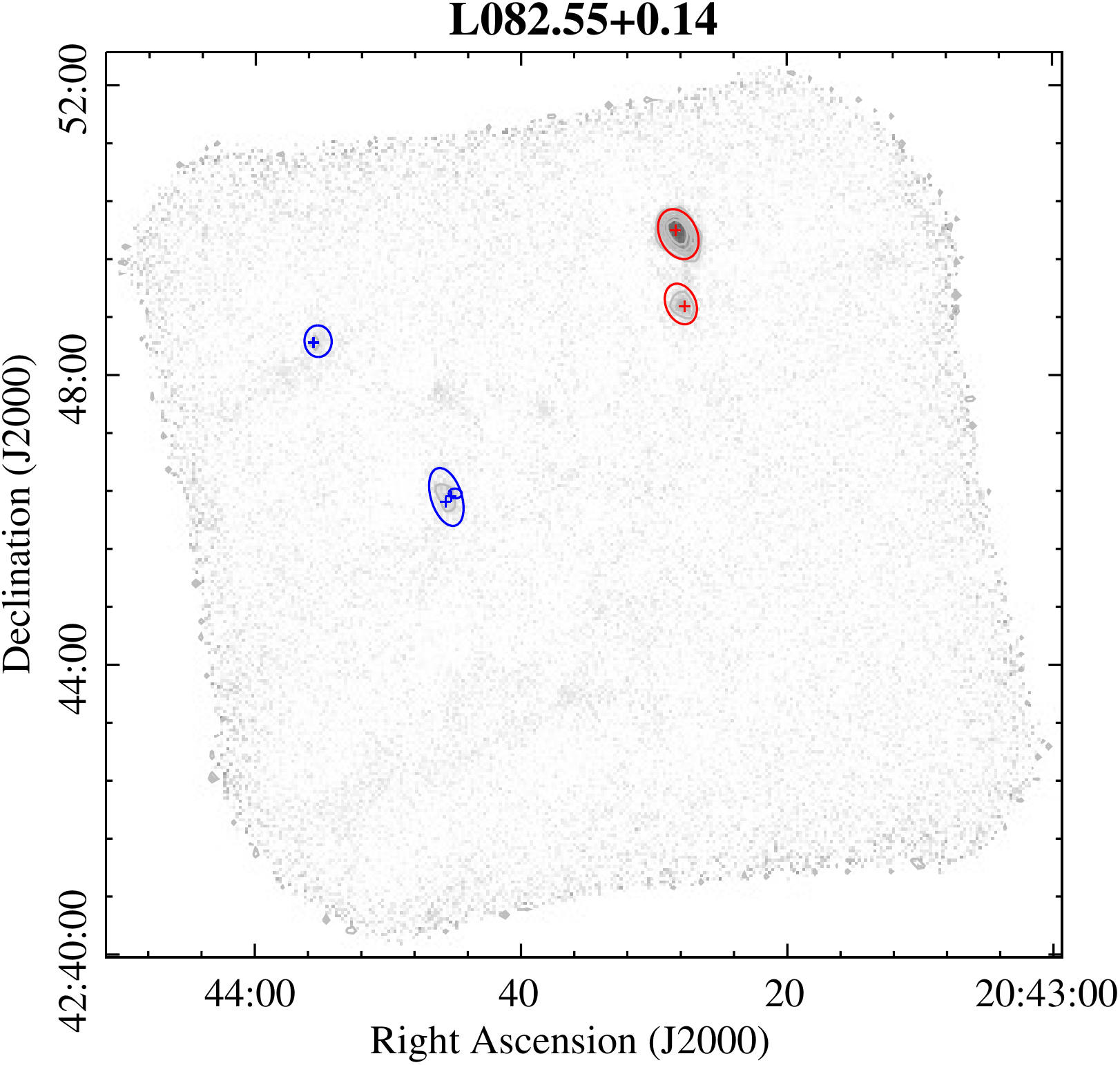}
}\\
\subfloat[L075.76+0.41 map, $\sigma_{rms}=652$ mJy beam$^{-1}$.]{
\includegraphics[scale=0.43]{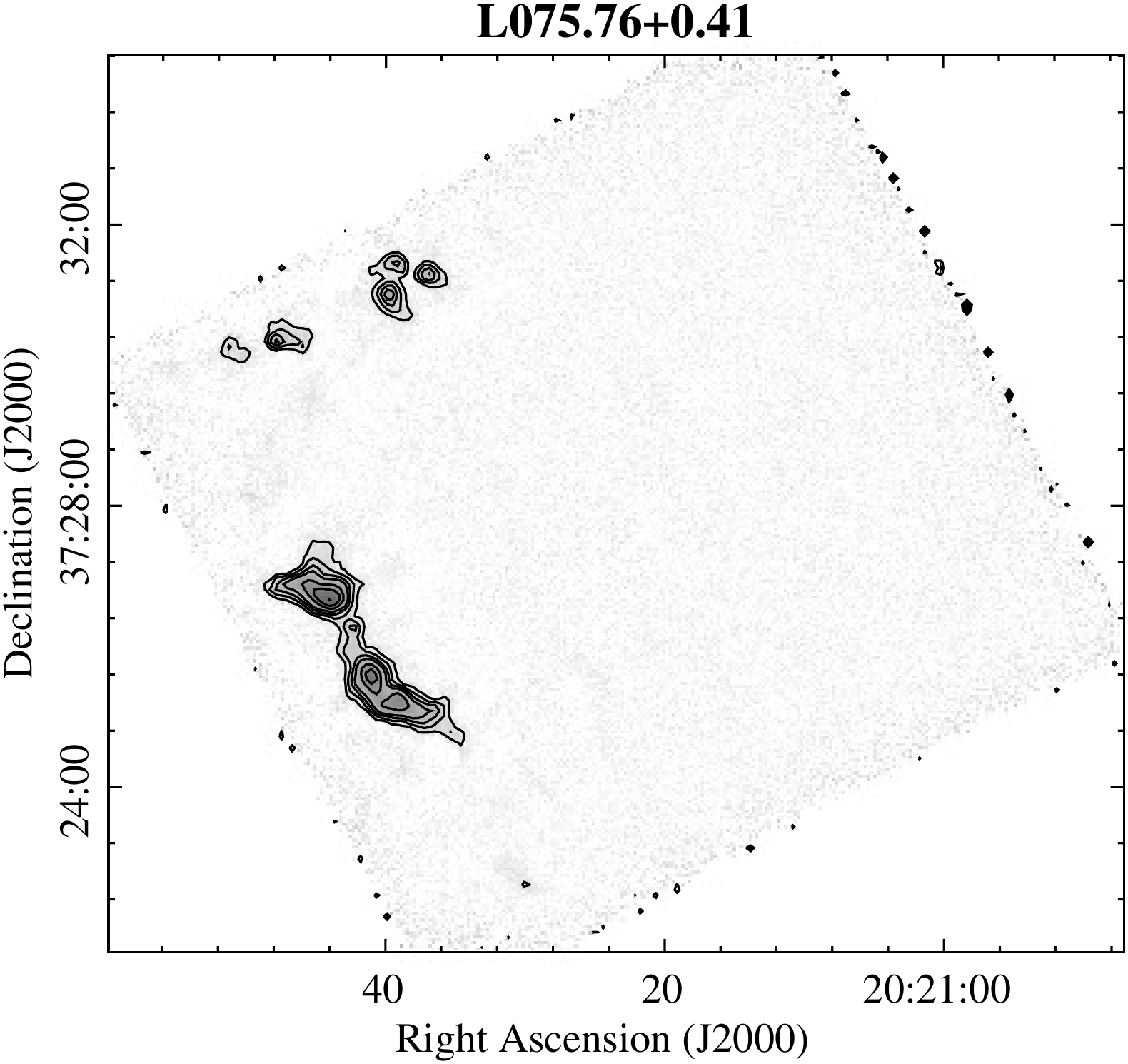}
\includegraphics[scale=0.43]{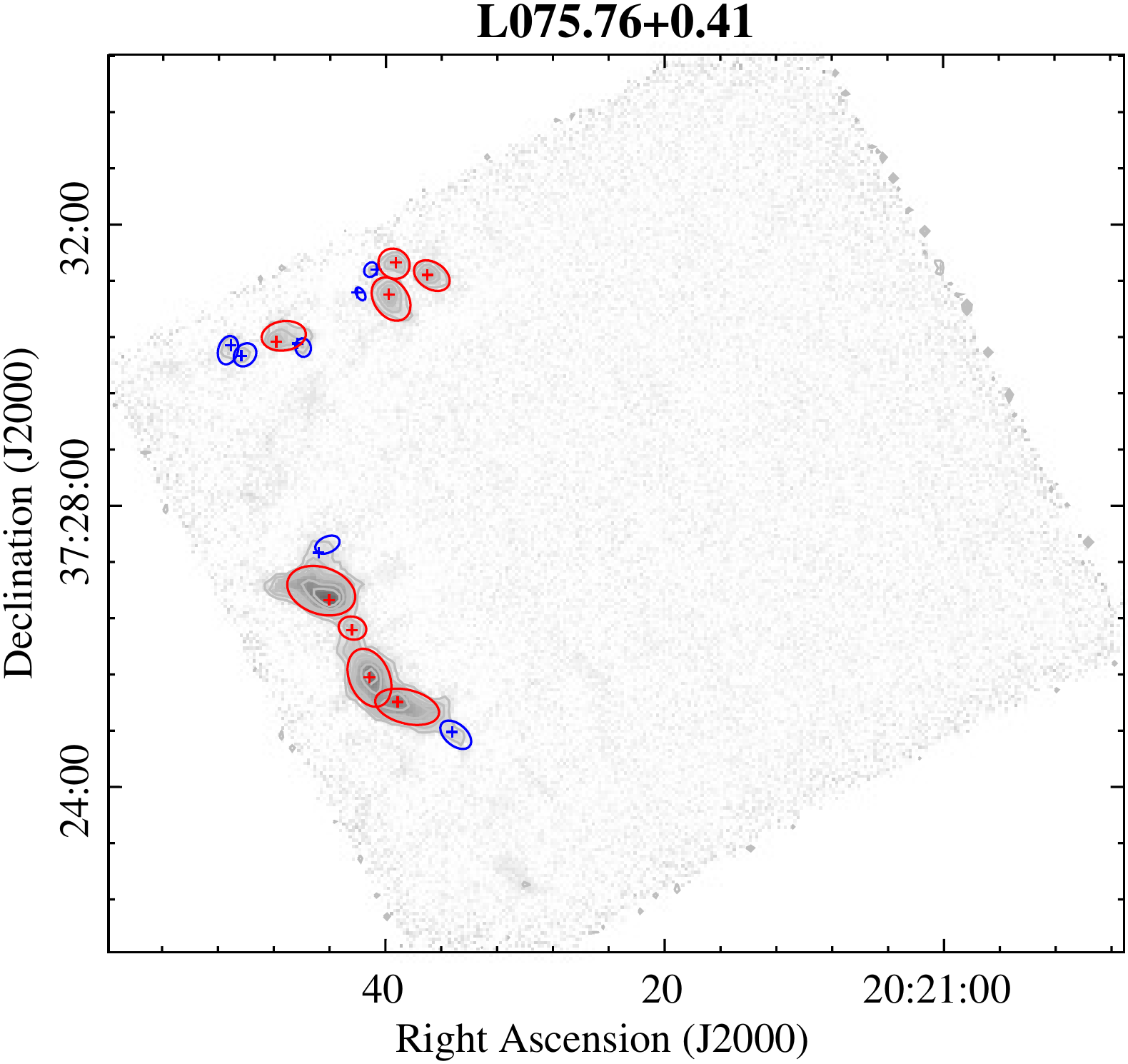}
}\\
\caption{Continuation}
\end{figure}

\clearpage
\begin{figure}\ContinuedFloat 
\center
\subfloat[L076.12-0.24 map, $\sigma_{rms}=388$ mJy beam$^{-1}$.]{
\includegraphics[scale=0.43]{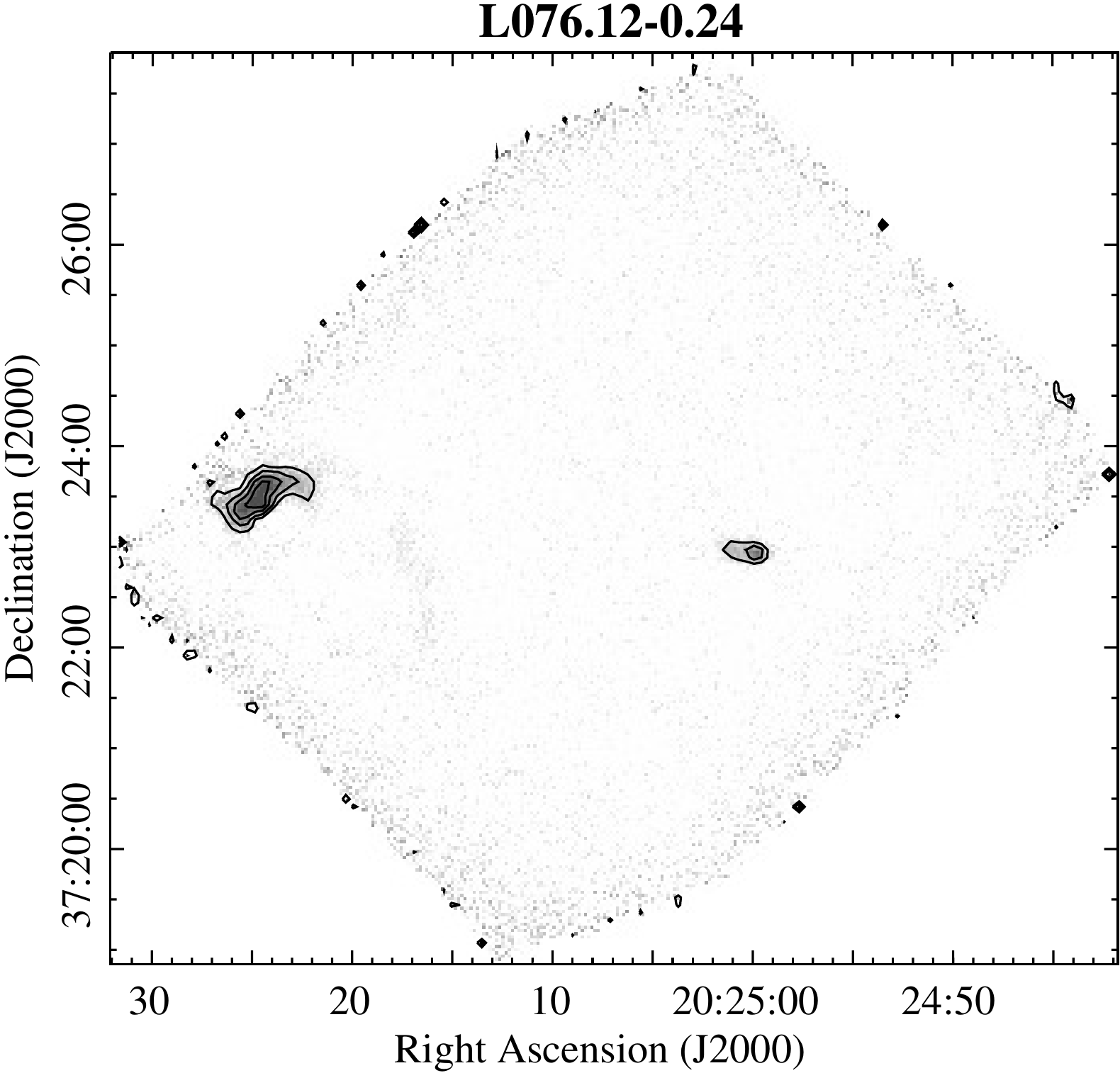}
\includegraphics[scale=0.43]{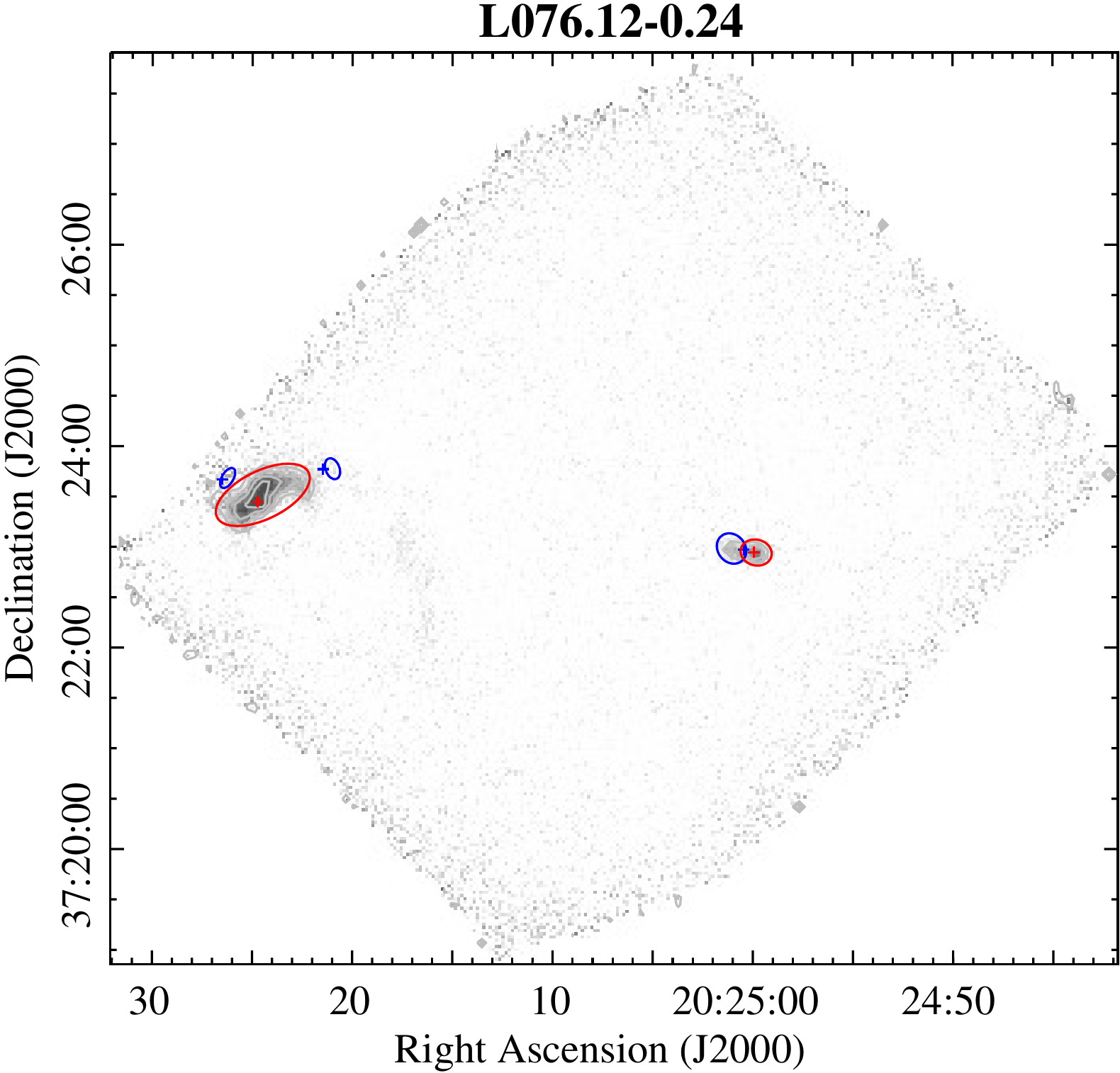}
}\\
\subfloat[L076.35-0.58 map, $\sigma_{rms}=430$ mJy beam$^{-1}$.]{
\includegraphics[scale=0.43]{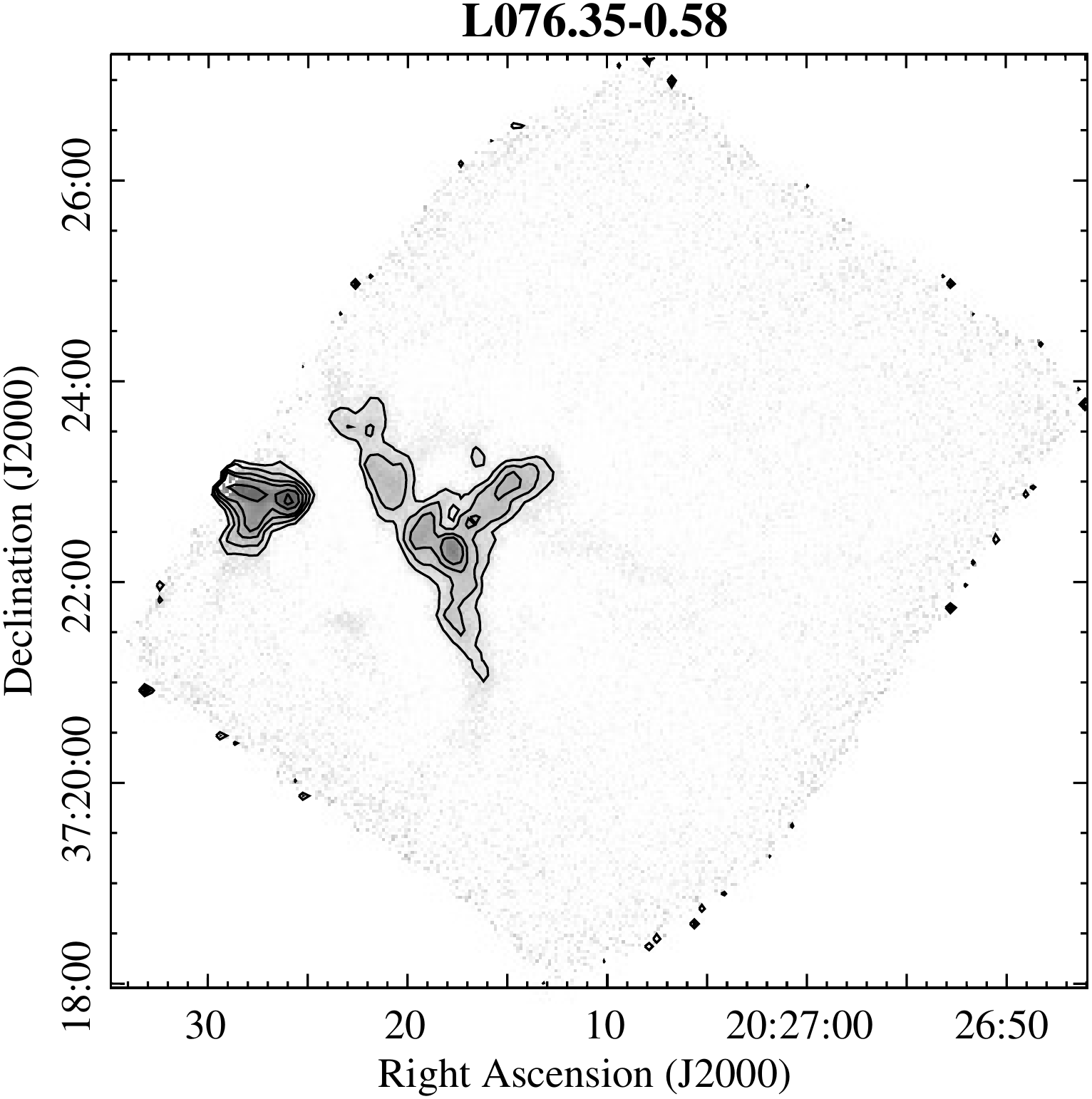}
\includegraphics[scale=0.43]{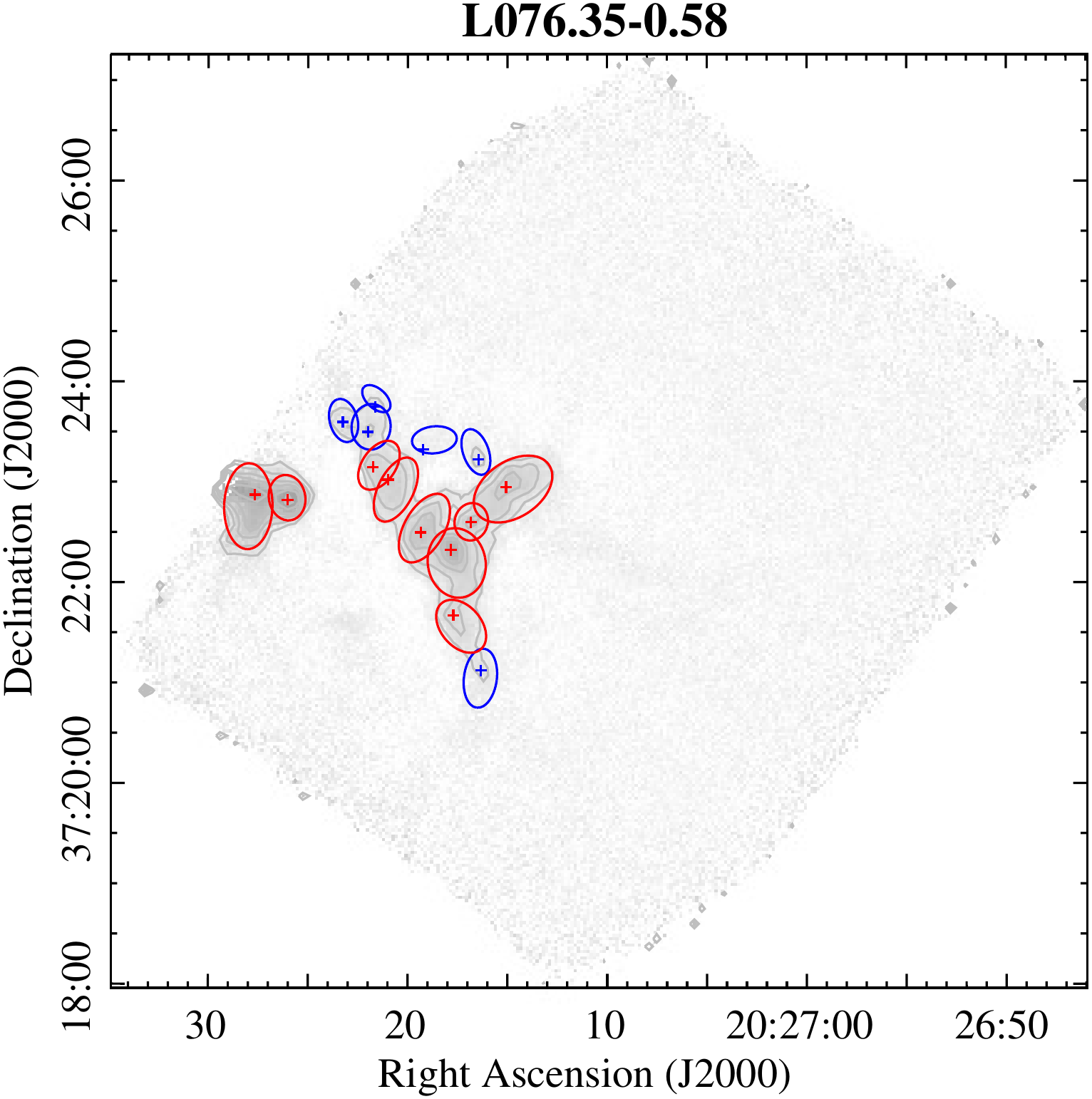}
}\\
\subfloat[L078.92-0.19 map, $\sigma_{rms}=689$ mJy beam$^{-1}$.]{
\includegraphics[scale=0.43]{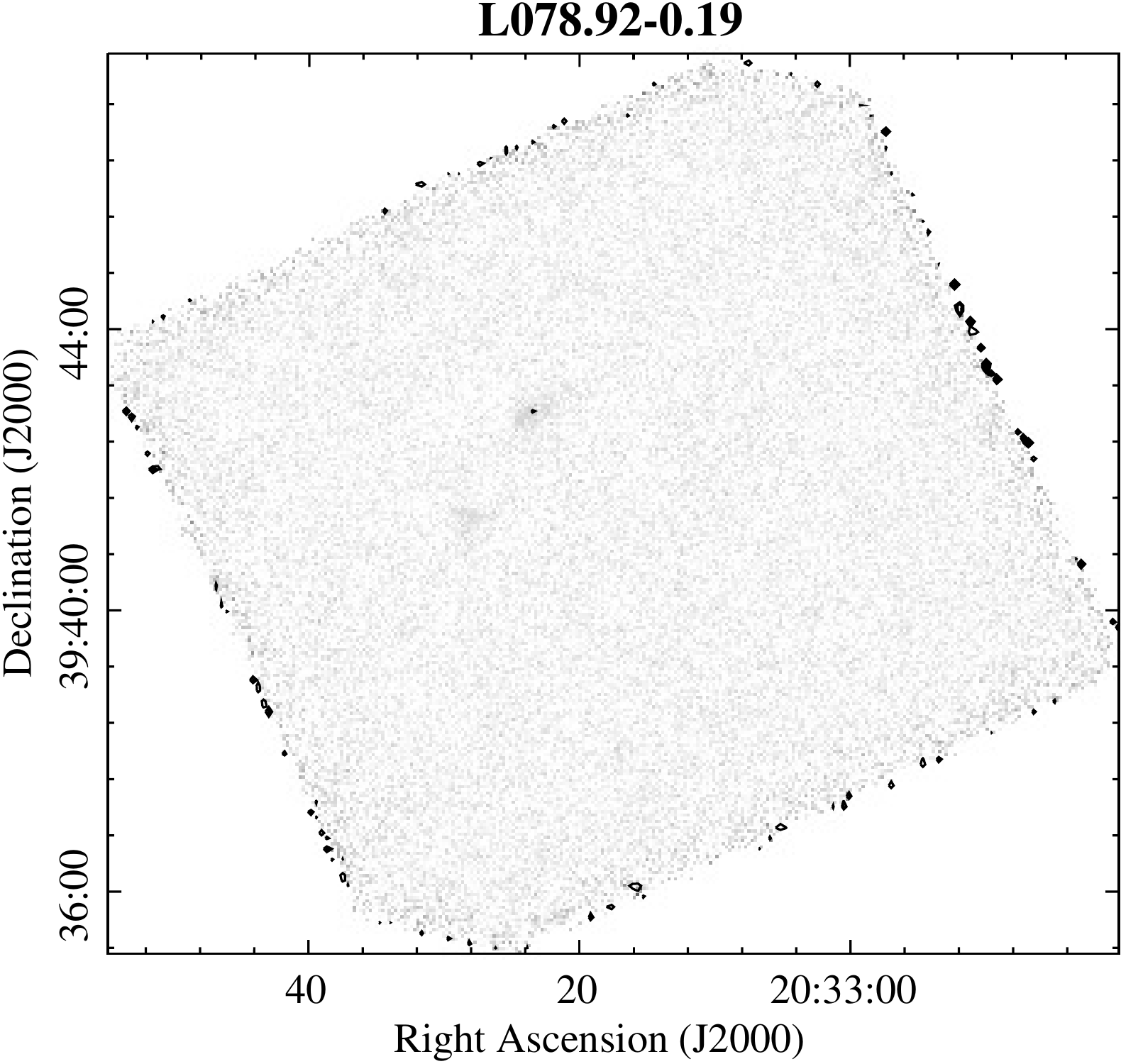}
\includegraphics[scale=0.43]{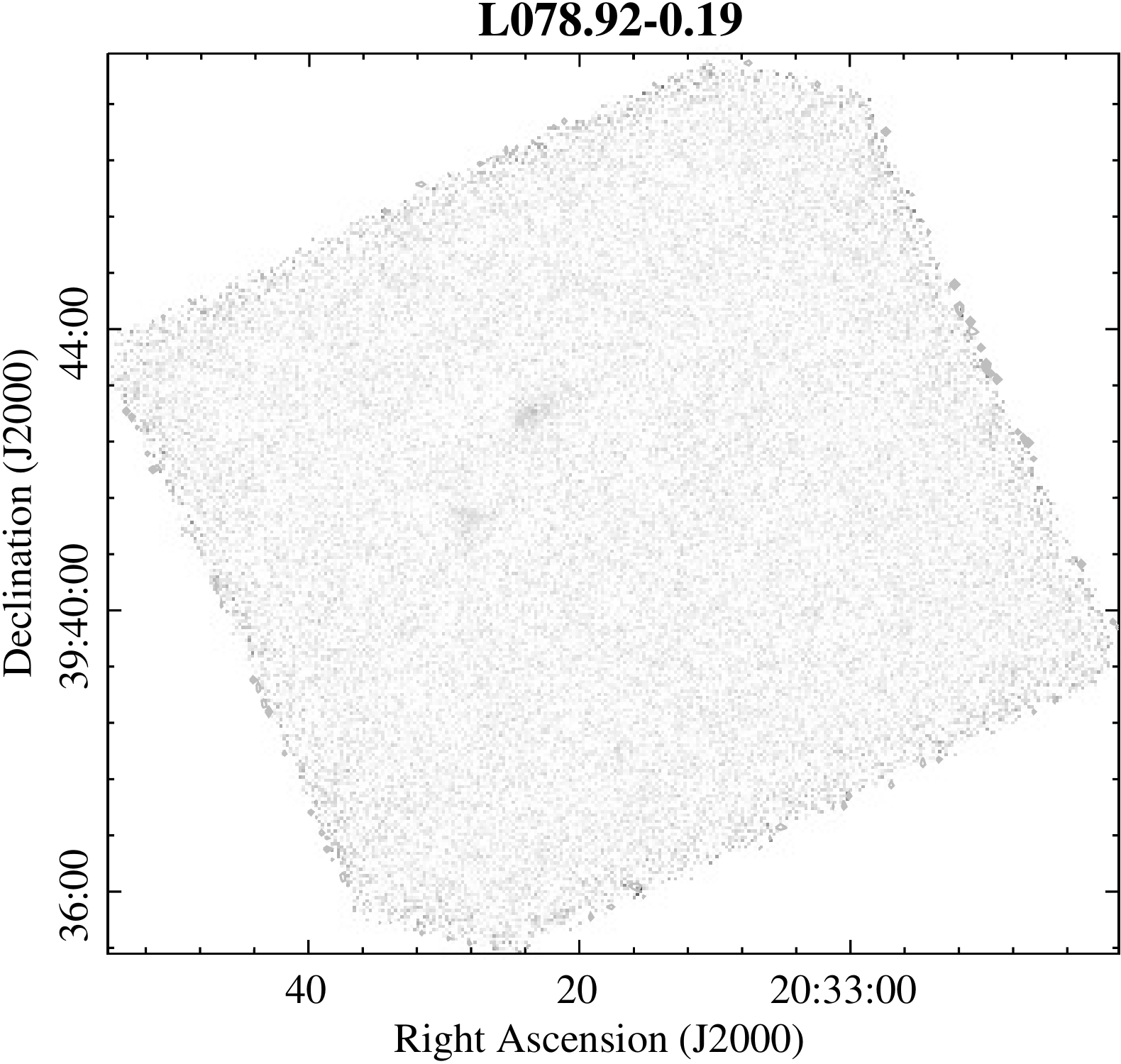}
}\\
\caption{Continuation}
\end{figure}

\clearpage
\begin{figure}\ContinuedFloat 
\center
\subfloat[L078.17-0.31 map, $\sigma_{rms}=800$ mJy beam$^{-1}$.]{
\includegraphics[scale=0.43]{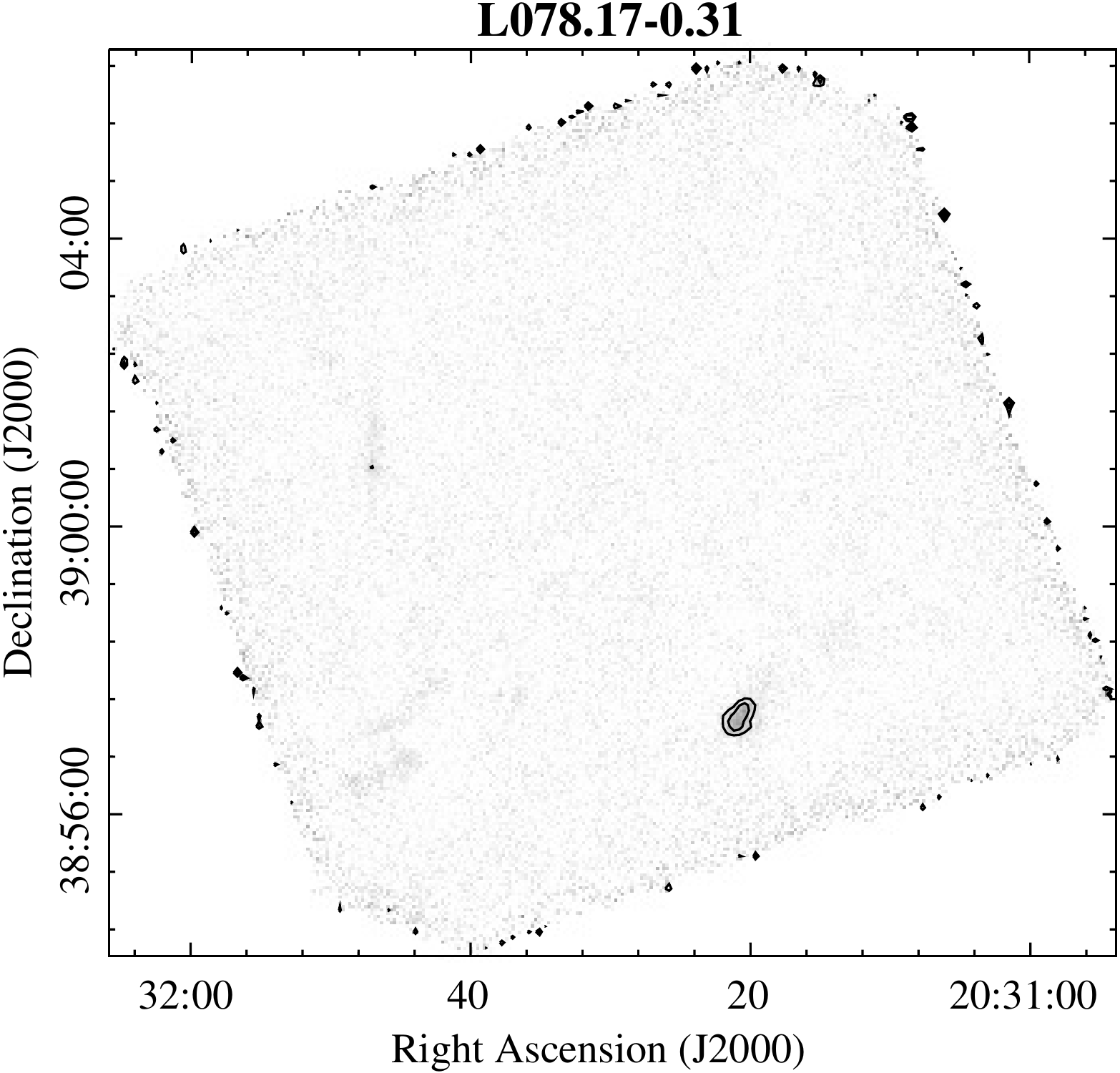}
\includegraphics[scale=0.43]{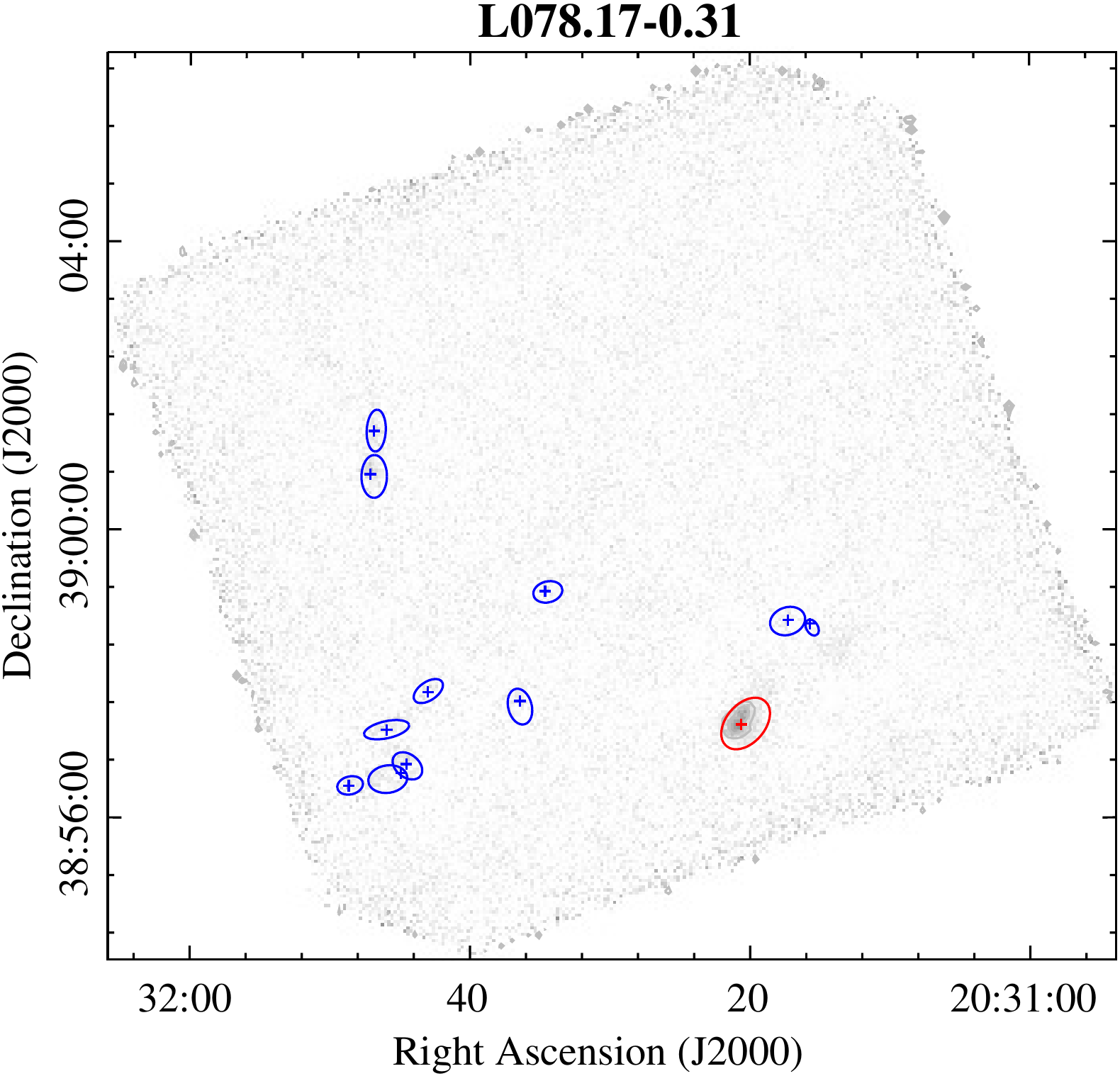}
}\\
\subfloat[L079.62+0.49 map, $\sigma_{rms}=1489$ mJy beam$^{-1}$.]{
\includegraphics[scale=0.43]{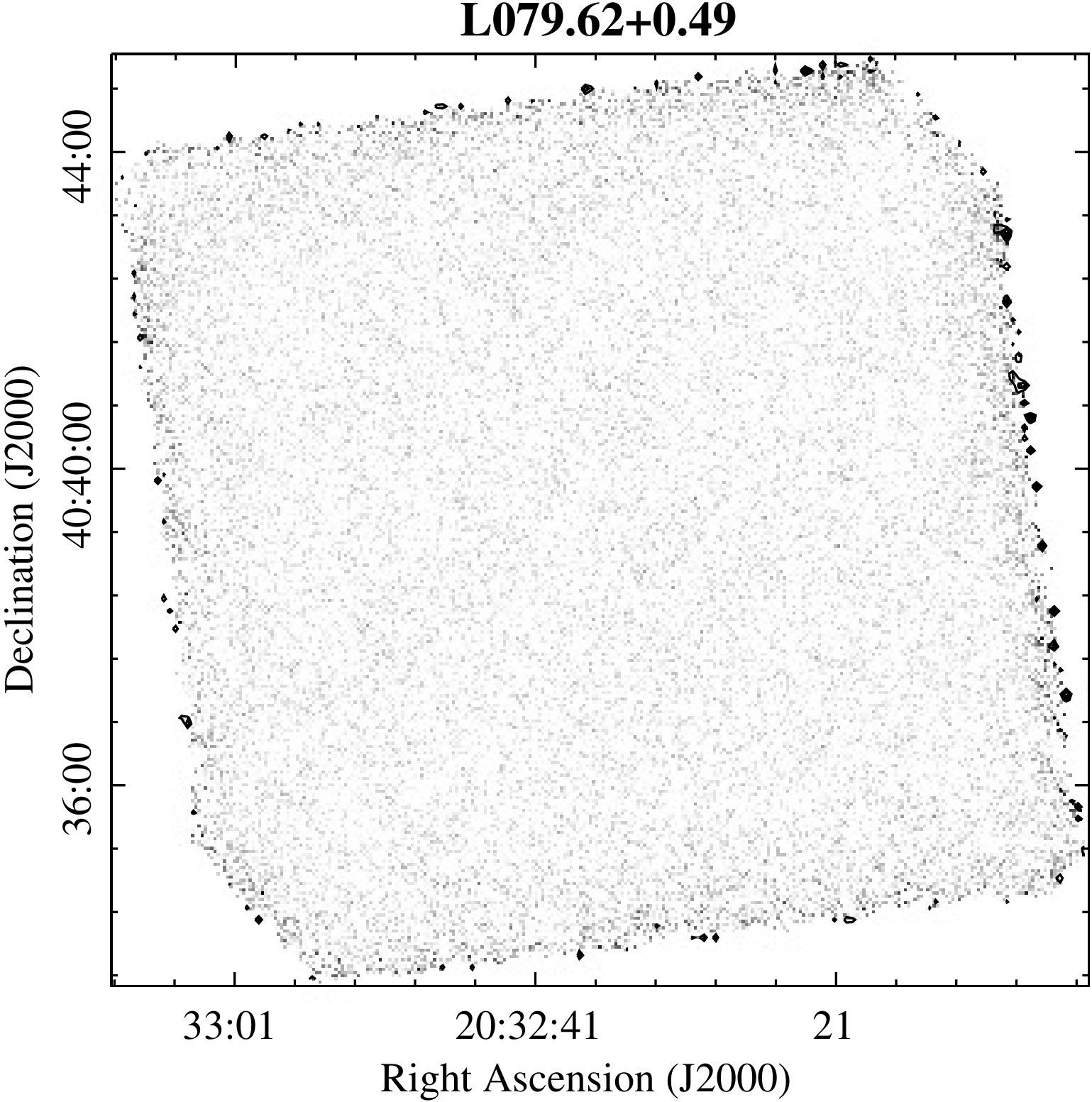}
\includegraphics[scale=0.43]{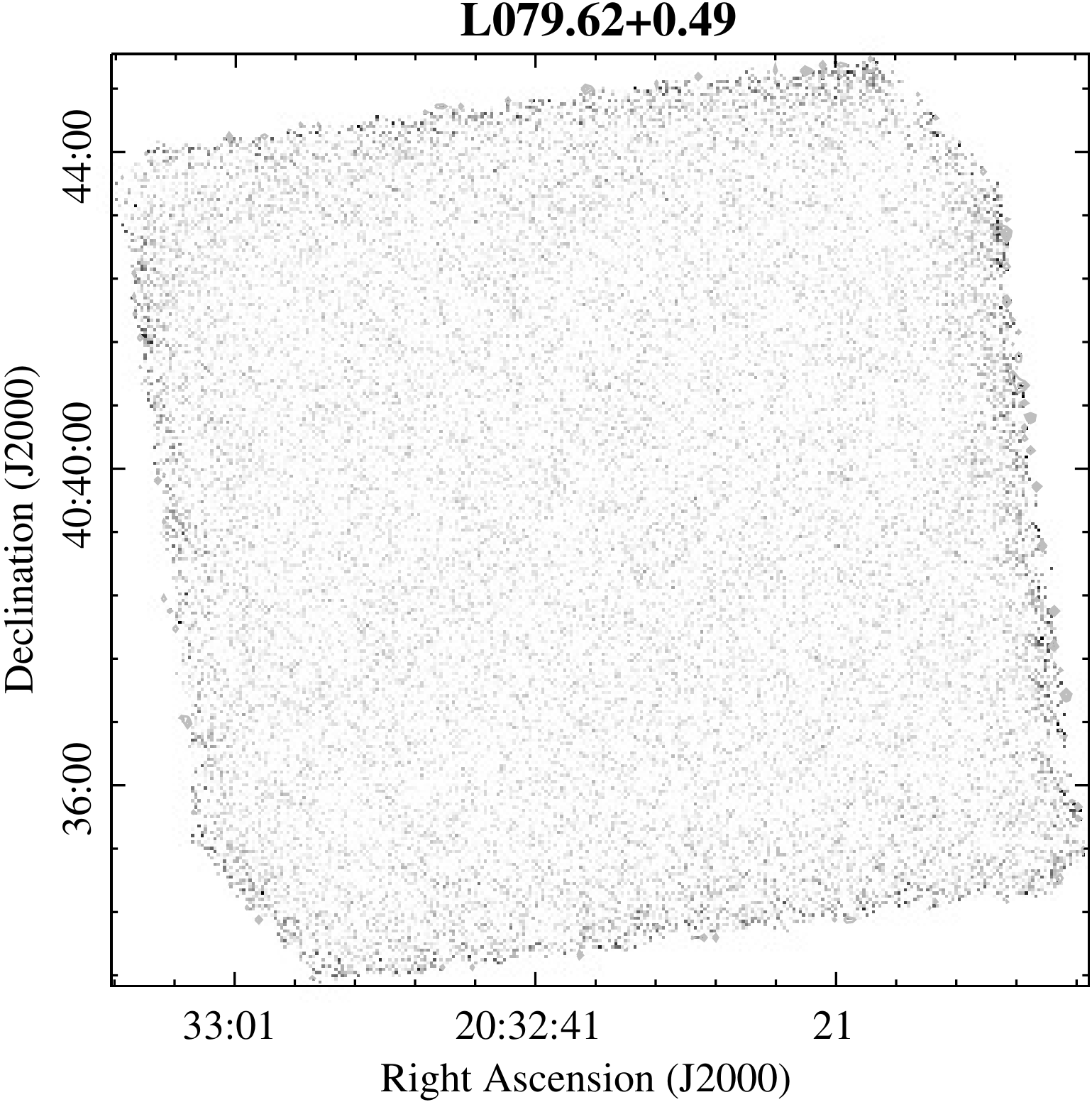}
}\\
\subfloat[L079.11-0.35 map, $\sigma_{rms}=1226$ mJy beam$^{-1}$.]{
\includegraphics[scale=0.43]{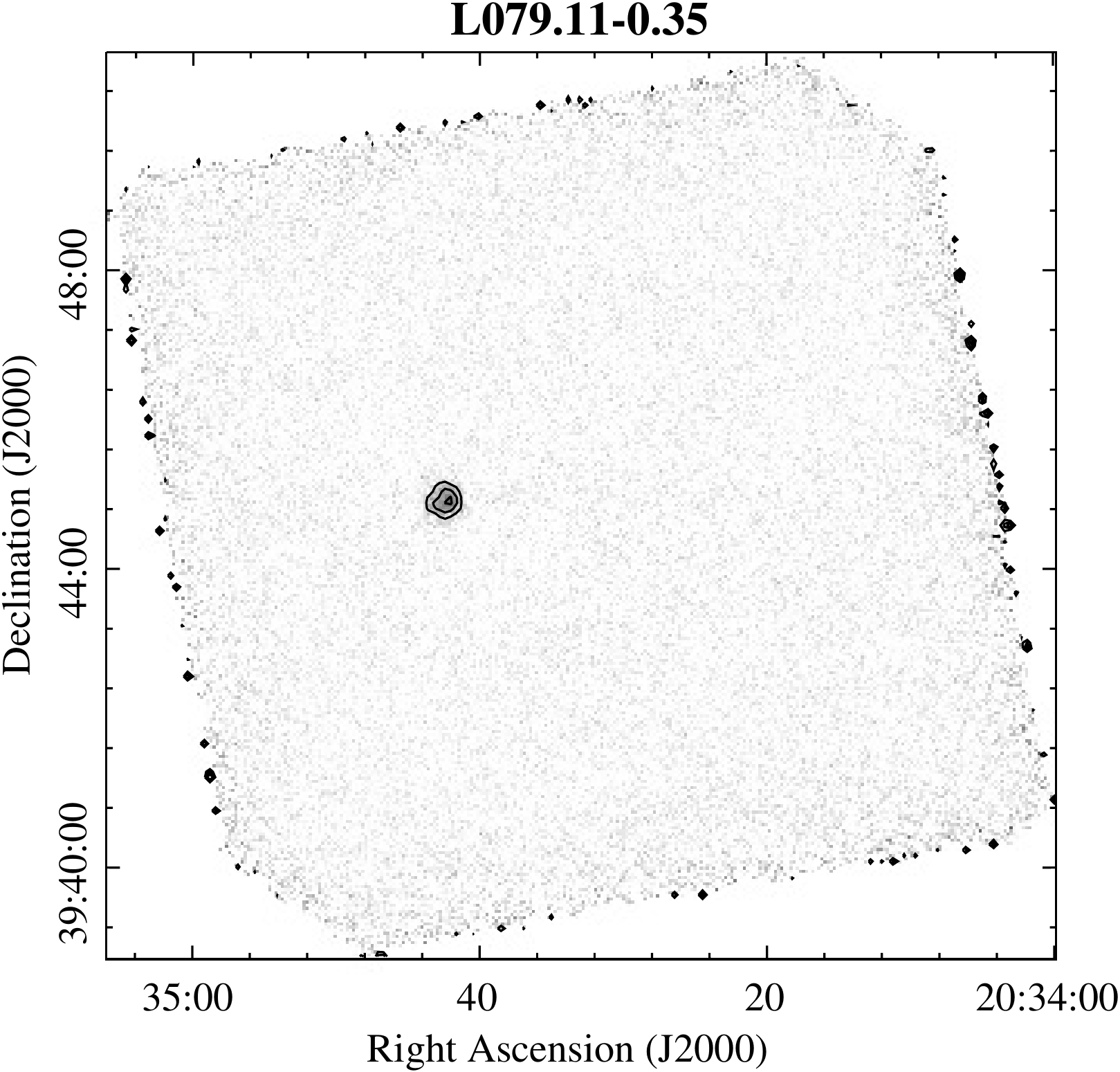}
\includegraphics[scale=0.43]{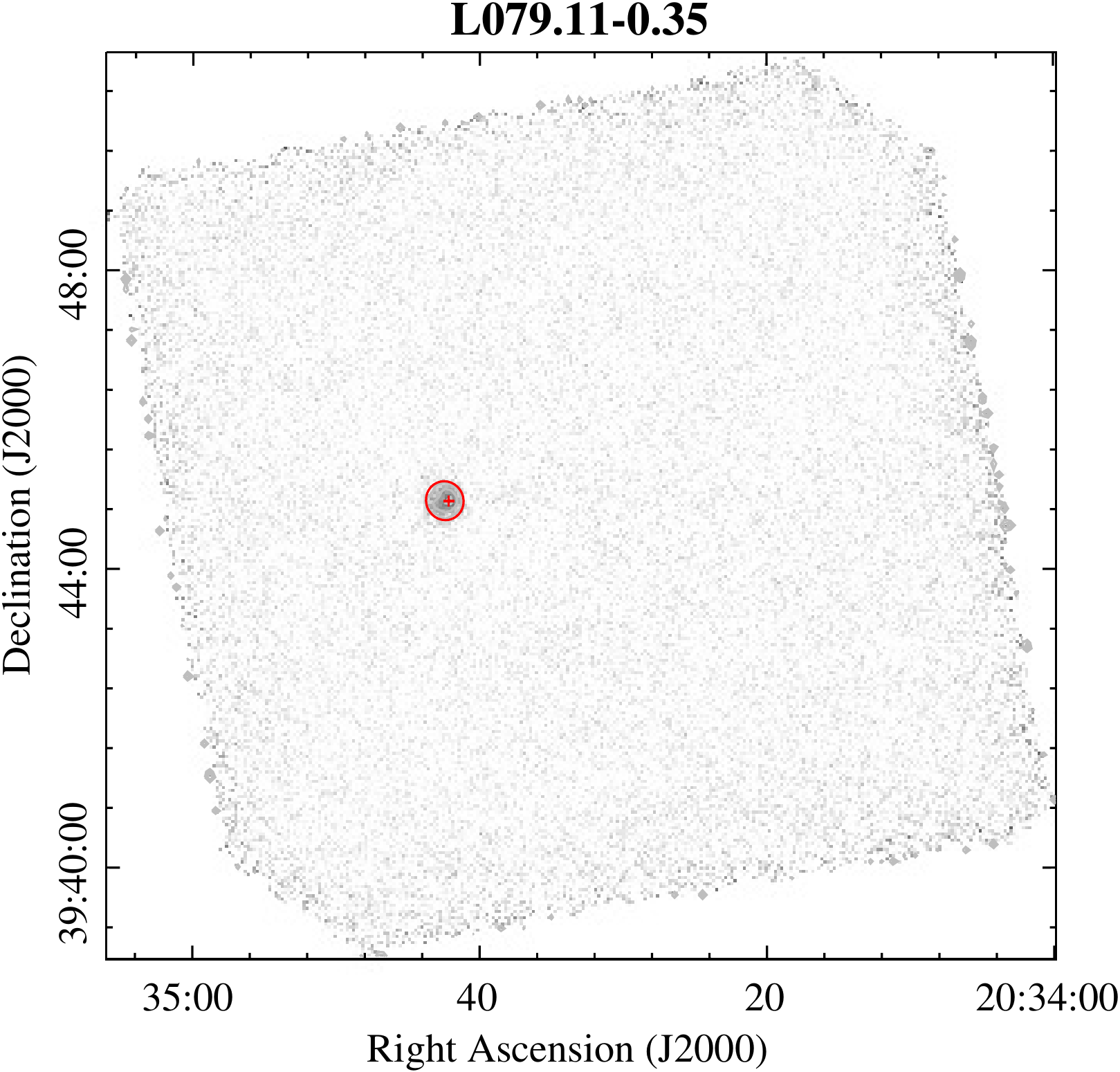}
}\\
\caption{Continuation}
\end{figure}

\clearpage
\begin{figure}\ContinuedFloat 
\center
\subfloat[L080.86+0.38 map, $\sigma_{rms}=1874$ mJy beam$^{-1}$.]{
\includegraphics[scale=0.43]{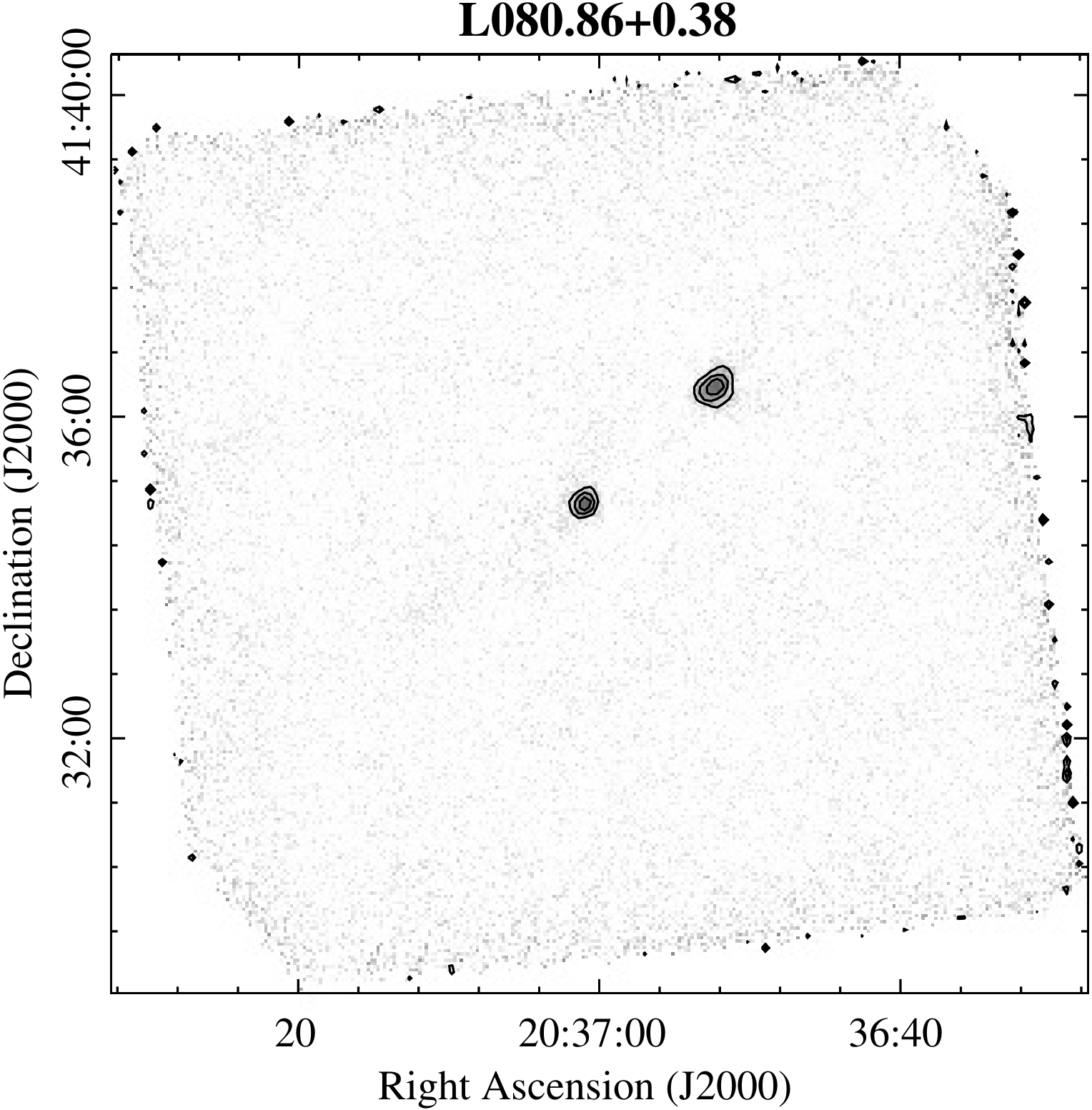}
\includegraphics[scale=0.43]{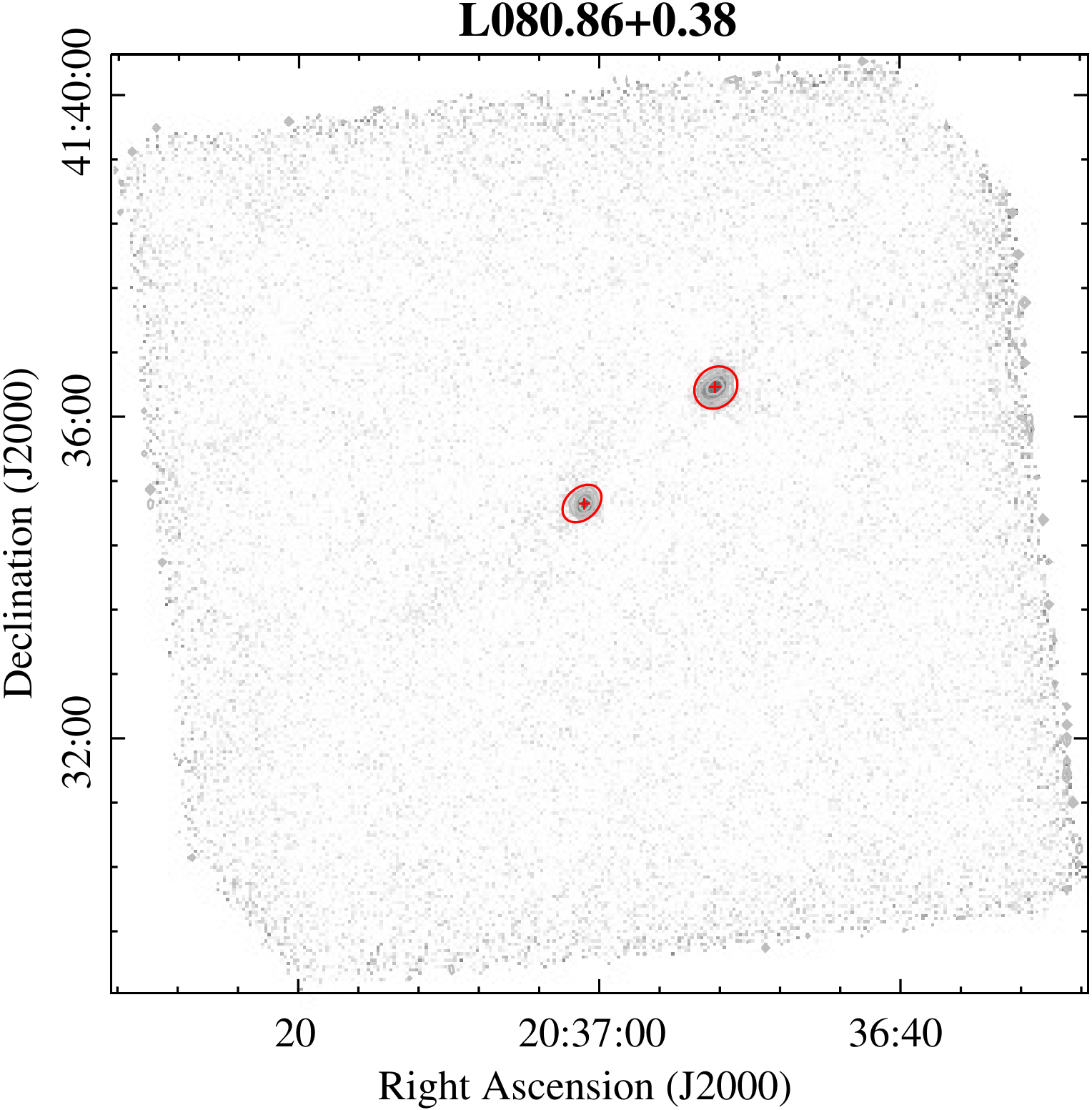}
}\\
\subfloat[L110.11+0.05 map, $\sigma_{rms}=288$ mJy beam$^{-1}$.]{
\includegraphics[scale=0.43]{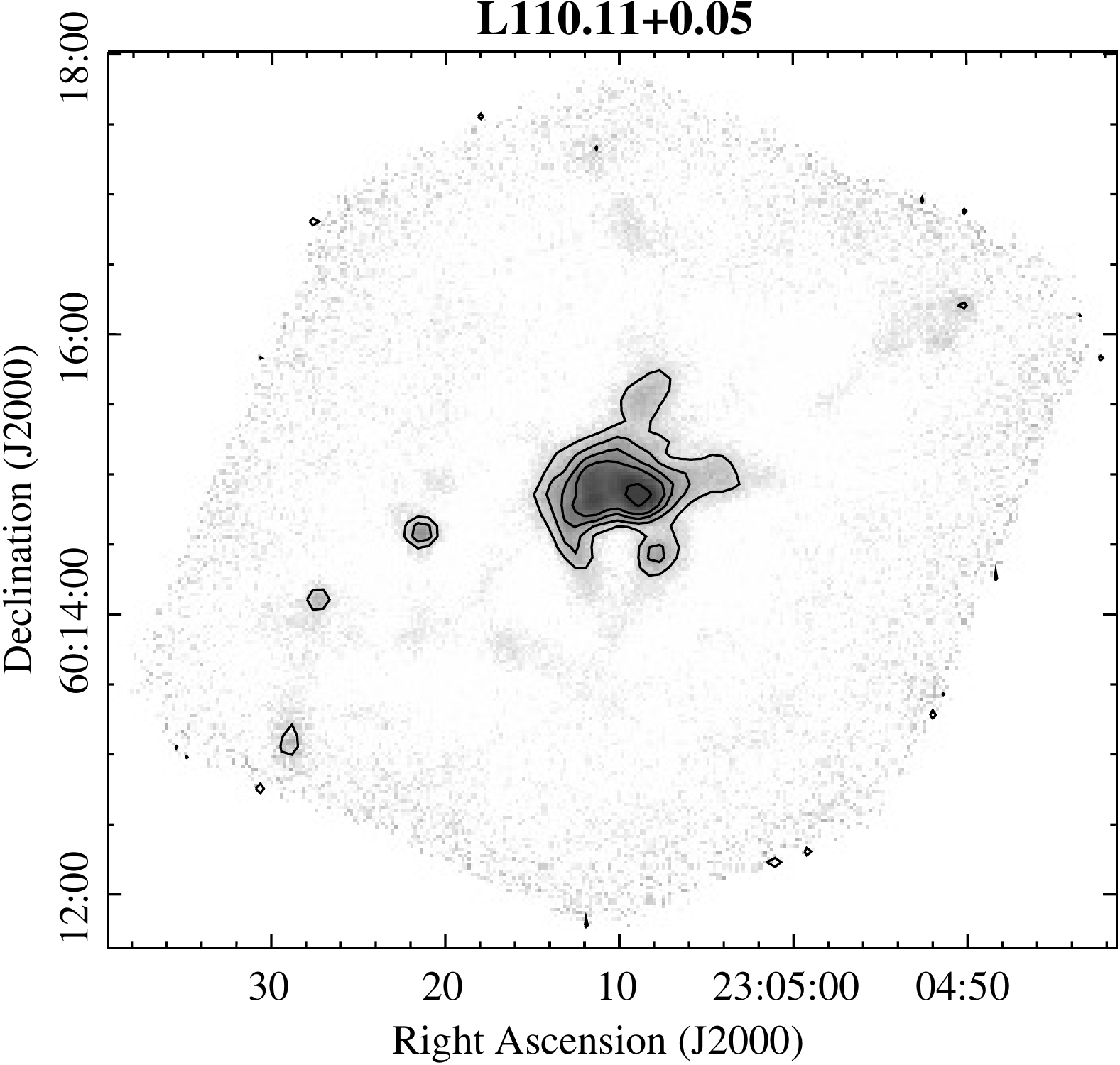}
\includegraphics[scale=0.43]{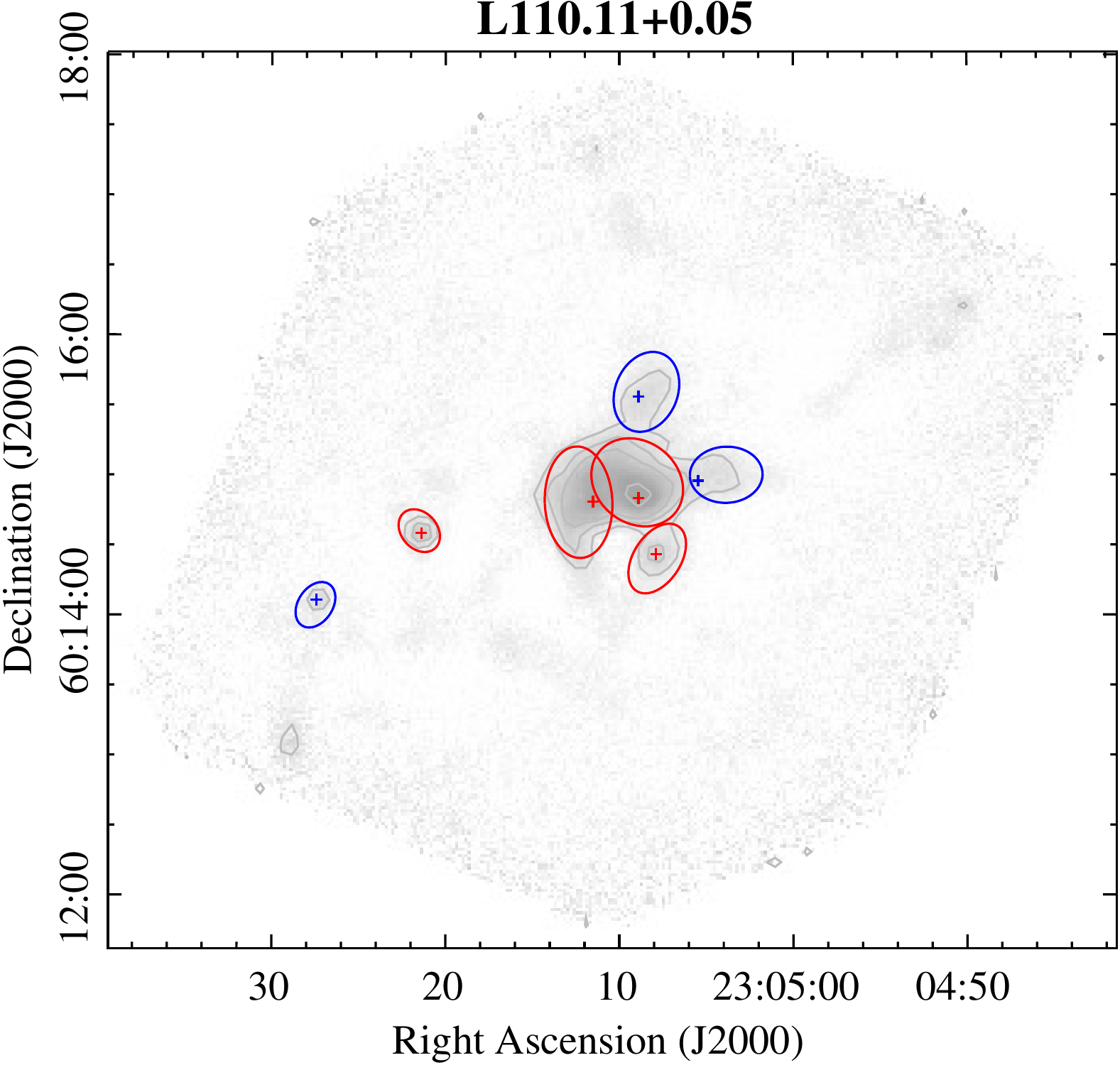}
}\\
\subfloat[L111.62+0.38 map, $\sigma_{rms}=342$ mJy beam$^{-1}$.]{
\includegraphics[scale=0.43]{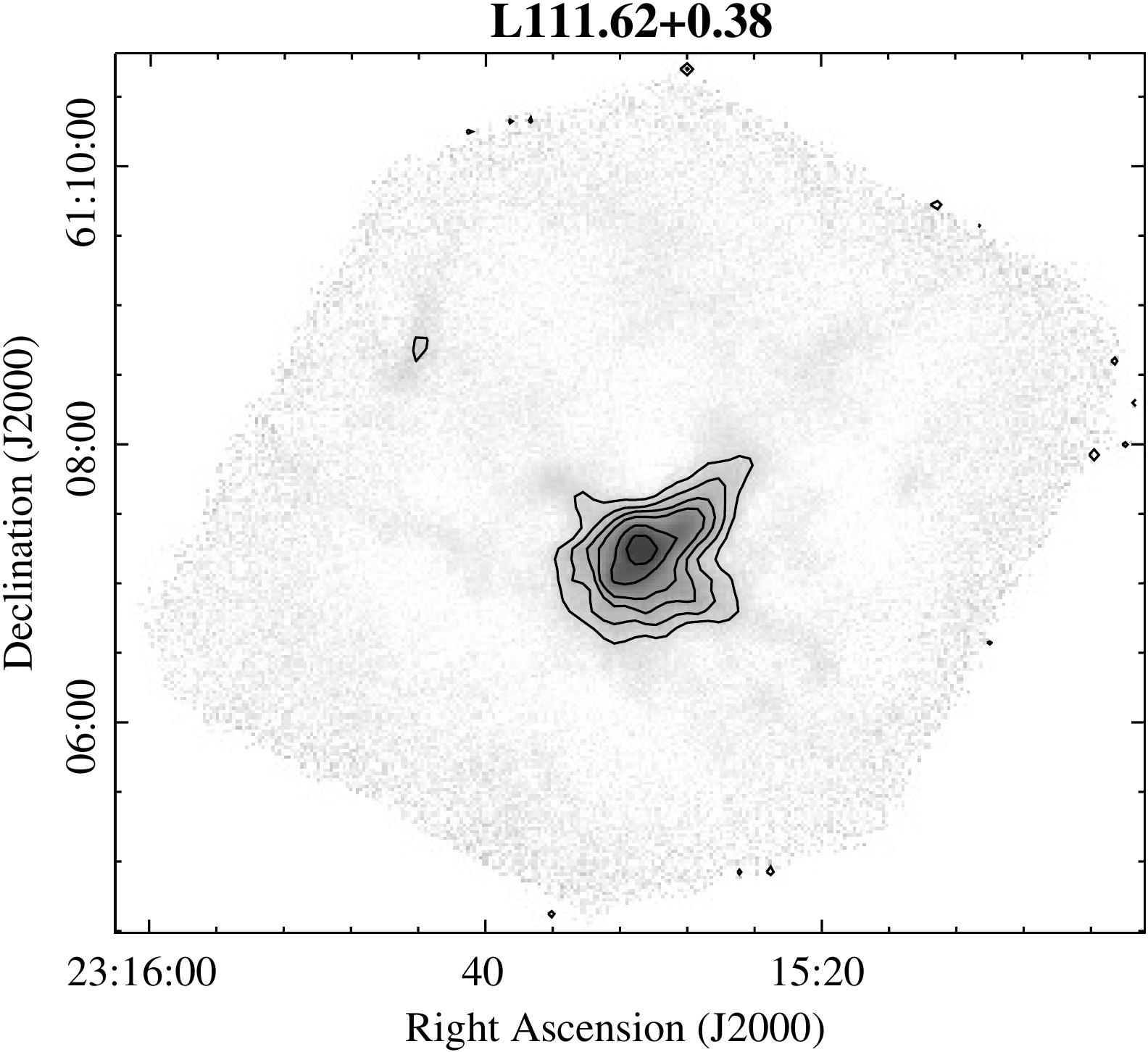}
\includegraphics[scale=0.43]{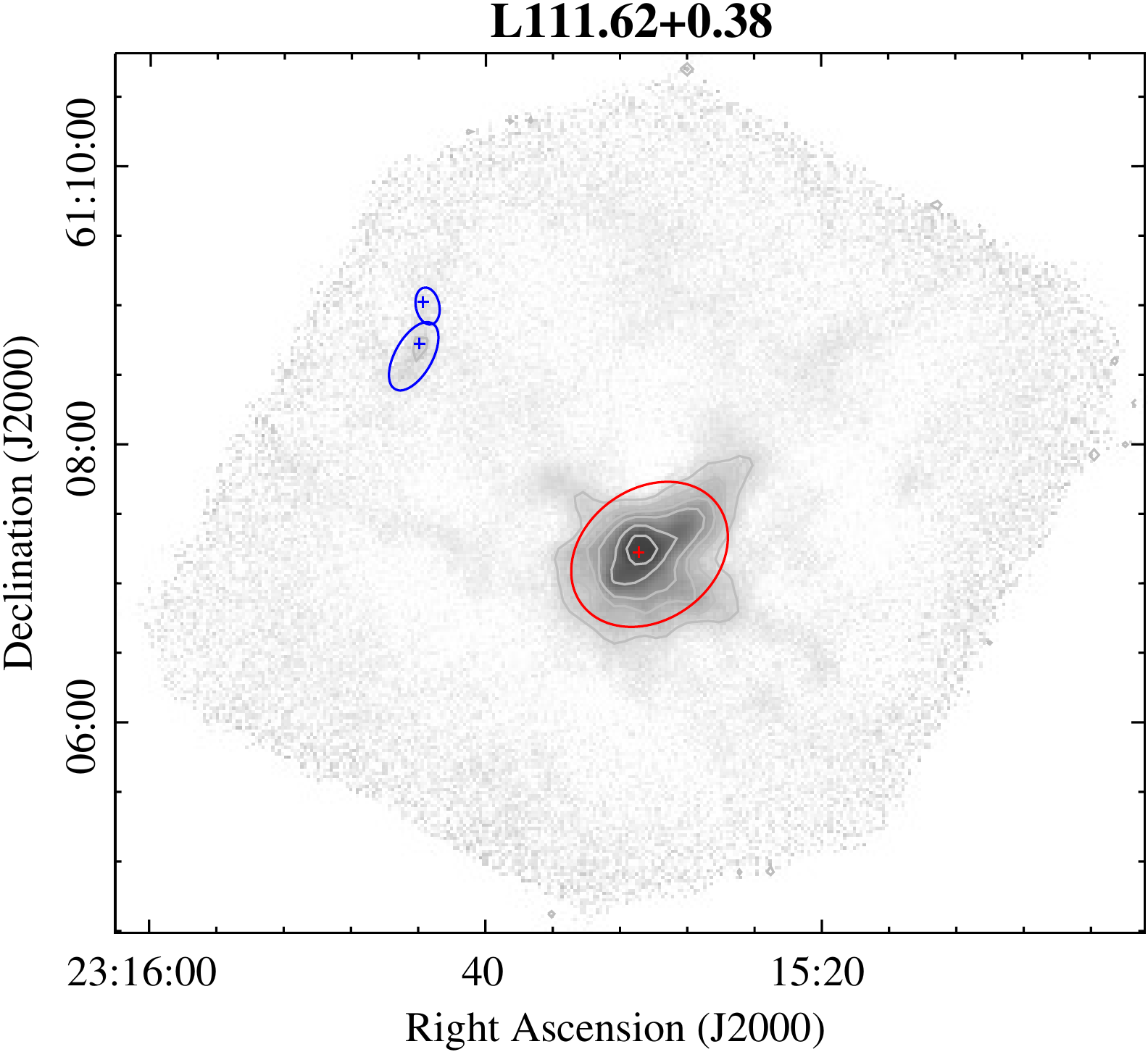}
}\\
\caption{Continuation}
\end{figure}

\clearpage
\begin{figure}\ContinuedFloat 
\center
\subfloat[L134.28+0.86 map, $\sigma_{rms}=275$ mJy beam$^{-1}$.]{
\includegraphics[scale=0.43]{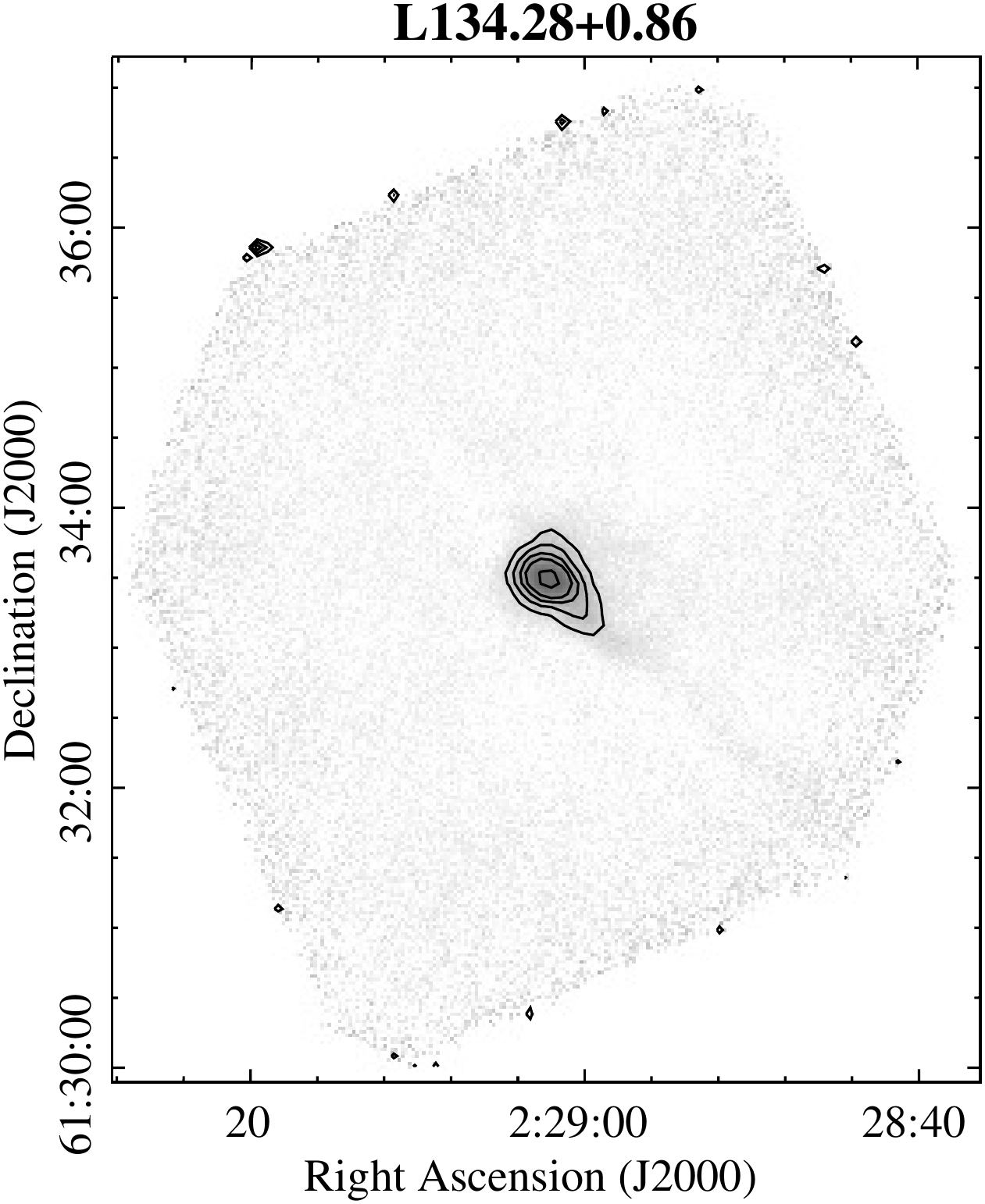}
\includegraphics[scale=0.43]{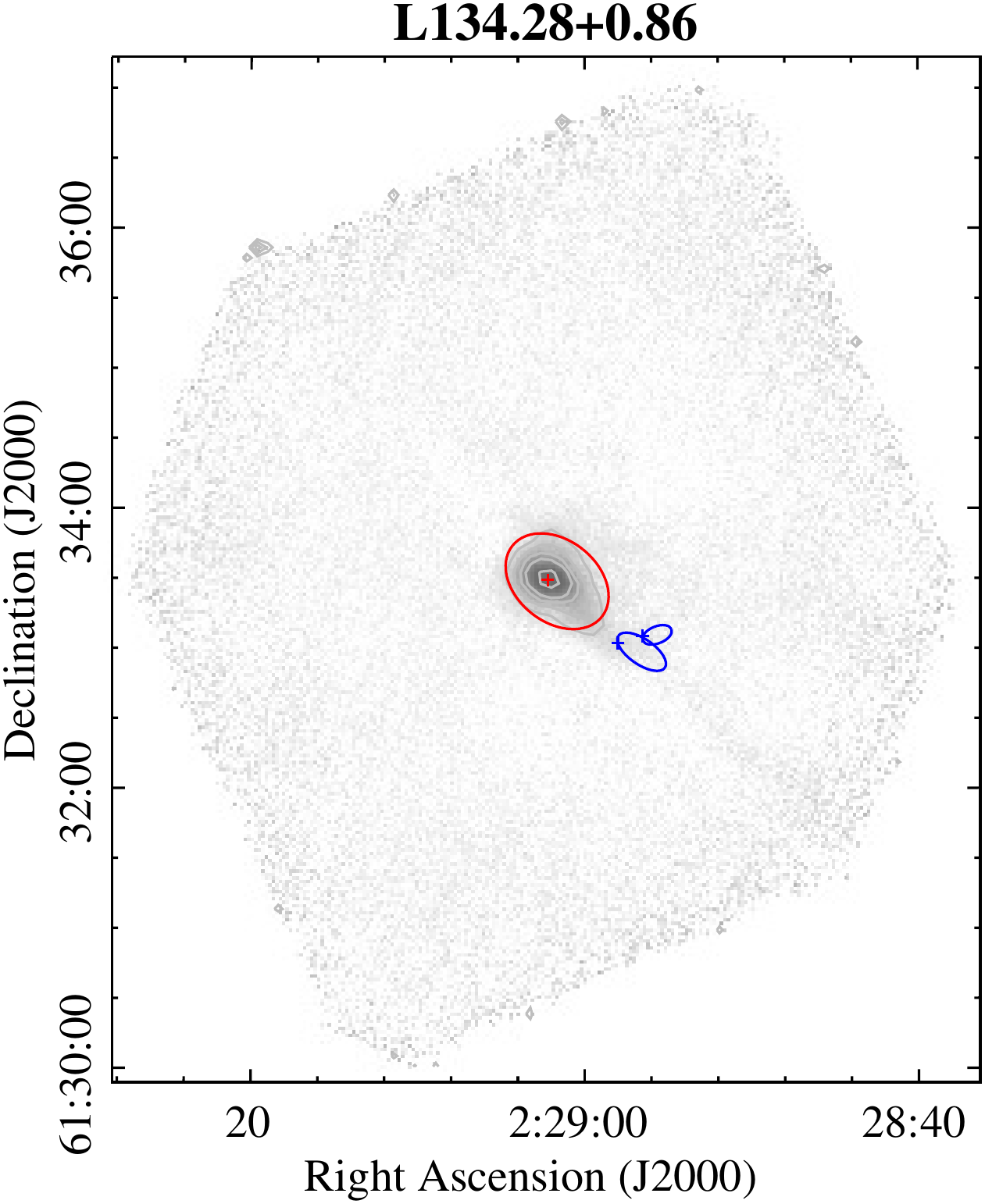}
}\\
\subfloat[L134.83+1.31 map, $\sigma_{rms}=296$ mJy beam$^{-1}$.]{
\includegraphics[scale=0.43]{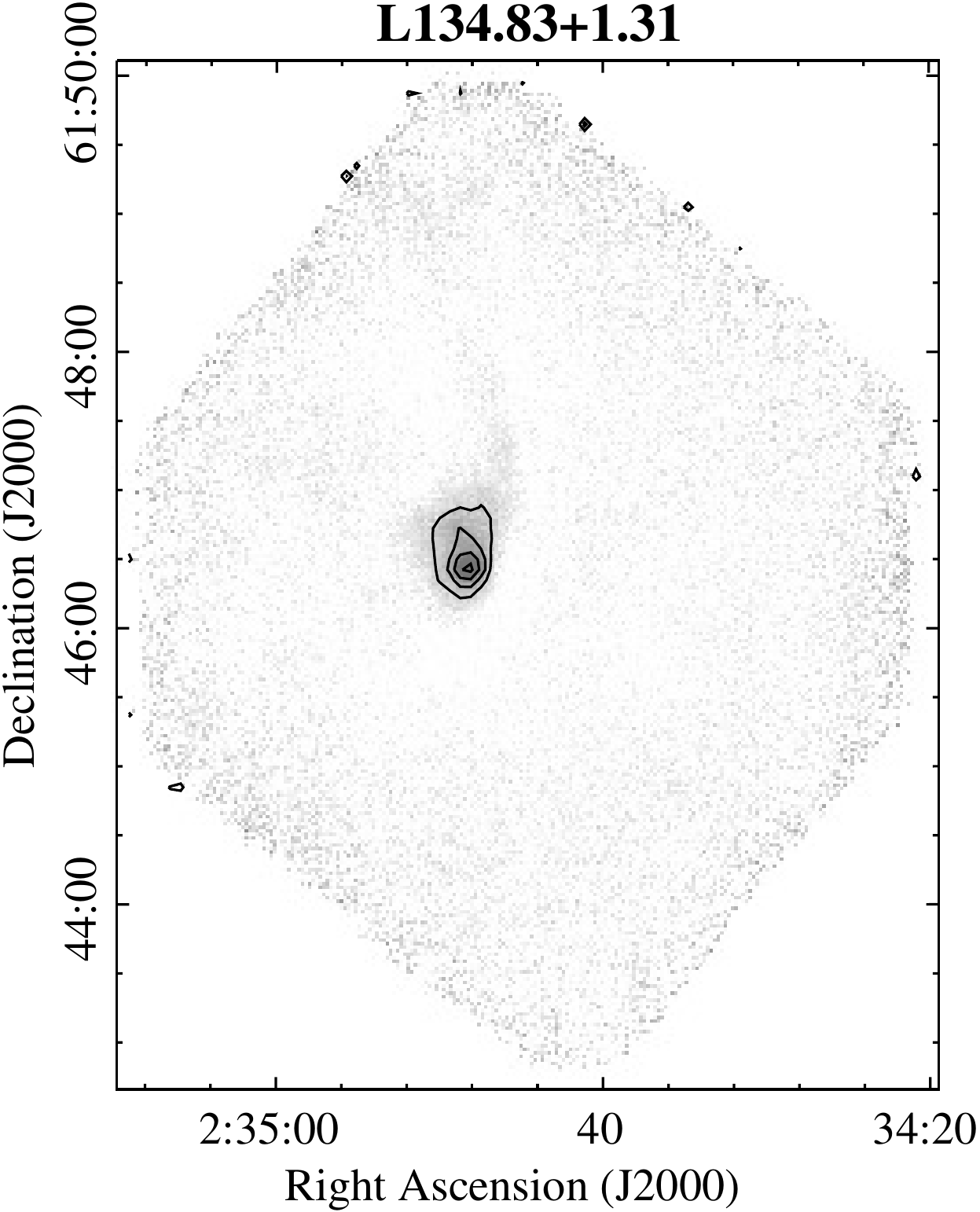}
\includegraphics[scale=0.43]{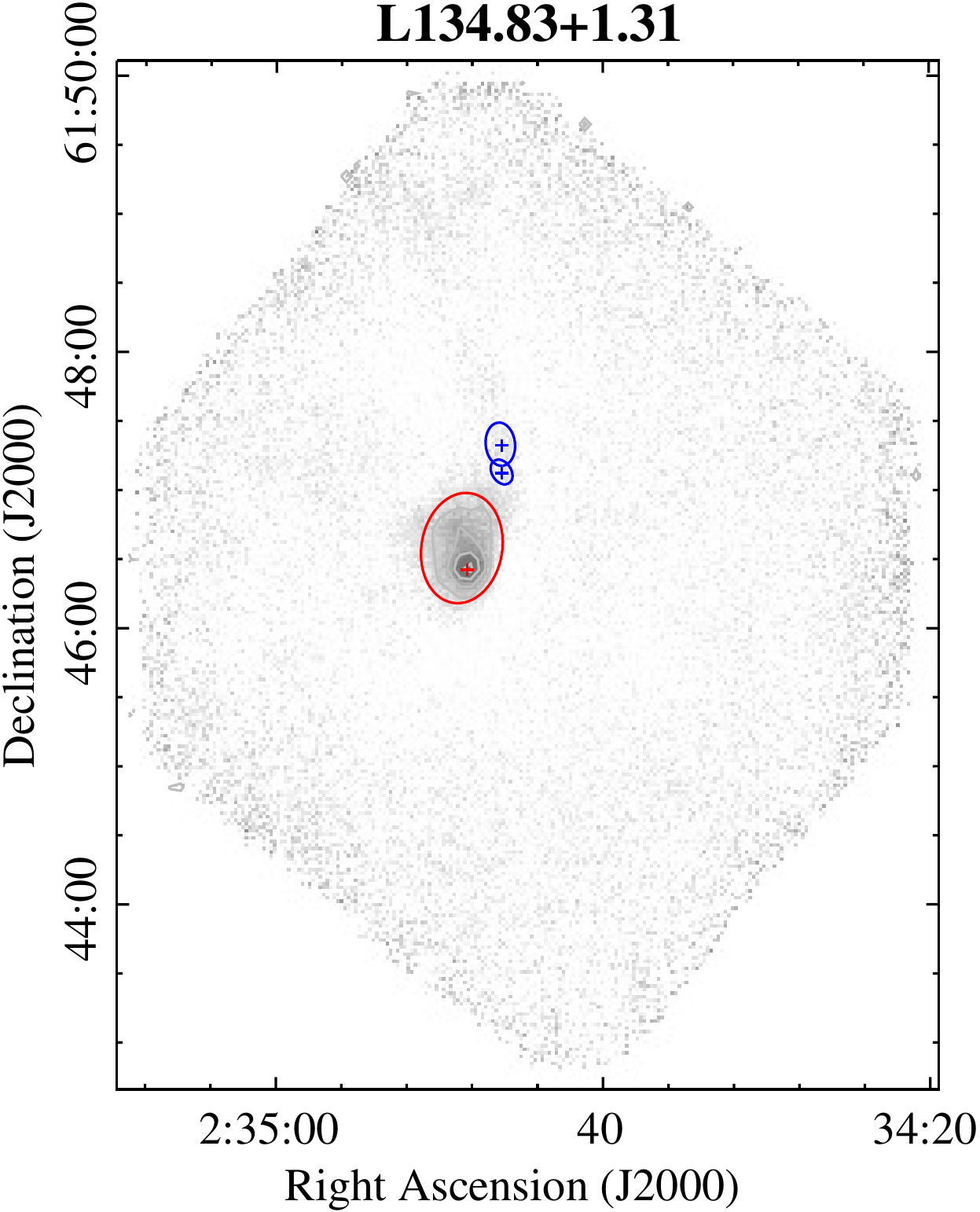}
}\\
\subfloat[L136.38+2.27 map, $\sigma_{rms}=257$ mJy beam$^{-1}$.]{
\includegraphics[scale=0.43]{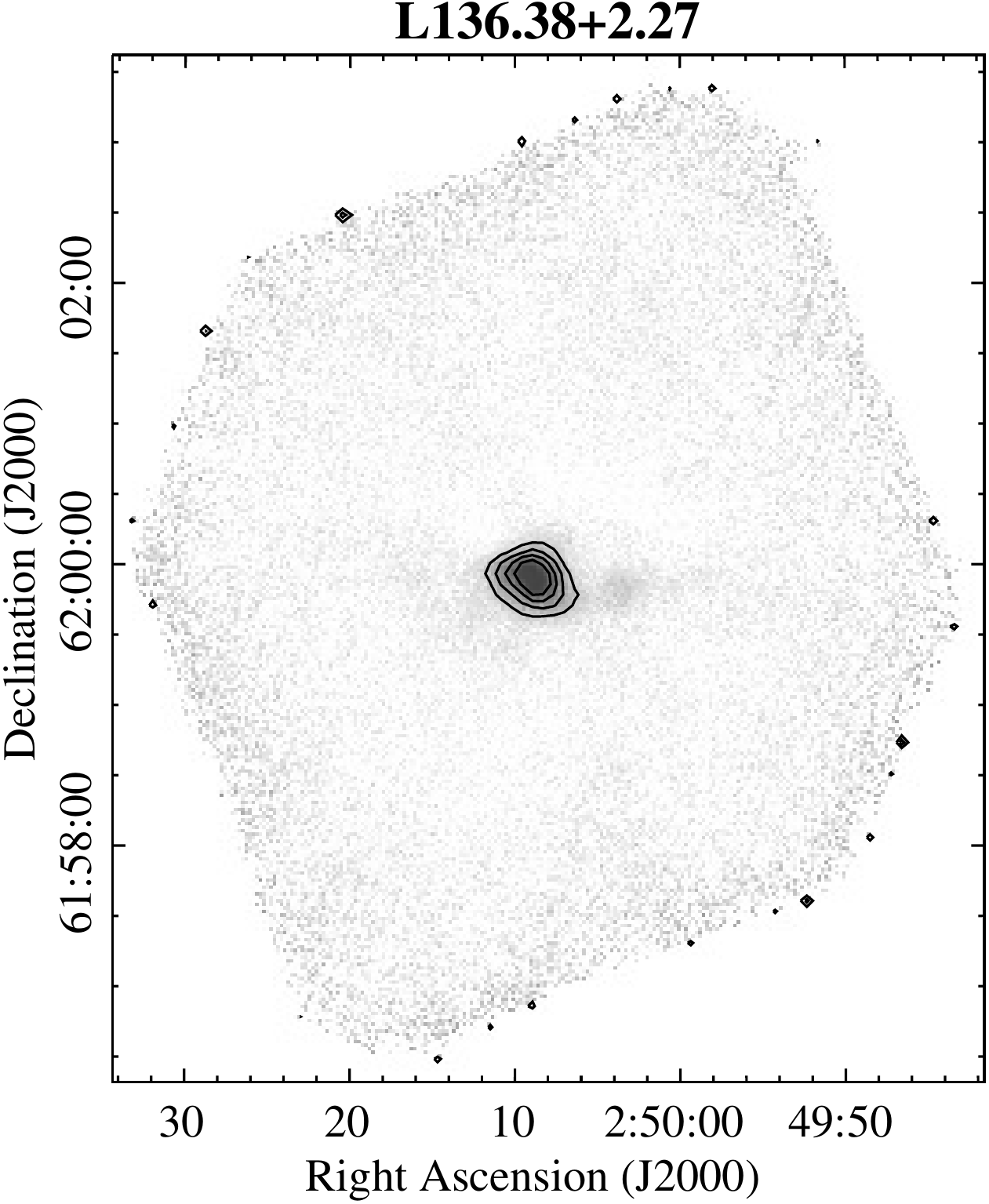}
\includegraphics[scale=0.43]{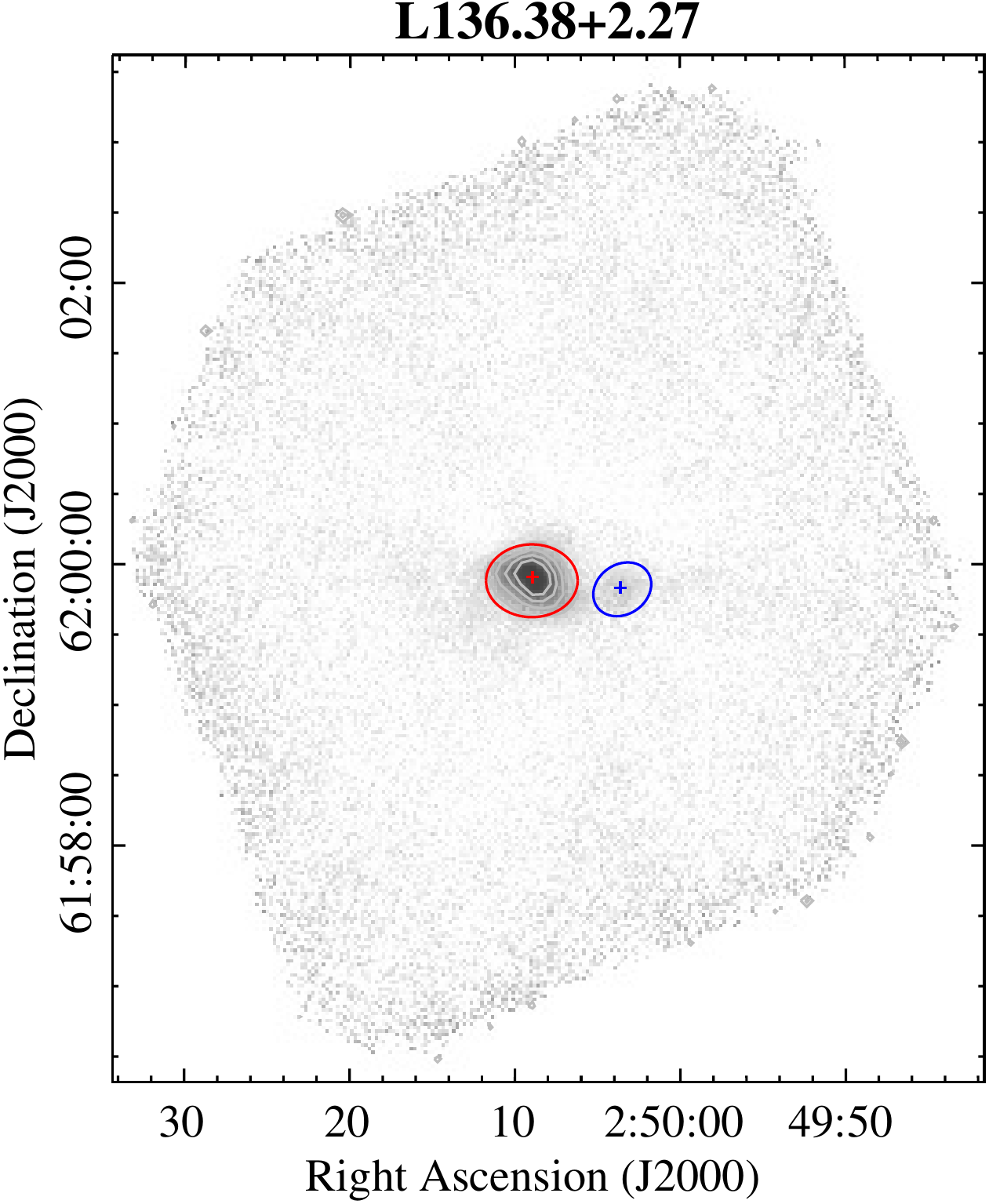}
}\\
\caption{Continuation}
\end{figure}

\clearpage
\begin{figure}\ContinuedFloat 
\center
\subfloat[L136.83+1.07 map, $\sigma_{rms}=250$ mJy beam$^{-1}$.]{
\includegraphics[scale=0.43]{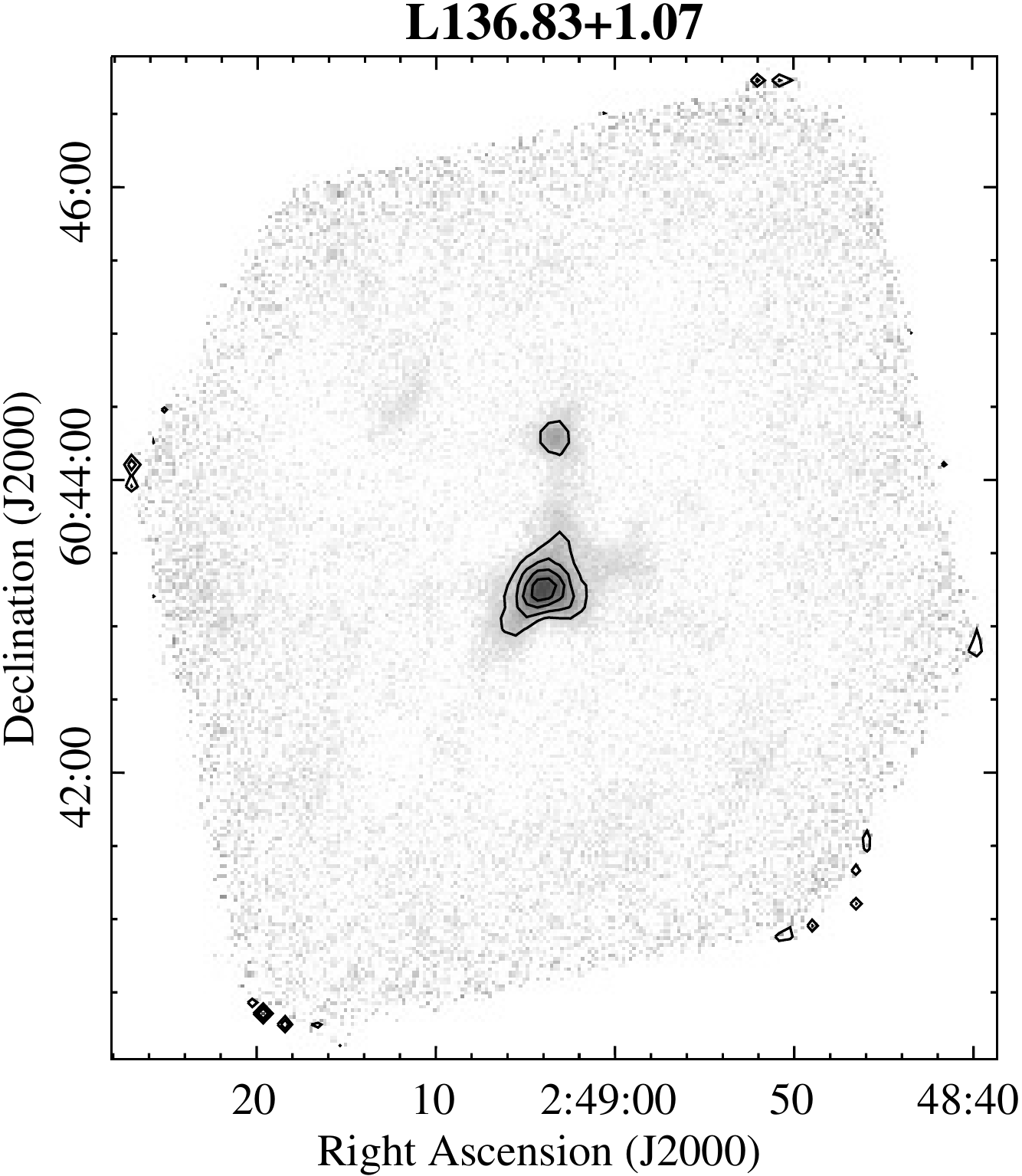}
\includegraphics[scale=0.43]{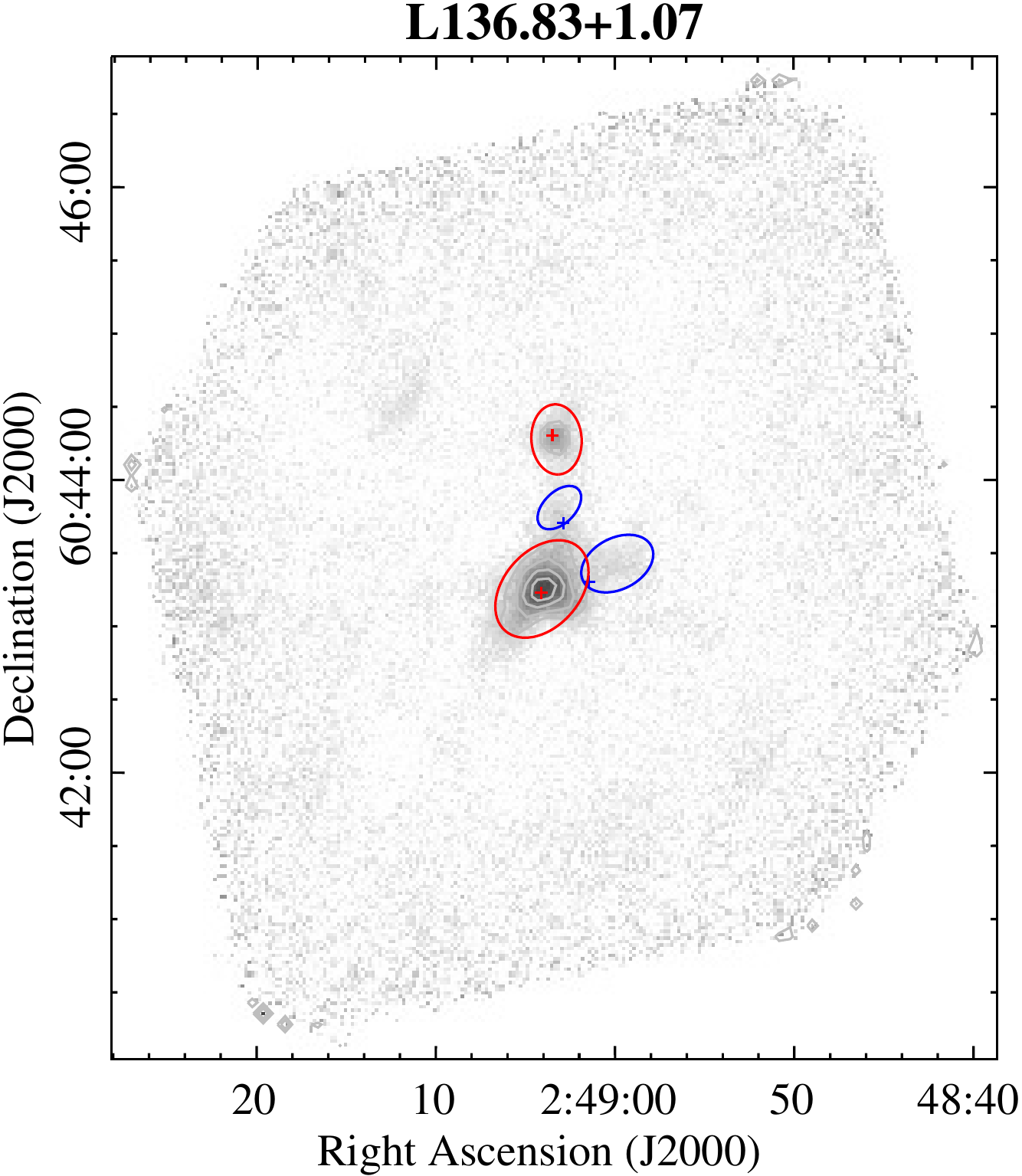}
}\\
\subfloat[L111.28-0.66 map, $\sigma_{rms}=286$ mJy beam$^{-1}$.]{
\includegraphics[scale=0.43]{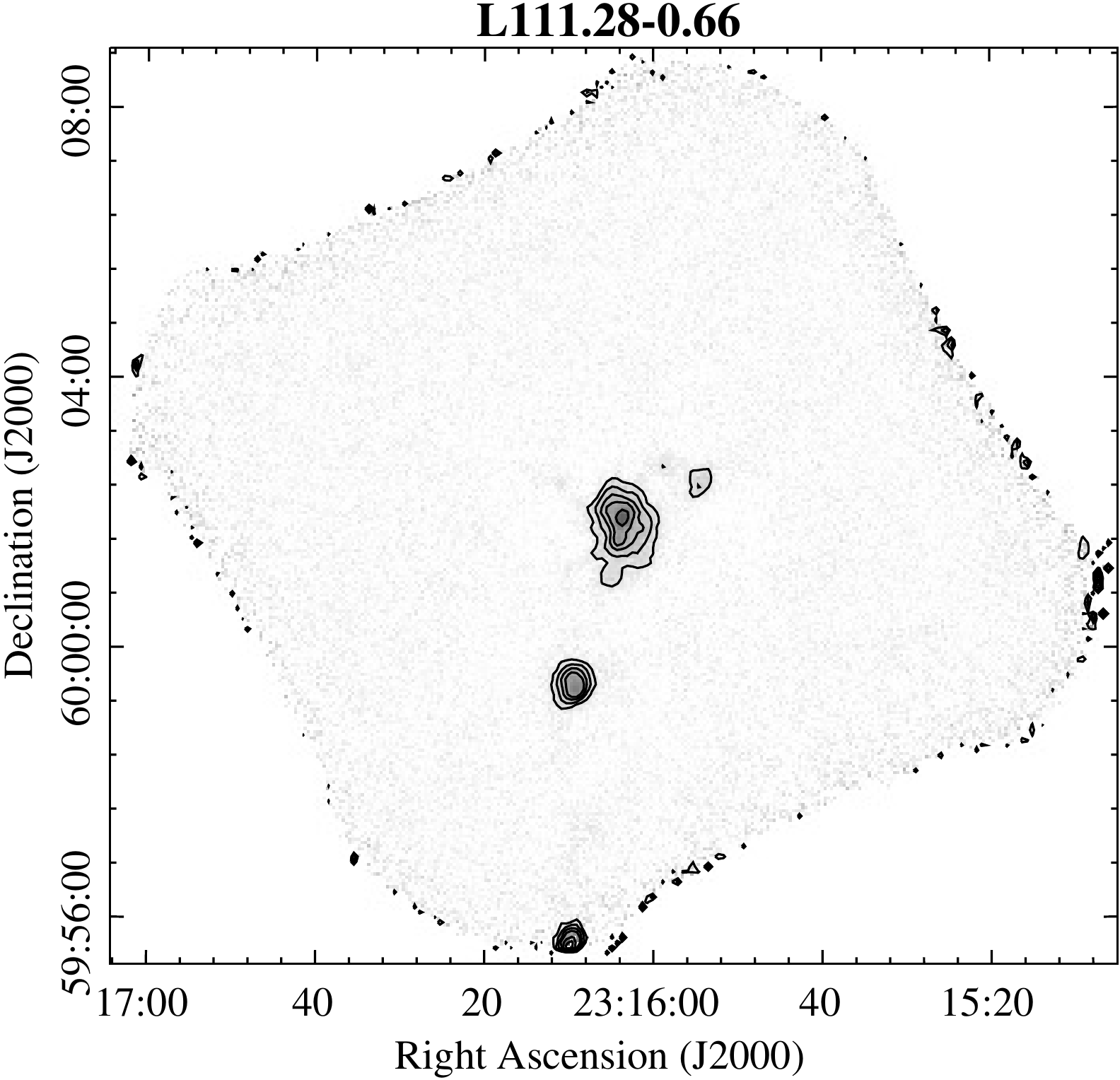}
\includegraphics[scale=0.43]{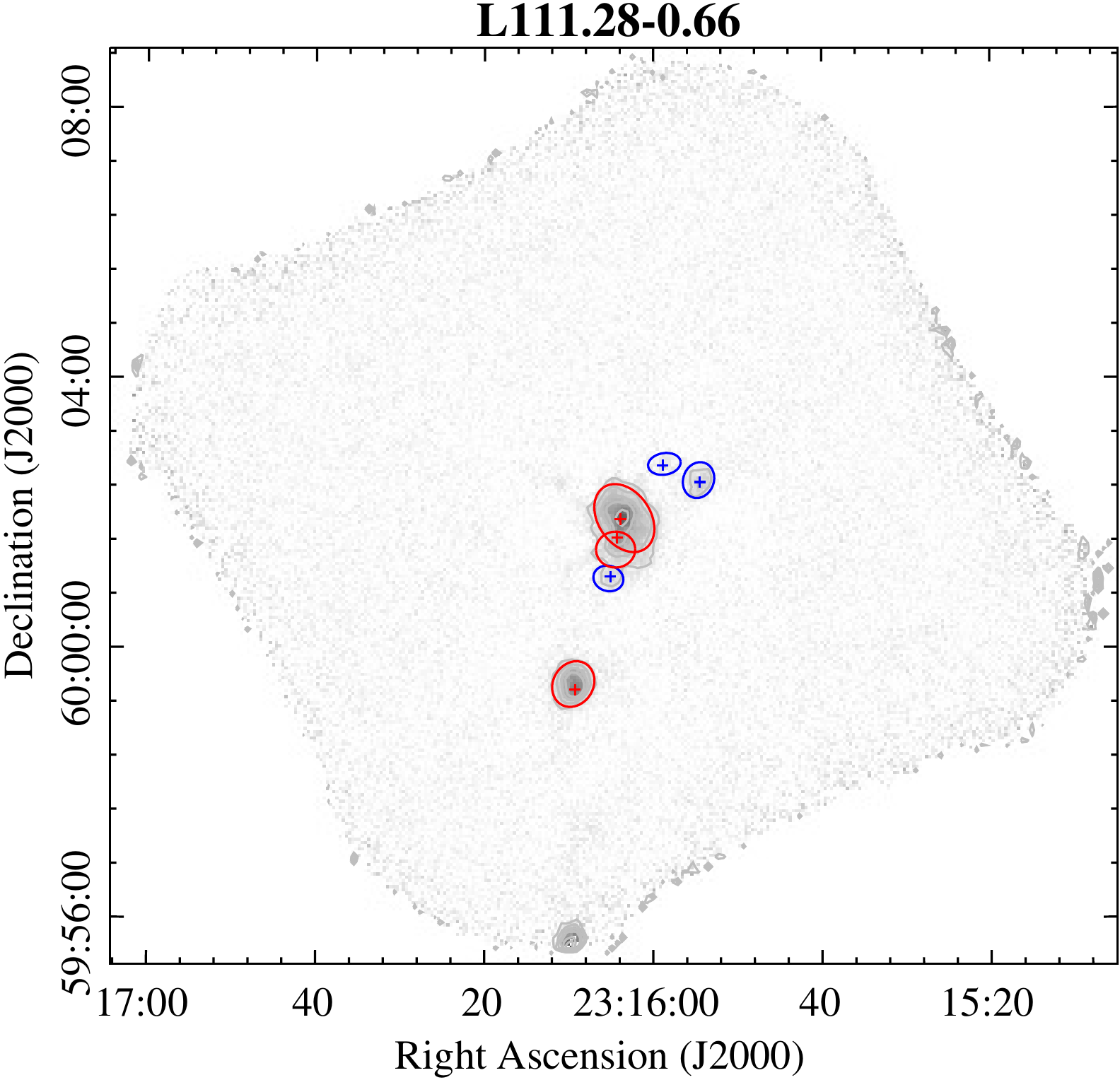}
}\\
\subfloat[L111.54+0.78 map, $\sigma_{rms}=496$ mJy beam$^{-1}$.]{
\includegraphics[scale=0.43]{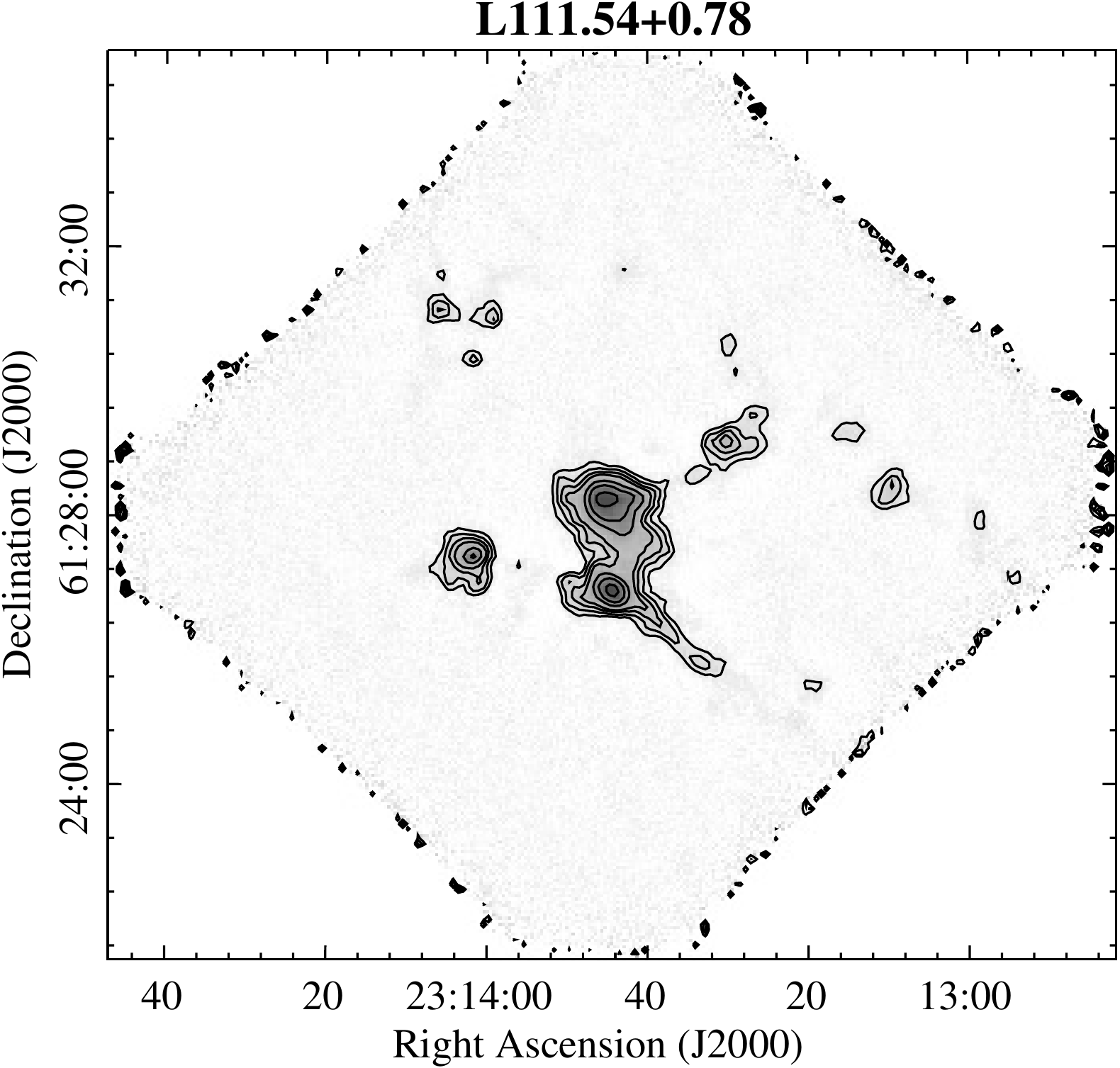}
\includegraphics[scale=0.43]{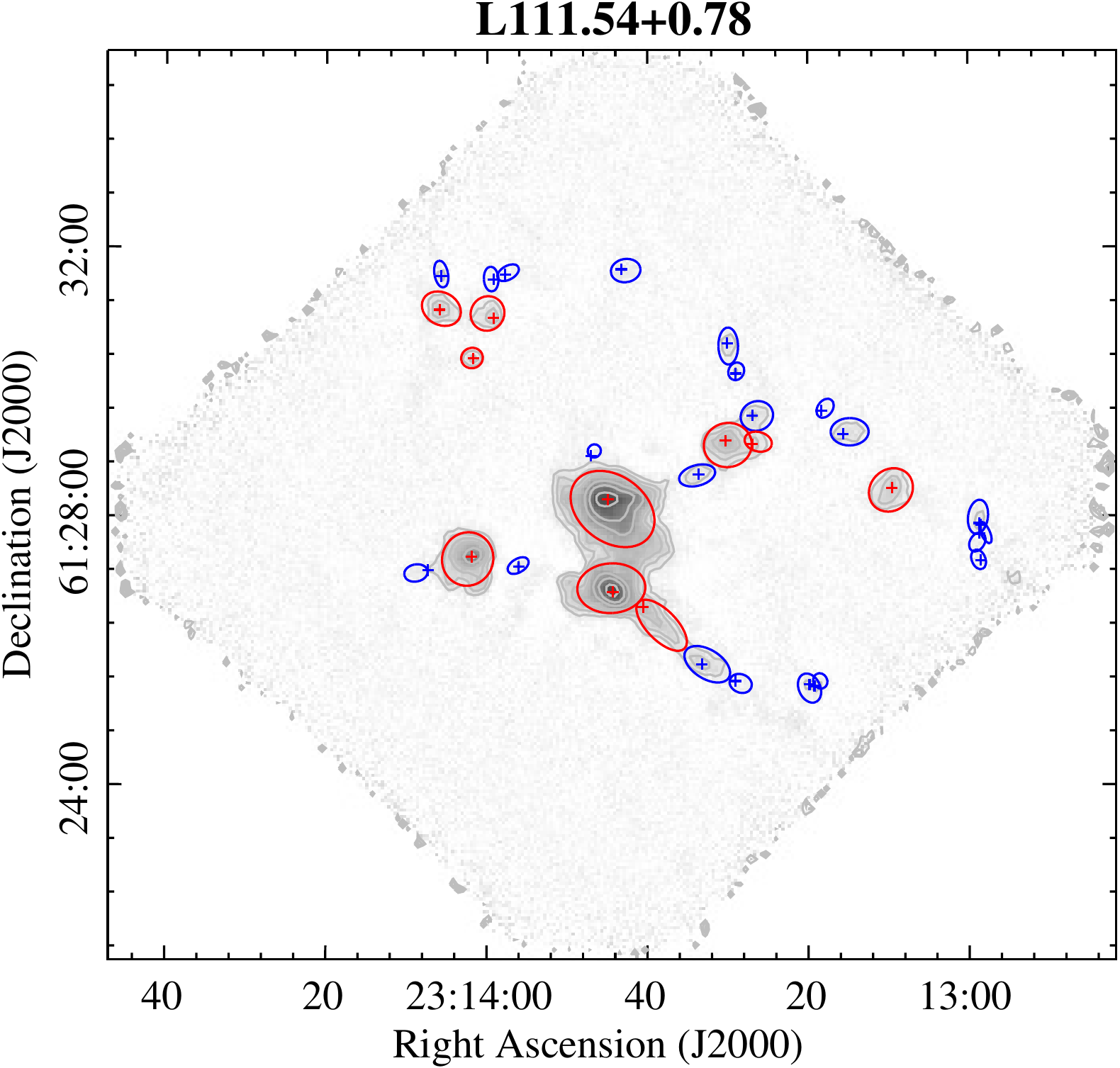}
}\\
\caption{Continuation}
\end{figure}

\clearpage
\begin{figure}\ContinuedFloat 
\center
\subfloat[L133.71+1.21 map, $\sigma_{rms}=338$ mJy beam$^{-1}$.]{
\includegraphics[scale=0.43]{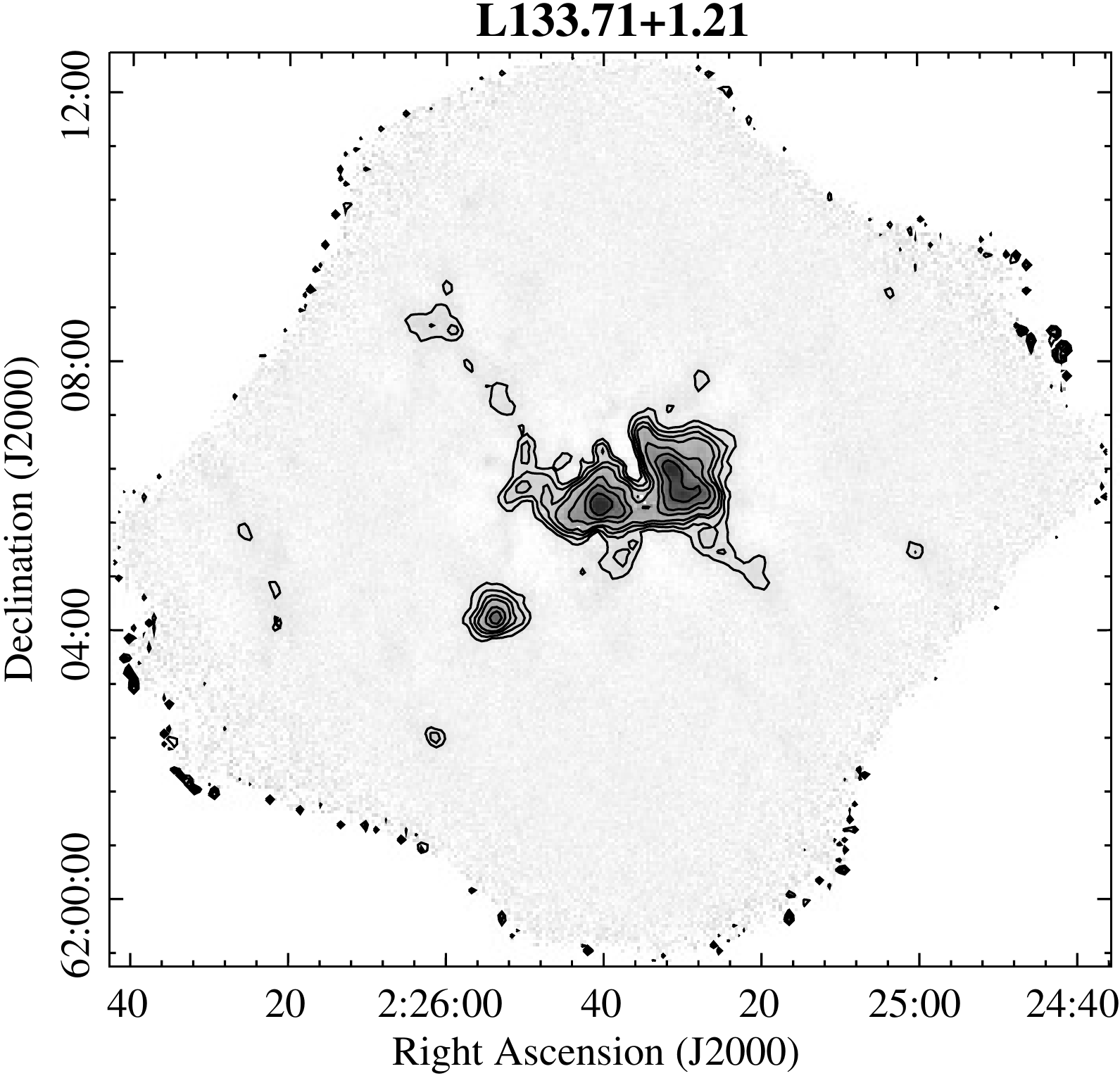}
\includegraphics[scale=0.43]{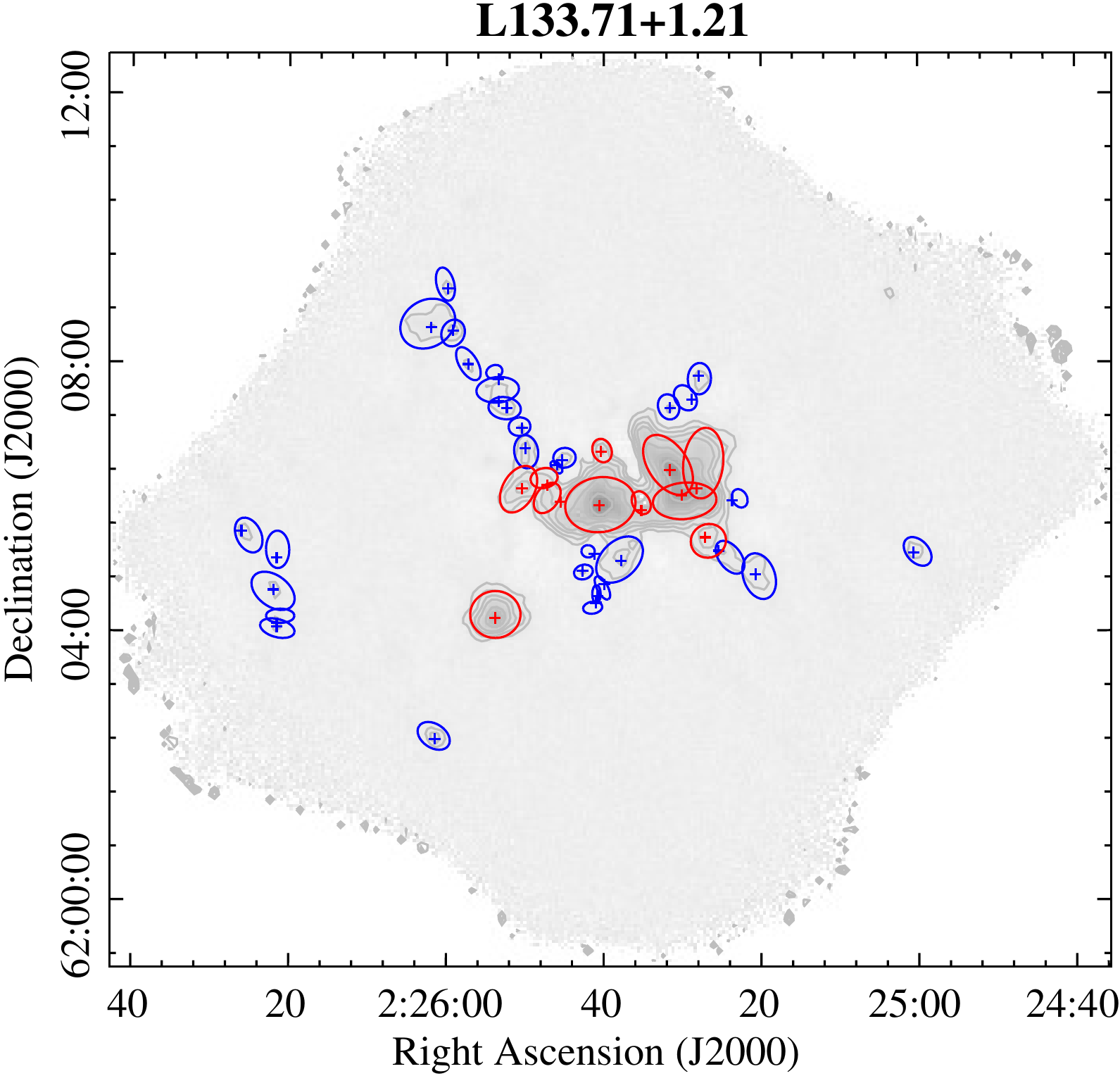}
}\\
\subfloat[L133.95+1.06 map, $\sigma_{rms}=338$ mJy beam$^{-1}$. Additional contour is drawn at 200$\sigma$.]{
\includegraphics[scale=0.43]{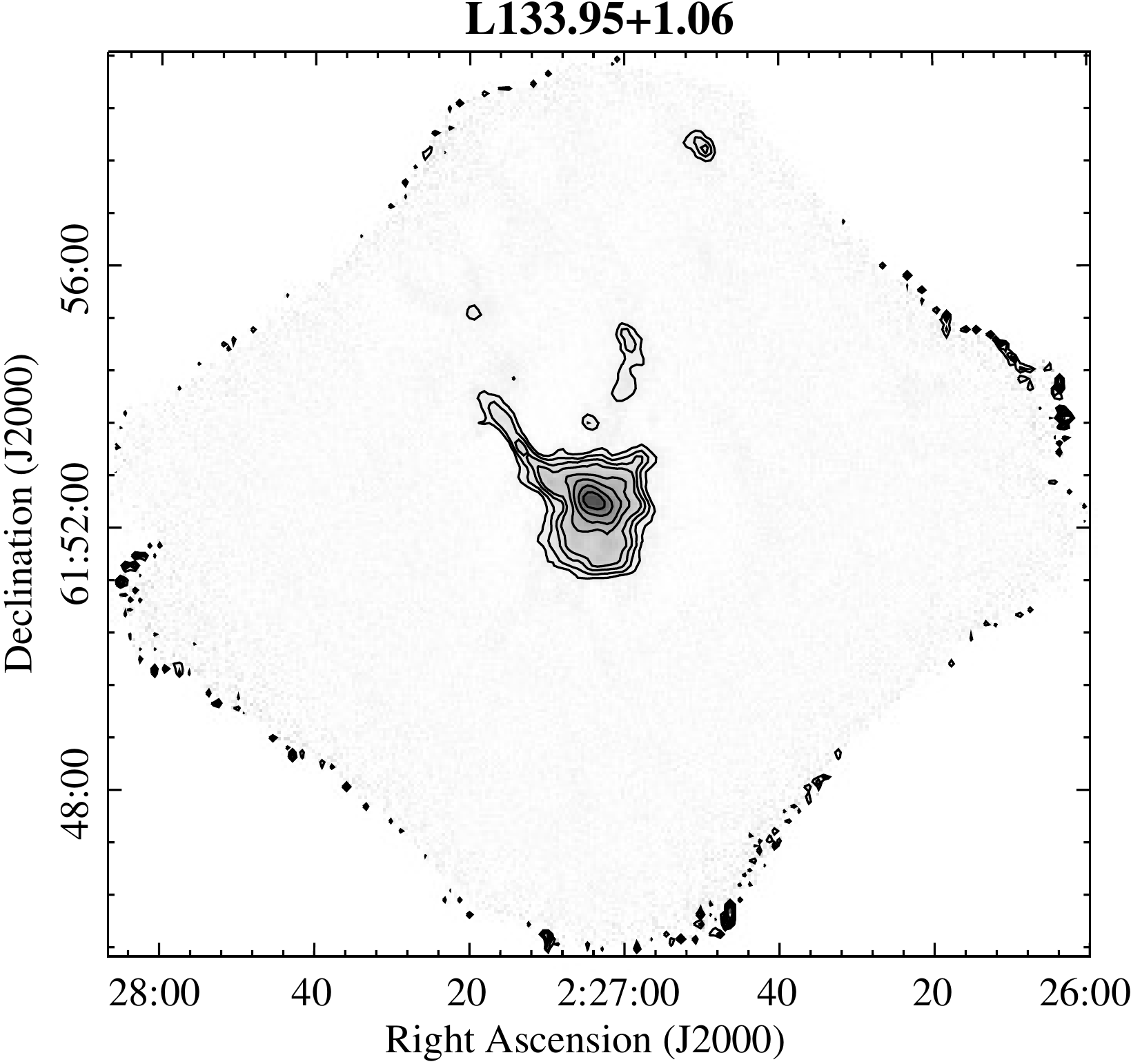}
\includegraphics[scale=0.43]{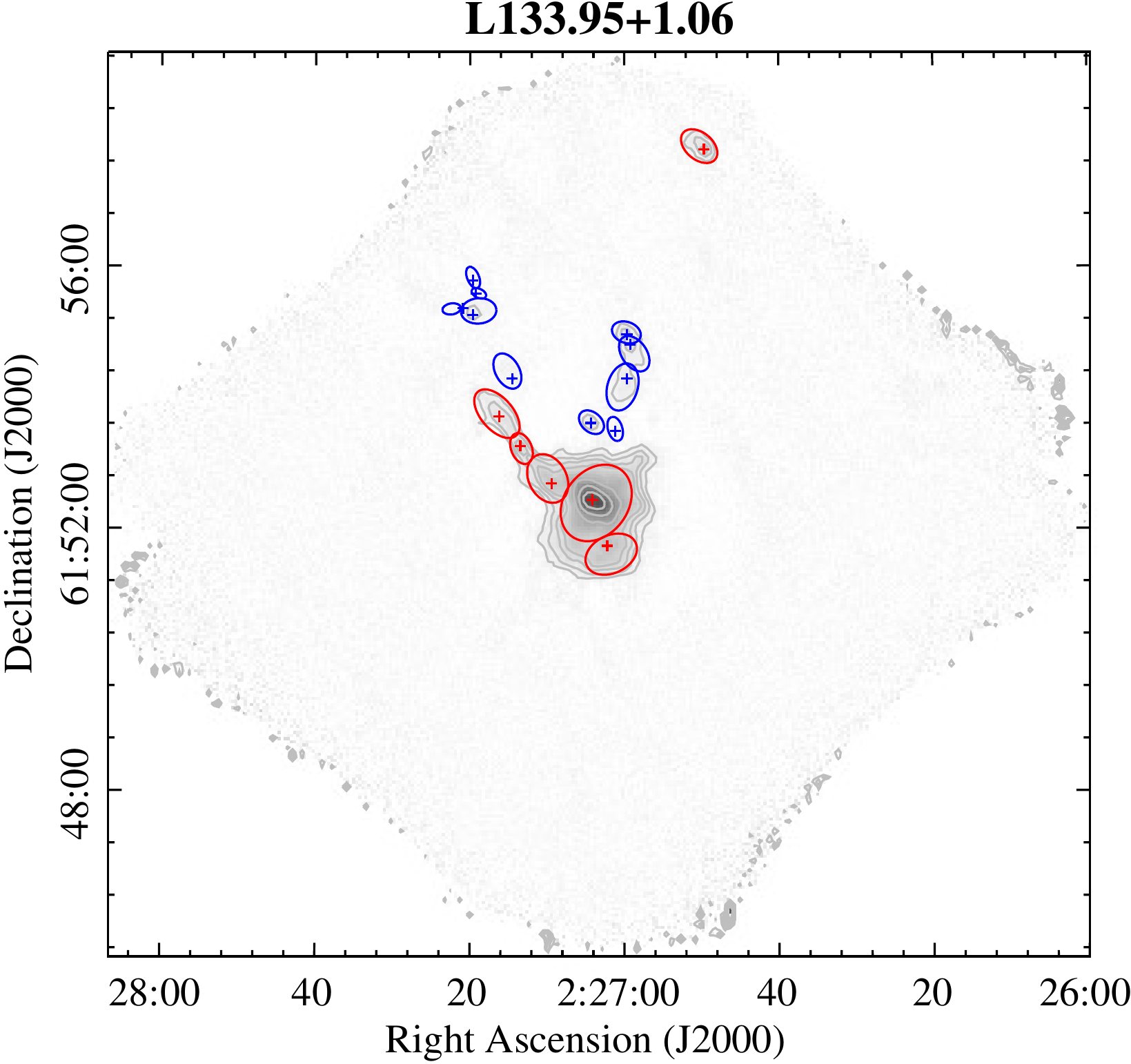}
}\\
\subfloat[L111.26-0.77 map, $\sigma_{rms}=538$ mJy beam$^{-1}$.]{
\includegraphics[scale=0.43]{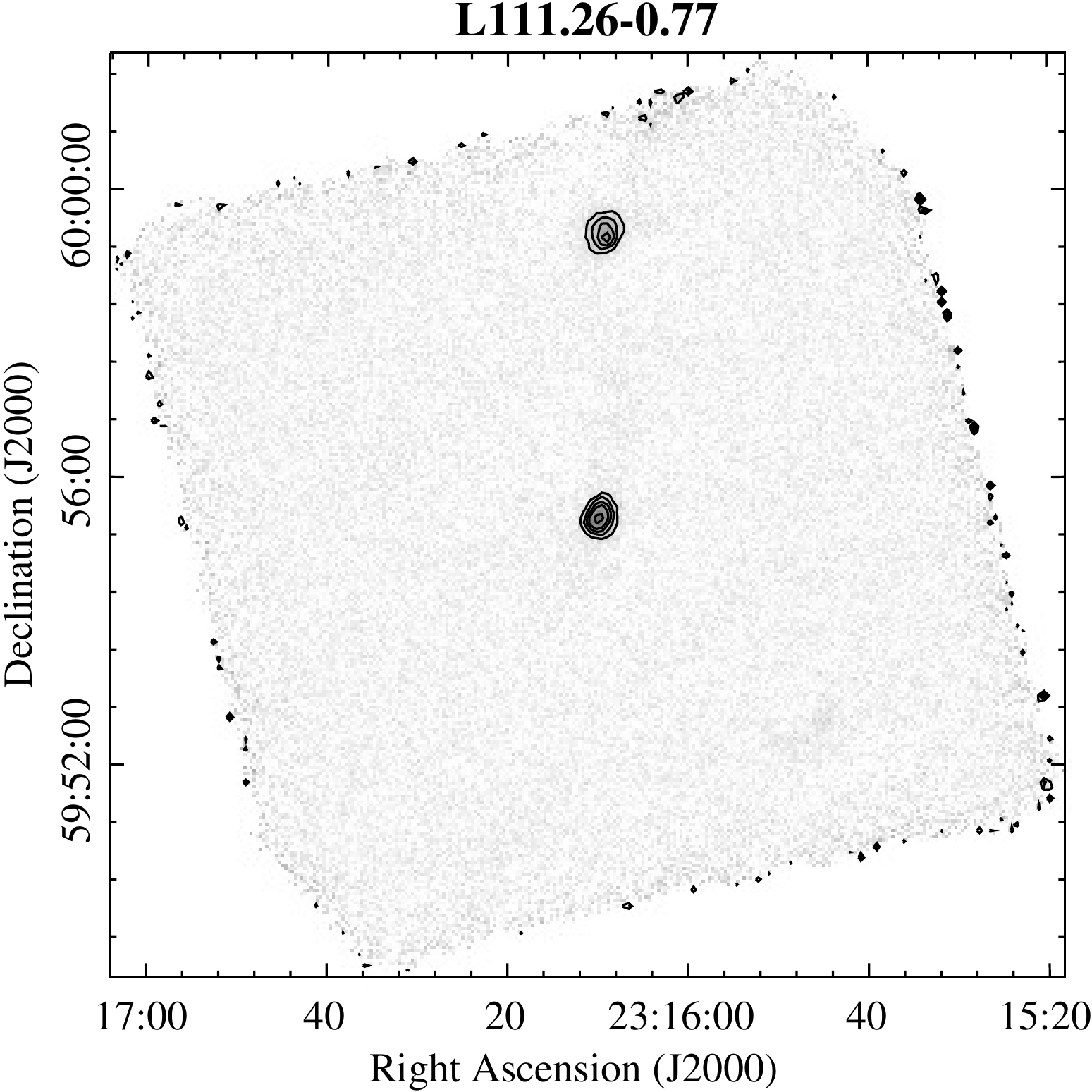}
\includegraphics[scale=0.43]{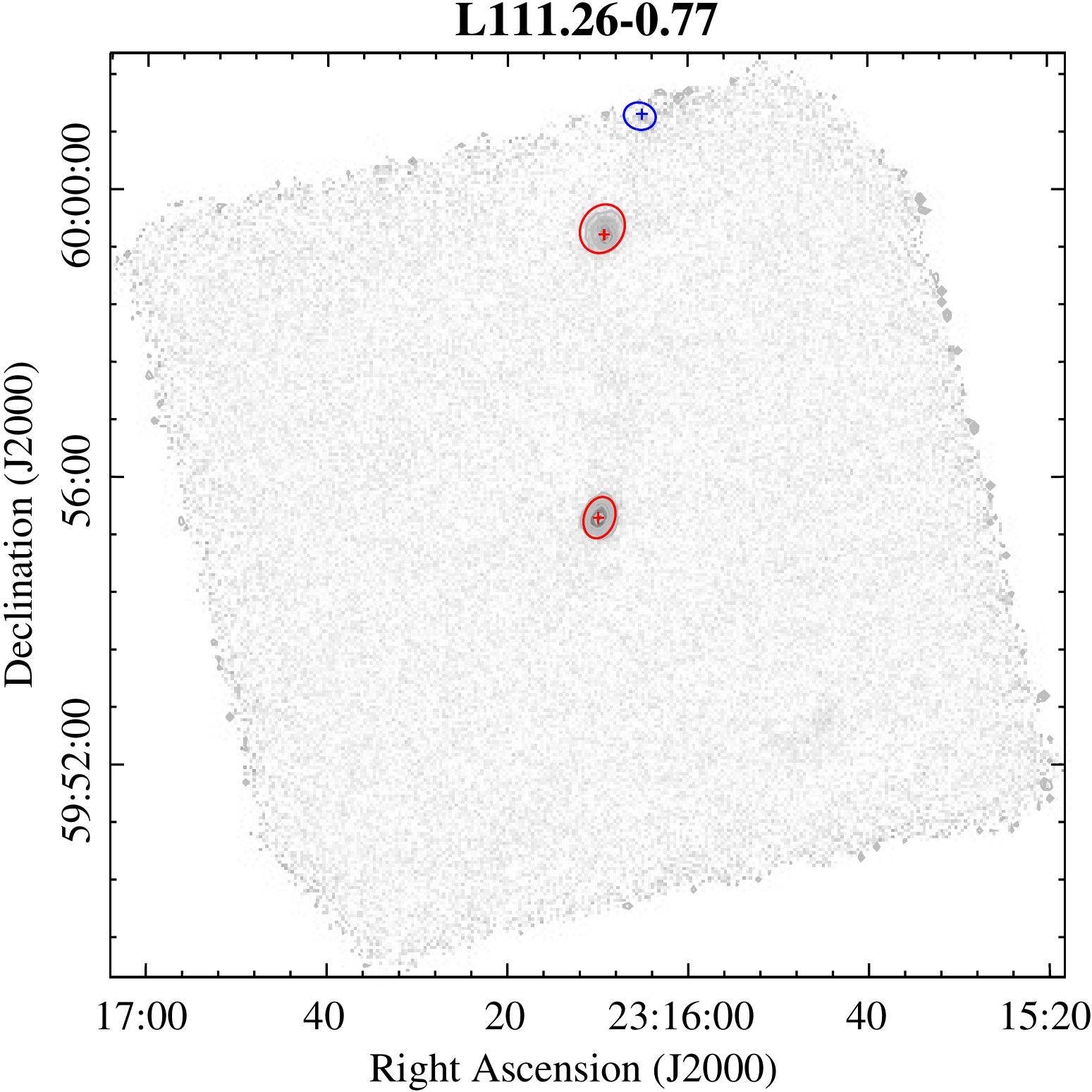}
}\\
\caption{Continuation}
\end{figure}

\clearpage
\begin{figure}\ContinuedFloat 
\center
\subfloat[L111.78+0.59 map, $\sigma_{rms}=412$ mJy beam$^{-1}$.]{
\includegraphics[scale=0.43]{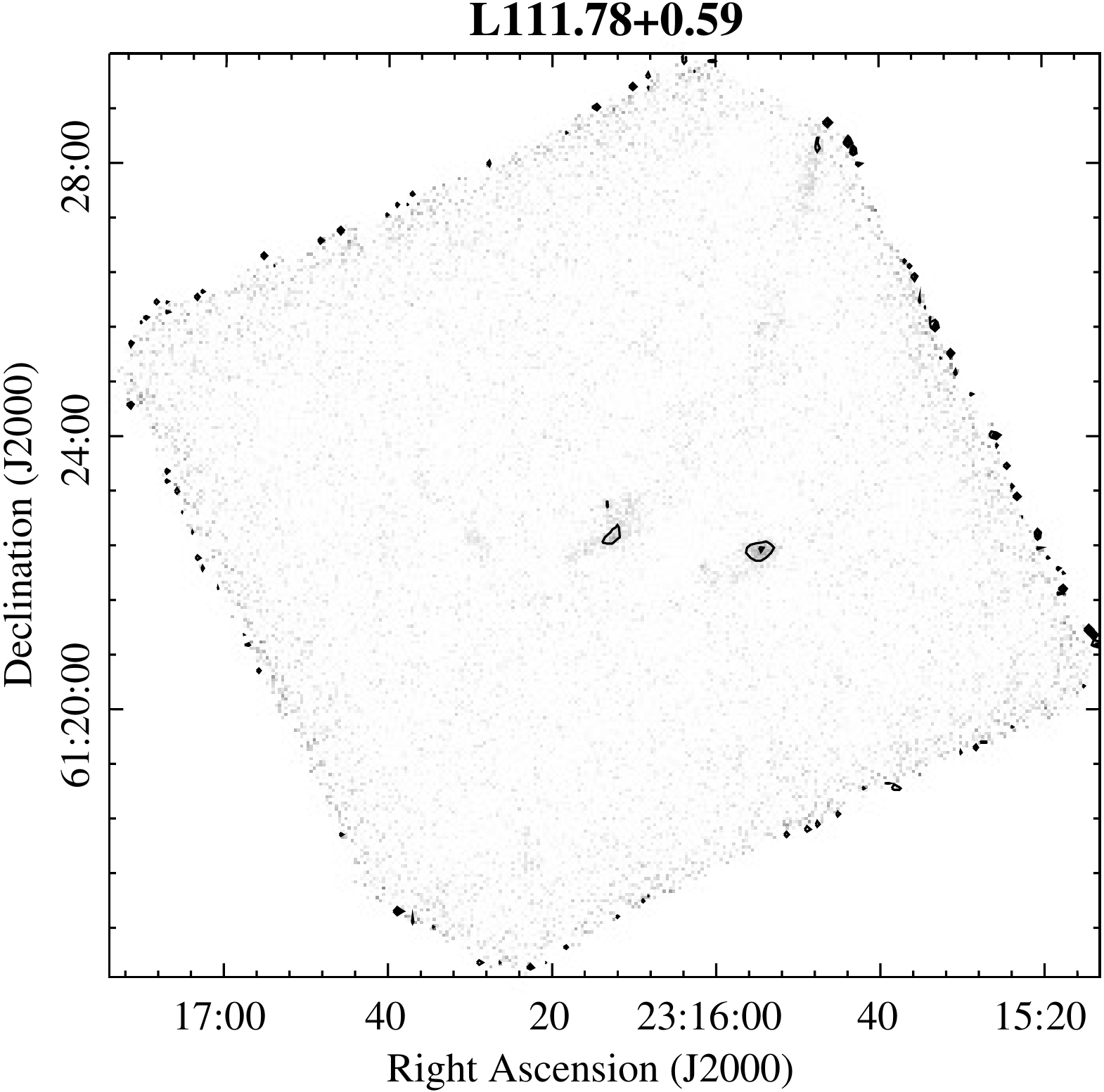}
\includegraphics[scale=0.43]{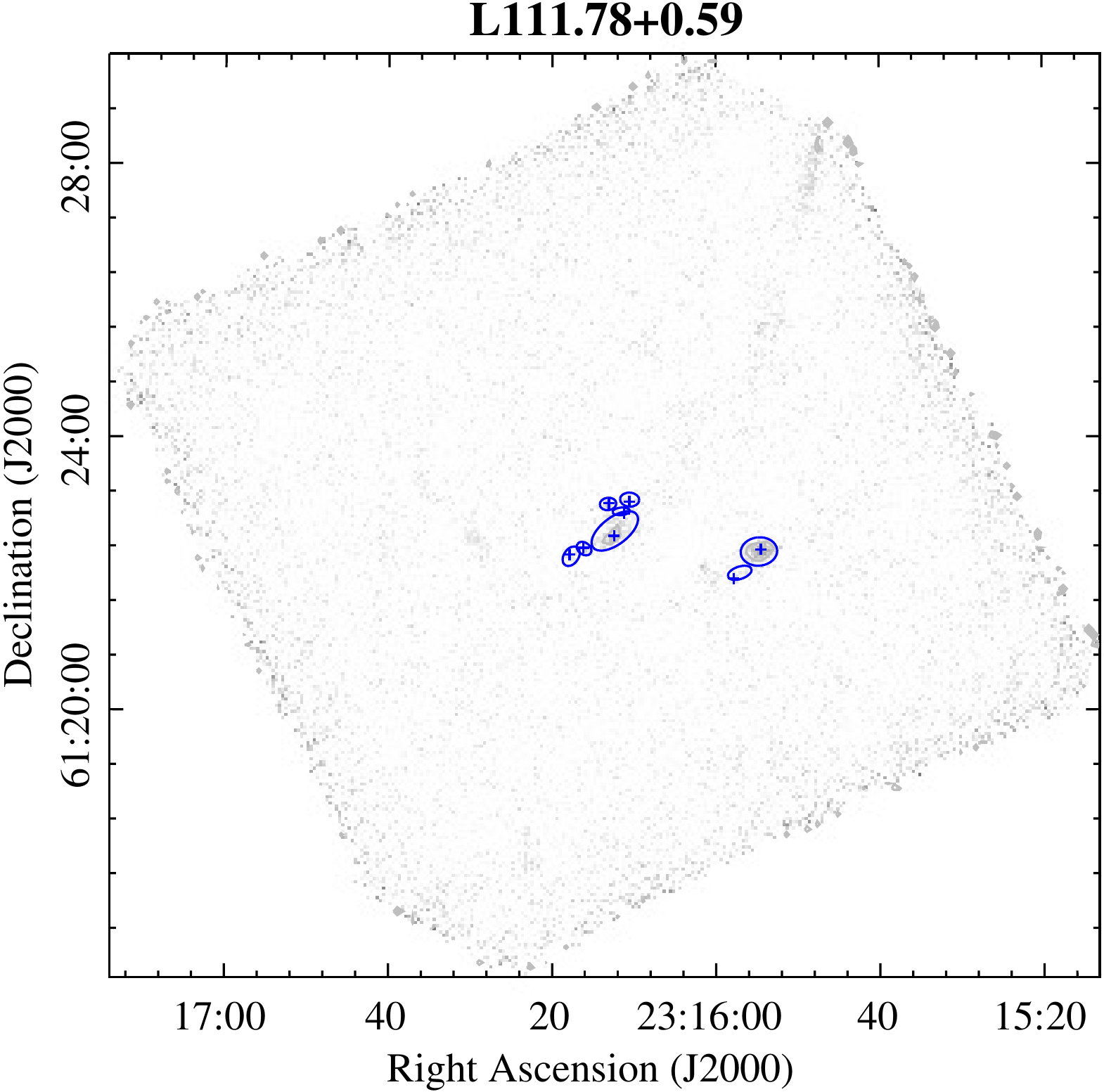}
}\\
\subfloat[L111.79+0.71 map, $\sigma_{rms}=307$ mJy beam$^{-1}$.]{
\includegraphics[scale=0.43]{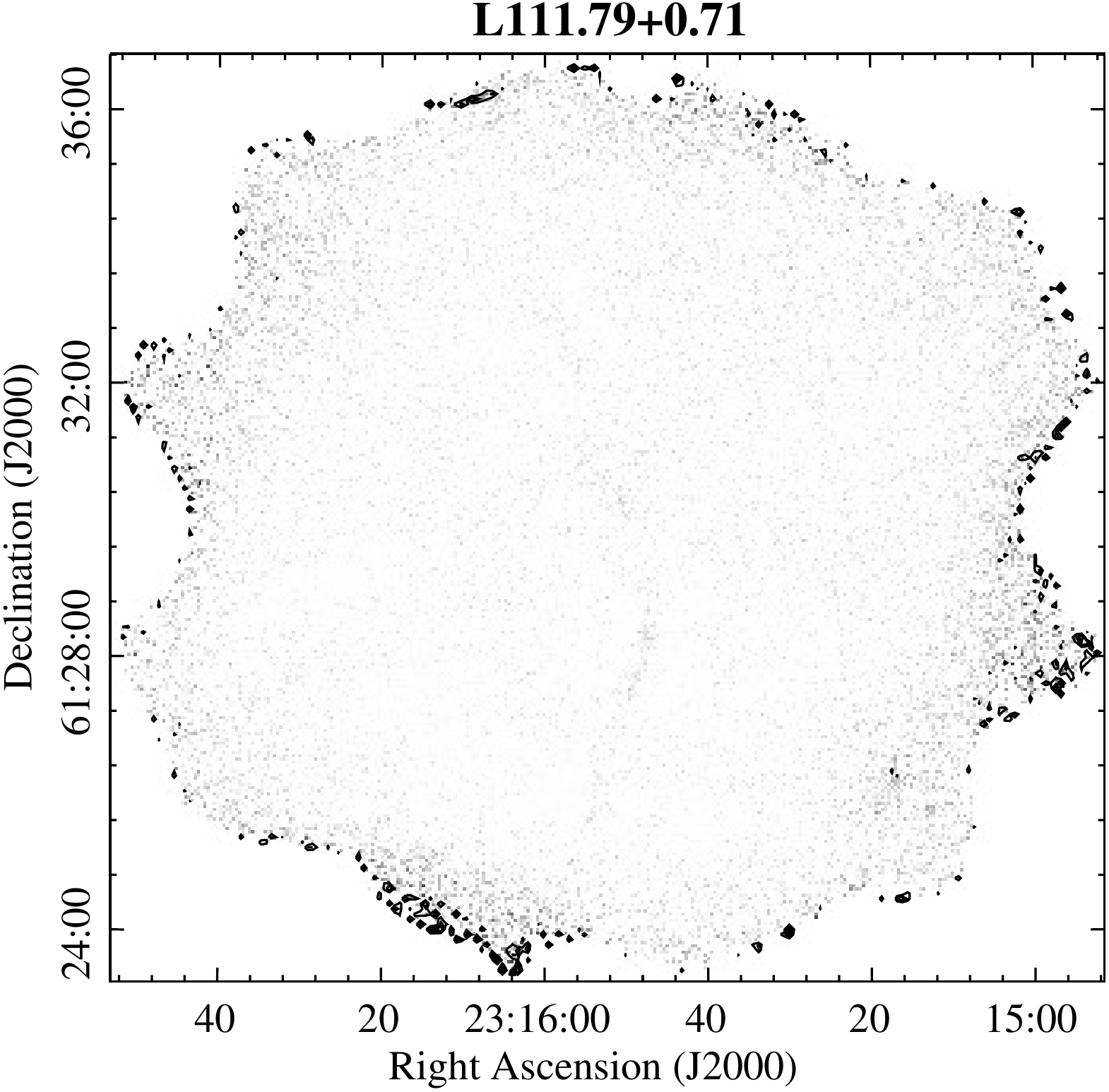}
\includegraphics[scale=0.43]{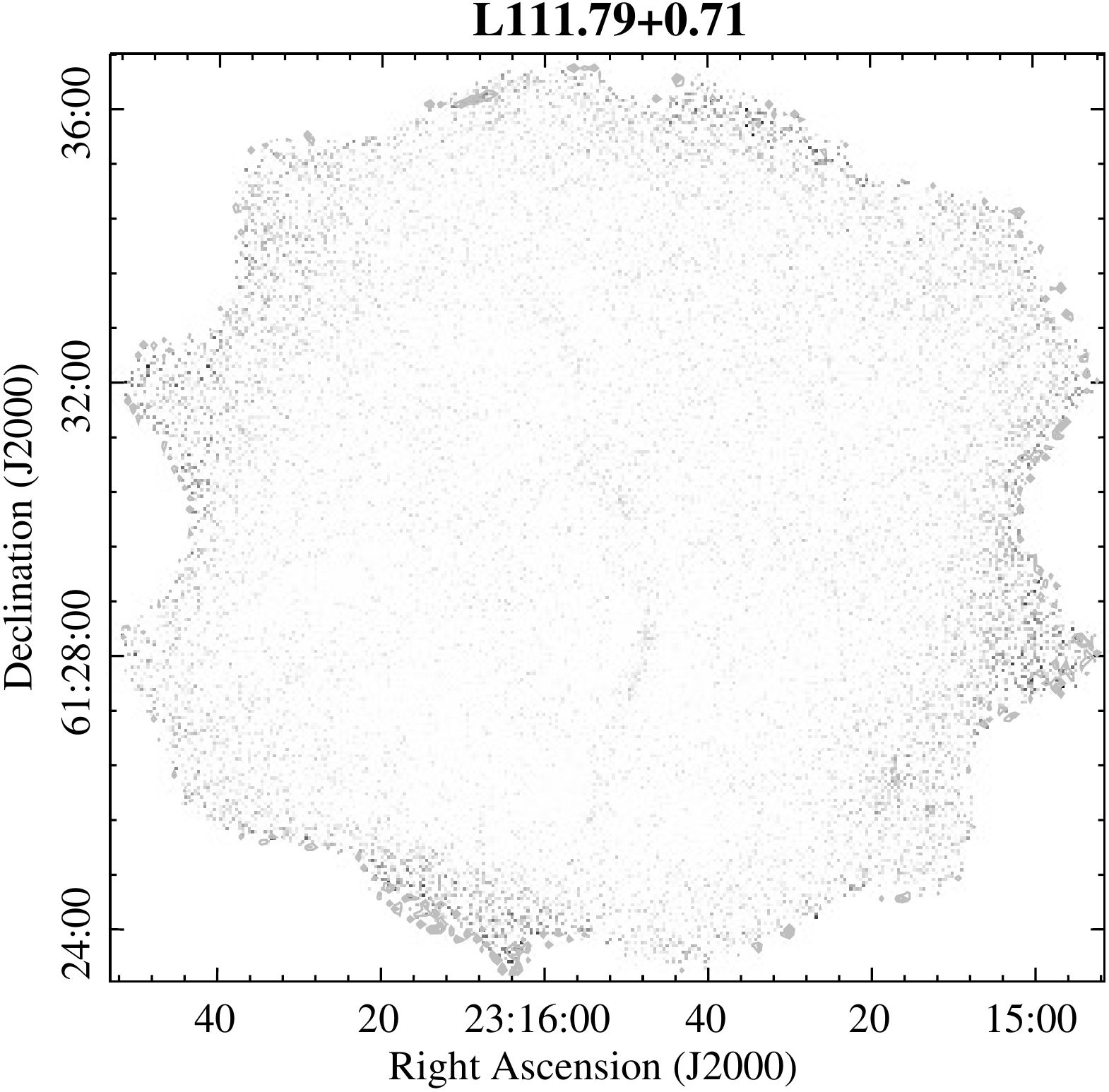}
}\\
\subfloat[L111.88+0.82 map, $\sigma_{rms}=473$ mJy beam$^{-1}$.]{
\includegraphics[scale=0.43]{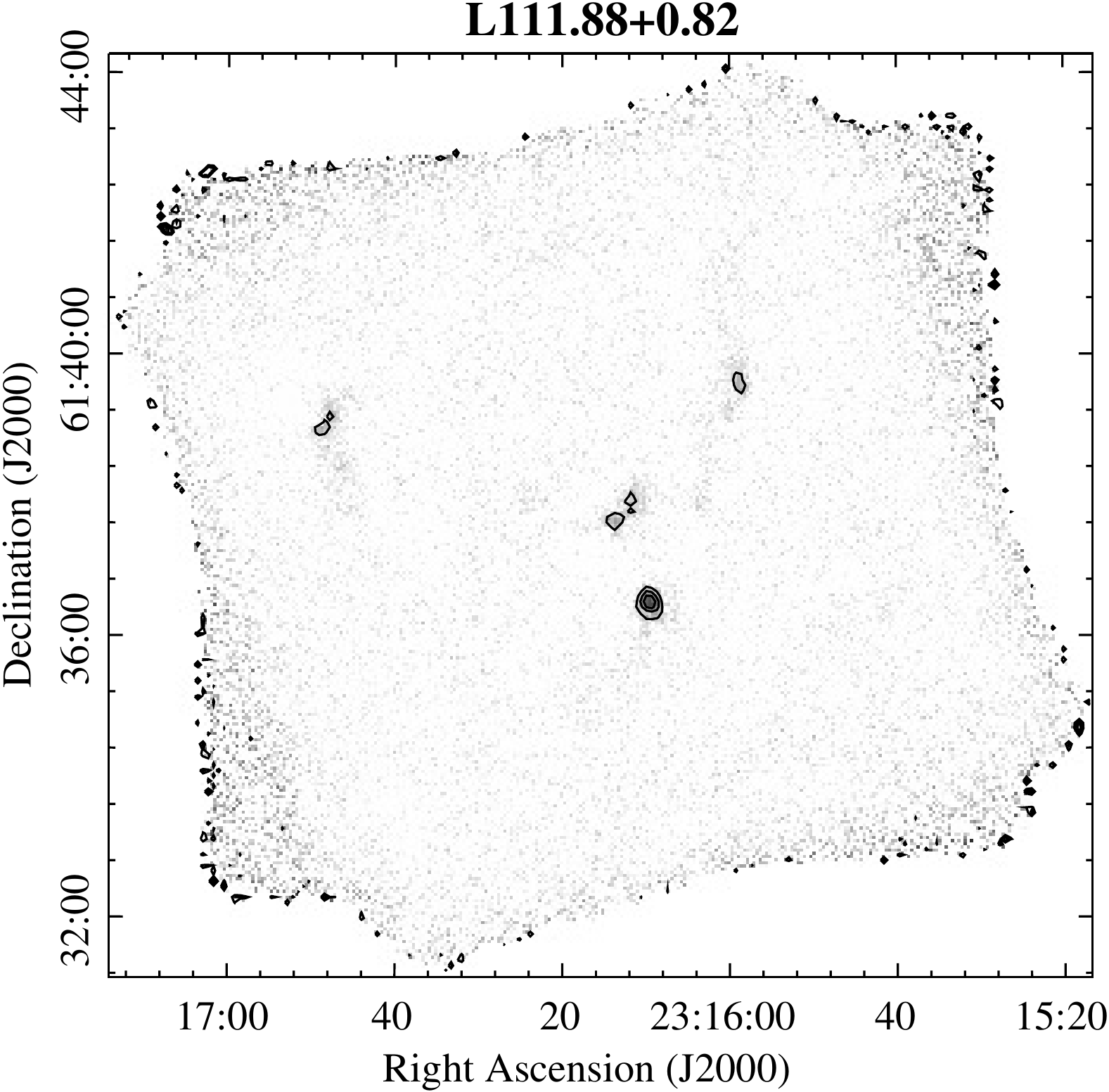}
\includegraphics[scale=0.43]{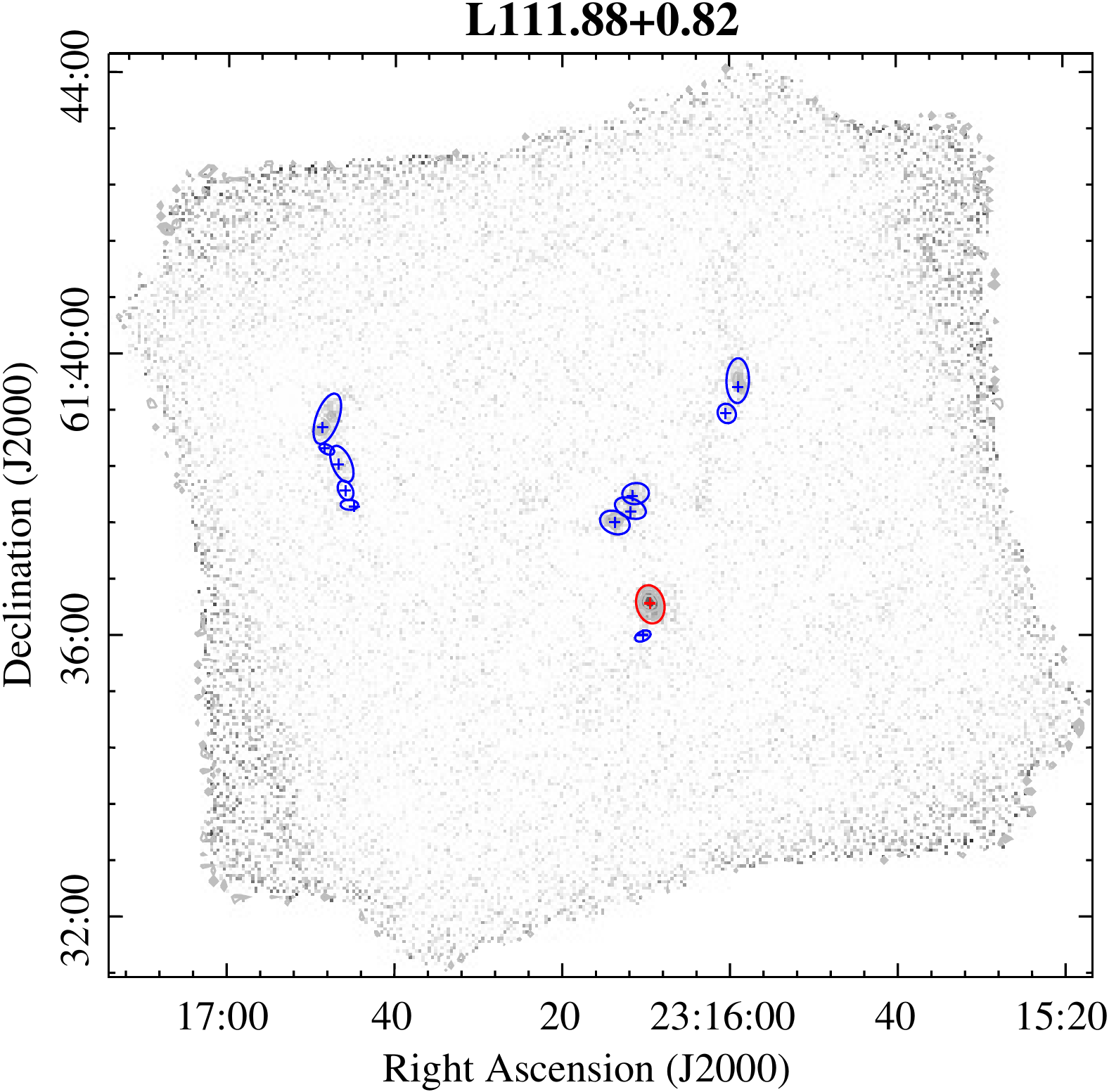}
}\\
\caption{Continuation}
\end{figure}

\clearpage
\begin{figure}\ContinuedFloat 
\center
\subfloat[L136.52+1.24 map, $\sigma_{rms}=260$ mJy beam$^{-1}$.]{
\includegraphics[scale=0.43]{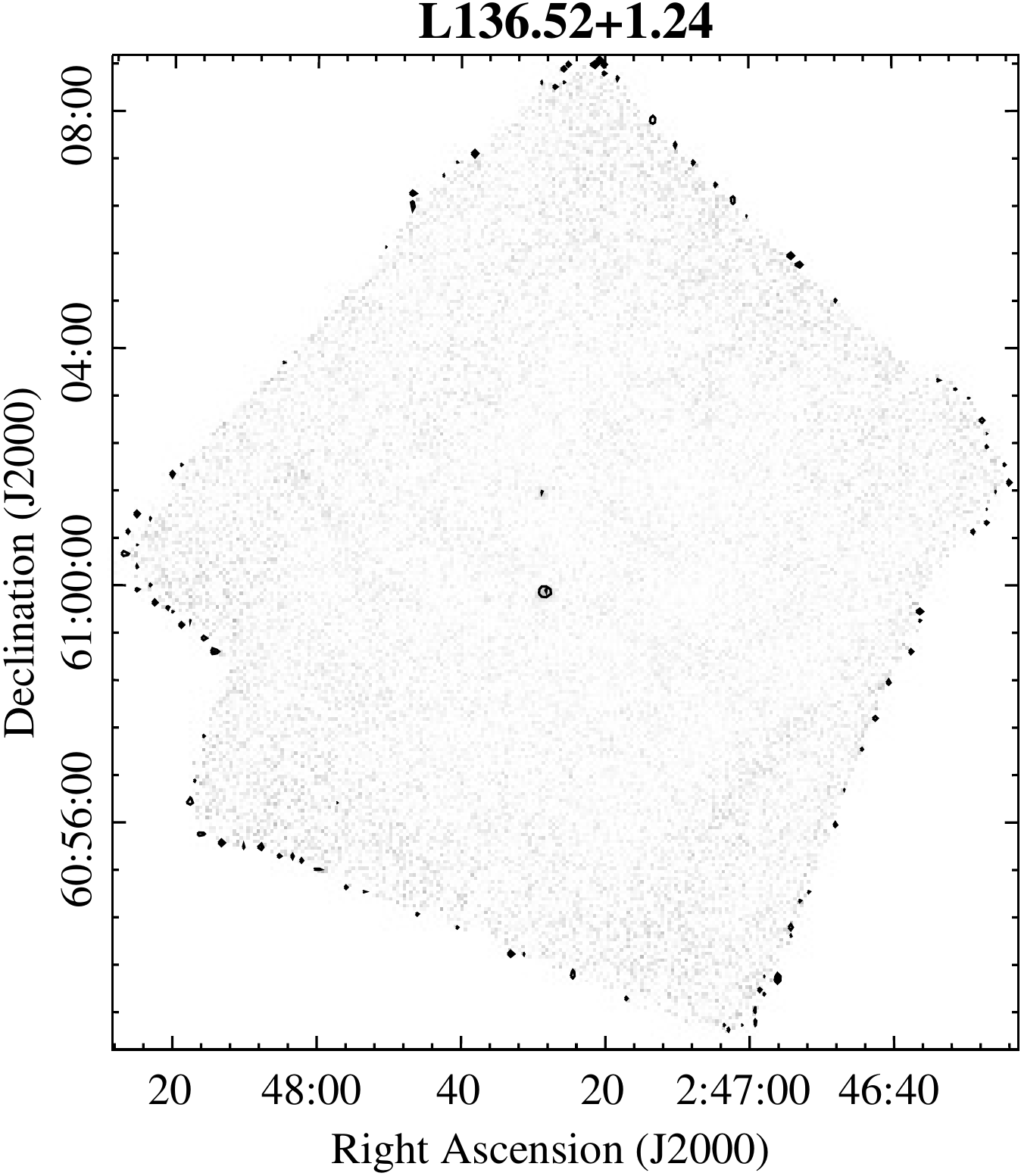}
\includegraphics[scale=0.43]{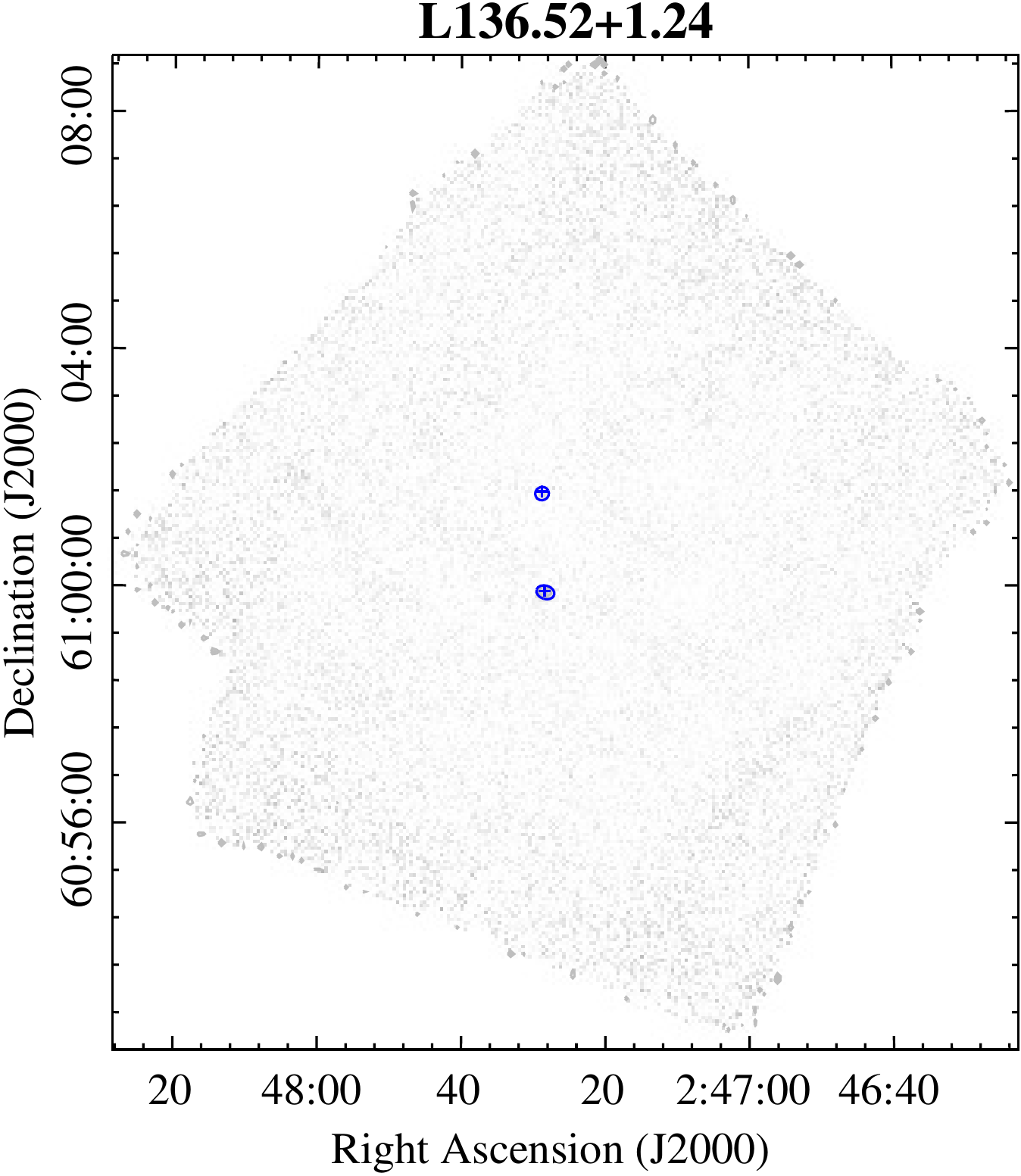}
}\\
\subfloat[L136.85+1.14 map, $\sigma_{rms}=337$ mJy beam$^{-1}$.]{
\includegraphics[scale=0.43]{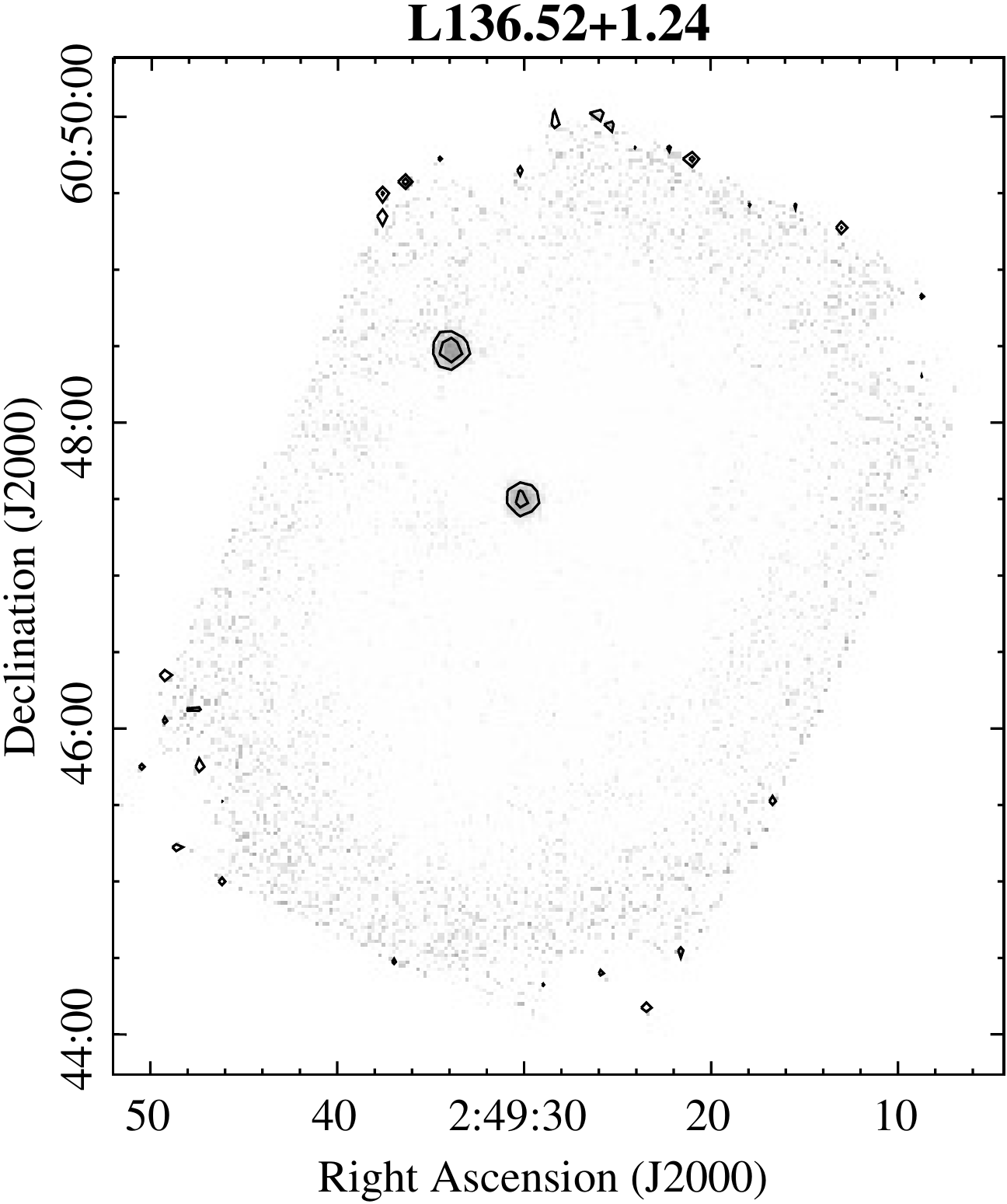}
\includegraphics[scale=0.43]{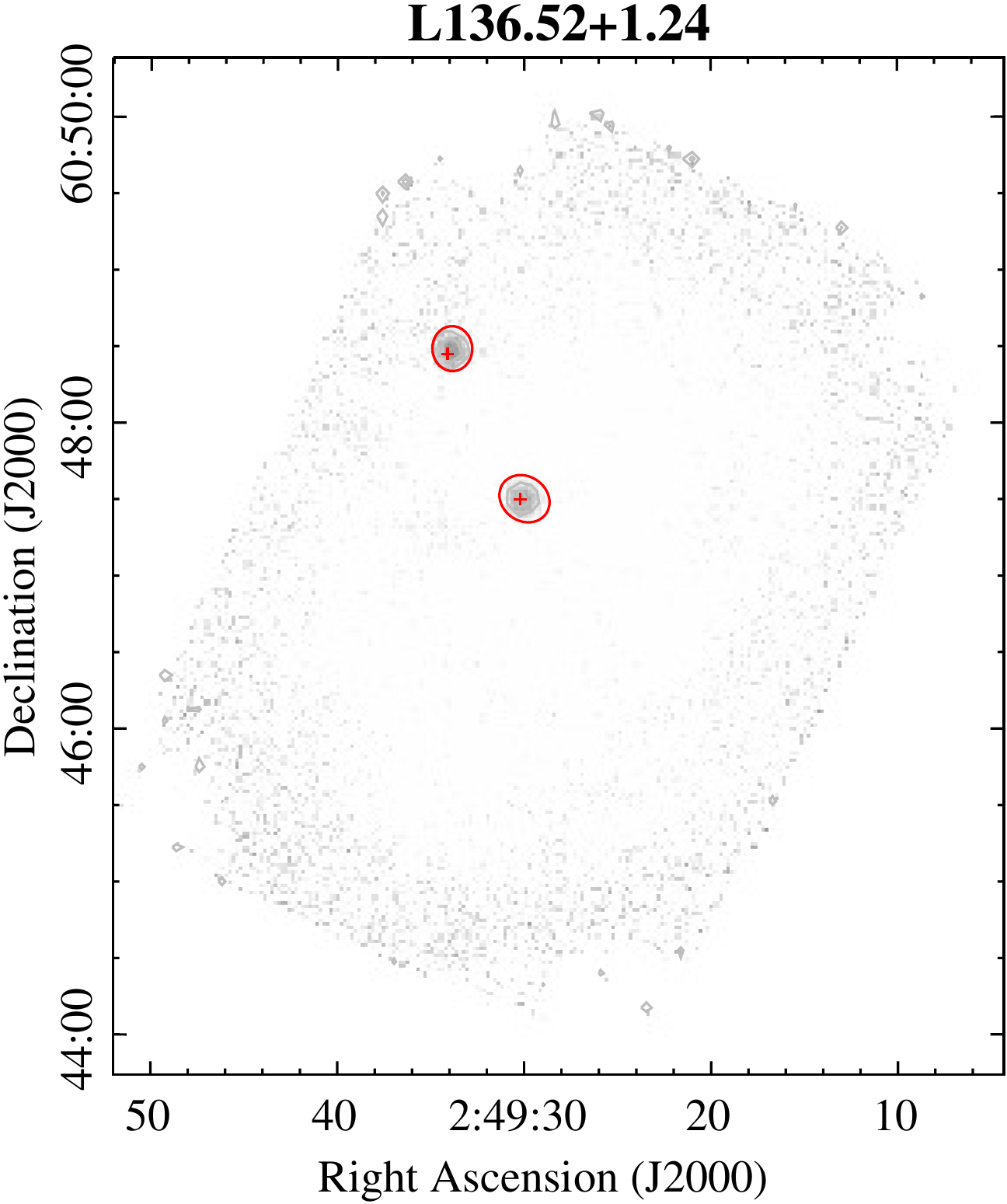}
}\\
\subfloat[L136.95+1.09 map, $\sigma_{rms}=475$ mJy beam$^{-1}$.]{
\includegraphics[scale=0.43]{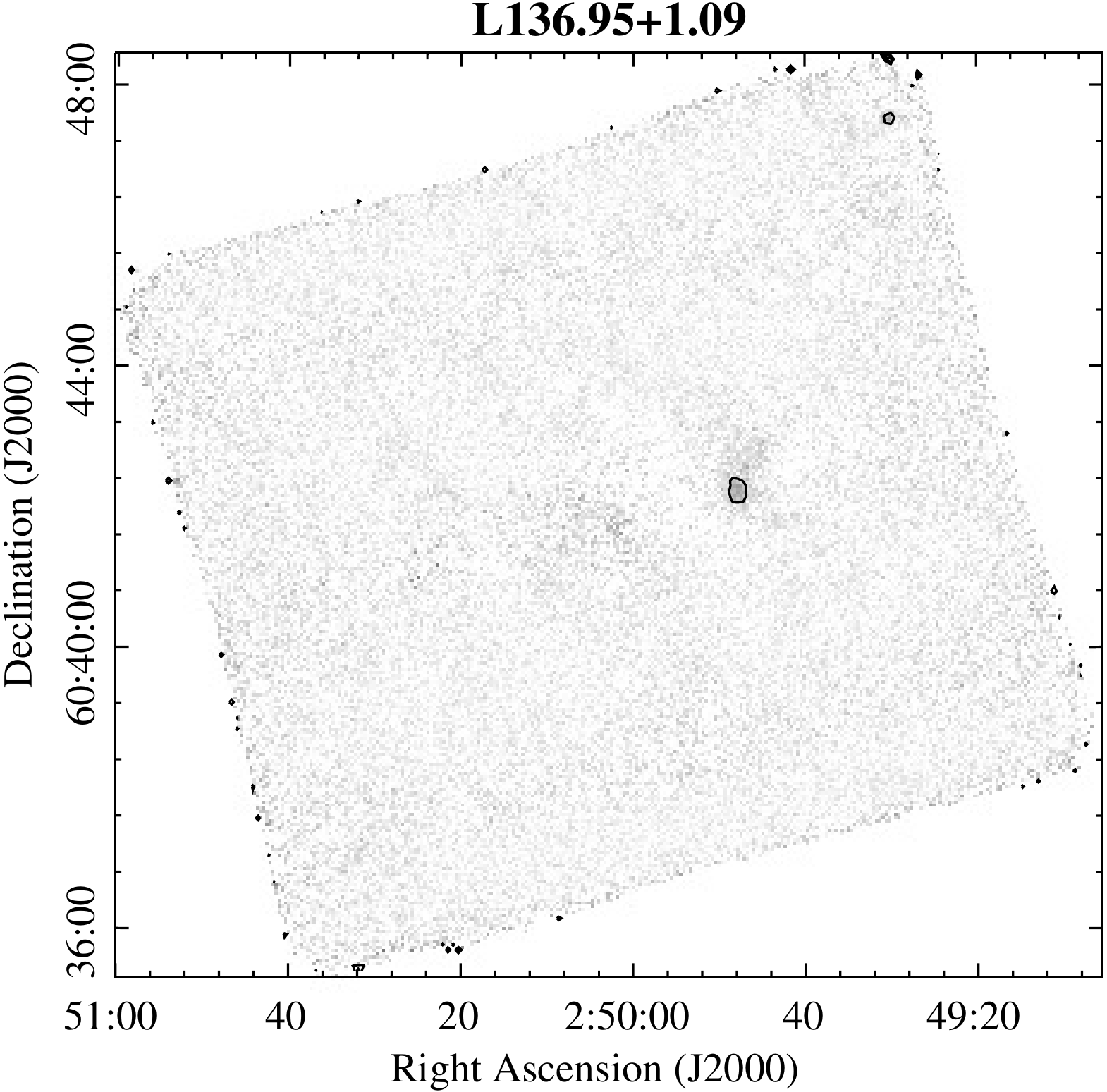}
\includegraphics[scale=0.43]{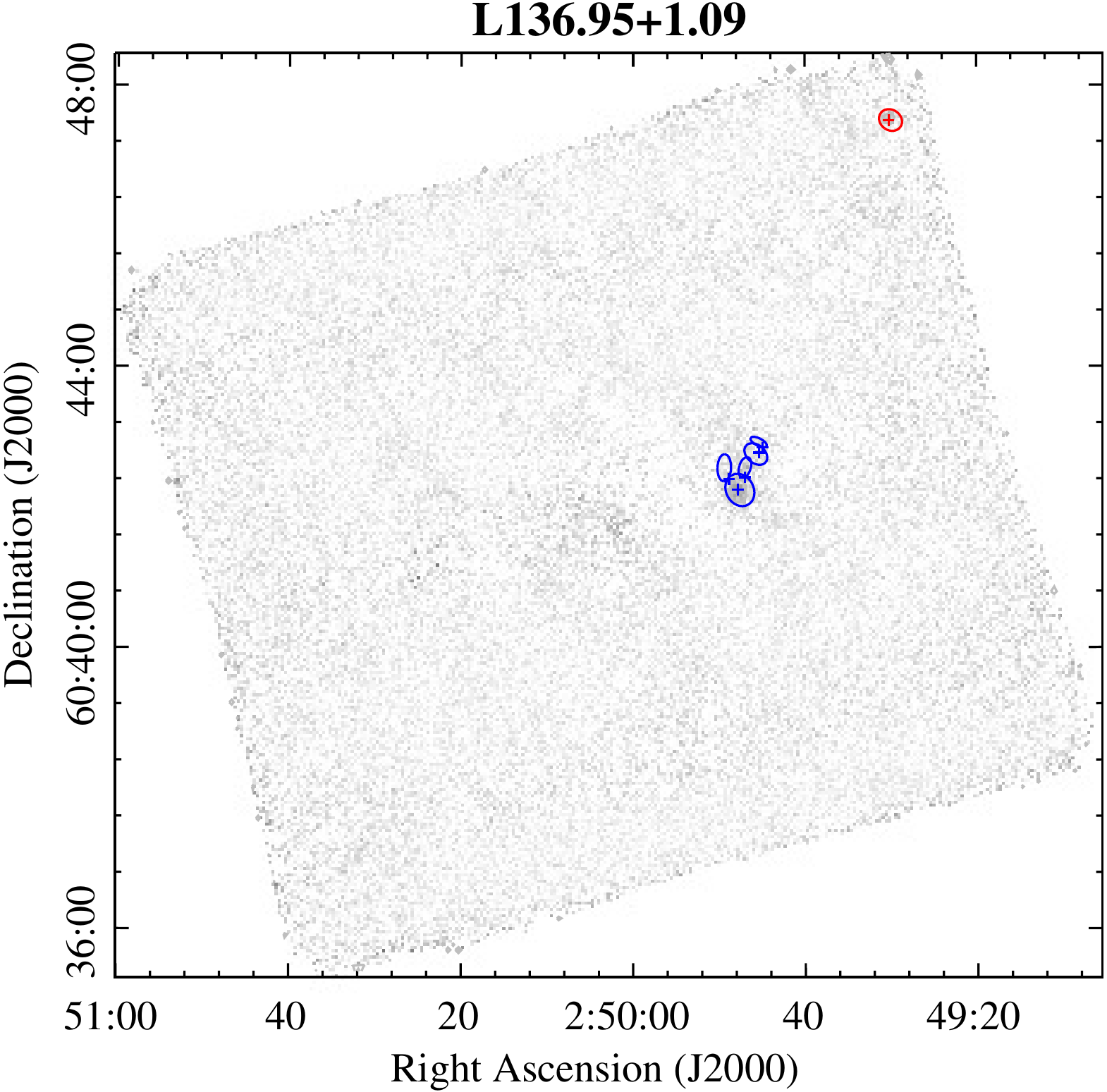}
}\\
\caption{Continuation}
\end{figure}

\clearpage
\begin{figure}\ContinuedFloat 
\center
\subfloat[L137.69+1.46 map, $\sigma_{rms}=667$ mJy beam$^{-1}$.]{
\includegraphics[scale=0.43]{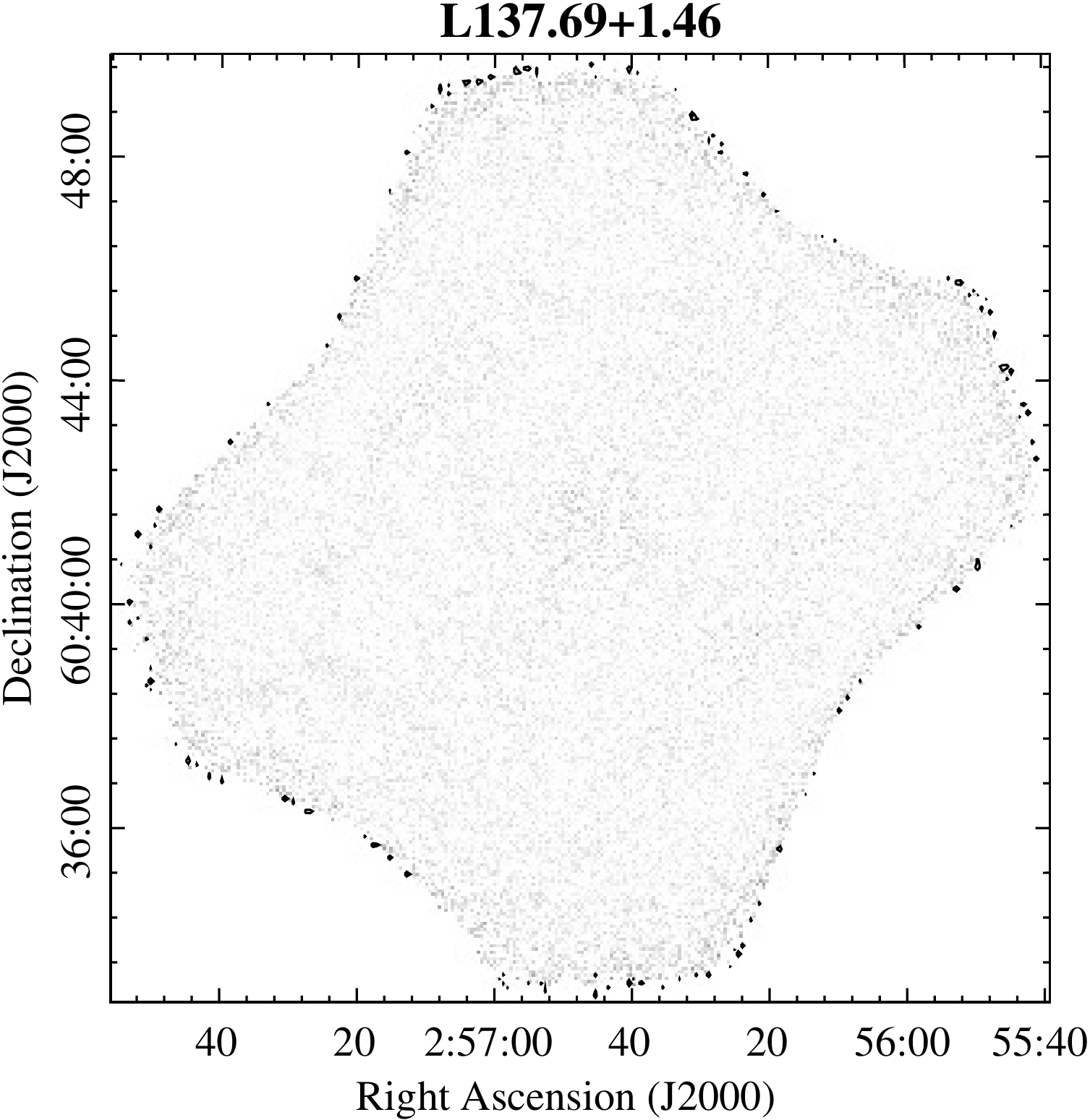}
\includegraphics[scale=0.43]{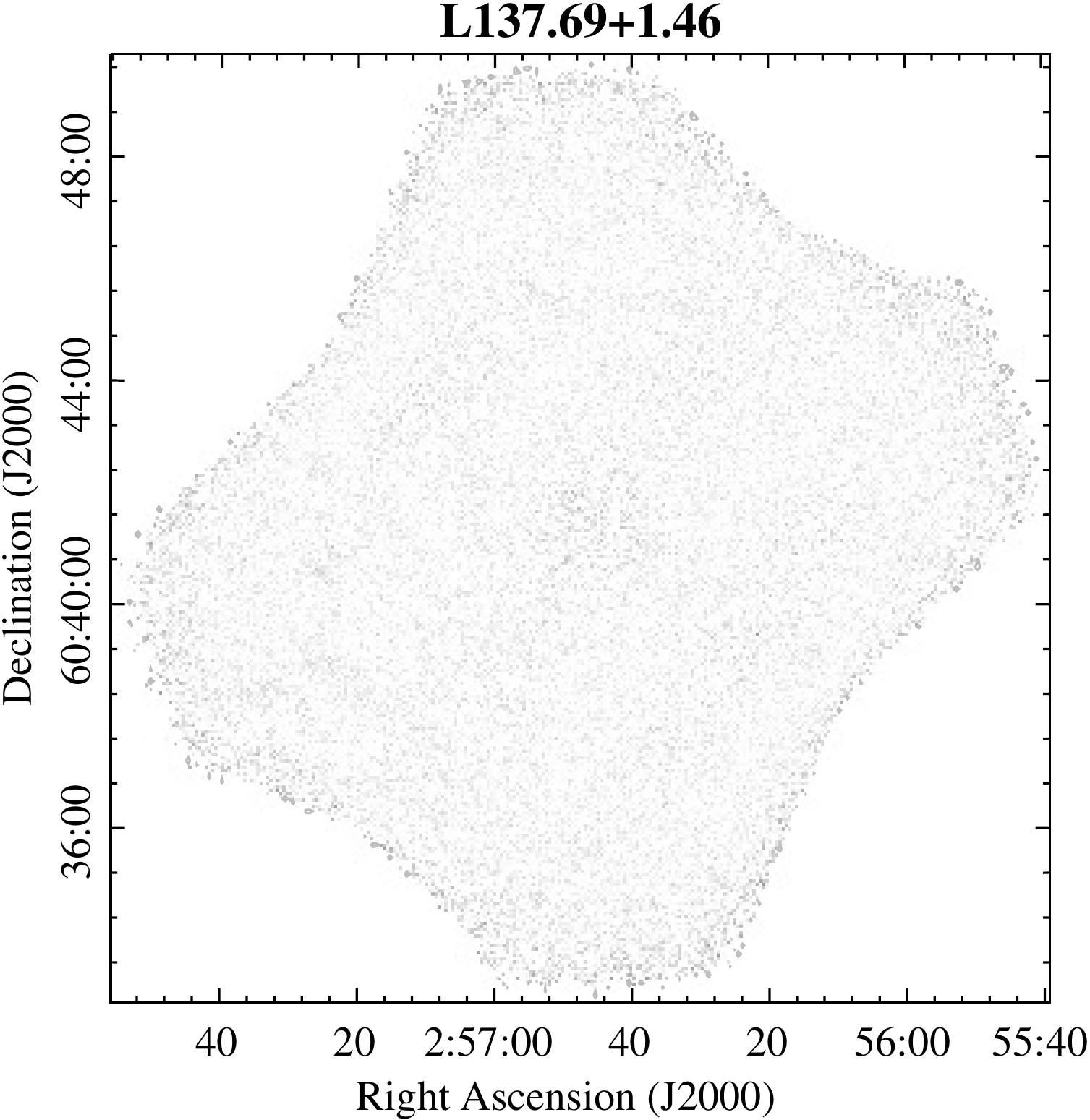}
}\\
\subfloat[L138.30+1.56 map, $\sigma_{rms}=205$ mJy beam$^{-1}$.]{
\includegraphics[scale=0.43]{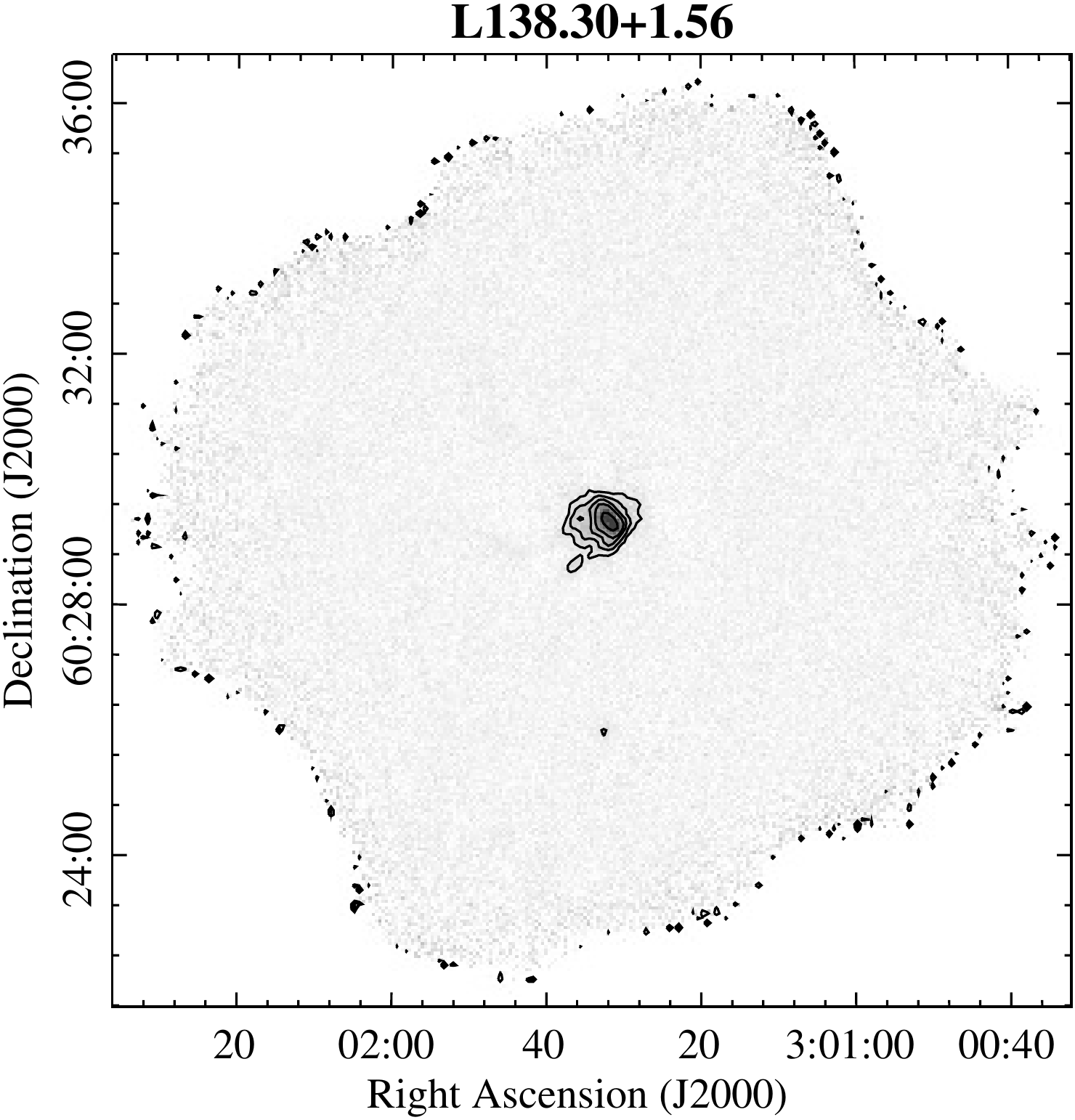}
\includegraphics[scale=0.43]{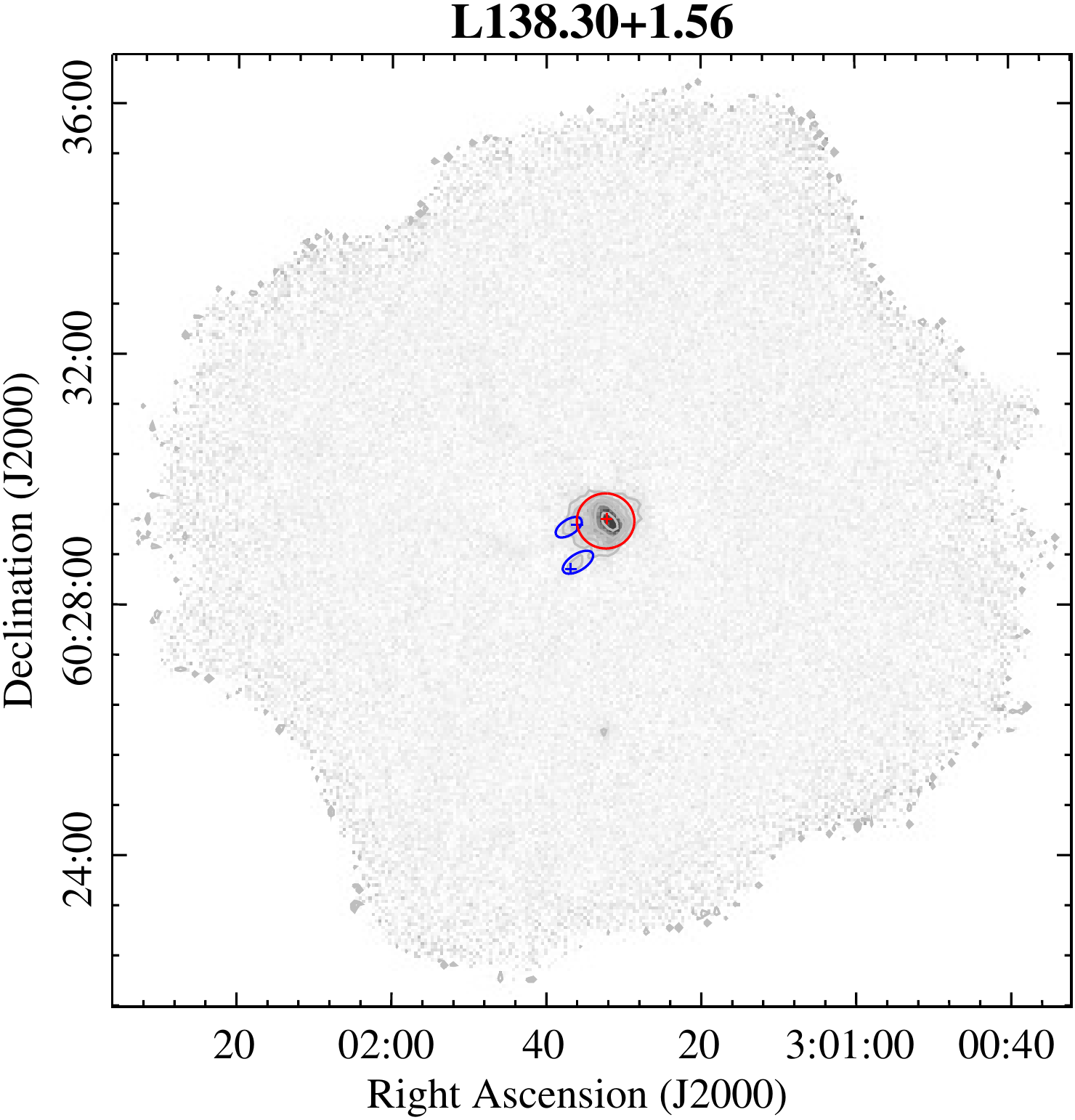}
}\\
\subfloat[L138.48+1.63 map, $\sigma_{rms}=266$ mJy beam$^{-1}$.]{
\includegraphics[scale=0.43]{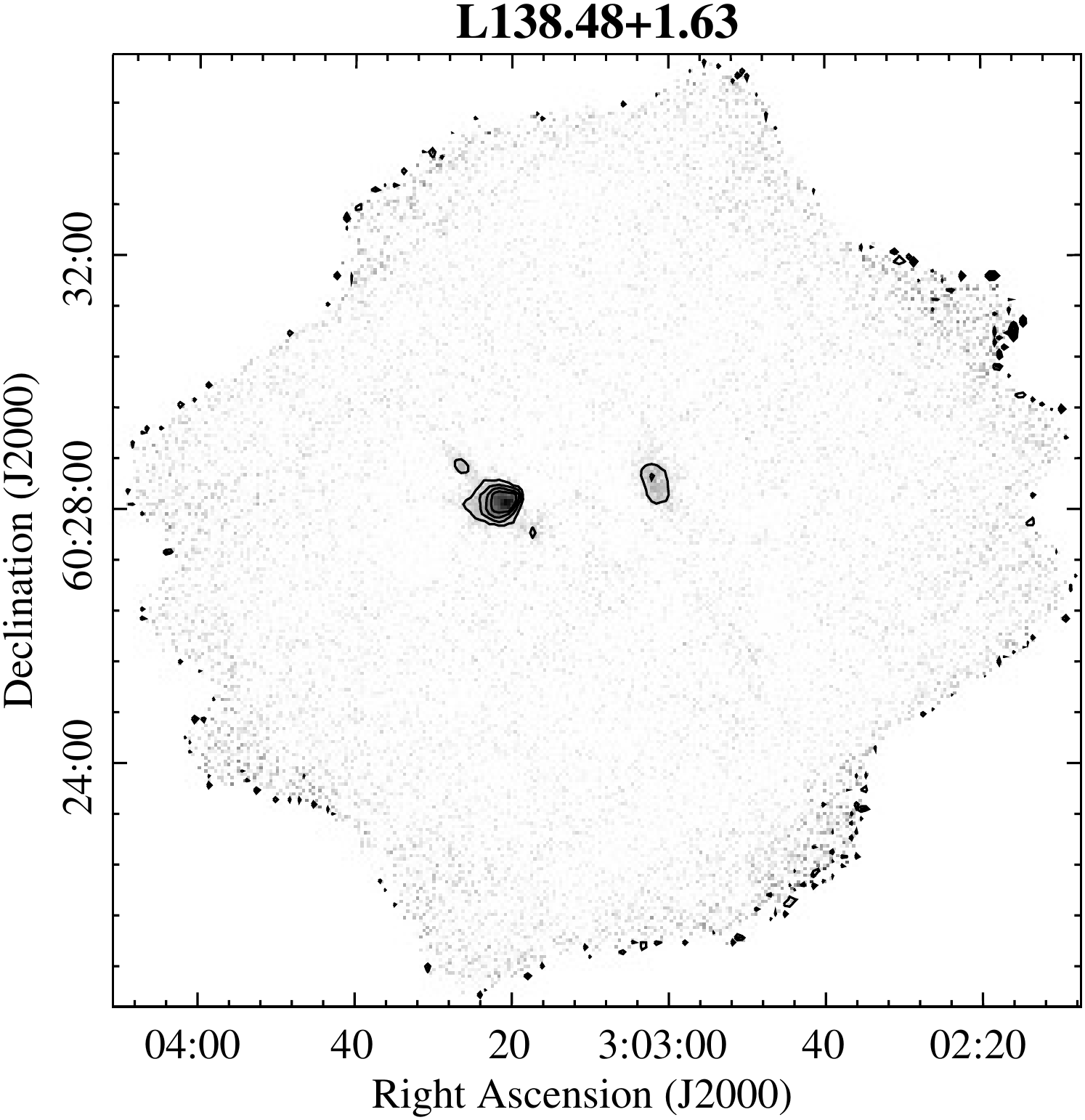}
\includegraphics[scale=0.43]{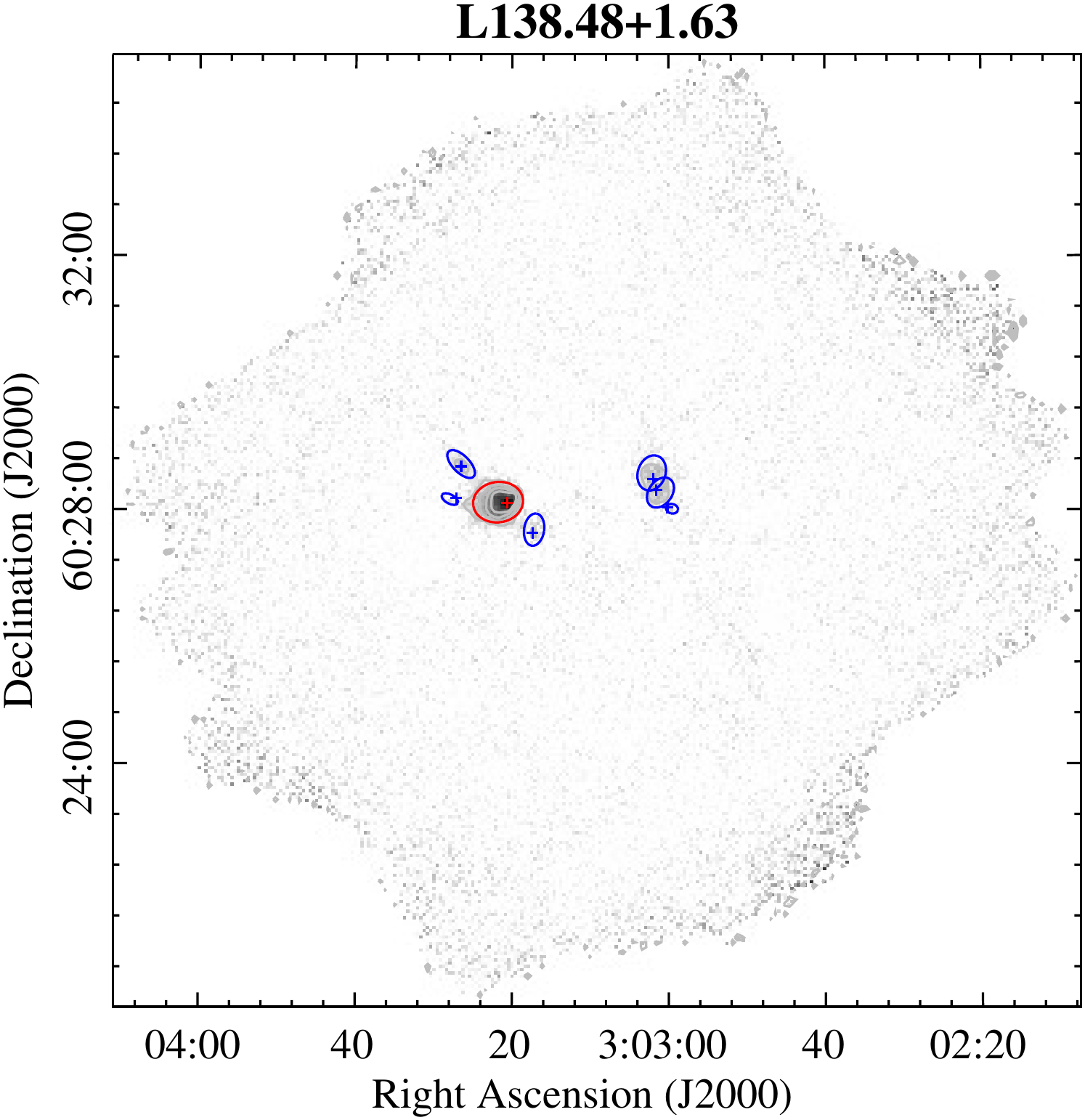}
}\\
\caption{Continuation}
\end{figure}

\clearpage
\begin{figure}\ContinuedFloat 
\center
\subfloat[L173.14+2.36 map, $\sigma_{rms}=777$ mJy beam$^{-1}$.]{
\includegraphics[scale=0.43]{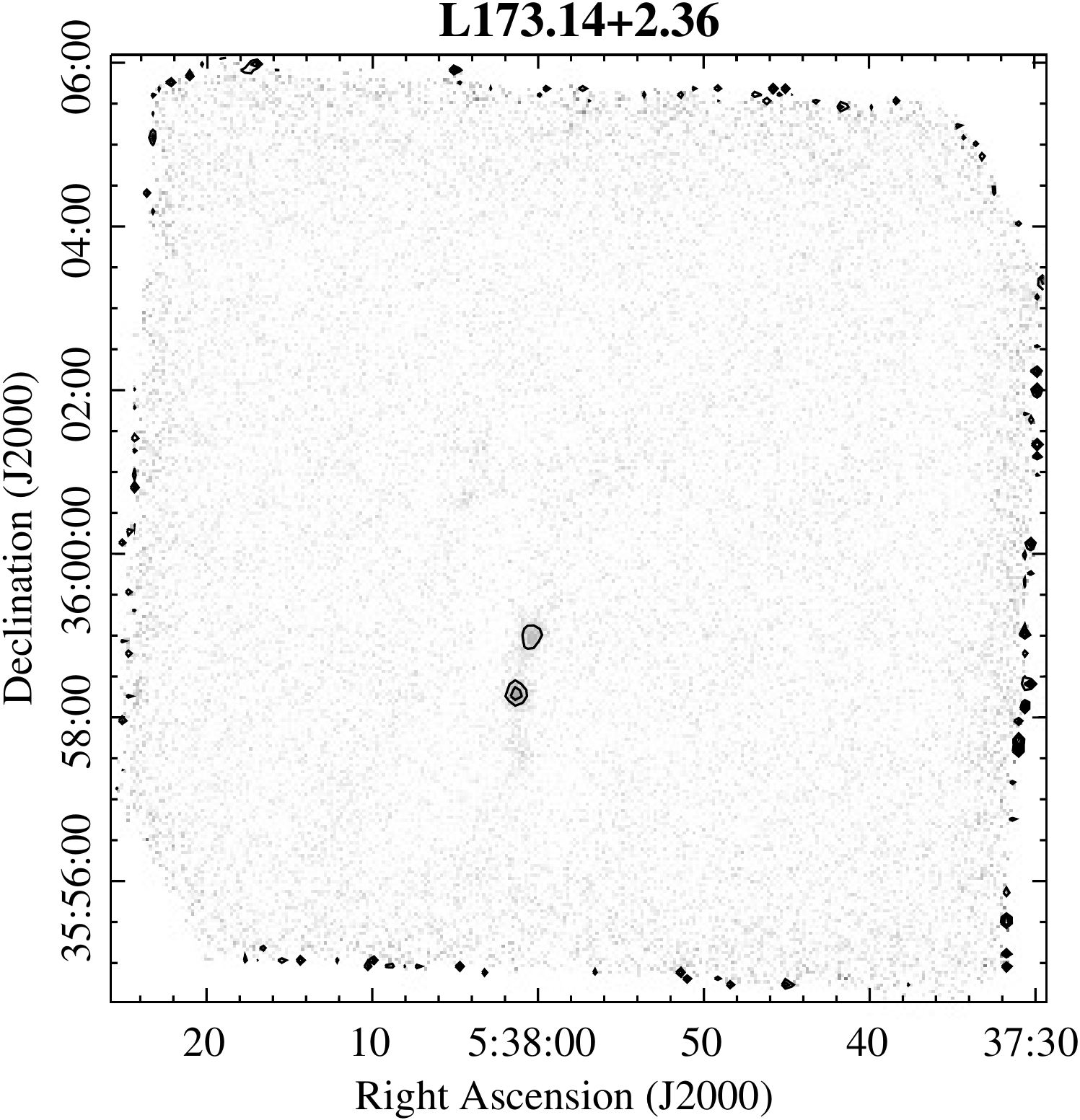}
\includegraphics[scale=0.43]{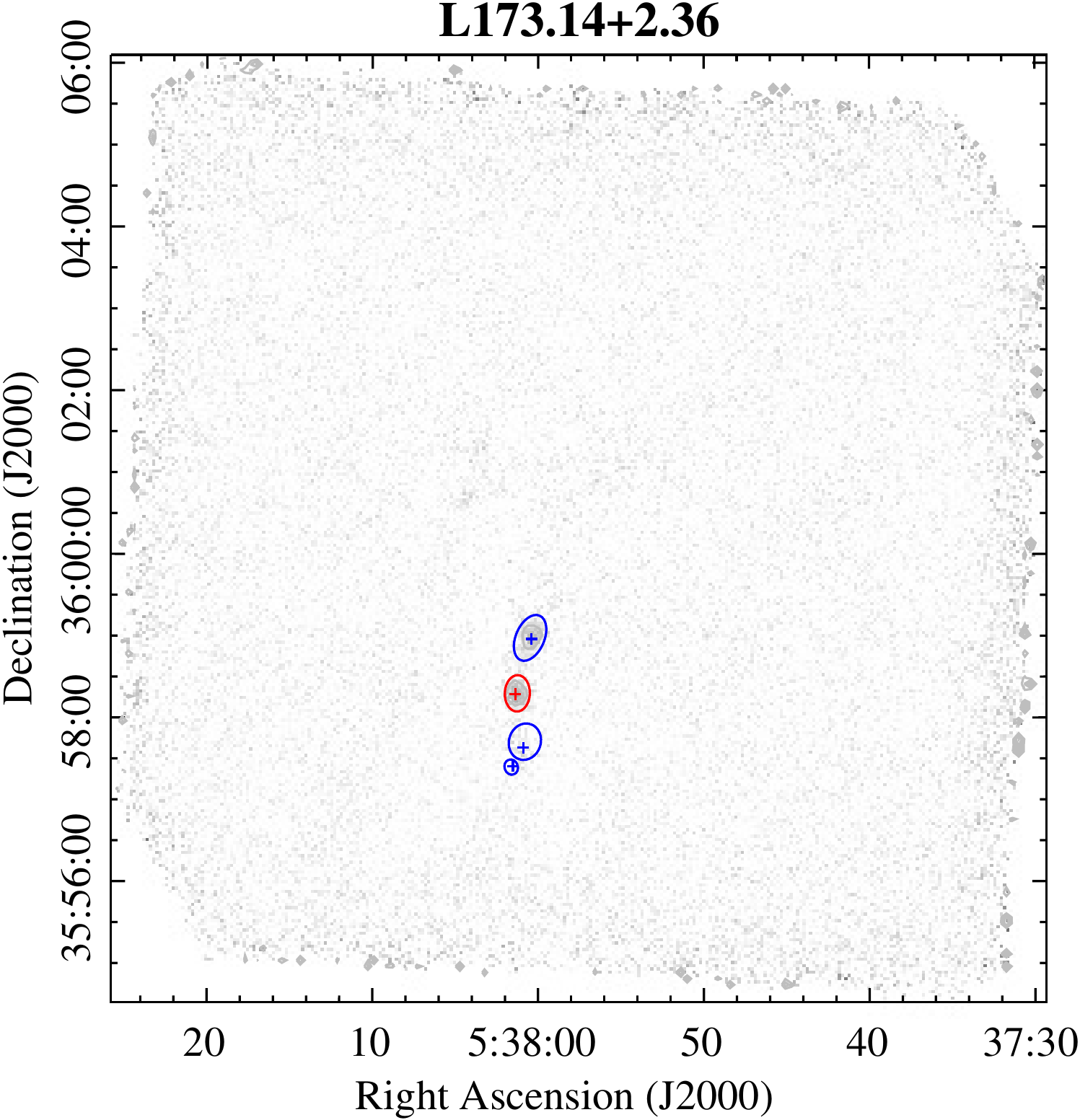}
}\\
\subfloat[L173.17+2.35 map, $\sigma_{rms}=320$ mJy beam$^{-1}$.]{
\includegraphics[scale=0.43]{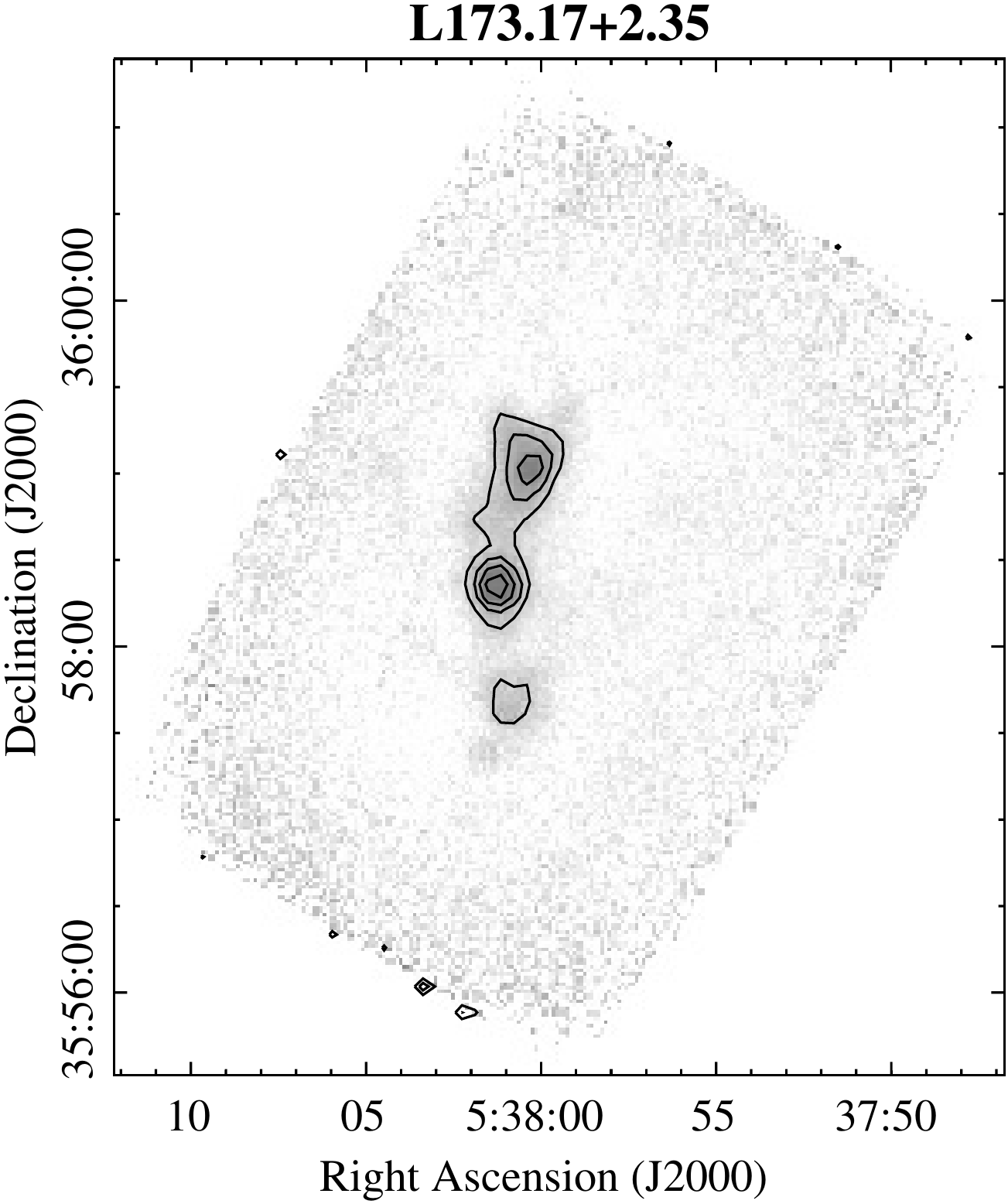}
\includegraphics[scale=0.43]{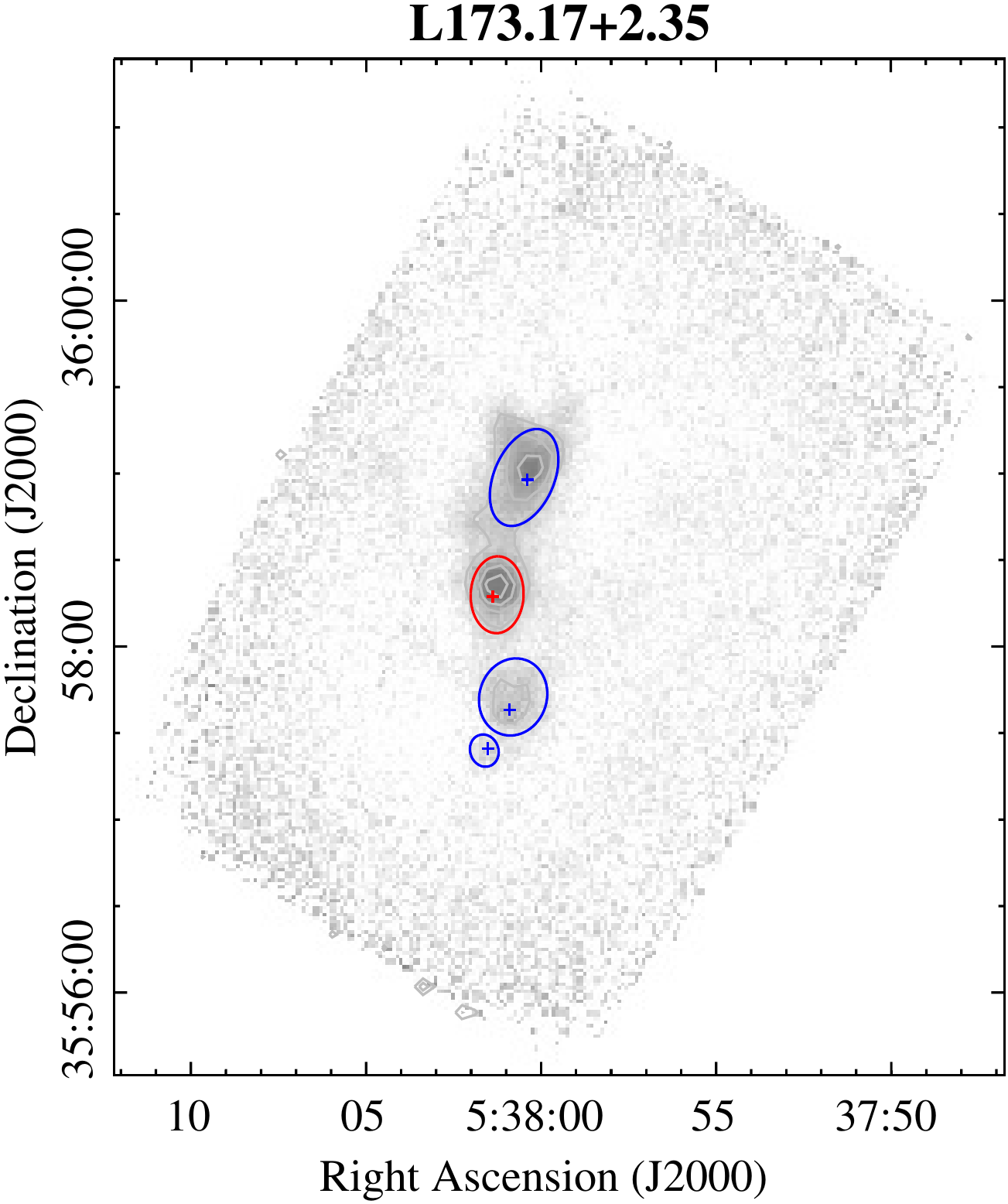}
}\\
\subfloat[L173.47+2.43 map, $\sigma_{rms}=487$ mJy beam$^{-1}$.]{
\includegraphics[scale=0.43]{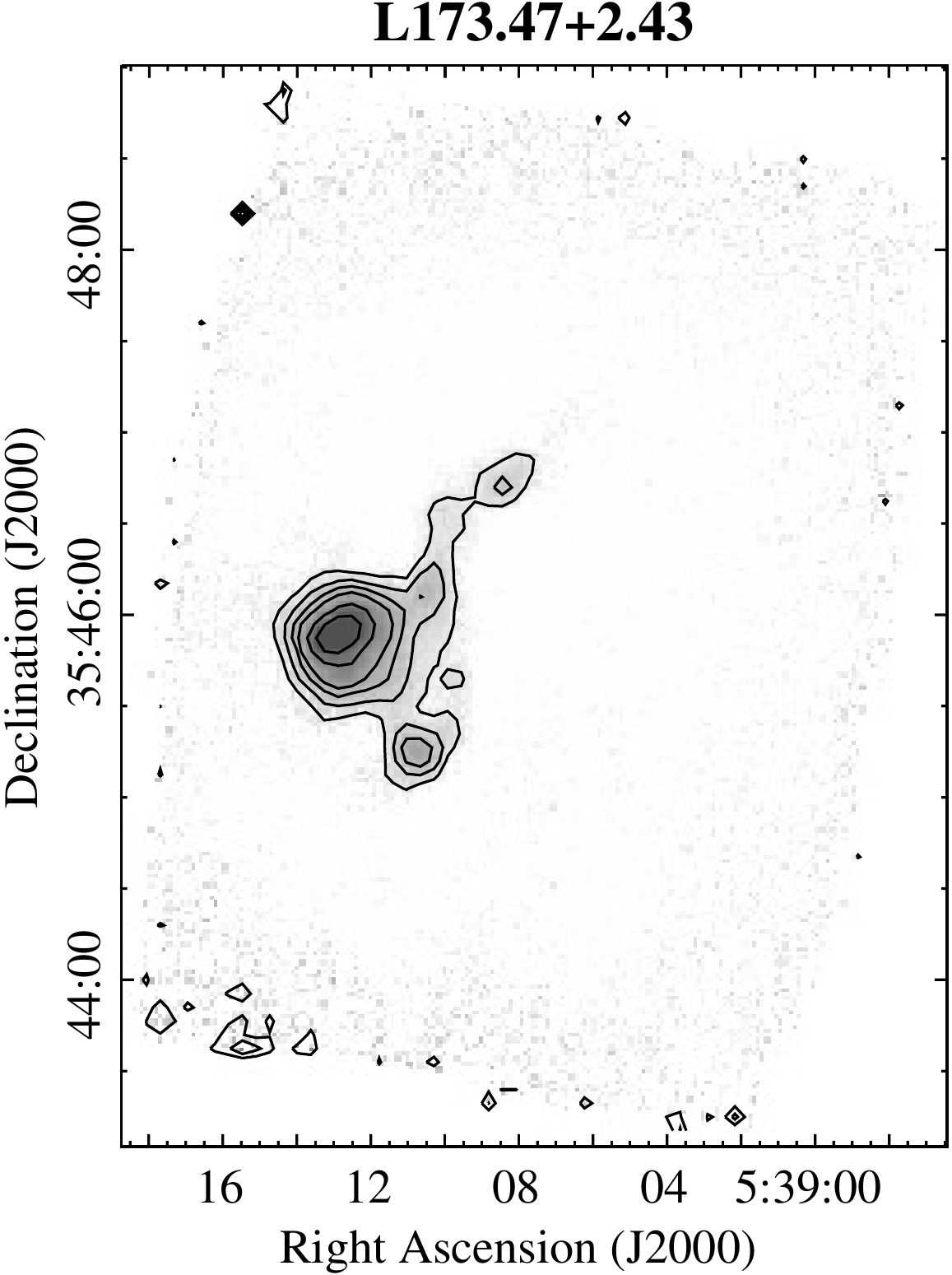}
\includegraphics[scale=0.43]{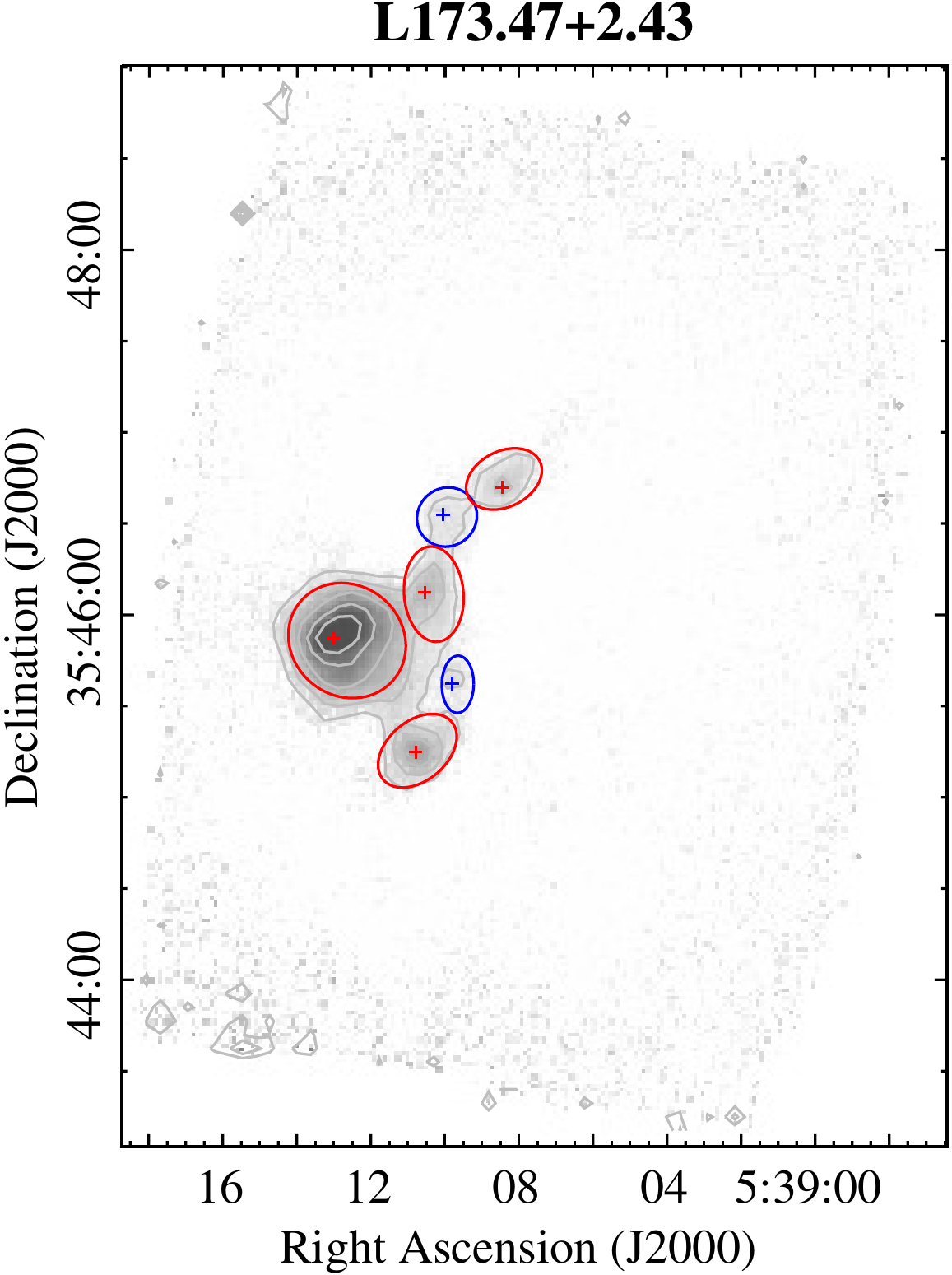}
}\\
\caption{Continuation}
\end{figure}

\clearpage
\begin{figure}\ContinuedFloat 
\center
\subfloat[L173.57+2.44 map, $\sigma_{rms}=421$ mJy beam$^{-1}$.]{
\includegraphics[scale=0.43]{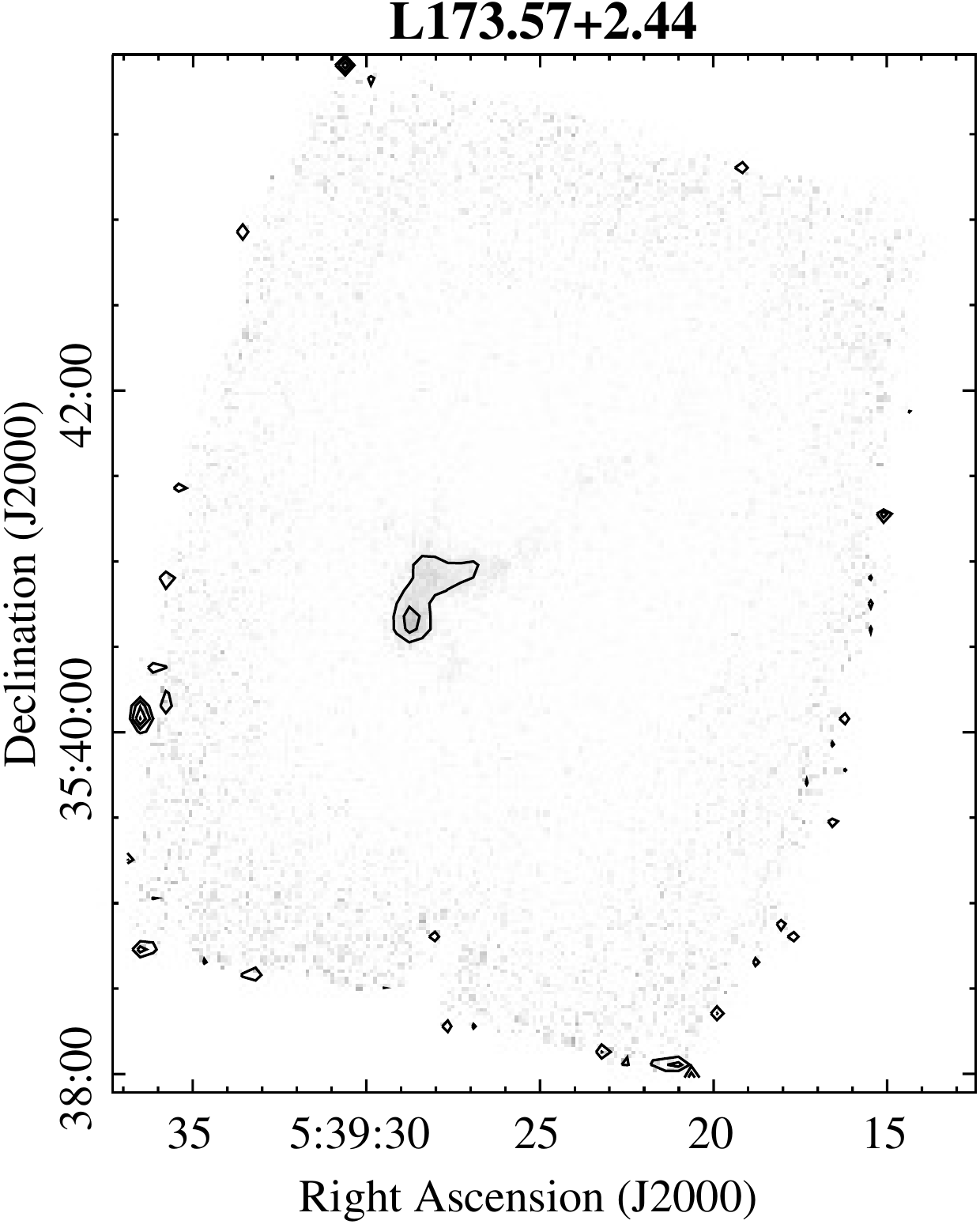}
\includegraphics[scale=0.43]{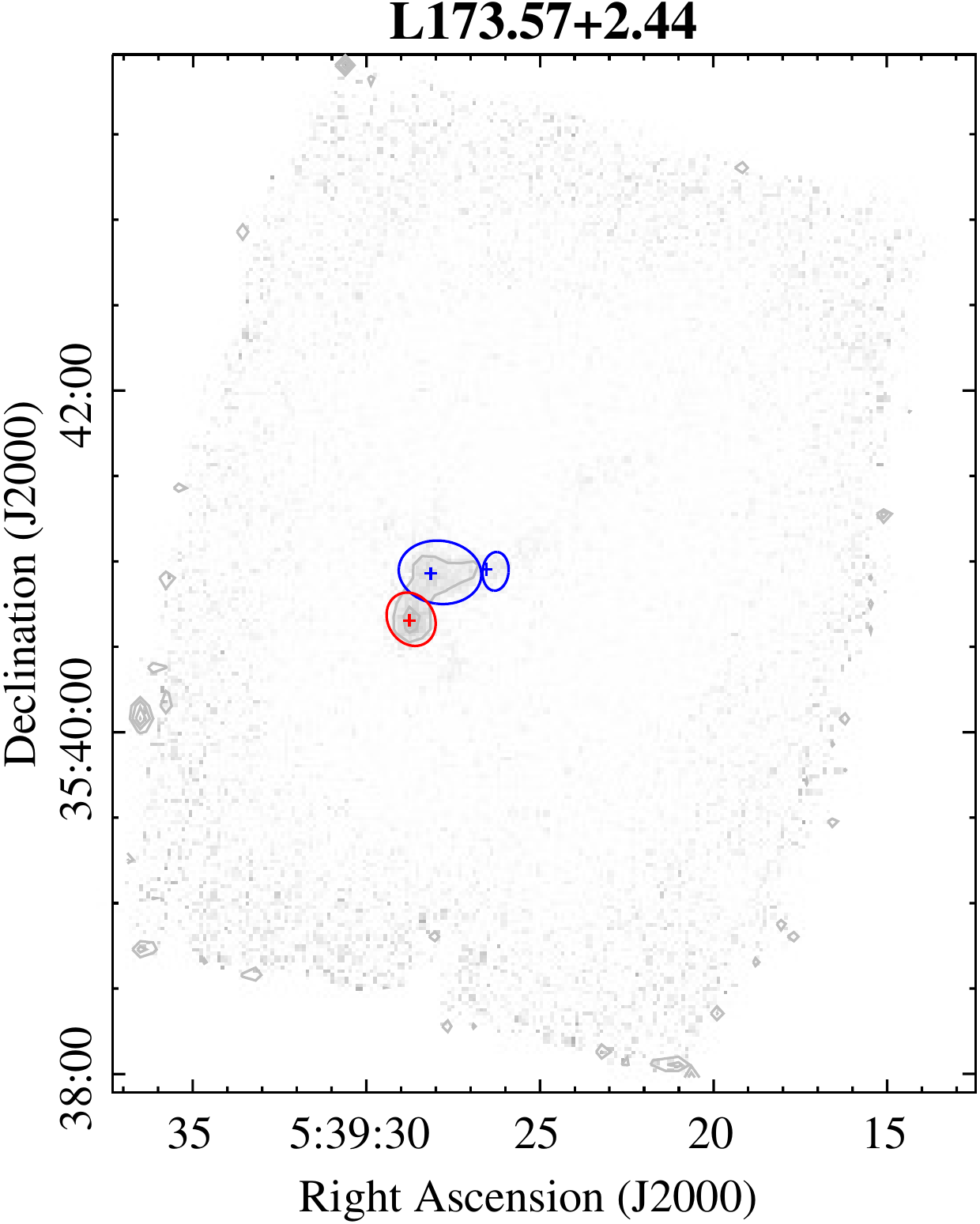}
}\\
\subfloat[L173.62+2.81 map, $\sigma_{rms}=359$ mJy beam$^{-1}$.]{
\includegraphics[scale=0.43]{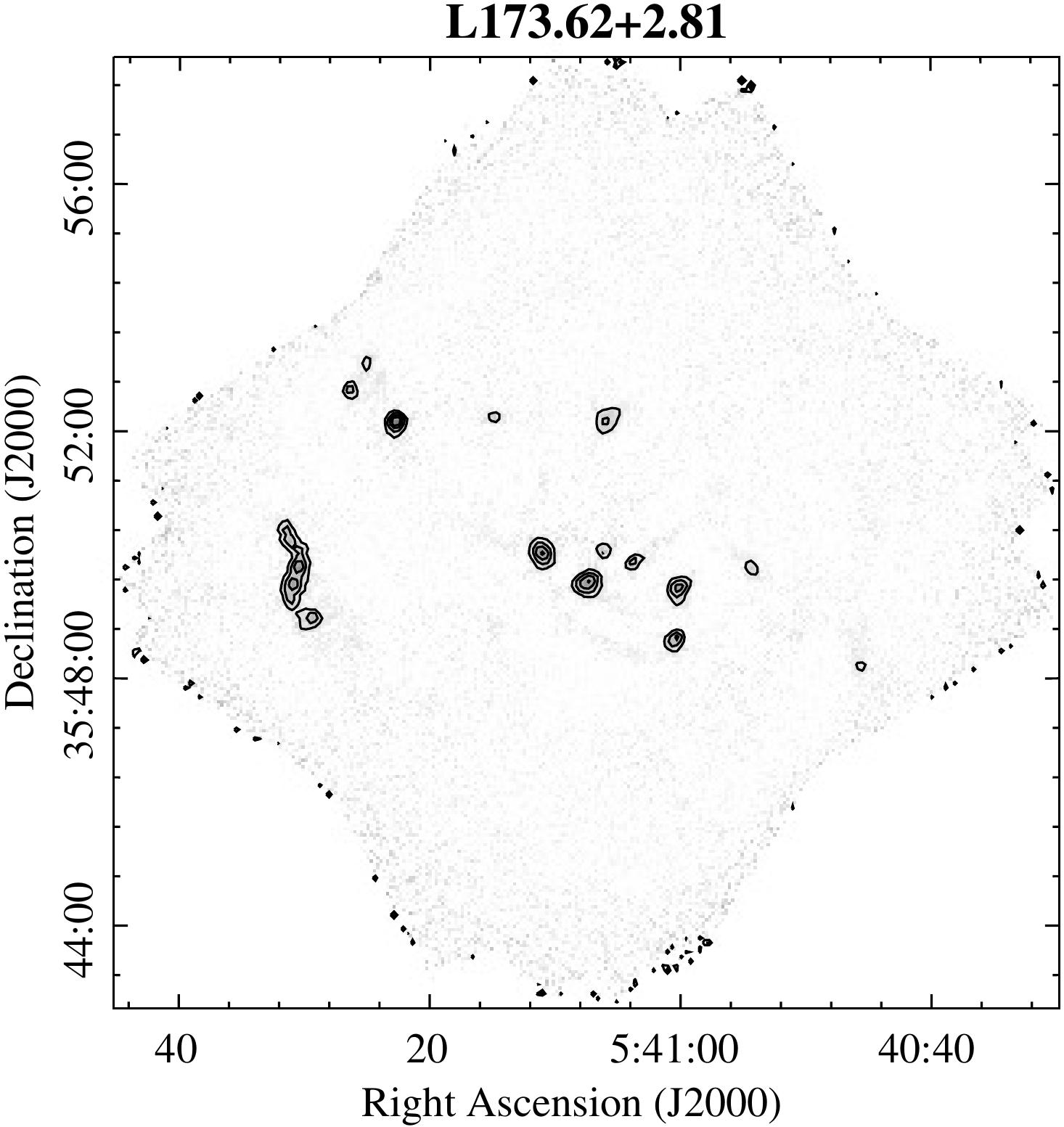}
\includegraphics[scale=0.43]{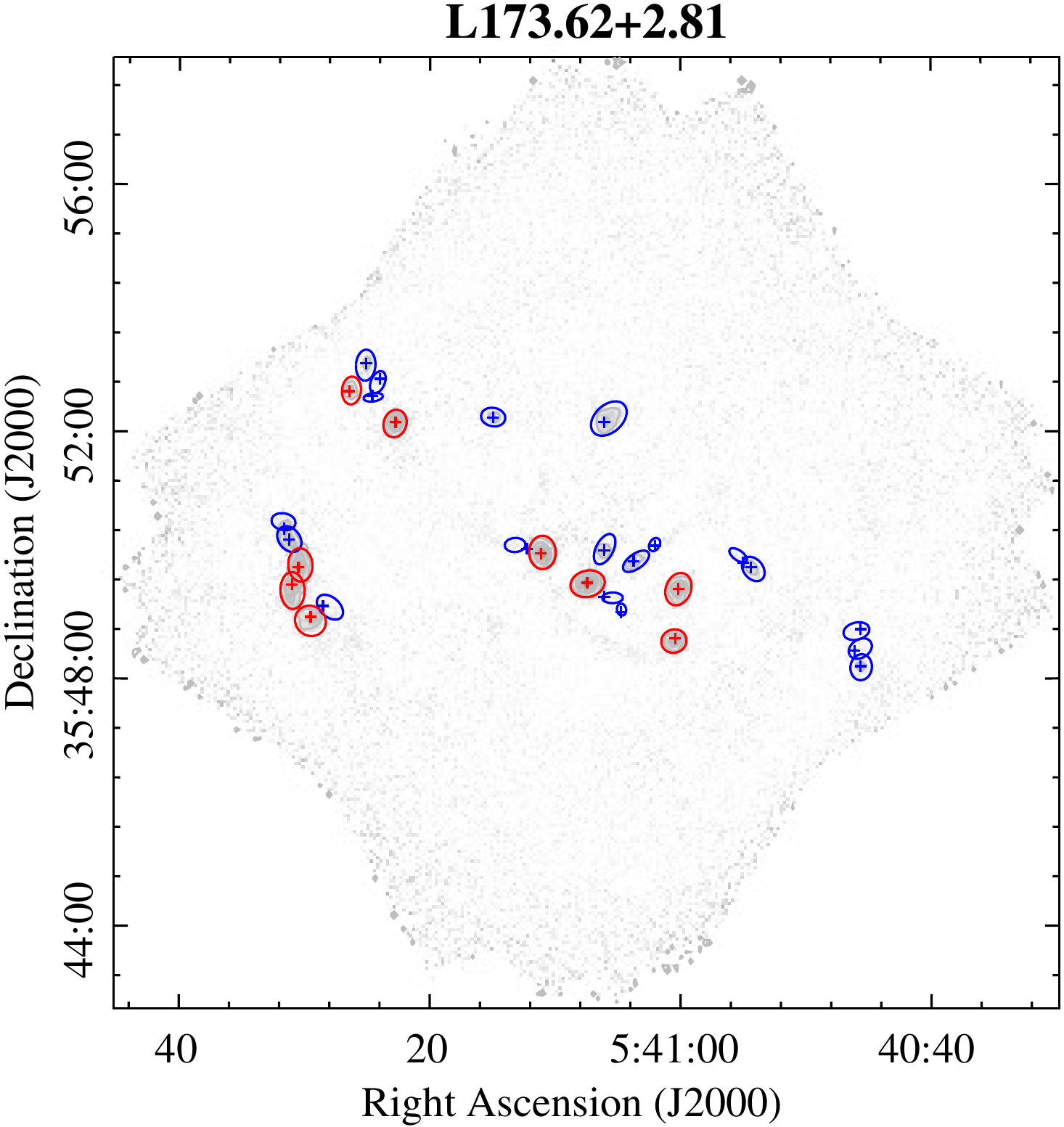}
}\\
\subfloat[L173.72+2.70 map, $\sigma_{rms}=387$ mJy beam$^{-1}$.]{
\includegraphics[scale=0.43]{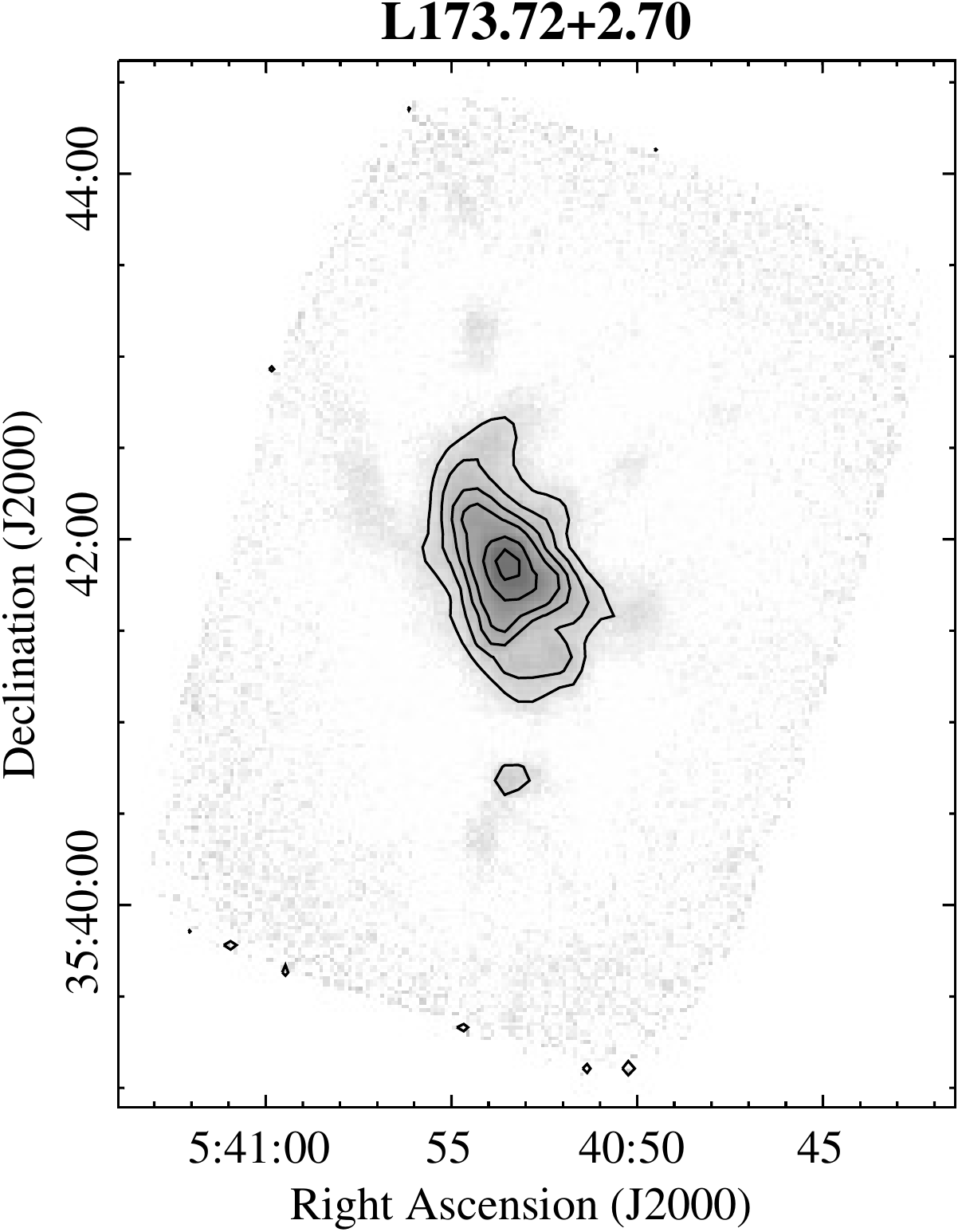}
\includegraphics[scale=0.43]{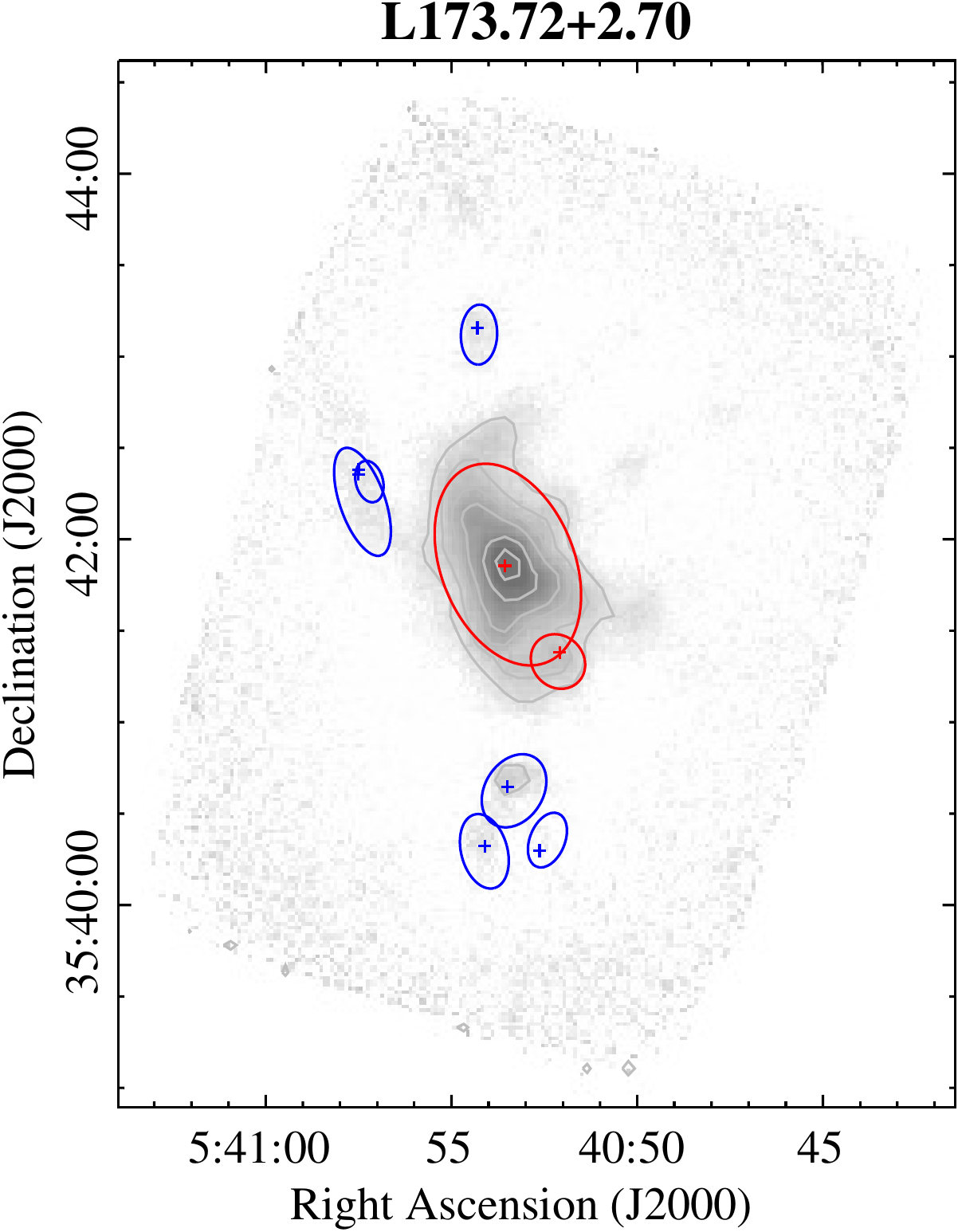}
}\\
\caption{Continuation}
\end{figure}

\clearpage
\begin{figure}\ContinuedFloat 
\center
\subfloat[L173.76+2.67 map, $\sigma_{rms}=301$ mJy beam$^{-1}$.]{
\includegraphics[scale=0.43]{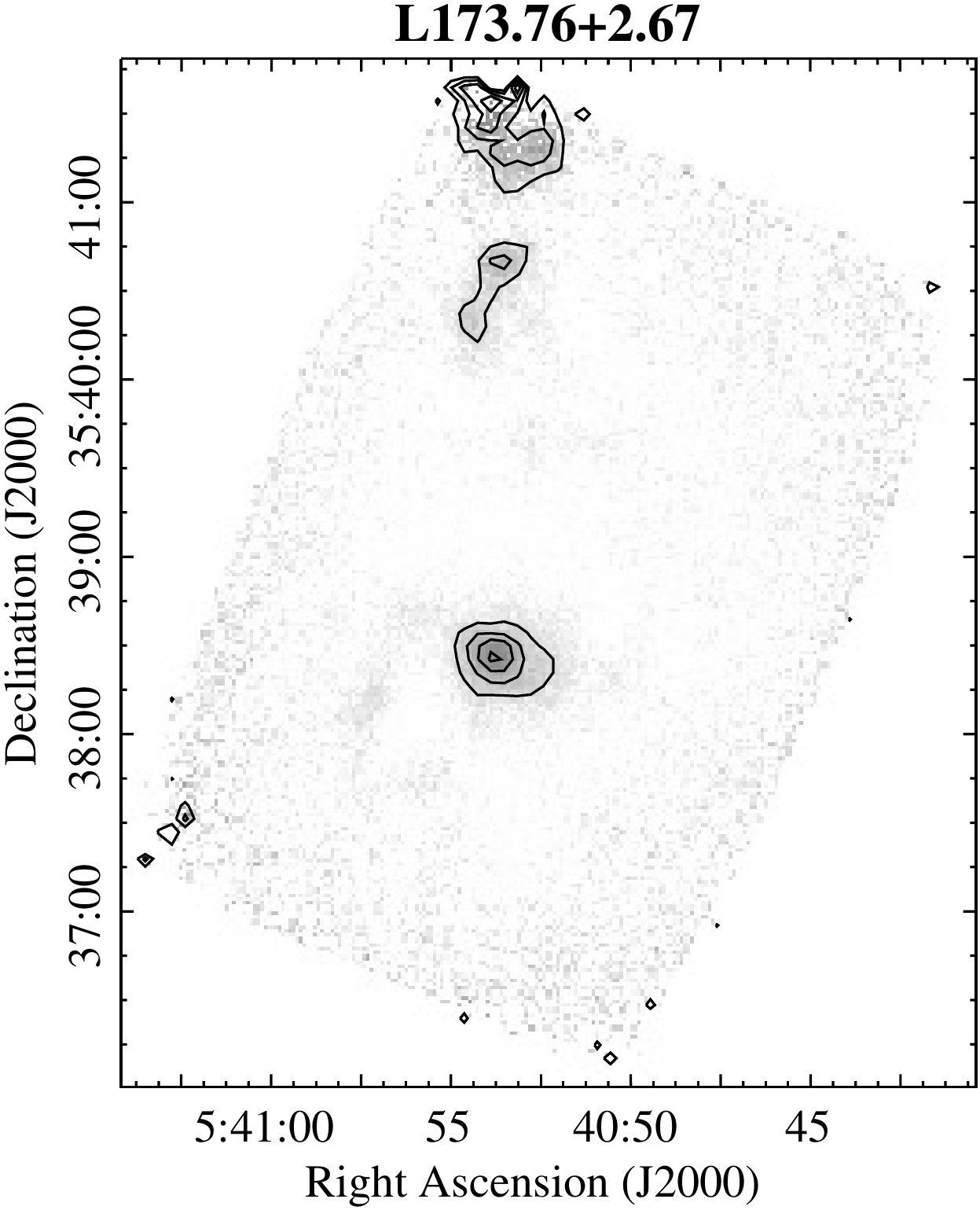}
\includegraphics[scale=0.43]{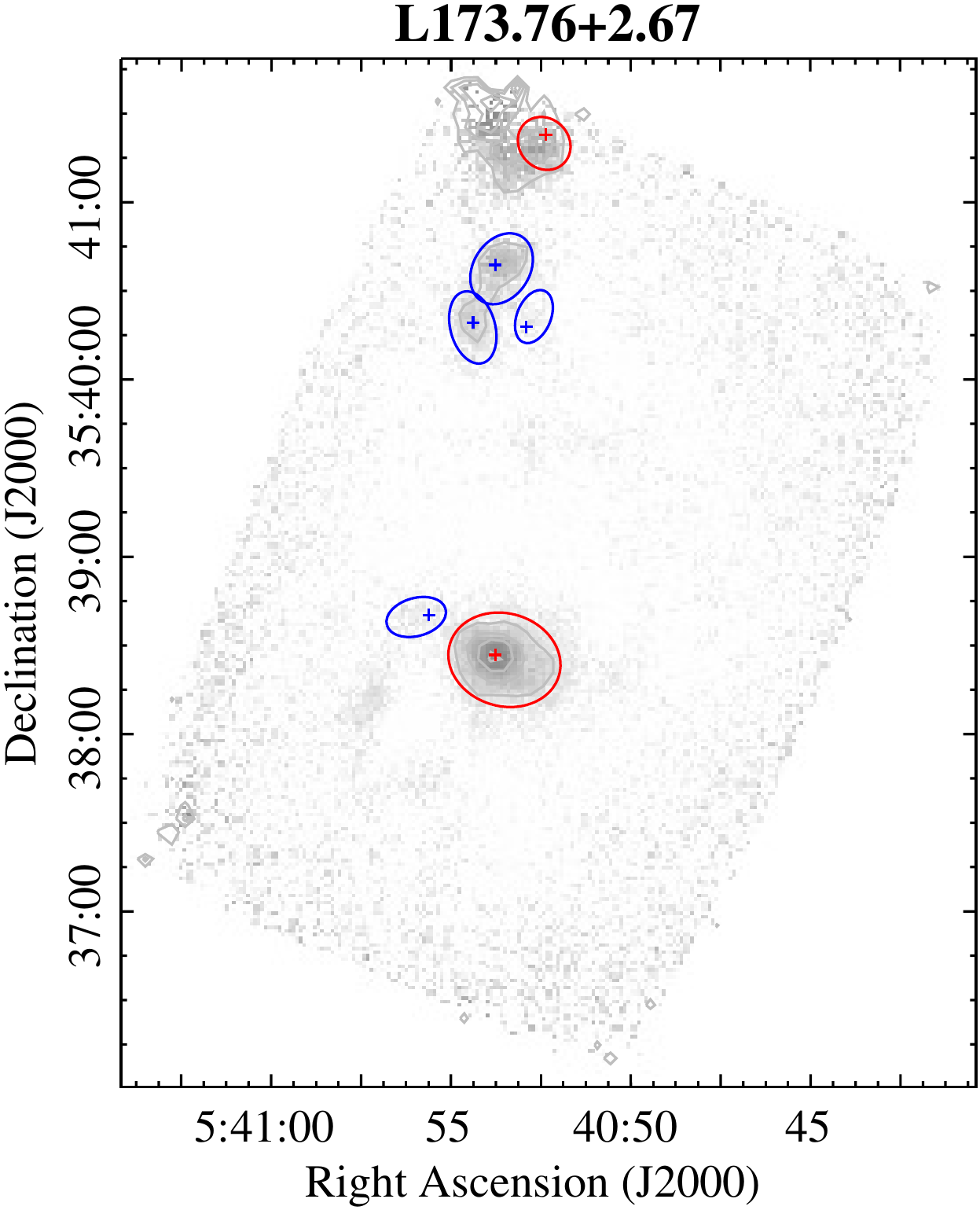}
}\\
\subfloat[L188.79+1.03 map, $\sigma_{rms}=259$ mJy beam$^{-1}$.]{
\includegraphics[scale=0.43]{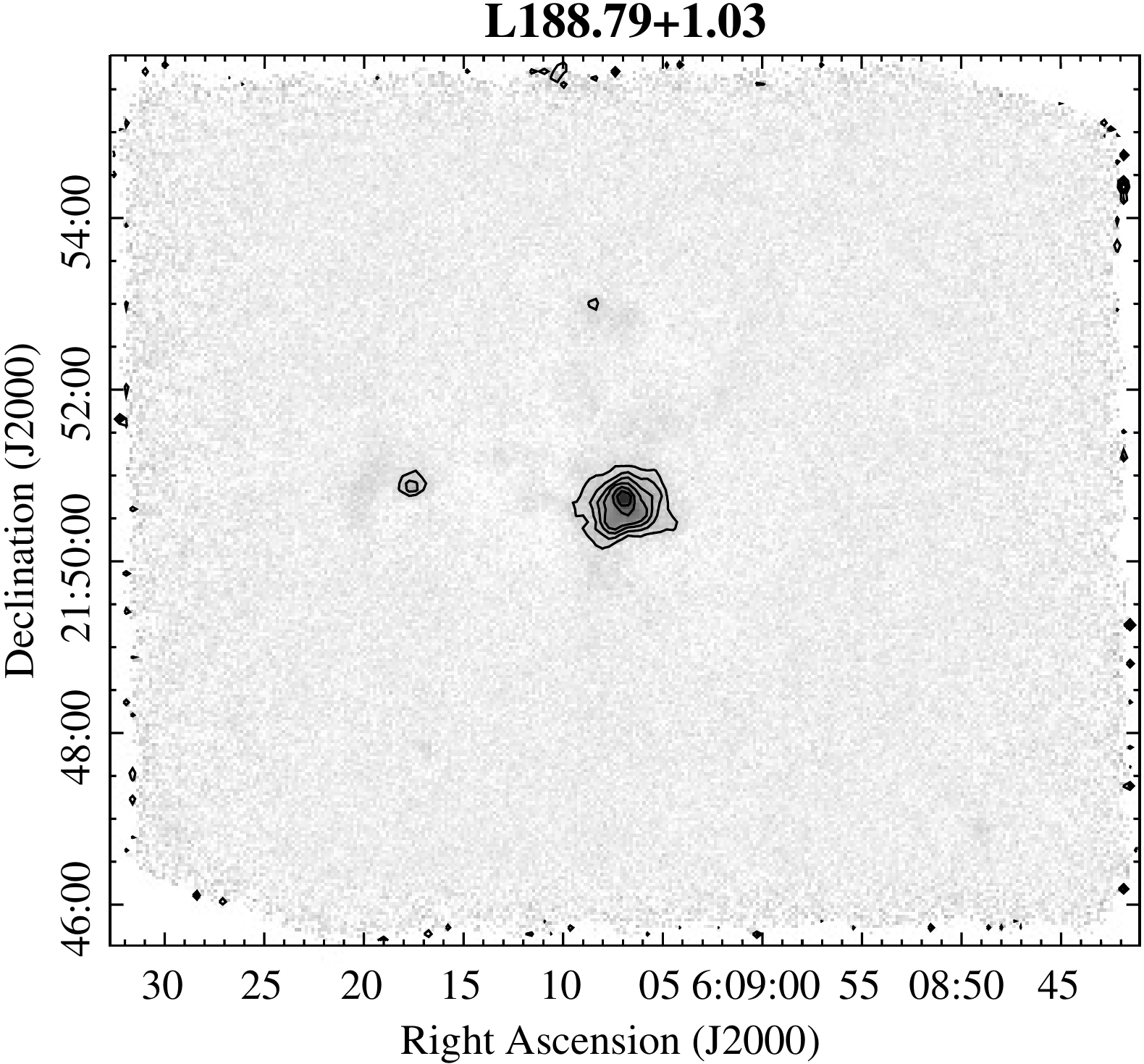}
\includegraphics[scale=0.43]{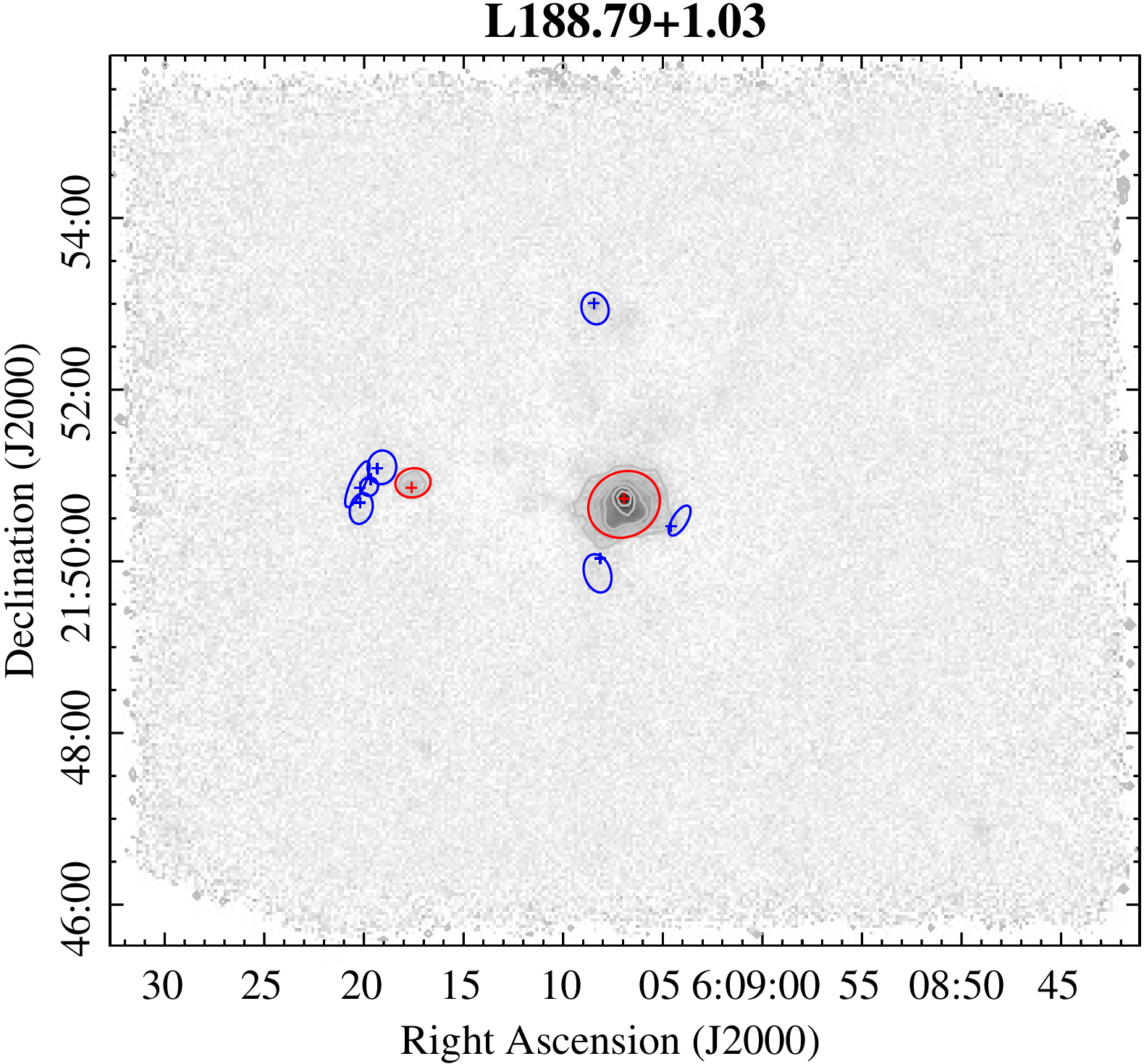}
}\\
\subfloat[L188.95+0.88 map, $\sigma_{rms}=268$ mJy beam$^{-1}$.]{
\includegraphics[scale=0.43]{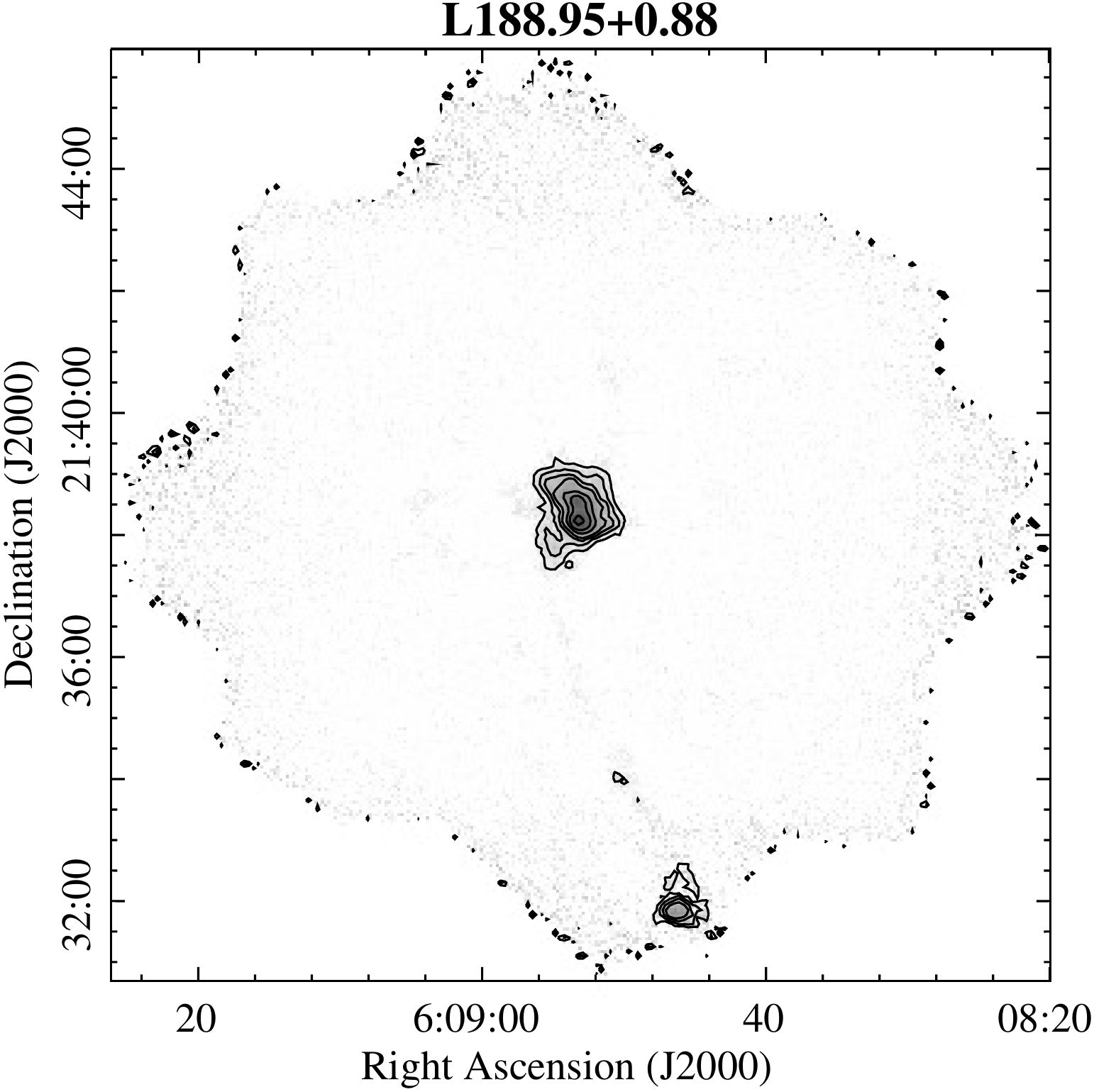}
\includegraphics[scale=0.43]{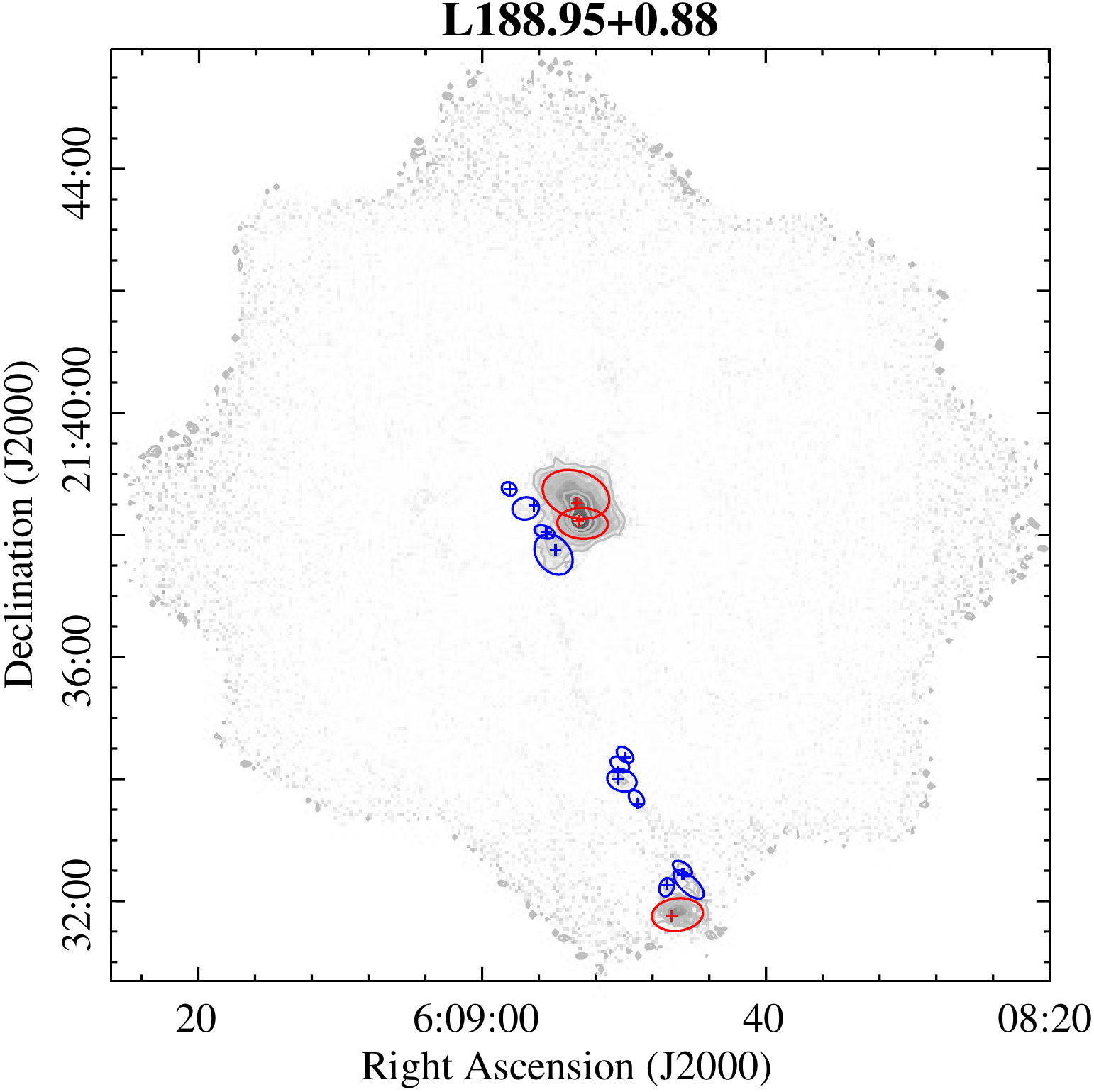}
}\\
\caption{Continuation}
\end{figure}

\clearpage
\begin{figure}\ContinuedFloat 
\center
\subfloat[L189.03+0.78 map, $\sigma_{rms}=290$ mJy beam$^{-1}$.]{
\includegraphics[scale=0.43]{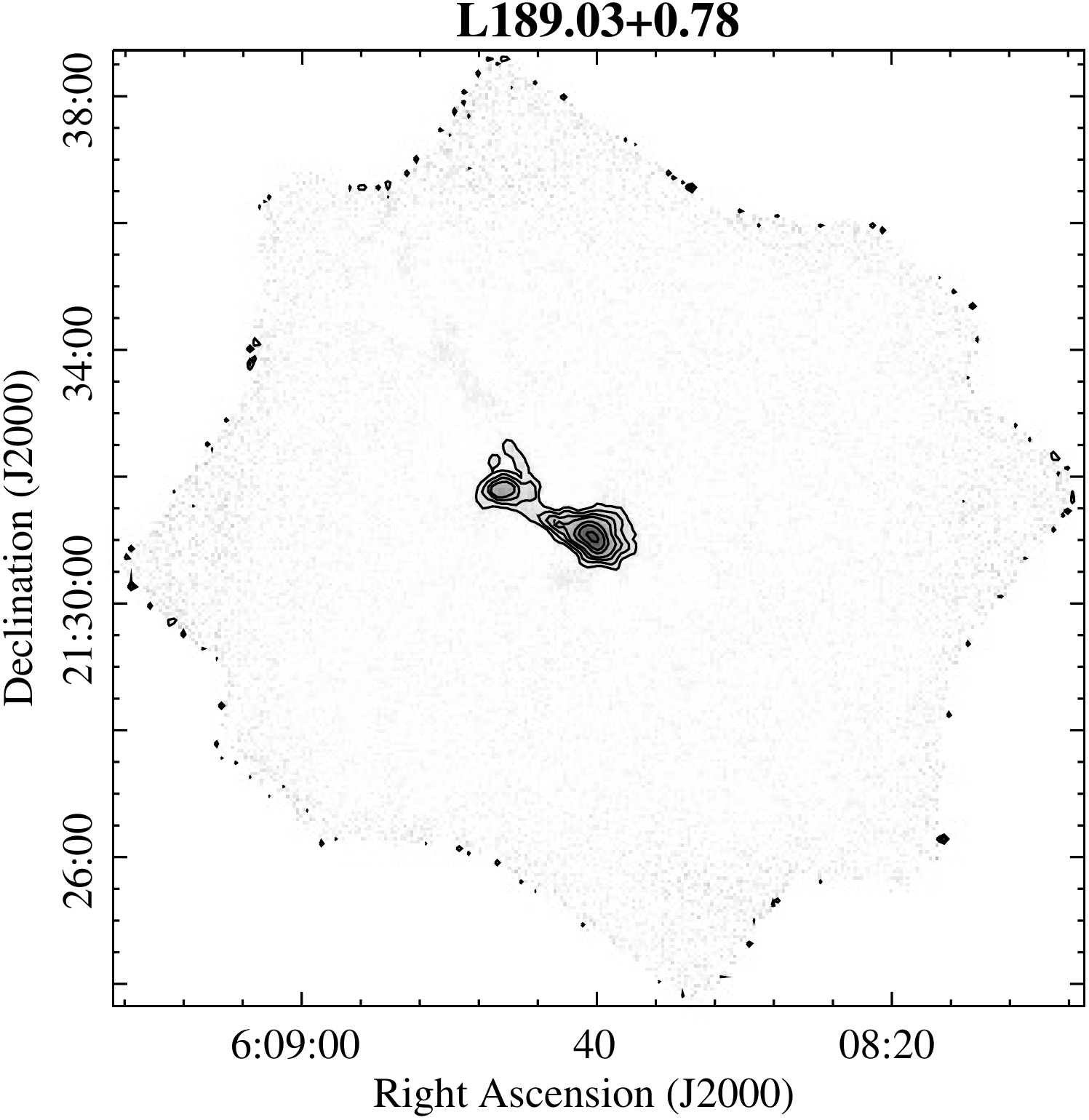}
\includegraphics[scale=0.43]{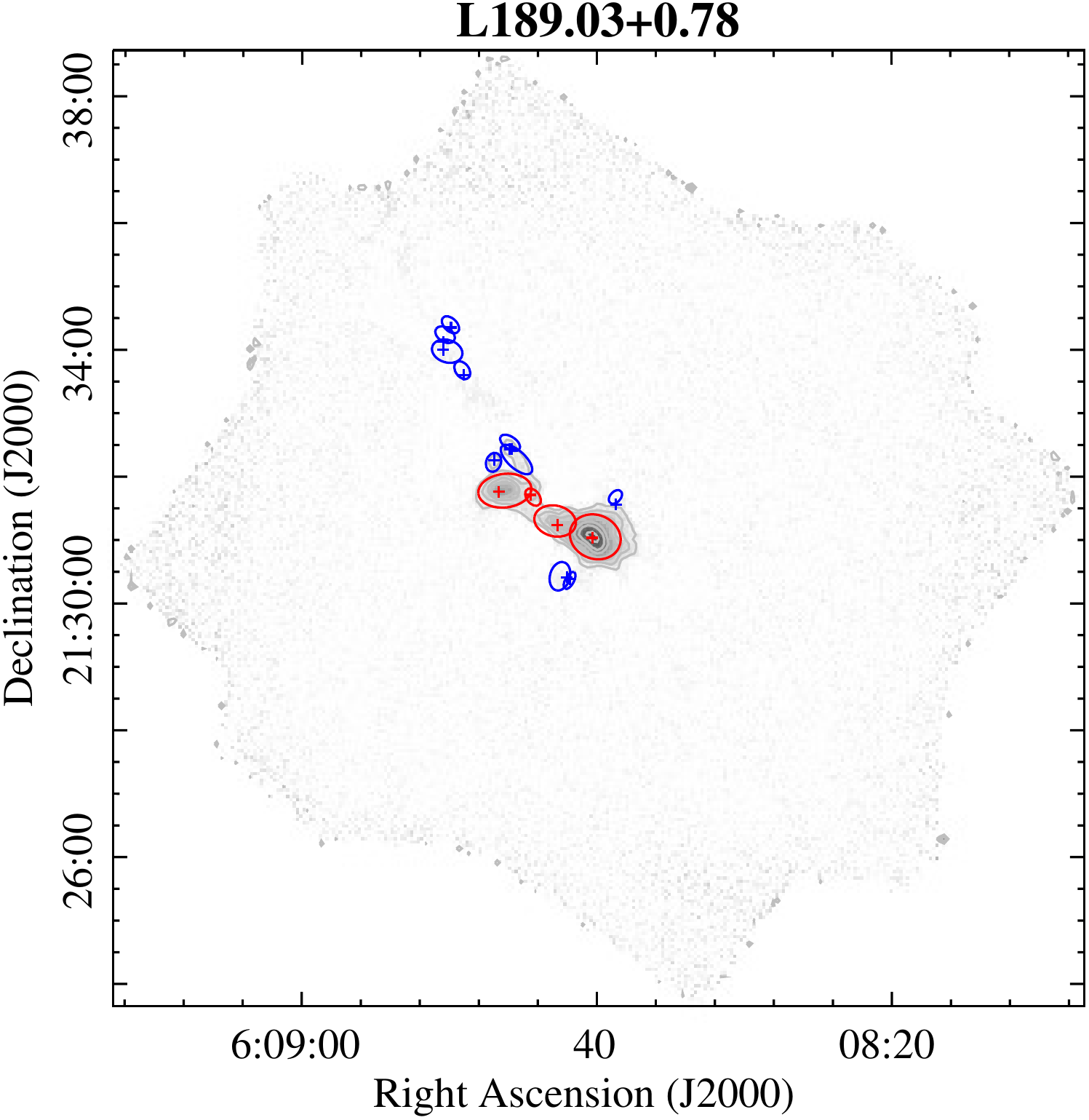}
}\\
\subfloat[L189.12+0.64 map, $\sigma_{rms}=403$ mJy beam$^{-1}$.]{
\includegraphics[scale=0.43]{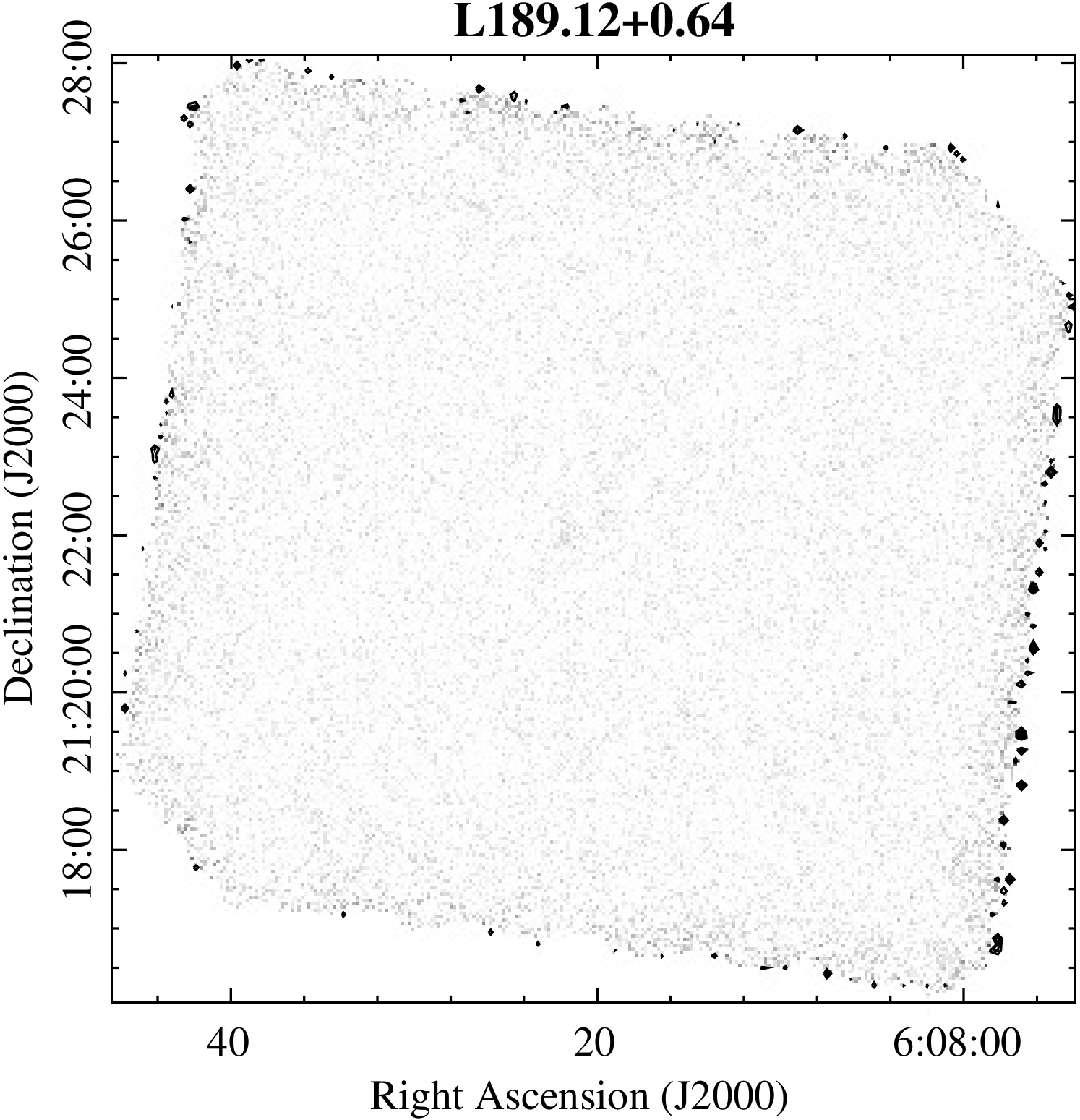}
\includegraphics[scale=0.43]{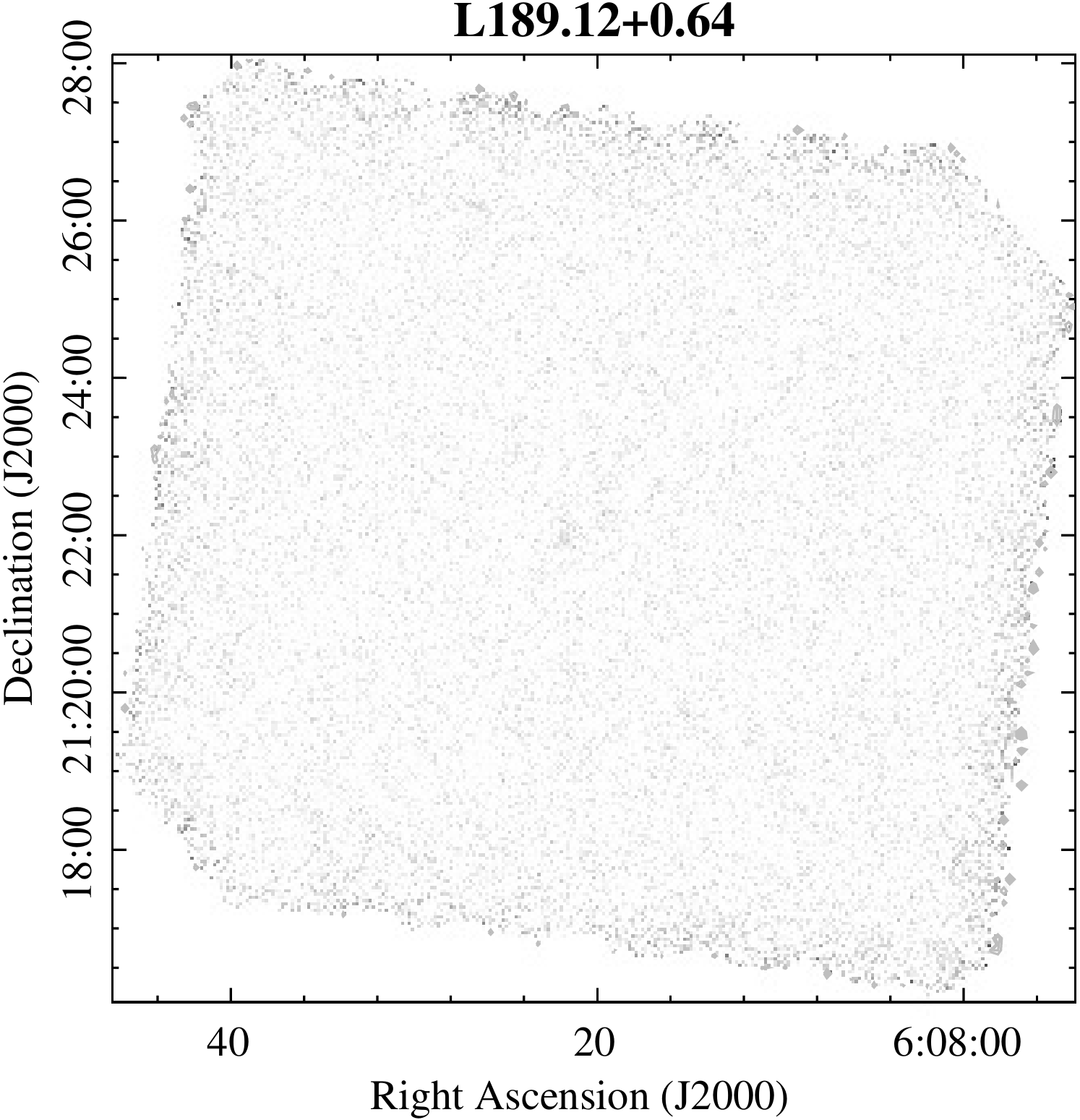}
}\\
\subfloat[L189.68+0.19 map, $\sigma_{rms}=155$ mJy beam$^{-1}$.]{
\includegraphics[scale=0.43]{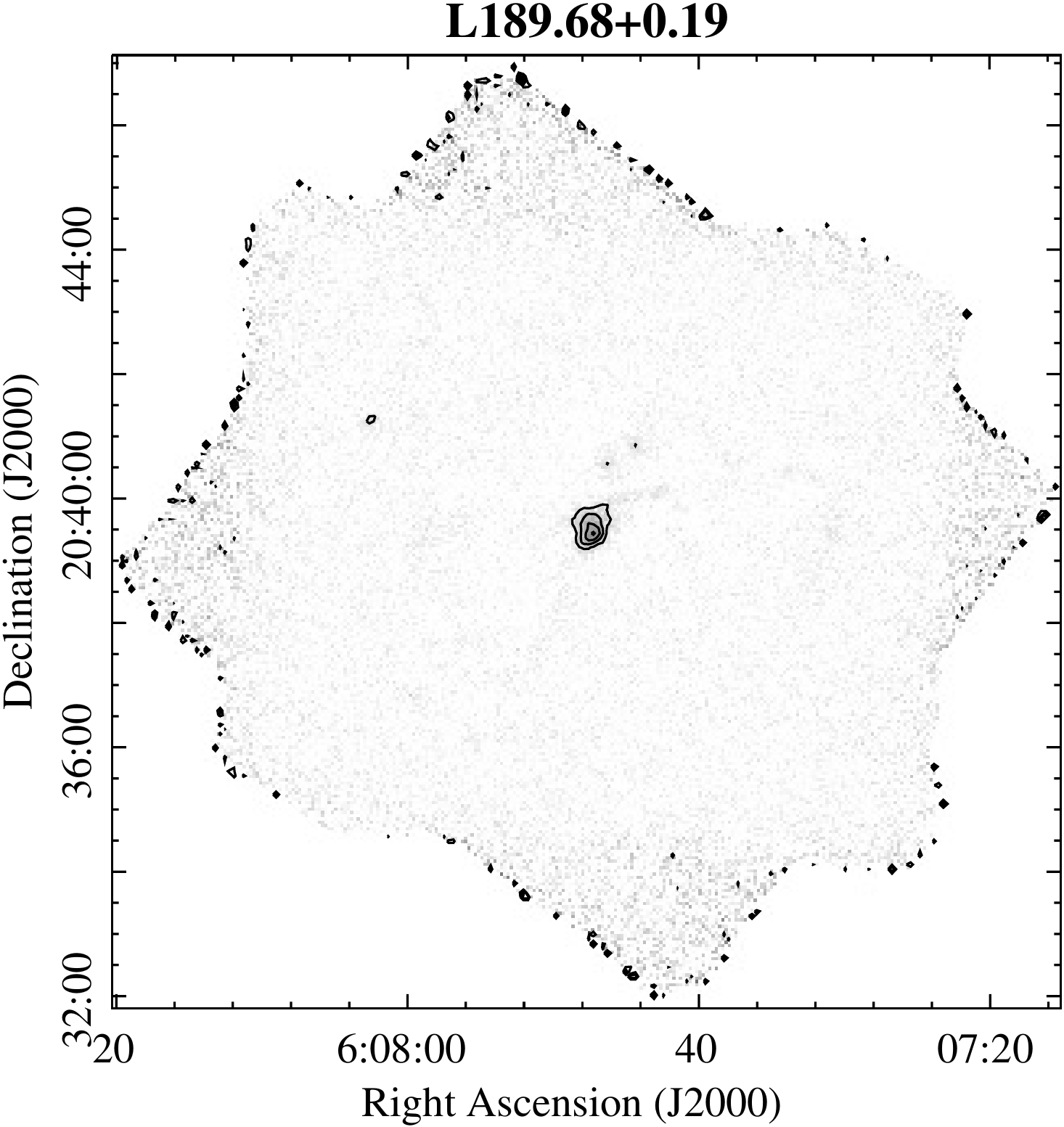}
\includegraphics[scale=0.43]{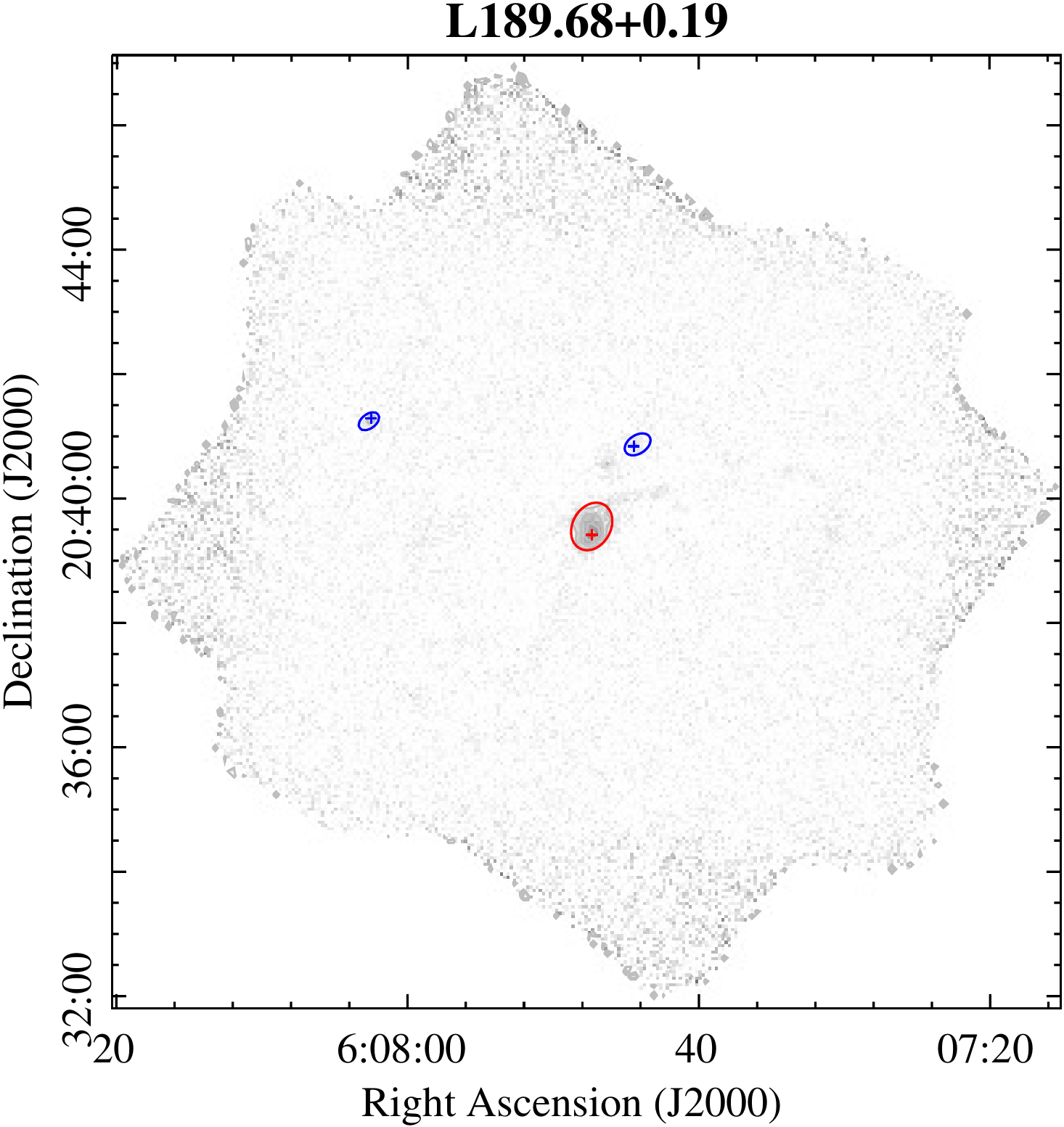}
}\\
\caption{Continuation}
\end{figure}

\clearpage
\begin{figure}\ContinuedFloat 
\center
\subfloat[L189.85+0.39 map, $\sigma_{rms}=505$ mJy beam$^{-1}$.]{
\includegraphics[scale=0.43]{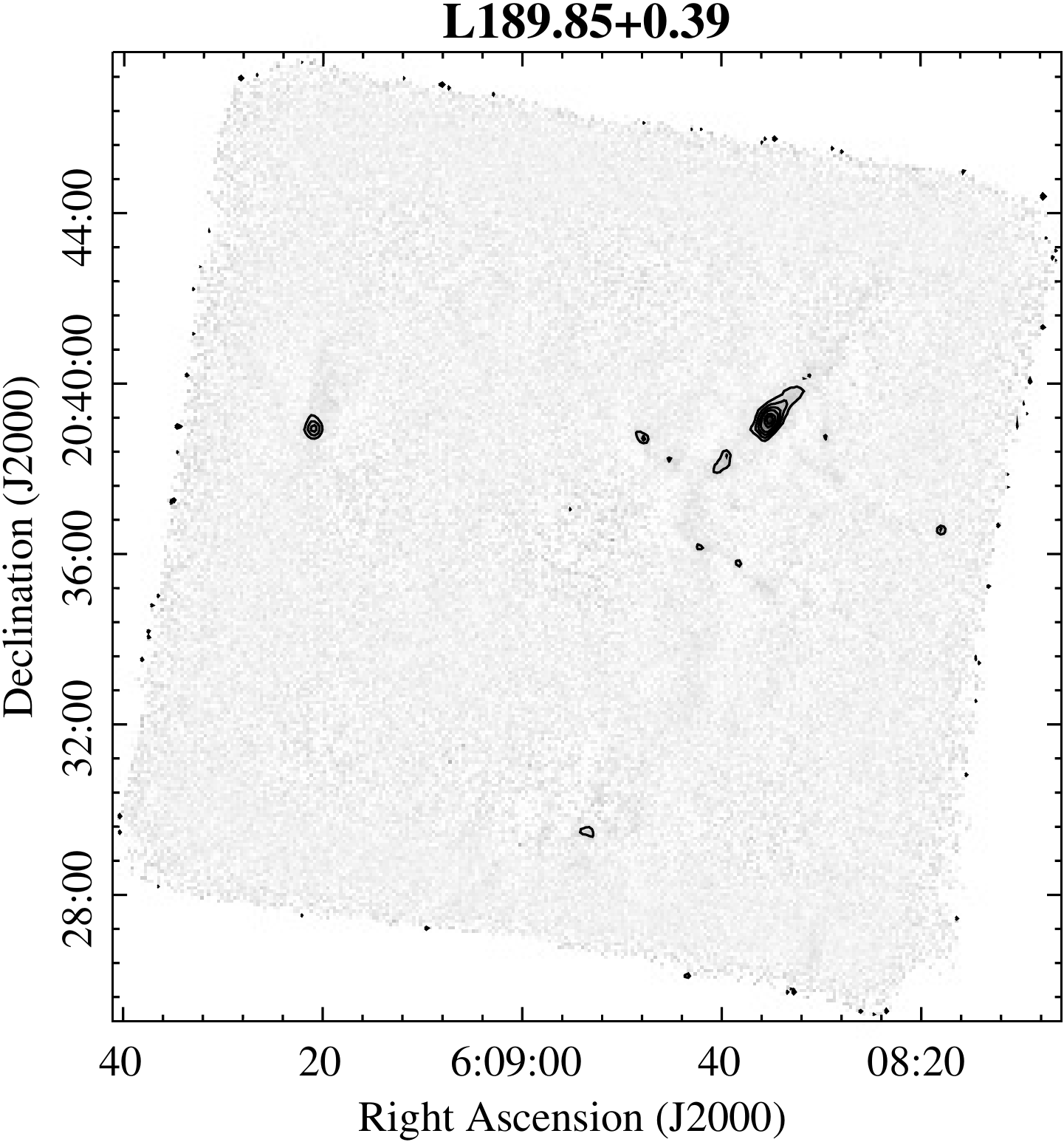}
\includegraphics[scale=0.43]{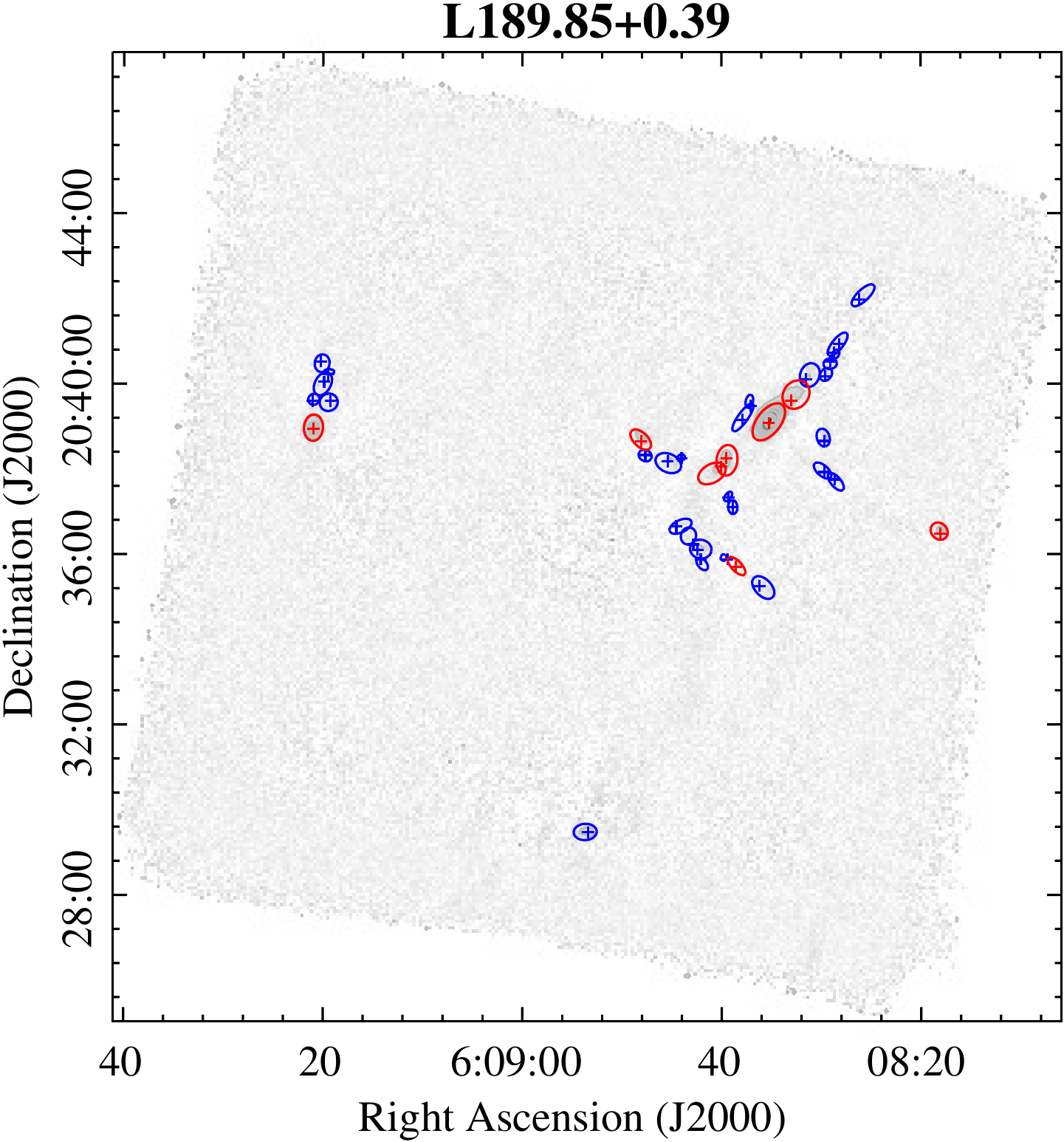}
}\\
\subfloat[L190.17+0.74 map, $\sigma_{rms}=184$ mJy beam$^{-1}$.]{
\includegraphics[scale=0.43]{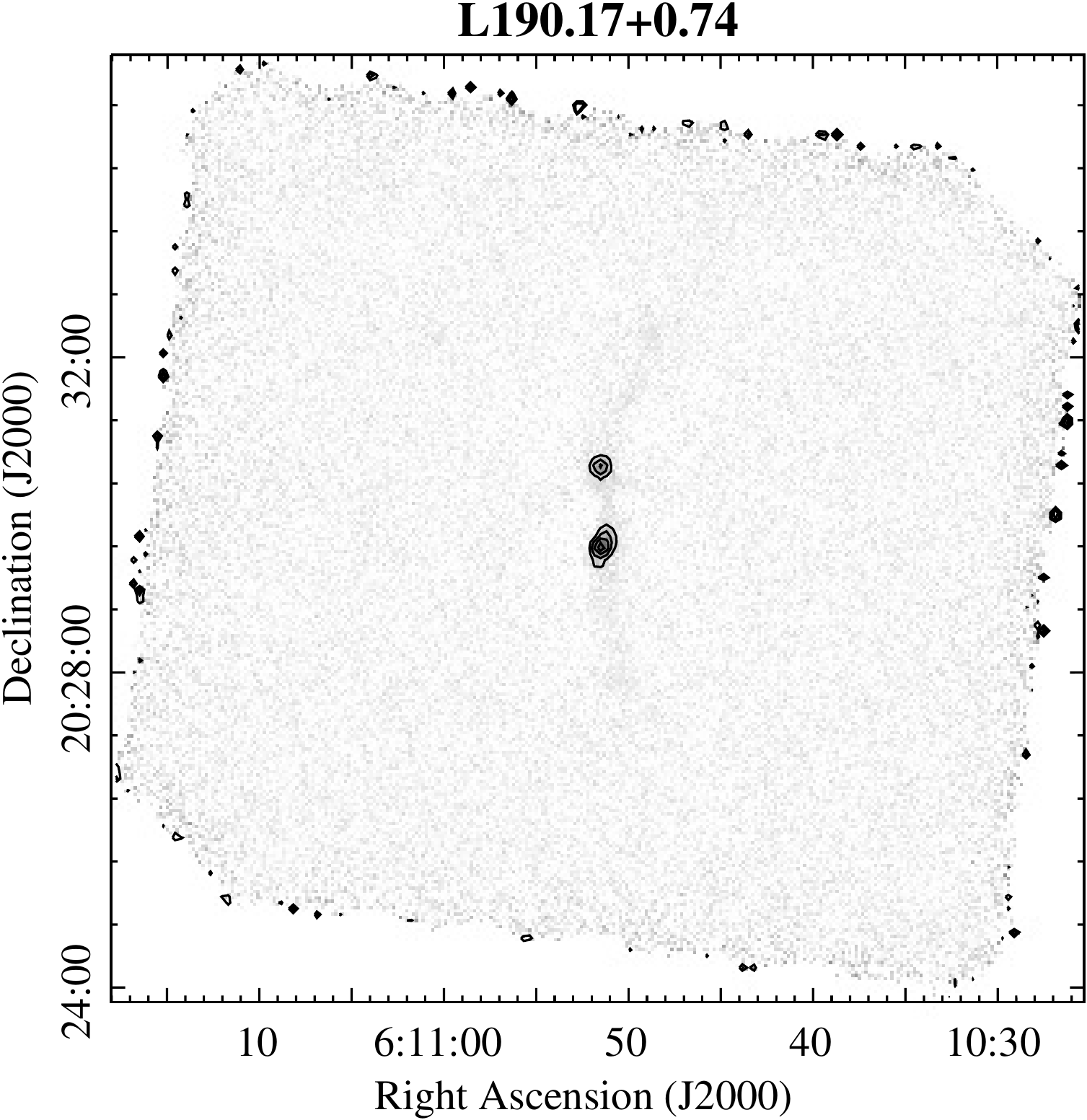}
\includegraphics[scale=0.43]{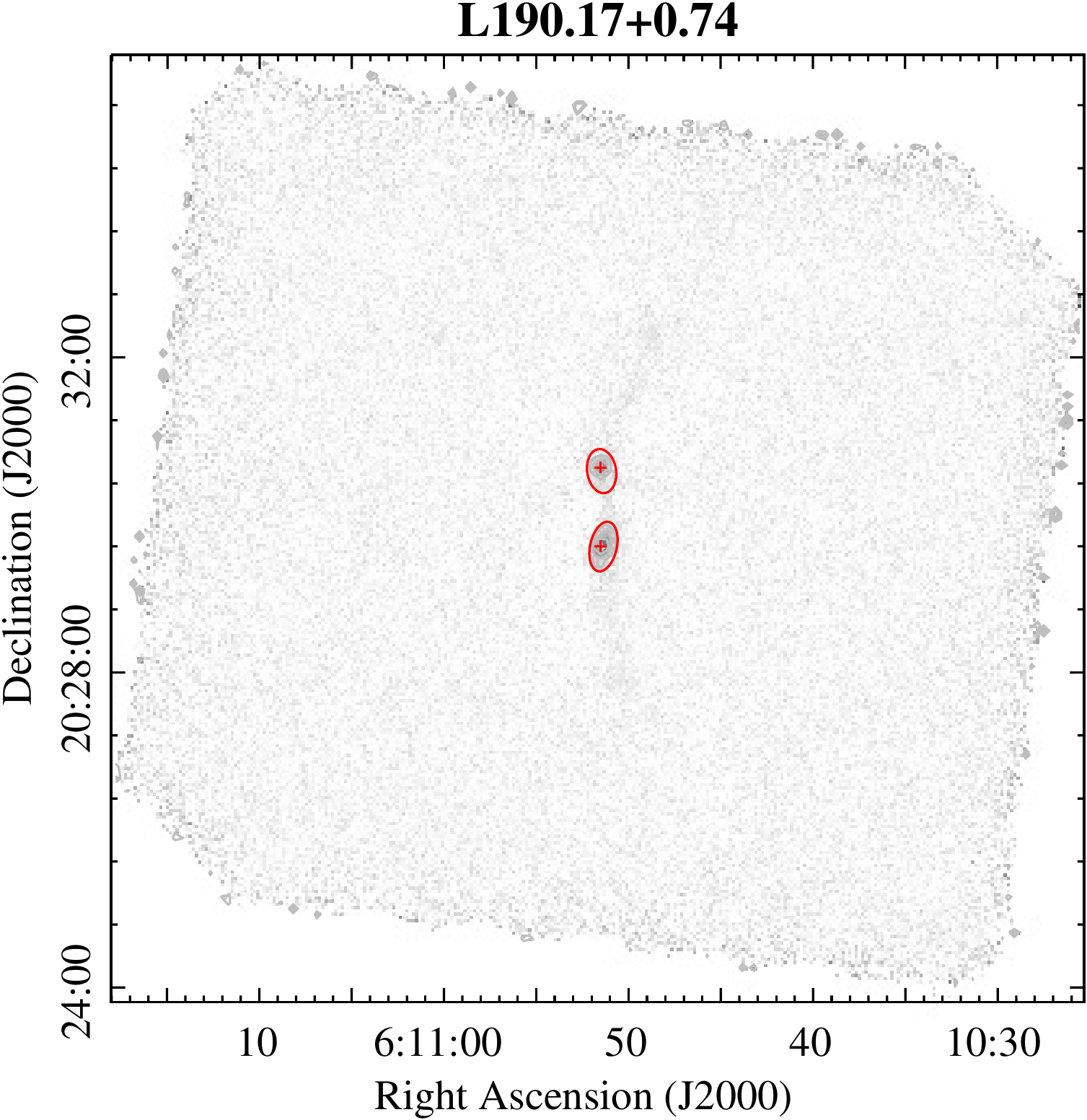}
}\\
\subfloat[L192.60-0.16 map, $\sigma_{rms}=340$ mJy beam$^{-1}$.]{
\includegraphics[scale=0.43]{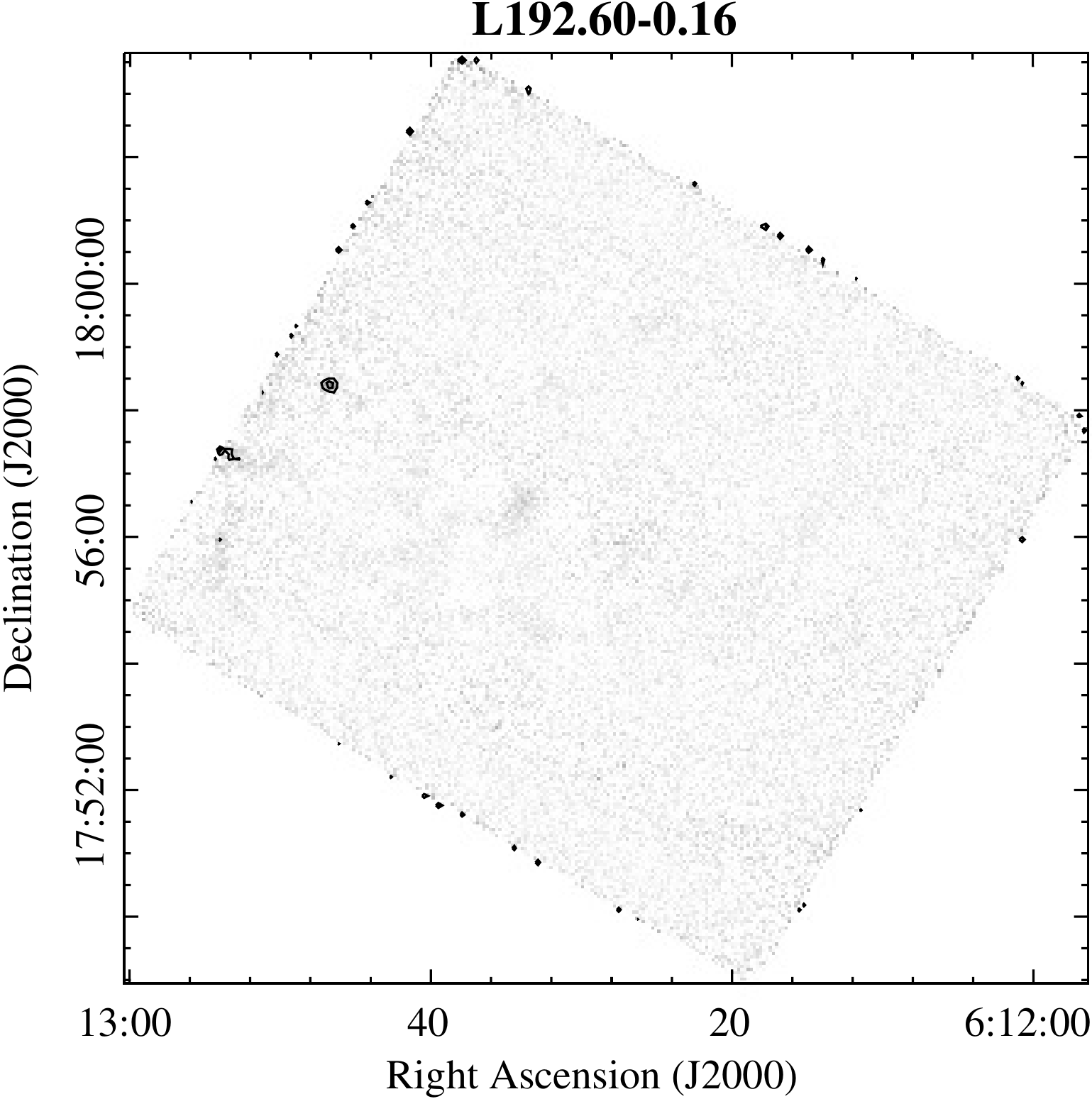}
\includegraphics[scale=0.43]{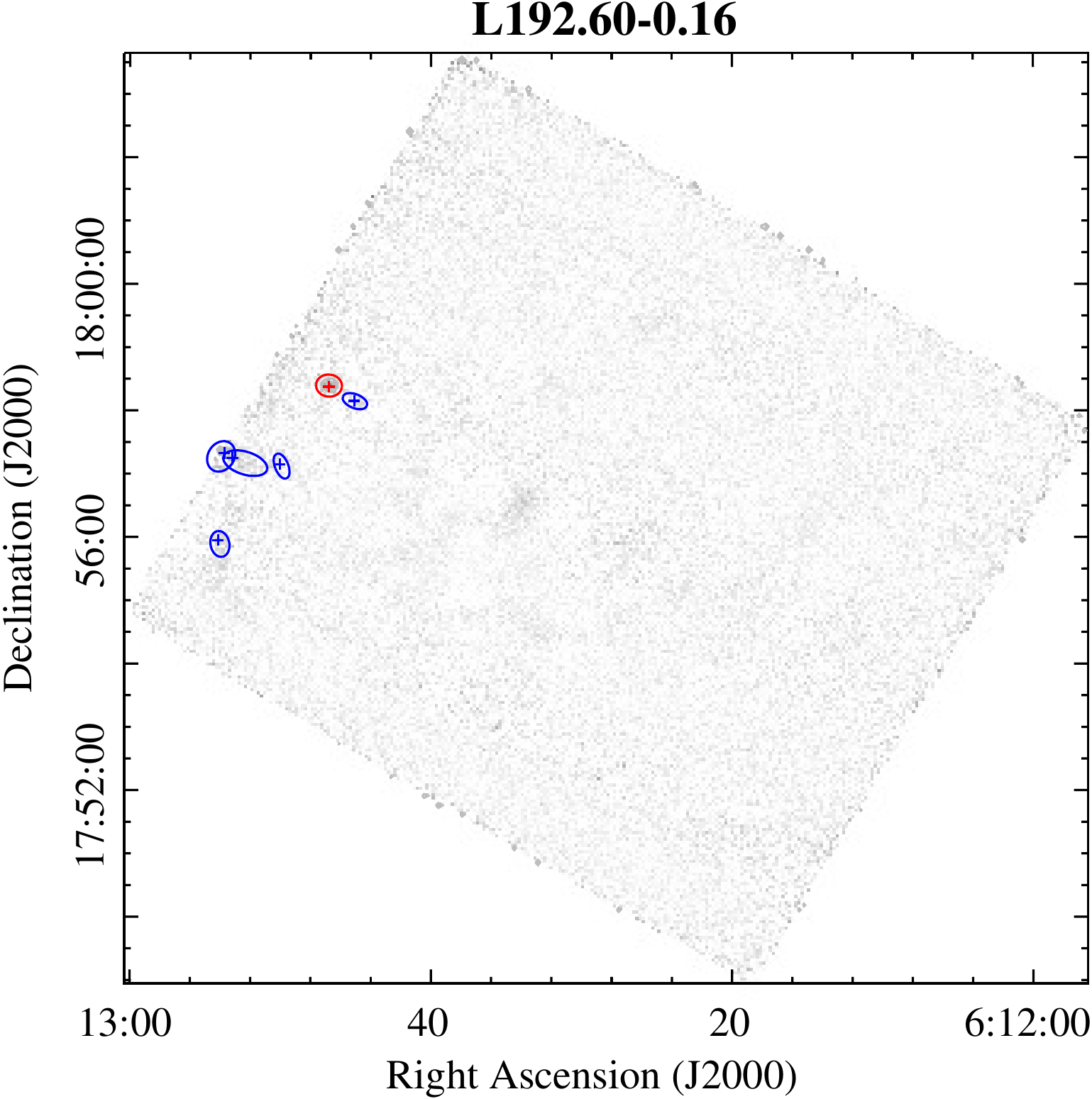}
}\\
\caption{Continuation}
\end{figure}

\clearpage
\begin{figure}\ContinuedFloat 
\center
\subfloat[L192.60-0.05 map, $\sigma_{rms}=271$ mJy beam$^{-1}$.]{
\includegraphics[scale=0.43]{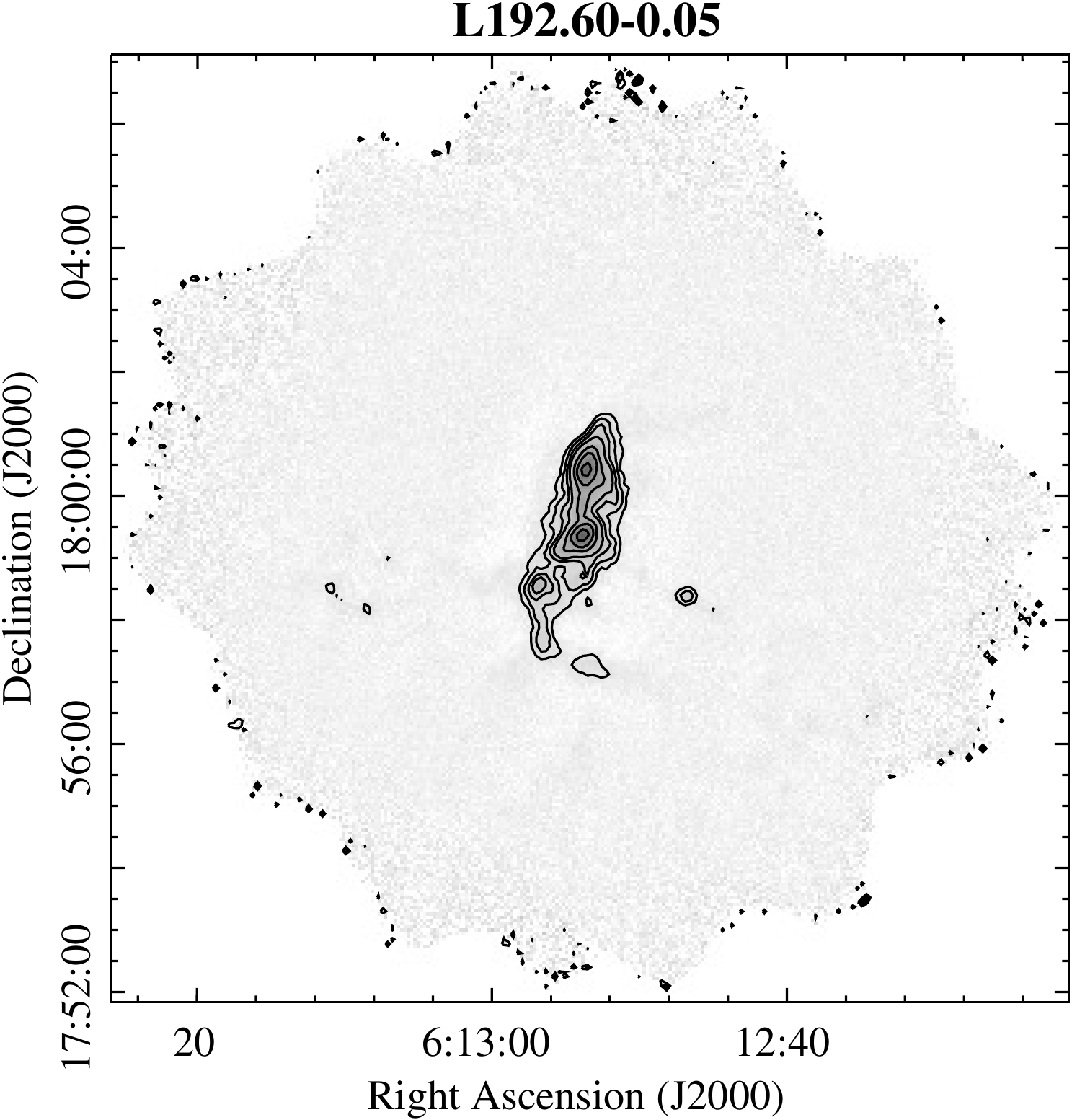}
\includegraphics[scale=0.43]{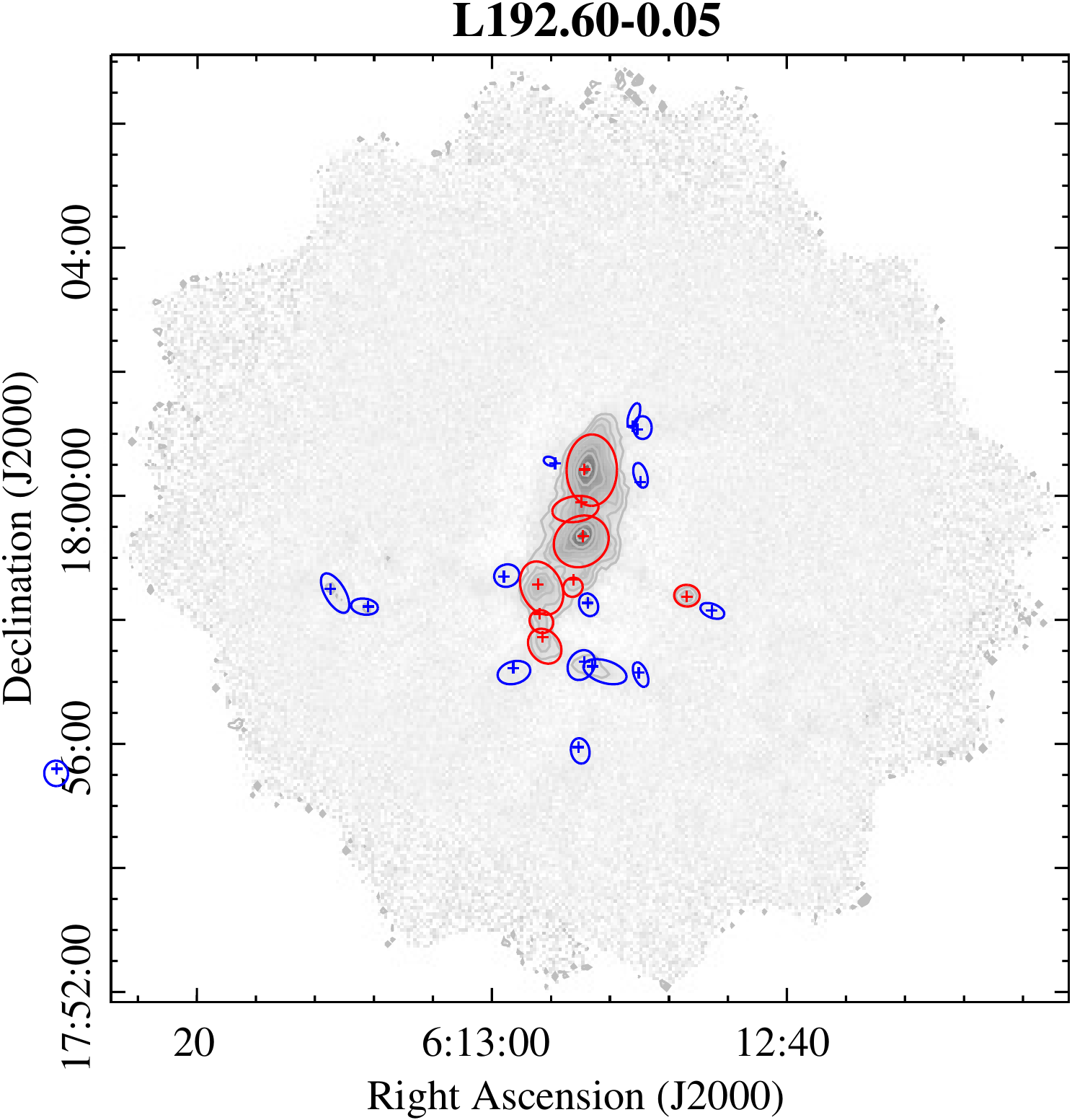}
}\\
\subfloat[L192.72+0.04 map, $\sigma_{rms}=335$ mJy beam$^{-1}$.]{
\includegraphics[scale=0.43]{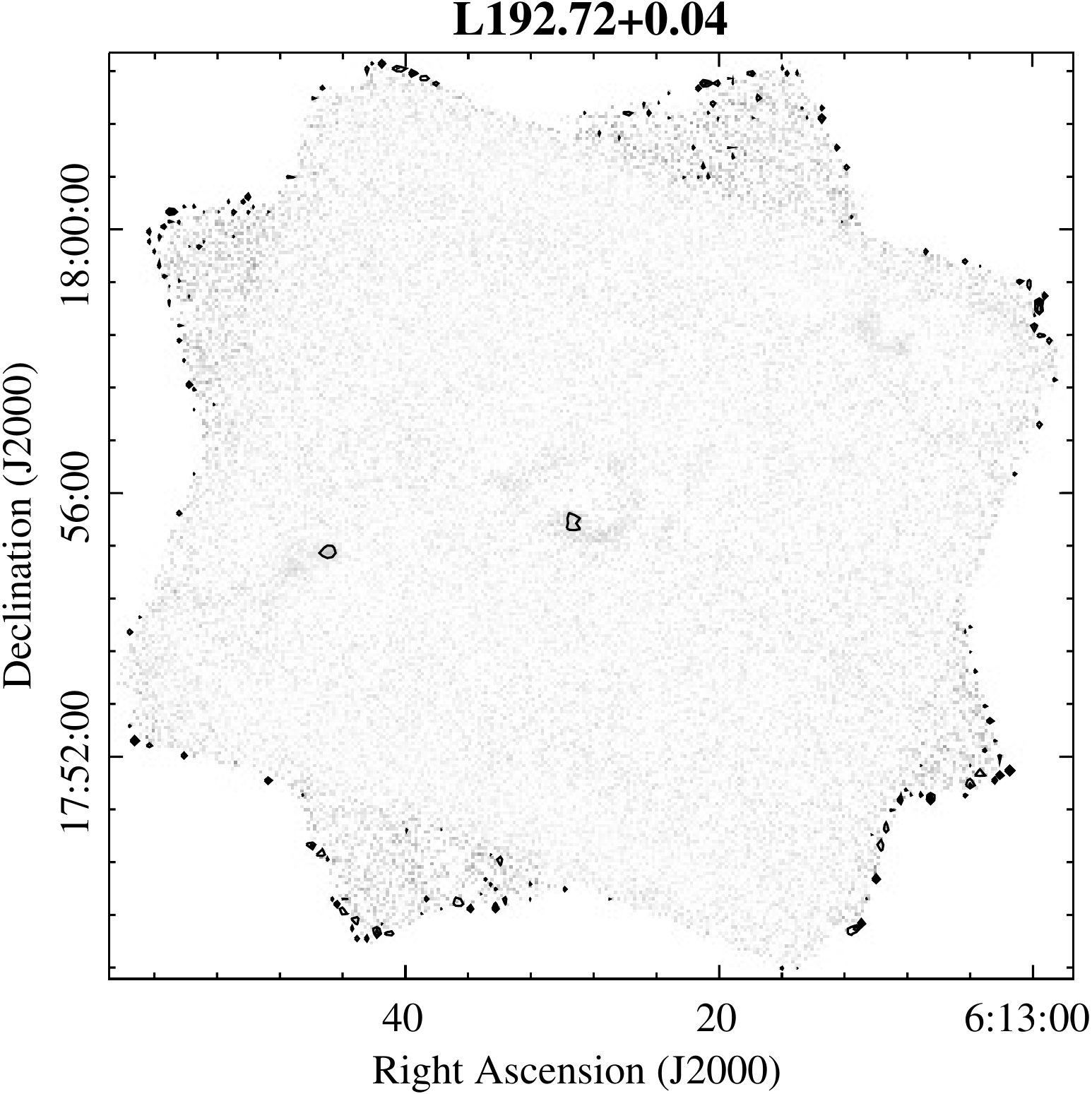}
\includegraphics[scale=0.43]{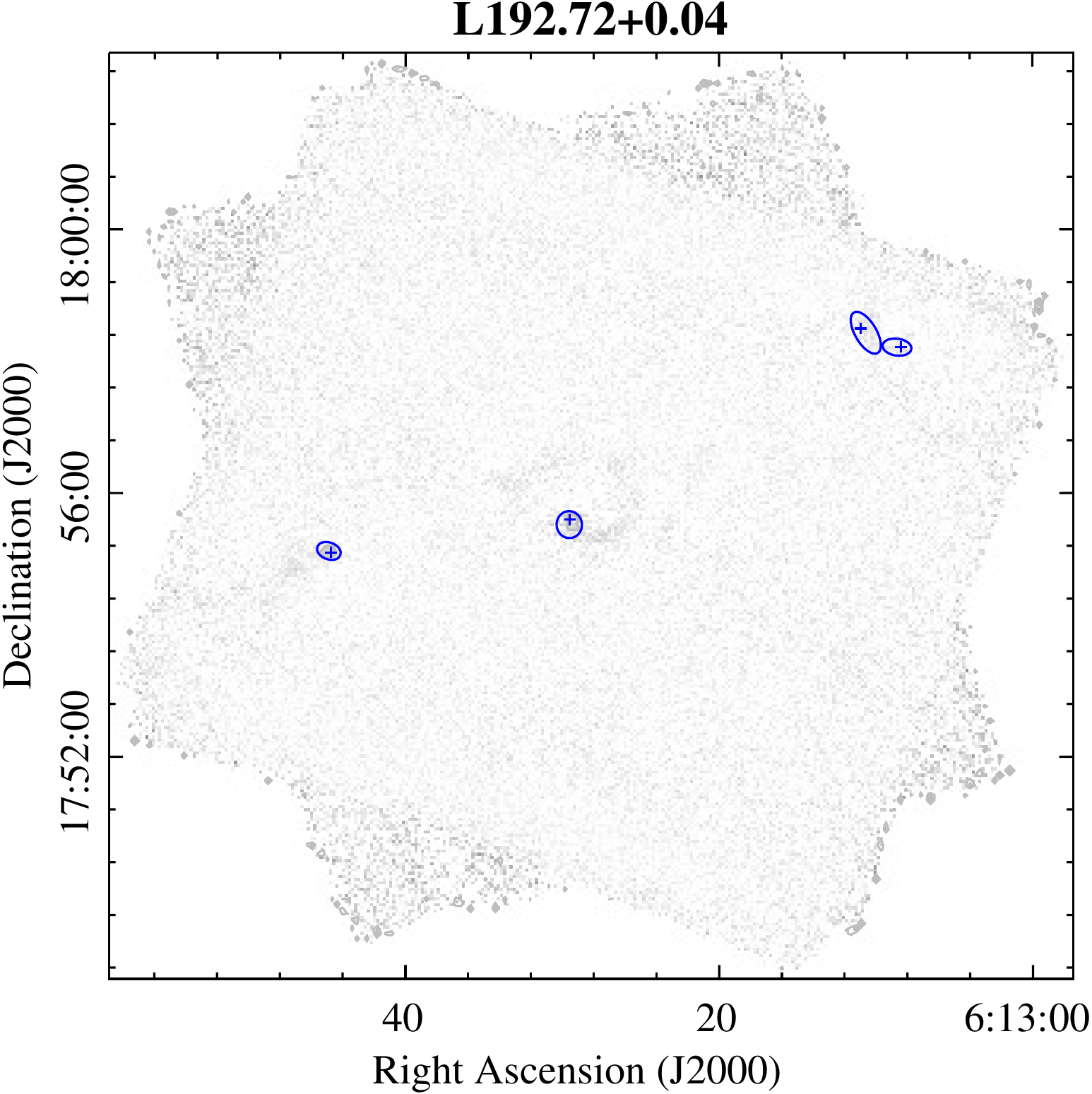}
}\\
\subfloat[L192.81+0.11 map, $\sigma_{rms}=328$ mJy beam$^{-1}$.]{
\includegraphics[scale=0.43]{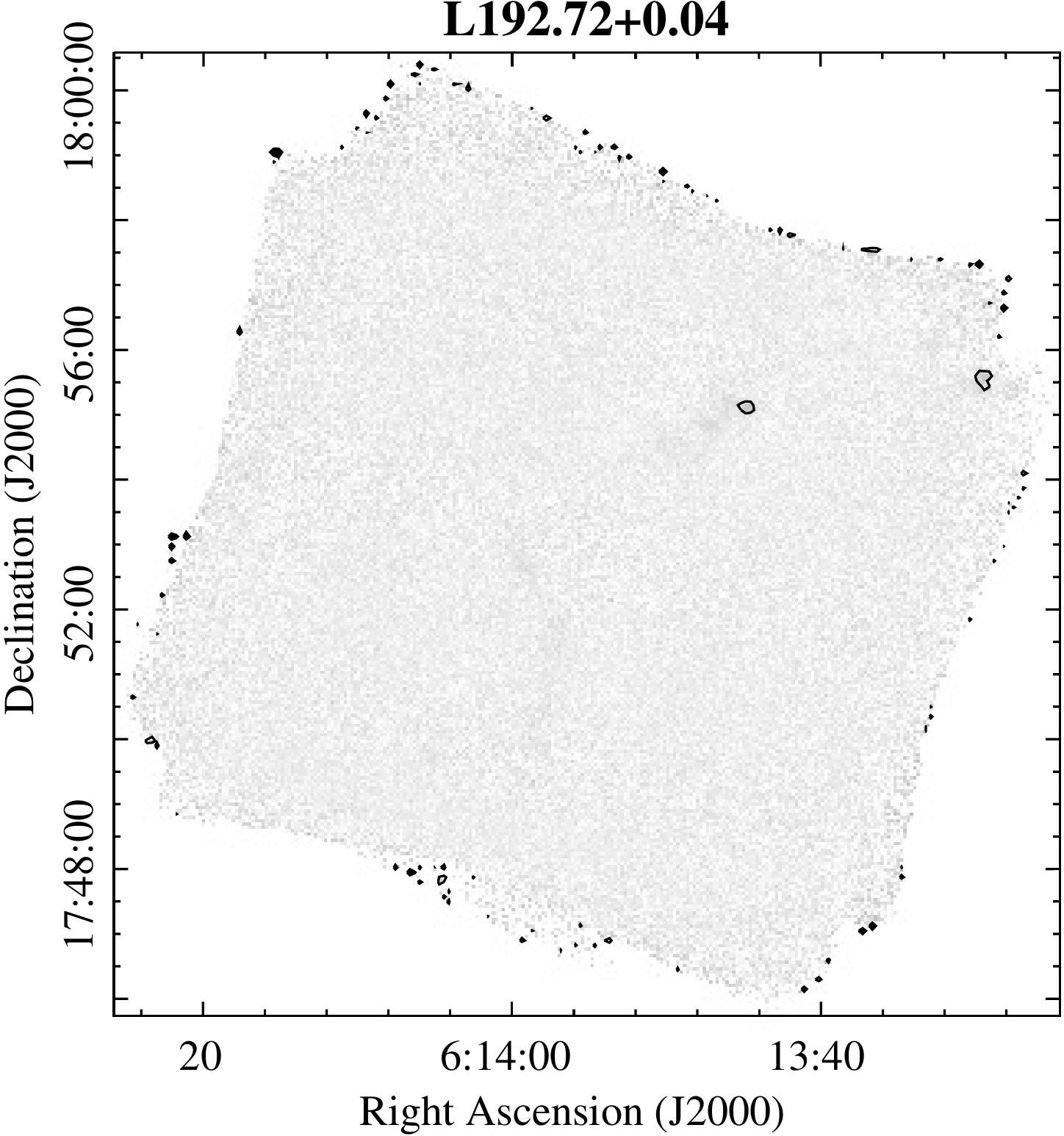}
\includegraphics[scale=0.43]{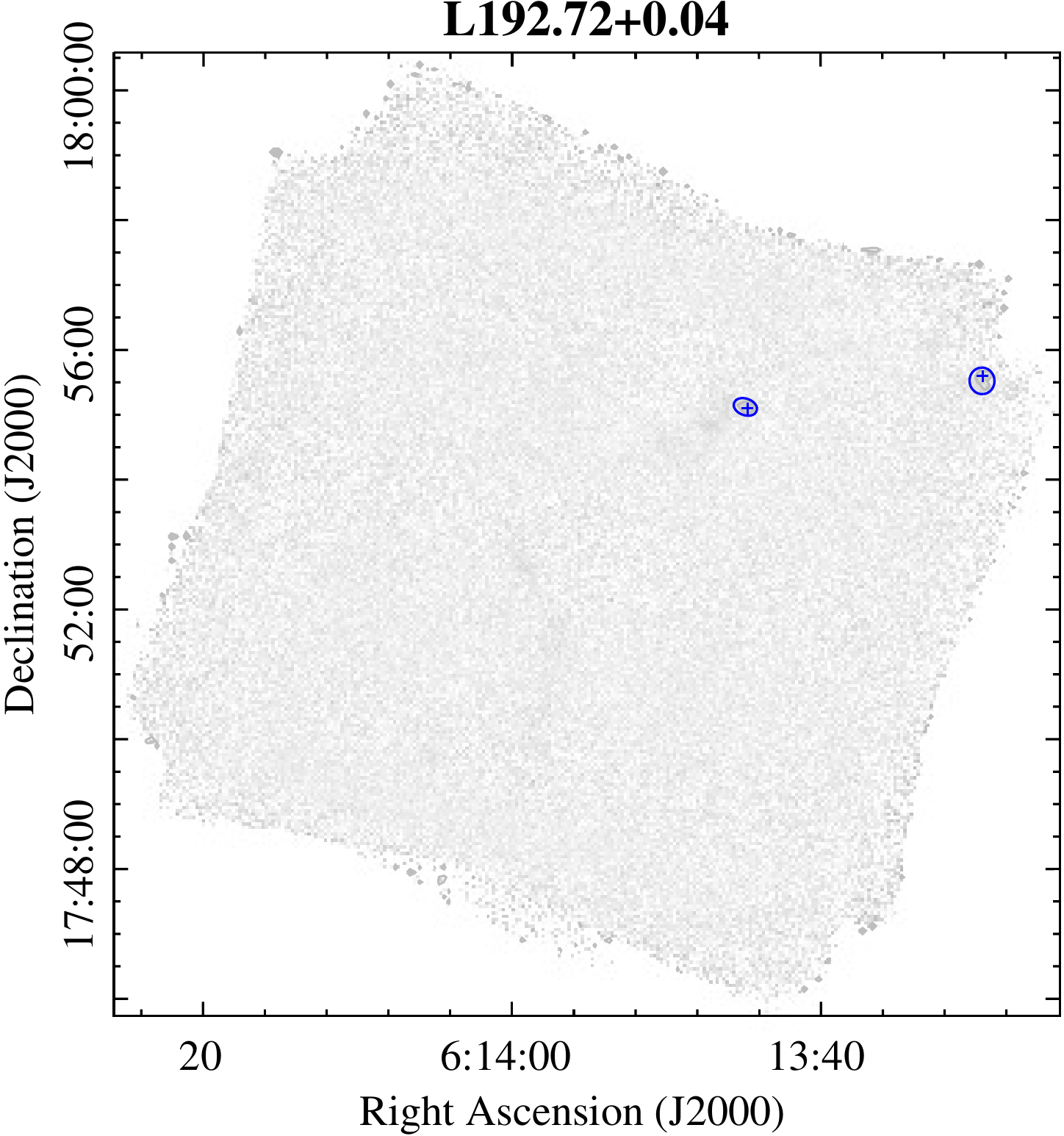}
}\\
\caption{Continuation}
\end{figure}

\clearpage
\begin{figure}\ContinuedFloat 
\center
\subfloat[L192.98+0.14 map, $\sigma_{rms}=383$ mJy beam$^{-1}$.]{
\includegraphics[scale=0.43]{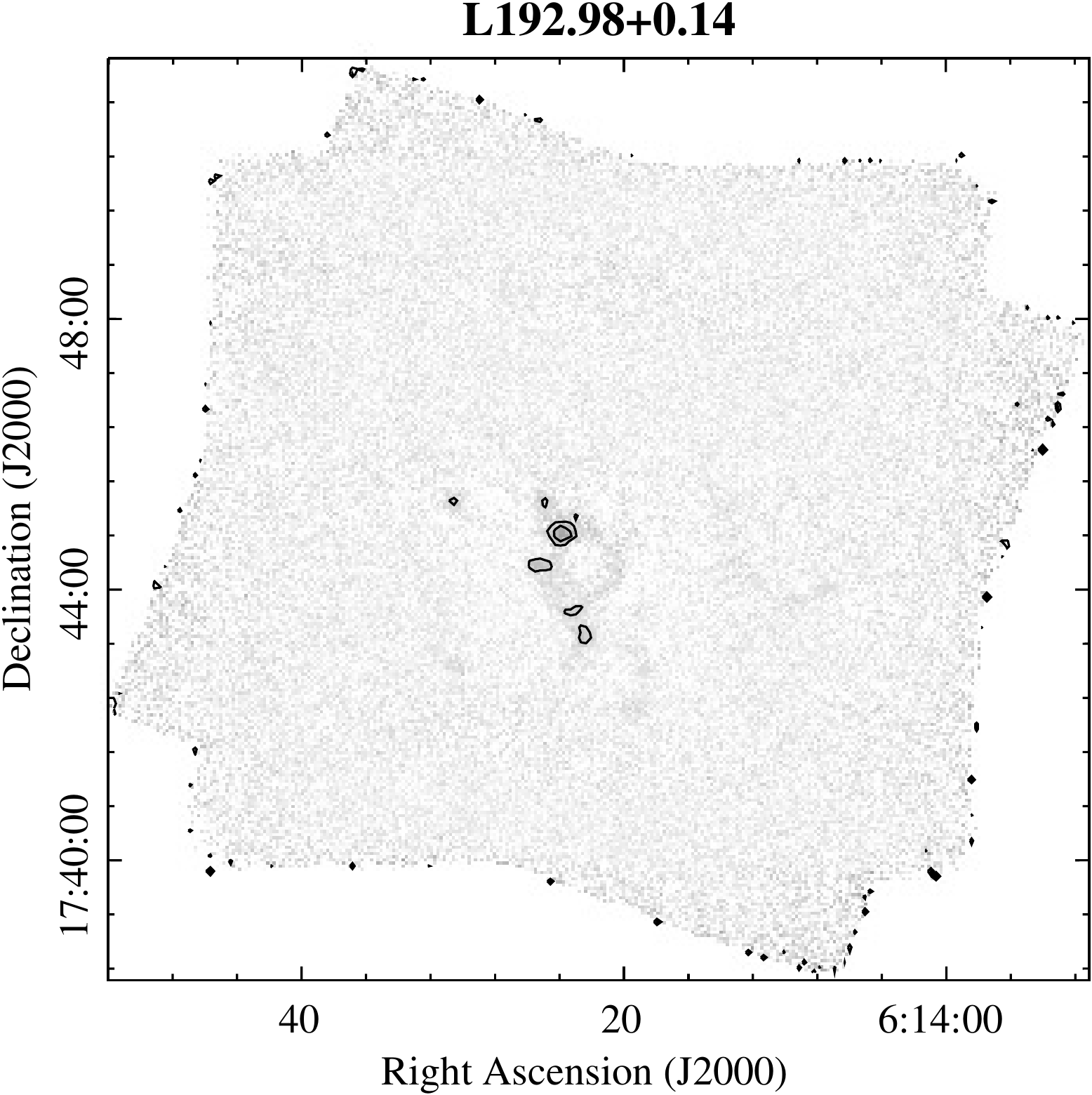}
\includegraphics[scale=0.43]{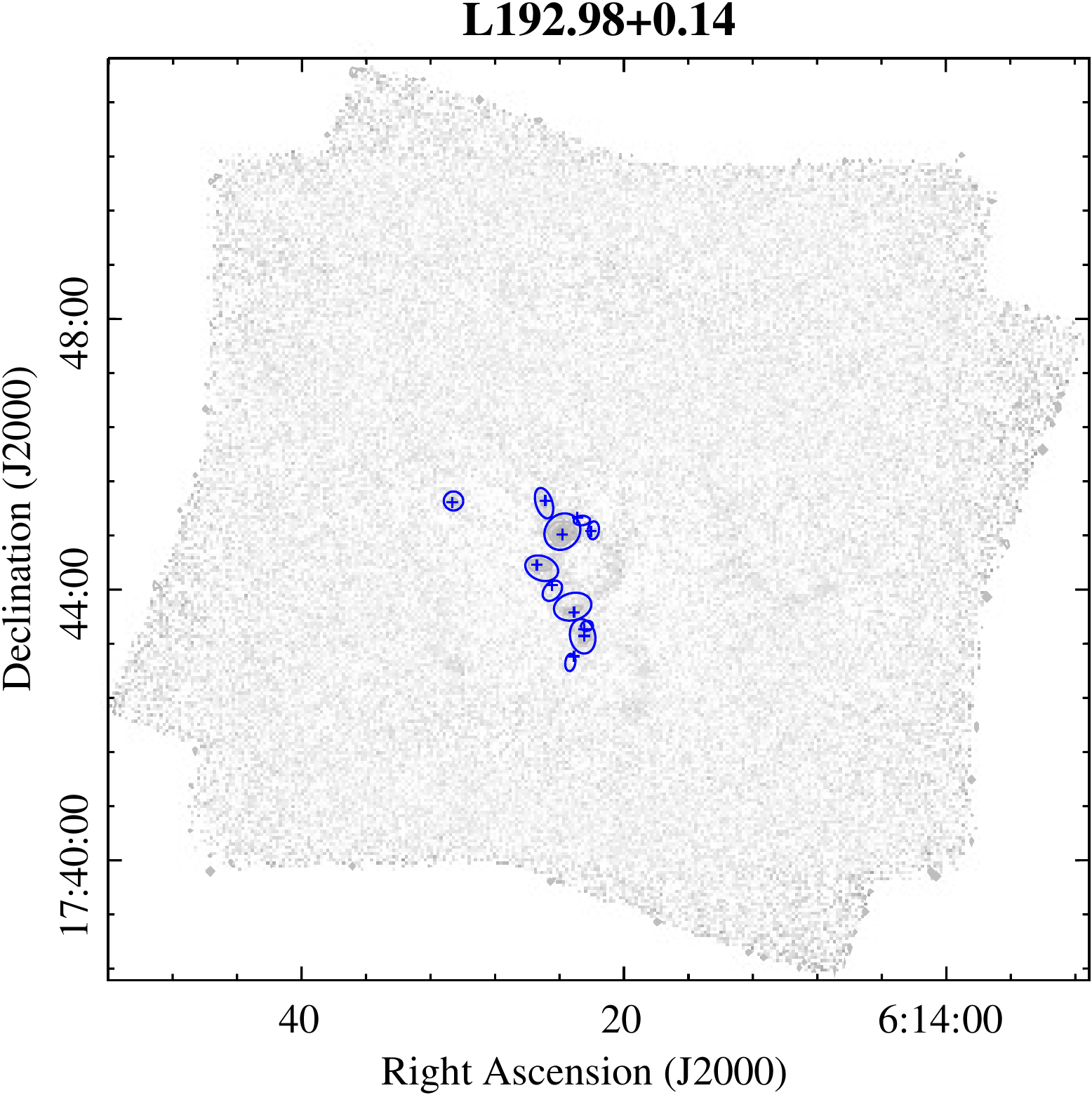}
}\\
\subfloat[L196.42-1.66 map, $\sigma_{rms}=470$ mJy beam$^{-1}$.]{
\includegraphics[scale=0.43]{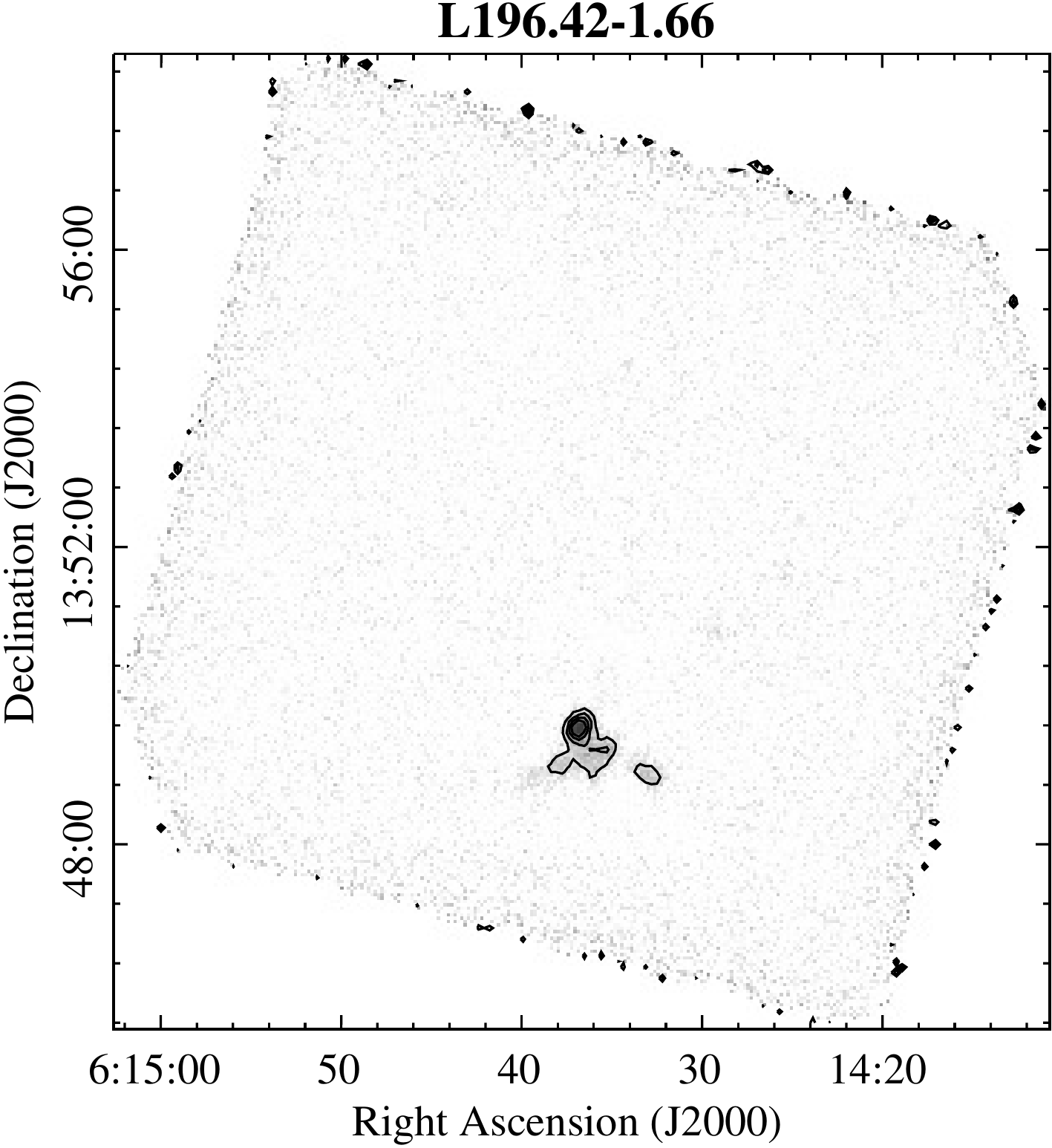}
\includegraphics[scale=0.43]{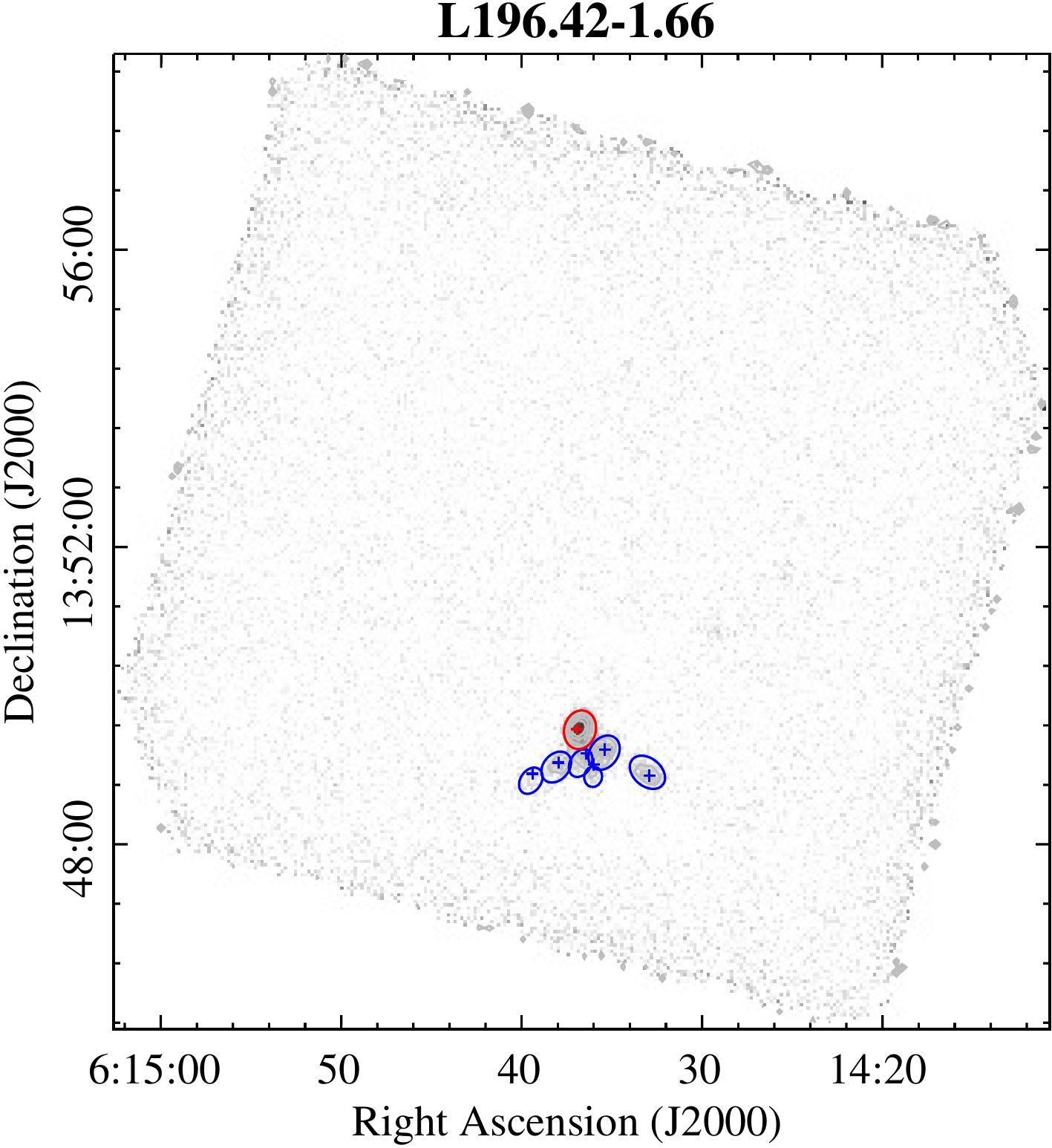}
}\\
\subfloat[L203.23+2.06 map, $\sigma_{rms}=269$ mJy beam$^{-1}$.]{
\includegraphics[scale=0.43]{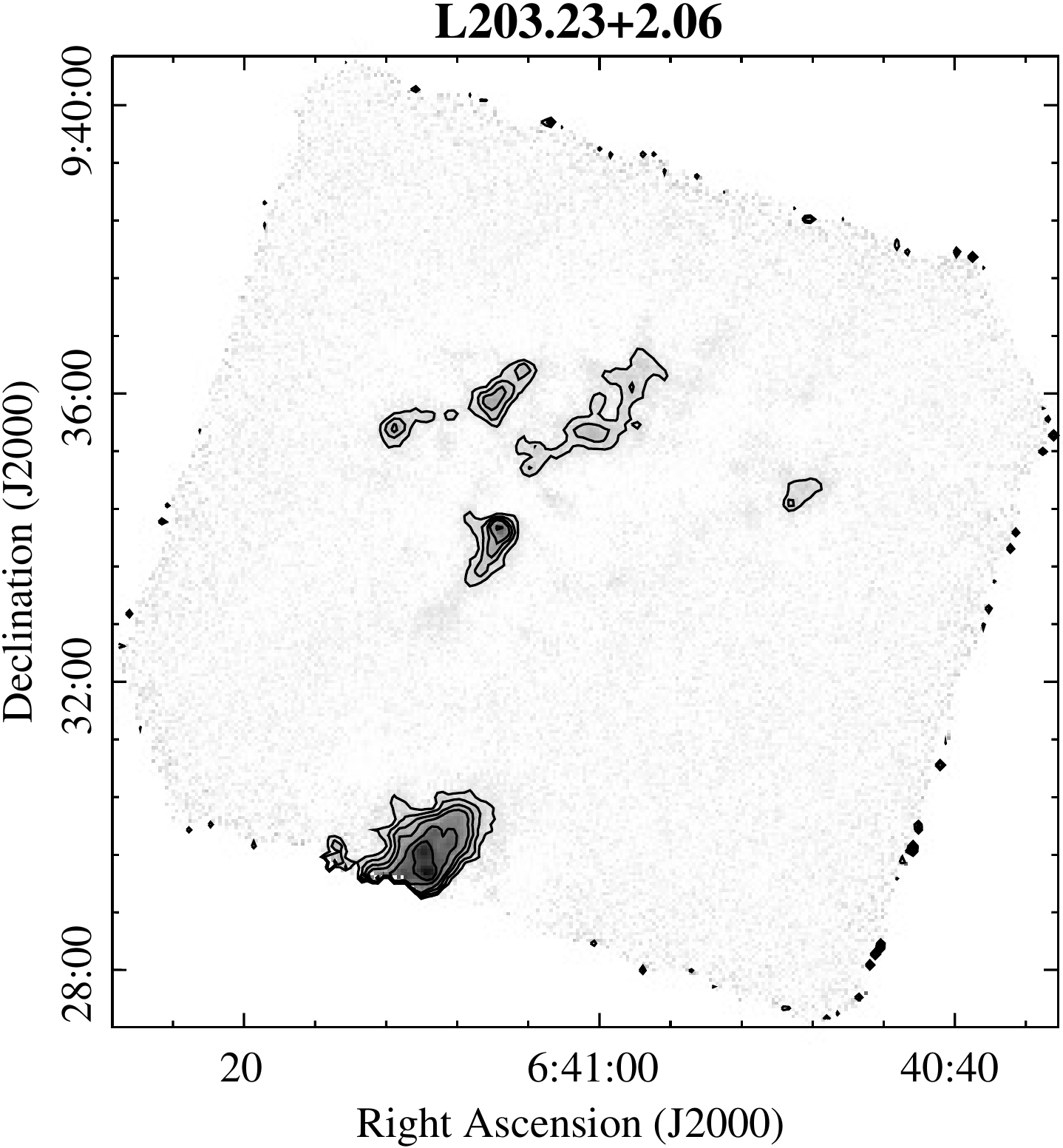}
\includegraphics[scale=0.43]{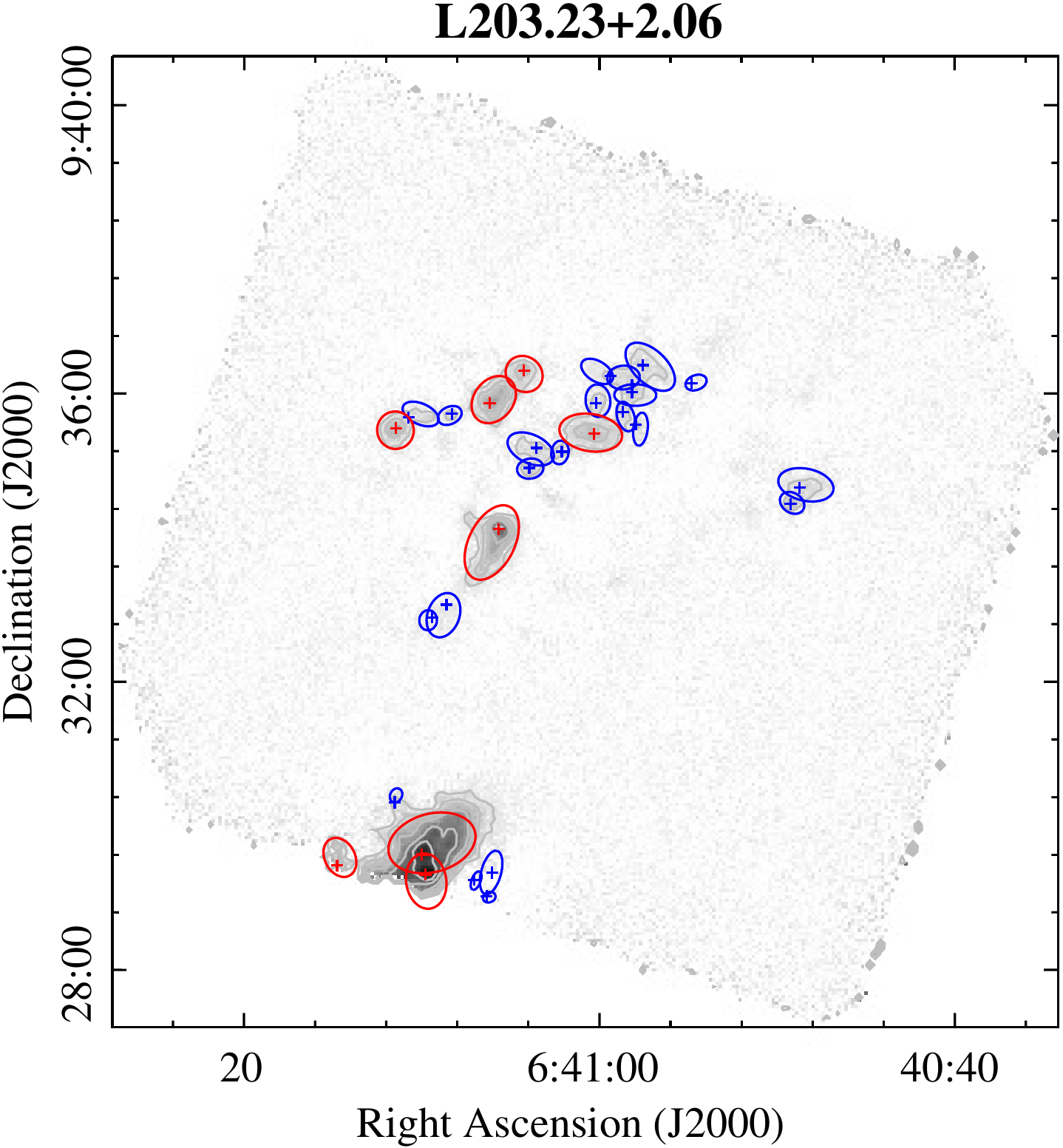}
}\\
\caption{Continuation}
\end{figure}

\clearpage
\begin{figure}\ContinuedFloat 
\center
\subfloat[L203.35+2.03 map, $\sigma_{rms}=314$ mJy beam$^{-1}$.]{
\includegraphics[scale=0.43]{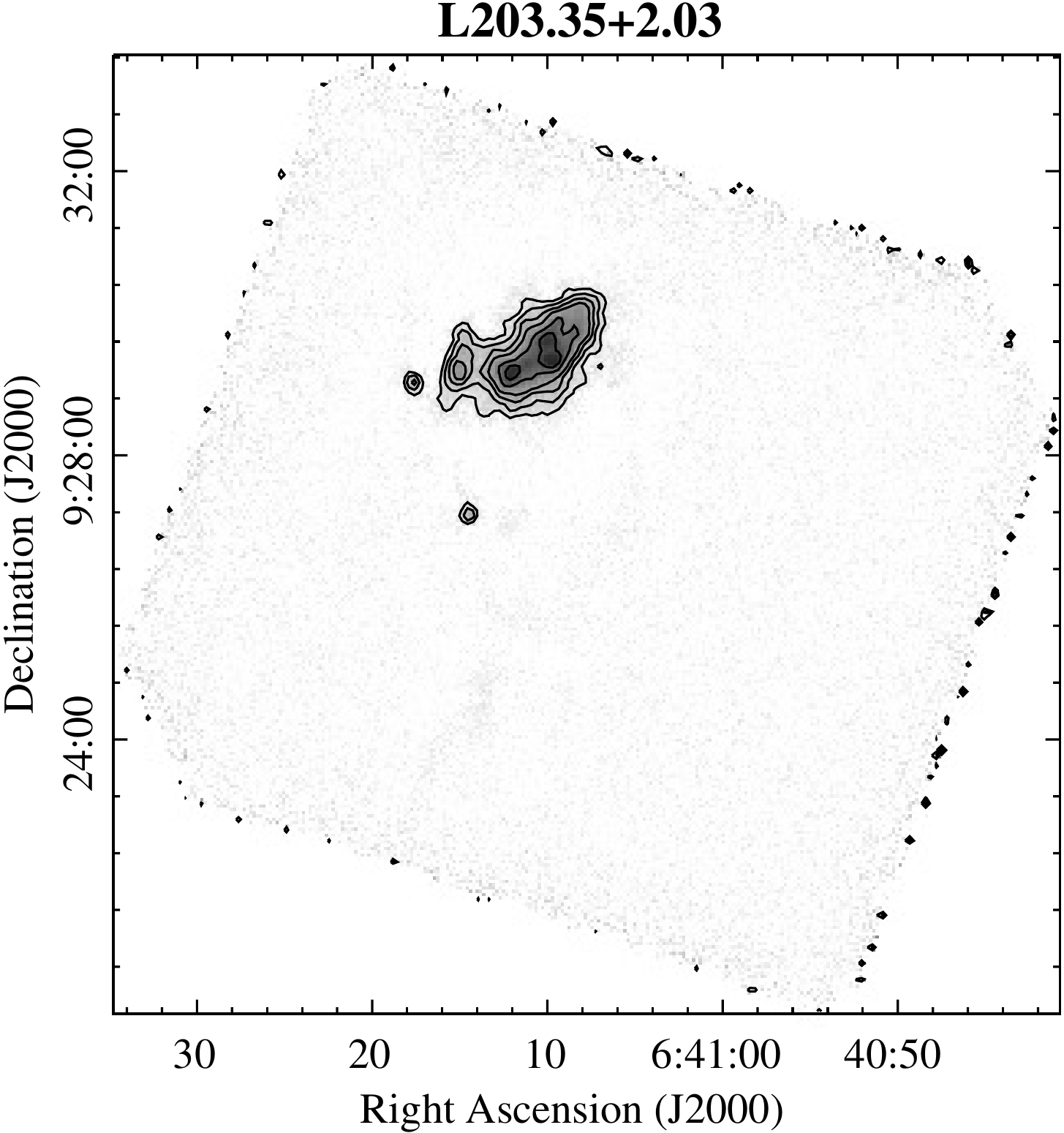}
\includegraphics[scale=0.43]{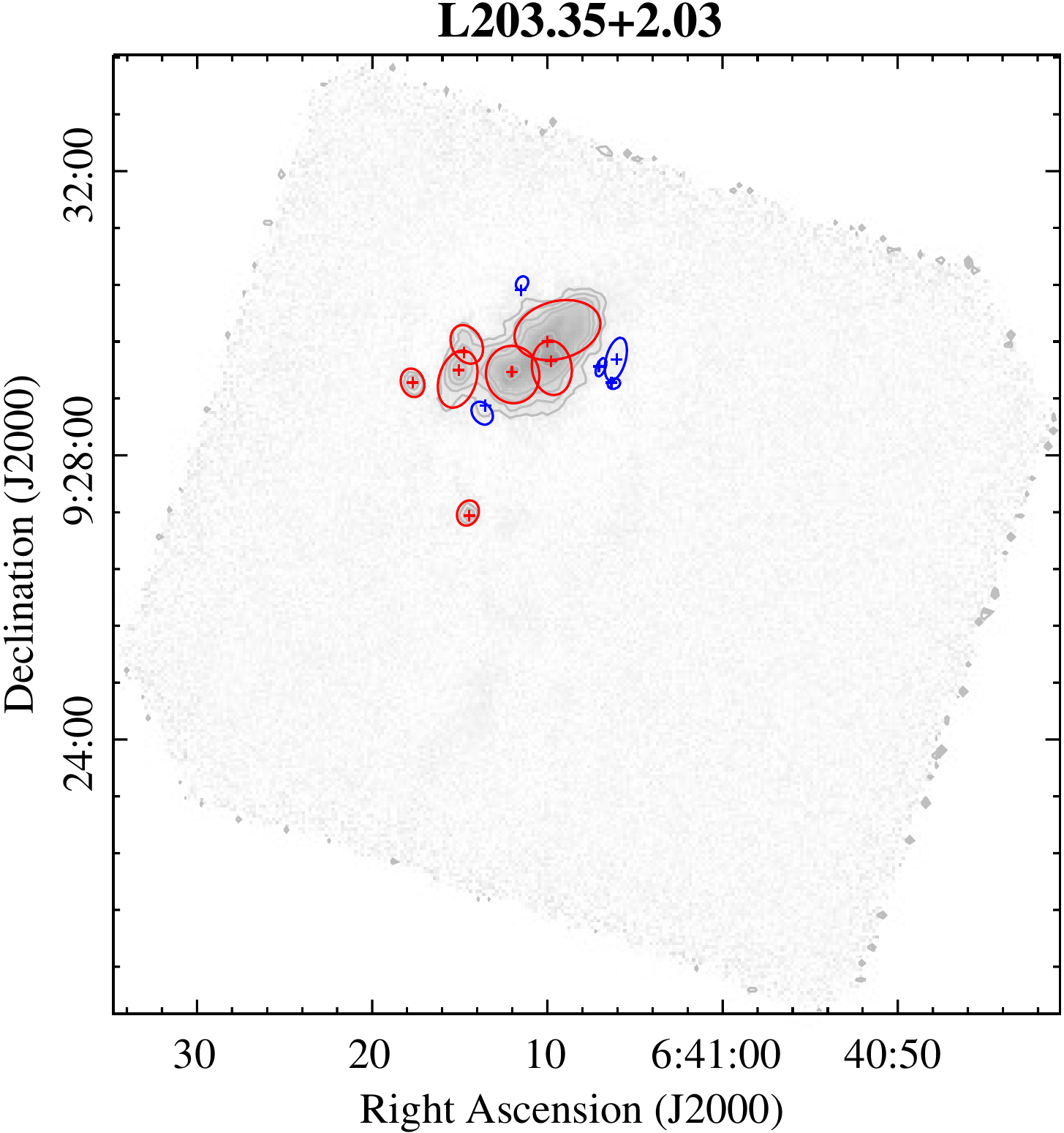}
}\\
\subfloat[L213.71-12.62 map, $\sigma_{rms}=421$ mJy beam$^{-1}$.]{
\includegraphics[scale=0.43]{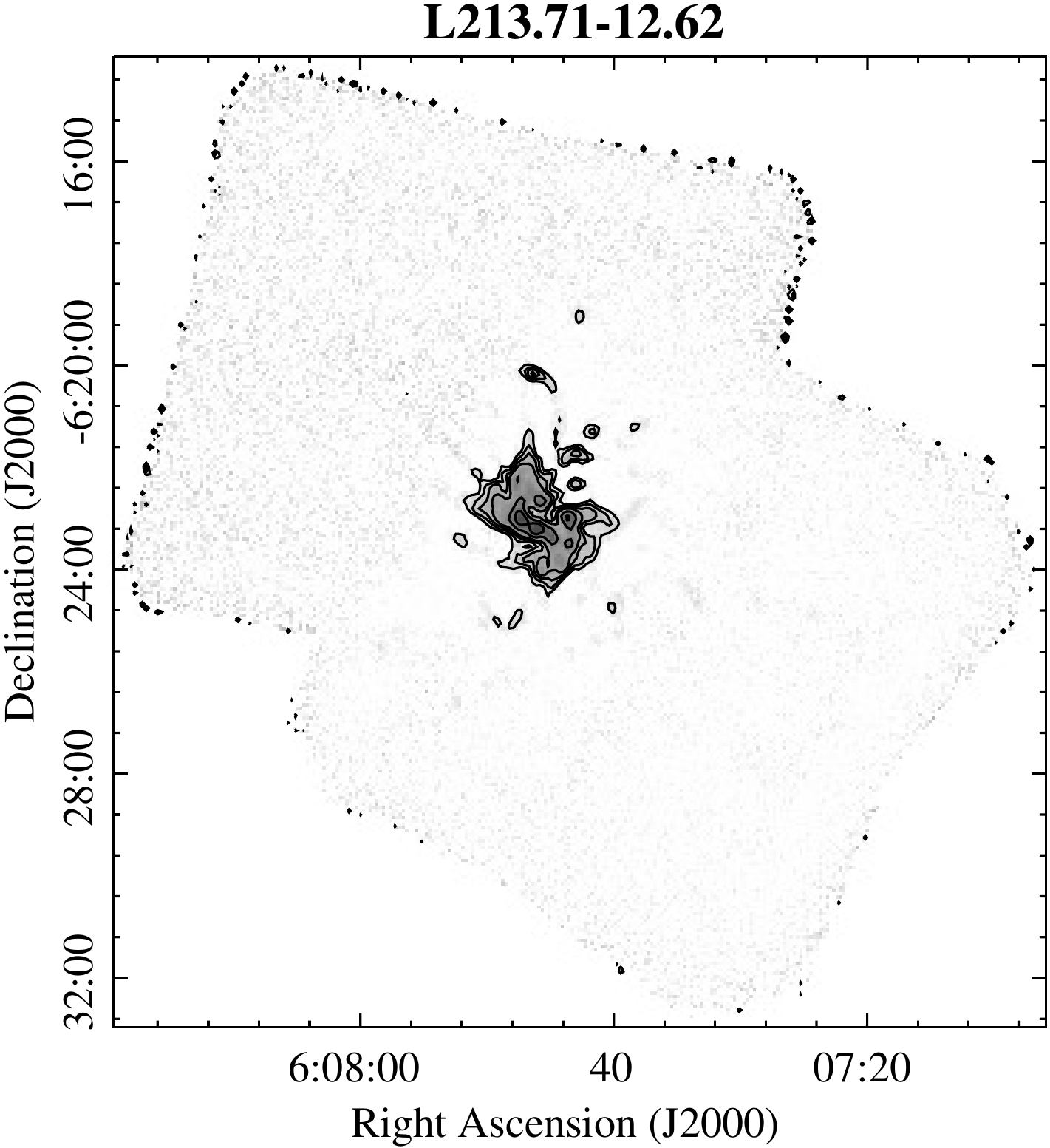}
\includegraphics[scale=0.43]{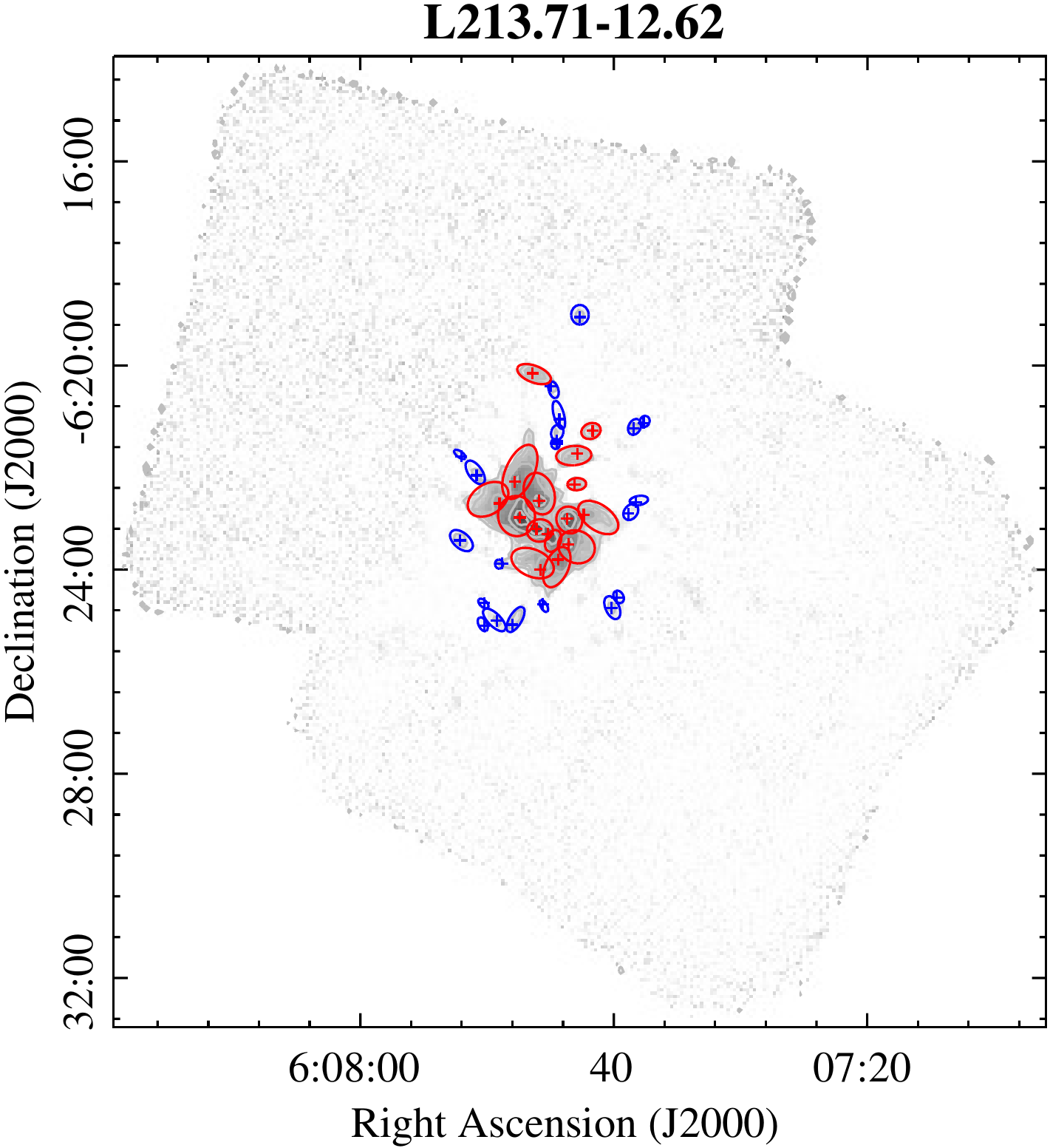}
}\\
\subfloat[L217.37-0.07 map, $\sigma_{rms}=704$ mJy beam$^{-1}$.]{
\includegraphics[scale=0.43]{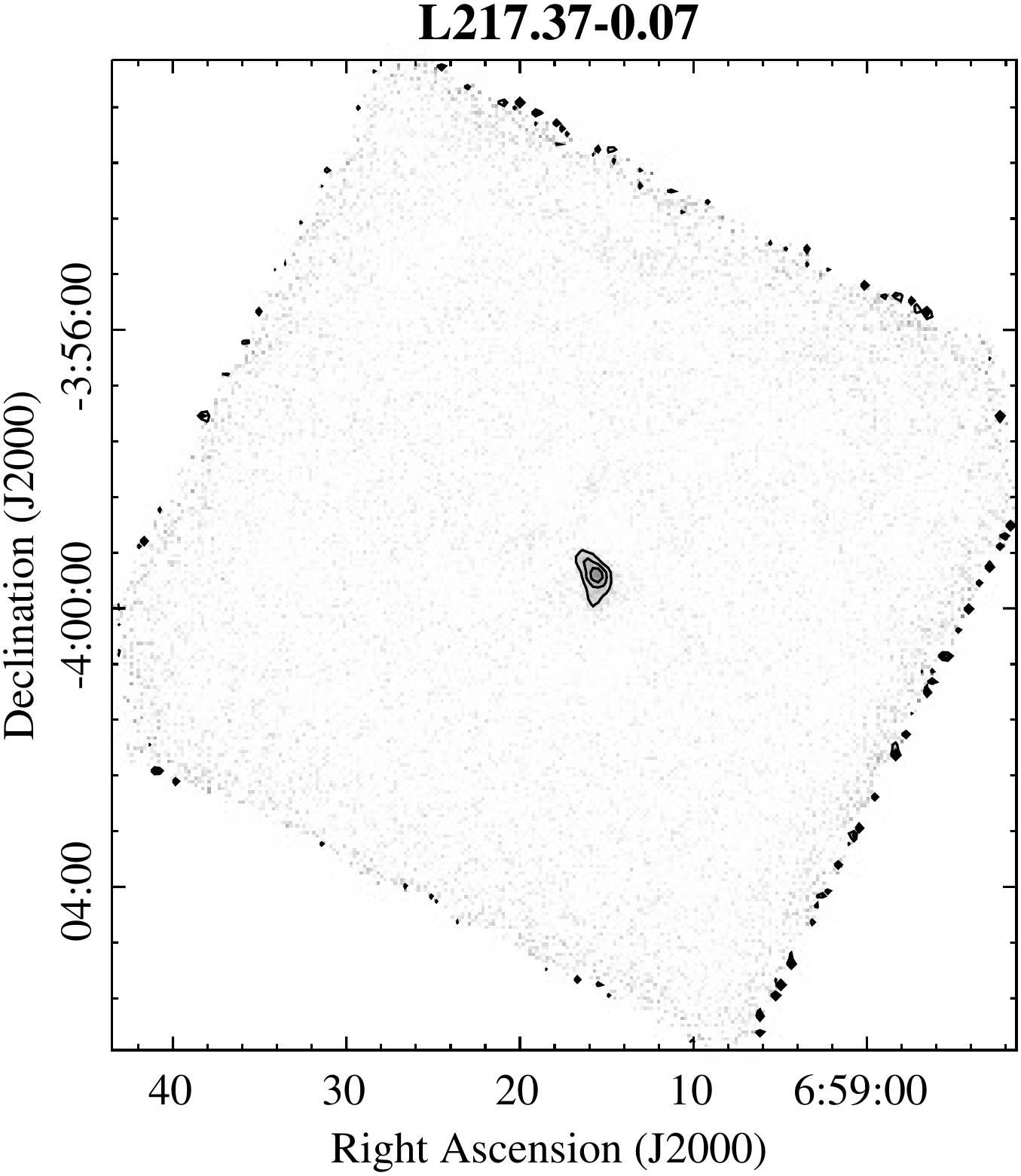}
\includegraphics[scale=0.43]{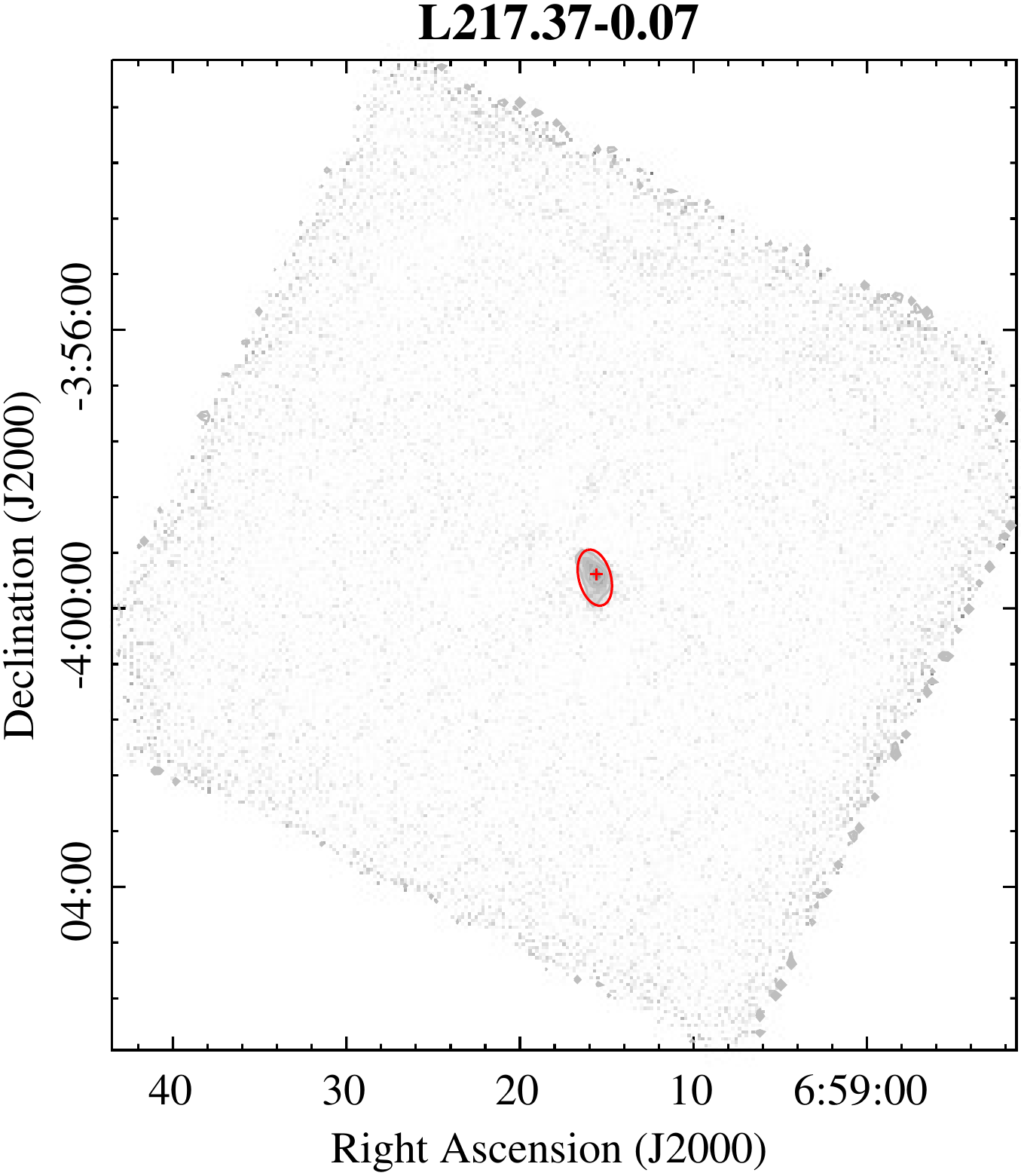}
}\\
\caption{Continuation}
\end{figure}

\clearpage
\begin{figure}\ContinuedFloat 
\center
\subfloat[L234.57+0.82 map, $\sigma_{rms}=1051$ mJy beam$^{-1}$.]{
\includegraphics[scale=0.43]{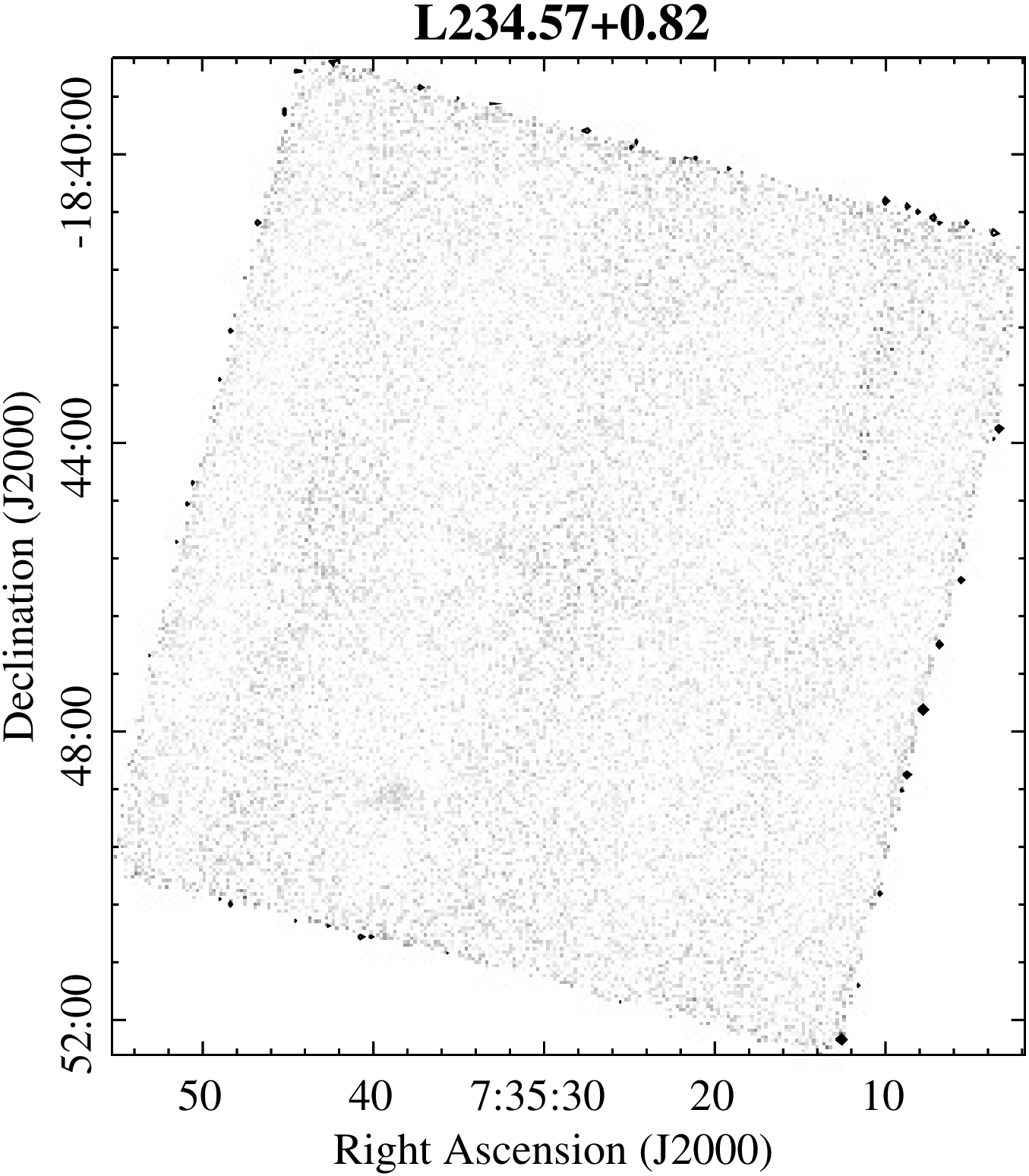}
\includegraphics[scale=0.43]{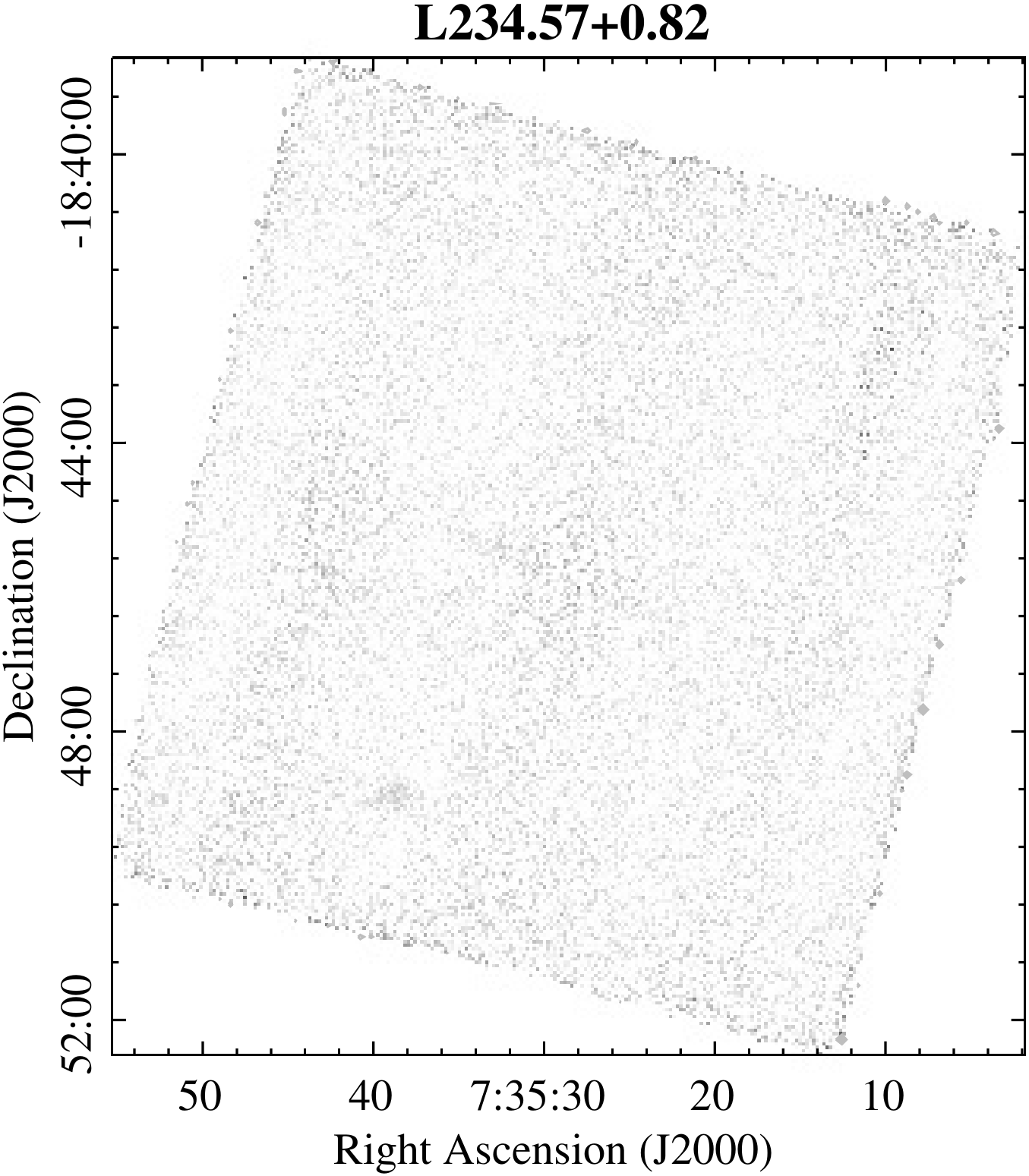}
}\\
\subfloat[L001.10-0.07 map, $\sigma_{rms}=849$ mJy beam$^{-1}$.]{
\includegraphics[scale=0.43]{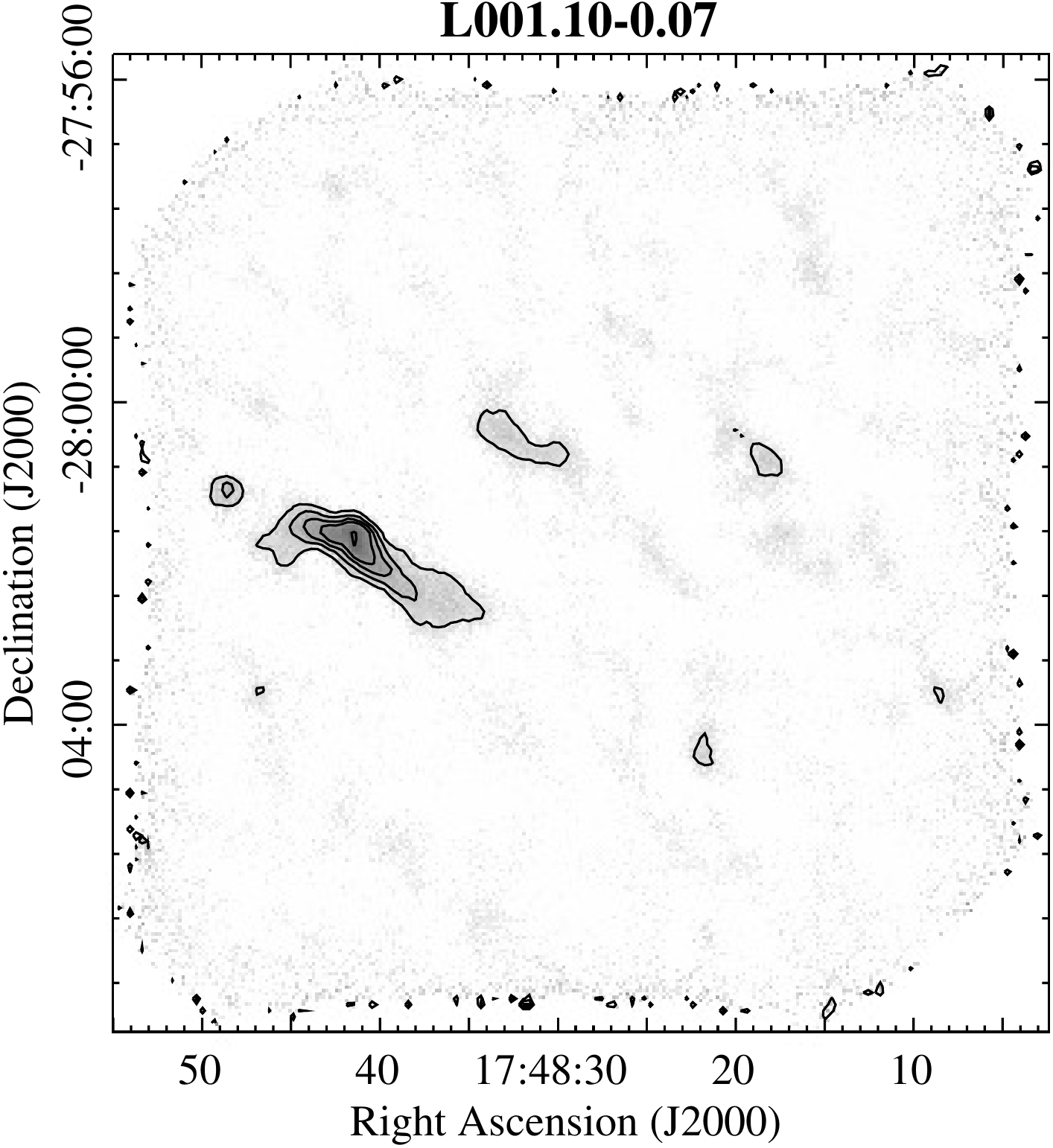}
\includegraphics[scale=0.43]{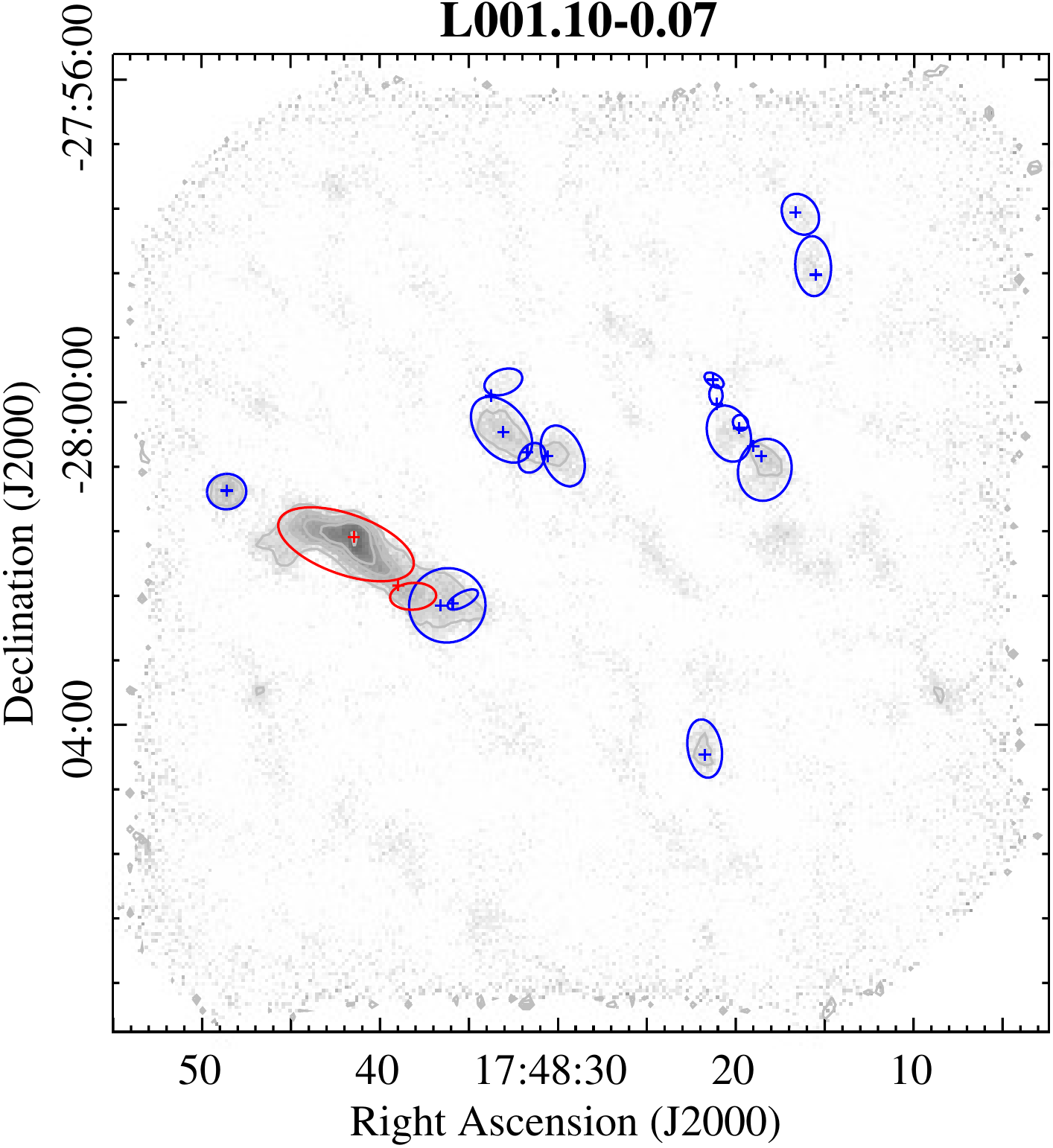}
}\\
\subfloat[L023.31-0.26 map, $\sigma_{rms}=379$ mJy beam$^{-1}$.]{
\includegraphics[scale=0.43]{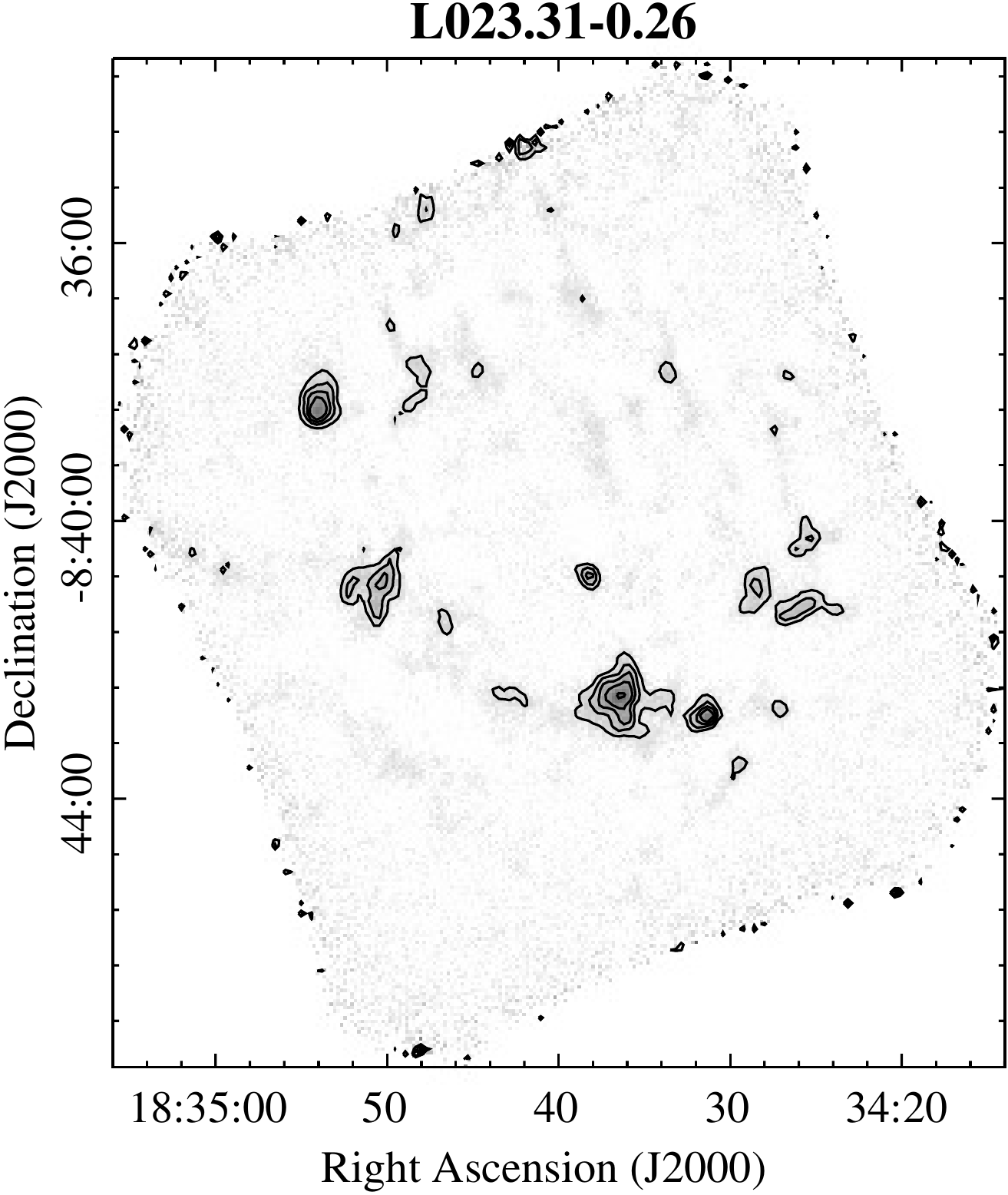}
\includegraphics[scale=0.43]{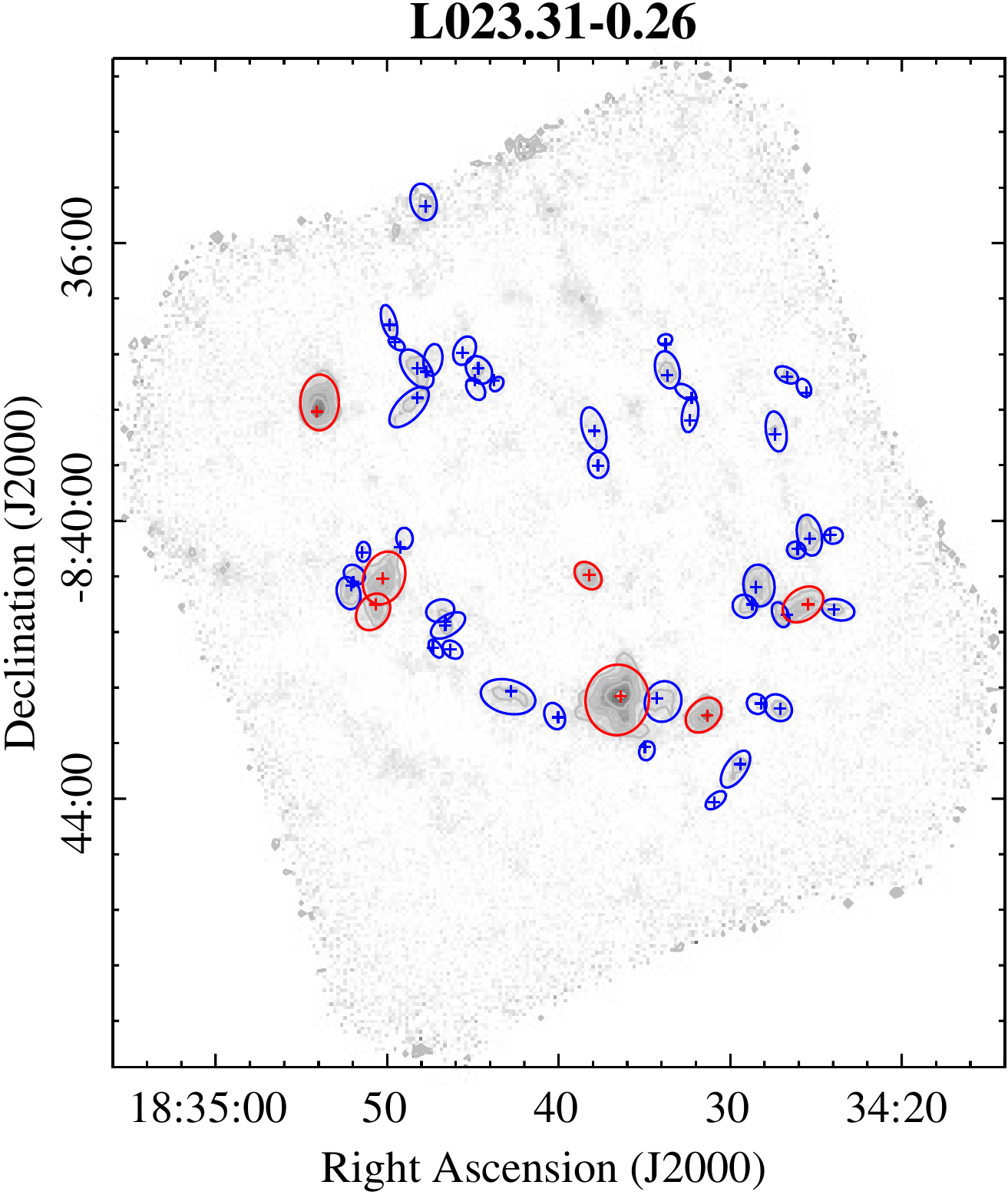}
}\\
\caption{Continuation}
\end{figure}

\clearpage
\begin{figure}\ContinuedFloat 
\center
\subfloat[L023.43-0.22 map, $\sigma_{rms}=375$ mJy beam$^{-1}$.]{
\includegraphics[scale=0.43]{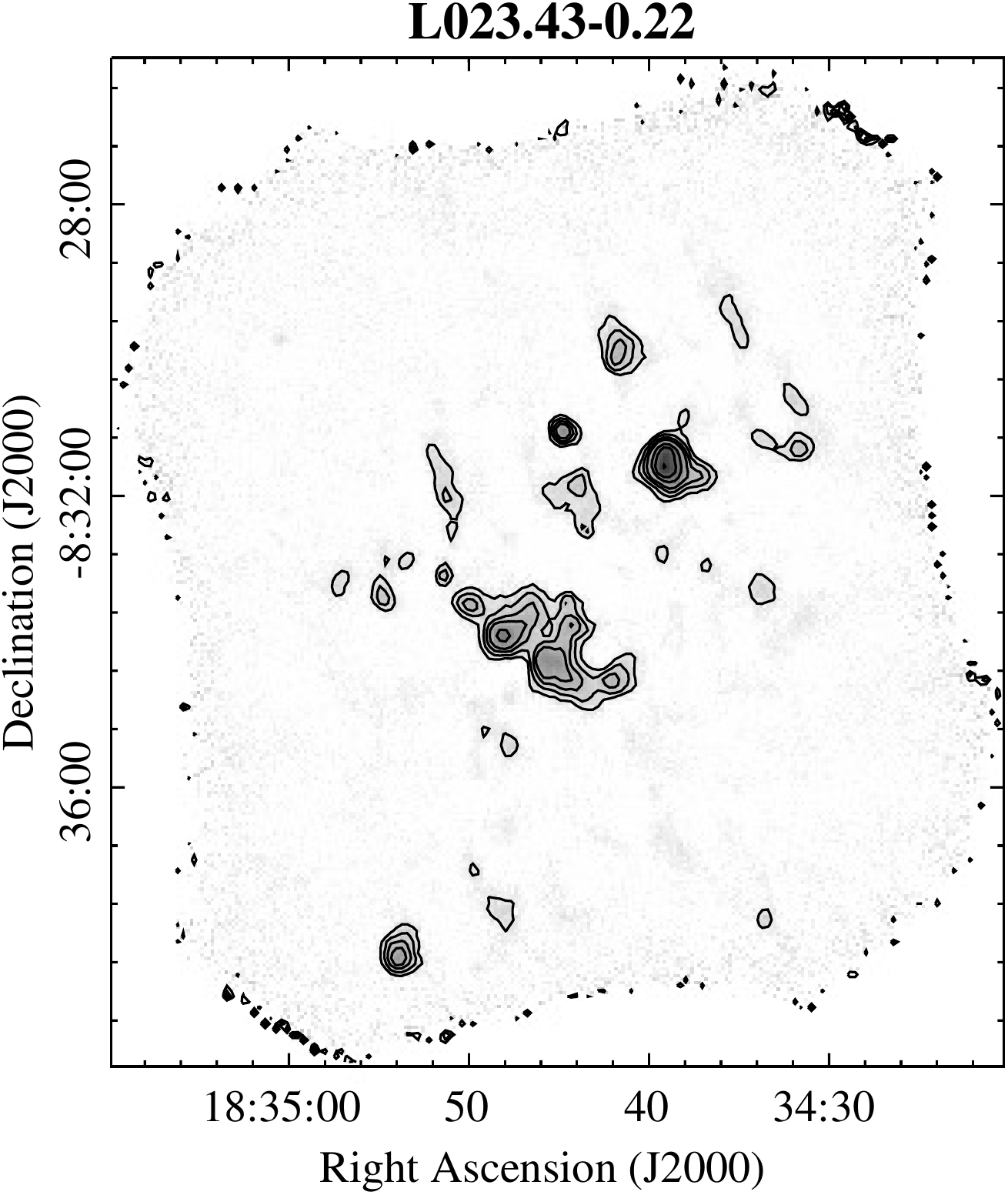}
\includegraphics[scale=0.43]{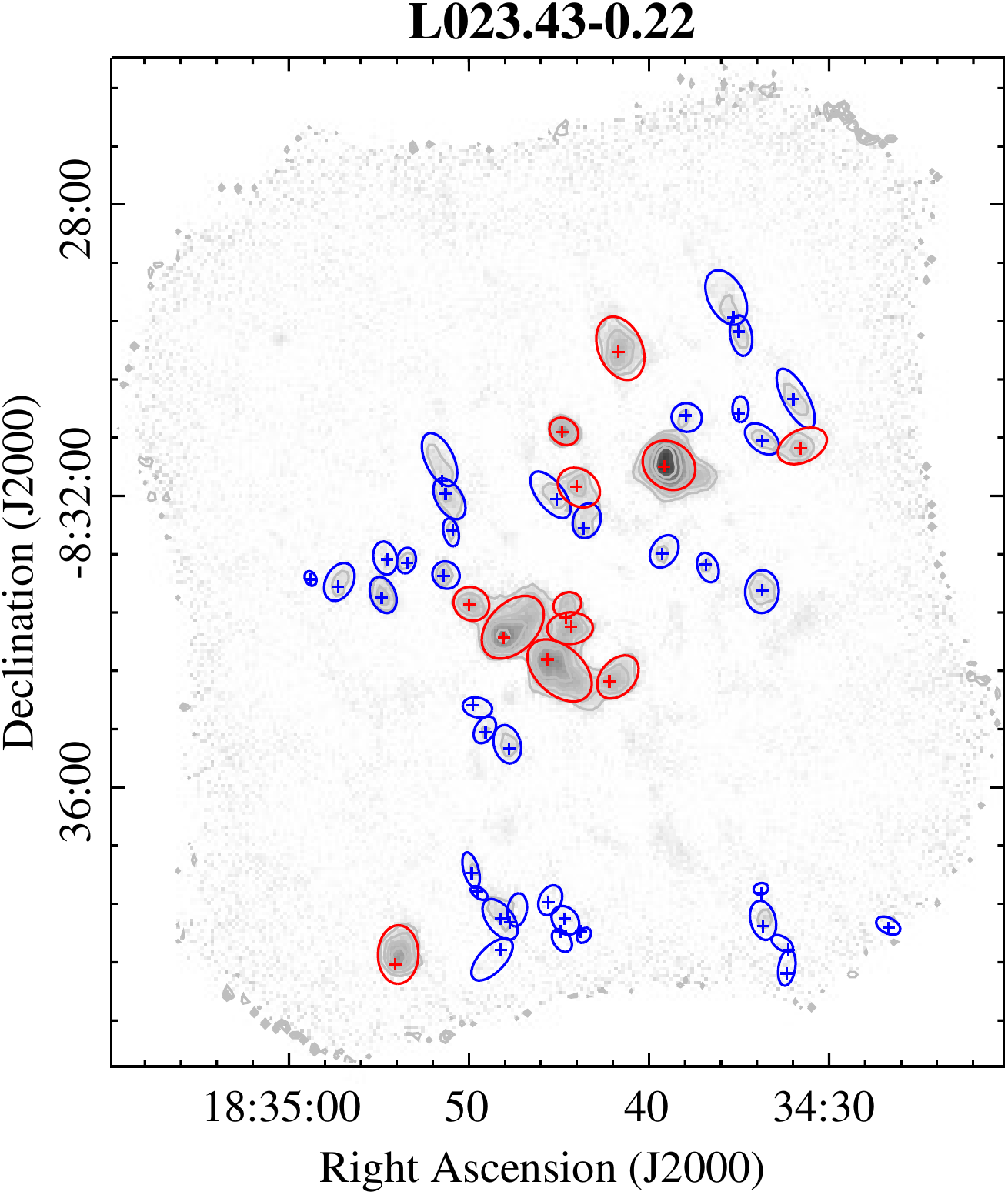}
}\\
\subfloat[L024.50-0.08 map, $\sigma_{rms}=306$ mJy beam$^{-1}$.]{
\includegraphics[scale=0.43]{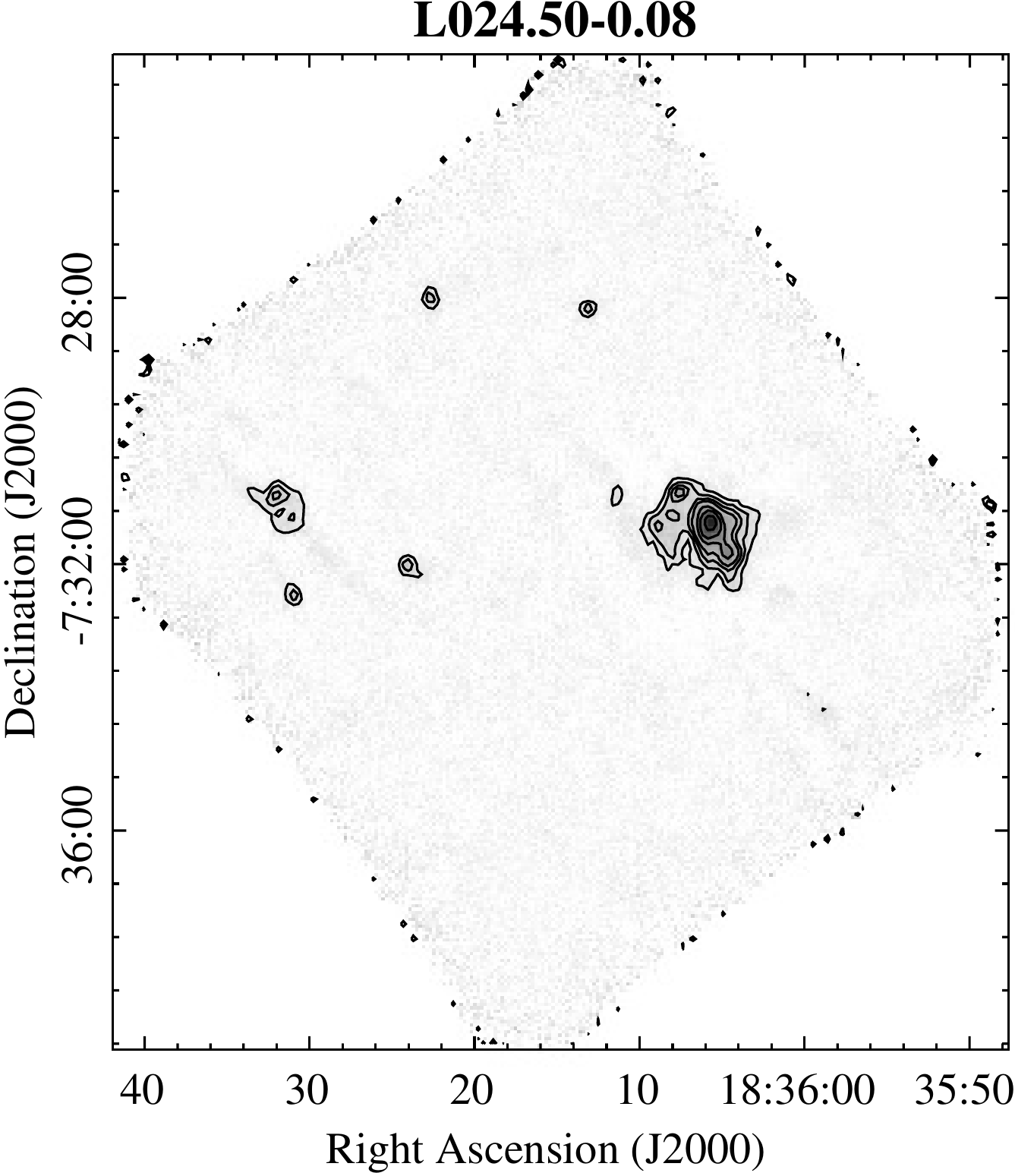}
\includegraphics[scale=0.43]{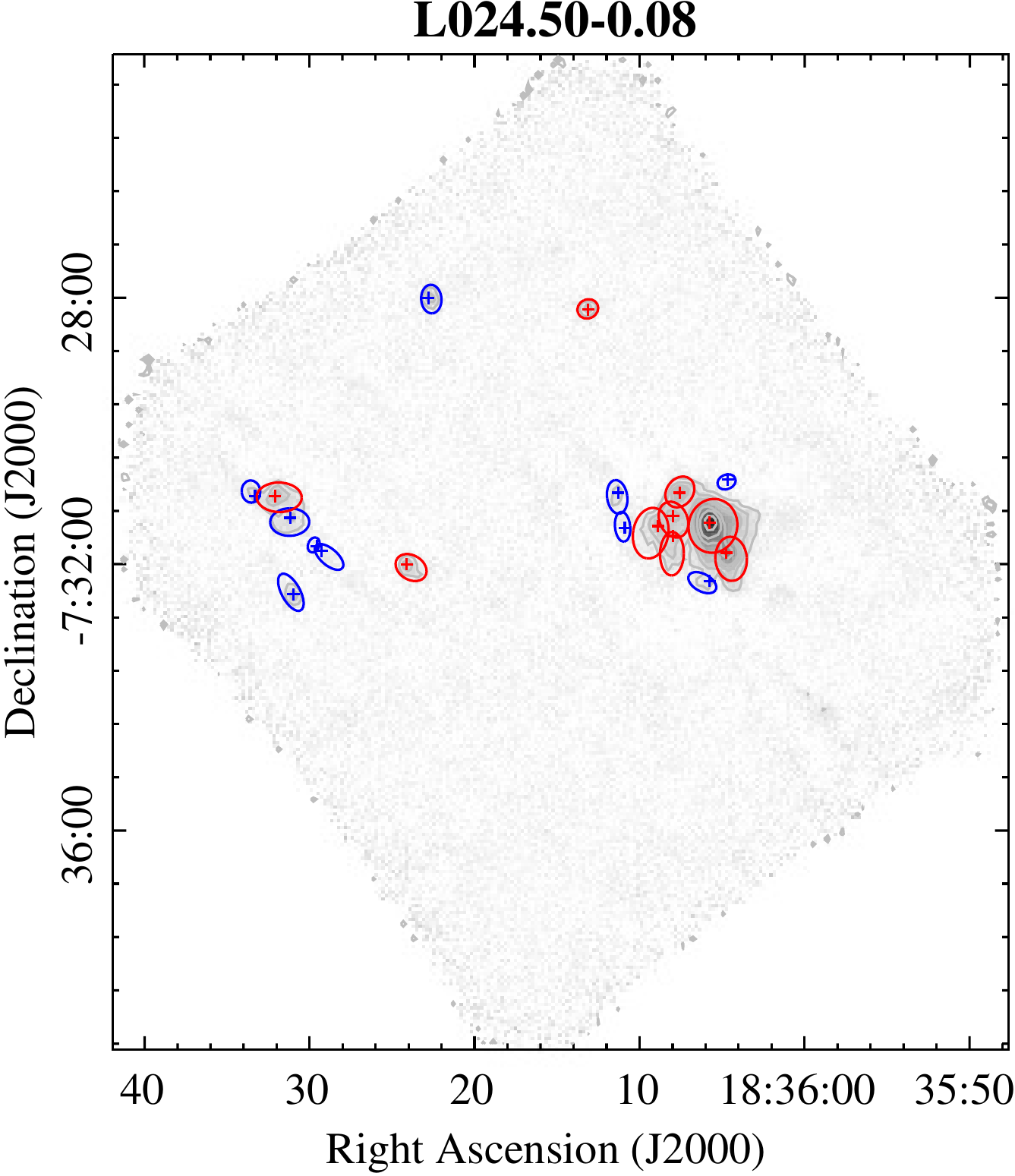}
}\\
\subfloat[L024.65-0.13 map, $\sigma_{rms}=357$ mJy beam$^{-1}$.]{
\includegraphics[scale=0.43]{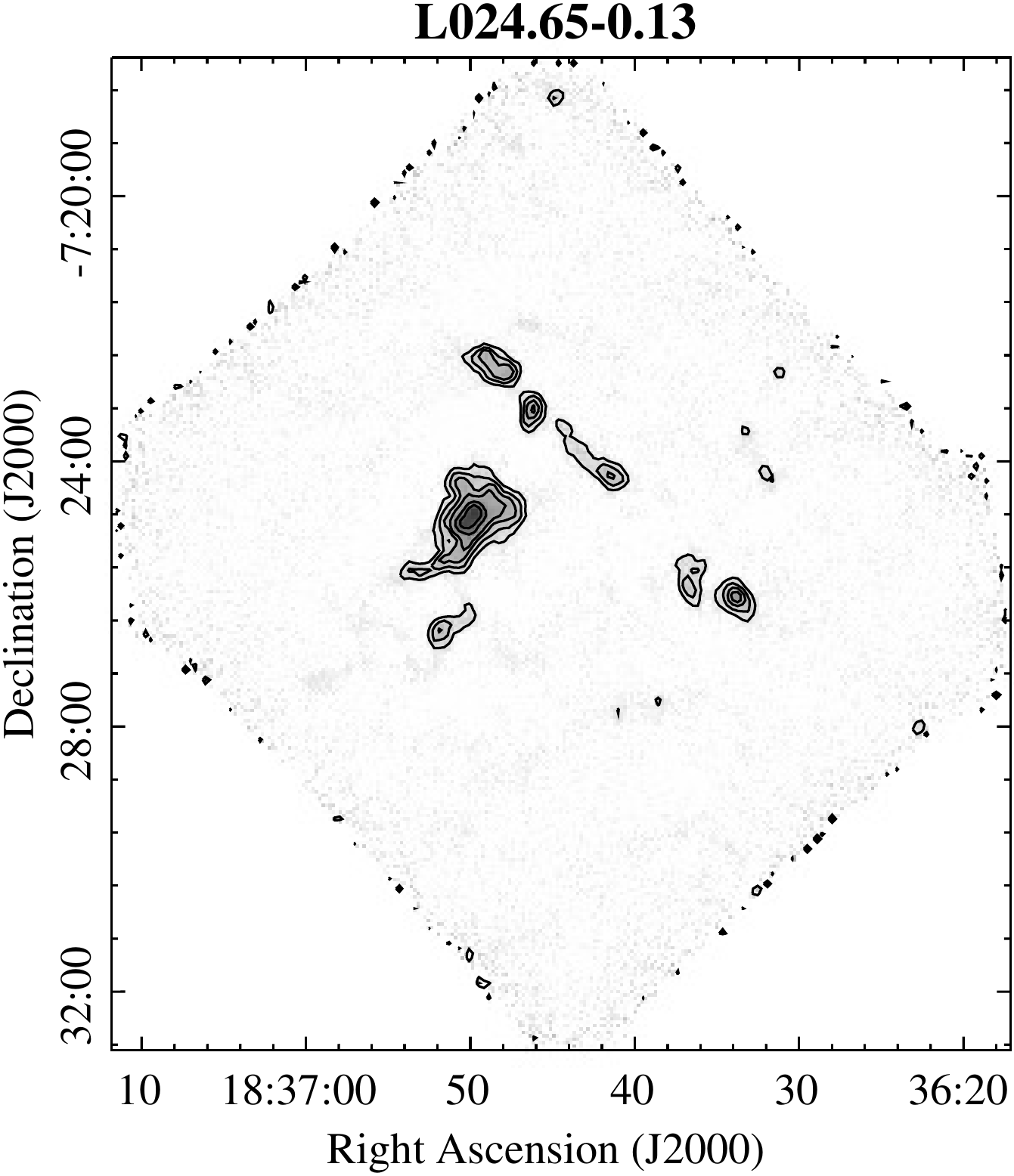}
\includegraphics[scale=0.43]{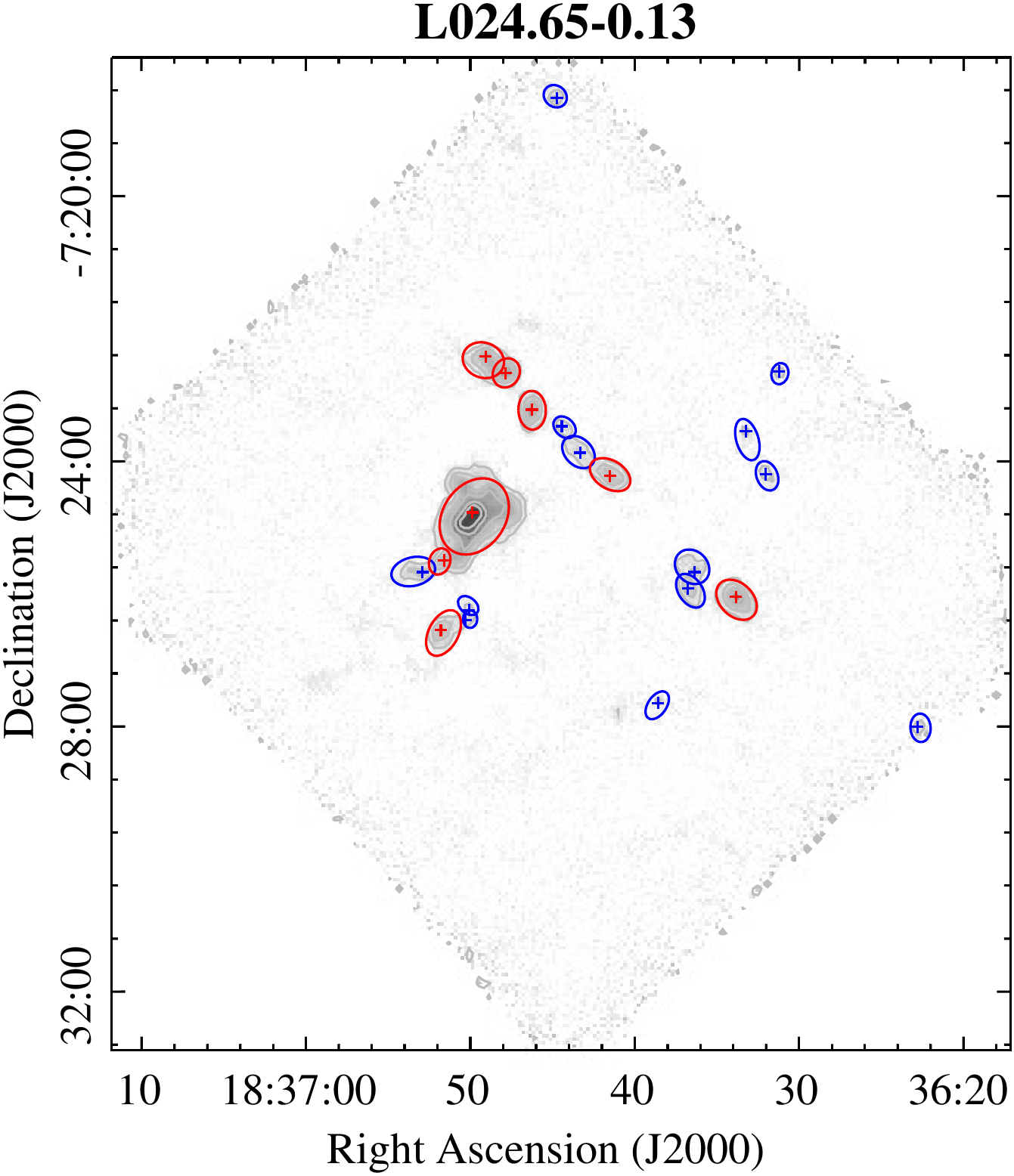}
}\\
\caption{Continuation}
\end{figure}

\clearpage
\begin{figure}\ContinuedFloat 
\center
\subfloat[L024.78+0.12 map, $\sigma_{rms}=363$ mJy beam$^{-1}$.]{
\includegraphics[scale=0.43]{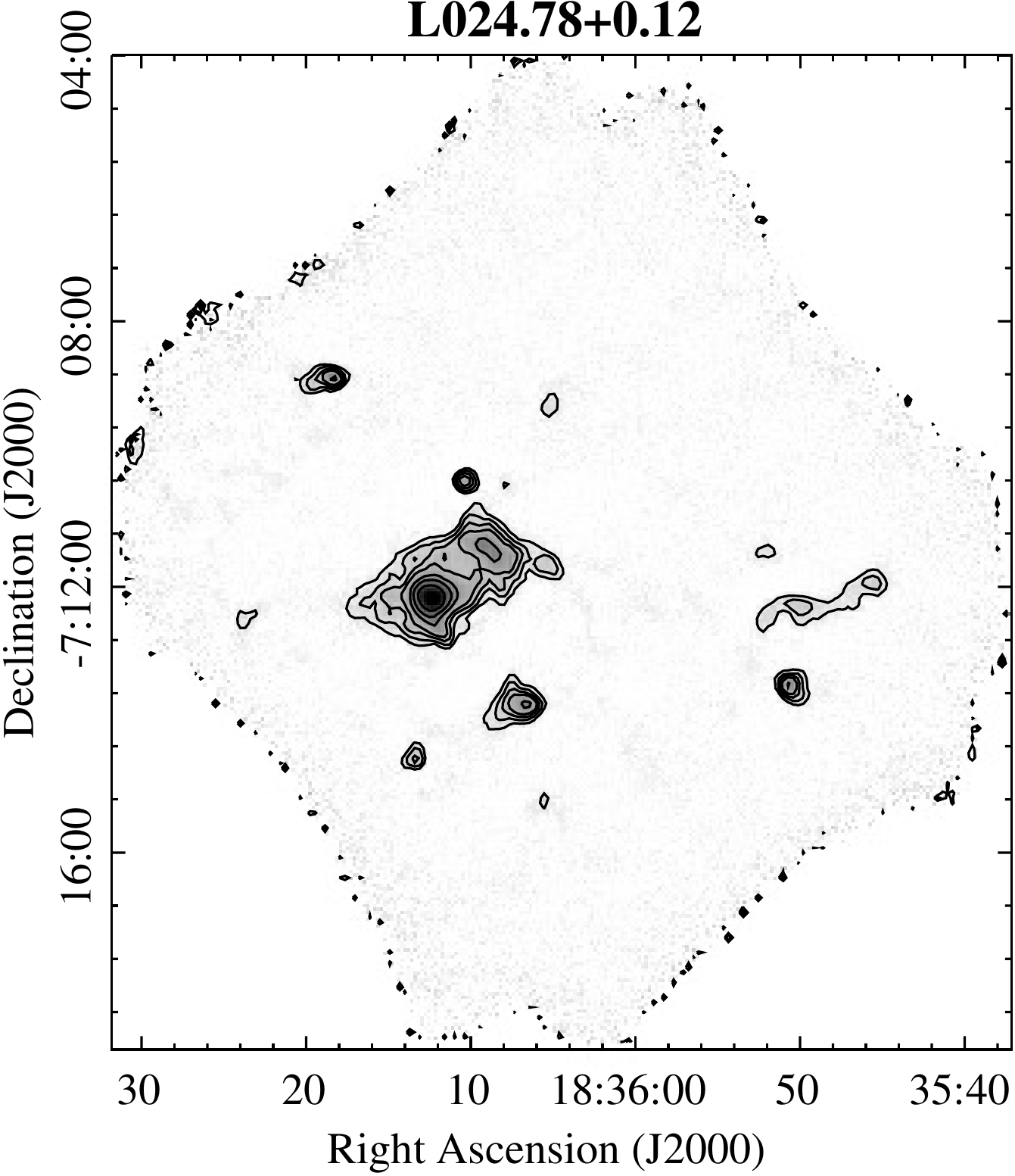}
\includegraphics[scale=0.43]{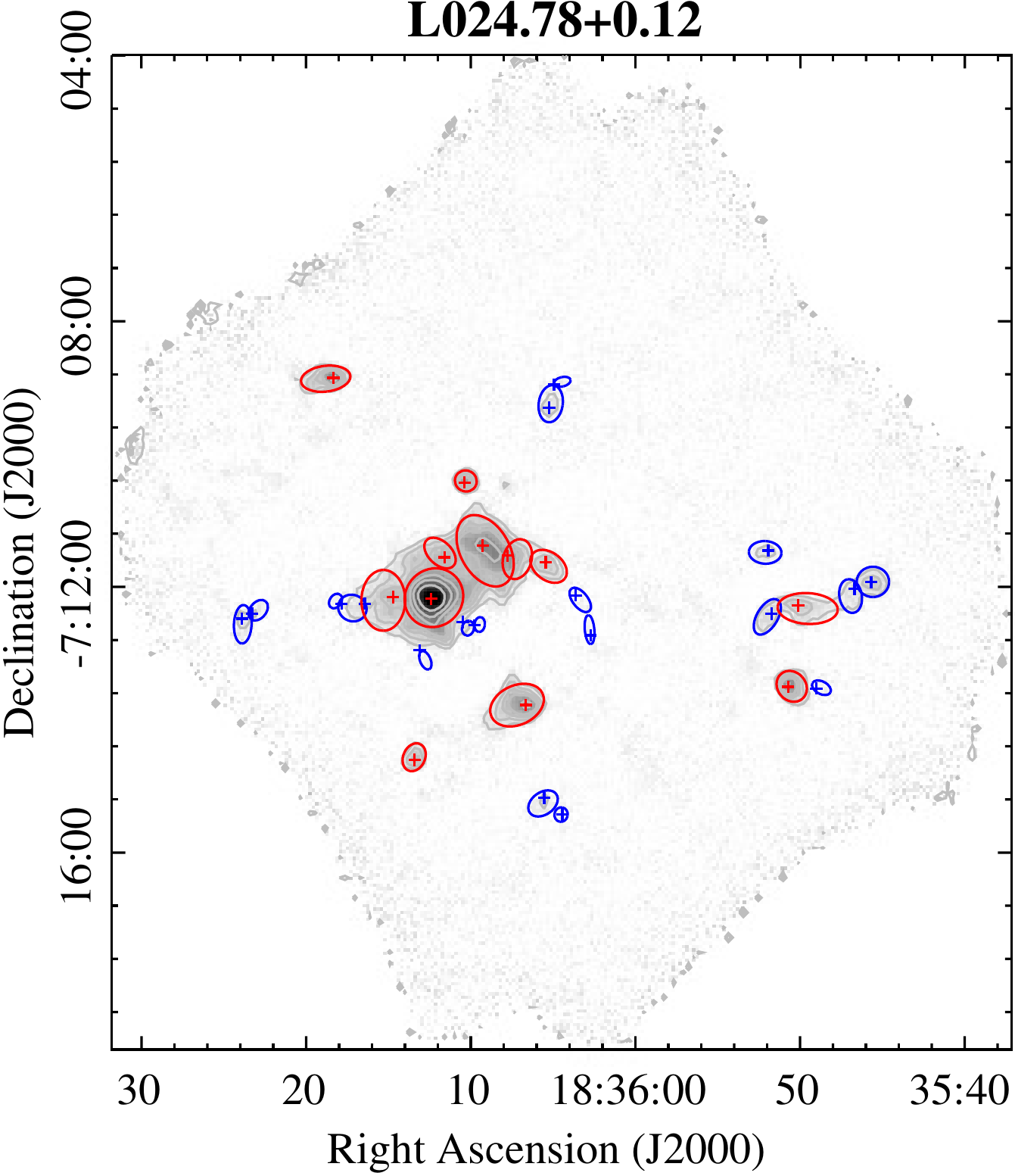}
}\\
\subfloat[L025.40-0.18 map, $\sigma_{rms}=482$ mJy beam$^{-1}$.]{
\includegraphics[scale=0.43]{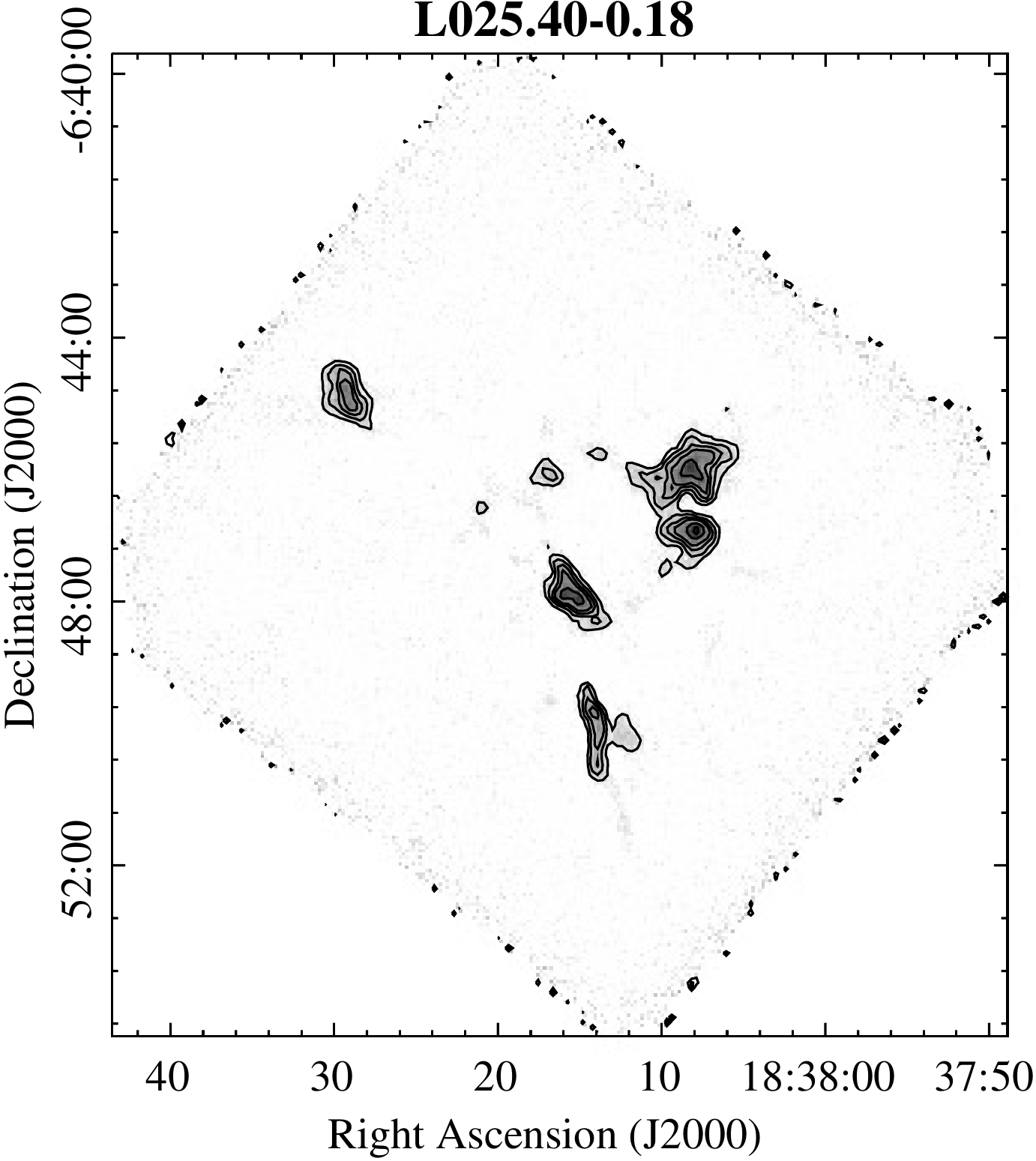}
\includegraphics[scale=0.43]{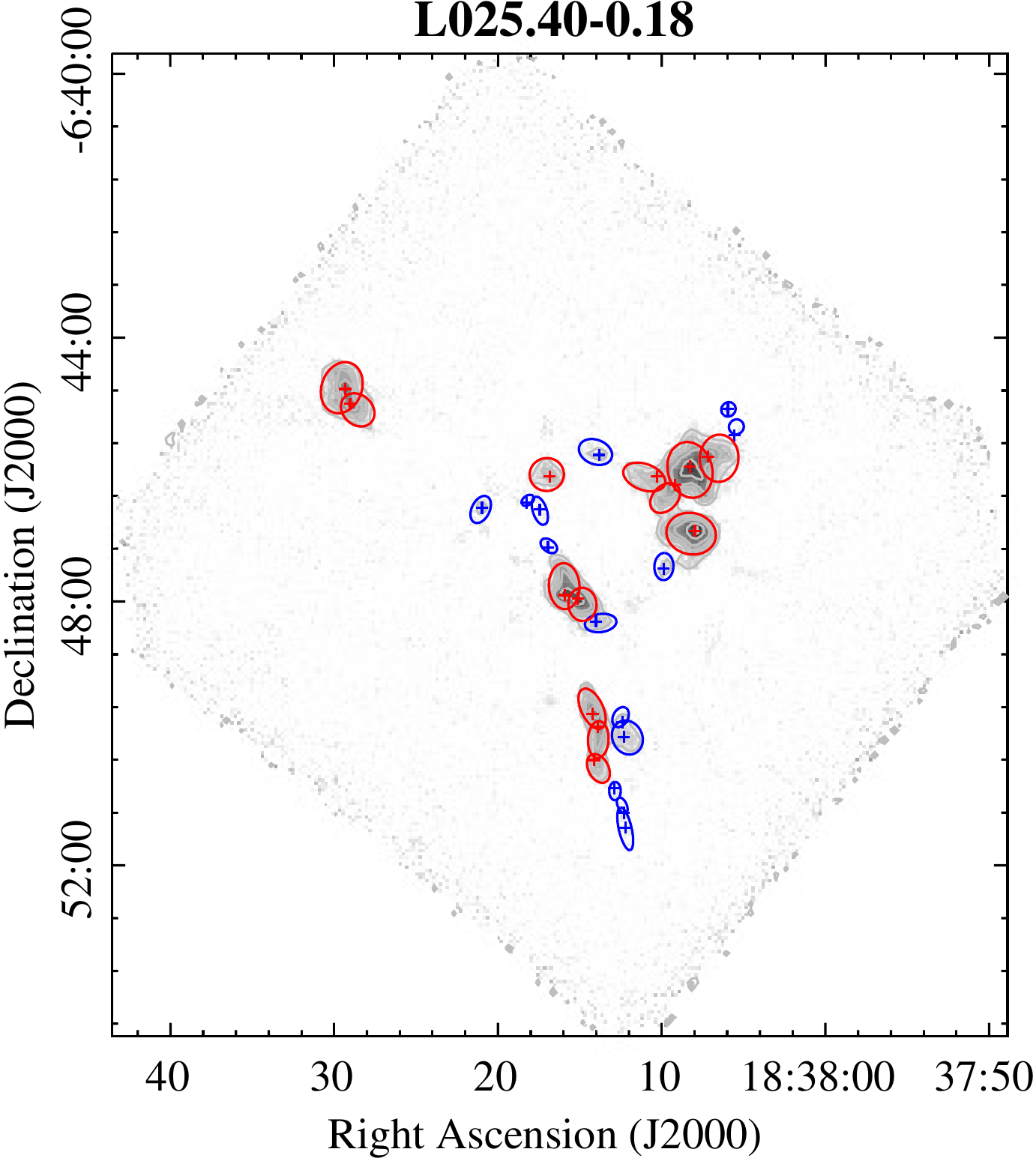}
}\\
\subfloat[L030.61+0.16 map, $\sigma_{rms}=349$ mJy beam$^{-1}$.]{
\includegraphics[scale=0.43]{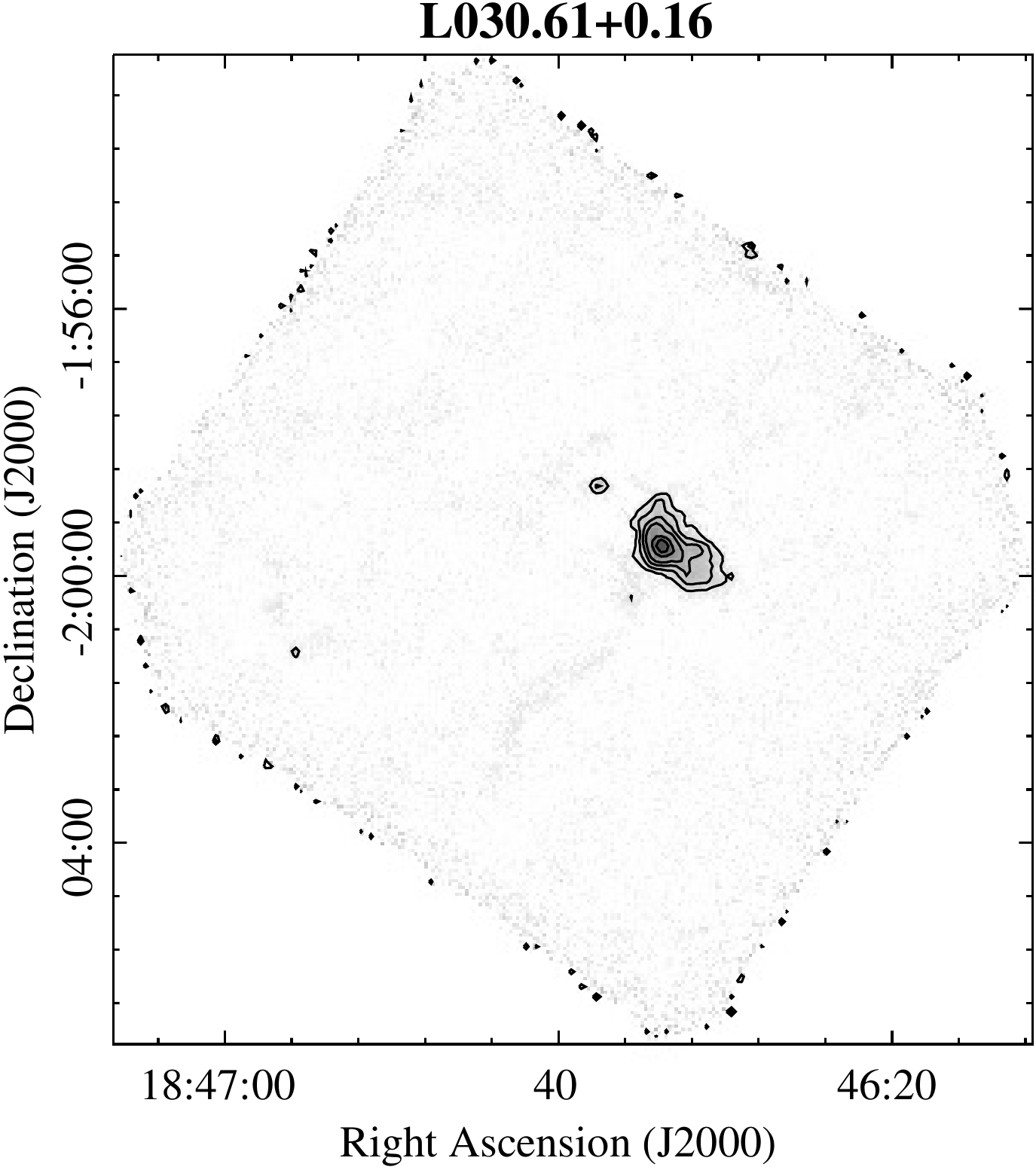}
\includegraphics[scale=0.43]{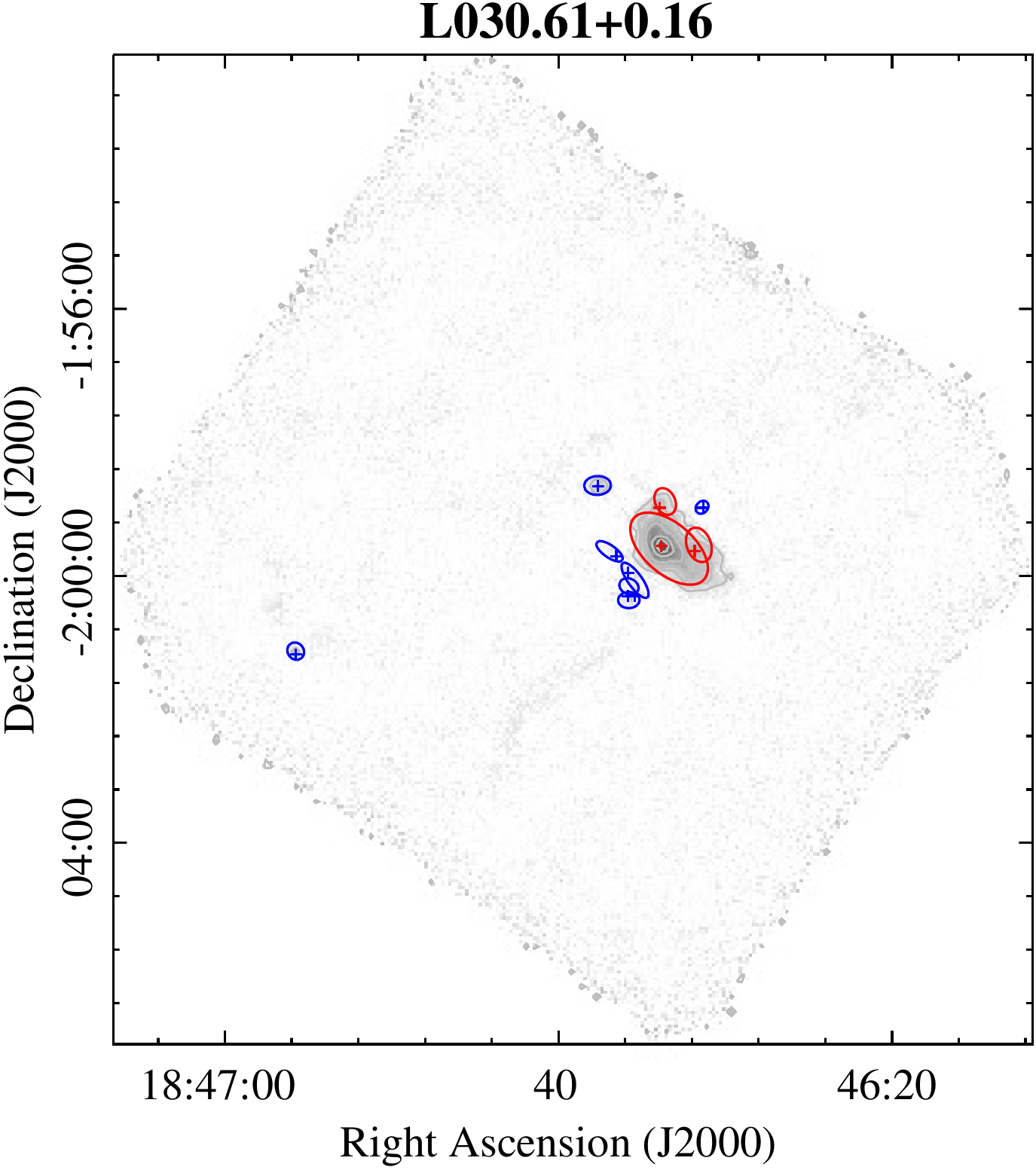}
}\\
\caption{Continuation}
\end{figure}

\clearpage
\begin{figure}\ContinuedFloat 
\center
\subfloat[L031.28+0.05 map, $\sigma_{rms}=303$ mJy beam$^{-1}$.]{
\includegraphics[scale=0.43]{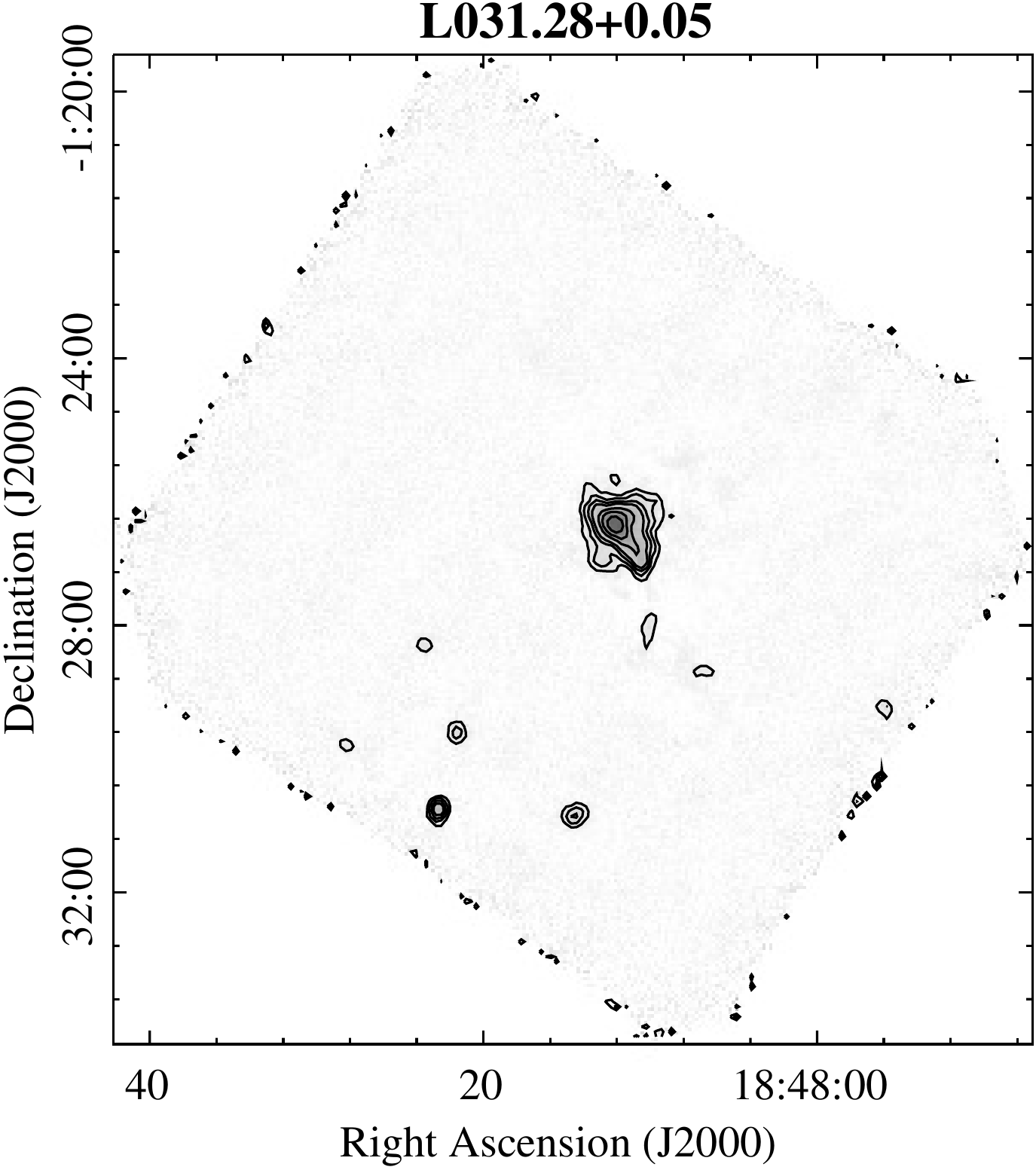}
\includegraphics[scale=0.43]{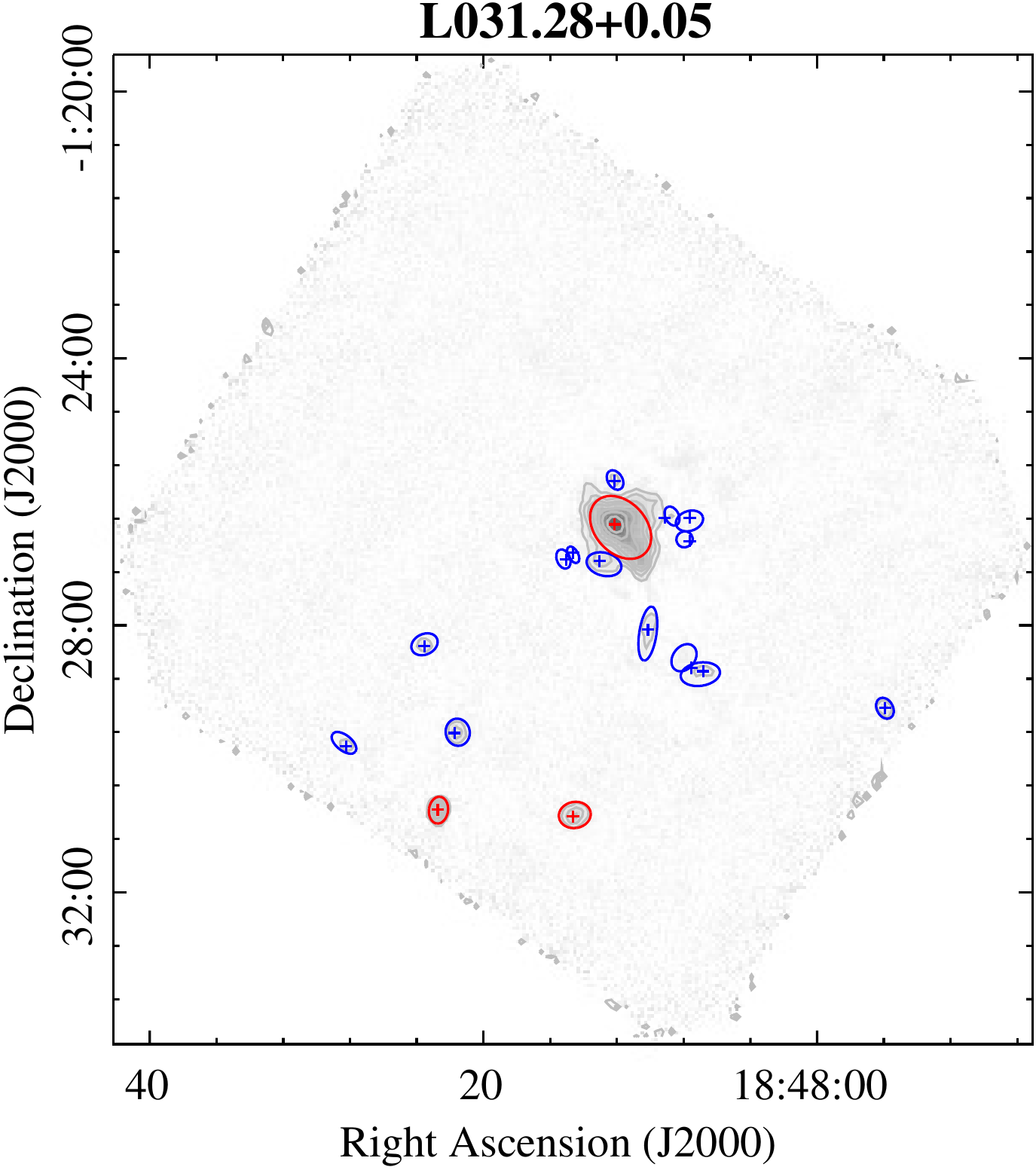}
}\\
\subfloat[L081.11-0.16 map, $\sigma_{rms}=341$ mJy beam$^{-1}$.]{
\includegraphics[scale=0.43]{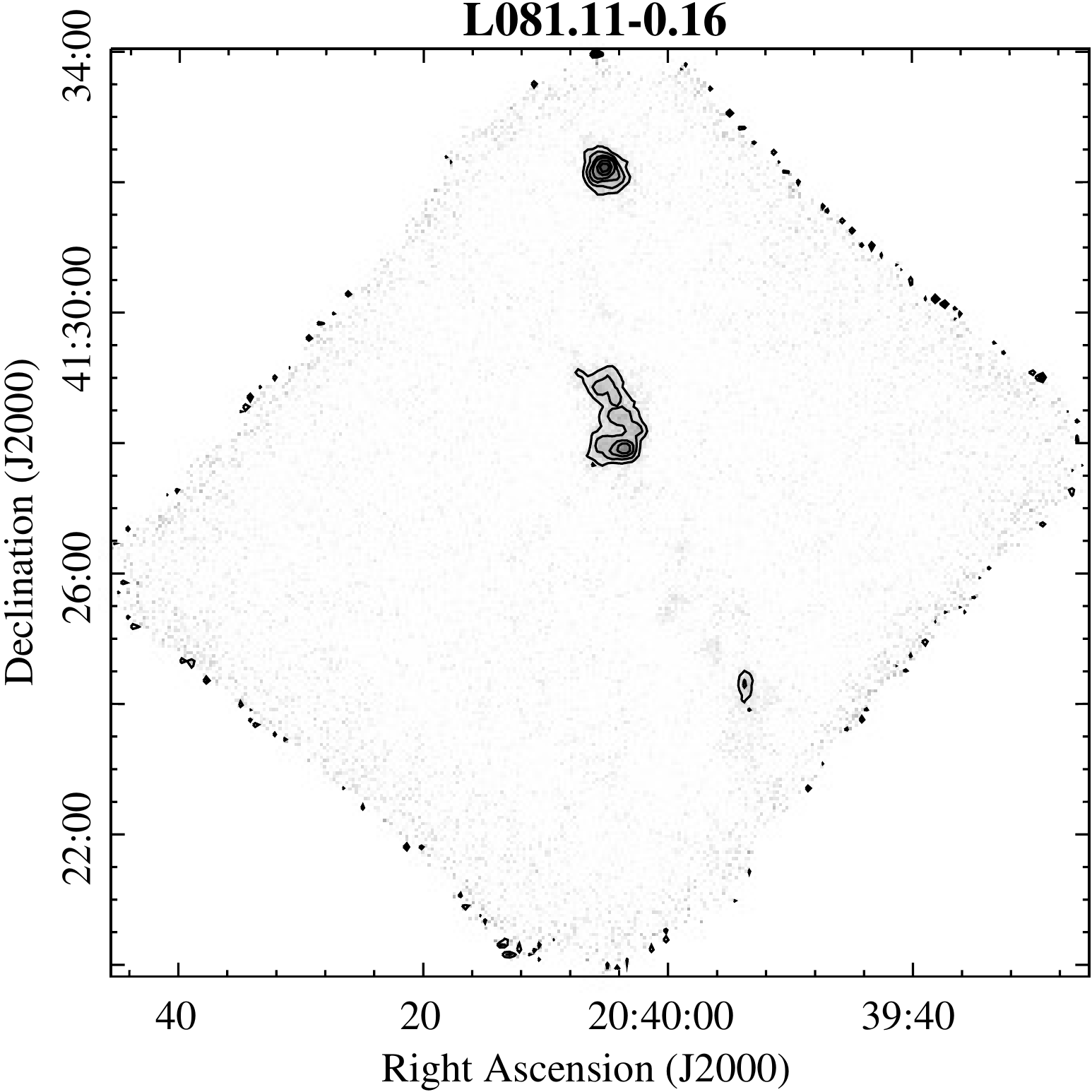}
\includegraphics[scale=0.43]{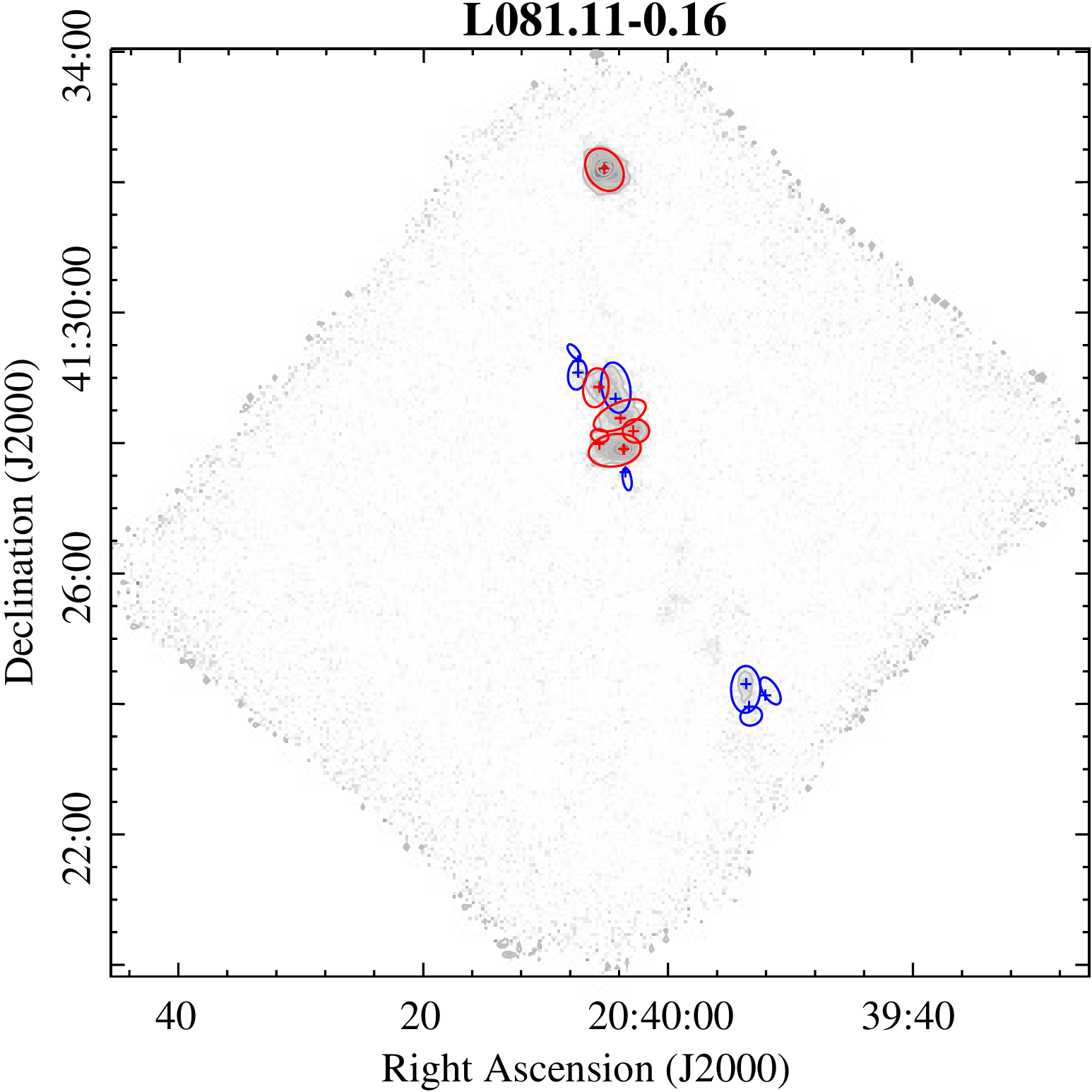}
}\\
\subfloat[L081.28+1.01 map, $\sigma_{rms}=250$ mJy beam$^{-1}$.]{
\includegraphics[scale=0.43]{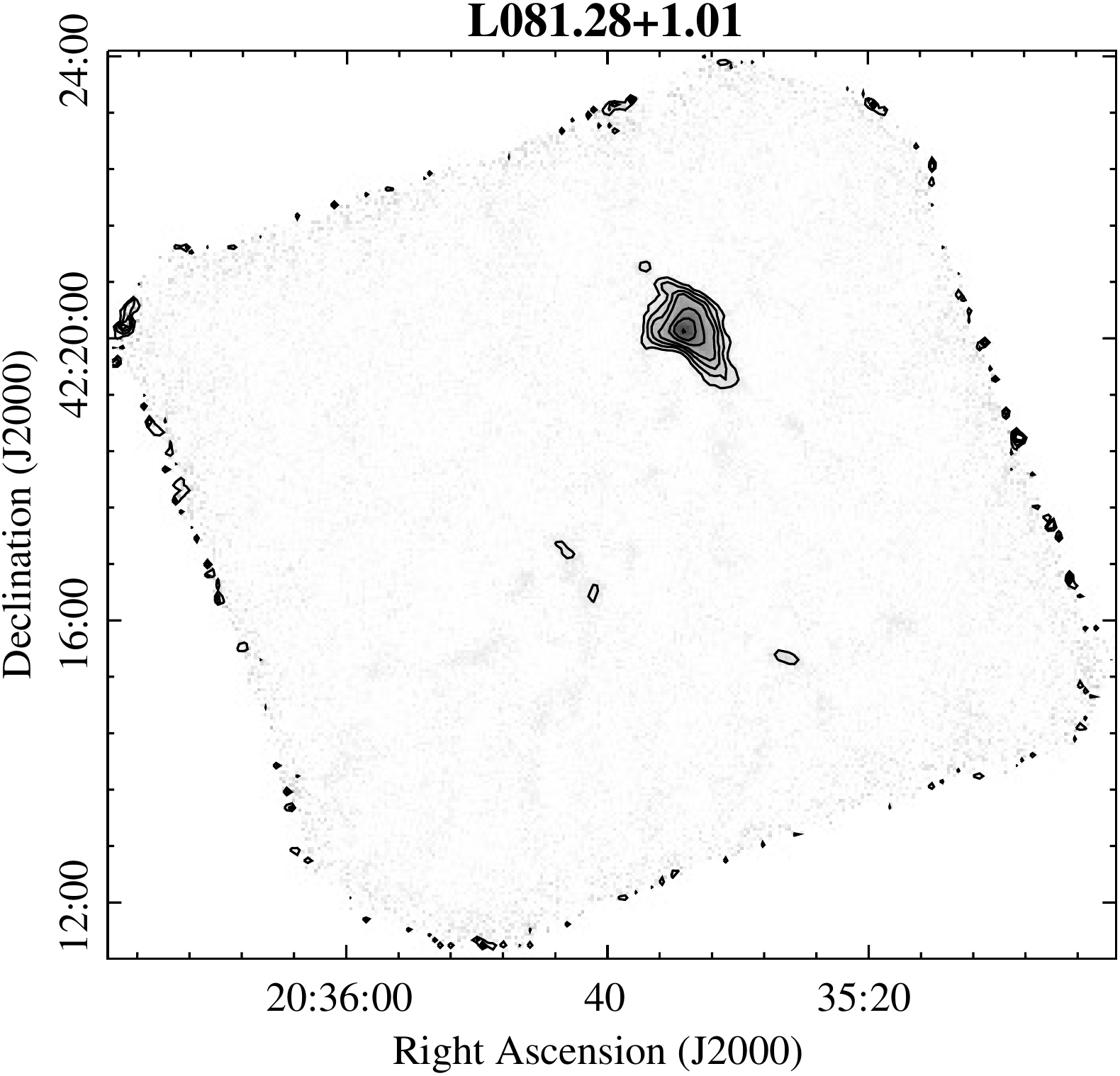}
\includegraphics[scale=0.43]{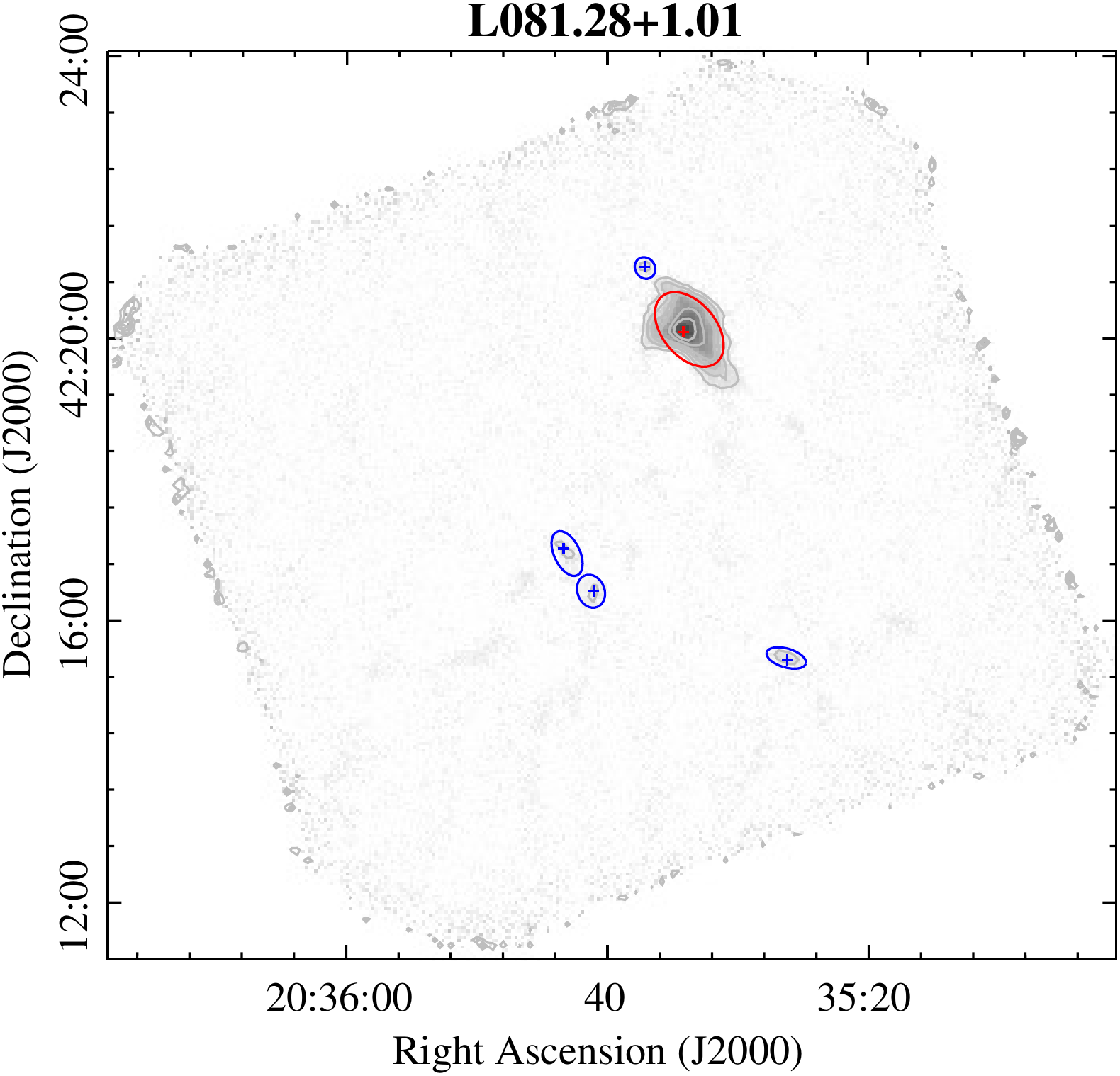}
}\\
\caption{Continuation}
\end{figure}

\clearpage
\begin{figure}\ContinuedFloat 
\center
\subfloat[L081.39+0.73 map, $\sigma_{rms}=341$ mJy beam$^{-1}$.]{
\includegraphics[scale=0.43]{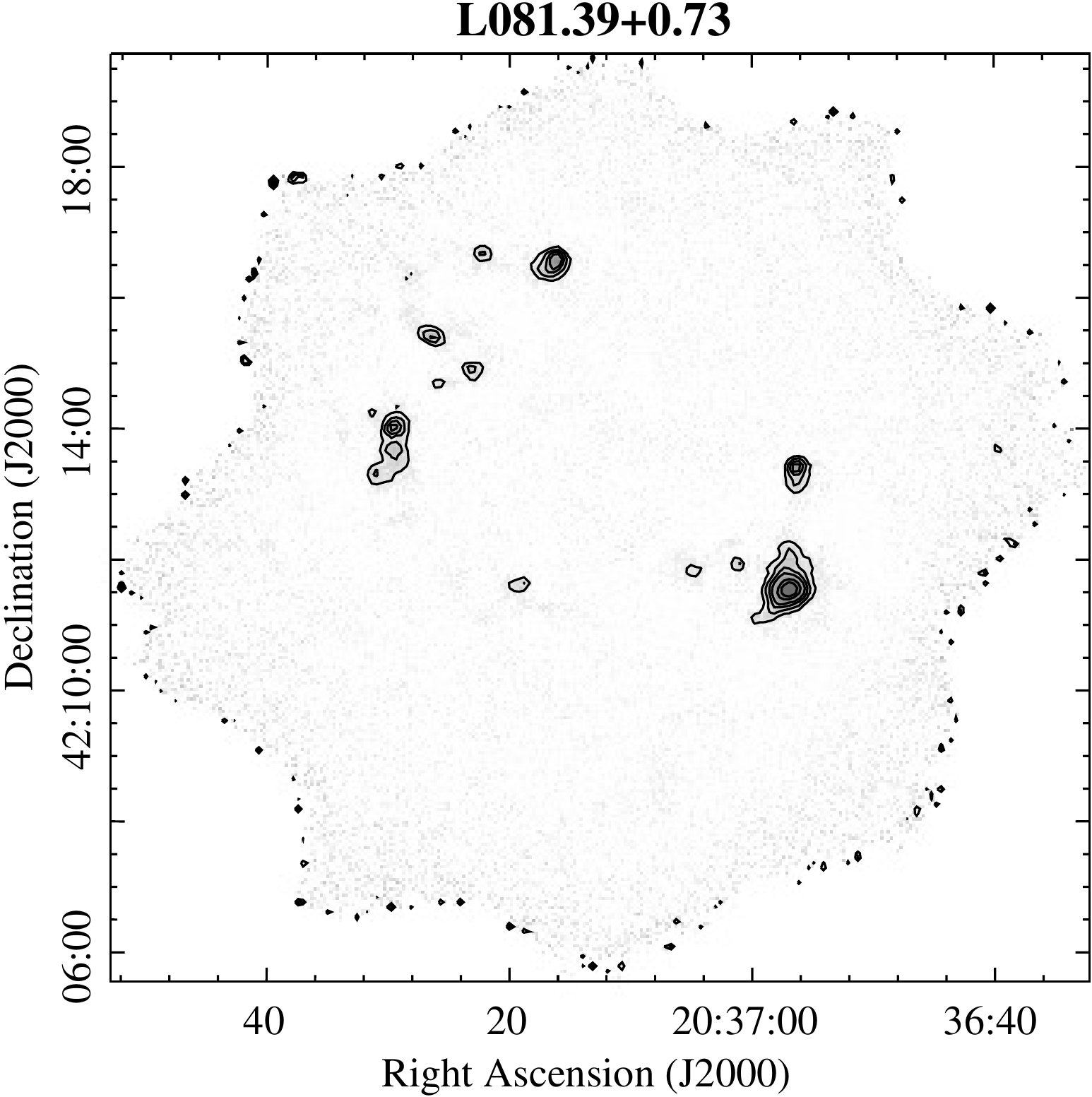}
\includegraphics[scale=0.43]{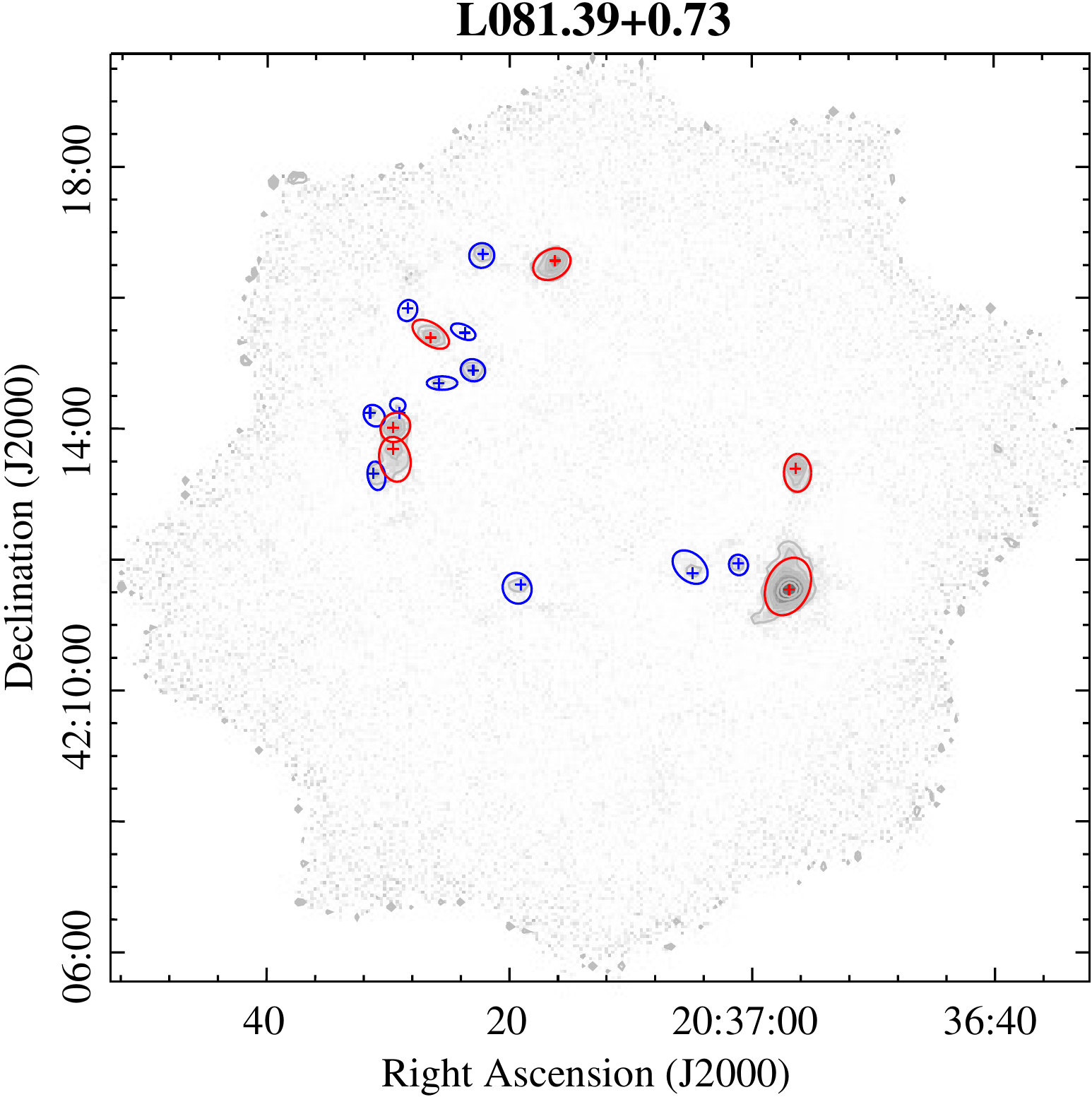}
}\\
\subfloat[L081.48+0.00 map, $\sigma_{rms}=353$ mJy beam$^{-1}$.]{
\includegraphics[scale=0.43]{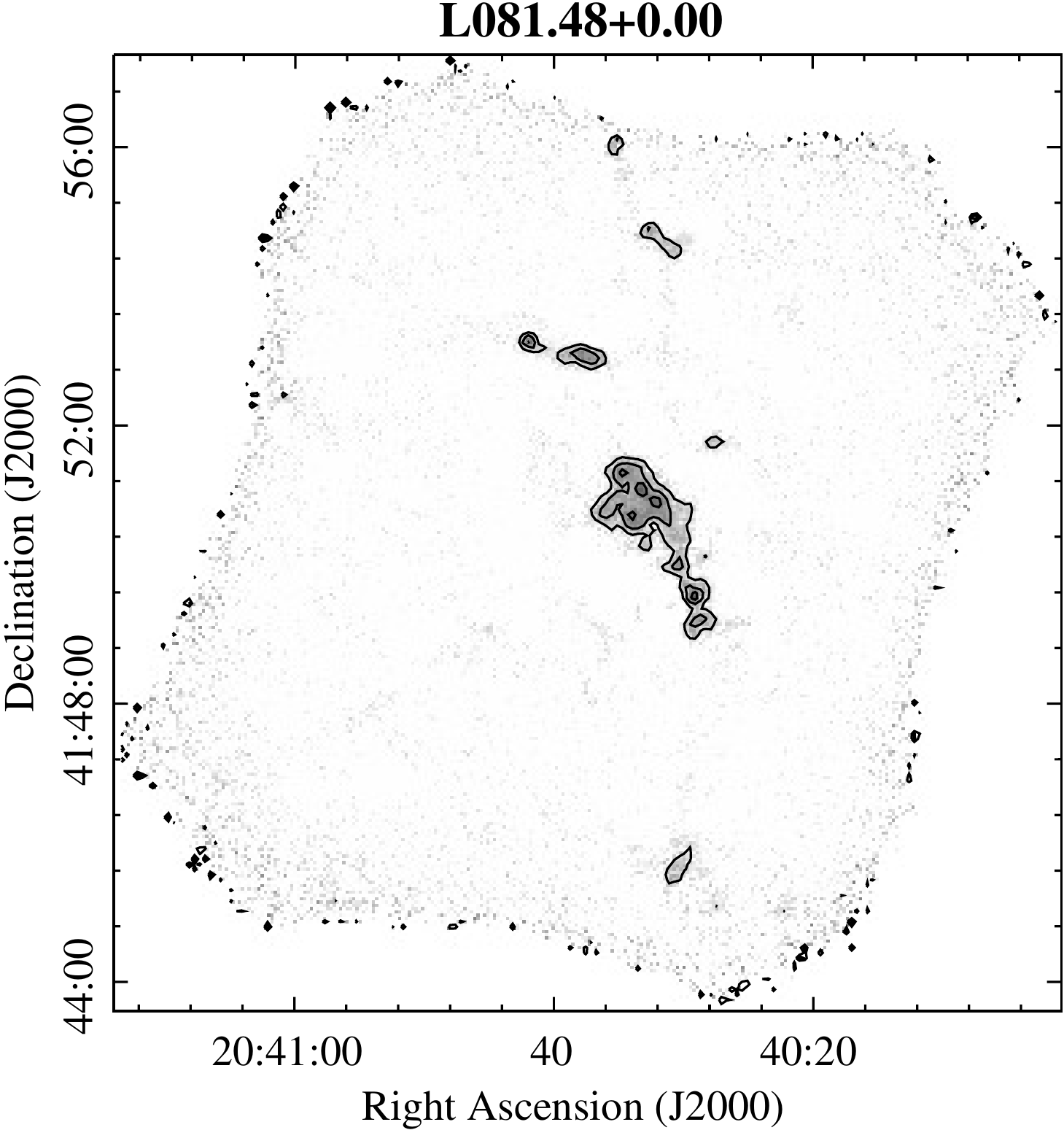}
\includegraphics[scale=0.43]{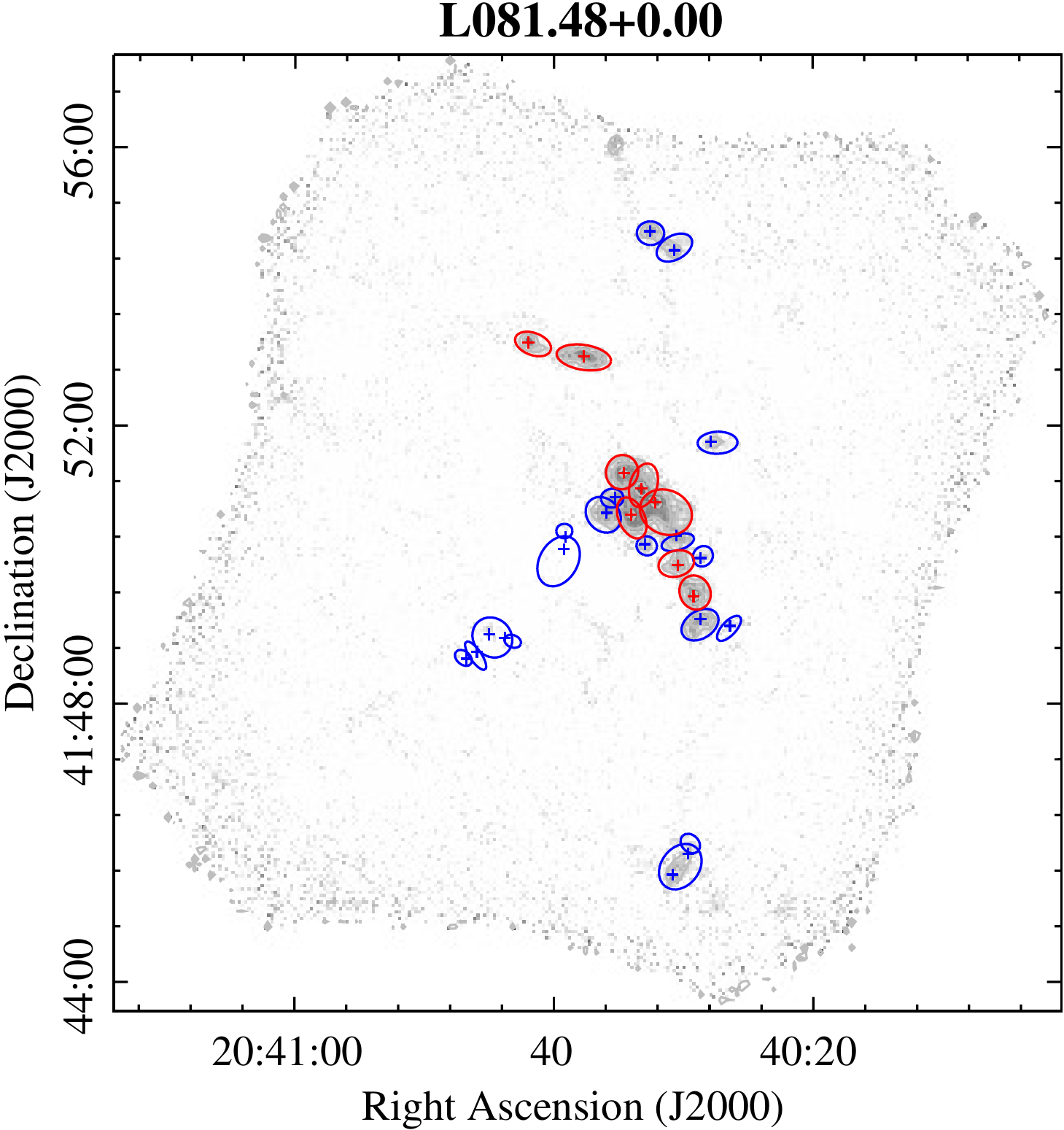}
}\\
\subfloat[L081.76+0.60 map, $\sigma_{rms}=329$ mJy beam$^{-1}$. Additional contour is drawn at 200$\sigma$.]{
\includegraphics[scale=0.43]{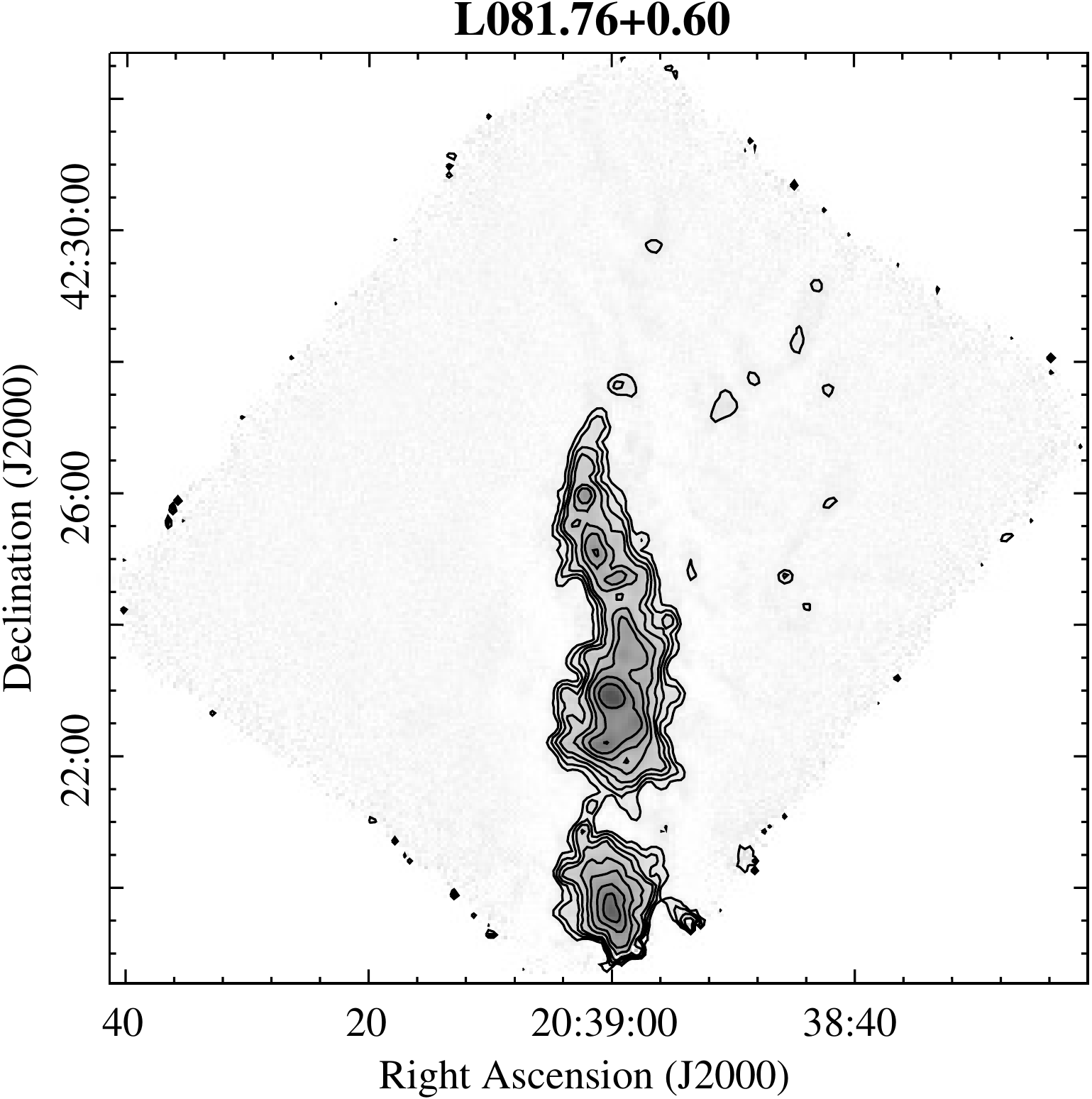}
\includegraphics[scale=0.43]{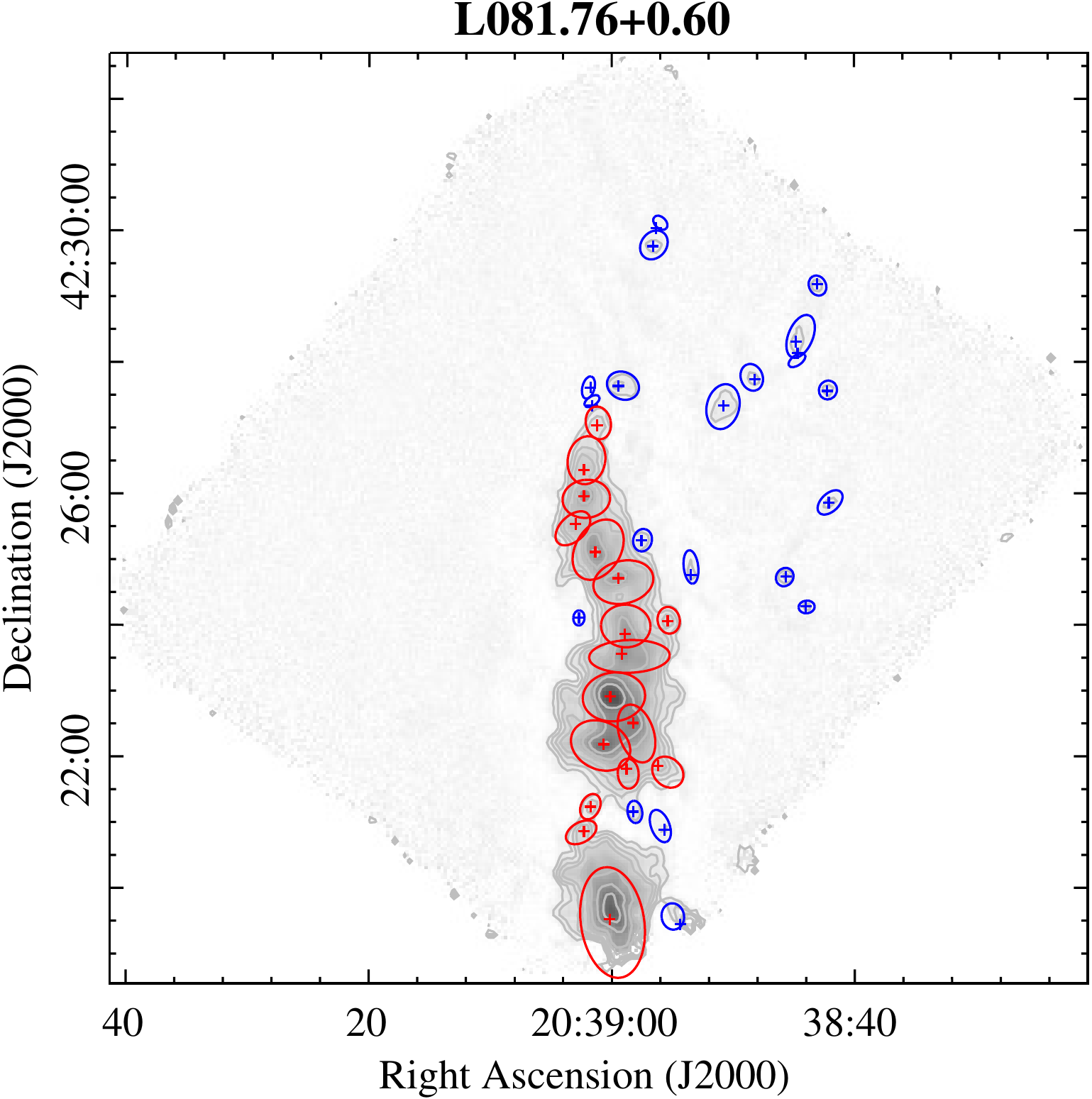}
}\\
\caption{Continuation}
\end{figure}

\clearpage
\begin{figure}\ContinuedFloat 
\center
\subfloat[L031.41+0.31 map, $\sigma_{rms}=791$ mJy beam$^{-1}$.]{
\includegraphics[scale=0.43]{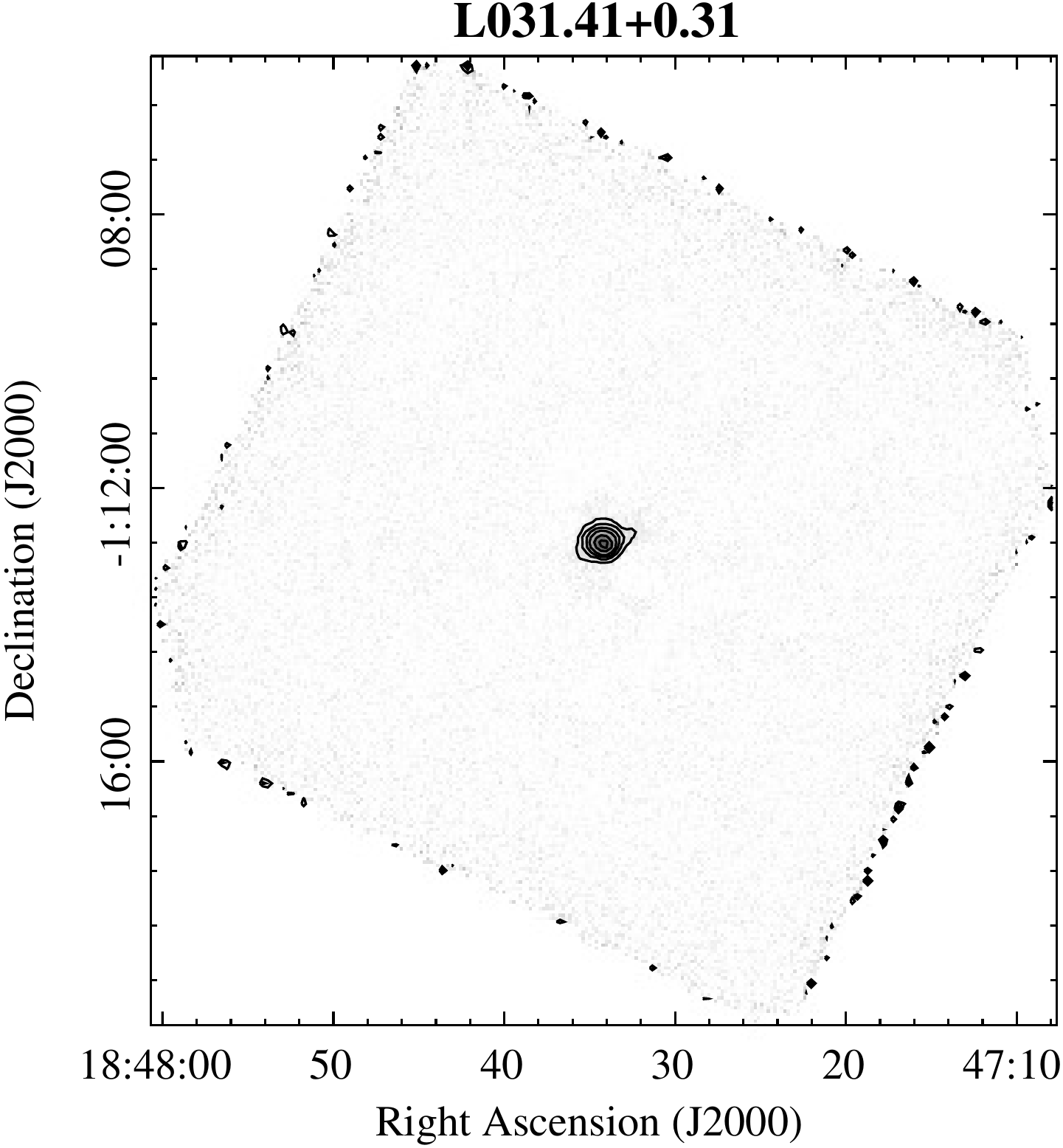}
\includegraphics[scale=0.43]{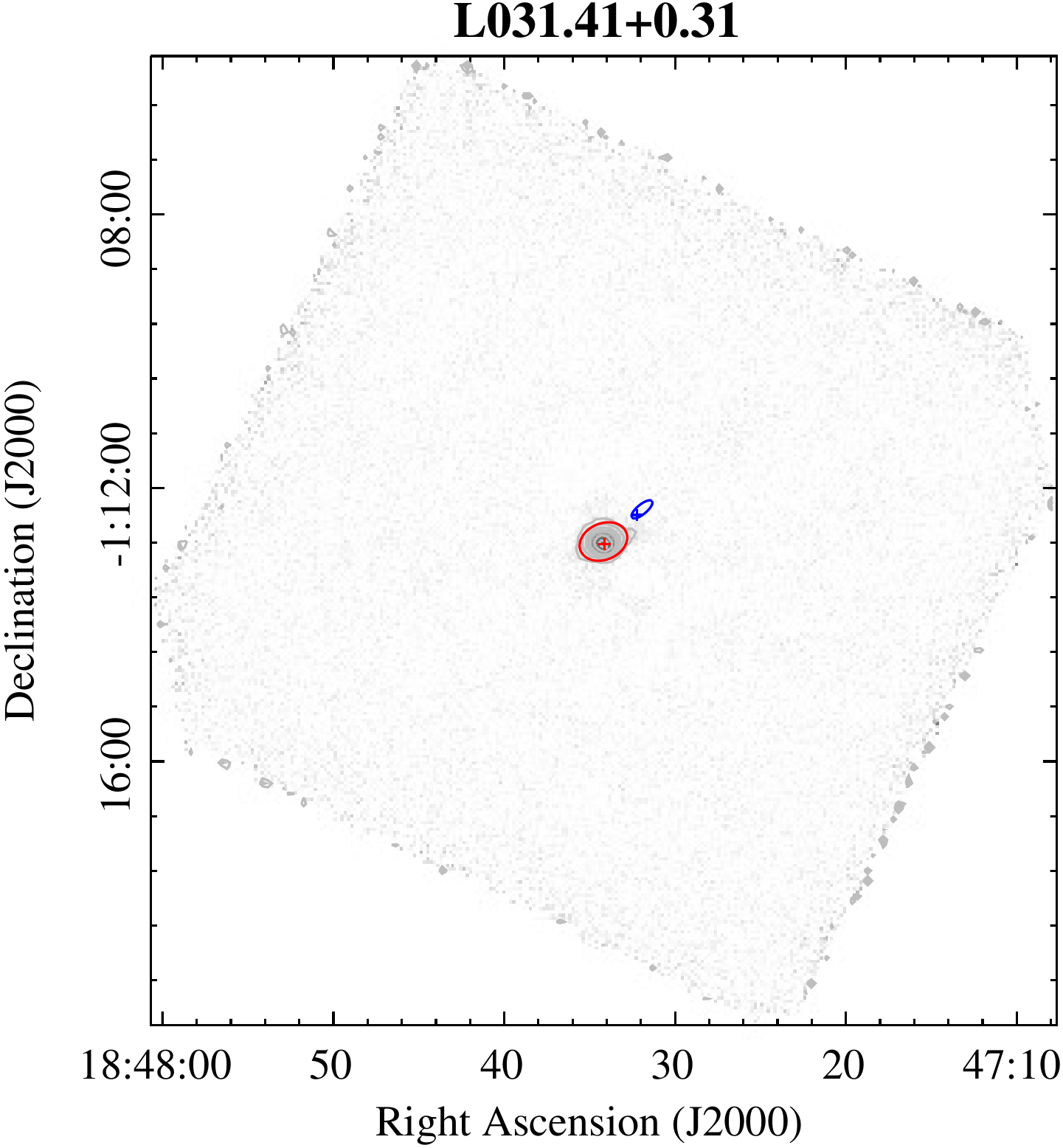}
}\\
\subfloat[L081.88+0.77 map, $\sigma_{rms}=717$ mJy beam$^{-1}$.]{
\includegraphics[scale=0.43]{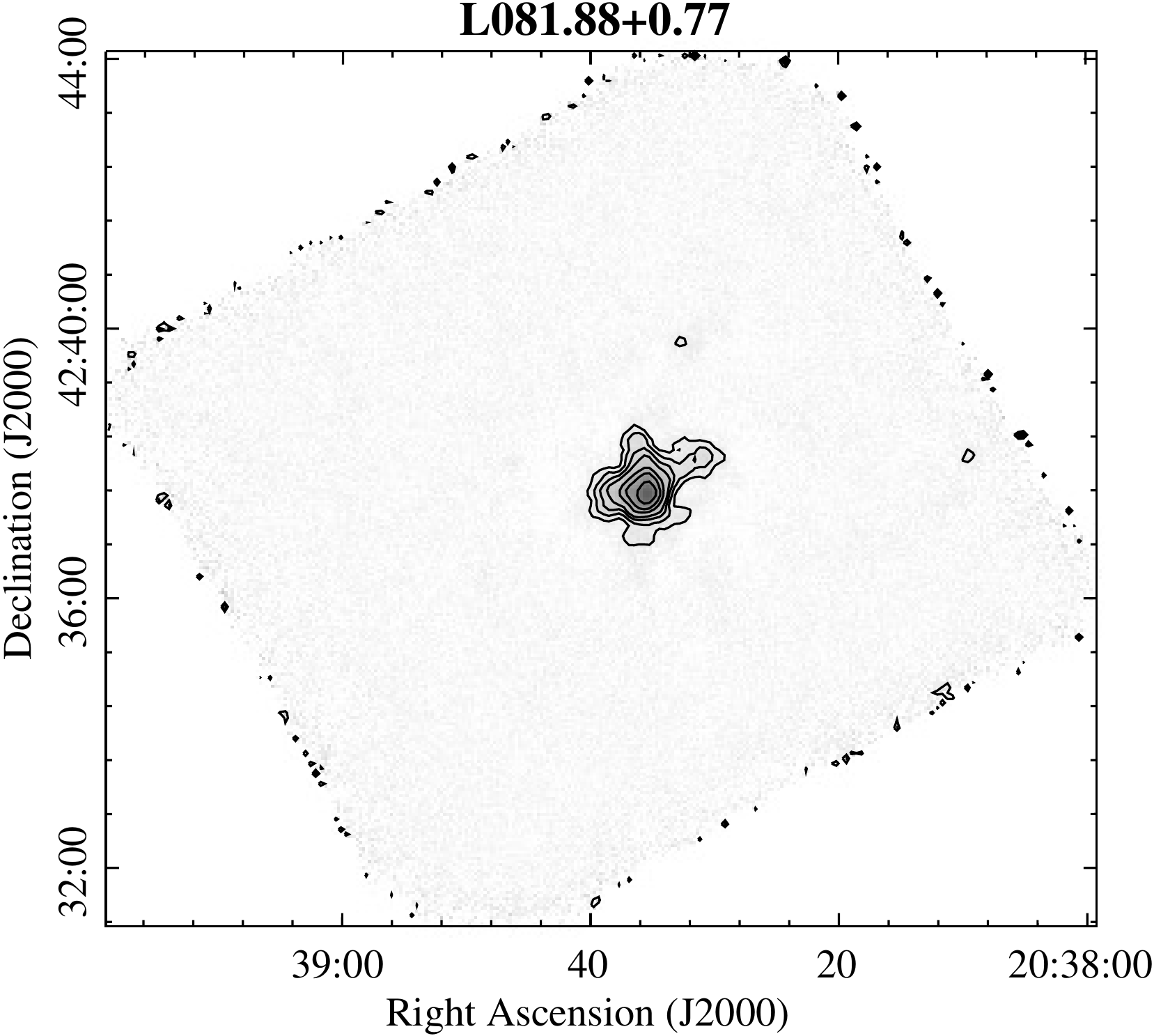}
\includegraphics[scale=0.43]{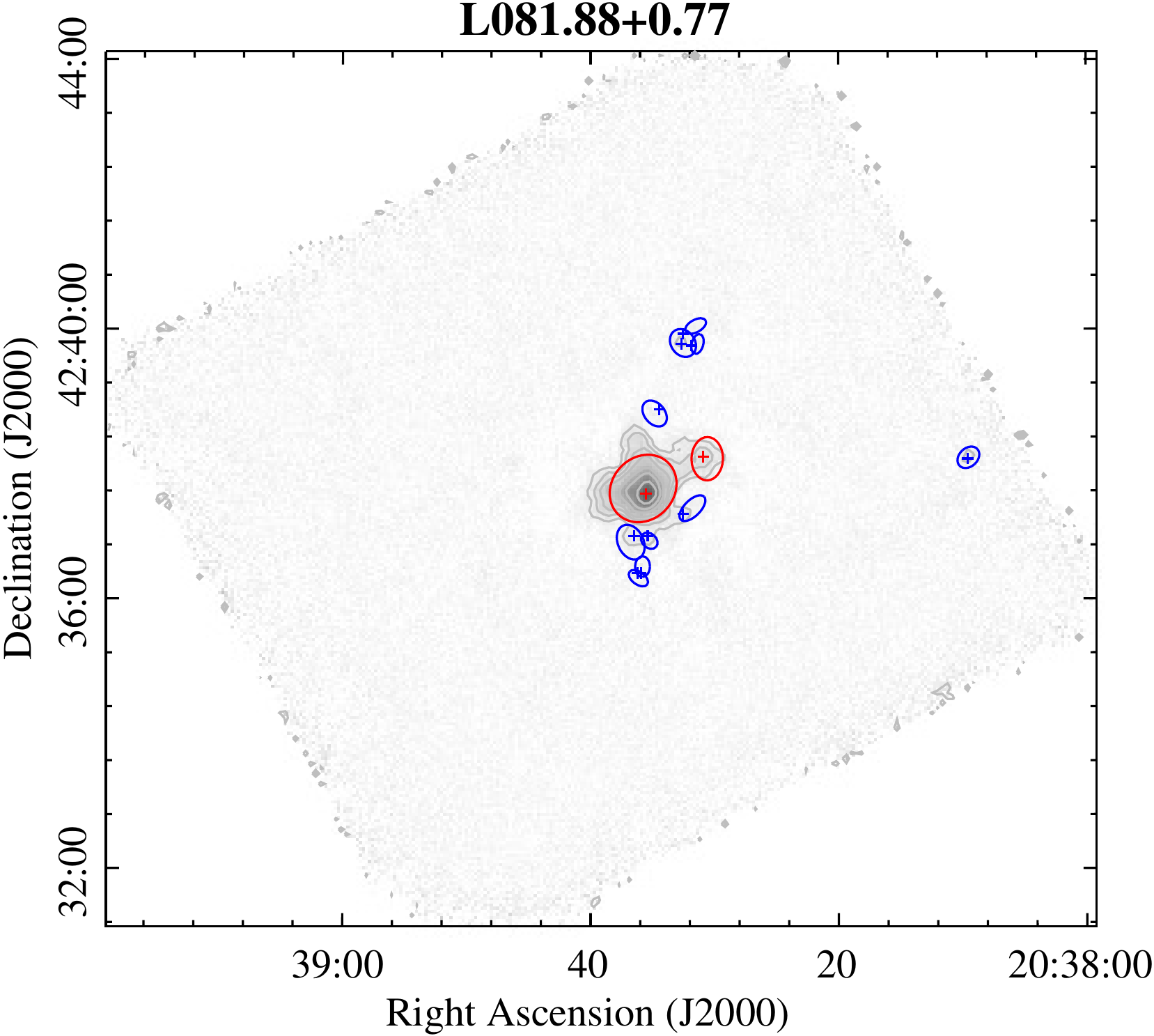}
}\\
\subfloat[L183.40-0.58 map, $\sigma_{rms}=255$ mJy beam$^{-1}$.]{
\includegraphics[scale=0.43]{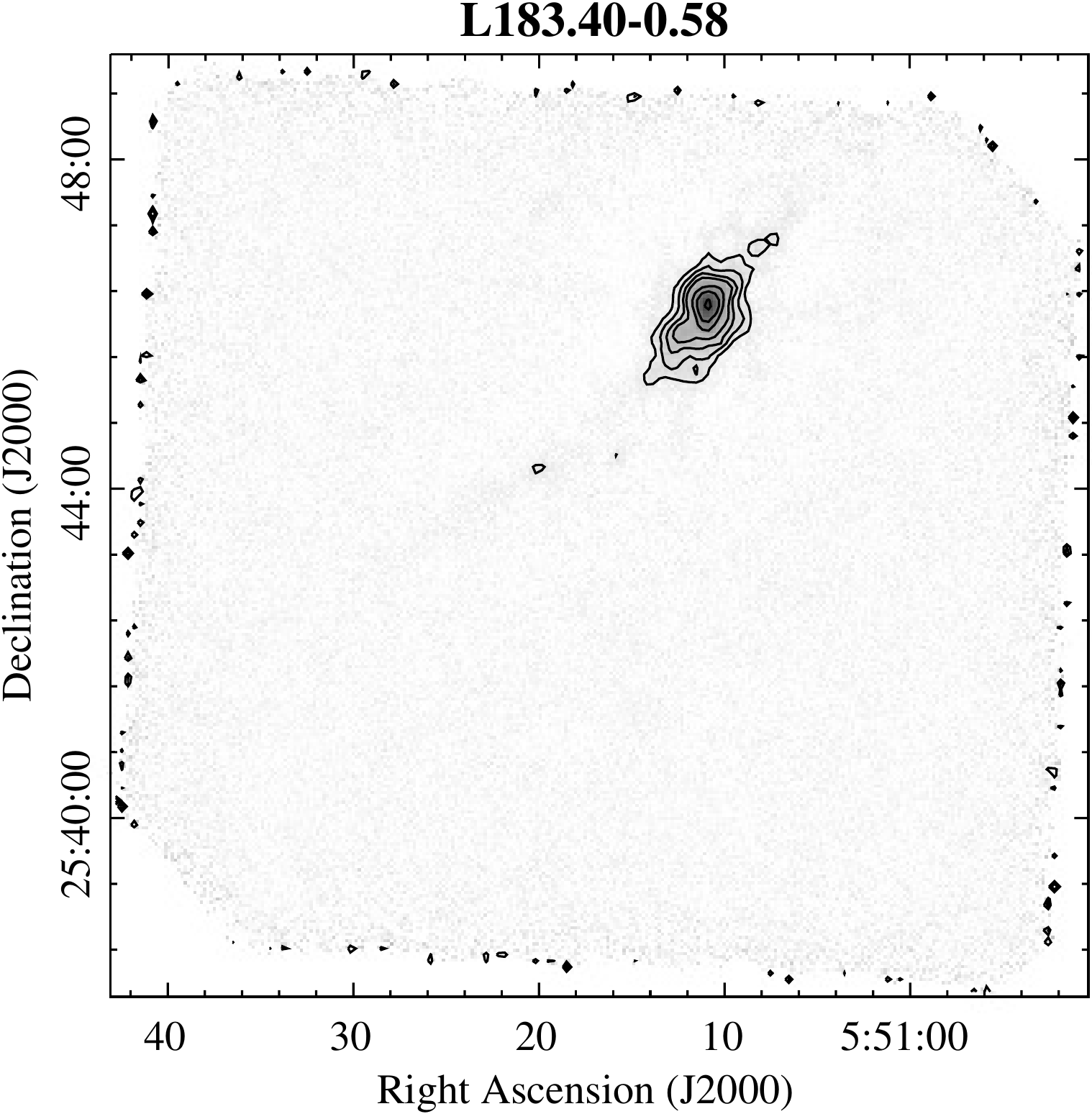}
\includegraphics[scale=0.43]{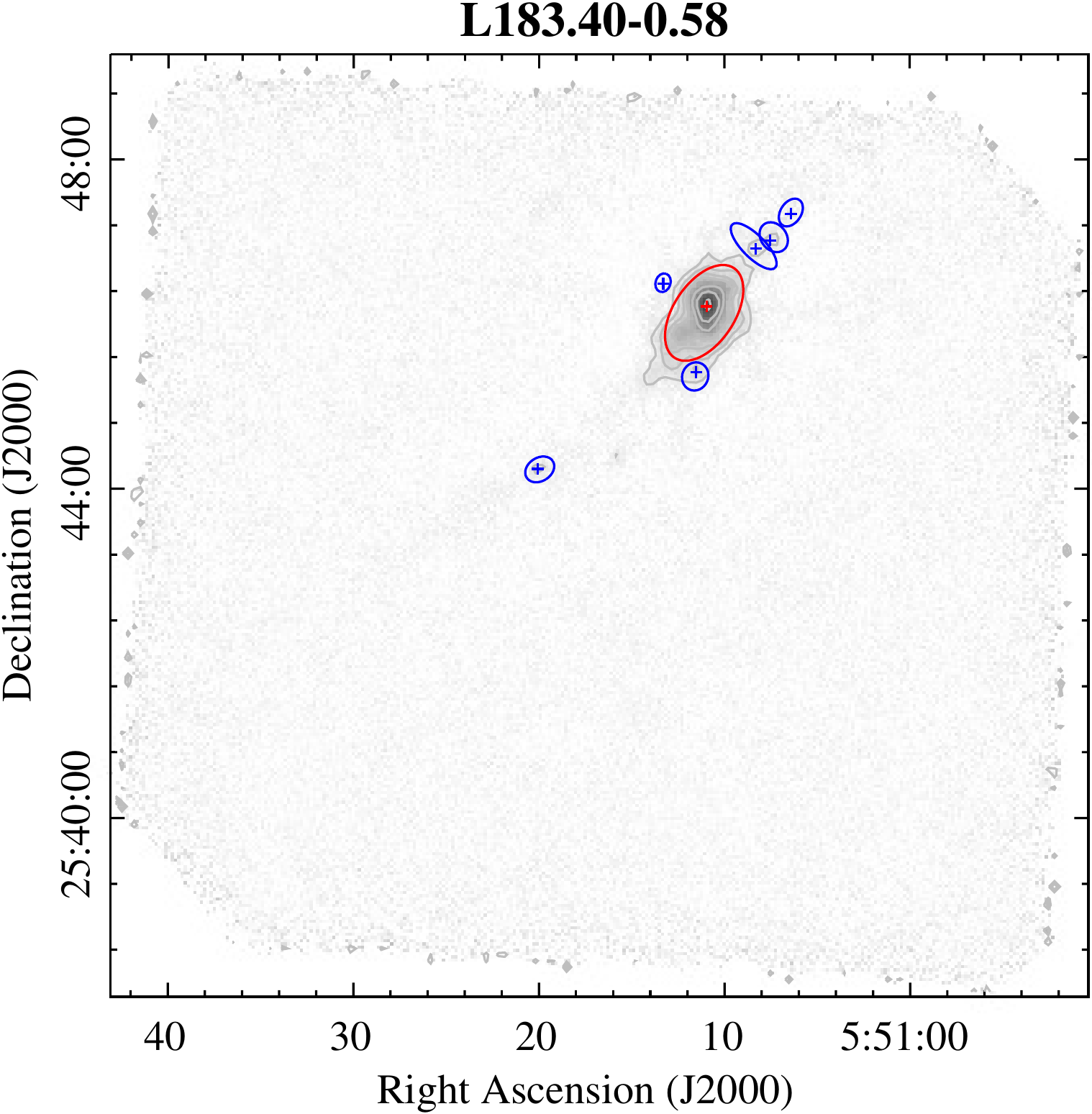}
}\\
\caption{Continuation}
\end{figure}

\clearpage
\begin{figure}\ContinuedFloat 
\center
\subfloat[L189.78+0.33 map, $\sigma_{rms}=294$ mJy beam$^{-1}$.]{
\includegraphics[scale=0.43]{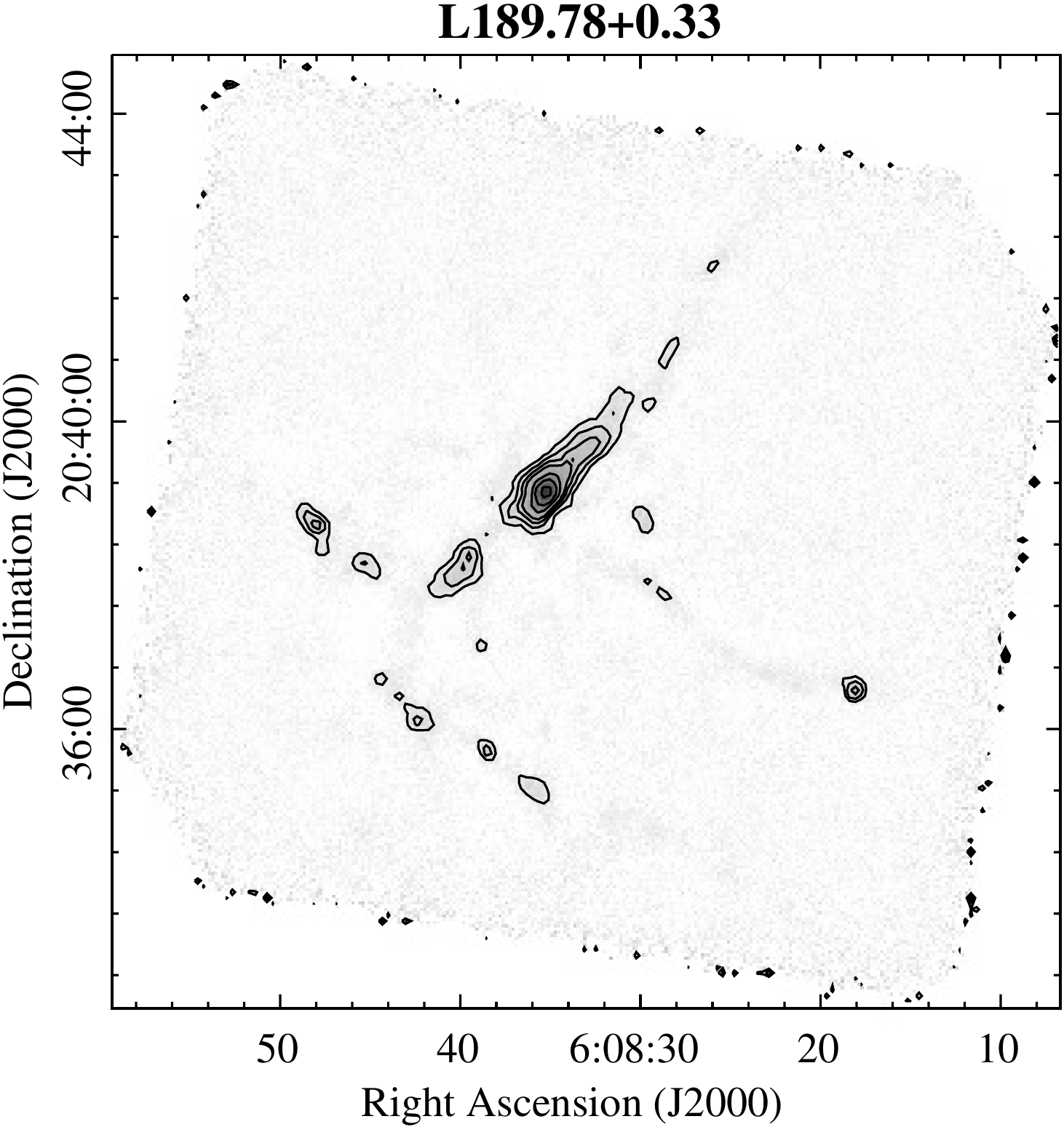}
\includegraphics[scale=0.43]{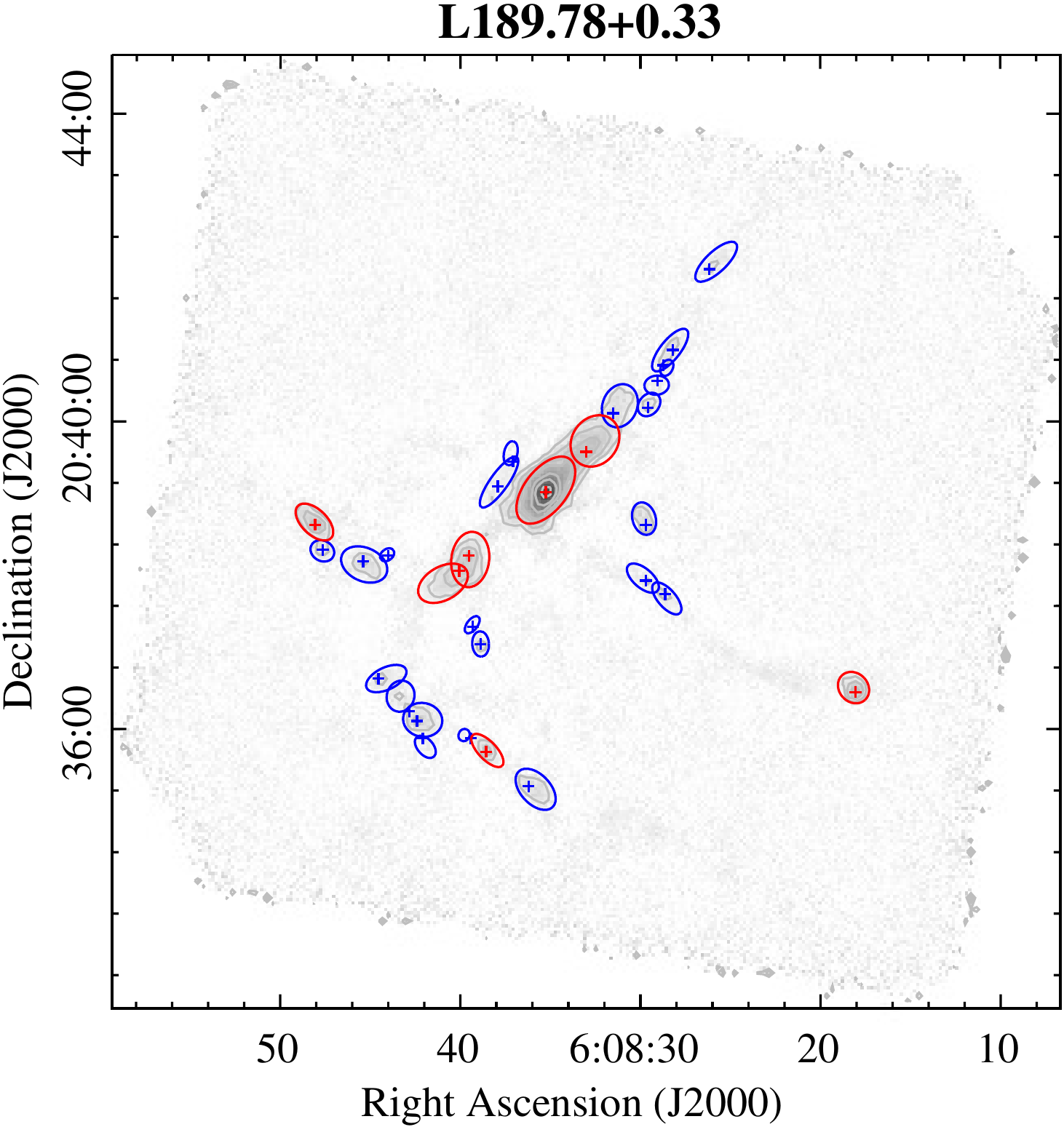}
}\\
\subfloat[L202.58+2.42 map, $\sigma_{rms}=370$ mJy beam$^{-1}$.]{
\includegraphics[scale=0.43]{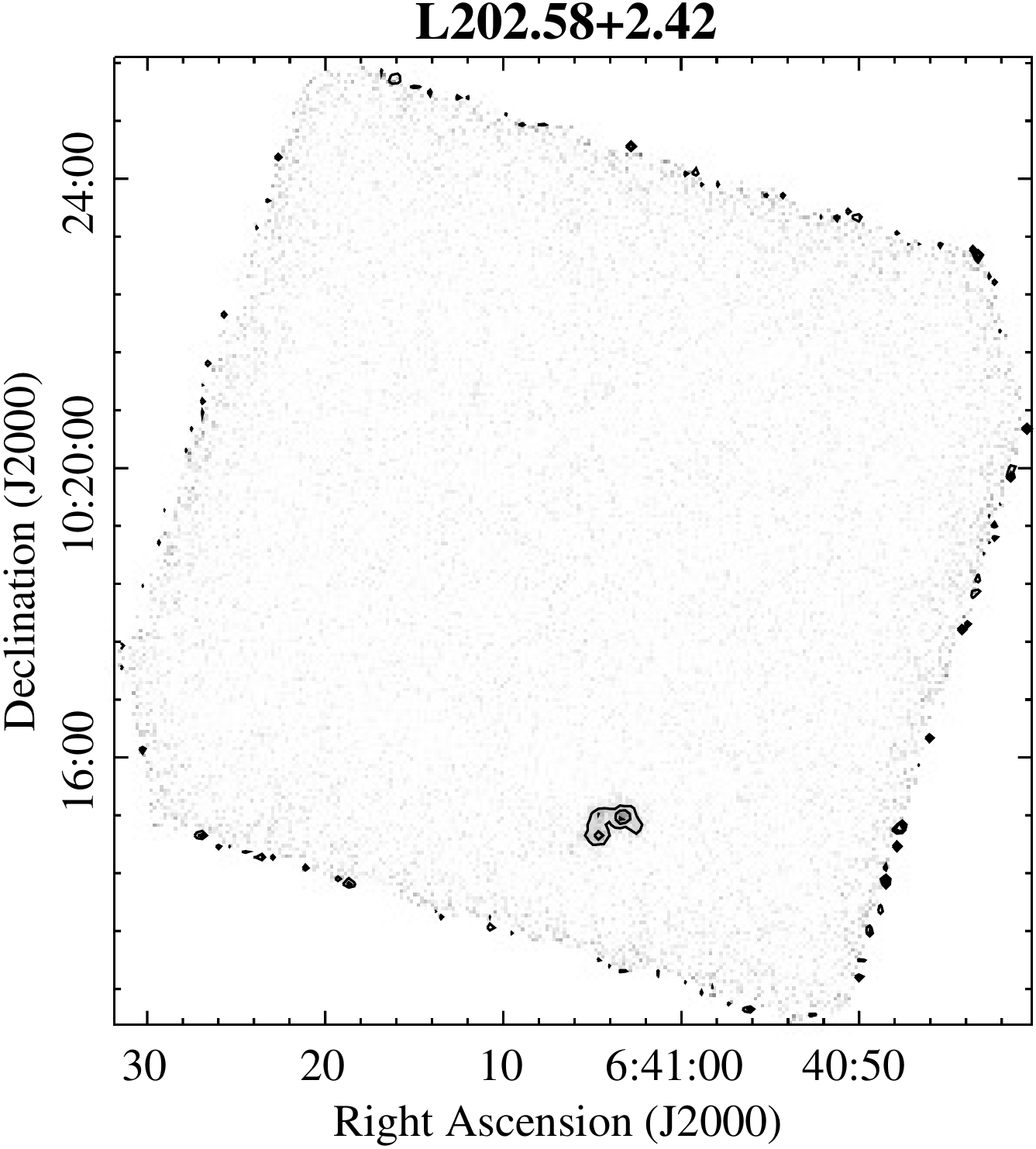}
\includegraphics[scale=0.43]{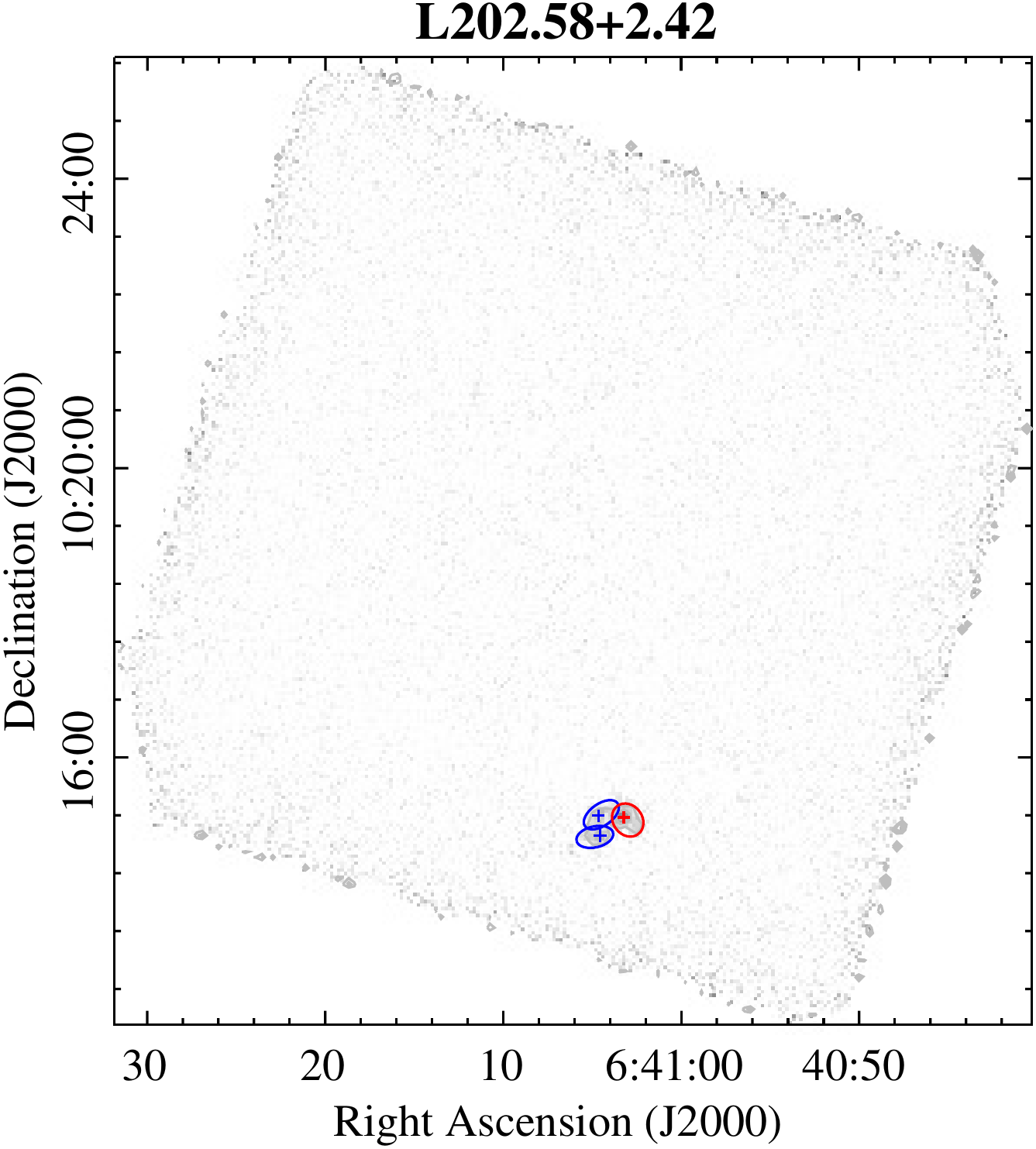}
}\\
\subfloat[L142.01+1.77 map, $\sigma_{rms}=305$ mJy beam$^{-1}$.]{
\includegraphics[scale=0.43]{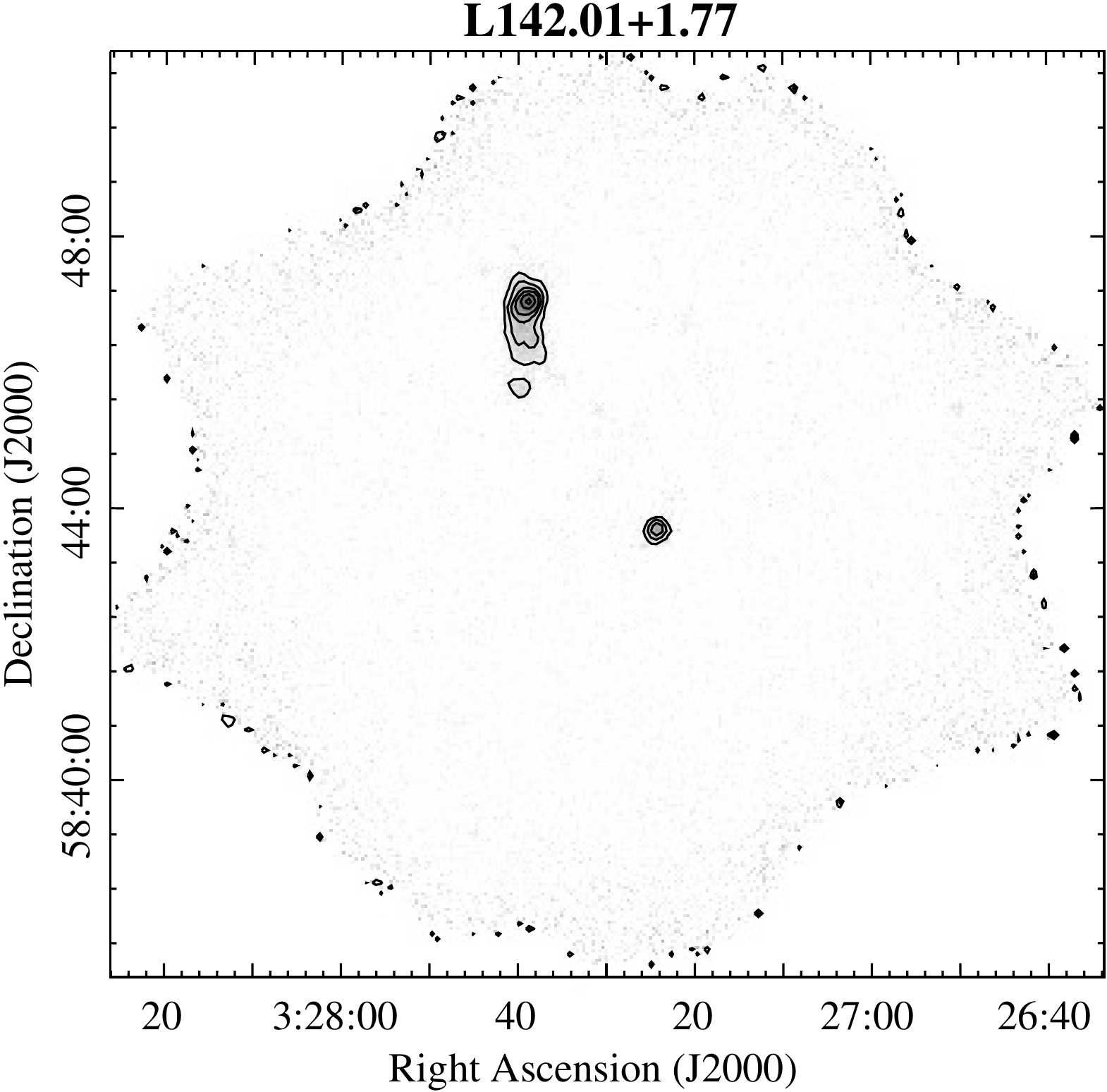}
\includegraphics[scale=0.43]{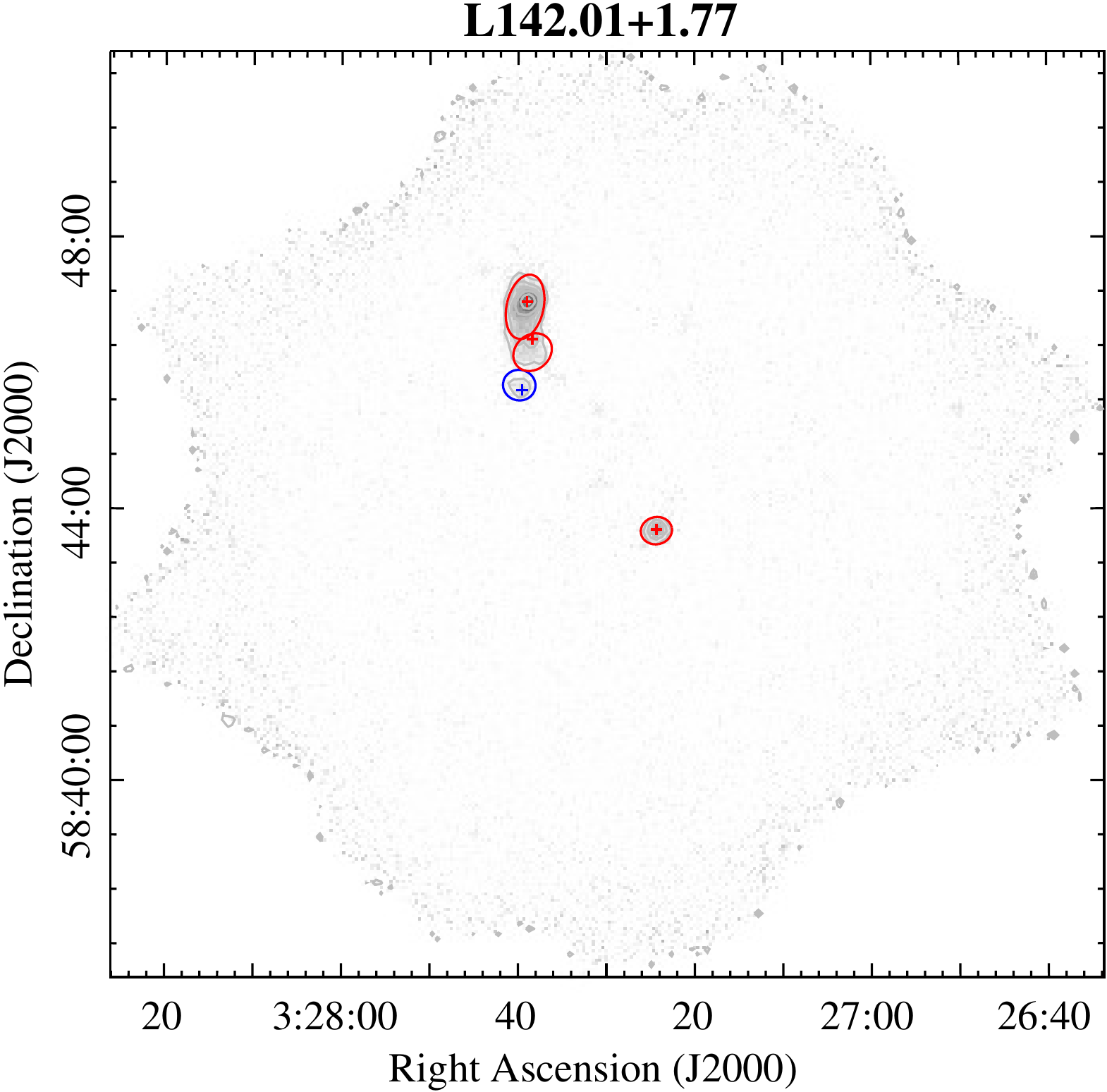}
}\\
\caption{Continuation}
\end{figure}

\clearpage
\begin{figure}\ContinuedFloat 
\center
\subfloat[L151.61-0.24 map, $\sigma_{rms}=266$ mJy beam$^{-1}$.]{
\includegraphics[scale=0.43]{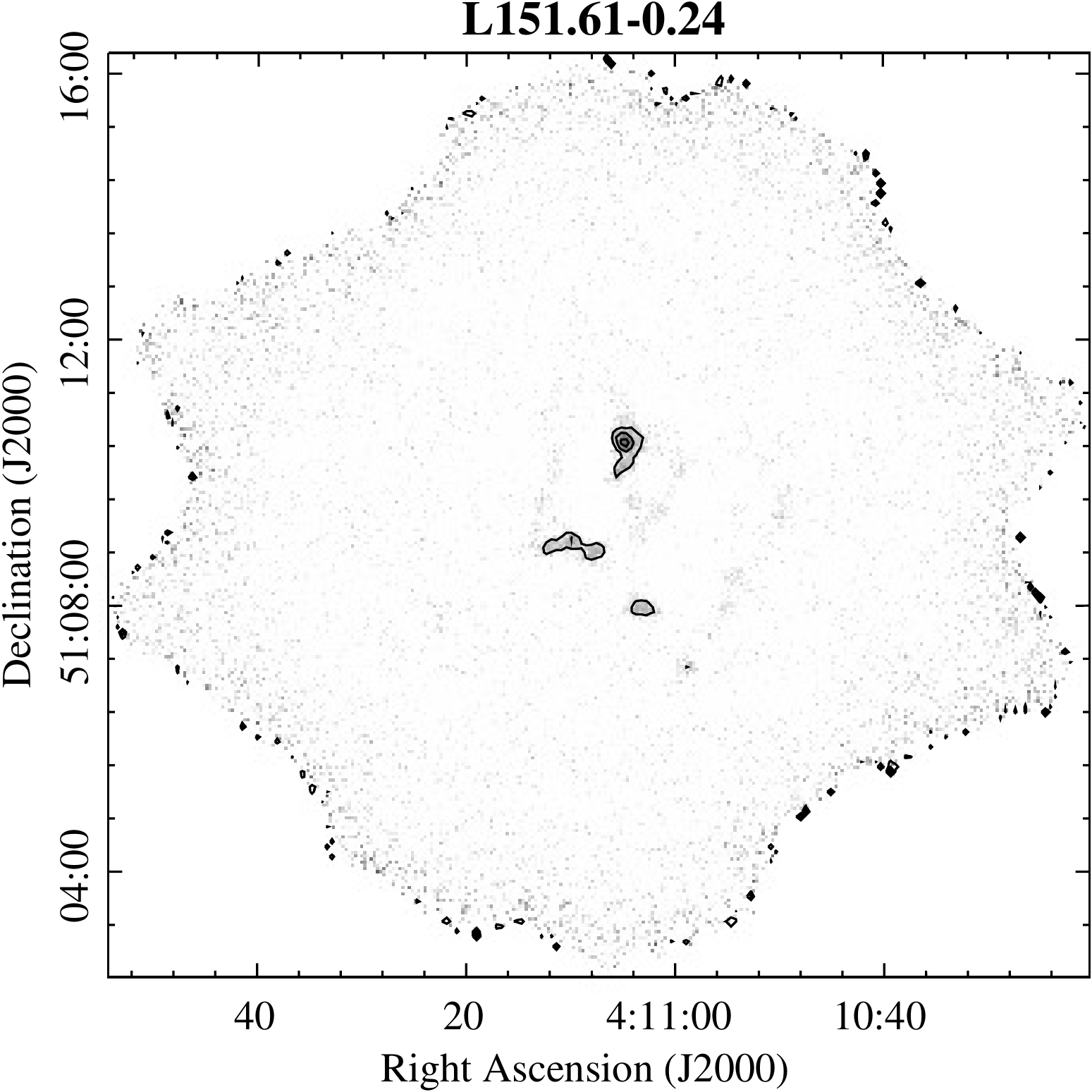}
\includegraphics[scale=0.43]{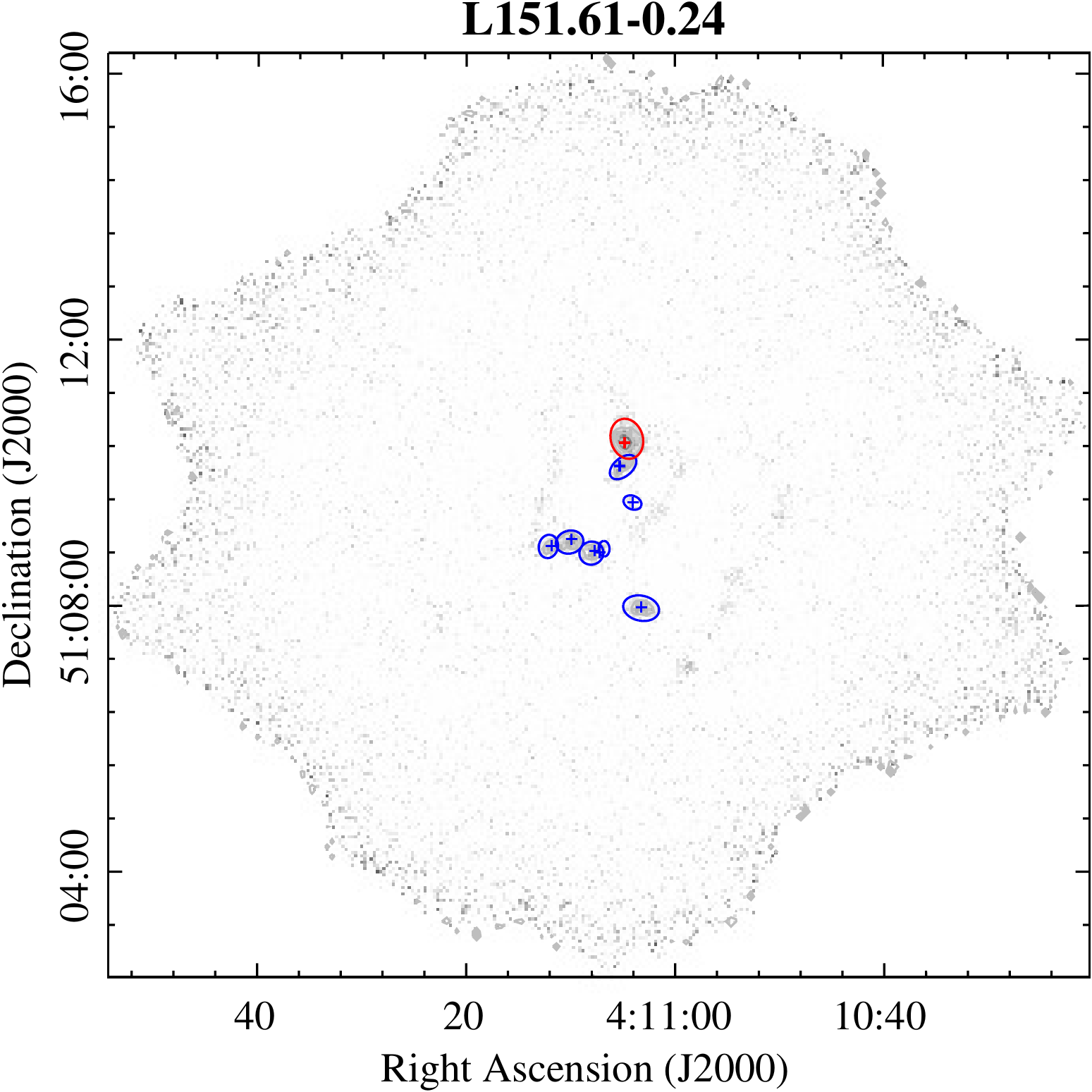}
}\\
\subfloat[L154.37+2.58 map, $\sigma_{rms}=266$ mJy beam$^{-1}$.]{
\includegraphics[scale=0.43]{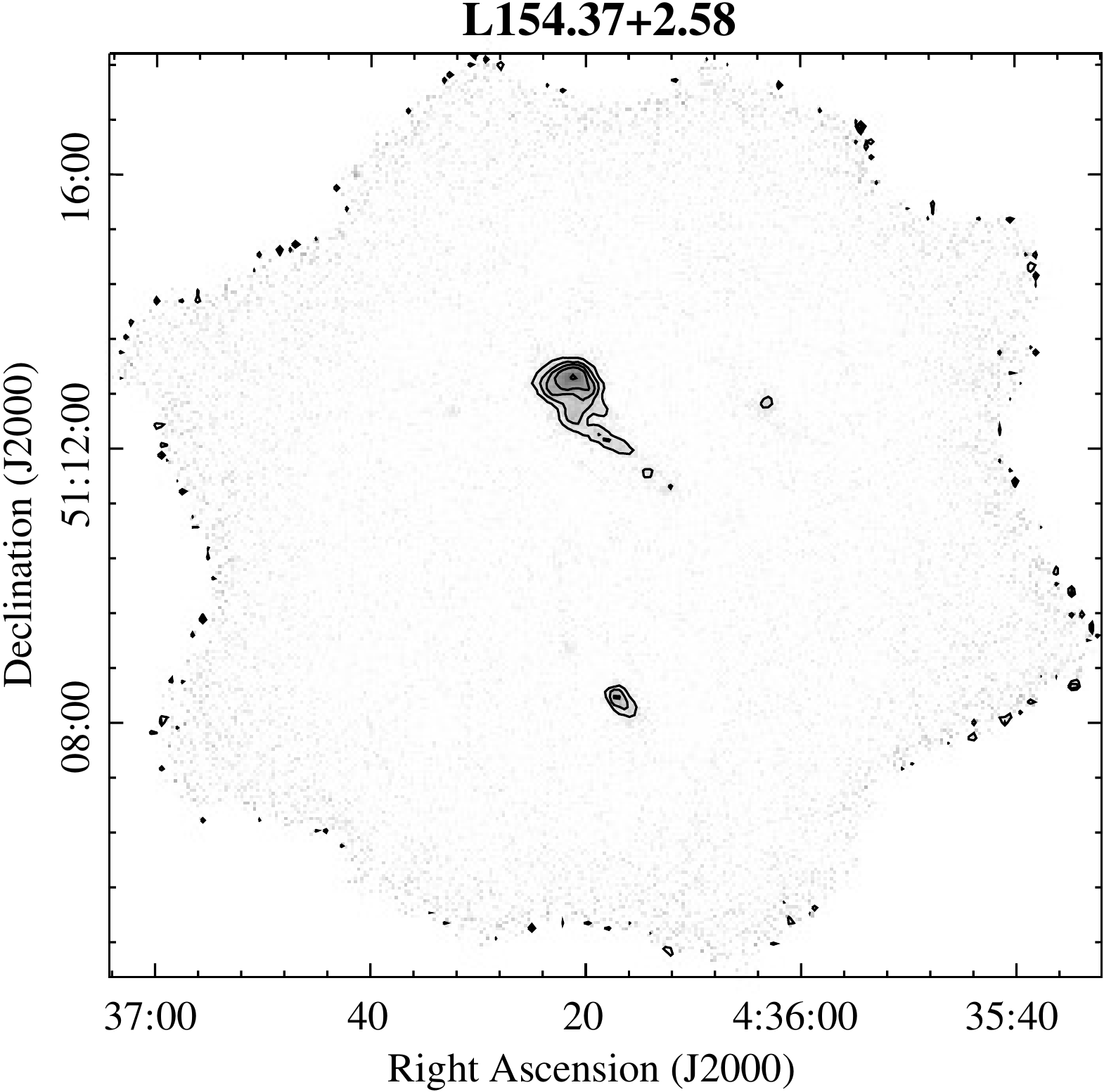}
\includegraphics[scale=0.43]{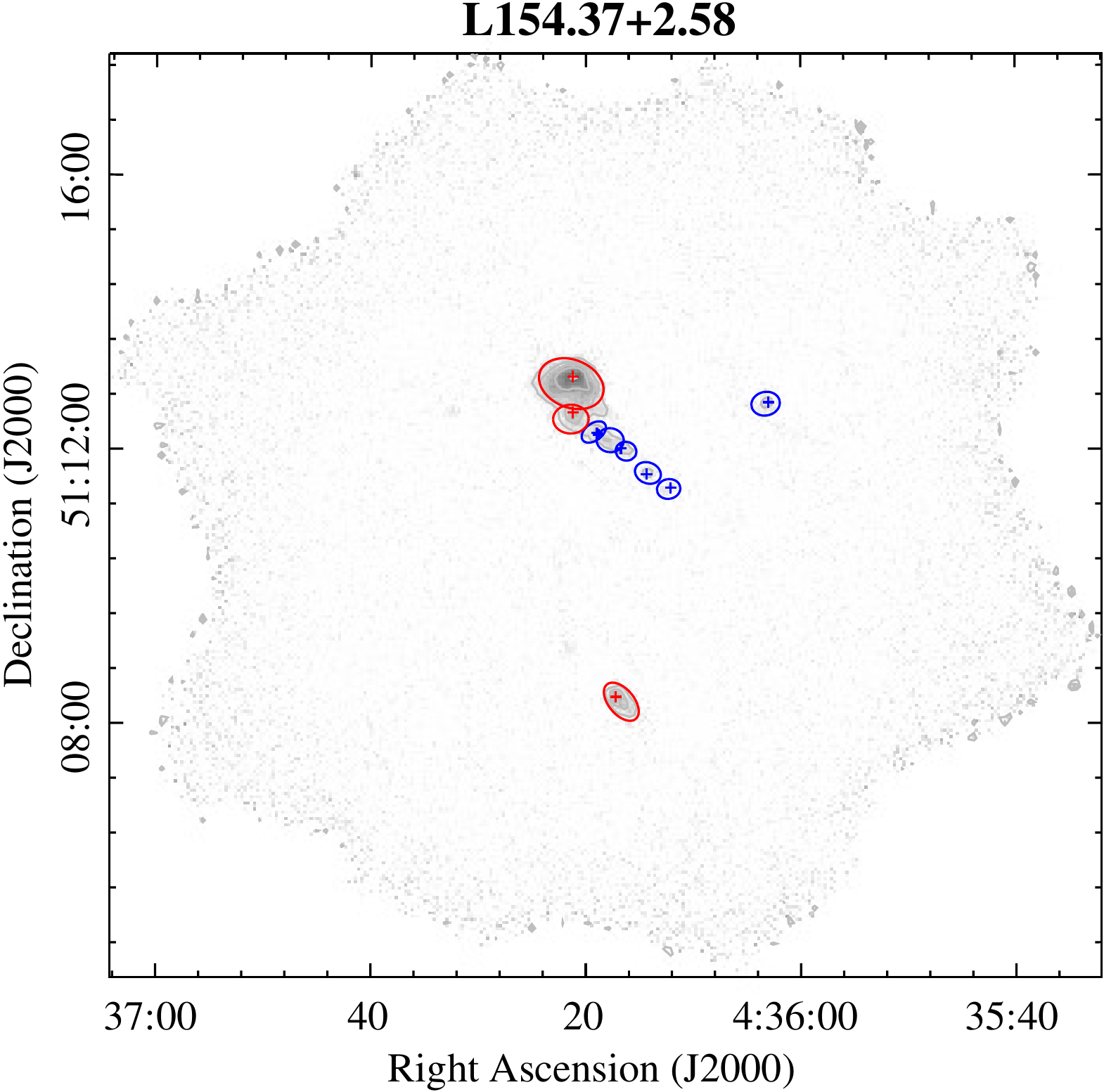}
}\\
\subfloat[L169.18-0.89 map, $\sigma_{rms}=232$ mJy beam$^{-1}$.]{
\includegraphics[scale=0.43]{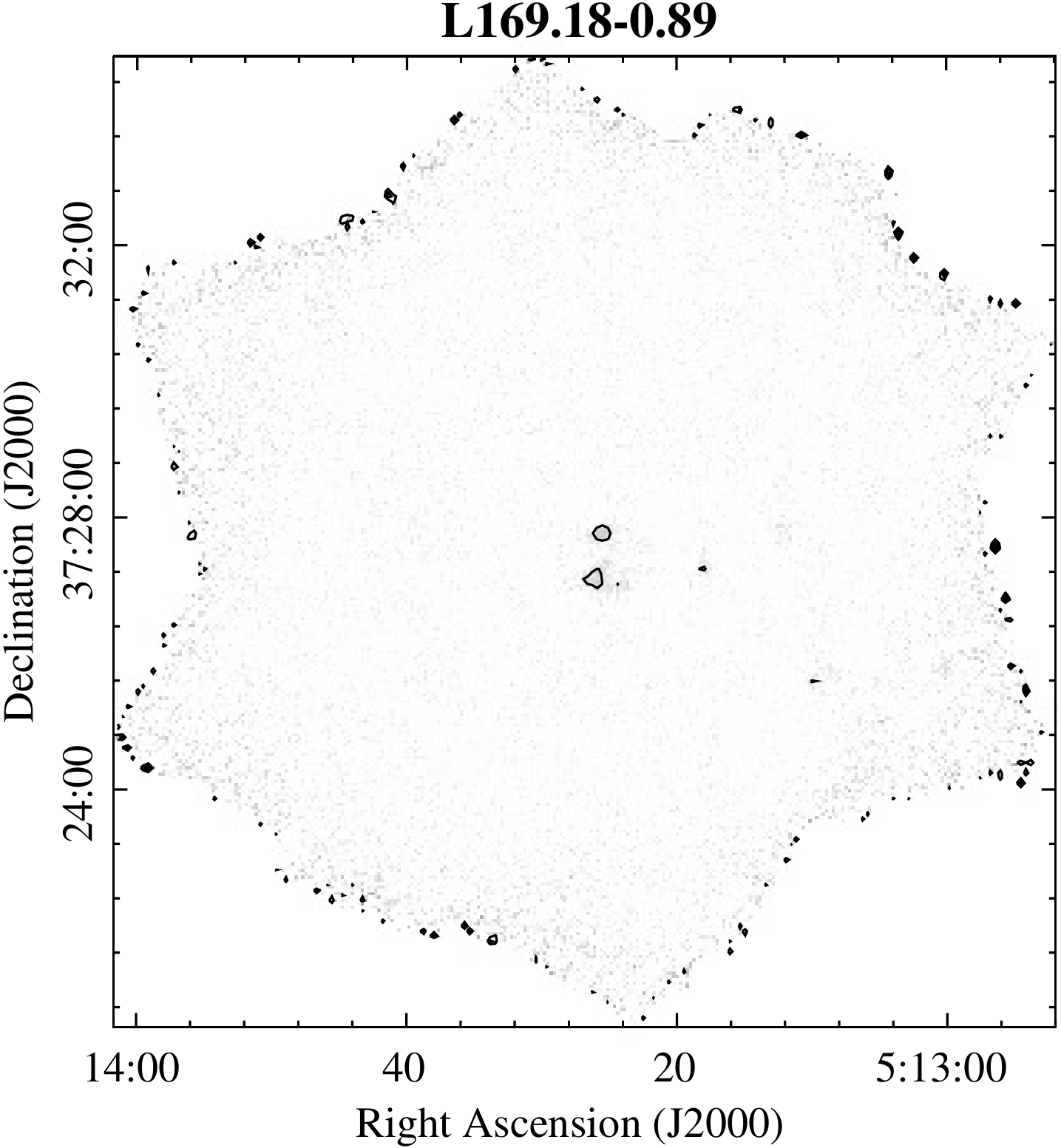}
\includegraphics[scale=0.43]{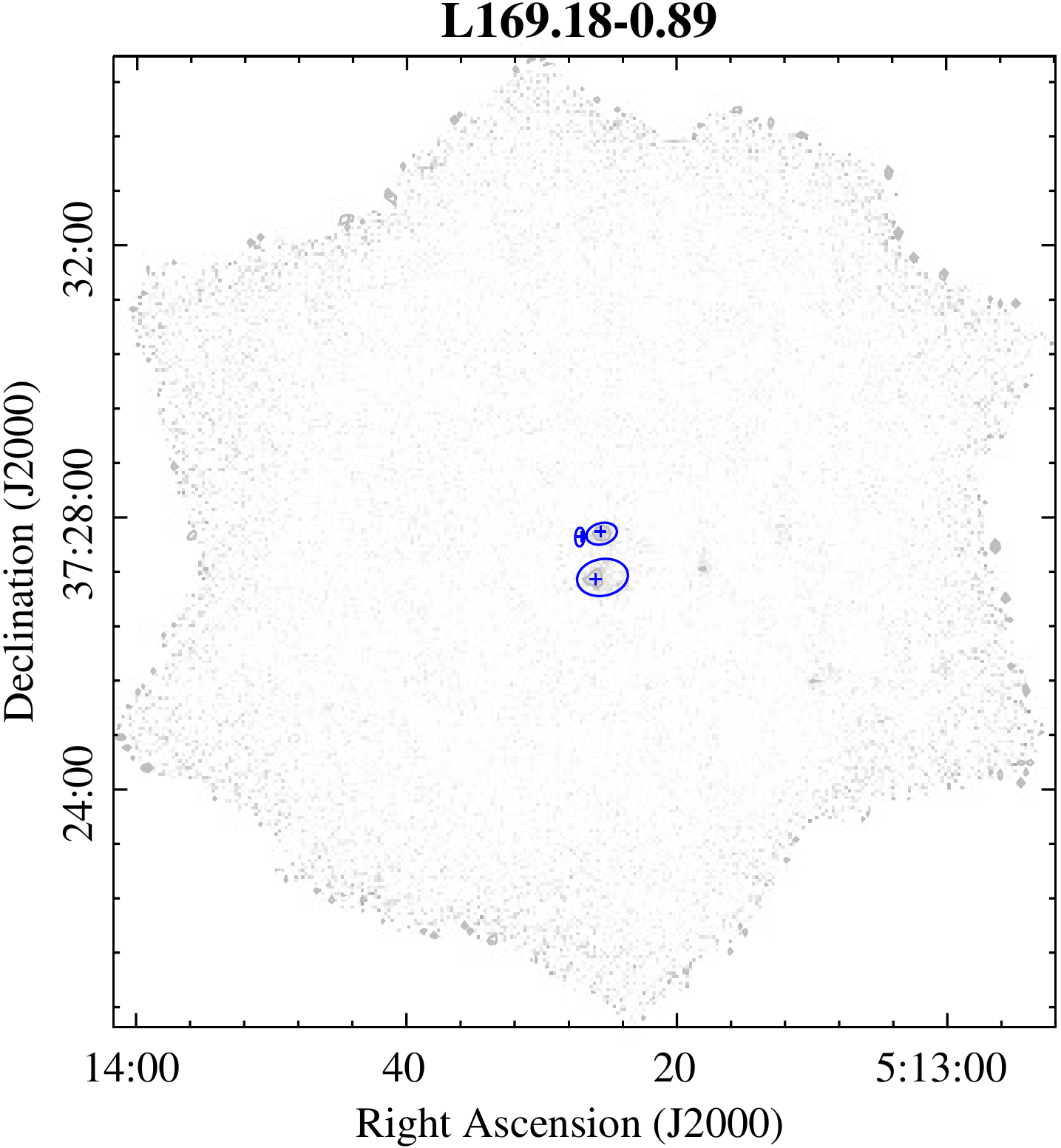}
}\\
\caption{Continuation}
\end{figure}

\clearpage
\begin{figure}\ContinuedFloat 
\center
\subfloat[L172.88+2.27 map, $\sigma_{rms}=315$ mJy beam$^{-1}$.]{
\includegraphics[scale=0.43]{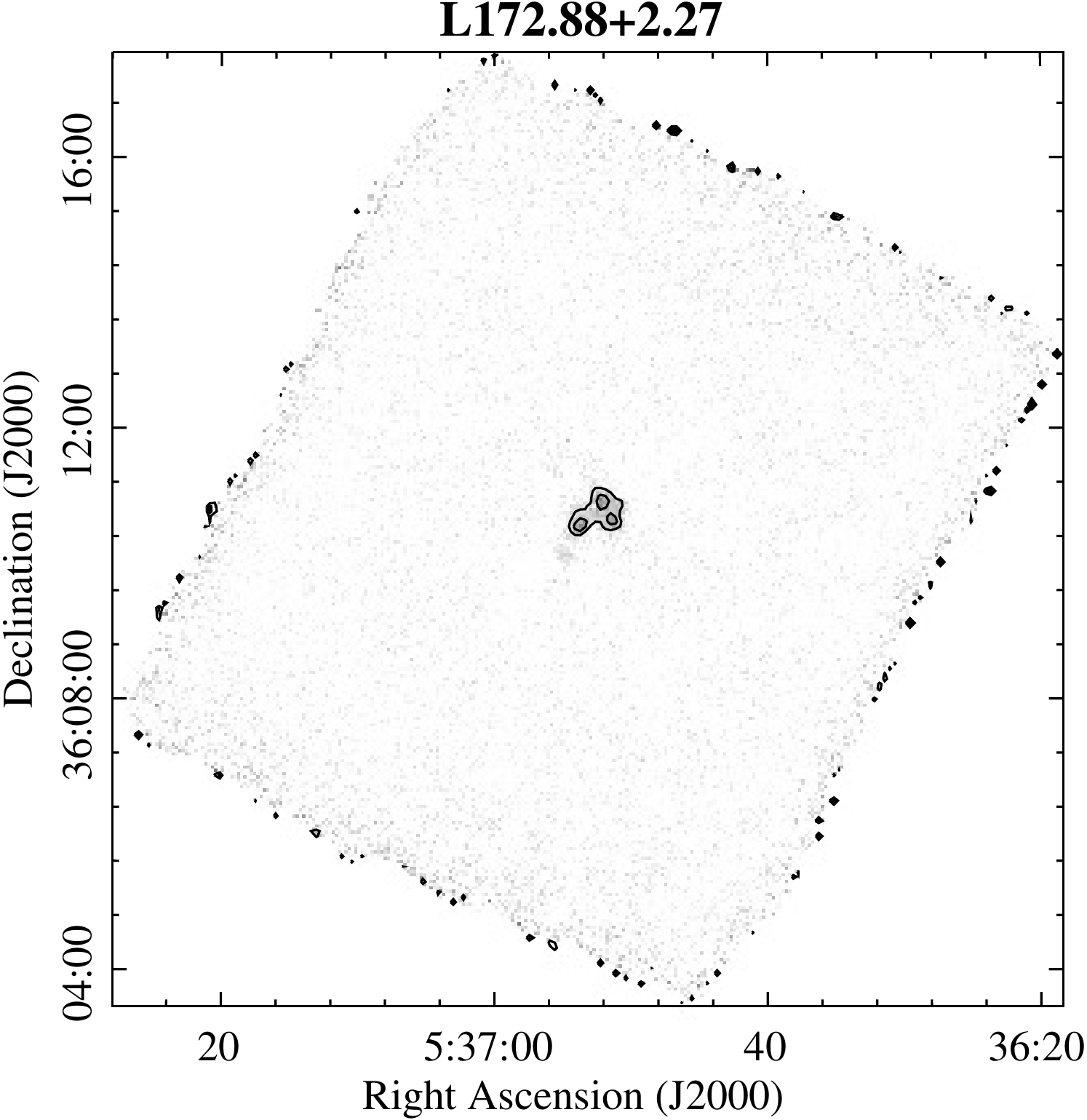}
\includegraphics[scale=0.43]{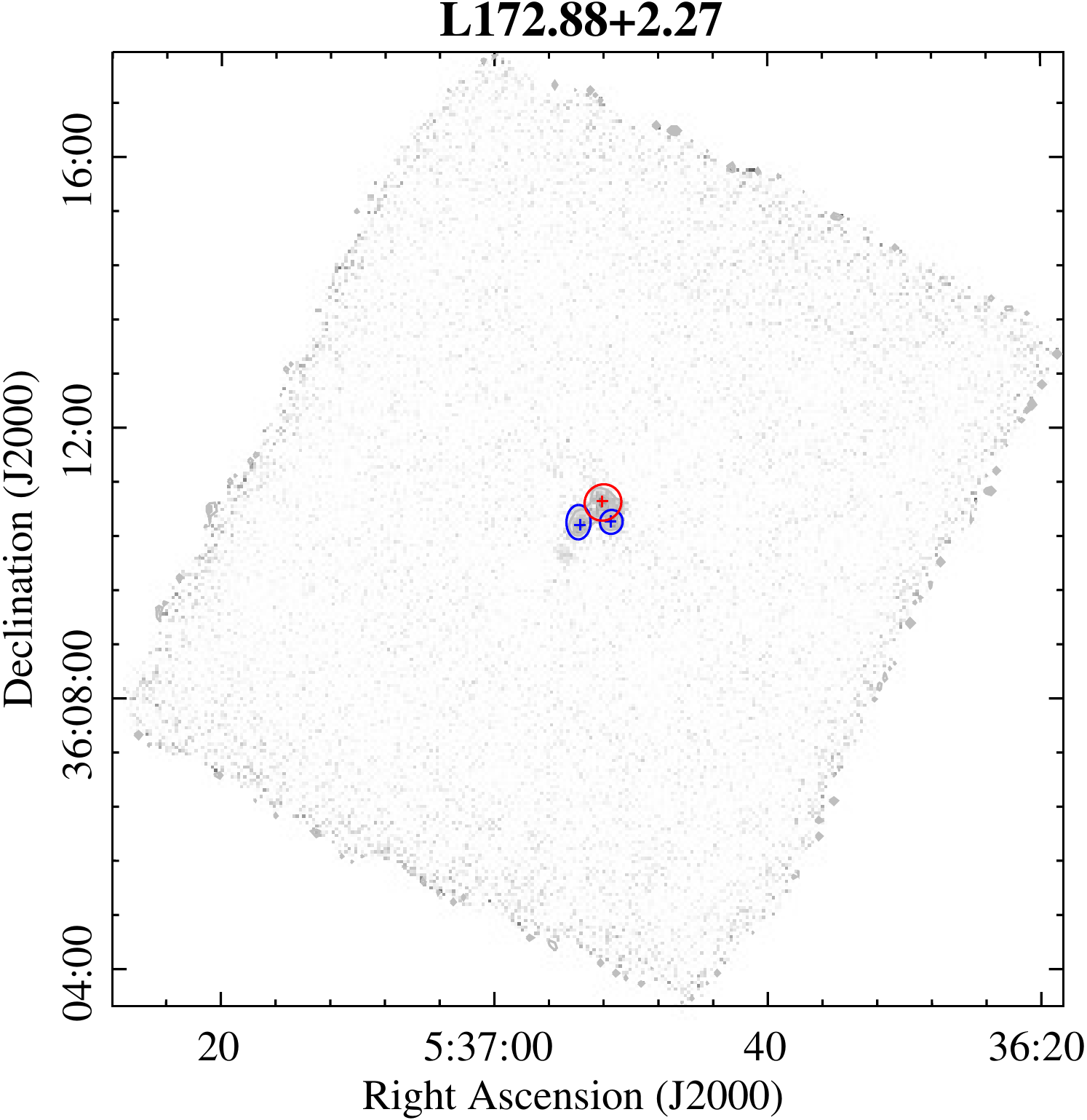}
}\\
\subfloat[L211.53-19.27 map, $\sigma_{rms}=254$ mJy beam$^{-1}$.]{
\includegraphics[scale=0.43]{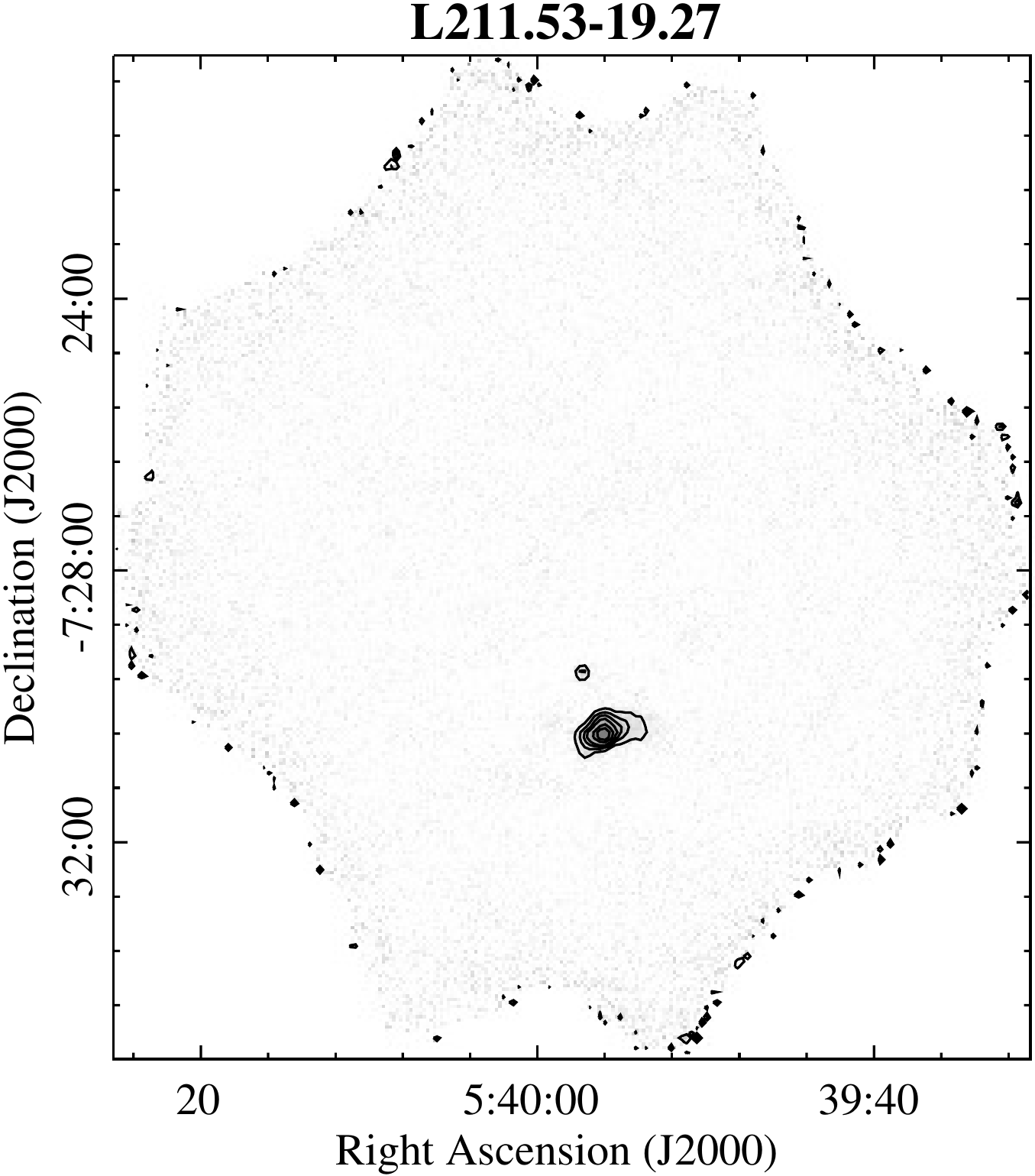}
\includegraphics[scale=0.43]{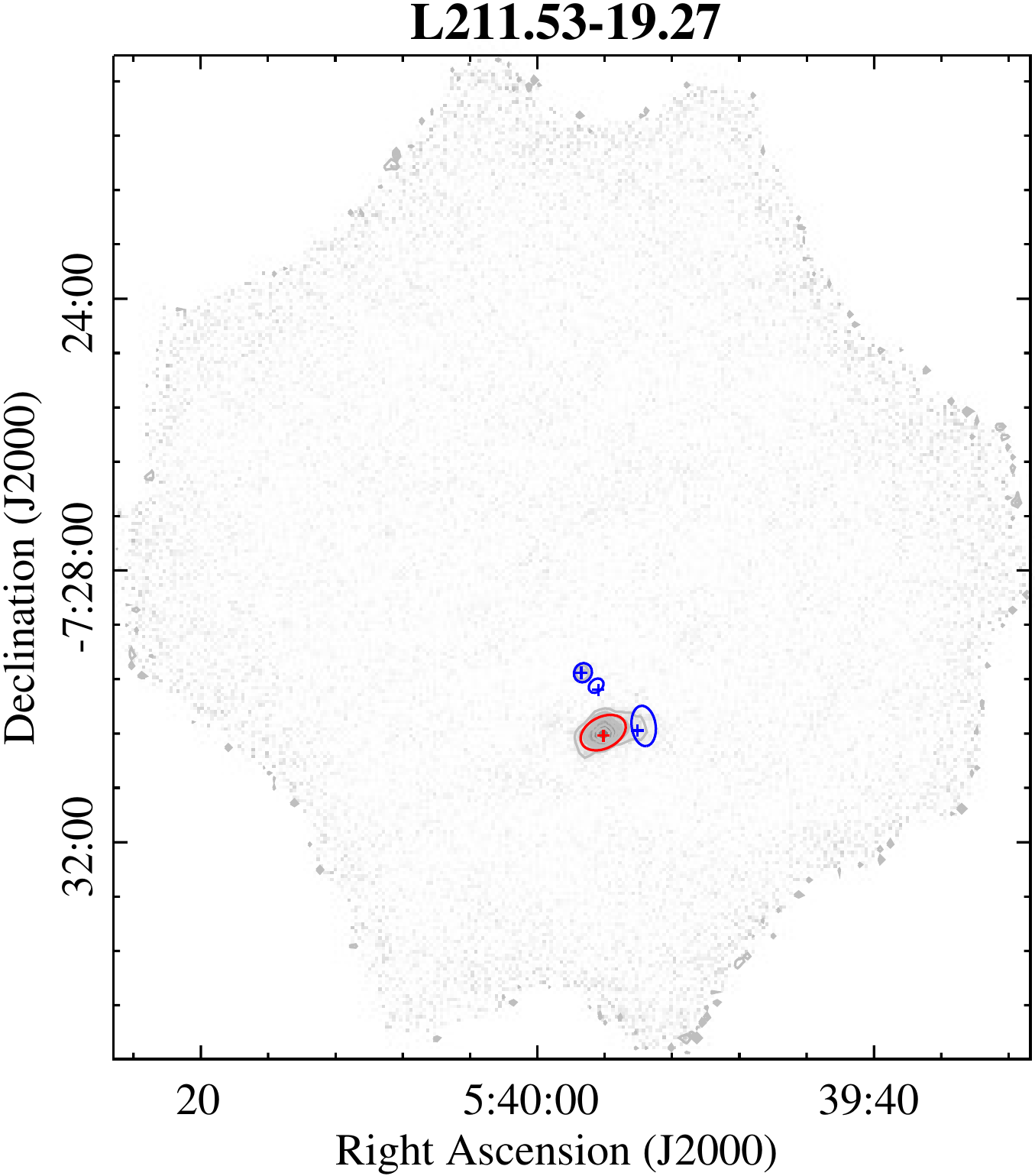}
}\\
\subfloat[L111.42+0.76 map, $\sigma_{rms}=375$ mJy beam$^{-1}$.]{
\includegraphics[scale=0.43]{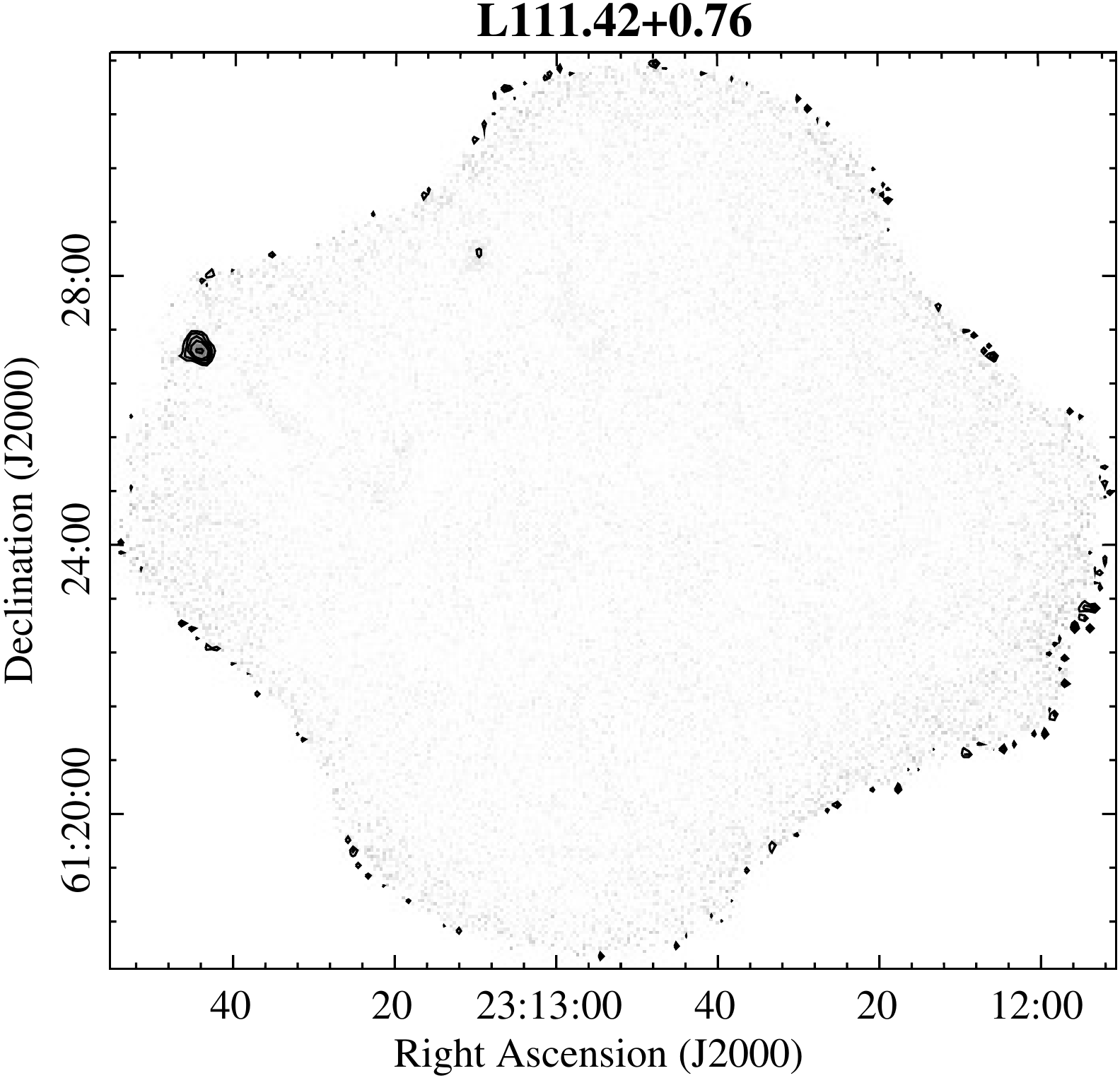}
\includegraphics[scale=0.43]{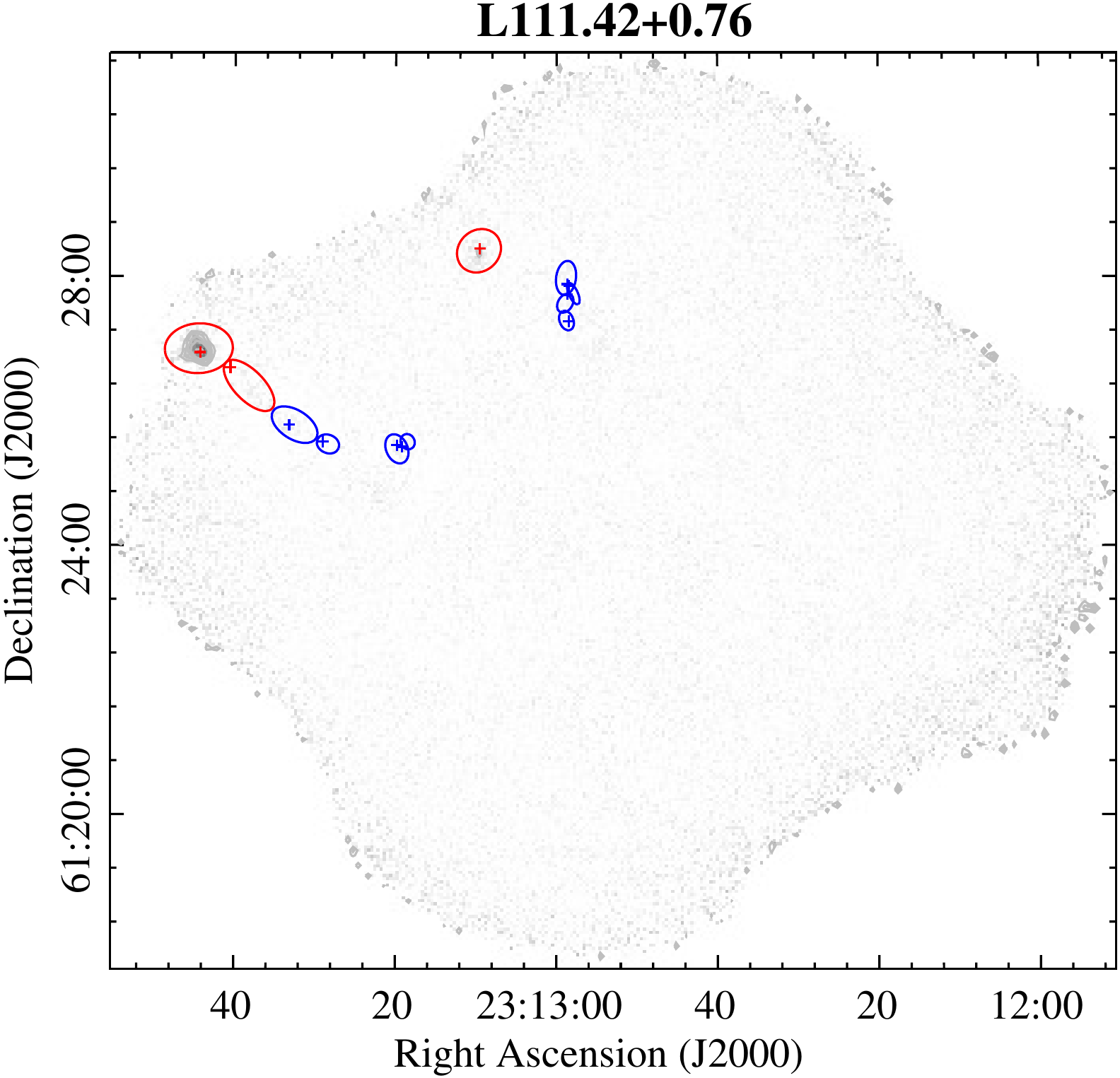}
}\\
\caption{Continuation}
\end{figure}

\clearpage
\begin{figure}\ContinuedFloat 
\center
\subfloat[L111.88+0.99 map, $\sigma_{rms}=226$ mJy beam$^{-1}$.]{
\includegraphics[scale=0.43]{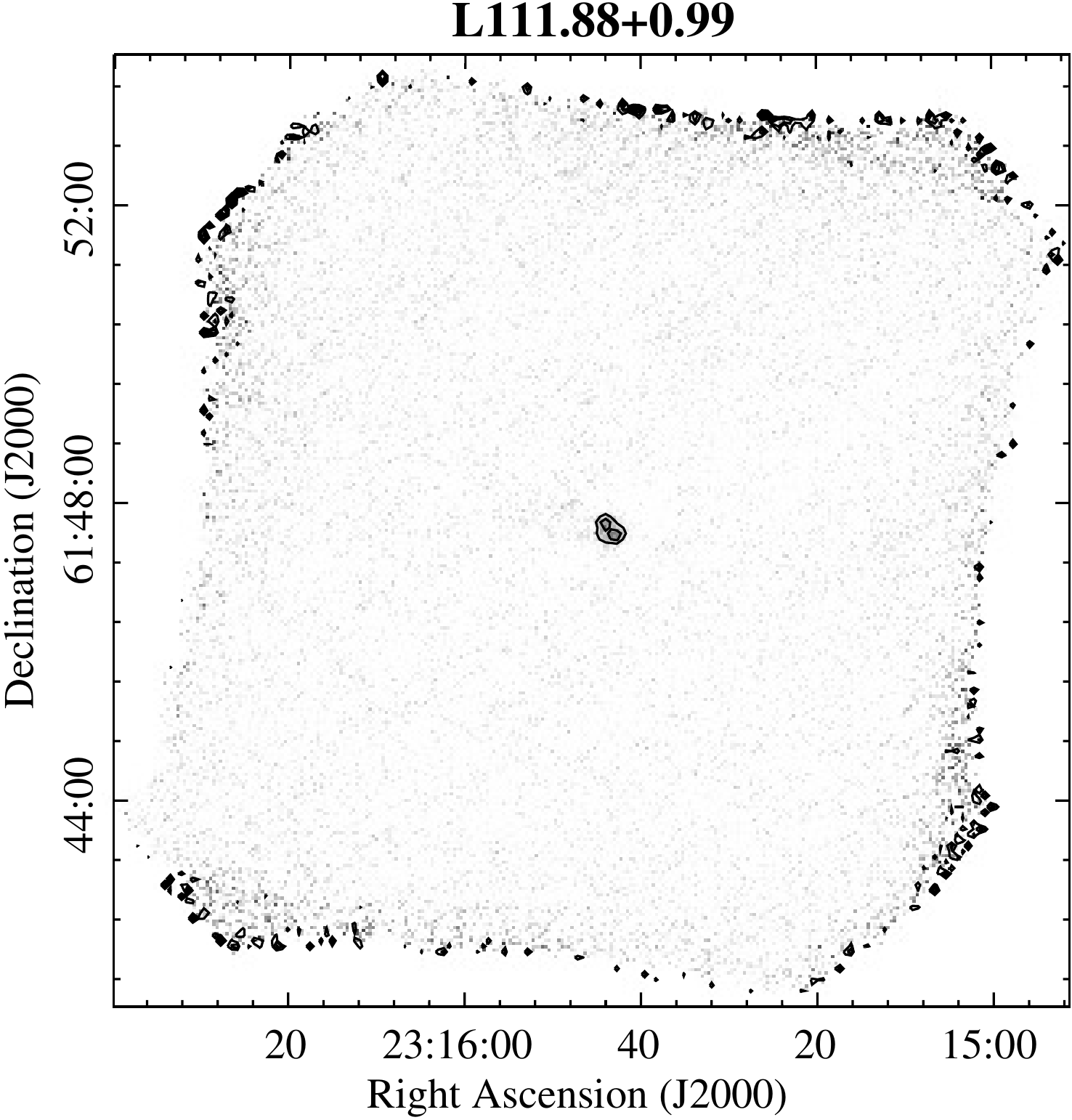}
\includegraphics[scale=0.43]{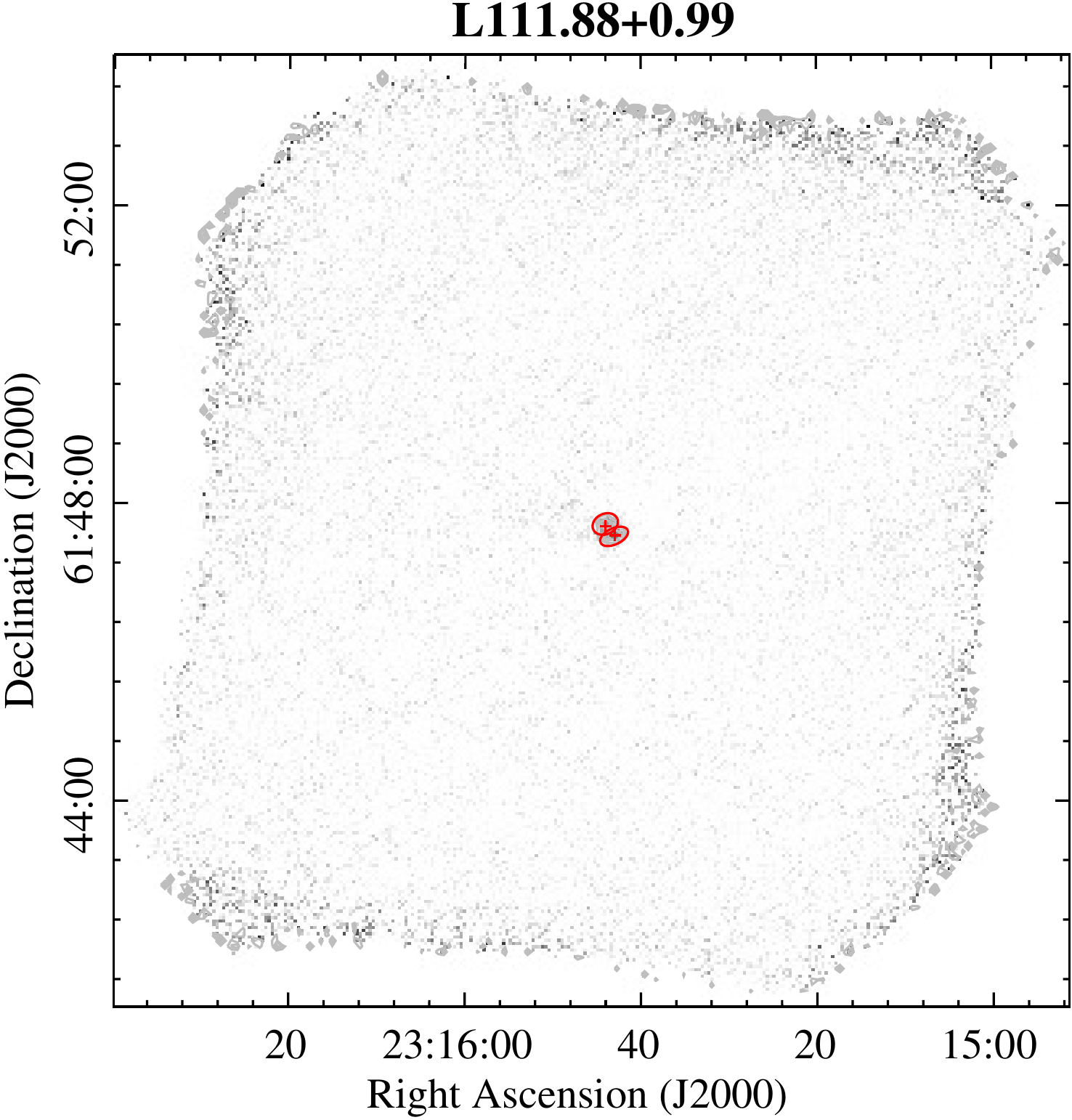}
}\\
\subfloat[L134.20+0.75 map, $\sigma_{rms}=525$ mJy beam$^{-1}$.]{
\includegraphics[scale=0.43]{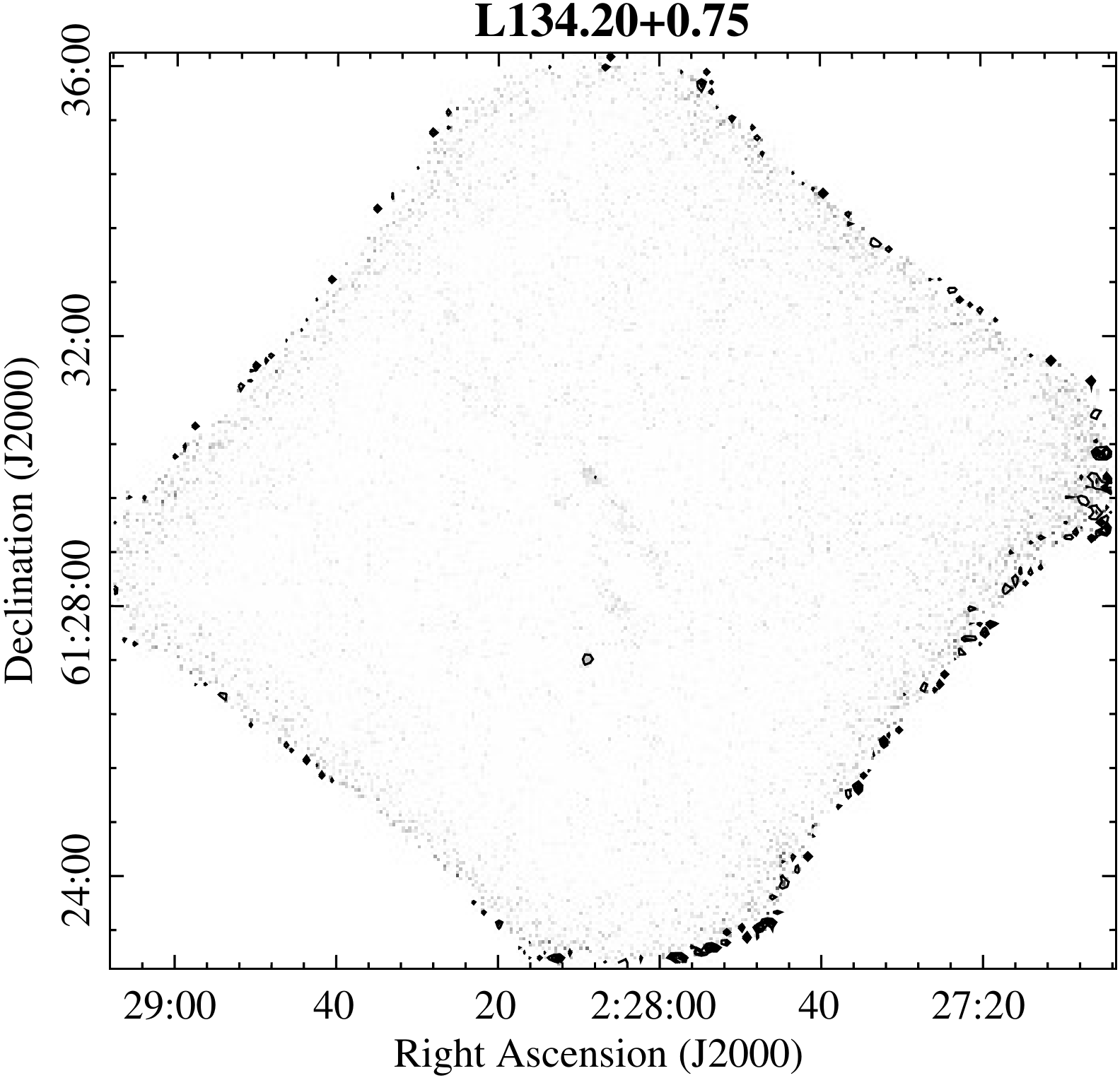}
\includegraphics[scale=0.43]{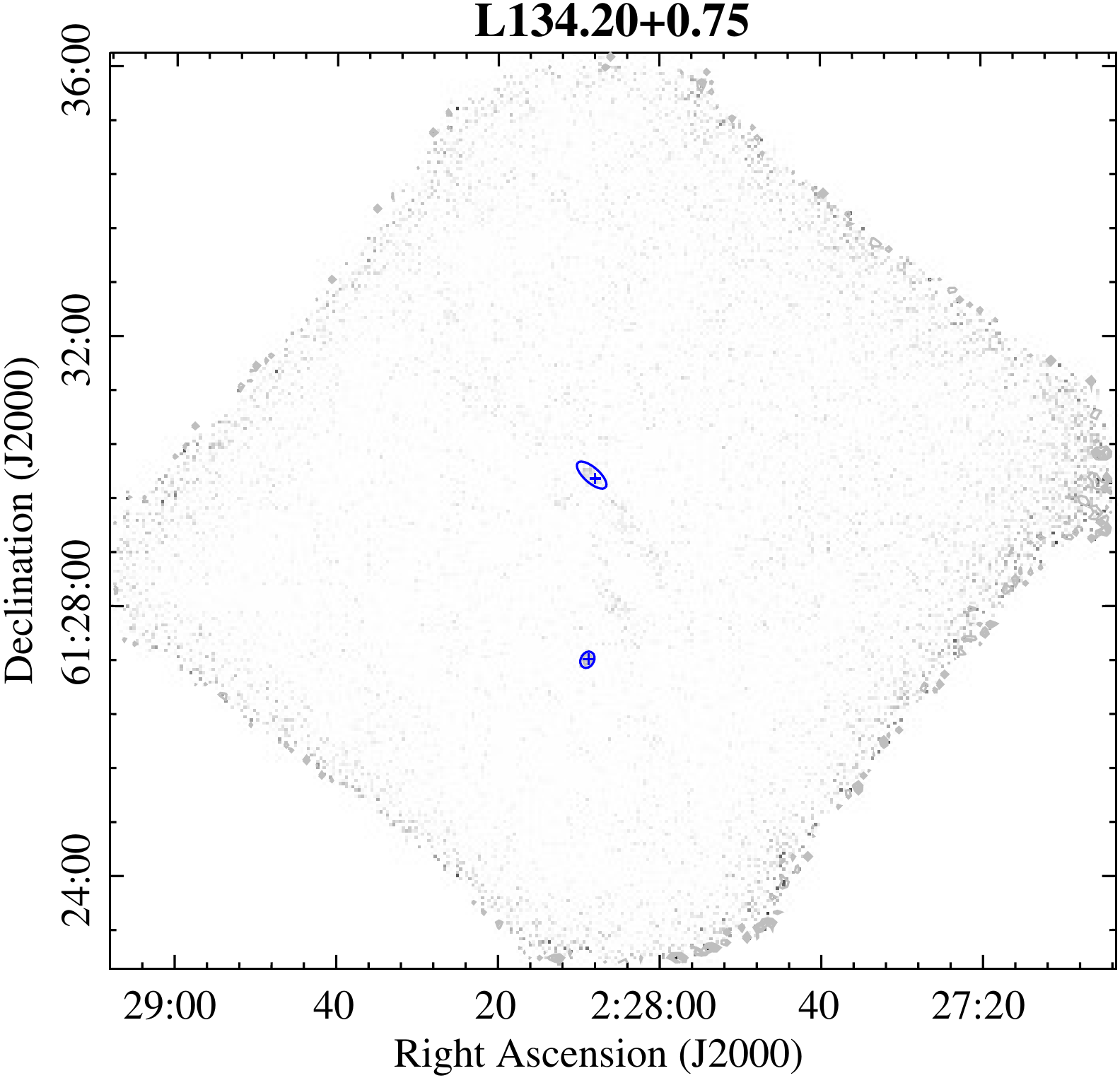}
}\\
\subfloat[L189.85+0.50 map, $\sigma_{rms}=506$ mJy beam$^{-1}$.]{
\includegraphics[scale=0.43]{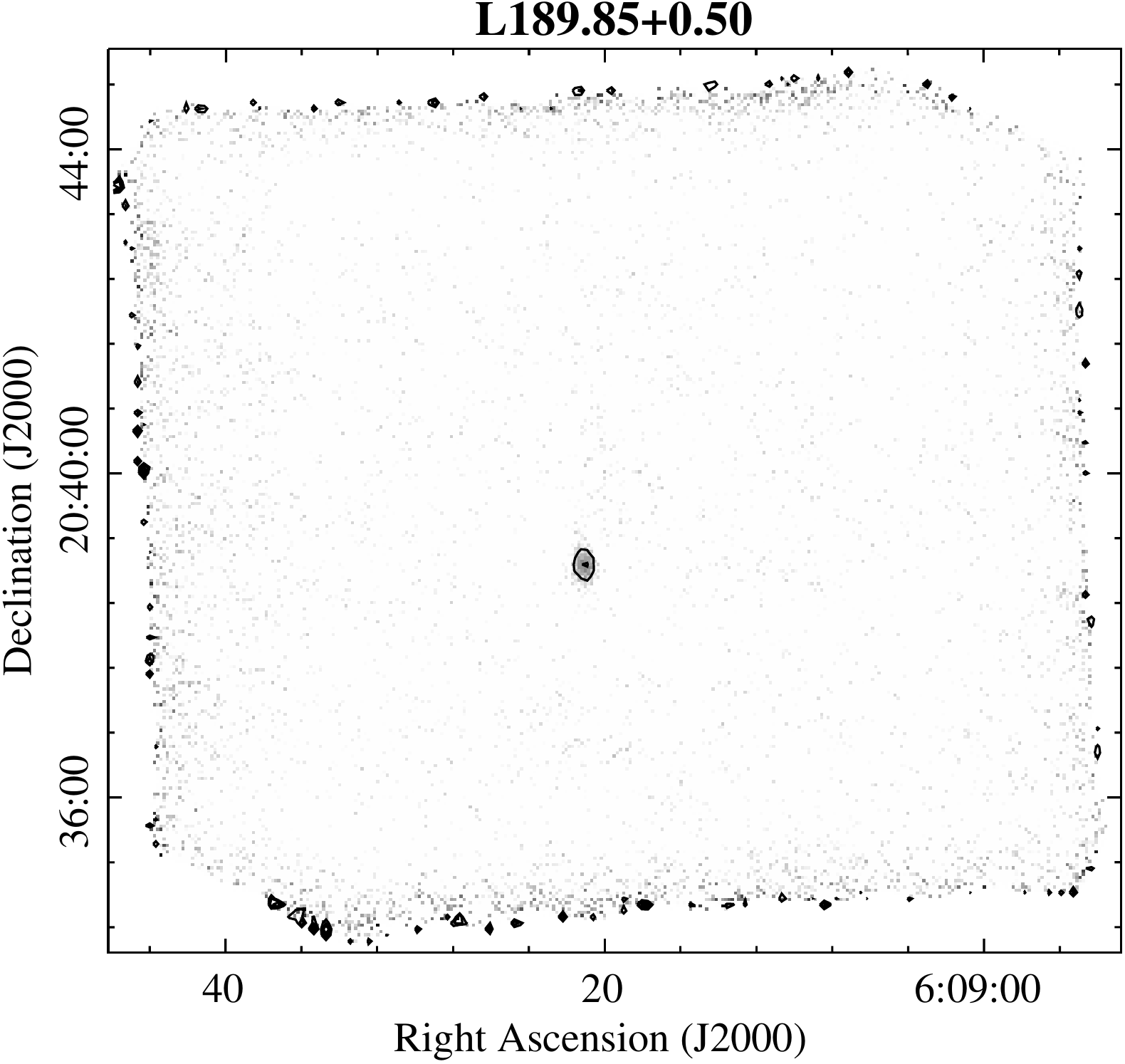}
\includegraphics[scale=0.43]{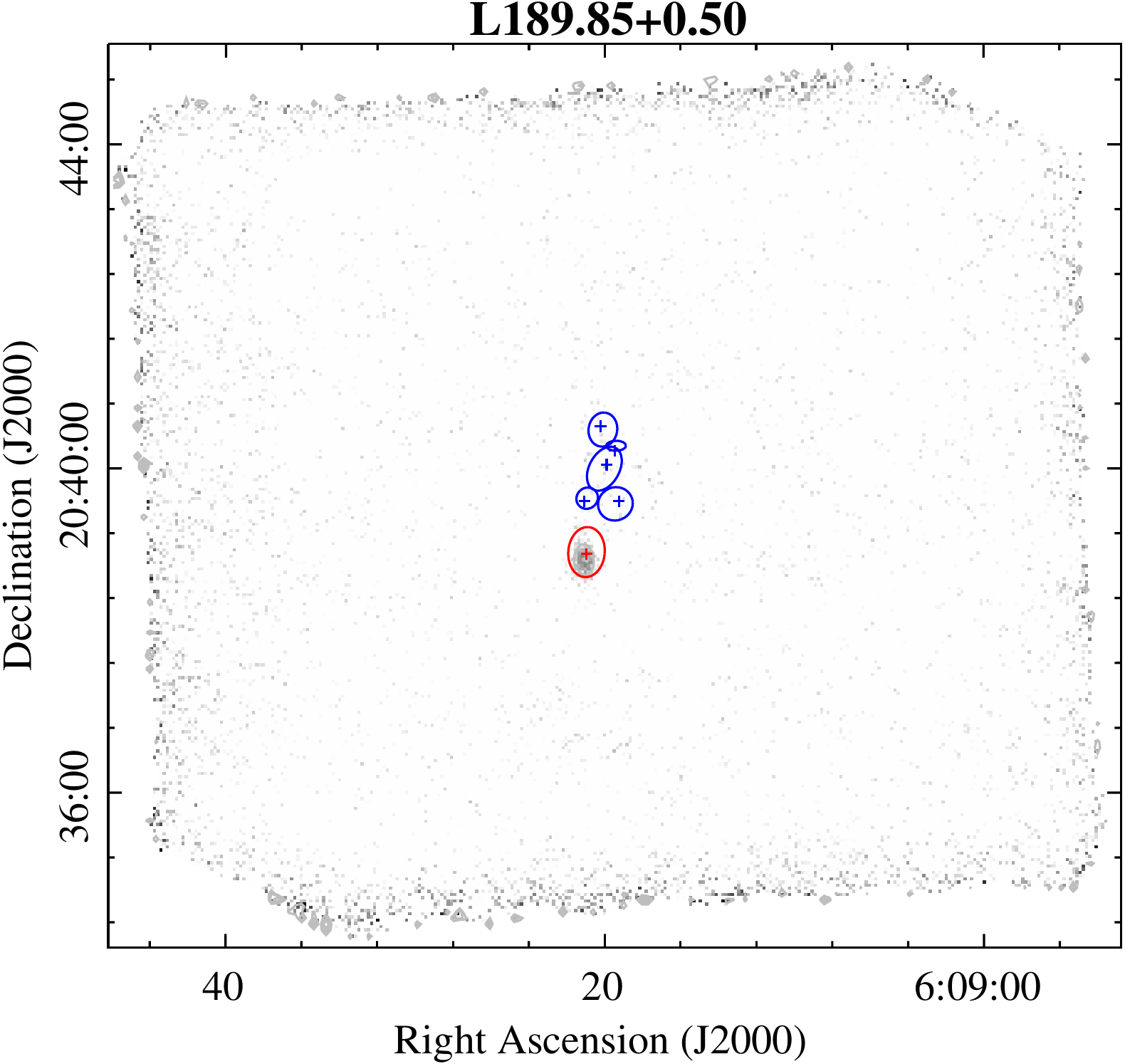}
}\\
\caption{Continuation}
\end{figure}

\clearpage
\begin{figure}\ContinuedFloat 
\center
\subfloat[L189.94+0.34 map, $\sigma_{rms}=590$ mJy beam$^{-1}$.]{
\includegraphics[scale=0.43]{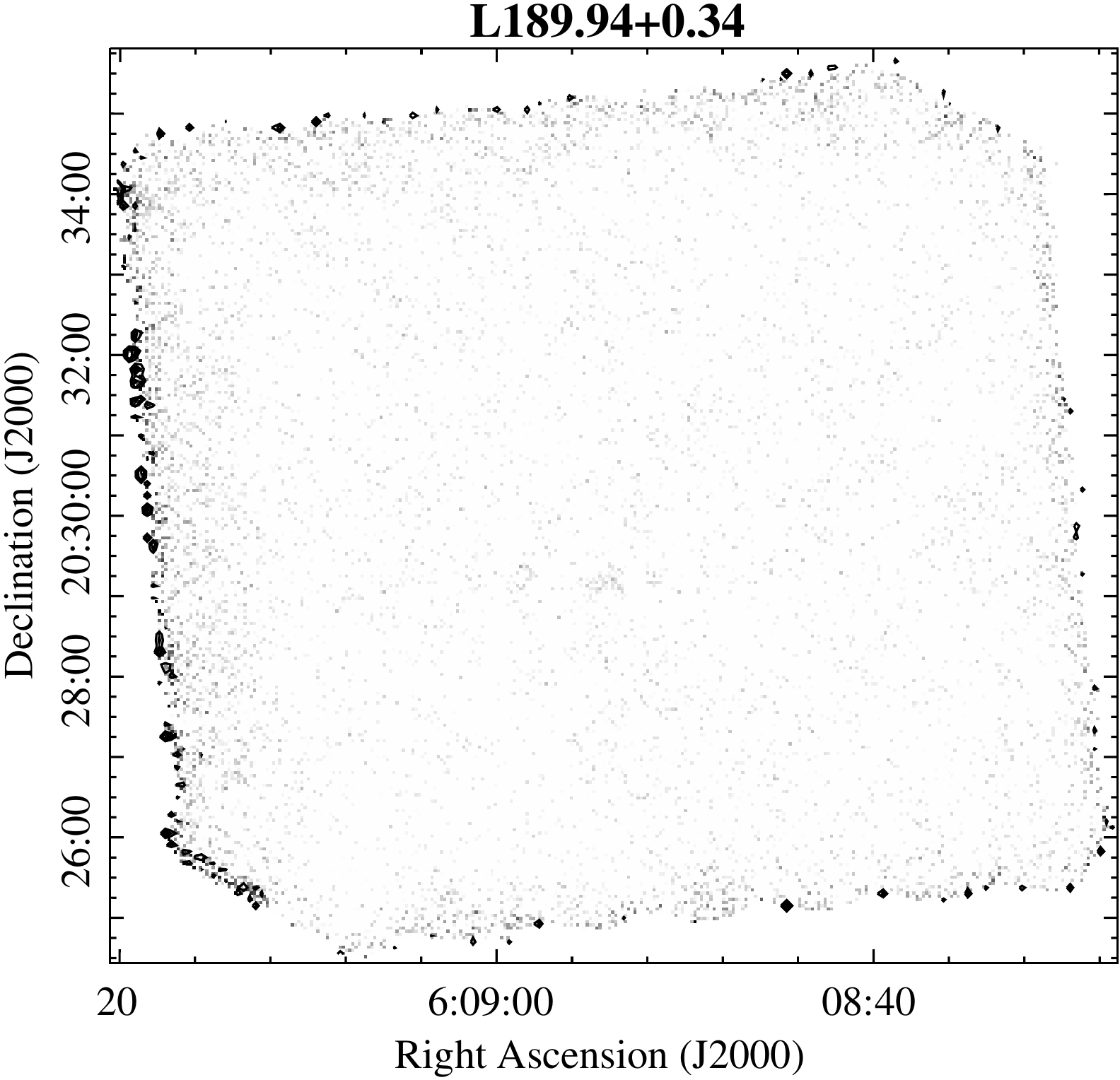}
\includegraphics[scale=0.43]{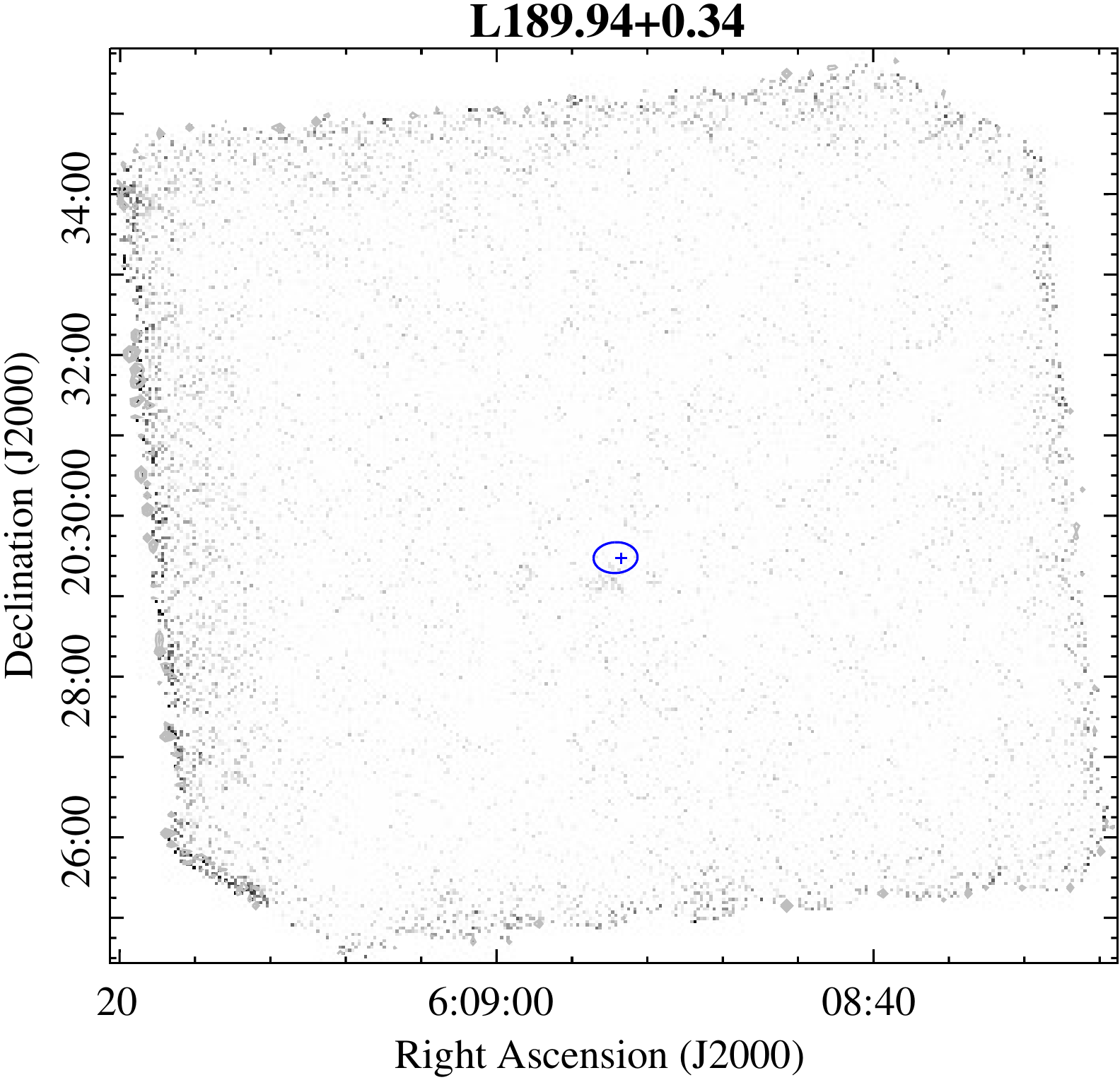}
}\\
\subfloat[L206.60-16.37 map, $\sigma_{rms}=563$ mJy beam$^{-1}$.]{
\includegraphics[scale=0.43]{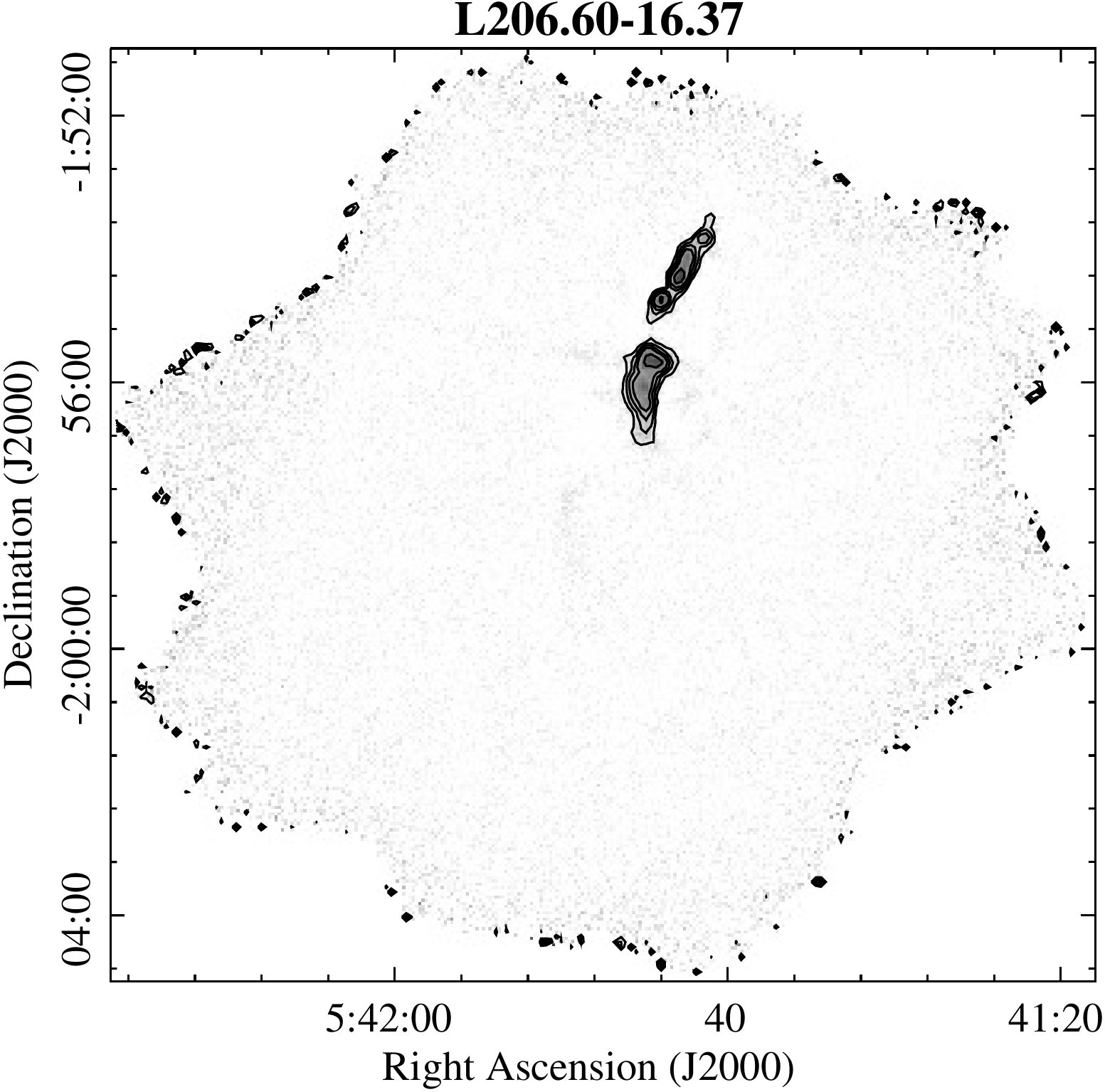}
\includegraphics[scale=0.43]{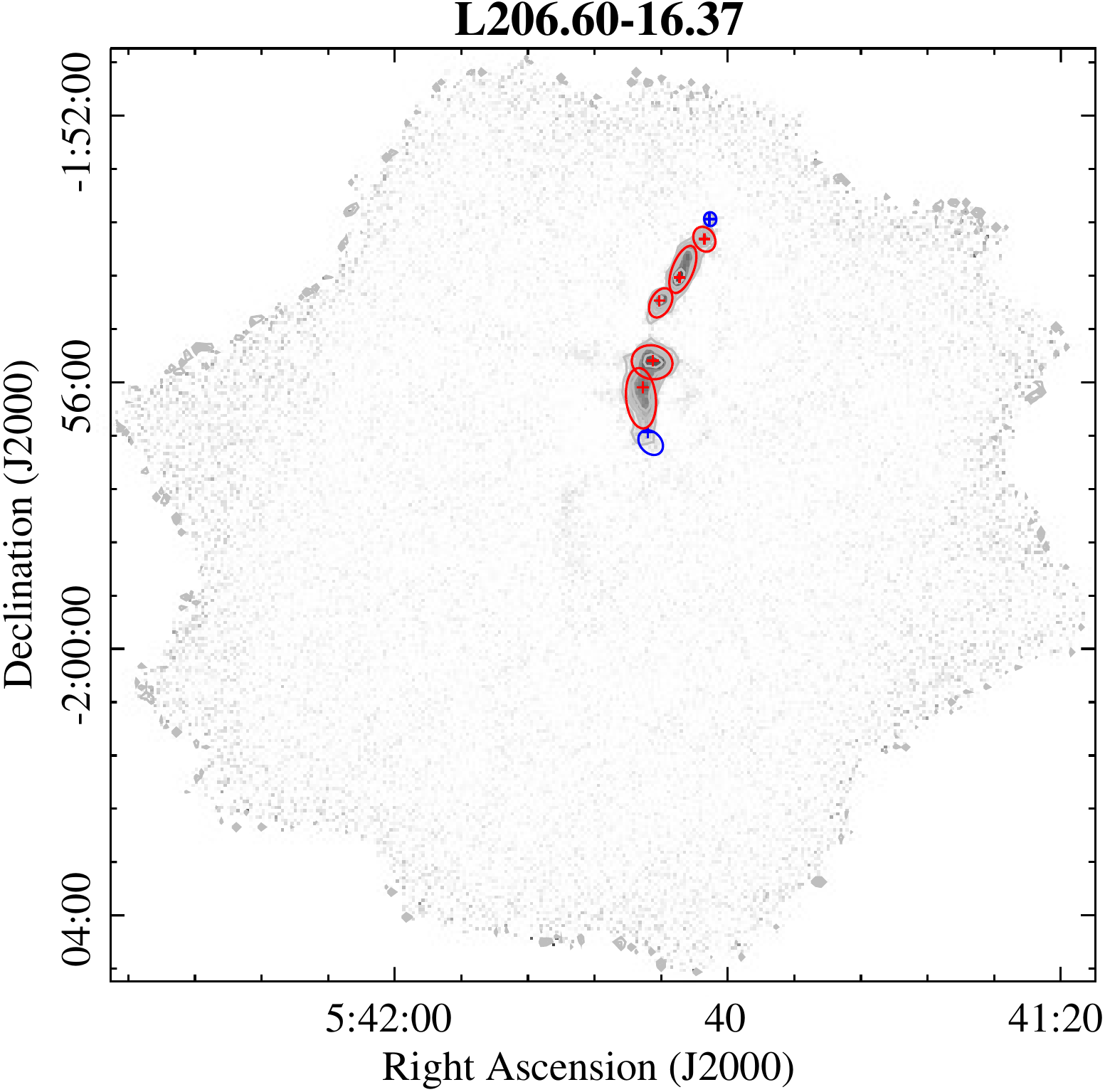}}\\
\caption{Continuation}
\end{figure}
\clearpage
\section{Hi-Gal SPIRE images at 350 $\mu$\textit{\MakeLowercase{m}} for $\textit{\MakeLowercase{l}}=30\arcdeg$}

In this section we indicate the locations of the high-resolution sources identified in the 350 \um\ SHARC-II maps (8.5\arcsec beam size) toward $l=30\arcdeg$, and their comparison with structures observed at lower resolution (24.9\arcsec) in the corresponding \textit{Herschel}/SPIRE maps at the same wavelength and BGPS maps at 1.1 mm at a resolution of 33\arcsec. Figure~\ref{fig:bgps_herschel1} shows 1.1 mm continuum gray scale maps with green contours representing sources from the BGPS catalog, and \textit{Herschel} grey scale maps with contours representing the emission at 10$\sigma$, and increasing steps of 5$\sigma$, with $\sigma$=164 MJy sr$^{-1}$. The thick black line on each map represent the area covered by the SHARC-II maps, with the name of those maps indicated for each region. Peak positions of detected high-resolution sources from SHARC-II maps are overlaid in red and blue crosses, representing sources with signal-to-noise above and below a value of 10, respectively. Most sources with high signal-to-noise are well associated with strong emission in the \textit{Herschel} maps, but several of the low signal-to-noise sources are not individually identified and they are part of extended background emission detected in \textit{Herschel} maps but not recovered in SHARC-II maps.

\begin{figure}
   \centering
\subfloat[Thick black contour in Herschel map represents the area covered by maps L029.95-0.05 and L030.00+00.]{%
   	\includegraphics[width=0.85\textwidth]{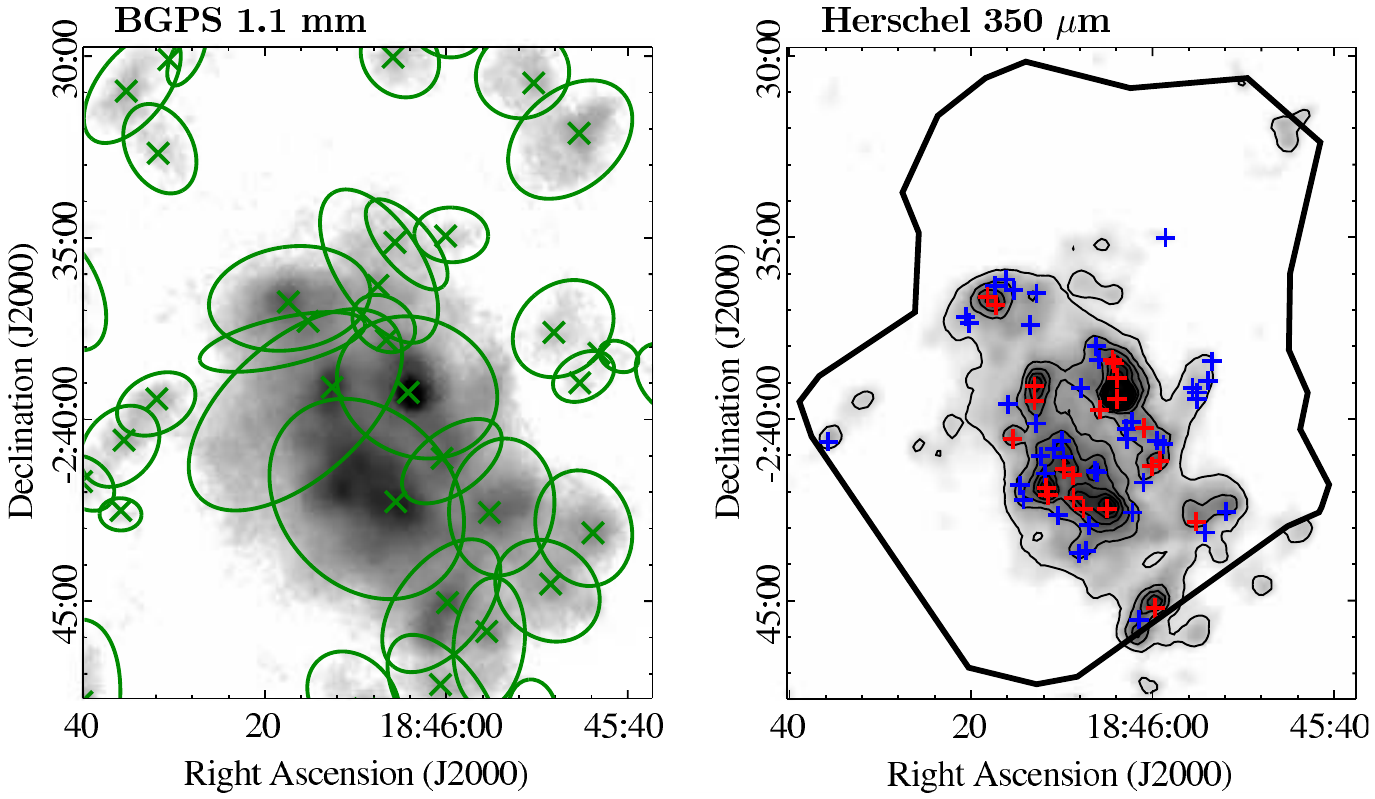}

    }\\%
        \subfloat[Thick black contour in Herschel map represents the area covered by maps L030.15+0.00, L030.30+0.00 and L030.45+0.00.]{%
      	\includegraphics[width=0.85\textwidth]{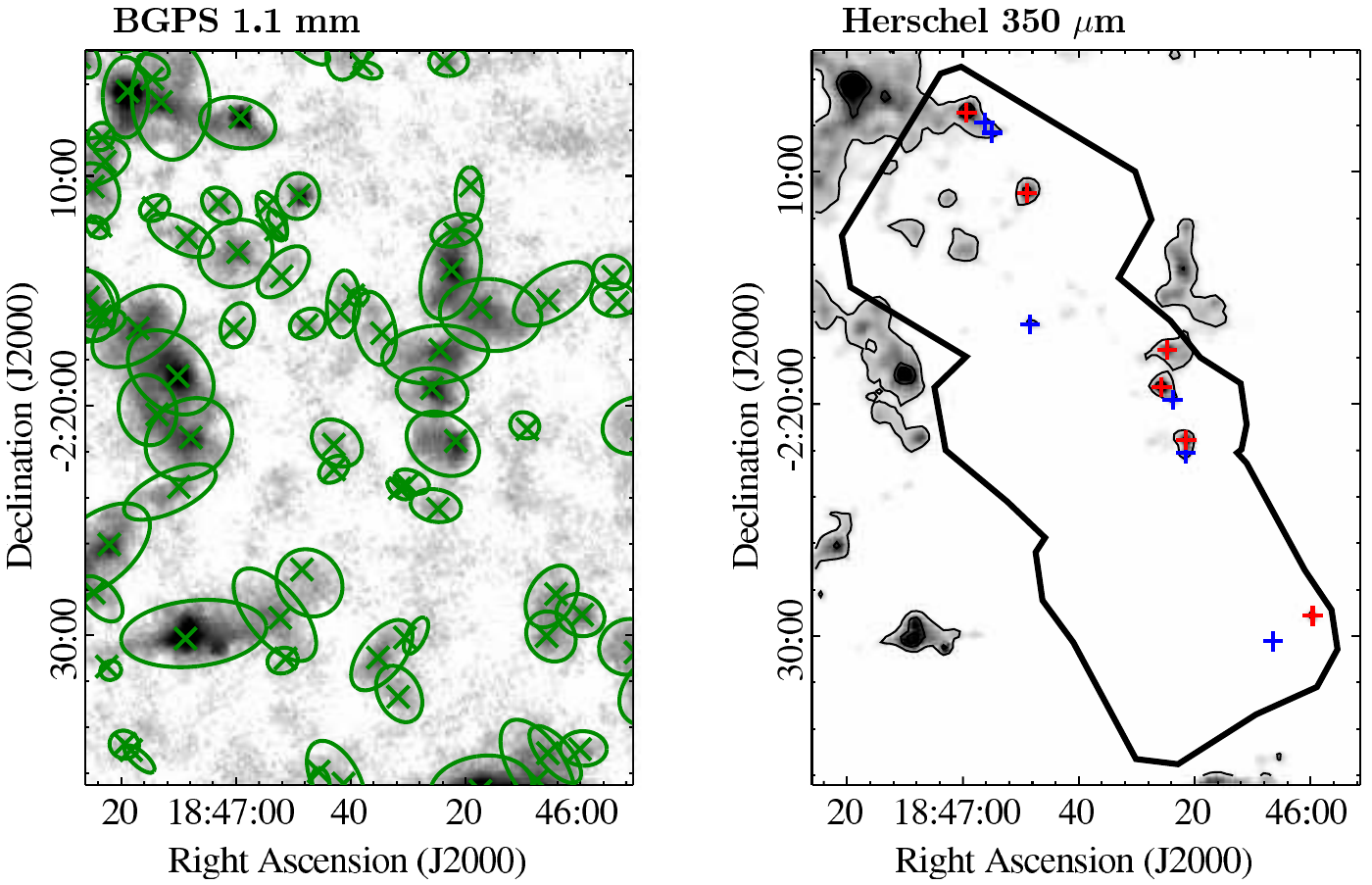}
    }%
   \caption{Left: BGPS map at 1.1 mm  toward $l=30\arcdeg$. The ellipses show the recovered sources from the BGPS catalog. Right: Corresponding 350 $\mu$m maps from \textit{Herschel} SPIRE. Contours start from 10$\sigma$, with increasing steps of 5$\sigma$ ($\sigma$=164 MJy sr$^{-1}$) The crosses show the peak position of the high-resolution sources recovered in SHARC-II map. Red and blue crosses show substructures with peak signal-to-noise above and below 10, respectively. The area covered by SHARC-II maps, indicated under each figure, is represented by the thick black contour. }
  \label{fig:bgps_herschel1}
\end{figure}

\clearpage

\begin{figure}\ContinuedFloat 
   \centering
   \subfloat[Thick black contour in Herschel map represents the area covered by maps L030.60+0.00, L030.70-0.07 and L030.80-0.05]{%
   \includegraphics[angle=90,width=0.95\textwidth]{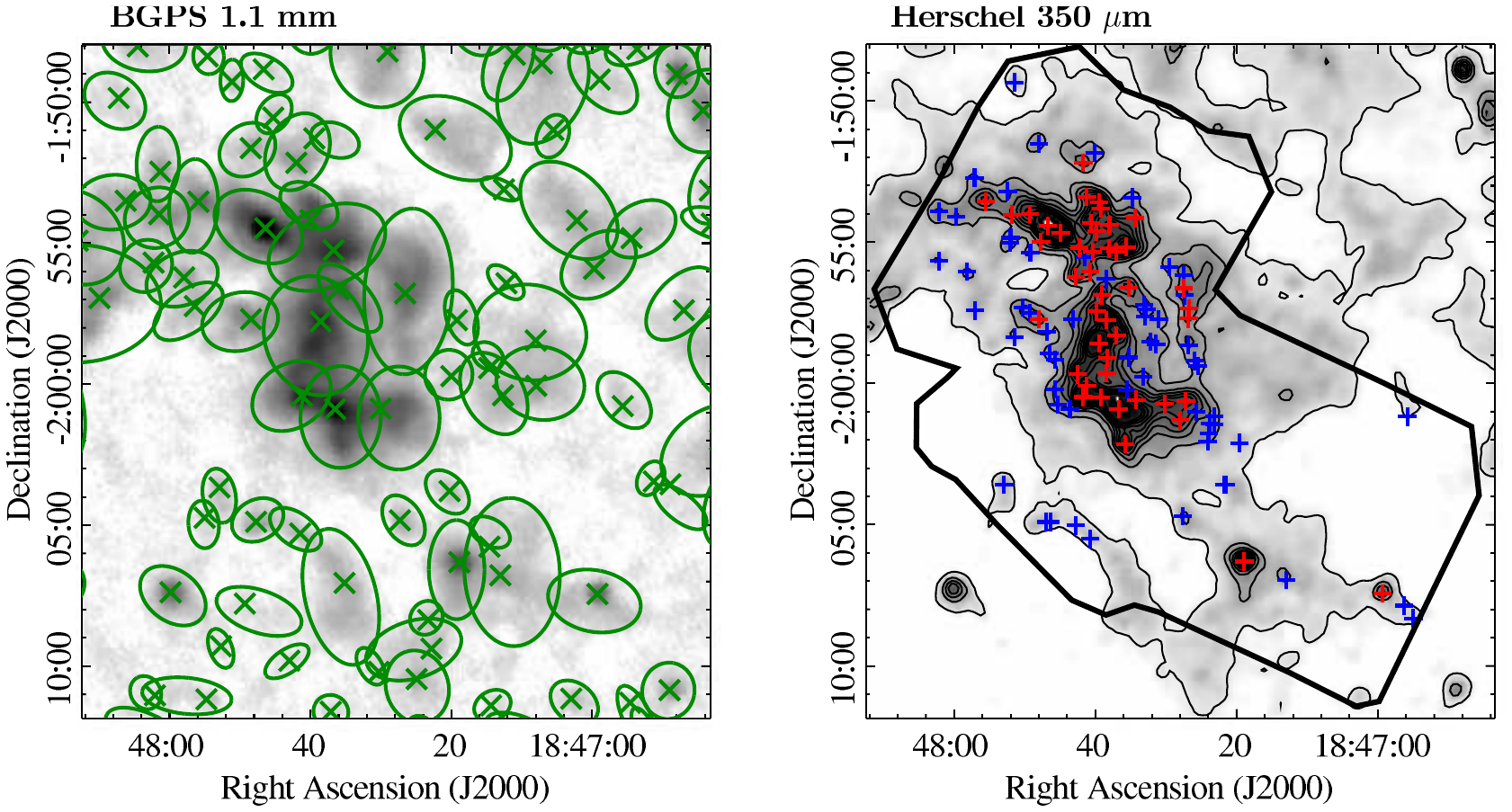}

    }\\%

\subfloat[Thick black contour in Herschel map represents the area covered by map L030.88+0.13.]{%
   \includegraphics[width=0.75\textwidth]{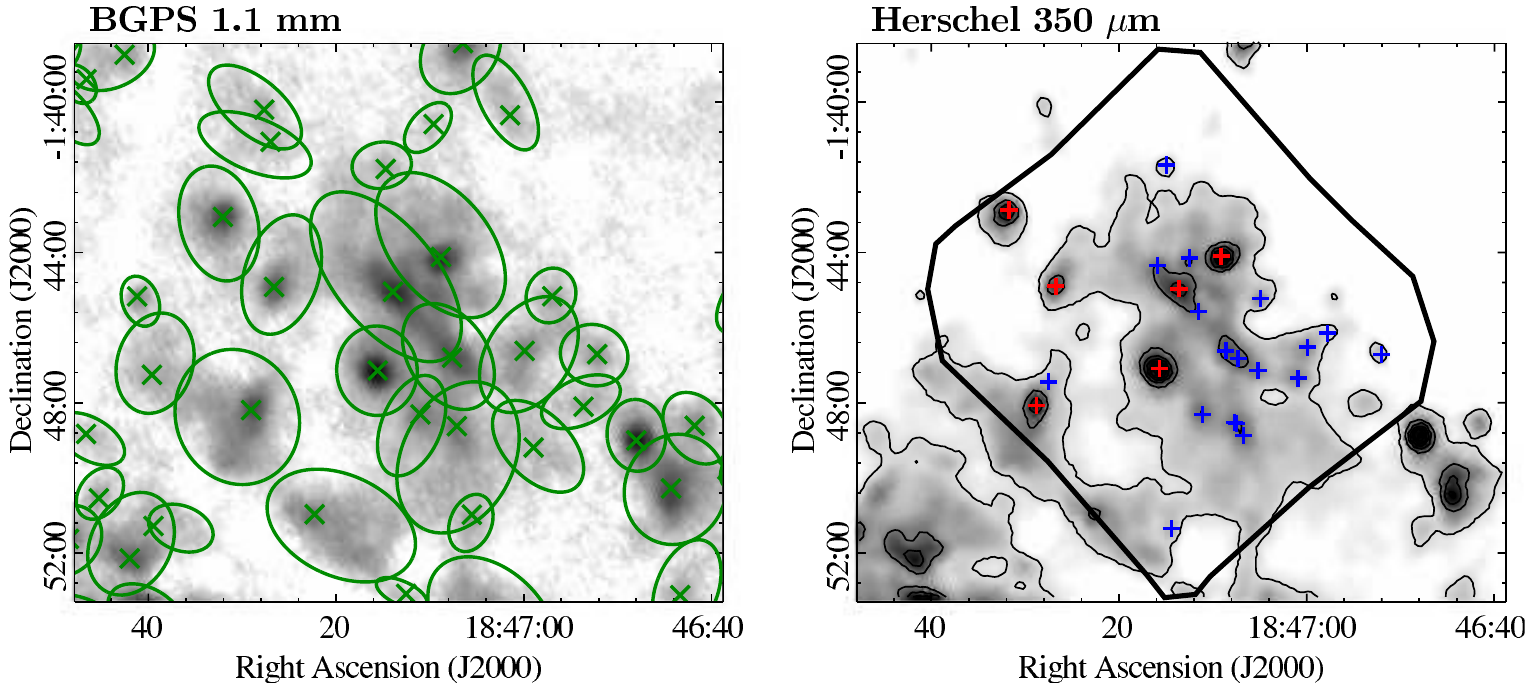}

    }\\%
        \subfloat[Thick black contour in Herschel map represents the area covered by map L030.61+0.16.]{%
   \includegraphics[width=0.75\textwidth]{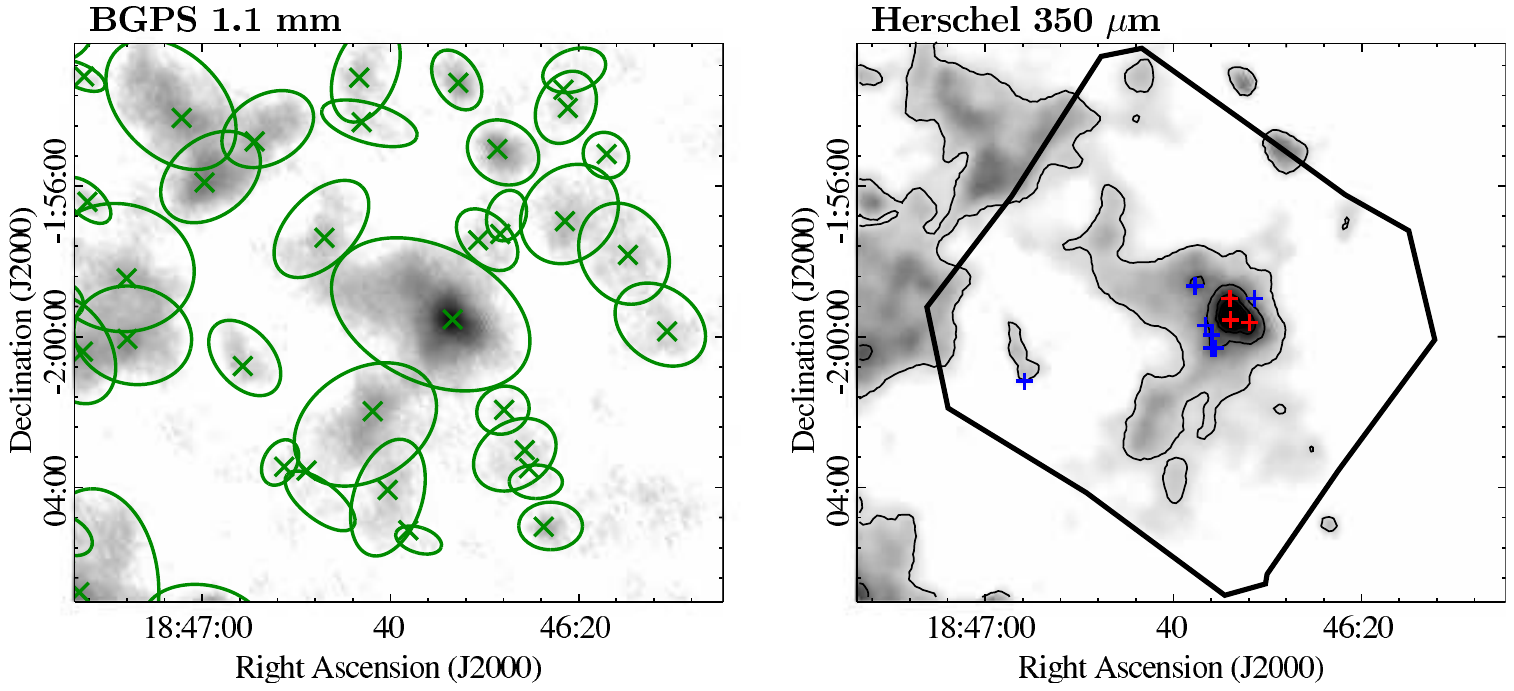}

    }%
   \caption{Continuation.}
\end{figure}

\clearpage



\begin{thebibliography}{}

\bibitem[Aguirre et al.(2011)]{agu11} Aguirre, J.~E., 
Ginsburg, A.~G., Dunham, M.~K., et al.\ 2011, \apjs, 192, 4


\bibitem[Bally et al.(2010)]{bal10} Bally, J., Aguirre, J.,
Battersby, C., et al.\ 2010, \apj, 721, 137

\bibitem[Bally
\& Zinnecker(2005)]{bal05} Bally, J., \& Zinnecker, H.\ 2005, \aj, 129, 2281

\bibitem[Battersby et al.(2010)]{bat10} Battersby, C., Bally, 
J., Jackson, J.~M., et al.\ 2010, \apj, 721, 222

\bibitem[Battersby et al.(2011)]{bat11} 
Battersby, C., Bally, J., Ginsburg, A., et al.\ 2011, \aap, 535, A128 

\bibitem[Beckwith 
\& Sargent(1991)]{bec91} Beckwith, S.~V.~W., \& Sargent, A.~I.\ 1991, \apj, 381, 250 

\bibitem[Beelen et al.(2006)]{bee06} Beelen, A., Cox, P., 
Benford, D.~J., et al.\ 2006, \apj, 642, 694

\bibitem[Benjamin et al.(2003)]{ben03} Benjamin, R.~A., 
Churchwell, E., Babler, B.~L., et al.\ 2003, \pasp, 115, 953 

\bibitem[Bertin 
\& Arnouts(1996)]{ber96} Bertin, E., \& Arnouts, S.\ 1996, \aaps, 117, 393

\bibitem[Bohlin et al.(1978)]{boh78} Bohlin, R.~C., Savage, 
B.~D., \& Drake, J.~F.\ 1978, \apj, 224, 132 

\bibitem[Butler 
\& Tan(2012)]{but12} Butler, M.~J., \& Tan, J.~C.\ 2012, \apj, 754, 5


\bibitem[Carey et al.(2009)]{car09} Carey, S.~J., 
Noriega-Crespo, A., Mizuno, D.~R., et al.\ 2009, \pasp, 121, 76

\bibitem[Cesaroni et
al.(1991)]{ces91} Cesaroni, R., Walmsley, C.~M., Koempe, C., \& Churchwell, E.\
1991, \aap, 252, 278

\bibitem[Cesaroni et al.(1994)]{ces94} Cesaroni, R., Olmi, 
L., Walmsley, C.~M., Churchwell, E., \& Hofner, P.\ 1994, \apjl, 435, L137

\bibitem[Churchwell et
al.(1990)]{chu90} Churchwell, E., Walmsley, C.~M., \& Cesaroni, R.\ 1990,
\aaps, 83, 119

\bibitem[Contreras et 
al.(2013)]{con13} Contreras, Y., Schuller, F., Urquhart, J.~S., et al.\ 2013, \aap, 549, A45 

\bibitem[Draine 
\& Lee(1984)]{dra84} Draine, B.~T., \& Lee, H.~M.\ 1984, \apj, 285, 89 

\bibitem[Dowell et al.(2003)]{dow03} Dowell, C.~D., et al.\
2003, \procspie, 4855, 73


\bibitem[Dunham et al.(2010)]{dun10} Dunham, M.~K., 
Rosolowsky, E., Evans, N.~J., II, et al.\ 2010, \apj, 717, 1157 

\bibitem[Dunham et al.(2011)]{dun11} Dunham, M.~K., 
Rosolowsky, E., Evans, N.~J., II, Cyganowski, C., 
\& Urquhart, J.~S.\ 2011, \apj, 741, 110 

\bibitem[Egan et al.(1998)]{ega98} Egan, M.~P., Shipman, 
R.~F., Price, S.~D., et al.\ 1998, \apjl, 494, L199 

\bibitem[Elia et al.(2013)]{eli13} Elia, D., Molinari, S., 
Fukui, Y., et al.\ 2013, \apj, 772, 45 

\bibitem[Ellsworth-Bowers et al.(2013)]{ell13} 
Ellsworth-Bowers, T.~P., Glenn, J., Rosolowsky, E., et al.\ 2013, \apj, 
770, 39 

\bibitem[Ellsworth-Bowers et al.(2015)]{ell15} 
Ellsworth-Bowers, T.~P., Rosolowsky, E., Glenn, J., et al.\ 2015, \apj, 
799, 29 

\bibitem[Evans(1999)]{eva99} Evans, N.~J., II 1999, \araa, 37, 311

\bibitem[Fa{\'u}ndez et 
al.(2004)]{fau04} Fa{\'u}ndez, S., Bronfman, L., Garay, G., et al.\ 2004, \aap, 426, 97

\bibitem[Foster 
\& Goodman(2006)]{fos06} Foster, J.~B., \& Goodman, A.~A.\ 2006, \apjl, 636, L105 

\bibitem[Garay
\& Lizano(1999)]{gar99} Garay, G., \& Lizano, S.\ 1999, \pasp, 111, 1049

\bibitem[Ginsburg et al.(2013)]{gin13} Ginsburg, A., Glenn, 
J., Rosolowsky, E., et al.\ 2013, \apjs, 208, 14

\bibitem[Groesbeck(1995)]{groesbeck95} Groesbeck, T.~D.\ 1995, PhD thesis, California Institute of Technology

\bibitem[Hatchell 
\& Dunham(2009)]{hatchell09} Hatchell, J., \& Dunham, M.~M.\ 2009, \aap, 502, 139



\bibitem[Johnstone
\& Bally(2006)]{joh06} Johnstone, D., \& Bally, J.\ 2006, \apj, 653, 383

\bibitem[Johnstone et al.(2006)]{joh06b} Johnstone, D., 
Matthews, H., \& Mitchell, G.~F.\ 2006, \apj, 639, 259 



\bibitem[Klein et al.(2005)]{kle05} Klein, R., Posselt, B., 
Schreyer, K., Forbrich, J., \& Henning, T.\ 2005, \apjs, 161, 361 

\bibitem[Kov\'acs(2006)]{kov06} Kov\'acs, A., {\it SHARC-2 350 \um\ Observations
of Distant Submillimeter-Selected Galaxies and Techniques for the Optimal
Analysis and Observing of Weak Signals}, PhD Thesis, Caltech, 2006.

\bibitem[Krugel et 
al.(1989)]{kru89} Krugel, E., Densing, R., Nett, H., et al.\ 1989, \aap, 211, 419

\bibitem[Krumholz 
\& McKee(2008)]{kru08} Krumholz, M.~R., \& McKee, C.~F.\ 2008, \nat, 451, 1082 

\bibitem[Longmore et al.(2012)]{lon12} Longmore, S.~N., 
Rathborne, J., Bastian, N., et al.\ 2012, \apj, 746, 117 

\bibitem[McKee 
\& Ostriker(2007)]{mck07} McKee, C.~F., \& Ostriker, E.~C.\ 2007, \araa, 45, 565

\bibitem[Megeath et al.(2008)]{meg08} Megeath, S.~T., 
Townsley, L.~K., Oey, M.~S., 
\& Tieftrunk, A.~R.\ 2008, Handbook of Star Forming Regions, Volume I: The Northern Sky
Astronomical Society of the Pacific Monograph Publications, Vol. 4. 
Edited by Bo Reipurth, p.264

\bibitem[Molinari et al.(2010)]{mol10} Molinari, S., 
Swinyard, B., Bally, J., et al.\ 2010, \pasp, 122, 314 

\bibitem[Molinari et 
al.(2011)]{mol11} Molinari, S., Schisano, E., Faustini, F., et al.\ 2011, \aap, 530, AA133 

\bibitem[Moore et al.(2007)]{moo07} Moore, T.~J.~T., 
Bretherton, D.~E., Fujiyoshi, T., et al.\ 2007, \mnras, 379, 663 

\bibitem[Motte et
al.(2007)]{mot07} Motte, F., Bontemps, S., Schilke, P., Schneider, N., Menten,
K.~M., \& Brogui{\`e}re, D.\ 2007, \aap, 476, 1243

\bibitem[Motte et 
al.(2010)]{mot10} Motte, F., Zavagno, A., Bontemps, S., et al.\ 2010, \aap, 518, LL77 

\bibitem[Mueller et al.(2002)]{mueller02} Mueller, K.~E., 
Shirley, Y.~L., Evans, N.~J., II, \& Jacobson, H.~R.\ 2002, \apjs, 143, 469e

\bibitem[Ossenkopf
\& Henning(1994)]{oss94} Ossenkopf, V., \& Henning, T.\ 1994, \aap, 291, 943

\bibitem[Padoan et al.(2006)]{pad06} Padoan, P., Juvela, M., 
\& Pelkonen, V.-M.\ 2006, \apjl, 636, L101 

\bibitem[Plume et al.(1992)]{plu92} Plume, R., Jaffe, D.~T.,
\& Evans, N.~J., II 1992, \apjs, 78, 505

\bibitem[Ragan et al.(2013)]{rag13} Ragan, S.~E., Henning, 
T., \& Beuther, H.\ 2013,  \aap, 559, AA79

\bibitem[Rathborne et al.(2006)]{rat06} Rathborne, J.~M., 
Jackson, J.~M., \& Simon, R.\ 2006, \apj, 641, 389 

\bibitem[Rathborne et al.(2010)]{rat10} Rathborne, J.~M., 
Jackson, J.~M., Chambers, E.~T., et al.\ 2010, \apj, 715, 310


\bibitem[Richards et al.(2012)]{ric12} Richards, E.~E., Lang, 
C.~C., Trombley, C., \& Figer, D.~F.\ 2012, \aj, 144, 89 

\bibitem[Rivera-Ingraham et al.(2013)]{riv13} 
Rivera-Ingraham, A., Martin, P.~G., Polychroni, D., et al.\ 2013, \apj, 
766, 85 

\bibitem[Rosolowsky et al.(2010)]{ros09} Rosolowsky, E.,
Dunham, M.~K., Ginsburg, A., et al.\ 2010, \apjs, 188, 123


\bibitem[Schuller et
al.(2009)]{sch09} Schuller, F., et al.\ 2009, \aap, 504, 415

\bibitem[Shirley et al.(2000)]{shi00} Shirley, Y.~L., Evans,
N.~J., II, Rawlings, J.~M.~C., \& Gregersen, E.~M.\ 2000, \apjs, 131, 249

\bibitem[Shirley et al.(2005)]{shi05} Shirley, Y.~L.,
Nordhaus, M.~K., Grcevich, J.~M., et al.\ 2005, \apj, 632, 982

\bibitem[Shirley et al.(2011)]{shirley11} Shirley, Y.~L., Huard, 
T.~L., Pontoppidan, K.~M., et al.\ 2011, \apj, 728, 143 


\bibitem[Shirley et al.(2013)]{shi13} Shirley, Y.~L., 
Ellsworth-Bowers, T.~P., Svoboda, B., et al.\ 2013, \apjs, 209, 2 

\bibitem[Stahler et al.(2000)]{sta00} Stahler, S.~W., Palla,
F., \& Ho, P.~T.~P.\ 2000, Protostars and Planets IV, 327

\bibitem[Schlingman et al.(2011)]{sch11} Schlingman, W.~M.,
Shirley, Y.~L., Schenk, D.~E., et al.\ 2011, \apjs, 195, 14



\bibitem[Simon et al.(2006)]{sim06} Simon, R., Jackson, 
J.~M., Rathborne, J.~M., \& Chambers, E.~T.\ 2006, \apj, 639, 227 

\bibitem[Tan et al.(2014)]{tan14} Tan, J.~C., Beltran, M.~T., Caselli, P., et al.\ 2014, arXiv:1402.0919 

\bibitem[Traficante et al.(2011)]{tra11} Traficante, A., 
Calzoletti, L., Veneziani, M., et al.\ 2011, \mnras, 416, 2932 

\bibitem[Urquhart et al.(2013)]{urq13} Urquhart, J.~S., 
Moore, T.~J.~T., Schuller, F., et al.\ 2013, \mnras, 431, 1752 

\bibitem[Wienen et 
al.(2012)]{wie12} Wienen, M., Wyrowski, F., Schuller, F., et al.\ 2012, \aap, 544, A146 

\bibitem[Williams et al.(1994)]{wil94} Williams, J.~P., de
Geus, E.~J., \& Blitz, L.\ 1994, \apj, 428, 693

\bibitem[Williams et al.(2000)]{williams2000} Williams, J.~P., 
Blitz, L., \& McKee, C.~F.\ 2000, Protostars and Planets IV, 97 

\bibitem[Wright et al.(2010)]{wri10} Wright, E.~L., 
Eisenhardt, P.~R.~M., Mainzer, A.~K., et al.\ 2010, \aj, 140, 1868

\bibitem[Wu et al.(2007)]{wu07} Wu, J., Dunham, M.~M.,
Evans, N.~J., II, Bourke, T.~L., \& Young, C.~H.\ 2007, \aj, 133, 1560


\bibitem[Wynn-Williams et al.(1972)]{wyn72} Wynn-Williams, 
C.~G., Becklin, E.~E., \& Neugebauer, G.\ 1972, \mnras, 160, 1

\bibitem[Zinnecker
\& Yorke(2007)]{zinn07} Zinnecker, H., \& Yorke, H.~W.\ 2007, \araa, 45, 481

\end{thebibliography}
\end{document}